\documentstyle[11pt,epsf]{article}
\begin{document}

\begin{titlepage}

\begin{center}

{\Large New Superconformal Field Theories in Four Dimensions and $N=1$ Duality}

\vspace{5mm}

{\large M. Chaichian${}^{1,2}$, 
W.F. Chen${}^3$\renewcommand{\thefootnote}{\dagger}\footnote{Address after September 1,
2000: Department of Mathematics and Statistics, University of Guelph, Guelph, Ontario,
Canada N1G 2W1}
 and C. Montonen${}^2$\renewcommand{\thefootnote}{\ddagger}\footnote{On 
leave from the Theoretical Physics Division, Department of Physics, 
University of Helsinki}}

\vspace{2mm}
${}^1$ High Energy Physics Division, Department of Physics\\
${}^2$ Helsinki Institute of Physics, P.O. Box 9 (Siltavuorenpenger 20 C)\\
FIN-00014 University of Helsinki, Finland\\
${}^3$ Winnipeg Institute of Theoretical Physics 
and Department of Physics\\
 University of Winnipeg, Winnipeg, 
Manitoba, Canada R3B 2E9

\end{center}

\vspace{1cm}

\begin{abstract}
{\noindent Recently, developments in the understanding of 
low-energy $N=1$ supersymmetric gauge theory have revealed 
two important phenomena: the appearance of new four-dimensional 
superconformal field theories and a non-Abelian generalization 
of electric-magnetic duality at the IR fixed point.
This report is a pedagogical introduction to these phenomena.
After presenting some necessary background material,
a detailed introduction to the low-energy non-perturbative dynamics
of $N=1$ supersymmetric $SU(N_c)$ QCD is given. The emergence of
new four-dimensional superconformal field theories and non-Abelian 
duality is explained. New non-perturbative phenomena in supersymmetric 
$SO(N_c)$ and $Sp(N_c=2n_c)$ gauge theories, such as the two 
inequivalent branches, oblique confinement and electric-magnetic-dyonic 
triality, are presented.  Finally, some new features of these 
four-dimensional superconformal field theories are exhibited: the 
universal operator product expansion, evidence for a possible
$c$-theorem in four dimensions and the critical behaviour of anomalous
currents. The concluding  remarks contain a brief history of 
electric-magnetic duality and a discussion on the possible
 applications of duality to ordinary QCD.}
\end{abstract}

\end{titlepage}

\newpage

\tableofcontents

\newpage
\section{Introduction}
\label{sect1}
\renewcommand{\theequation}{1.\arabic{equation}}
\setcounter{equation}{0}

During the past several years there has been a great progress
in understanding the non-perturbative dynamics of supersymmetric 
gauge theories. The breakthrough has been made in two aspects. 
One is in $N=2$ supersymmetric gauge theories pioneered by Seiberg and 
Witten \cite{ref1p1,ref1p1a}. Based on the low energy effective action 
obtained by Seiberg earlier from the non-perturbative $\beta$ function
 analysis \cite{ref1p2}, which has now been verified up to 
at most two derivatives and not more than four-fermion coupling 
by various calculation methods \cite{ref1p3,ref1p4,ref1p5}, 
Seiberg and Witten found that in the Coulomb phase $N=2$ supersymmetric
Yang-Mills theory can admit a self-dual electric-magnetic duality.
This kind of duality was originally proposed by Montonen and 
Olive \cite{ref1p6} and previously it was believed that it could only exist 
in an $N=4$ supersymmetric Yang-Mills theory \cite{ref1p7}. Furthermore,
with this duality Seiberg and Witten explicitly showed that the confinement
mechanism is indeed provided by the condensation of magnetic monopoles
and justified in supersymmetric gauge theory 
the original conjecture by 't Hooft 
and Mandelstam \cite{ref1p8,ref1p8a}. 
In the presence of $N=2$ matter multiplets, they also
showed that the chiral symmetry breaking is driven by the condensation
of magnetic monopoles.

Another direction in which progress has been made 
is $N=1$ supersymmetric gauge theory,
mainly based on ideas by Seiberg. From
an analysis of the quantum moduli space,  a series of exact results
has been obtained in $N=1$ supersymmetric gauge theory, and an almost
complete phase diagram of $N=1$ supersymmetric gauge theory 
including the dynamical features and the particle
spectrum in each phase has been worked out 
\cite{se2,ref525,ref1p9,ref1p10,ref51,ref1p11}. In particular, Seiberg
found that the electric-magnetic duality can also exist
in the IR fix point of $N=1$ supersymmetric gauge theory, where
the theory is described by an interacting four-dimensional
superconformal field theory. This report concentrates on
these new four-dimensional superconformal field theories and $N=1$ duality.

It is not accidental that so many non-perturbative results
can be obtained in supersymmetric gauge theories. Supersymmetric 
quantum field theory
is much more tractable than usual quantum field theory. One of the
most remarkable characteristics of supersymmetric gauge theory is 
the non-renormalization theorem \cite{ref1p12,ref1p12a,ref1p13}, 
which claims that an interaction vertex
of a supersymmetric gauge theory is not spoiled by quantum corrections.
If the theory is expressed in superfield form, this immediately
implies that the classical superpotential withstands 
quantum corrections and thus
remains holomorphic. This property imposes a very strict constraint
on the form of the interaction vertex at the quantum level.

In addition, there is another direct consequence of the 
non-renormalization theorem. Since the vertex receives no quantum 
corrections, one can always define the vertex renormalization constant
to be one. Thus the renormalization of the coupling constants is only 
related to the wave function renormalization constants. This means that the
beta function of the theory depends only on the anomalous dimensions
of the fields. In supersymmetric gauge theory with gauge group
$G$ and $N_f$ species of matter field in the representation $T_i$
($i=1,2,{\cdots},N_f)$, this fact is quantitatively manifested
in the NSVZ (Novikov-Shifman-Vainshtein-Zakharov) 
beta function \cite{ref1p14,ref1p14a}:
\begin{eqnarray}
\beta^{({\rm NSVZ})}(g)=-\frac{g^3}{16\pi^2}\frac{3T(G)\mbox{dim}G
-\sum_{i=1}^{N_f} T(R_i)\left[1-\gamma_i\right]}{1-T(G)g^2/(8\pi^2)},
\label{eq1p1}
\end{eqnarray}
where $\gamma_i$ are the anomalous dimensions of the matter fields, 
and the group invariants $T(G)$ and $T(R)$ are defined as 
follows: 
\begin{eqnarray}
\mbox{Tr}_R(T^aT^b)=T(R)\delta^{ab}, 
~~~T(G)=T(R=\mbox{adjoint representation)}. 
\end{eqnarray} 
The explicit form of this beta function provides a possibility to
determine the non-trivial IR fixed points. It should be emphasized
 that the NSVZ beta function
(\ref{eq1p1}) is a non-perturbative result and is valid to all
orders. Furthermore, recalling that there exists a direct 
connection between the trace anomaly and the beta function, 
for example in a pure Yang-Mills theory \cite{ref1p15},
\begin{eqnarray} 
{\theta}_{~\mu}^{\mu}=\frac{{\beta}(g)}{2g}F_{\mu\nu}F^{\mu\nu},
\end{eqnarray}
one can see that at fixed points, the trace anomaly vanishes
and hence conformal symmetry emerges since ${\theta}_{~\mu}^{\mu}$
is the measure of quantum conformal symmetry \cite{ref1p16}. 
Therefore, an interacting superconformal field theory can
arise at the IR fixed point. 
The NSVZ beta function says that for an asymptotically 
free supersymmetric gauge theory, such an IR fixed point must exist.

The origin of the non-renormalization theorem 
lies in the supersymmetry, which ensures the 
cancellation of quantum fluctuations
from the bosonic and fermionic modes, since in a supersymmetric field
theory the same number of bosonic and fermionic 
degrees of freedom occur in a supermultiplet, and
they contribute to a virtual process
with the same amplitude but with opposite sign. The basic
building blocks of a quantum field theory are the Green functions.
As manifested in the Green 
functions supersymmetry, just like a usual local internal symmetry,
leads to relations between various Green functions, which 
take the form of Ward-Takahashi (WT) identities. With the assumption 
that the supersymmetry is not spontaneously broken, these WT identities
impose strong constraints on the form of the Green functions. For example,
the chiral supersymmetric WT identities not only 
determine that the Green function
of the lowest component field in a chiral multiplet is space-time 
independent, but also, together with the internal symmetries,
renormalization group invariance and other physical requirements, 
specify the explicit dependence of the Green functions on the 
parameters of the theory such as
masses and coupling constants \cite{ref1p17,akmrv}. 
This is another important reason why 
supersymmetric gauge theories are under better control.

The algebraic foundation of a supersymmetric field theory,  
 superalgebra, is an extension of the usual Poincar\'{e}
symmetry. It unifies some of the most fundamental conserved currents
such as the energy-momentum tensor, the supercurrent and the axial type
$R$-current into a supermultiplet. Consequently, the quantum
anomalies of these currents should 
also fit into a supermultiplet \cite{fz,so}.
Furthermore, there exists  a new type of anomaly, which was originally
found in Ref.\,\cite{ref1p19} 
and independently re-derived by Konishi \cite{ref1p20}. This
anomaly gives a connection between the squark condensation and the 
gluino condensation. All these new features provide possible 
ways to explore the non-perturbative aspects of a supersymmetric 
gauge theory.

The vacuum structure of a supersymmetric gauge theory and
the relevant non-perturbative dynamics have a rich physical content.
Unbroken supersymmetry requires that there exists at least one ground 
state with zero energy \cite{wit1,wit2}. 
In a four-dimensional supersymmetric gauge
theory there usually exists a continuous degeneracy of inequivalent
ground states \cite{ref1p22}. Classically, these ground states correspond to
the flat directions of the scalar potential and thus form a classical
moduli space. Along these flat directions, some of the squarks can acquire 
expectation values which break the gauge symmetry. As a consequence,
some fields will acquire masses due to the super-Higgs mechanism and 
the moduli space will be characterized by the light degrees of freedom.
At the origin of the classical moduli space the gauge symmetry will 
be restored fully. Quantum mechanically, a dynamically 
generated superpotential can arise and 
dynamical supersymmetry breaking occurs.
Consequently, the vacuum degeneracy will be lifted \cite{ref1p22}. 
Note that the non-renormalization theorem only refers to the perturbative 
quantum correction and it does not pose any restriction on the 
non-perturbative quantum effects. The vacuum degeneracy
may still persist after inclusion of non-perturbative effects,
and then the theory has a quantum moduli space of vacua.
The holomorphy of the quantum superpotential makes it possible 
to determine the light degrees of freedom and hence the 
quantum moduli space \cite{ref1p23}. By analyzing the structure 
of the quantum moduli space, one can get a handle on some of the 
important non-perturbative features such as confinement and chiral 
symmetry breaking since many non-perturbative physical phenomena 
are related closely to the vacua such as mass generation and
fermion condensates etc. 
If the dynamically generated superpotential has erased all  
vacua, then dynamical supersymmetry breaking occurs.
Supersymmetry breaking induced by non-perturbative quantum effects 
was proposed by Witten nearly two decades ago \cite{wit1,wit2}. 
Only in recent years have its possible physical applications been 
considered. This kind of supersymmetry breaking mechanism introduces 
very few physical parameters and thus has a great advantage over 
the soft supersymmetry breaking mechanism. At present dynamical 
supersymmetric breaking is a popular topic in supersymmetry 
phenomenology \cite{ref1p24}.

The non-perturbative aspects of an $N=1$ supersymmetric gauge theory
exhibit a rich phase structure, depending heavily on
the choice of gauge group and the matter contents 
\cite{ref1p10,ref51,ref1p11}. The most remarkable non-perturbative 
dynamical phenomenon is that in a special range of colour number 
and flavour number, i.e. in the so-called conformal window \cite{ref1p10},
 the theory may have a non-trivial infrared fixed point implied 
from by NSVZ beta function (\ref{eq1p1}), at which the 
theory becomes a superconformal field theory. In particular, it now 
admits a physically equivalent dual description but with the strong
and weak coupling exchanged \cite{ref1p10}.
 
 The emergence of $N=1$ superconformal symmetry in the IR region
has a great significance. In addition to $N=4$ supersymmetric 
Yang-Mills theory and $N=2$ supersymmetric
gauge theory with vanishing beta function, this is a new non-trivial conformal field
 theory in four dimensional space-time. Especially,  
the $N=1$ supersymmetric gauge theory has in general the property
of asymptotic freedom, at high energy it can be regarded as
a free field theory. Thus the existence of  the IR fixed point
means that an off-critical quantum field theory can be regarded 
as a radiative interpolation between a pair of four dimensional conformal
field theories \cite{afgj}. A conventional method to observe the physical
phenomena at different energy scales is to investigate the flow
of some physical quantity along the renormalization group trajectory
from the UV region to the IR region. In two-dimensional
quantum field theory, there is a famous "$c$-theorem" on the renormalization
group flow proposed by Zamolodchikov \cite{zamo}. It states that there
exists a $c$-function $C(g)$ of the coupling $g$ associated
with the two-point function of the energy-momentum tensor,
which decreases monotonically along the renormalization group
flow from the UV region to the IR region and becomes stationary
at the fixed point, where the theory is a two-dimensional
conformal invariant quantum field theory and the $c$-function coincides
exactly with the central charge ( the coefficient of the conformal
anomaly). Since the central charge actually counts the number
of dynamical degrees of freedom of the theory \cite{jack,duff}, 
this theorem shows precisely how the information about the 
dynamical degrees of freedom
at short distance is lost in the renormalization group flow to
long distance. Therefore, it would be helpful for the study of
some non-perturbative dynamical phenomena in four dimensional
quantum field theory if a four dimensional analogue of the $c$-theorem
could be found. For example, one could get insight in a number of
non-perturbative phenomena such as confinement, chiral symmetry
breaking, supersymmetry breaking and even the Higgs mechanism.

However, unexpectedly  the search and verification of
a four dimensional $c$-theorem turned out to be extremely 
difficult. Soon after the proposal of the two dimensional 
$c$-theorem, much effort was put into
looking for a candidate for a $c$-function and checking its
evolution along the renormalization group flow 
was made \cite{car,osb1,cfl}, but
a general proof of the existence of a
monotonically decreasing $c$-function is still lacking.
Recently, some progress has been made in this direction \cite{fola}.  
The availability of a series of exact results in $N=1$ 
supersymmetric gauge theory and especially the discovery of
superconformal symmetry in the IR fixed point has provided  
useful tools for testing $c$-theorems and finding the
right $c$-function. As in the two-dimensional case, a candidate for a
$c$-function should be related to the central function \cite{jhep}, which 
is the coefficient function of the most singular term in the operator
product expansion of the energy-momentum tensor and should coincide 
with the conformal anomaly coefficients at the fixed points.
A qualitative analysis using the exact results give the first indications of
the possible existence of a four dimensional $c$-theorem
in $N=1$ supersymmetric gauge theory \cite{bast}. Furthermore, asymptotic
freedom implies that various quantities at the UV region 
can be determined in the framework of a free field theory, while
the superconformal invariance at the IR fixed point allows 
an exact calculation on the anomaly coefficients at the IR region
\cite{afgj}: the trace anomaly of the energy-momentum tensor and the chiral
anomaly of $R$-current lie in the same supermultiplet and
the latter can be exactly calculated through the 't Hooft anomaly 
matching. An explicit calculation suggests that the central function
coinciding with the coefficient of the Euler term of in the trace anomaly
at the IR fixed point is the most appropriate candidate for
the $c$-function \cite{afgj,fros}.

 The $N=1$ four dimensional superconformal field theory arising
at the IR fixed point has also some remarkable features in comparison
with the two dimensional case \cite{ans,ans1}. The operator product 
expansion of two energy-momentum
tensor operators or of an energy-momentum tensor and
a conserved current does not close, another current called the Konishi
current must be introduced to ensure closure. Consequently,   
a four dimensional superconformal field theory is characterized
by two central charges and a conformal dimension.
The central charges  are four dimensional superconformal
invariants in the sense that they receive no higher order
(more than two-loop) quantum correction and remain  
at their one-loop values (i.e. always proportional
to the number of dynamical degrees of freedom), while the
conformal dimension does not, it can receive quantum corrections
from every order \cite{ans1}.

 The dual theory arising at the IR fixed point 
is of significance since it gives an equivalent
physical description to the low energy dynamics of the original
theory but in terms of a fundamental Lagrangian of a supersymmetric
gauge theory with one (or several) additional superpotential(s).
The dynamics of the mesons and baryons in the original theory
can be equivalently replaced by the dynamics of dual fundamental
dynamical degrees of freedom and one (or several) gauge singlet 
particle(s). Especially, the gauge couplings of the original
and the dual theories are inversely related. This provides
a possibility to study the low-energy non-perturbative dynamics 
of the original theory through the perturbation theory of the
dual theory.

There have already appeared several excellent reviews on 
$N=1$ duality with different emphasis such as those by Intriligator 
and Seiberg \cite{ref1p11},
Shifman \cite{ref1p25}, Shifman and Vainshtein \cite{ref1p27} 
and Peskin \cite{ref1p26}.
The present review mainly emphasizes superconformal
field theory. It is written a pedagogical manner, 
giving detailed mathematical derivations
and explanations of these new developments and including much
of the preliminary knowledge needed to understand the new progress
such as the representation of conformal algebra, the renormalization
group equation, anomaly matching, phases of gauge theory and the Wilson
effective action, etc. This report is based on a series of discussions and
seminars in the theoretical physics group of the University of Helsinki
and Helsinki Institute of Physics. We hope that it
may prove helpful for beginners interested in this topic.

The organization of this report is as follows: Sect.\,\ref{sect2} contains 
some background material. We first review 
the definition of conformal transformations, the derivation of the conformal 
algebra and the field representation of conformal algebra in terms of
Wigner's little group method. 
In a renormalizable relativistic quantum field theory, 
scale symmetry implies conformal symmetry, 
the anomalous breaking of scale symmetry
means  the violation of conformal symmetry at quantum level.
We thus introduce scale symmetry breaking in the context of
a massless scalar field theory and the renormalization group 
equation, which plays the role of anomalous Ward identity
of scale symmetry.
Since new developments occur in supersymmetric QCD,
the supersymmetric extension of the non-supersymmetric theory,  
we introduce the chiral symmetry of ordinary QCD and its breaking.
We discuss the possible (external) chiral anomaly in QCD and the 't Hooft anomaly matching.
We shall see that the 't Hooft anomaly matching is
an important tool to investigate electric-magnetic duality.
The non-perturbative dynamical structure of a quantum field theory
is quite rich and the theory can present several phases. The  
discovery of new non-perturbative phenomena in supersymmetric gauge theory
has verified this. Hence it is necessary to introduce 
the various phase structures and their dynamical behaviour. The 
$N=1$ superconformal algebra is the algebraic foundation of constructing 
a superconformal field theory, and thus based on the current supermultiplet 
and the Jacobi identities, we give the full superconformal algebra. 
Further, we review the representation of the superconformal algebra in field 
operator space. 
It is much more complicated than the ordinary conformal algebra:
the ordinary conformal algebra can be realized on one type of
fields, whereas the superconformal algebra must
be realized on a supermultiplet. In particular, 
the representations realized on different
supermultiplets, such as the chiral and the vector ones,
 have their own specific features. In particular, the representation
of the superconformal algebra on the chiral multiplet yields
a simple relation between the conformal dimension of the chiral superfield 
and its $R$-charge, which has played a key role in determining Seiberg's 
conformal window.
 
From Sect.\,\ref{sect6}, we begin to introduce the 
low-energy dynamics of supersymmetric
QCD. Supersymmetric gauge theory exhibits some new characteristics
compared with the non-supersymmetric case,
the most important of which are $R$-symmetry and holomorphicity. Thus
we first give a detailed introduction to these two aspects. 
We explain in detail how to combine the anomalous 
$R$-symmetry and the axial vector
$U_A(1)$ symmetry to get an anomaly-free $R$-symmetry, and then 
introduce the low-energy dynamics of supersymmetric QCD.
Since supersymmetric QCD is a theory sensitive to 
flavour number $N_f$ and colour number $N_c$, we analyze the
theory with respect to different ranges of $N_f$ and $N_c$.
Using the Georgi-Glashow model to illustrate the general definition
of a classical moduli space, we explain how to describe
the classical moduli space of supersymmetric QCD in the cases of $N_f<N_c$ and
$N_f{\geq}N_c$, respectively. Especially for $N_f=N_c$ and
$N_f=N_c+1$ the constraint equations 
characterizing the classical moduli space are
explicitly given. A large part of this section is
devoted to describing the quantum moduli space 
and low energy  dynamics of supersymmetric QCD. 
In the case $N_c<N_f$, we give a detailed derivation of the 
ADS (Affleck-Dine-Seiberg) superpotential and argue the reasonableness of 
this non-perturbative superpotential
by considering its various limits and the physical consequences 
obtained from this superpotential. When $N_f=N_c$, we show 
how the quantum corrections have modified the classical moduli space
and what physical effects are produced, and check the
reasonableness of these physical pictures using the anomaly matching
condition. For $N_f=N_c+1$, we introduce the effective
superpotential to determine the quantum moduli space and the
physical effects resulting from the quantum moduli space,
't Hooft anomaly matching providing a strong support. For the case 
$N_f>N_c$,  we show that the NSVZ beta function
 implies a nontrivial IR fix point in the
conformal window $3N_c/2<N_f<3N_c$.
Then we briefly explain
the situation in the range $N_f>3N_c$, where the asymptotic freedom of
the theory is lost. 

In Sect.\,\ref{sect7n}, we concentrate on non-Abelian 
electric-magnetic duality in the conformal window
by first introducing the dual description of supersymmetric QCD and 
showing the reasonableness of this duality conjecture using 't Hooft's anomaly
matching. Then we explain what non-Abelian electric-magnetic duality is
and how the duality arises in the conformal window and show
how the duality remains valid in various limits.
The duality in the Kutasov-Schwimmer model is briefly reviewed
since it provides a possibility to study the non-perturbative aspects 
of supersymmetric extensions of the Standard Model. This model also gives 
a deep understanding to the duality of $N=1$ supersymmetric QCD.   
Sect.\,\ref{sect5} mainly summarizes the non-perturbative phenomena 
of $N=1$ supersymmetric $SO(N_C)$ gauge theory. In contrast to the
$SU(N_c)$ case, the gauge group $SO(N_c)$ is not simply connected
and has only real representations. Depending thus heavily on the
number of colours and flavours, the theory has richer and more novel 
non-perturbative dynamical phenomena. When $N_f{\leq}N_c-5$, 
a dynamical superpotential is generated by gaugino condensation. In the cases
 $N_f=N_c-3$ or $N_c-4$, the theory has two inequivalent ground states, and some
exotic particle states appear. When $N_f=N_c-2$, the theory is in
a Coulomb phase and the particle spectrum contains magnetic monopoles and 
dyons, and the oblique confinement conjectured by 't Hooft occurs naturally.
As in the $SU(N_c)$ case, in the range $N_f{\geq}4$, $N_f{\geq}N_c-1$,
the theory has a dual magnetic description, a supersymmetric
$SO(N_f-N_c+4)$ gauge theory. The most novel dynamical phenomenon
is that the $SO(3)$ gauge theory exhibits electric-magnetic-dyonic triality.
In the one-flavour and two-flavour cases, a quantum symmetry with a non-local
action on the matter field arises. If the theory has three flavours,
it was surprisingly  found that the $N=1$ duality in the $SO(3)$ theory
can actually be identified as the electric-magnetic duality
of $N=4$ supersymmetric Yang-Mills theory. Further, the concrete
form of the dyonic dual of the $SO(N_c)$ gauge theory with $N_f=N_c-1$
is introduced since its dual magnetic theory is an $SO(3)$ gauge theory,
while the $SO(3)$ gauge theory has a dual dyonic theory. This means
that the $SO(N_c=N_f+1)$ theory should also admit a dual dyonic theory.
Another typical supersymmetric model showing duality, the $Sp(N_c)$ gauge 
theory, is briefly reviewed.
In Sect.\,\ref{sect7} we introduce some of new progress in exploring 
four-dimensional superconformal
field theory and non-Abelian electric-magnetic duality including
't Hooft anomaly matching in the presence of
higher quantum corrections, 
the universality of the operator product expansion in four-dimensional superconformal
field theory and the explicit 
evidence supporting a possible four-dimensional $c$-theorem
in supersymmetric gauge theory.
In the concluding remarks, Sect.\,\ref{sect8n}, we briefly recall the history
of searching for electric-magnetic duality symmetry in relativistic
quantum field theory and explore 
the possible applications of the non-perturbative
results of $N=1$ supersymmetric gauge theory to ordinary QCD by softly
breaking the supersymmetry.


\section{Some background knowledge}
\label{sect2}
\setcounter{equation}{0}

\subsection{Four-dimensional conformal algebra and its representation}
\label{subsect21}
\renewcommand{\theequation}{2.1.\arabic{equation}}
\setcounter{equation}{0}
\renewcommand{\thetable}{2.1.\arabic{table}}
\setcounter{table}{0}

Conformal transformations preserve angles but change the scale. 
In a flat space-time, they are defined by the following transformation 
of the line element \cite{ref21},
\begin{eqnarray}
ds'^2={\eta}_{\mu\nu}dx'^{\mu}dx'^{\nu}=f(x)ds^2=
f(x){\eta}_{\alpha\beta}dx^{\alpha}dy^{\beta},
\label{eq1}
\end{eqnarray}
where $f(x)$ is a scalar function and $\eta_{\mu\nu}$ 
is the space-time metric.
The concrete form of the conformal transformation can be found by
considering infinitesimal transformations
\begin{eqnarray}
x'^{\mu}=x^{\mu}-{\epsilon}^{\mu}(x).
\label{eq3}
\end{eqnarray} 
Eqs.\,(\ref{eq1}) and (\ref{eq3}) give 
\begin{eqnarray}
{\partial}_{\mu}{\epsilon}_{\nu}(x)+{\partial}_{\nu}{\epsilon}_{\mu}(x)
=-(f(x)-1){\eta}_{\mu\nu}{\equiv}-h(x){\eta}_{\mu\nu}.
\label{eq5}
\end{eqnarray}
and further, after contracting $\eta^{\mu\nu}$ with (\ref{eq5}),
\begin{eqnarray}
h(x)&=&-\frac{2}{n}{\partial}_{\mu}{\epsilon}^{\mu}(x), \label{hequ}\\
{\partial}_{\mu}{\epsilon}_{\nu}(x)+{\partial}_{\nu}{\epsilon}_{\mu}(x)
&=&\frac{2}{n}{\partial}_{\alpha}{\epsilon}^{\alpha}(x){\eta}_{\mu\nu}, 
\end{eqnarray}
where $n$ is the dimension of space-time. Thus
\begin{eqnarray}
(n-2){\partial}_{\mu}{\partial}_{\nu}h(x)&=&-\frac{2}{n}(n-2){\partial}_{\mu}
{\partial}_{\nu}{\partial}^{\rho}{\epsilon}_{\rho}=0.
\label{eq6}
\end{eqnarray}
 For $n>2$, Eq.\,(\ref{eq6}) implies that $h(x)$ is at most linear 
in $x$ and from (\ref{hequ}) that
${\epsilon}_{\mu}(x)$ is at most quadratic in $x^{\mu}$. The general form
of ${\epsilon}_{\mu}(x)$ is
\begin{eqnarray}
{\epsilon}^{\mu}(x)=a^{\mu}+{\epsilon}x^{\mu}+{\omega}^{\mu\nu}x_{\nu}+
2b{\cdot}xx^{\mu}-b^{\mu}x^2,
\label{eq7}
\end{eqnarray}
where $a$ and $b$ are constant $n$-dimensional vectors, ${\epsilon}$ is
an infinitesimal constant and ${\omega}_{\mu\nu}=-{\omega}_{\nu\mu}$.  
The finite form of the above infinitesimal conformal transformations 
is listed in Table (\ref{ta2.1}). Counting the number of the generators,
the dimension of the conformal group is $(n+1)(n+2)/2$. It is isomorphic 
to the orthogonal group $O(n,2)$ in Minkowski space. 
For $n=4$, it is a 15-dimensional space-time symmetry group.

\begin{table}
\begin{center}
\begin{tabular}{|c|c|c|} \hline
Translations  & $x'^{\mu}=x^{\mu}-a^{\mu} $ & $h(x)=0$ \\ \hline 
Scale transformation  & $x'^{\mu}=e^{-\epsilon}x^{\mu}$ 
& $h(x)=-2{\epsilon}$\\ \hline
Lorentz transformation & $x'^{\mu}
={\Lambda}^{\mu}_{~\nu}x^{\nu} $ & $h(x)=0$ \\ \hline
Special conformal transformation 
& $\displaystyle x'^{\mu}=\frac{x^{\mu}+b^{\mu}x^2}{1+2
b{\cdot}x+b^2 x^2} $ & $h(x)=-4b{\cdot}x $ \\ \hline
\end{tabular}
\caption{\protect\small Conformal transformations. \label{ta2.1}}
\end{center}
\end{table}

The infinitesimal version of the conformal transformation clearly shows 
that the conformal group is a generalization of the Poincar\'{e} group.
Thus, in addition to $P_{\mu}$ and $M_{\mu\nu}$, 
the set of generators consists of $n+1$ new ones, $D$ and $K_{\mu}$,
 which correspond to scale and special conformal transformations, 
respectively. 

 According to the relation between symmetry and conservation law
(Noether's theorem), corresponding to each continuous global 
invariance, there exists a conserved current $j_{\mu}^k$. The space
integral of the time component gives a conserved charge, 
$Q^k=\int d^3x j_0^k$,
and the conserved charges yield a representation
of the generators of the symmetry group. 
It is well known that the conserved current for $P_{\mu}$ and $M_{\mu\nu}$
are, respectively, the 
energy-momentum tensor $\theta_{\mu\nu}$ and the 
moment ${\cal M}_{\mu,\nu\rho}$ of $\theta_{\mu\nu}$,
\begin{eqnarray}
{\cal M}_{\mu,\nu\rho}=\theta_{\mu\nu}x_{\rho}-{\theta}_{\mu\rho}x_{\nu}.
\label{eq2a8}
\end{eqnarray}
The generators of the Poincar\'{e} group are hence 
\begin{eqnarray}
P_{\mu}=\int d^3x \theta_{0\mu},
~~~~M_{\mu\nu}=\int d^3x{\cal M}_{0,\mu\nu}({\bf x},t).
\label{eq2a10}
\end{eqnarray}

In a quantum field theory with conformal symmetry,
the energy-momentum tensor $\theta_{\mu\nu}$ 
is traceless,
\begin{eqnarray}
\theta^{\mu}_{~\mu}=0.
\label{eq2a17}
\end{eqnarray}
Using this condition of tracelessness, one can construct the
conserved currents for the scale and special conformal
transformations,
\begin{eqnarray}
d_{\mu}=x^{\nu}\theta_{\nu\mu}, ~~~
k_{\mu\nu}=2x_{\nu}x^{\rho}\theta_{\rho\mu}-x^2\theta_{\nu\mu},
\label{eq2a18}
\end{eqnarray} 
and the corresponding generators
\begin{eqnarray}
D={\int}d^3x\, d_0({\bf x},t), ~~~K_{\mu}={\int}d^3x \, k_{0\mu}({\bf x},t).
\label{eq2a19x}
\end{eqnarray}
The full conformal algebra can be worked out in a model
independent way,
\begin{eqnarray}
[M_{\mu\nu}, M_{\alpha\beta}]&=&i\left({\eta}_{\mu\beta}M_{\nu\alpha}+
{\eta}_{\nu\alpha}M_{\mu\beta}-{\eta}_{\mu\alpha}M_{\nu\beta}-
{\eta}_{\nu\beta}M_{\mu\alpha}\right),\nonumber \\
\left[ M_{\mu\nu}, P_{\rho}\right]
&=&-i{\eta}_{\mu\rho}P_{\nu}+i{\eta}_{\nu\rho}P_{\mu},
\nonumber \\
\left[M_{\mu\nu}, K_{\rho}\right]
&=&-i{\eta}_{\mu\rho}K_{\nu}+i{\eta}_{\nu\rho}K_{\mu},
\nonumber \\
\left[ D, P_{\mu} \right]&=&-iP_{\mu},~~ 
\left[ D, K_{\mu} \right]=iK_{\mu},   \nonumber \\
\left[ K_{\mu}, P_{\nu} \right]&=&-2i({\eta}_{\mu\nu}D+M_{\mu\nu}), 
\nonumber \\
\left[ K_{\mu}, K_{\nu} \right] &=&[M_{\mu\nu},D]=[P_{\mu},P_{\nu}]=0.
\label{eq2a29}
\end{eqnarray}

The four-dimensional conformal group is isomorphic 
to the pseudo-orthogonal group $O(4,2)$, whose
covering group is $SU(2,2)$. The conformal algebra (\ref{eq2a29})
can be brought into a form which exhibits the $O(4,2)$ (or $SU(2,2)$) 
structure by the identification 
\begin{eqnarray}
M_{ab}&=&(M_{\mu\nu},M_{{\mu}5},M_{{\mu}6},M_{56}),\nonumber\\[2mm]
M_{{\mu}5}&=&\frac{1}{2}(P_{\mu}+K_{\mu}), 
~~M_{{\mu}6}=\frac{1}{2}(P_{\mu}-K_{\mu}), 
~~M_{56}=D.
\label{eq2.43}
\end{eqnarray}
Then $M_{ab}$ satisfy the algebraic relation of 
the group $O(4,2)$ (or $SU(2,2)$)
\begin{eqnarray}
[M_{ab},M_{cd}]=-i{\eta}_{ac}M_{bd}+i{\eta}_{ad}M_{bc}+i{\eta}_{bc}M_{ad}
-i{\eta}_{bd}M_{ac},
\label{eq2.42}
\end{eqnarray}
where ${\eta}_{ab}=\mbox{diag}(+1,-1,-1,-1,+1,-1)$, $a,b,c,d=0,{\cdots}3,5,6$.
Therefore, the representations 
of the conformal algebra can be obtained through
those of $O(2,4)$ (or $SU(2,2)$).

The conformal group in a quantum field theory  is
realized through unitary operators $T(g)$, which are 
exponential functions of the generators and which 
transform the field operators as
\begin{eqnarray}
T(g)\varphi_r(x)T(g)^{-1}=S_{rs}(g,x)\varphi_s(g^{-1}x).
\label{eq2p144}
\end{eqnarray}
Here $r$ is a generic index labelling the fields 
and the matrices $S_{rs}$ form a representation of the 
conformal group. 
The infinitesimal form of Eq.\,(\ref{eq2p144}) involves the commutators of the 
generators with the fields, and thus the action of the 
conformal group on the fields is determined by these 
commutation relations. We can deduce the form of these commutators using 
the method of induced representations.

The starting point are the translations, generated by the momentum 
operators $P_\mu$,
\begin{eqnarray}
\left[P_\mu,\varphi_r(x)\right] = -i\partial_\mu\varphi_r(x),
\label{Pcomm}
\end{eqnarray}
or, equivalently,
\begin{eqnarray}
e^{iP\cdot x}\varphi_r(0)e^{-iP\cdot x} = \varphi_r(x).
\end{eqnarray}
From (\ref{eq2p144}) we see that if $g$ is a 
transformation belonging to the stability group or little 
group of $x = 0$, i.e. leaves $x = 0$ invariant, 
then the commutator of a generator of the little group 
with a field operator at $x = 0$ will only involve the 
field at that point. Let  $\{S^a\}$ be the generators 
of the little group obeying the algebra
\begin{eqnarray}
\left[S^a,S^b\right] = if^{abc}S^c.
\label{Scomm}
\end{eqnarray}
If we now posit the commutation relations
\begin{eqnarray}
\left[S^a,\varphi_r(0)\right] = -(\sigma^a)_{rs}\varphi_s(0),
\label{Sphicomm}
\end{eqnarray}
the matrices $\{\sigma^a\}$ will form a representation of 
the little group algebra (\ref{Scomm}):
\begin{eqnarray}
\left[\sigma^a,\sigma^b\right] = if^{abc}\sigma^c,
\end{eqnarray}
as can be easily checked from the Jacobi identities involving two 
generators $S^a, S^b$ and the field
$\varphi_r(0)$. Translating the relations (\ref{Sphicomm}) 
to a general point $x$,
\begin{eqnarray}
\left[e^{iP\cdot x}S^ae^{-iP\cdot x},\varphi_r(x)\right] = -(\sigma^a)_{rs}\varphi_s(x),
\label{eiPSphicomm}
\end{eqnarray}
and evaluating
\begin{eqnarray}
\mbox{exp}[ix^\mu P_{\mu}]S^a\mbox{exp}[-ix^\mu P_{\mu}]
=\sum_{n=0}^{\infty}\frac{i^n}{n!}x^{\mu_1}{\cdots}x^{\mu_n}
[P_{\mu_1},[\cdots ,[P_{\mu_n},S^a]\cdots ]]],
\label{eq2a37}
\end{eqnarray} 
the equations for the commutators $[S^a,\varphi_r(x)]$ are obtained, i.e.
a representation of the conformal algebra on the field operators is induced.

From Table \ref{ta2.1} we see that for the conformal group, 
the generators of the little group are $M_{\mu\nu}$, $D$ and 
$K_\mu$. Thus we can write
\begin{eqnarray}
\left[M_{\mu\nu},{\varphi_r}(0)\right]&=&-
({\Sigma}_{\mu\nu})_{rs}{\varphi_s}(0),\nonumber\\[2mm]
\left [D,{\varphi_r}(0) \right]&=&-i\Delta_{rs}\varphi_s(0),\nonumber\\[2mm]
\left [K_{\mu},{\varphi_r}(0) \right]&=&-({\kappa}_{\mu})_{rs}{\varphi_s}(0),
\label{eq2a34}
\end{eqnarray}  
and the matrices $\Sigma_{\mu\nu}, \Delta$ and $\kappa_\mu$ obey the same
algebra as the corresponding generators, viz.
\begin{eqnarray}
&&[\kappa_{\mu},\kappa_{\nu}]=[\Delta,\Sigma_{\mu\nu}]=0,\label{ekv12}\\
&&[\Delta,\kappa_{\mu}]=i\kappa_{\mu},\label{ekv11}\\
&&[\kappa_{\rho},\Sigma_{\mu\nu}]
=i(\eta_{\rho\mu}\kappa_{\nu}-\eta_{\rho\nu}\kappa_{\mu}),\\
&&[\Sigma_{\rho\sigma},\Sigma_{\mu\nu}]=i(\eta_{\sigma\mu}\Sigma_{\rho\nu}
-\eta_{\rho\mu}\Sigma_{\sigma\nu}-\eta_{\sigma\nu}\Sigma_{\rho\mu}
+\eta_{\rho\nu}\Sigma_{\sigma\mu}).
\label{eq2a35}
\end{eqnarray} 
Note that if the $\Sigma_{\mu\nu}$, which according 
to (\ref{eq2a35}) form a representation
of the Lorentz algebra, generate an irreducible 
representation, Schur's lemma implies,
by virtue of (\ref{ekv12}), that the matrix $\Delta$ is 
proportional to the unit matrix:
\begin{eqnarray}
\Delta_{rs} = id\delta_{rs}.
\end{eqnarray}
Here $d$ is called the scale or conformal 
dimension of the field. In this case it follows
from (\ref{ekv11}) that $\kappa_\mu = 0$.

The relations (\ref{eq2a34}) can now be translated to a general $x$. 
We need to evaluate
\begin{eqnarray}
&&e^{iP\cdot x}M_{\mu\nu} e^{-iP\cdot x}=M_{\mu\nu}-
x_{\mu}P_{\nu}+x_{\nu}P_{\mu},\nonumber\\
&& e^{iP\cdot x}D e^{-iP\cdot x}=D-x\cdot P,\nonumber\\
&& e^{iP\cdot x}K_{\mu} e^{-iP\cdot x}=K_{\mu}-
2x_{\mu}D+2x^{\nu}M_{\nu\mu}+2x_{\mu}(x\cdot P)-x^2P_{\mu}.
\label{translconfalg}
\end{eqnarray}
Inserting (\ref{translconfalg}) into (\ref{eiPSphicomm}) and
using (\ref{Pcomm}), we finally get
\begin{eqnarray}
\left[ M_{\mu\nu},{\varphi}_r(x)\right]&=&-i(x_{\mu}{\partial}_{\nu}-
x_{\nu}{\partial}_{\mu}){\varphi}_r(x)-({\Sigma})_{rs}{\varphi}_s(x),
\nonumber\\[2mm]
\left[ D,{\varphi}_r(x)\right]&=&-i(x{\cdot}{\partial}){\varphi}_r(x)-
{\Delta}_{rs}{\varphi}_s(x),\nonumber\\[2mm]
\left[ K_{\mu},{\varphi}_r(x)\right]&=& i(x^2{\partial}_{\mu}-
2x_{\mu}(x{\cdot}{\partial})){\varphi}_r(x)-2x_{\mu}{\Delta}_{rs}
{\varphi}_s(x) \nonumber\\[2mm]
&&+ 2x^{\nu}({\Sigma}_{\mu\nu})_{rs}{\varphi}_s(x)
-({\kappa}_{\mu})_{rs}{\varphi}_s(x).
\label{eq2p1.31}
\end{eqnarray}
Together with (\ref{Pcomm}), these give the action of the 
conformal group on the fields.

Finally, let us briefly mention the finite dimensional representations of the conformal
algebra on the state space of a quantum field theory. Usually, to 
find a representation of a Lie algebra, one should first find the 
lowest (or highest) weight state, then use the raising (or lowering) 
operators to construct the whole irreducible representation. Each 
irreducible representation is labelled by the quantum numbers associated 
with the eigenvalues of the Casimir operators. For the conformal group,
the finite-dimensional irreducible representations of its subgroup, 
the Lorentz group, are labelled by the angular momentum quantum numbers 
$(j_1,j_2)$, since the Lorentz
group is locally isomorphic to $SU(2){\times}SU(2)$. 
The lowest weight states are 
$(-j_1, -j_2)$. Since $P_{\mu}$ and
$M_{\mu\nu}$ constitute the Poincar\'{e} algebra, the
quantum numbers classifying its representations, the particle mass $m$ and
spin $s$ for $m^2>0$ or
the helicity $\lambda$ for $m^2=0$, also play a role in characterizing
the representations. In particular,  
the commutation relations $[D,K_{\mu}]=iK_{\mu}$ and  $[D,P_{\mu}]=-iP_{\mu}$
imply that the conformal dimension $d$, the eigenvalue
associated with the scale transformation generator $D$, is another (magnetic) 
quantum number labelling the representations of the conformal algebra 
and $P_{\mu}$, $K_{\mu}$ are the raising and lowering operators, 
respectively. This can be easily seen as follows.
Assume that $|\varphi (0)\rangle$ is an eigenstate of $-iD$,
\begin{eqnarray}
-iD|\varphi (0)\rangle=d|\varphi (0)\rangle.
\end{eqnarray} 
Then
\begin{eqnarray}
DK_{\mu} |\varphi (0)\rangle &=& \left([D,K_{\mu}]+ K_{\mu}D\right)
|\varphi (0)\rangle = i(d+1)K_{\mu}|\varphi (0)\rangle ,\nonumber\\
DP_{\mu} |\varphi (0)\rangle &=& \left([D,P_{\mu}]+ P_{\mu}D\right)
|\varphi (0)\rangle = i(d-1)P_{\mu}|\varphi (0)\rangle.
\end{eqnarray}         
Thus, given a conformal dimension $d$, the lowest weight is
\begin{eqnarray}
\lambda = (d,-j_1,-j_2).
\end{eqnarray}  
It was found that there are only five classes of unitary irreducible
representations of the four-dimensional conformal algebra. They are
listed in Table (\ref{ta2p2}). They
differ in their Poincar\'{e} content $(m,s)$ or $(m,\lambda)$ \cite{ref25}.

\begin{table}
\begin{center}
\begin{tabular}{|c|c|}\hline

Weight (magnetic) quantum numbers &  Poincar\'{e} quantum numbers \\
   $(d,j_1,j_2)$                  &    $(m,s)$ or $(m,\lambda)$ \\  \hline
   $d=j_1=j_2=0$                  & Trivial 1-dimensional representation
\\ \hline
$j_1{\neq}0$, $j_2{\neq}0$, $d>j_1+j_2+2$  & $m>0$, 
$s=|j_1-j_2|,{\cdots},j_1+j_2 $ (integer steps) \\  \hline
$j_1j_2=0$, $d>j_1+j_2+1$  & $m >0$, $s=j_1+j_2$ \\  \hline
$j_1{\neq}0$, $j_2{\neq}0$, $d=j_1+j_2+2$ & $m >0$, $s=j_1+j_2$;\\ \hline
$j_1j_2=0$, $d=j_1+j_2+1$ & $m =0$, $\lambda =j_1-j_2$.\\ \hline
\end{tabular}
\caption{\protect\small Unitary representation of conformal algebra.
\label{ta2p2} }
\end{center}
\end{table}

A state in the Hilbert space is generated by the action
of an operator on the vacuum, 
\begin{eqnarray}
|{\cal O}\rangle ={\cal O}|0\rangle.
\end{eqnarray}
Without spontaneous (conformal) symmetry breaking, the quantum 
states and the quantum operators are in one-to-one correspondence. 
An operator generating a quantum state with the lowest weight 
$(d,-j_1,-j_2)$ is called a primary field. An operator with conformal
dimension $d+n$ is called an $n$th-stage descendant field.

Although the conformal symmetry must be broken in physics, 
we see that these unitary representations of the
conformal algebra have imposed highly nontrivial constraints 
on the conformal dimensions of the fields. 

\subsection{Scale symmetry breaking and renormalization group equation}
\label{subsect22}
\renewcommand{\theequation}{2.2.\arabic{equation}}
\setcounter{equation}{0}
\renewcommand{\thetable}{2.2.\arabic{table}}
\setcounter{table}{0}

The scale symmetry is at the heart of conformal symmetry. In fact,  
in a renormalizable relativistic quantum field theory, scale invariance 
implies conformal symmetry. However, scale symmetry cannot be an exact 
symmetry in the nature, since in a field theory with exact scale 
symmetry the mass spectrum must be continuous or massless.
To break scale invariance, three roads are open to us. The first would be to
break the symmetry explicitly by introducing terms containing
dimensional parameters into the Lagrangian. We shall not consider
this case. A second way is spontaneous breaking,
$D|0{\rangle}{\neq}0$.
In this case, Goldstone's theorem implies that there will be a corresponding
massless Goldstone boson, which is awkward from a phenomenological
point of view. There remains the third alternative, anomalous symmetry
breaking. This means that although the classical theory is symmetric, there
is no quantization scheme that would respect the symmetry. In a quantum
field theory, renormalization introduces a scale into the theory and scale
symmetry is broken.

To discuss the anomalous breaking of scale symmetry, 
we shall first derive the naive Ward identity
corresponding to scale transformation and then compare it with 
the renormalization group equation. The renormalization group equation
can in fact be thought of as a kind of anomalous scaling Ward 
identity since it reflects the scale dependence of a physical amplitude. 

We take the massless scalar field theory 
\begin{eqnarray}
{\cal L}=\frac{1}{2}\partial_{\mu}\phi \partial^{\mu}\phi-
\frac{1}{4!} \lambda\phi^4,
\end{eqnarray}
as an example. Consider a general Green function with the dilatation current,
\begin{eqnarray}
{\langle}0|T[d_{\mu}(y){\phi}(x_1){\cdots}{\phi}(x_n)]|0{\rangle}.
\end{eqnarray}
Taking the derivative with respect to $y^\mu$ gives
\begin{eqnarray}
&&\frac{\partial}{{\partial}y^{\mu}}{\langle}0|T[d^{\mu}(y){\phi}(x_1){\cdots}
{\phi}(x_n)]|0{\rangle}=
{\langle}0|T[{\partial}_{\mu}d^{\mu}(y){\phi}(x_1){\cdots}
{\phi}(x_n)]|0{\rangle}\nonumber\\[2mm]
&&+\sum_{i=1}^n{\delta}(x_i^0-y^0)
{\langle}0|T[{\phi}(x_1){\cdots}[d^0(y),{\phi}(x_i)]{\cdots}
{\phi}(x_n)]|0{\rangle}.
\label{eq57}
\end{eqnarray}
Integrating over $y$, using the scale transformations
\begin{eqnarray}
i[D,{\phi}(x)]= {\delta}{\phi}(x)=(d_{\phi}+x^{\mu}{\partial}_{\mu}){\phi},
\end{eqnarray}
the classical relation ${\theta}^{\mu}_{~\mu}={\partial}^{\mu}d_{\mu}$=0
and discarding the surface term, we get the Ward identity
\begin{eqnarray}
&&{\int}d^4y{\langle}0|T\left[{\theta}^{\mu}_{~\mu}(y){\phi}(x_1){\cdots}
{\phi}(x_n)\right]|0{\rangle}
=i\sum_{i=1}^n{\langle}0|T\left[{\phi}(x_1){\cdots}{\delta}{\phi}(x_i){\cdots}
{\phi}(x_n)\right]|0{\rangle}\nonumber\\[2mm]
&&=i\sum_{i=1}^n{\langle}0|T\left[{\phi}(x_1){\cdots}\left(d_{\phi}+x_i^{\mu}
\frac{\partial}{{\partial}x_i^{\mu}}\right){\phi}(x_i){\cdots}
{\phi}(x_n)\right]|0{\rangle}\nonumber\\[2mm]
&&=i\left(nd_{\phi}+x_1^{\mu}\frac{\partial}{{\partial}x_1^{\mu}}+{\cdots}
+x_n^{\mu}\frac{\partial}{{\partial}x_n^{\mu}}\right)
{\langle}0|T\left[{\phi}(x_1){\cdots}{\phi}(x_n)\right]|0{\rangle}=0.
\label{eq60}
\end{eqnarray}
In momentum space, this gives
\begin{eqnarray}
\left(-\sum_{i=1}^{n-1}p_i{\cdot}\frac{\partial}{\partial p_i}+
D\right)G^{(n)}(p_1,{\cdots},p_{n-1})=0,
\end{eqnarray}
where $D{\equiv}nd_{\phi}-4n+4$ is just the canonical dimension
of the Fourier transform of the Green function
${\langle}0|T[{\phi}(x_1){\cdots}{\phi}(x_n)]|0{\rangle}$.
Parameterizing the momenta $p_i=e^tp_i^{(0)}$ with $p_i^{(0)}$ being
certain fixed momenta,                                           
and considering the corresponding 1PI Green function 
${\Gamma}^{(n)}(e^tp_i^{(0)})$, we have
\begin{eqnarray}
\left(-\frac{\partial }{\partial t}+D\right)
{\Gamma}^{(n)}(e^tp_i^{(0)})=0,
\end{eqnarray}                                              
and hence
\begin{eqnarray}
{\Gamma}^{(n)}(e^tp_i^{(0)})=e^{Dt}{\Gamma}^{(n)}(p_i^{(0)}).
\label{eq66}
\end{eqnarray}
This means that the Green function has canonical scaling dimension. However,
this is not correct, since the naive scaling Ward identity ignores
quantum effects. As a consequence of renormalization, anomalous
dimensions have to be added to the canonical ones.

Renormalization is a necessary procedure to deal with UV divergences.
Its basic idea is to absorb the divergences into a redefinition
of the parameters and a rescaling of the fields. To extract the UV 
divergences, one should impose renormalization conditions
on the renormalized Green functions.
Under different renormalization prescriptions, the renormalized 
Green functions can differ by a finite quantity. Physical
results should, however, be independent of the renormalization
prescription. Since the renormalized parameters depend on the renormalization
prescription, a change in the prescription is compensated
by simultaneous changes of the renormalized parameters of the theory.
Hence the physical amplitude can remain invariant and this
is described by renormalization group equations (RGE). 
For massive $\lambda\phi^4$ theory, the renormalization group equation
for the 1PI part of the renormalized Green function
$G^{(n)}(p_1,{\cdots},p_{n-1})$ is
\begin{eqnarray}
\left[{\mu}\frac{\partial}{{\partial}{\mu}}+{\beta}(\lambda_R)
\frac{\partial}{{\partial}{\lambda}_R}+{\gamma}_m({\lambda}_R)
m_R\frac{\partial}{{\partial}m_R}-n{\gamma}({\lambda}_R)\right]
{\Gamma}_R^{(n)}(p_i,{\lambda}_R,m_R,\mu)=0,
\label{eq74}
\end{eqnarray}
where $\beta (\lambda_R)$, $\gamma (\lambda_R)$ and $\gamma_m (\lambda_R)$
are the $\beta$-function of the scalar self-coupling, the anomalous dimensions
of wave function and mass renormalization. However,
this equation is of little practical use since it only
 describes the dependence on $\mu$. It allows us, however, to derive
an equation describing the behaviour of Green
functions under a variation of the external momenta. We rescale
the momenta,
\begin{eqnarray}
p_i{\equiv}{\rho}p_i^{(0)}, ~i=1,2,{\cdots},n,
\end{eqnarray}
with $p_i^{(0)}$ being certain fixed momenta, and get
\begin{eqnarray}
\left({\rho}\frac{\partial}{{\partial}{\rho}}
+{\mu}\frac{\partial}{{\partial}{\mu}}+
m_R\frac{{\partial}}{{\partial}m_R}-D_{\Gamma}\right){\Gamma}_R^{(n)}
({\rho}p_i^{(0)}, m_R,{\mu},{\lambda}_R)=0.
\label{eq4.29}
\end{eqnarray}
The combination of (\ref{eq4.29}) and (\ref{eq74}) gives the
RGE we prefer,
\begin{eqnarray}
\hspace{-6mm}\left[{\rho}\frac{\partial}{{\partial}{\rho}}-{\beta}({\lambda}_R)
\frac{\partial}{{\partial}{\lambda}_R}+(1-{\gamma}_m({\lambda}_R))
m_R\frac{{\partial}}{{\partial}m_R}  
+n{\gamma}({\lambda}_R)-D_{\Gamma}\right]
{\Gamma}_R^{(n)}({\rho}p_i^{(0)}, m_R,{\mu},{\lambda}_R)=0.
\label{eq78}
\end{eqnarray}

The solution to the RGE (\ref{eq78}) can be worked out by defining
the running functions ${\lambda}(\rho)$ and $m({\rho})$
with the boundary condition,
\begin{eqnarray}
\lambda(1)={\lambda}_R, ~~m(1)=m_R,
\label{eq79}
\end{eqnarray}
and the running $\beta$-functions and the anomalous dimension 
${\gamma}_m$ for the mass renormalization,
\begin{eqnarray}
&&{\beta}(\lambda(\rho))={\rho}\frac{d}{d{\rho}}{\lambda}(\rho),
\label{eq2p12}
\nonumber\\[2mm]
&&\left[{\gamma}_m({\lambda}(\rho))-1\right]m(\rho)
={\rho}\frac{d}{d{\rho}}{m}(\rho).
\end{eqnarray}
With the replacement of the renormalized parameters
$m_R$ and $\lambda_R$ by the corresponding running coupling, the
RGE (\ref{eq78}) is converted into an integrable differential
equation,
\begin{eqnarray}
\left[\frac{d}{d\rho}+n\gamma[\lambda (\rho)]-D\right]
{\Gamma}_R^{(n)}(\rho p_i^{(0)},{\lambda}(\rho),m(\rho),{\mu})=0. 
\end{eqnarray}
The solution can be easily written out
\begin{eqnarray}
{\Gamma}_R^{(n)}(p_i,{\lambda}_R,m_R,{\mu})&=&
{\Gamma}_R^{(n)}({\rho}p_i^{(0)},{\lambda}_R,m_R,{\mu})\nonumber\\[2mm]
&=&{\rho}^{D_{\Gamma}}\mbox{exp}\left[-n{\int}_{\lambda_R}^{\lambda(\rho)}
\frac{{\gamma}(\lambda')}{\beta(\lambda')} d\lambda'\right]
{\Gamma}_R^{(n)}(p_i^{(0)},{\lambda}(\rho),m(\rho),{\mu}) 
\nonumber\\[2mm]
&=&{\rho}^{D_{\Gamma}}\mbox{exp}\left[-n{\int}_1^{\rho}
{\gamma}(\lambda(\rho')) \frac{d\rho'}{\rho'}\right]
{\Gamma}_R^{(n)}(p_i^{(0)},{\lambda}(\rho),m(\rho),{\mu}).
\label{eq84}
\end{eqnarray}
The solution (\ref{eq84}) shows why ${\gamma}$ is called the 
anomalous dimension.

Comparing with the naive scale Ward identity (\ref{eq66}),
one can see that the scale symmetry is broken by renormalization
effects. Only when in a massless theory the $\beta$-function  
and the anomalous dimensions vanish, i.e. the theory is 
finite, can the Green function have canonical scaling behaviour
\begin{eqnarray}
&&{\beta}={\gamma}=0,~~~({\rho}\frac{\partial}{{\partial}{\rho}}
-D_{\Gamma}){\Gamma}_R^{(n)}({\rho}p_i^{(0)},{\mu},\lambda_R)
=0,\nonumber\\[2mm]
&&{\Gamma}_R^{(n)}({\rho}p_i^{(0)},\mu, \lambda_R)={\rho}^{D_{\Gamma}}
{\Gamma}_R^{(n)}(p_i^{(0)},\mu, \lambda_R).
\end{eqnarray}

A main application of the RGE
is to discuss the large or small momentum behaviour of
quantum field theory, giving information about the physics
at different energy scales. According to the definition 
$p_{\mu}{\equiv}{\rho}p^{(0)}_{\mu}$, the case ${\rho}{\rightarrow}{\infty}$
is called the UV limit
and ${\rho}{\rightarrow}0$ is called the IR limit. 
We assume Eq.\,(\ref{eq2p12})
\begin{eqnarray}
\ln\rho={\int}_{\lambda_R}^{\lambda(\rho)}\frac{d\lambda^{\prime}}
{\beta(\lambda^{\prime})}
\label{eq89}
\end{eqnarray}
is valid in the whole range $0<{\rho}<{\infty}$.
Otherwise, the renormalization scale $\mu$ could not vary 
arbitrarily and the
theory would need a cut-off.
(\ref{eq89}) is divergent when ${\rho}{\rightarrow}{\infty}$ or 
${\rho}{\rightarrow}0$.
If an integral over a finite interval is divergent, 
the integrand must be singular at 
either endpoint (or both). Then one must have $\beta (\lambda){\to}0$
when ${\rho}{\to}{\infty}$ or ${\rho}{\to}0$. Thus we have
\begin{eqnarray}
\lim_{\rho{\to}{\infty}~{\rm or}~0}\lambda(\rho)=\lambda_f, 
~~~{\beta}(\lambda_f)=0,
\end{eqnarray}
 that is, $\lambda (\rho)$ must approach a zero of the $\beta$ function.
The zeroes of $\beta$ are called fixed points. 
If $\beta^{\prime}(\lambda_f)<0$ 
with $\beta^{\prime}(\lambda)=d{\beta}/d\lambda$,
and $\lim_{\rho{\to}{\infty}}\lambda (\rho)=\lambda_f$,
$\lambda_f$ is called an UV stable fixed point. 
If $\beta^{\prime}(\lambda_f)>0$ and 
$\lim_{\rho{\to}0}\lambda (\rho)=\lambda_f$,
$\lambda_f$ is called an IR stable fixed point.

For theories with only one coupling constant,
$\lambda_f=0$ must be one of the fixed points. 
If $\lambda_f=0$ is a UV stable fixed point,
the theory is called asymptotically free. 
If $\lambda_f=0$ is a IR stable fixed point, 
the theory is said to be IR stable.
For example, QCD is asymptotically free, while
 $\lambda {\phi}^4$ theory and QED are IR stable. 

One can now see that some theories have
 asymptotic scale invariance at high energy. From the solution
to the RGE of ${\lambda}{\phi}^4$ theory,
\begin{eqnarray}
{\Gamma}_R^{(n)}(e^tp_i^{(0)},\lambda,m,\mu)=e^{D_{\Gamma}t}\,
{\Gamma}^{(n)}[p_i^{(0)},\lambda(t),m(t),\mu]\,\mbox{exp}\left[-n{\int}_0^t
{\gamma}(\lambda(t'))dt'\right].
\end{eqnarray}
Near the UV fixed point ${\lambda}_f$, with the definition ${\rho}
{\equiv}e^t$, we can approximately write
\begin{eqnarray}
\mbox{exp}\left[-n{\int}_0^t{\gamma}(\lambda(t'))dt'\right]=
{\rho}^{-n{\gamma}({\lambda}_f)+{\epsilon}(t)},
~~ {\epsilon}(t)=-\frac{1}{t}{\int}_0^tdt'[\gamma({\lambda}(t'))-{\gamma}
(\lambda_f)].
\end{eqnarray}
If ${\epsilon}{\propto}{\cal O}(1/t)$ as $t{\to}{\infty}$, the theory is
called asymptotically scale invariant. The asymptotic scale behaviour
of ${\Gamma}^{(n)}_R(e^tp_i^{(0)},\lambda,\mu)$ can be obtained by expanding
$\lambda (t)$ around $\lambda_f$. The leading term is
\begin{eqnarray}
{\Gamma}^{(n)}_R(e^tp_i^{(0)},\lambda,m,\mu){\sim}
{\rho}^{D_{\Gamma}
-n{\gamma}(\lambda_f)}{\Gamma}^{(n)}_R(p_i^{(0)},\lambda,m,\mu).
\end{eqnarray}
Rigorously speaking, only ${\gamma}({\lambda}_f)$ is called
the anomalous dimension of the field. 

\subsection{Chiral symmetry in massless QCD}
\label{subsect24}
\renewcommand{\theequation}{2.3.\arabic{equation}}
\setcounter{equation}{0}
\renewcommand{\thetable}{2.3.\arabic{table}}
\setcounter{table}{0}

\subsubsection{Global symmetries of massless QCD}
\label{subsub241}

The Lagrangian of massless QCD with $N_f$ flavours and $N_c$ colours
(colour gauge group $G=SU(N_c)$) reads as  follows:
\begin{eqnarray}
{\cal L}=\sum_{\alpha\beta}
\sum_{rs}\sum_{ij}\overline{\psi}_{\alpha ri}i{\gamma}^{\mu}_{\alpha\beta}D_{\mu rs}
{\delta}_{ij}{\psi}_{\beta sj}
-\frac{1}{4}F_{\mu\nu}^aF^{\mu\nu a}, ~
D_{\mu rs}={\partial}_{\mu}{\delta}_{rs}-igA_{\mu}^aT^a_{rs},
\label{eq97}
\end{eqnarray}
where for clarity we explicitly write the various indices; 
${\alpha},{\beta}=1,{\cdots},4$ are the spinor indices, 
$a=1,{\cdots},\mbox{dim}G=N_c^2-1$ are group indices, 
$i,j=1,{\cdots},N_f$ are the flavour
indices and $r,s=1,{\cdots},N_c$ are the colour  indices. 
The Lagrangian (\ref{eq97}) has an explicit $SU_V(N_f){\times}U_B(1)$
flavour symmetry,
\begin{eqnarray}
{\Psi}{\longrightarrow}e^{i{\alpha}^At^A}{\Psi}~ &,&~
\overline{\Psi}{\longrightarrow}\overline{\Psi}e^{-i{\alpha}^At^A}, \nonumber\\[2mm]
{\Psi}{\longrightarrow}e^{i{\alpha}}{\Psi} ~&,&~
\overline{\Psi}{\longrightarrow}e^{-i{\alpha}}\overline{\Psi}, 
\end{eqnarray}
where $A=1,{\cdots},N_f^2-1$ are the $SU_V(N_f)$ flavour group indices and
$t^A$ are $N_f{\times}N_f$ matrices.
The Noether theorem gives the conserved vector current, 
baryon number current and the corresponding charges:
\begin{eqnarray}
j_{\mu}^A=\overline{\Psi}{\gamma}_{\mu}t^A{\Psi}&,&~~~
j_{\mu}=\overline{\Psi}{\gamma}_{\mu}{\Psi},\nonumber\\[2mm]
Q_V^A={\int}d^3xj_0^A={\int}d^3x{\Psi}^{\dagger}t^A{\Psi}&,&~~~
Q_B={\int}d^3xj_0={\int}d^3x{\Psi}^{\dagger}{\Psi}. 
\end{eqnarray}
Since $\{{\gamma}_5,{\gamma}_{\mu}\}=0$ and there is no quark mass term,
the Lagrangian (\ref{eq97}) possesses yet other global flavour symmetry
$SU_A(N_f)\times U_A(1)$:
\begin{eqnarray}
\Psi{\rightarrow}e^{i{\alpha}^At^A{\gamma}_5}{\Psi} &,&
\overline{\Psi}{\rightarrow}\overline{\Psi}e^{i{\alpha}^At^A{\gamma}_5},
\nonumber\\[2mm]
\Psi{\rightarrow}e^{i{\alpha}{\gamma}_5}{\Psi} &,&
\overline{\Psi}{\rightarrow}\overline{\Psi}e^{i{\alpha}{\gamma}_5}.
\end{eqnarray}
The corresponding conserved axial vector current, $U(1)$
axial current and charges are as follows,
\begin{eqnarray}
j_{\mu}^{5A}=\overline{\Psi}{\gamma}_{\mu}{\gamma}_5t^A{\Psi} &,&~~~
j_{\mu}^5=\overline{\Psi}{\gamma}_{\mu}{\gamma}_5{\Psi},
\nonumber\\[2mm]
Q_5^A={\int}d^3x {\Psi}^{\dagger}{\gamma}_5t^A{\Psi} &,&~~~
Q_5={\int}d^3x {\Psi}^{\dagger}{\gamma}_5{\Psi}. 
\end{eqnarray}
The vector and axial vector conserved charges form representations
of the Lie algebra of the $SU(N_f)$ group 
\begin{eqnarray}
[Q^A , Q^B]=[Q^A_5, Q^B_5]=if^{ABC}Q^C,~~[Q_5^A,Q^B]=
if^{ABC}Q_5^C,
\end{eqnarray}
where $f^{ABC}$ are the structure constants
of the Lie algebra of $SU(N_f)$.
The Lagrangian (\ref{eq97}) explicitly exhibits the chiral symmetry
\begin{eqnarray}
SU_L(N_f){\times}SU_R(N_f){\times}U_L(1){\times}U_R(1)
\label{eq105m}
\end{eqnarray}
if it is rewritten by means of chiral spinors,
\begin{eqnarray}
{\cal L}=\overline{\Psi}_Li{\gamma}^{\mu}D_{\mu}{\Psi}_L+
\overline{\Psi}_Ri{\gamma}^{\mu}D_{\mu}{\Psi}_R-\frac{1}{4}F_{\mu\nu}^aF^{\mu\nu a},
\label{eq105}
\end{eqnarray}
where the left- and right- handed chiral spinors are associated with
the Dirac spinor as follows:
\begin{eqnarray}
\Psi_L{\equiv}\frac{1}{2}(1-{\gamma}_5){\Psi},~~~ 
\Psi_R{\equiv}\frac{1}{2}(1+{\gamma}_5){\Psi}.  
\end{eqnarray}
The chiral transformation under (\ref{eq105m}) are:
\begin{eqnarray}
{\Psi'}_{L(R)}&=&e^{i{\alpha}^At^A}{\Psi}_{L(R)},~~~
\overline{\Psi}'_{L(R)}=\overline{\Psi}_{L(R)}e^{-i{\alpha}^At^A},\nonumber\\[2mm]
{\Psi'}_{L(R)}&=&e^{i{\alpha}}{\Psi}_{L(R)},~~~
\overline{\Psi}'_{L(R)}=\overline{\Psi}_{L(R)}e^{-i{\alpha}} .
\end{eqnarray}
and the corresponding Noether currents and charges are, respectively,
\begin{eqnarray}
&&j_{L(R)\mu}^A=\overline{\Psi}_{L(R)}{\gamma}_{\mu}t^A{\Psi}_{L(R)},~~
j_{L(R)\mu}=\overline{\Psi}_{L(R)}{\gamma}_{\mu}{\Psi}_{L(R)},\nonumber\\[2mm]
&&Q_{L(R)}^A={\int}d^3xj_{L(R)0}^A
={\int}d^3x\Psi^{\dagger}_{L(R)}t^A{\Psi}_{L(R)}, \nonumber\\[2mm]
&&Q_{L(R)}={\int}d^3xj_{L(R)0}
={\int}d^3x\Psi^{\dagger}_{L(R)}{\Psi}_{L(R)}.
\end{eqnarray}
These conserved charges give the representations of the generators 
of the symmetry groups (\ref{eq105m}), for example,   
\begin{eqnarray}
[Q^A_{L(R)} , Q^B_{L(R)}]=if^{ABC}Q^C_{L(R)}, ~~~~
[Q_L^A,Q_R^B]=0,
\end{eqnarray}
and their relations with the the vector and axial vector conserved charge
are
\begin{eqnarray}
Q^A_R = \frac{1}{2}(Q^A + Q^A_5)&,& Q^A_L = \frac{1}{2}(Q^A - Q^A_5);
\nonumber\\
Q_R = \frac{1}{2}(Q + Q_5)&,& Q_L = \frac{1}{2}(Q - Q_5).
\end{eqnarray}

Note that there are  dynamical vector conserved currents,  
which correspond to global gauge transformation in colour space,
\begin{eqnarray}
J_{\mu}^a=\overline{\Psi}{\gamma}_{\mu}T^a{\Psi}, ~\widetilde{Q}^a=
{\int}d^3x J_0^a,~~[\widetilde{Q}^a,\widetilde{Q}^b]=iC^{abc}\widetilde{Q}^c,
\end{eqnarray}
where $T^a$ is the $N_c{\times}N_c$ matrix representation 
of the generators of the $SU(N_c)$ colour gauge group and $C^{abc}$
are the structure constants of $SU(N_c)$.

\subsubsection{Chiral symmetry breaking}
\label{subsub242}

Although the Lagrangian (\ref{eq97}) possesses a 
large global flavour symmetry
$SU_L(N_f){\times}$ $SU_R(N_f)$ ${\times}U_B(1)$ ${\times}U_A(1)$, the observed
symmetry is only $SU_V(N_f){\times}U_B(1)$. This means that a chiral
symmetry breakdown must occur:
$SU_L(N_f){\times}SU_R(N_f){\longrightarrow}SU_V(N_f)$,
and that the $U_A(1)$ symmetry also breaks. Unfortunately, up to now
the mechanisms for both breakdowns have not been 
understood  completely.

In fact, the observed hadron spectrum tells us that chiral symmetry should be broken. 
Otherwise there would naturally exist parity degenerate states, whereas this
is not a feature of the hadron spectrum. 
The argument runs briefly as follows.
First, according to Coleman's theorem, if there is no spontaneous breakdown 
of symmetry, the symmetry of the vacuum should be that of the world,
i.e. if $Q|0{\rangle}=0$, then $[H,Q]=0$, $H$ being the Hamiltonian
of the theory. Then consider a state $|\Psi{\rangle}$, which is an
eigenstate  of both the Hamiltonian $H$ and the parity operator $P$,
\begin{eqnarray}
H|\Psi{\rangle}=E |\Psi{\rangle}, ~P|\Psi{\rangle}=|\Psi{\rangle}.
\end{eqnarray}
If chiral symmetry were not spontaneously  broken, 
we would have
\begin{eqnarray}
&& HQ_L|\Psi{\rangle}=E Q_L|\Psi{\rangle}, ~~
HQ_R|\Psi{\rangle}=E Q_R|\Psi{\rangle}, \nonumber\\[2mm]
&& PQ_{L(R)}|\Psi{\rangle}=PQ_{L(R)}P^{\dagger}P|\Psi{\rangle}=
Q_{R(L)}|\Psi{\rangle}.
\end{eqnarray}
Defining the state
\begin{eqnarray}
|{\Psi}'{\rangle}=\frac{1}{\sqrt{2}}(Q_R-Q_L)|\Psi{\rangle},
\end{eqnarray}
we obtain
\begin{eqnarray}
H|{\Psi}'{\rangle}=E|{\Psi}'{\rangle},~P|{\Psi}'{\rangle}=-|{\Psi}'{\rangle}.
\end{eqnarray}
Therefore, we come to the conclusion 
that $|{\Psi}'{\rangle}$  and $|{\Psi}{\rangle}$
describe parity degenerate particles, hence the assumption
$Q_R|0{\rangle}=Q_L|0{\rangle}=0$ is not correct.
The breaking pattern should be 
\begin{eqnarray}
Q_5^a|0{\rangle}{\neq}0, ~~Q^a|0{\rangle}=0,
\label{eq5.23}
\end{eqnarray}
i.e. the axial symmetry is spontaneously broken. 
Although the mechanism of the spontaneous breaking of chiral symmetry is not
yet understood, the consequences of the breaking
pattern (\ref{eq5.23}) agrees with experimental observations.
The first consequence of the spontaneous chiral symmetry breaking
which agrees with experimental data is  the Goldberg-Trieman 
relation \cite{ref27,ref217}.  The second piece of evidence is 
that some pseudoscalar mesons
can be explained as the Goldstone bosons of the spontaneous breakdown
of the chiral symmetry. For example, if 
we consider the spontaneous breaking of $SU_L(2){\times}SU_R(2)$ in QCD, the
triplet of pions has the right quantum numbers and are candidates
for Goldstone bosons. A concrete phenomenological
model which manifests chiral symmetry breaking 
is the $\sigma$ model describing the interactions 
between nucleons and mesons \cite{ref27,ref217}.

Usually, a spontaneous symmetry breaking
is described by a non-vanishing vacuum expectation value. However, 
unlike in electroweak theory, there is no scalar field in $QCD$.  
The characteristic of chiral symmetry
breaking is the appearance of a non-vanishing  quark condensate
\begin{eqnarray}
{\langle}0|\overline{\Psi}\Psi|0{\rangle}{\neq}0, 
\label{5.24}
\end{eqnarray}          
or equivalently,
\begin{eqnarray}
{\langle}0|\overline{\Psi}_L\Psi_R|0{\rangle}{\neq}0, 
~{\langle}0|\overline{\Psi}_R\Psi_L|0{\rangle}{\neq}0. 
\label{eq5.25}
\end{eqnarray}          
Roughly speaking, the dynamical reason for this condensation  
could be that the coupling constant of QCD becomes strong 
at low energy. Thus it is possible that in the ground state
of QCD  there is an indefinite number of massless fermion pairs
which can be created and annihilated due to the strong coupling.
These condensed fermion pairs have zero total momentum
and angular momentum and hence make the ground state
Lorentz invariant. Therefore, the QCD vacuum has
the property that the operators which annihilate or create such fermion
pairs can have non-zero vacuum expectation values (\ref{5.24})
or (\ref{eq5.25}).  This can be regarded as a qualitative explanation
of the chiral symmetry breaking in QCD.

\subsubsection{Anomaly in QCD}
\label{subsub243}

QCD is a vector gauge theory. The couplings of left-handed 
and right-handed fermions with gauge field are parity-symmetric,
thus no dynamical chiral anomaly arises. This is unlike a
chiral gauge theory such as the electroweak model, where the left-handed and
right-handed fermions can be either in different representations of
the gauge group or coupled to different gauge groups, and the axial
vector currents or the chiral currents are dynamical currents.
At the quantum level, if we require that the vector current is conserved,
the conservation of the axial vector currents is violated and this is
reflected in the violation of a Ward identity. One typical amplitude 
is given by the triangle diagram consisting of one axial vector current
and two vector currents or three axial vector currents.
An anomaly will make the quantum chiral gauge theory non-renormalizable.
Thus one must choose appropriate fermion species to make the anomaly
cancel, otherwise we have no way to quantize a chiral gauge theory.

In QCD, there are two kinds of axial vector flavour currents, the non-singlet
one, $j_{\mu}^{5A}$, and the singlet one, $j^5_\mu$. The singlet
axial vector current usually becomes anomalous, while the non-singlet
one remains conserved. However, if the quarks participate in other interactions, 
the non-singlet axial vector current may become anomalous. 
Like the dynamical chiral anomaly, 
these non-dynamical anomalies still reflect the violation
of Ward identities.  We consider the
triangle diagrams ${\langle}j^5_{\mu}J^a_{\nu}J^b_{\rho}{\rangle}$
and ${\langle}j_{\mu}^{5A}J^b_{\nu}J^c_{\rho}{\rangle}$. 
At the classical level all the currents are conserved
$$ {\partial}^{\mu}j_{\mu}^{5A}={\partial}^{\mu}j_{\mu}^5=
\partial^{\mu}J_{\mu}^a=0.$$
In terms of the Fourier transforms of the triangle diagrams:
\begin{eqnarray}
{\Gamma}^{ab}_{\mu\nu\rho}(p,q,r)(2\pi)^4\delta^{(4)}(p+q+r)={
\int}d^4xd^4yd^4ze^{i(r{\cdot}x+p{\cdot}y+q{\cdot}z)}
{\langle}j^5_{\mu}(x)J^a_{\nu}(y)J^b_{\rho}(z){\rangle},\nonumber\\
\hspace{-10mm}{\Gamma}^{Abc}_{\mu\nu\rho}(p,q,r)(2\pi)^4\delta^{(4)}(p+q+r)
={\int}d^4xd^4yd^4ze^{i(r{\cdot}x
+p{\cdot}y+q{\cdot}z)}
{\langle}j^{5A}_{\mu}(x)J^b_{\nu}(y)J^c_{\rho}(z){\rangle}, 
\end{eqnarray}
the na\"{\i}ve Ward identities read as follows,
\begin{eqnarray}
(p+q)^{\mu}{\Gamma}^{ab}_{\mu\nu\rho}(p,q,r)
&=&p^{\nu}{\Gamma}^{ab}_{\mu\nu\rho}(p,q,r)
=q^{\rho}{\Gamma}^{ab}_{\mu\nu\rho}(p,q,r)=0,\nonumber\\[2mm]
(p+q)^{\mu}{\Gamma}^{Abc}_{\mu\nu\rho}(p,q,r)&=&
p^{\nu}{\Gamma}^{Abc}_{\mu\nu\rho}(p,q,r)
=q^{\rho}{\Gamma}^{Abc}{\Gamma}_{\mu\nu\rho}(p,q,r)=0.
\end{eqnarray} 
Usually the gauge symmetry is required to be preserved,
\begin{eqnarray} 
p^{\nu}{\Gamma}^{ab}_{\mu\nu\rho}(p,q,r)
&=&q^{\rho}{\Gamma}^{ab}_{\mu\nu\rho}(p,q,r)=0 ,\nonumber\\[2mm]
p^{\nu}{\Gamma}^{Abc}_{\mu\nu\rho}(p,q,r)
&=&q^{\rho}{\Gamma}^{Abc}_{\mu\nu\rho}(p,q,r)=0.
\label{eq5.102}
\end{eqnarray} 
(\ref{eq5.102}) are actually
the renormalization conditions for evaluating the triangle diagrams.
With these (physical) renormalization conditions, explicit calculations
yield
\begin{eqnarray}
(p+q)^{\mu}{\Gamma}^{ab}_{\mu\nu\rho}(p,q,r)&=& \frac{i}{2\pi^2}
{\epsilon}_{\nu\rho\alpha\beta}p^{\alpha}q^{\beta}
\mbox{Tr}(T^aT^b)\nonumber\\[2mm]
(p+q)^{\mu}{\Gamma}_{\mu\nu\rho}^{Abc}(p,q,r)&=&\frac{i}{2\pi^2}
{\epsilon}_{\nu\rho\alpha\beta}p^{\alpha}q^{\beta}\mbox{Tr}(t^A\{T^b,T^c\})=0,
\end{eqnarray}
where the trace is taken over both colour and flavour indices.
The corresponding operator equations in coordinate space are
\begin{eqnarray}
{\partial}^{\mu}j_{\mu}^5&=&
-\frac{g^2}{16\pi^2}{\epsilon}^{\mu\nu\alpha\beta}
F_{\mu\nu}^aF_{\alpha\beta}^b\mbox{Tr}(T^aT^b),
~~~\partial^{\mu}j_{\mu}^{5A}=0.
\end{eqnarray}                                     
Therefore, only the singlet axial current $j^5_{\mu}$ is anomalous. 
However,  if the quarks interact electromagnetically, 
an anomaly for the non-singlet current may exist. As an example, consider the case of
two flavours ($N_f=2$),
\begin{eqnarray}
{\psi}=\left(\begin{array}{c} u\\d\end{array}\right).
\end{eqnarray}
Correspondingly, $t^a={\sigma}^a/2$ and the $T$´s are replaced by the
electric charge matrix $Q$
\begin{eqnarray}
Q=\left(\begin{array}{cc}2/3 & \\ & -1/3\end{array}\right), ~
{\partial}^{\mu}j_{\mu}^{5A}=
-\frac{e^2}{16\pi^2}{\epsilon}^{\mu\nu\alpha\beta}
F_{\mu\nu}F_{\alpha\beta}\mbox{Tr}(Q^2{\sigma}^A).
\end{eqnarray}                                     
If we choose $A=3$, we obtain the operator anomaly equation,
\begin{eqnarray}
{\partial}^{\mu}j_{\mu}^{5(3)}=-
\frac{e^2}{32\pi^2}{\epsilon}^{\mu\nu\alpha\beta}
F_{\mu\nu}F_{\alpha\beta}.
\end{eqnarray}
It is well known that this anomaly contributes to the decay 
$\pi^0{\longrightarrow}2{\gamma}$.

\subsubsection{Anomaly matching}
\label{subsub244}

Anomaly matching \cite{ref218} 
is a basic tool in testing non-Abelian duality conjectures
of $N=1$ supersymmetric QCD. 
Roughly speaking, the matching condition means
that in a confining theory like QCD, the anomaly equations
should survive confinement \cite{ref218,ref219,ref220}.  
Concretely, let us consider $SU(N_c)$ QCD with $N_f$ flavours.
The fundamental building blocks are quarks
represented by ${\psi}_{ir}$ with $i$ and $r$ being flavour 
and colour indices, respectively. The general 
form of the conserved flavour singlet current is
\begin{eqnarray}
j_{\mu}=\overline{\psi}_{ir}{\gamma}_{\mu}[A_{ij}(1-{\gamma}_5)
+B_{ij}(1+{\gamma}_5)]{\delta}_{rs}{\psi}_{js},
\end{eqnarray}
where $A$ and $B$ are Hermitian matrices in flavour space.  
This singlet flavour current will suffer from a chiral anomaly. 
Since the observed particles in QCD  are colourless bound states of 
quarks --- mesons and baryons, we should consider the matrix elements
of the above current between these particle states. Let $|u,p,\alpha{\rangle}$
denote a massless baryon state, where $u$ is a solution of the massless
Dirac equation, $p$ is the four-momentum of the particle and $\alpha$ labels
the other quantum numbers of the baryon. The matrix elements
of the current operator between these hadron states are
\begin{eqnarray}
{\langle}u',p,\alpha |j_{\mu}|u,p, \beta{\rangle}=
\overline{u}'{\gamma}_{\mu}[C_{\alpha\beta}(1-{\gamma}_5)
+D_{\alpha\beta}(1+{\gamma}_5)]u.
\end{eqnarray}
If the symmetry associated with $j_{\mu}$ does not suffer spontaneous
breakdown, then the following relation should hold: 
\begin{eqnarray}
\mbox{Tr}(C-D)=N_c\mbox{Tr}(A-B).
\label{eq5.123}
\end{eqnarray}
This is the matching condition suggested by 't Hooft. Obviously,
the matching condition is connected with the current 
triangle anomaly. Recalling the Fourier transformation of the triangle 
Green function, 
\begin{eqnarray}
{\Gamma}_{\mu\nu\rho}(p,q,r)(2\pi)^{4}{\delta}^{(4)}(p+q+r){\equiv}
{\int}d^4xd^4yd^4ze^{i(p{\cdot}x+q{\cdot}y+r{\cdot}z)}{\langle}
0|T\left[J_{\mu}(x)J_{\nu}(y)j_{\rho}(z)\right]|0{\rangle},
\end{eqnarray} 
$J$ being the gauge symmetry current, we obtain the 
anomalous Ward identity, 
\begin{eqnarray}
r^{\rho}{\Gamma}_{\mu\nu\rho}(p,q,r)=\frac{N_c}{2\pi^2}\mbox{Tr}(A-B)
{\epsilon}_{\mu\nu\alpha\beta}p^{\alpha}q^{\beta}
=\frac{1}{2\pi^2}\mbox{Tr}(C-D)
{\epsilon}_{\mu\nu\alpha\beta}p^{\alpha}q^{\beta}.
\end{eqnarray}
It is well known that this anomaly equation is true to all orders
of perturbation theory and it even survives non-perturbative
effects such as instanton correction \cite{ref220}. Although 
${\Gamma}_{\mu\nu\rho}$ can receive contributions
from every order of perturbation theory, the anomaly
equation remains the same as given by the zeroth order triangle diagram. 
Therefore, the anomaly matching can be formulated in a stricter
way: {\it If one treats the massless baryons as if they were fundamental
spin $1/2$ particles, i.e. quarks, and 
ignores all other particles, one still
gets the correct anomaly}. This is 
the reason why (\ref{eq5.123}) is satisfied. In fact,
this matching condition reveals the deep origin of 
the anomaly \cite{ref221}:
the anomaly is connected with the IR singularity of the amplitude, which
gets contributions only from the massless spin $1/2$ particles.

\subsection{ Various phases of gauge theories}
\label{subsect27}
\renewcommand{\theequation}{2.4.\arabic{equation}}
\setcounter{equation}{0}

The possible phases of a field theory model are associated with 
symmetries of the theory. Phase transitions are usually 
associated with changes of the symmetry. In different phases, the 
particle spectrum and the dynamics of the theory can be greatly 
different. The quantity characterizing the phase is the order 
parameter.

  In a gauge theory the phases are classified by the symmetry 
that is realized in the phase. The order parameter should be 
gauge and Lorentz-invariant. There are several ways to
define an order parameter. The most common choice is the Wilson loop, which 
was proposed by Wilson in 1974 \cite{wloop}. 
The definition of the Wilson loop
can be illustrated by a simple example of an Abelian gauge field:
Assume that two charges $\pm e$ are created at some point in the
Euclidean plane $({\bf x},\tau)$, then 
separated to a distance $R$ and kept static. Finally they come
together and annihilate at another point after some Euclidean
time $T$. The world lines of these two charges will form a contour. The 
interaction of these two external charges with the gauge field is
\begin{eqnarray}
S_{\rm int}=e\int d^4xj^{\mu}A_{\mu}=e\oint A_{\mu}dx^{\mu}.
\end{eqnarray}
The current density for two point charges moving on the perimeter
of a loop is
\begin{eqnarray}
j^{\mu}=e\frac{dx^{\mu}}{ds}\delta^{(4)} (x-x(s)), ~~~ 0 <s \leq 2\pi, 
\end{eqnarray}
with $s$ being the parameter describing the contour.
In the path integral formalism, this source will 
introduce into the vacuum functional integral an extra factor, 
the Wilson loop
\begin{eqnarray}
W_w{\equiv}\exp\left[iS_{\rm int}\right]
=\exp\left[ie\oint A_{\mu}dx^{\mu}\right].
\end{eqnarray} 
The generating functional with this point electric charge 
interaction is just the ``quantum Wilson loop",
\begin{eqnarray}
Z=\langle W_w \rangle.
\end{eqnarray}
In the non-relativistic limit, the quantum Wilson loop is associated with
the static potential of two particles with opposite charges. To see this, 
we calculate the quantum Wilson loop over a rectangle with width $R$ and length
$T$. Choosing the gauge condition $A_0({\bf x},t)=0$, 
we then have \cite{chanel}
\begin{eqnarray}
\langle W_w\rangle \equiv \langle W_w(R,T)\rangle 
=\langle \psi (0) \psi^{\dagger}(T)\rangle,
\label{eq27p5}
\end{eqnarray}
where
\begin{eqnarray}
\psi (0)=P\exp\left[ie\int_0^R ds \frac{d{\bf x}}{ds}{\cdot}
 {\bf A}({\bf x},0)\right],
~~~\psi (T)=P\exp\left[ie\int_0^Rds 
\frac{d{\bf x}}{ds}{\cdot} {\bf A}({\bf x},T)\right], 
\end{eqnarray}
$P$ denoting the path ordering.
Performing the sum over the intermediate states in Eq.\,(\ref{eq27p5}) 
and using the
translation invariance (in Euclidean space)
\begin{eqnarray}
\psi (T) =e^{-HT}\psi (0) e^{HT},
\end{eqnarray}
we get
\begin{eqnarray}
\langle W_w(R,T)\rangle=\sum_n \langle \psi (0)|n\rangle\langle n|
 \psi^{\dagger}(T)\rangle =\sum_n |\langle \psi (0)|n\rangle|^2e^{-E_nT},
\end{eqnarray}
where $H$ is the Hamiltonian and $\lbrace E_n \rbrace$ is the energy
spectrum of the system. In the limit $T\rightarrow \infty$, only
the ground state with the lowest energy $E_0$ contributes to
$\langle W_w(R,T)\rangle $, 
\begin{eqnarray}
\langle W_w(R,T)\rangle \stackrel{T\to\infty}{\longrightarrow}e^{-E_0(R) T}. 
\end{eqnarray}
Wilson's prescription for obtaining the static potential 
between the 
pair of charges as a function of their distance is the following:
\begin{eqnarray}
V(R){\equiv}\lim_{T\to\infty}\left[-\frac{1}{T}\ln \langle W (R,T)
\rangle\right].
\end{eqnarray}
Thus, the phase can be characterized by the static potential
defined in this way. Since the vacuum expectation value of the Wilson
loop operator is determined by the full quantum theory, the general form
of the static potential at large $R$ should be
\begin{eqnarray}
V(R)\sim \frac{\alpha (R)}{R}.
\label{eq27p11}
\end{eqnarray}
The generalization of the above discussion 
to the non-Abelian case is straightforward. The two 
test charges should be in conjugate representations $r$ and $\overline{r}$
of the gauge group and the Wilson loop operator is the trace of the holonomy 
operator in the representation $r$,
\begin{eqnarray}
W_w=\mbox{Tr}_rP\exp\left[\oint A_{\mu}dx^{\mu}\right].
\end{eqnarray}
In Eq.\,(\ref{eq27p11}), 
$\alpha$ is classically a constant, $\alpha =g^2/(4\pi)$, with
$g$ being the gauge coupling constant, but the quantum corrections make 
$\alpha$ be a function of $R$, since $\alpha$ runs due to 
renormalization effects. Depending on the functional form of $\alpha (R)$ at
large $R$ (up to a non-universal additive constant), the phases are classified
as listed in Table (\ref{ta2p71}).

\begin{table}
\begin{center}
\begin{tabular}{|c|c|}\hline 
 phase & static potential $V(R)$ \\
       & at large $R$ \\ \hline
 Coulomb &  $ 1/R $ \\ \hline
 free electric & $ 1/[R\ln\left(R\Lambda\right)]$ \\ \hline
 free magnetic & $\ln\left(R\Lambda\right)/R$\\ \hline
 Higgs & constant \\ \hline
 confining & $\kappa R$ \\ \hline
\end{tabular}

\caption{\protect\small Various phases in gauge theory 
($\kappa$ is the string tension and $\Lambda$ is the renormalization scale).}
\end{center}
\label{ta2p71}

\end{table}
In the following we shall give a detailed explanation of 
each phase and mention the known field theories possessing such a phase.

\vspace{4mm}
\begin{flushleft}
{\it Coulomb phase}
\end{flushleft}
\vspace{4mm}

This dynamical regime has long-distance interaction behaviour. One typical
theory presenting this phase is massive QED, where 
the running of the coupling constant is given by the Landau formula, which  
in momentum space reads: 
\begin{eqnarray}
\alpha_{\rm eff}(q^2)=\frac{\alpha_0}{1-\alpha_0/(3\pi)
\ln\left(|q^2|/m^2\right)},
\end{eqnarray}
where $m$ is the mass of the charged particle.
Thus $\alpha$ decreases logarithmically at large distances. However, 
it stops running at $R{\sim} m^{-1}$, the corresponding 
limiting value of $\alpha$ being
\begin{eqnarray}
\alpha_{\rm eff}^*=\alpha_{\rm eff} (R{\sim}m^{-1})
\end{eqnarray}
and the static potential at large $R$ being
\begin{eqnarray}
V(R)\sim \frac{\alpha_{\rm eff}^*}{R}\propto\frac{1}{R}.
\end{eqnarray}

The Coulomb phase also appears in a non-Abelian theory 
with massless interacting quarks and gluons, and is 
then called the non-Abelian Coulomb phase. 
The Coulomb potential can emerge at a non-trivial infrared fixed 
point of the renormalization group. Thus, such a theory is a non-trivial
interacting four dimensional conformal field theory. 
One known field theory having this
feature is supersymmetric QCD with an appropriate choice of flavours and
colours. We shall give a detailed discussion of this theory later.

\vspace{4mm}
\begin{flushleft}
{\it Free electric phase}
\end{flushleft}
\vspace{4mm}

This dynamical feature is also familiar. 
It occurs in massless QED. From the running of the coupling constant 
in momentum space,
\begin{eqnarray}
\alpha_{\rm eff}(q^2)=\frac{\alpha_0}{1-\alpha_0/(3\pi)
\ln\left(|q^2|/\Lambda^2\right)},
\end{eqnarray} 
one can see that the electric charge  is renormalized to zero at large
distance and thus there will be a factor 
$\ln\left(R\Lambda\right)$ in the static
potential. An intuitive physical reason is that a strong screening
occurs due to quantum effects, and
the photon propagator is dressed by virtual pairs of electrons. 
This dressing makes the running coupling constant 
behave as follows at large $R$:
\begin{eqnarray}
\alpha (R) \sim \frac{1}{\ln (R\Lambda )}. 
\end{eqnarray}
Note that this is greatly different from the massive case, where
the running of the effective charge is frozen at the distance $R=m^{-1}$.
In the massless case,  the logarithmic falloff continues indefinitely. Thus 
the asymptotic limit of massless QED is a free photon plus massless electrons  
whose charge is completely screened. This is why this phase is called a free
phase. Strictly speaking, the model with this phase is ill-defined at
short distances, since the effective coupling grows continuously and
finally hits the Landau pole. Usually, this kind of theory must be
embedded into an asymptotically free theory. A free electric phase also occurs
in the IR region of a non-Abelian gauge theory which is not asymptotically 
free, and it is then called a non-Abelian free electric phase. Ordinary QCD 
with $N_f >16$ can have this phase.

\vspace{4mm}
\begin{flushleft}
{\it Free magnetic phase}
\end{flushleft}
\vspace{4mm}

The free magnetic phase owes its existence to the occurrence of
 magnetic monopole states. Assume that a magnetic monopole 
behaves like a point particle and participates in
interactions like an electron. Analogous to the free electric phase, 
the free magnetic phase occurs when there are massless magnetic monopoles, 
which renormalize the magnetic coupling to zero at large distance,
\begin{eqnarray}
\alpha^{\rm m}_{\rm eff} (R) = 
{\frac{g^2(R)} {4\pi}} \sim \frac{1}{\ln (R\Lambda )}.
\end{eqnarray}
Due to the Dirac condition $e(R)g(R) \sim 1$, 
the electric coupling constant is 
correspondingly renormalized to infinity at large distance,
\begin{eqnarray}
\alpha^{\rm e}_{\rm eff} (R) \sim \ln (R\Lambda ).
\end{eqnarray}
There also exists a non-Abelian free magnetic phase. The known examples
are $N=2$ $SU(2)$ Supersymmetric Yang-Mills theory at the massless monopole
points \cite{ref1p1,ref1p1a} and the dual magnetic theory of $N=1$ supersymmetric $SU(N_c)$
gauge theory with $N_f$ flavours when $N_c+2{\leq}N_f{\leq}3/2N_c$ \cite{ref1p11}.        

\vspace{4mm}
\begin{flushleft}
{\it Higgs phase}
\end{flushleft}
\vspace{4mm}

In a Higgs phase, the gauge group $G$ is spontaneously 
broken to a subgroup $H$ by a scalar field or by the condensation
of a fermionic field. The gauge bosons corresponding to the broken 
generators will become massive due to the Higgs mechanism. One 
typical model is the Georgi-Glashow model, which describes the interaction 
of an $SU(2)$ gauge field with the scalar field in 
the adjoint representation of the gauge group,
\begin{eqnarray}
{\cal L}=-\frac{1}{4}G^a_{\mu\nu}G^{\mu\nu a}+\frac{1}{2}
(D_{\mu}\phi)^a(D^{\mu}\phi)^a-\frac{\lambda}{4}
\left(\phi^a\phi^a-v^2\right)^2.
\end{eqnarray}
The non-vanishing expectation value $\langle |\phi|\rangle =v$ leads to
the spontaneous breaking of the gauge symmetry, $SU(2){\longrightarrow} U(1)$. 
Corresponding to the two broken generators, two gauge bosons become 
massive with mass $M_W=|gv|$ due to the Higgs mechanism, and one 
remains massless, corresponding to the unbroken generator. 
The dynamics is a little complicated: at a distance
less than $M_W^{-1}$, the static potential is the Coulomb potential
$\sim 1/R$; at a distance larger than $M_W^{-1}$, the interaction
force is short-range and the potential  
behaves as a Yukawa potential, $V(R)\sim {\exp\left(-M_W R\right)}/{R}$,
 i.e. the electric charge
is exponentially screened. The gauge coupling constant 
runs according to the Landau formula 
at distances shorter than $M_W^{-1}$, since the
remaining theory is an Abelian gauge theory, and the running is frozen
at the distance $M_W^{-1}$. Thus, in the Higgs phase the static potential 
between two test charges should tend to a constant 
value. This can also be seen from an explicit computation 
of the Wilson loop in lattice gauge theory, 
where the quantum Wilson loop obeys a ``perimeter law",
\begin{eqnarray}
\langle W_w\rangle \sim \exp\left[-\Lambda\times (\mbox{perimeter})\right]. 
\end{eqnarray}

\vspace{4mm}
\begin{flushleft}
{\it Confining phase}
\end{flushleft}
\vspace{4mm}

Confinement means that the particles corresponding to the fields appearing 
in the fundamental Lagrangian are absent in the observed particle
spectrum. One well-known example of a confining phase 
is low-energy QCD. In QCD, the microscopic dynamical variables are
quarks and gluons, they carry colour charges responsible 
for the strong interactions. However, all the observed particles are
colourless hadrons, i.e. bound states of quark-antiquark pairs (mesons)
or three quarks (baryons) or gluons (glueballs). 
The microscopic degrees of freedom are
always confined. This kind of dynamical feature is called a 
confining phase. Strictly speaking, the dynamical mechanism for colour
confinement is not quite clarified yet. However, 
a phenomenological picture is believed to be as follows. 

 If we place a static colour charge and a conjugate charge 
at a large distance from each other, they will create 
a chromoelectric field formed like a thin flux tube between
the two charges, in contrast to
the electric-magnetic field in the Abelian case, 
which is dispersed. The flux tube is a
 string-like object with constant string
tension $\kappa$. Both the cross section and the string tension are
determined by the energy scale at which confinement occurs. 
Thus, the potential between quarks 
grows linearly with the distance since the string tension is constant:
\begin{eqnarray}
V(R)=\kappa R.
\end{eqnarray}
The quarks cannot leave the hadron since this would need an
infinite energy. This dynamical feature manifests itself in the Wilson
loop as the ``area law". 

This physical picture is reminiscent of the Meissner effect
in type II superconductivity. From the viewpoint of field theory,
superconductivity can be thought of as the spontaneous breakdown
of electromagnetic gauge symmetry \cite{weinII} created by the 
Bose condensation
of the Cooper electron pairs in the vacuum state. One of the most prominent
features in the superconductor is the exclusion of magnetic fields,
the Meissner effect. However, if we put two static magnetic charges inside
the superconductor, since the magnetic flux is conserved, the magnetic 
field cannot vanish everywhere. The magnetic flux will be 
pressed into narrow tubes connecting these two magnetic charges.
A phenomenological model describing this dynamics is the Landau-Ginzburg 
theory,
\begin{eqnarray}
S=\int d^3x \left[-\frac{1}{2} \left(\partial_i\psi_p
-2ie t_{pq}A_i\psi_q\right)^2+\frac{1}{2}m^2\psi_p\psi_p
-\frac{1}{4}g\left(\psi_p\psi_p\right)^2\right],
\end{eqnarray} 
where $i=1,2,3$; $p,q=1,2$; $g,m^2 >0$ and $t$ is the Hermitian $U(1)$
generator,
\begin{eqnarray}
t=\left(\begin{array}{cc} 0 & -i\\ i & 0\end{array}\right).
\end{eqnarray} 
This model is the non-relativistic analogue
of a $U(1)$ gauge field interacting with a scalar field. Defining
\begin{eqnarray}
\psi_1+i\psi_2=\rho\exp \left(2ie\phi\right),
\end{eqnarray}
we can rewrite the above action as
\begin{eqnarray}
S=\int d^3x \left[-\frac{1}{2} \left(\partial_i\rho\right)^2-
2e^2\rho^2\left(\partial_i\phi+A_i\right)^2+
\frac{1}{2}m^2\rho^2-\frac{1}{4}g\rho^4\right].
\end{eqnarray}
The Landau-Ginzburg action shows
 spontaneous symmetry breakdown due to the condensation of electron
pairs. The flux tubes are just the vortex solutions
of the classical equations of motion derived from the above action. 
One can calculate the energy carried by the vortex per unit length 
at a distance far away from the magnetic 
charge source with the result being a constant \cite{weinII}. Thus, 
the energy between two magnetic charges in a superconductor 
grows linearly with the separation between them. 

We expect a similar dynamical mechanism to exist in QCD resulting
in colour confinement. However, there are two 
difficulties to overcome.

First, in QCD, it is the chromoelectric field
that forms flux tubes. The vacuum medium should expel the
chromoelectric field, and this can only be achieved by the condensation 
of particles carrying magnetic charge. Thus, if the colour confinement
is produced in this way, it should be a dual Meissner effect.
It is well known that monopole solutions had been found
in the Georgi-Glashow model independently by 't Hooft \cite{monopole} and 
Polyakov \cite{monopolea}.
In this model there exists a scalar field and the Higgs mechanism can break
the original $SU(2)$ gauge group to $U(1)$, leading to the 
emergence of monopoles as soliton solutions
to the classical equations of motion. But QCD, whose gauge group $SU(3)$
remains unbroken, does not have classical magnetic monopole solutions.

A possible scheme to solve this problem was proposed by 't Hooft. His
idea is quite simple \cite{maxag}: choose an appropriate 
non-propagating gauge
condition to reduce the original $SU(N_c)$ QCD to a multiple Abelian theory
with gauge group $U(1)^{N_c-1}$, i.e. the maximal Cartan subgroup of $SU(N_c)$.
This procedure is called Abelian projection, which can artificially create
monopole configurations. Note that the QCD monopoles obtained in this 
way are not physical objects, since they are gauge dependent. Nevertheless,
they may play an important role in implementing the dual Meissner effect
in QCD.

A non-propagating gauge 
means a gauge that is fixed in such a way that no unphysical 
degrees of freedoms such as Faddeev-Popov ghosts emerge.
A familiar example is the unitary gauge. Usually, as 
indicated by 't Hooft, this kind of gauge choice can render 
 the theory non-manifestly renormalizable and therefore, 
in concrete calculations one should go over to 
a nicer ``approximately non-propagating" gauge 
condition. Another more important 
feature of the non-propagating gauge is that it can induce singularities in 
space-time. 't Hooft interpreted these singularities as the ``monopoles", 
additional physical dynamical variables.  
     
A concrete method to implement this gauge is as follows. 
Let us consider some operator that transforms
 covariantly under the local gauge transformation $U(x)$,
\begin{eqnarray}
X \longrightarrow U X U^{-1}.
\end{eqnarray}  
For instance, $X= G_{\mu\nu}G^{\mu\nu}$, $X= G_{\mu\nu}D^2G^{\mu\nu}$,  
or $X=\overline{\psi}_r\psi_s$ and so on, $r,s=1,\cdots,N_c$ being colour indices. 
Then by choosing $U(x)$ appropriately $X$ can be diagonalized,
\begin{eqnarray}
X=\left(\begin{array}{ccc} \lambda_1 & \cdots & 0 \\
                          \vdots & \ddots &\vdots \\
                           0 &\cdots & \lambda_{N_c} \end{array}\right),
\end{eqnarray} 
where the eigenvalues are ordered,
\begin{eqnarray}
\lambda_1 \geq \lambda_2 \geq \cdots \geq \lambda_{N_c}.
\end{eqnarray}
The gauge is not yet fixed completely, as any diagonal gauge rotation
\begin{eqnarray}
U(x)=\left(\begin{array}{ccc} e^{i\theta_1} &\cdots & 0 \\
                      \vdots   & \ddots & \vdots \\
                           0 &\cdots & e^{i\theta_{N_c}} \end{array}\right)
\end{eqnarray} 
leaves $X$ invariant. These gauge transformations form the group
\begin{eqnarray}
U(1)^{N_c-1},
\end{eqnarray} 
i.e. the maximal Abelian subgroup of the gauge group. Thus, 
with this gauge choice 
 $SU(N_c)$ gluodynamics is reduced 
to a dynamics of $N_c-1$ ``photons" 
(the gluon components corresponding to diagonal generators) and $N_c(N_c-1)$
``matter fields" (the gluon components corresponding to all off-diagonal
generators) charged with respect to 
the ``photon" fields.

Let us see how the QCD monopole configurations emerge. As mentioned above,
the monopoles correspond to singularities in space induced by the gauge
choice. Singularities arise when two consecutive eigenvalues of 
$X$ coincide. Without loss of generality, we 
consider the case in which $\lambda_1$ and $\lambda_2$
coincide at some space point ${\bf x}_0$, 
while the other eigenvalues remain distinct. Close to
${\bf x}_0$, a nonsingular gauge transformation brings $X$ 
into a form comprising a $2\times 2$ Hermitian matrix
\begin{eqnarray}
\lambda {\bf 1}+\epsilon^a ({\bf x})\sigma^a,
\end{eqnarray}   
where ${\bf 1}$ is the $2\times 2$ unit matrix 
and $\sigma^a$ are the Pauli matrices, in the upper 
left hand corner; otherwise, $X$
is diagonal. With respect to the $SU(2)$ 
subgroup rotating the 1,2-components into
each other, the fields $\epsilon^a ({\bf x})$ behave as the components
of an isovector. The coincidence $\lambda_1 =\lambda_2$ 
at ${\bf x}_0$ means that 
\begin{eqnarray}
\epsilon^a ({\bf x}_0)=0, ~~~a=1,2,3.
\end{eqnarray}
The vanishing of these three real functions
 can only happen in isolated 
points in three-dimensional space. 
Thus, $\epsilon^a ({\bf x})$ carries the characteristic features 
of the Higgs field of the 't Hooft-Polyakov monopole. 
Indeed, performing the final step
of the gauge fixing, making $\epsilon^1 = \epsilon^2= 0$ and $\epsilon^3 > 0$,
generates Dirac strings in the photon fields $A_{\mu}^1$ and $A_{\mu}^2$.

If the $U(1)$ charges of the ``matter" field $A_{\mu}^{ij}$ are
\begin{eqnarray}
q_i=-q_j=g; q_k=0, k\neq i,j,
\end{eqnarray}
then the magnetic charges of the singularity are
\begin{eqnarray}
(h_1,h_2,h_3,\cdots,h_{N_c}) = (\frac{2\pi} {g},-\frac{2\pi} {g},0,\cdots,0).
\label{eq27101}
\end{eqnarray}
Note that the ``elementary monopoles" only have adjacent magnetic
charges different from zero. Monopoles with nonadjacent
magnetic charges can only arise as bound states of these ``elementary" poles.

Although we have been able to identify states that could give rise
to a dual Meissner effect through condensation, there remains
the formidable problem of showing that this actually is what happens
in QCD. Lattice simulations have given some quite encouraging results,
 but they are still far from complete and many aspects still 
remain unclear \cite{kswetc}. It was thus very encouraging 
that a demonstration of this confinement mechanism could be given 
for $N=2$ supersymmetric Yang-Mills theory \cite{ref1p1} and 
QCD \cite{ref1p1a}. Seiberg and Witten were able to deduce an 
exact low-energy effective action using the electric-magnetic 
duality conjecture, and showed that the dual Meissner 
effect indeed happens.

\vspace{4mm}
\begin{flushleft}
{\it Oblique confinement}
\end{flushleft}
\vspace{4mm}

In addition to the various phases 
introduced above, there still exists
another peculiar dynamical scenario in which states with fractional
baryon charges can emerge. This phase is called the oblique confinement phase. 
Since this phase indeed emerges in supersymmetric $SO(N_c)$ gauge theory,
we shall in the following give a detailed explanation. 

Roughly speaking, oblique confinement is produced by dyon 
condensation. A dyon is a state carrying both electric and magnetic
charges. As a soliton solution in the Georgi-Glashow
model it was first found by Julia and Zee \cite{dyons}
in the gauge choice of non-zero $A_0^a$, the     
time component of the gauge field. The magnetic monopole is the static
solution of the Georgi-Glashow model in the gauge 
condition $A_0^a=0$ \cite{monopole}. The 
non-zero $A_0$ should give the static solution both electric 
and magnetic charges since the electric field does not vanish. 
It was further shown by Witten that a non-vanishing vacuum 
angle $\theta$ \cite{wit79},
\begin{eqnarray}
{\cal L}=\frac{g^2\theta}{32\pi^2}G_{\mu\nu}^a\widetilde{G}^{a\mu\nu}
\end{eqnarray}
affects the electric charge of magnetically charged states.
 Although this term is a surface term and does not affect
the classical equations of motion, it makes a particle with the magnetic 
charge $h$ necessarily acquire an electric charge,
\begin{eqnarray}
q=\frac{\theta g^2}{4\pi^2}h.
\label{eq27p37}
\end{eqnarray}

Before discussing the effects of the $\theta$-vacuum, we shall show that
for QCD monopole configurations the ``electric" charges of off-diagonal 
 gluons and the magnetic charges of singularities
form a $(2N-2)$-dimensional lattice. 

Consider two particles, (1) and (2),
with magnetic charges $h_i^{(1)}$ and $h_i^{(2)}$ and electric charges
$q_i^{(1)}$ and $q_i^{(2)}$, $i=1,\cdots, N_c$ labelling the $U(1)$
group. The charges have to obey the Schwinger-Zwanziger
quantization condition,
\begin{eqnarray}
\sum_{i=1}^{N_c}\left(h_i^{(1)}q_i^{(2)}-q_i^{(1)}h_i^{(2)}\right)=2\pi n
\label{eq27p44}
\end{eqnarray}
with 
\begin{eqnarray}
\sum_{i=1}^{N_c} h_i=\sum_{i=1}^{N_c} q_i=0.
\label{eq27p45}
\end{eqnarray}
In 't Hooft's terms ``the particle $(1)$ has a Dirac quantum $n$ 
with respect to particle $(2)$". 

It is easy to show that a particle, with electric and magnetic charges
being integer coefficient linear combinations of the charges of 
the particles $(1)$ and $(2)$, still satisfies the Schwinger-Zwanziger
quantization condition with respect to both others. Therefore, the 
whole particle spectrum satisfying (\ref{eq27p44}) and (\ref{eq27p45}) 
can be constructed by finding a basis of $2(N_c-1)$ particles with 
charges $h_i^{(A)}$ and $q_i^{(A)}$, $A=1, \cdots, 2N_c-2$, and 
then all allowed sets of charges are 
\begin{eqnarray}
h_i=\sum_{A=1}^{2N_c-2}k_Ah_i^{(A)}, ~~~~~ q_i=\sum_{A=1}^{2N_c-2}k_Aq_i^{(A)}.
\label{eq27p46}
\end{eqnarray}   
So they form $(2N_c-2)$-dimensional lattice.

Since in each of the $N-1$ $U(1)$ groups, the dynamics is described 
by Maxwell's equations, which are invariant under rotations of the electric 
and magnetic charges,
\begin{eqnarray}
h_i &\rightarrow& h_i\cos\phi_i+q_i\sin\phi_i,\nonumber\\
q_i &\rightarrow &-h_i\sin\phi_i+q_i\cos\phi_i,
\end{eqnarray}    
we can define a ``standard" basis by rotating away the 
magnetic part of the $(N_c-1)$ basic charges:
\begin{eqnarray}
h_i^{(A)}=0, ~~~~A=1,\cdots, N_c-1.
\label{eq27100}
\end{eqnarray}   
The electrically charged gluons provide us with a set of $N-1$ basic charges
obeying (\ref{eq27100}):
\begin{eqnarray}
q_i^{(A)}=g\delta_i^A-g\delta_i^{A+1}, ~~~
A=1,\cdots, N_c-1.
\label{eq27p49}
\end{eqnarray}
The standard basis is completed by the magnetic monopoles with magnetic
charges as in (\ref{eq27101}):
\begin{eqnarray}
h_i^{(A)}=\frac{2\pi}{g}\delta_i^{A+1-N_c}-\frac{2\pi}{g}\delta_i^{A+2-N_c},
~~~ A=N_c,\cdots, 2N_c-2.
\label{eq27p50} 
\end{eqnarray} 
Quarks, if they occur, would have only electric charges,
\begin{eqnarray}
q_{i}^{(i_0)}=g\delta_{ii_0}-\frac{g}{N_c}, ~~~~i=1,\cdots,N_c,
\label{eq27p51}
\end{eqnarray}
where $i_0$ labels a fixed $U(1)$ group and the last term 
is necessary in order to satisfy the 
constraint (\ref{eq27p45}). The lattice for the case $N_c=2$ is 
sketched in Fig.\,1.a. The electrically charged particles
lie on the horizontal axis. From (\ref{eq27p49}) the gluons have
charges $0$ (corresponding to diagonal generators), $\pm g$ 
(corresponding to off-diagonal generators). Other particles composed 
of gluons, according to Eq.\,(\ref{eq27p46}), have the charge $\pm 2g$ 
and so on. The quarks, according to Eq.\,(\ref{eq27p51}), have 
charges $\pm g/2$ in the case $N_c=2$. The monopoles lie on the vertical
axis with magnetic charge $h=4\pi /g$ and the electric 
charge $q=0$. Other states on the
vertical axis are the anti-monopole, a pair of monopoles and so on.
All the points which do not belong to the horizontal and vertical axis     
are bound states of the electric and magnetic quanta.

Now we switch on the $\theta$ vacuum term. According to 
Eq.\,(\ref{eq27p37}), the QCD monopoles acquire fractional $U(1)$ charges,
\begin{eqnarray}
q_i^{(A)}=\frac{\theta g^2}{4\pi^2}h_i^{(A)}, ~~~~A=N_c,\cdots, 2N_c-2.
\end{eqnarray}
Consequently, if $\theta\neq 0$ or $2\pi$, the square lattice
shown in Fig.\,1.a. will become oblique (this is why the name 
oblique confinement is applied). Fig.\,1.b shows the lattice corresponding to
$\theta =\pi+\epsilon$, $0 < \epsilon \ll \pi$. 

 From the above discussion, we infer that there are three physical dynamical
scenarios. If one of the purely electrically charged objects is a Lorentz 
scalar, it can develop a non-vanishing vacuum expectation value and then 
the theory is in the Higgs phase. If the field representing a monopole
develops a non-vanishing vacuum expectation value, the quarks will be
confined, and the theory is in the confining phase. If no condensation
occurs, the particle spectrum is given by the lattice, 
and the theory is in the so-called Coulomb phase.
Therefore, all the phases, if they emerge in $SU(N)$ gauge 
theory, can be characterized by designating those points on the charge 
lattice that develop vacuum expectation values.
The quantum in the Schwinger-Zwanziger quantization 
condition for every pair of these points must vanish,
\begin{eqnarray}
 h_i^{(1)}q_i^{(2)}-q_i^{(1)}h_i^{(2)}=0,
\label{eq27p53}
\end{eqnarray}
For the case of $SU(2)$ sketched in Fig.\,1, 
Eq.\,(\ref{eq27p53}) implies that these
points should lie on a straight line passing through the origin.
In the general $SU(N_c)$ case, they can span a linear subspace with the dimension
not larger than $N_c-1$. All particles whose charges lie in this subspace
show only short-range interactions, and their gauge fields are screened by 
the Higgs mechanism. All the particles that have a non-vanishing 
Schwinger-Zwanziger quantum with respect to one of the points on this
subspace are confined.  

Overall, all the possible physical phases of the theory can be 
characterized by the linear spaces spanned by at most
$N$ points on the electric-magnetic charges lattice. (In particular, the pure
Coulomb phase corresponds to choosing only the origin.) Phase transitions 
correspond to moving from one linear subspace to another.

With this electric-magnetic lattice and its phase description, we can
explain what the oblique confinement looks like. Increasing 
$\theta$, we can deform the $\{q,h\}$ lattice in a continuous way.
Suppose we have a confinement mode corresponding to the line $I$ sketched
in Fig.\,1.b., i.e. monopoles in this direction develop expectation
values. If $\theta$ runs from $0$ to $\pi$, 
this line will become continuously more tilted. At $\theta=\pi$, the
other line $II$, corresponding to the parity image of $I$ with respect
to the vertical axis, is also a confinement mode. When $\theta$ is 
very near $\pi$ but not equal to $\pi$, vacuum expectation values of the
monopoles cannot be developed in the direction represented by the 
line $I$ or $II$. This follows from the Schwinger-Zwanziger
quantization condition. It seems that the monopole particles have to 
move collectively and carry large electric charges. However, the monopoles
corresponding to $I$ and $II$ carry opposite electric charges. There is
a possibility that they form a tight bound state, which in turn condenses
and hence develops a vacuum expectation value. The possible direction
in which this condensation happens is the line $III$ as shown in Fig.\,1.b.
This alternative is referred to as oblique confinement 
by 't Hooft \cite{maxag, cardrab}. However, one can see that 
this confinement mode is 
very peculiar. Some of the particles with the external quantum numbers of
fundamental quarks exist in the observable spectrum. Of course, 
the fundamental quarks are confined, since they do not lie on the line
$III$. But the bound state of a quark and a dyon can lie on this line and 
hence is not confined. Since the dyon has no baryon charge, or any other
external quantum numbers, this kind of composite particle has exactly 
the same baryon charge as the fundamental quarks, i.e. 
fractional. In usual QCD, the vacuum angle is empirically
very close to zero, and oblique confinement cannot occur. 
However, in supersymmetric $SO(N_c)$ gauge theory, this dynamical
phenomenon indeed exists.     


\begin{figure}
\begin{center}
\leavevmode
{\epsfxsize=3.00truein \epsfbox{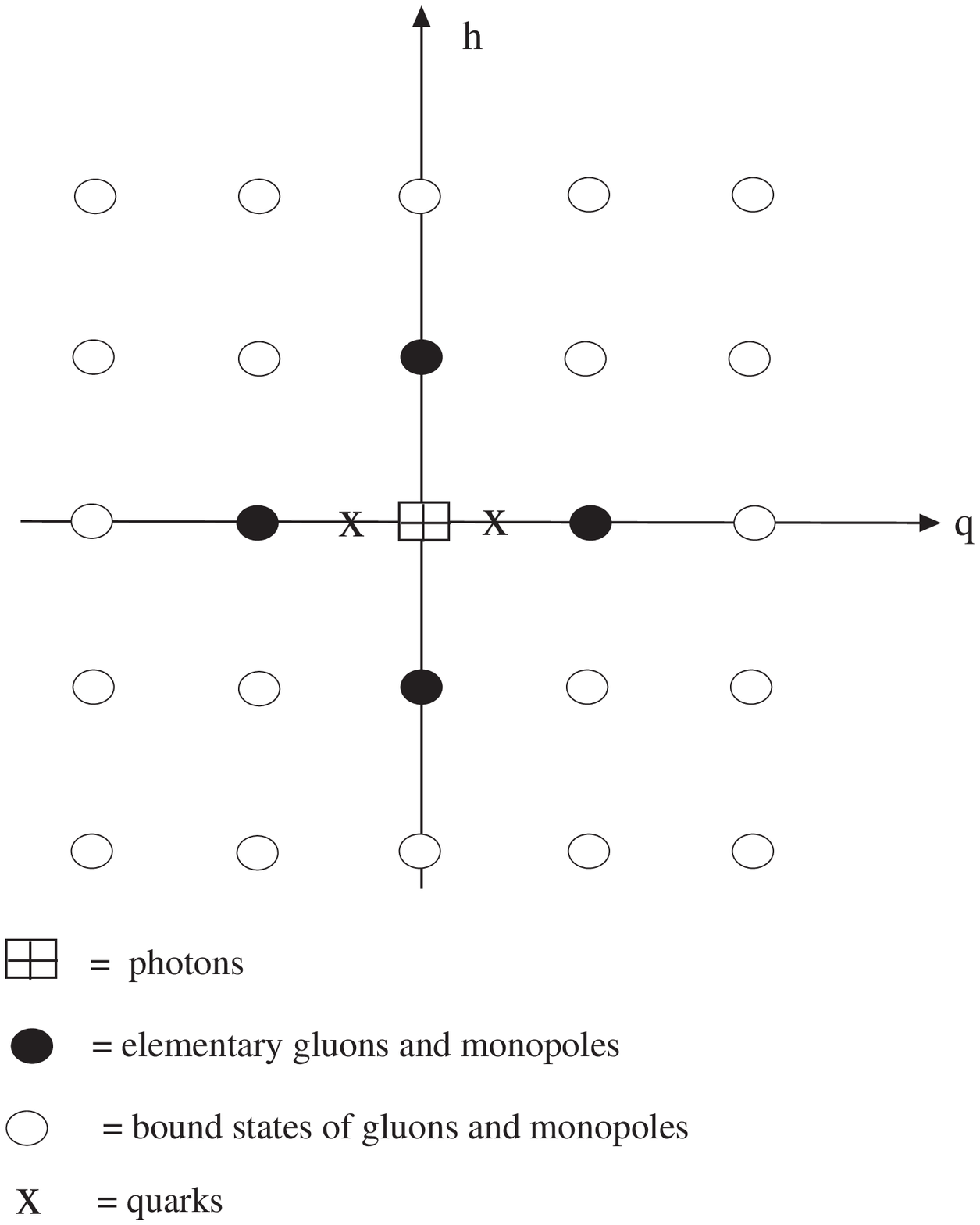}}\hspace{5mm}
{\epsfxsize=3.00truein \epsfbox{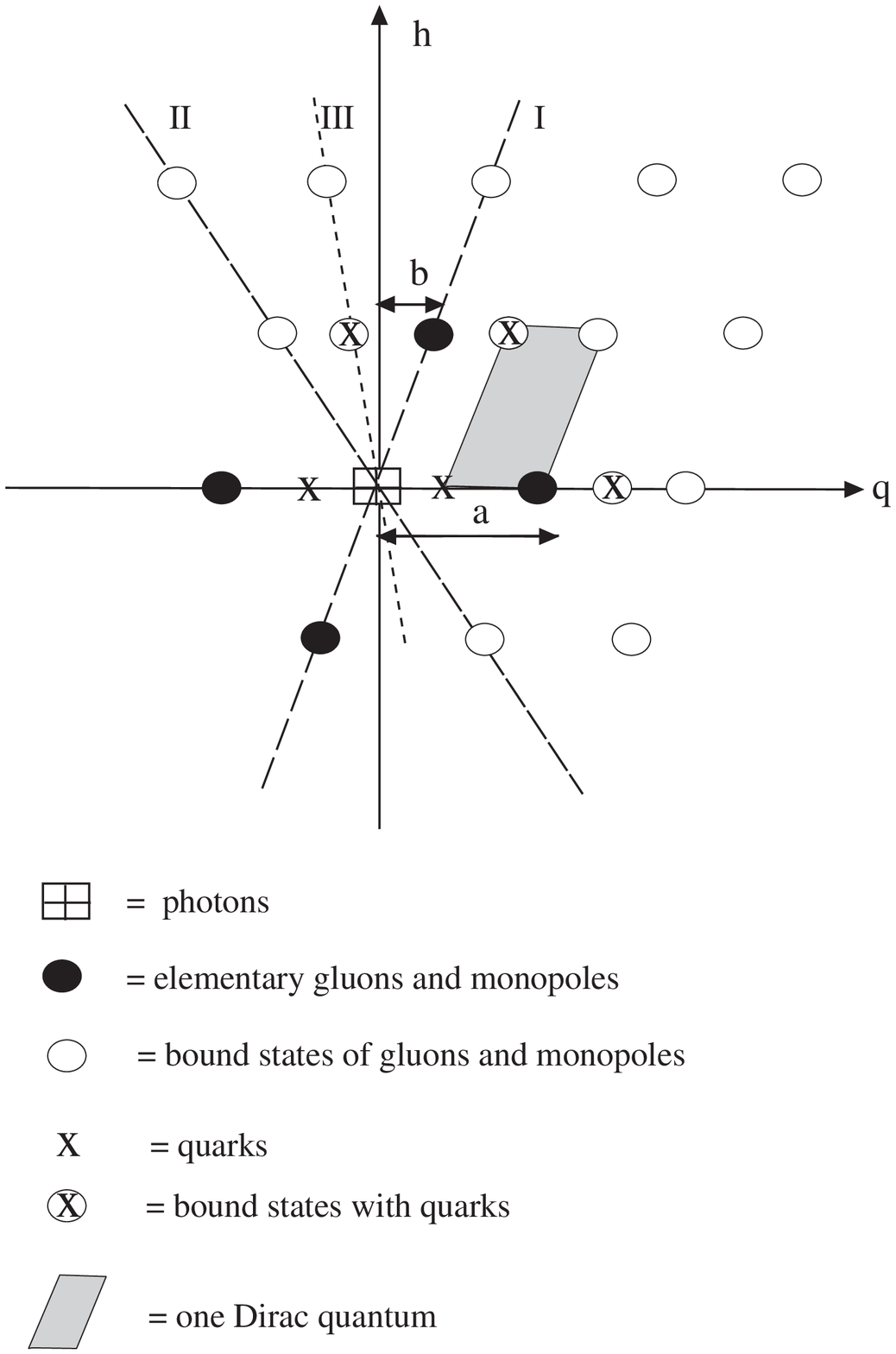}}
\caption{\protect\small Electric-magnetic charge lattice for the
$SU(2)$ case, $q$ = electric charge, $h$ = magnetic charge.}
\end{center}
\end{figure}


\vspace{4mm}
\begin{flushleft}
{\it More order parameters}
\end{flushleft}
\vspace{4mm}

 In the following, we shall introduce two other order parameters, which are
very convenient for describing monopole condensation (quark confinement)
and dyon condensation (oblique confinement).   

The first order parameter is the 't Hooft loop $W_t$, 
which is defined by cutting 
a contour out of the space-time and considering non-trivial 
boundary conditions 
around it. This definition itself has geometric meaning.   
In a fashion similar to the Wilson loop, it can be interpreted as 
the creation and annihilation of a monopole and anti-monopole pair. With
the choice of a rectangular contour, one can define the static potential
between the monopole and anti-monopole at large distance $R$,
\begin{eqnarray}
\lim_{T\to\infty}\langle W_t\rangle =e^{-TV_t(R)}.
\end{eqnarray}
It can be argued from some concrete calculations \cite{ref1p11}  
that the potential
between the monopoles can be classified as that listed in Table \ref{ta2p72}. 

\begin{table}
\begin{center}
\begin{tabular}{|c|c|} \hline
 phase & static potential $V(R)$ \\
       & at large $R$ \\ \hline
 Coulomb &  $ 1/R $ \\ \hline
 free electric & $\ln\left(R\Lambda\right)/R$\\ \hline
 free magnetic & $ 1/[R\ln\left(R\Lambda\right)]$ \\ \hline    
 Higgs & $\rho R$ \\ \hline
 confining & constant \\ \hline
\end{tabular}

\caption{\protect\small Static potential obtained from the 't Hooft loop and 
corresponding phases in gauge theory
($\rho$ is the string tension and $\Lambda$ is the renormalization scale).
\label{ta2p72}}
\end{center}

\end{table}

Comparing the static potential between electrically charged particles obtained
from the Wilson loop with that of magnetic monopoles 
from the 't Hooft loop, one can see that the 
dynamical behaviour of free electric and free magnetic phases are
exchanged. The Higgs phase and the confining phase are also
exchanged. This is in fact a reflection of electric-magnetic duality:
the Wilson loop and the 't Hooft loops are exchanged with the exchange of 
electrically charged particles with magnetic monopoles. In fact, 
the exchange of the Higgs phase and confinement phase is a conjecture 
by Mandelstam and 't Hooft \cite{ref1p8,ref1p8a} based on the 
the exchange of the free electric phase and
the free magnetic phase. As mentioned above, this conjecture provides a 
conceivable mechanism for colour confinement. It can be understood as a
dual Meissner effect produced by the condensation of monopoles, since the Higgs
phase is associated with the condensation of electrically charged particles,
and the magnetically charged particles are confined.

 The Coulomb phase goes into itself under the electric-magnetic duality
transformation. This is the unique self-dual phase. For an Abelian Coulomb 
phase with free photons, this can be easily seen from a standard 
duality transformation. The search for duality in non-Abelian Coulomb 
phase has become very popular in recent years. The original 
conjecture for the existence of this duality was made by Montonen 
and Olive \cite{ref1p6}. Osborn \cite{ref1p7} found that this 
duality can exist in $N=4$ supersymmetric
Yang-Mills theory. Some years ago a major progress was made by 
Seiberg and Witten \cite{ref1p1,ref1p1a}. 
They found that the electric-magnetic duality in the low-energy $N=2$
supersymmetric Yang-Mills theory play a crucial role in understanding
the non-perturbative dynamics.  The electric-magnetic duality
transformation turned out to be the monodromy, i.e. the transformation
of a complex function relevant to the coupling around the singularity
in the Riemann surface of the moduli space of the theory, while the singularities
represent the various particles implied from the duality such as
monopole and dyon etc. In this way an exact coupling of low-energy $N=2$ supersymmetric 
Yang-Mills theory is thus determined and consequently, the full
low-energy quantum effective action (including the non-perturbative contribution)
is given. This effective action is confirmed by some explicit
instanton calculations \cite{ref512}. Furthermore, 
Seiberg has shown that the non-Abelian
electric-magnetic duality can emerge in the infrared fixed point of 
$N=1$ supersymmetric QCD \cite{se2,ref525,ref1p9}. This is the main topic 
we shall discuss in the following sections.

 With the Wilson and the 't Hooft loops, we formally define another
gauge invariant order parameter as the product of them,  
\begin{eqnarray}
W_d=W_w W_t,
\end{eqnarray} 
which is called a dyon loop since it can describe the dynamics of
the particles with both electric and magnetic charges.
This order parameter is particularly suitable for describing 
the oblique confinement
phase. To compare the dynamical behaviour of each phase reflected in 
these parameters, we collect them in Table \ref{ta2p73}.

\begin{table}
\begin{center}
\begin{tabular}{|c|c|c|c|} \hline
 phase &  & dynamical behaviour & \\ \hline 
       & $W$ & $W_t$ & $W_d$ \\ \hline   
Coulomb &   perimeter law & area law & area law\\ \hline
Confinement  &  area law & perimeter law & area law \\ \hline
Oblique confinement & area law & area law &  perimeter law \\ \hline
\end{tabular}

\caption{\protect\small Dynamical law of the phases described by the
order parameters.  \label{ta2p73}}
 \end{center}
\end{table}

\subsection{$N=1$ superconformal algebra and its representation} 
\label{sect8}

\renewcommand{\theequation}{2.5.\arabic{equation}}
\setcounter{equation}{0}

\subsubsection{Current supermultiplet}
\label{subsect8.1}

The algebraic foundation of a superconformal invariant quantum 
field theory is superconformal algebra. To introduce superconformal 
symmetry, a natural way is to start from the supercurrent 
supermultiplet. Like in the derivation of ordinary conformal 
algebra in Subsect. \ref{subsect21}, we exhibit the structure
of the supercurrent multiplet without resorting to 
a particular model and only by demanding that the currents 
and their supersymmetric transformations yield the supersymmetry algebra.
 
According to the Noether theorem, corresponding to supersymmetry
invariance of a relativistic quantum field theory
we have the conserved supersymmetry current $j_{\mu\alpha}(x)$, 
$\partial_{\mu}j^{\mu}{_\alpha}(x)=0$, and the supercharge
\begin{eqnarray}
Q=\int d^3x j_0.
\label{eq9p1}
\end{eqnarray}
The supercharge generates the supersymmetric transformation,
$\delta \phi (x)=i[\overline{\zeta}Q,\phi (x)]$. Here we use four-component 
notation, 
\begin{eqnarray}
Q=\left(\begin{array}{c} Q_\alpha \\ \overline{Q}^{\dot{\alpha}}\end{array} \right),
~~~\overline{Q}=\left(Q^{\alpha},\overline{Q}_{\dot{\alpha}}\right),
~~~\overline{\zeta}Q=\overline{Q}\zeta.
\end{eqnarray} 
In two-component notation,
$\overline{\zeta}Q=\zeta^{\alpha}Q_{\alpha}+\overline{Q}_{\dot{\alpha}}
\overline{\zeta}^{\dot{\alpha}}$. 
The supersymmetric variation of the supersymmetric current
gives
\begin{eqnarray}
\delta j_{\mu}(x)=-i[j_{\mu}(x), \overline{\zeta}Q].
\label{eq9p2}
\end{eqnarray} 
Taking the space integral of the time component of the above supersymmetry
variation and requiring that the $N=1$ supersymmetry algebra 
$\{Q,\overline{Q}\}=2\gamma^{\mu}P_{\mu}$ is reproduced, we have
\begin{eqnarray}
\int d^3x\delta j_{0}(x)=-i[Q, \overline{\zeta}Q]
=-i[Q, \overline{Q}{\zeta}] =-2i\gamma^{\mu}{\zeta}P_{\mu}=
-2i\gamma^{\mu}{\zeta}\int d^3x \theta_{0\mu}.
\label{eq9p4}
\end{eqnarray}
In view of Lorentz covariance, the supersymmetric transformation 
of supersymmetry current $j_{\mu}$ must have the following general form:
\begin{eqnarray}
\delta j_{\mu}=2 \gamma^{\nu}\theta_{\nu\mu}+\partial^{\rho}R_{\mu\rho},
\label{eq9p8}
\end{eqnarray}
where $R_{\rho\mu}=-R_{\mu\rho}$. The second term of (\ref{eq9p8}) 
vanishes
when taking $\mu=0$ and integrating over $\int d^3x$. 
Furthermore, the chiral $U_R(1)$ rotation of the supercharge yields
\begin{eqnarray}
[Q, R] &=& \gamma_5 Q, ~~~R=\int d^3x j_0^{5}, \nonumber\\[2mm]
[\overline{\zeta}Q, \int d^3x j_0^{5}] &=& -i\int d^3x \delta j_0^{5}
=\overline{\zeta}{\gamma}_5 Q=\overline{\zeta}{\gamma}_5 \int d^3x j_0.
\label{eq9p}
\end{eqnarray}
Hence
\begin{eqnarray}
\delta j_\mu^{5}=i\gamma_5j_{\mu}+\partial^{\nu}r_{\nu\mu},
\label{eq9p9}
\end{eqnarray}
with $r_{\nu\mu}=-r_{\mu\nu}$.  
(\ref{eq9p8}) and (\ref{eq9p9}) imply that the $R$-current 
$j_\mu^{5}$, the supersymmetry current $j_{\mu\alpha}$ and the energy-momentum 
tensor $\theta_{\mu\nu}$ belong to a supermultiplet
\begin{eqnarray} 
(\theta_{\mu\nu},j_{\mu},j_\mu^{5}). 
\label{eq9px}
\end{eqnarray}
At quantum level, each member of the above supermultiplet will become 
anomalous. It was shown that the trace anomaly $\theta^{\mu}_{~\mu}$,
the $\gamma$-trace anomaly of the supersymmetry current, $\gamma^{\mu}j_\mu$,
and the chiral anomaly of the $R$-current, $\partial^{\mu}j_\mu^{5}$, 
lie in a chiral supermultiplet, $\Phi{\equiv}\left(\gamma^{\mu}j_\mu,
\theta^{\mu}_{~\mu}, \partial^{\mu}j_\mu^{5}\right)$ \cite{so, ref521}. 

In a concrete classical superconformal invariant field theory 
-- the massless Wess-Zumino model -- the explicit but model independent
supersymmetry transformations among the members of the current supermultiplet 
(\ref{eq9px}) are as follows:
\begin{eqnarray}
\delta j_{\mu}&=&-2i\gamma^{\nu} {\zeta}\theta_{\mu\nu}+i\gamma^{\nu}\gamma_5
{\zeta} \left(\partial_{\nu}j^{5}_{\mu}
-\eta_{\mu\nu}\partial^{\rho}j^{5}_{\rho}\right)
+\frac{1}{2}i\epsilon_{\mu\nu\rho\sigma}\gamma^{\nu}{\zeta}\partial^{\rho}
j^{5{\sigma}},
\nonumber\\
\delta j^{5}_{\mu}&=&i\overline{\zeta}\gamma_5j_{\mu}
-\frac{i}{3}\overline{\zeta}\gamma_5\gamma_{\mu}\gamma^{\nu}j_{\nu},\nonumber\\
\delta \theta_{\mu\nu}&=&\frac{1}{4}i\overline{\zeta}
(\sigma_{\mu\rho}\partial^{\rho}
j_{\nu}+\sigma_{\nu\rho}\partial^{\rho}j_{\mu}).
\label{eq9p30}
\end{eqnarray}
Consequently, the supersymmetry transformation of  
the members of the anomaly multiplet is
\begin{eqnarray}
\delta\left(-i\frac{1}{3}\gamma^{\mu}j_{\mu}\right)&=&-\left(\frac{2}{3}
\theta_{~\mu}^{\mu}-i\gamma_5 \partial^{\mu}j_{\mu}^{5}\right){\zeta},
\nonumber\\
\delta \left(\frac{2}{3}\theta_{~\mu}^{\mu}\right)&=&i\overline{\zeta}
\gamma^{\nu}\partial_{\nu}\left(-i\frac{1}{3}\gamma^{\mu}j_{\mu}\right),
\nonumber\\
\delta (\partial^{\mu}j_{\mu}^{5})&=&\overline{\zeta}\gamma_5
\gamma^{\nu}\partial_{\nu}\left(-i\frac{1}{3}\gamma^{\mu}j_{\mu}\right).
\label{eq9p30a}
\end{eqnarray} 
Like in the non-supersymmetric case discussed in 
Sect.\,\ref{subsect21}, if we require that the conformal symmetry is preserved
at the quantum level, $\Phi=0$, i.e.  
\begin{eqnarray}
\theta_{~\mu}^{\mu}=0, ~~~\gamma^{\mu}j_{\mu}=0,~~~
\partial^\mu j_\mu^{5}=0,
\label{eq9p9a}
\end{eqnarray}
then the supersymmetric transformation (\ref{eq9p30}) 
reduces to the form of the naive supersymmetry
transformations (\ref{eq9p8}) and (\ref{eq9p9}),
\begin{eqnarray}
\delta j_{\mu}&=&-2i \gamma^{\nu}{\zeta}\theta_{\mu\nu}
+\gamma^{\nu}\gamma_5{\zeta}\partial_{\nu}j_{\mu}^{5}+\frac{i}{2}
\epsilon_{\mu\nu\rho\sigma}\gamma^{\nu}{\zeta}\partial^{\rho}j^{5{\sigma}},
\nonumber\\
\delta \theta_{\mu\nu}&=&\frac{1}{4}i\overline{\zeta}
(\sigma_{\mu\rho}\partial^{\rho}j_{\nu}+\sigma_{\nu\rho}
\partial^{\rho}j_{\mu}),
~~~\delta j_{\mu}^{5}=i\overline{\zeta}\gamma_5 j_{\mu}.
\label{eq9p33}
\end{eqnarray}
In particular, with the vanishing of the quantum anomalies (\ref{eq9p9a}),
we have not only the conserved currents $d_{\mu}$ and $k_{\mu\nu}$ and their
charges $D$ and $K_{\mu}$ shown in (\ref{eq2a19x}),  
but also a new conserved  fermionic current
\begin{eqnarray} 
s_{\mu}&{\equiv}&ix^{\nu}\gamma_{\nu}j_{\mu}, ~~
\partial^{\mu}s_{\mu}=i\gamma^{\mu}j_{\mu}+ix^{\nu}\gamma_{\nu}\partial^{\mu}
j_{\mu}=0,
\label{eq9p10a}
\end{eqnarray}
and the corresponding supercharge
\begin{eqnarray} 
S{\equiv}\int  d^3x s_0.
\label{eq9p10}
\end{eqnarray}
$S$ is called the generator of special supersymmetry 
transformations. Like the supersymmetric charge, $S$ is a Majorana spinor, 
\begin{eqnarray} 
S=\left(\begin{array}{c} S_{\alpha} \\ 
\overline{S}^{\dot{\alpha}}\end{array}\right), ~~~
\overline{Q}=\left(S^{\alpha},\overline{S}_{\dot{\alpha}}\right).
\end{eqnarray}

\subsubsection{$N=1$ superconformal algebra }
\label{subsub8.3}

In this section, we first derive 
the whole superconformal algebra and then 
work out in detail the representation of the superconformal 
algebra following Ref.\,\cite{so}.
In particular, we shall introduce the relation between
the conformal dimension and the $R$-charge of a chiral
superfield, which plays a crucial role in determining
the conformal window in $N=1$ supersymmetric gauge theory.   

Like the ordinary conformal algebra, the superconformal 
algebra can be derived directly from the transformation 
property of the current given in (\ref{eq9p33}). We first get
\begin{eqnarray}
\left[Q,\overline{Q}\zeta \right]&=&[\int d^3x j_0, \overline{Q}\zeta]
=i \int d^3x \delta j_0\nonumber\\
&=&2\gamma^{\mu}\zeta \int d^3x\theta_{0\mu}+i\gamma^i\gamma_5\zeta\int d^3x
\partial^i j_{0}^{5}\nonumber\\
&&-i\gamma^0\gamma_5\zeta\int d^3x\partial^i j_{i}^{5}
-\frac{1}{2}{\epsilon}^{ijk}\gamma_i\zeta\int d^3x\partial_j j_{k}^{5}
\nonumber\\
&=&2{\gamma}^{\mu}{\zeta} \int d^3x{\theta}_{0\mu}=2{\gamma}^{\mu}{\zeta} 
P_{\mu}, \label{eq9p55} \\[2mm]
[P_{\mu},Q]&=&[\int d^3x{\theta}_{0\mu},\overline{\zeta}Q]=\int d^3x{\delta}
{\theta}_{0\mu}
=\frac{1}{4}i\overline{\zeta}\int d^3x\left(\sigma_{0\rho}\partial^{\rho}j_{\mu}
+\sigma_{\mu\rho}\partial^{\rho}j_0\right)
\nonumber\\
&=&\frac{1}{4}i\overline{\zeta}\int d^3x\left(\sigma_{0i}\partial^{i}j_{\mu}
+\sigma_{\mu i}\partial^{i}j_{0}-\sigma_{\mu 0}\partial^{i}j_{i}\right)=0,
\label{eq9p56}\\[2mm]
[M_{\mu\nu},\overline{\zeta} Q]&=& i\int d^3x \left(x_{\mu}\delta\theta_{\nu 0}
-x_{\nu}\delta\theta_{\mu 0}\right)\nonumber\\
&=& -\frac{1}{4}\overline{\zeta}\int d^3x\left[ (\sigma_{\nu 0}j_\mu-
\sigma_{\nu\mu}j_0-\sigma_{0\mu}j_\nu -(\mu{\longleftrightarrow}\nu)\right]
=-\frac{1}{2}\overline{\zeta}\sigma_{\nu\mu} Q,
\label{eq9p59}\\[2mm]
[R,\overline{\zeta} Q]&=&[\int d^3x j_0^{5},\overline{\zeta} Q]
=i\int d^3x \delta j_0^{5}=-\overline{\zeta}\gamma_5 \int d^3x j_0
=-\overline{\zeta}\gamma_5 Q,
\label{eq9p62}
\end{eqnarray}
where we have used
\begin{eqnarray}
&&\int d^3x \partial^i(\mbox{anything})=0,~~i=1,2,3; ~~\sigma_{\mu\nu}
=-\sigma_{\nu\mu}, ~~\partial^{\mu}j_{\mu}=\partial^{\mu}j_{\mu}^{5}= 0,
\nonumber\\
&&x_{\mu}\sigma_{\nu\rho}\partial^{\rho}j_0
=x_{\mu}\left(\sigma_{\nu 0}\partial^{0}j_0+\sigma_{\nu i}\partial^{i}j_0
\right)
=x_{\mu}\partial^{i}\left(\sigma_{\nu i}j_0-\sigma_{\nu 0}j_i\right).
\label{eq9p60}
\end{eqnarray}
Eqs.\,(\ref{eq9p55})-(\ref{eq9p62}) give the fermionic part of the
super-Poincar\'{e} algebra:
\begin{eqnarray}
[Q, M_{\mu\nu}]=\frac{1}{2}\sigma_{\mu\nu}Q,~~[Q, P_{\mu}]=0,~~
\{Q,\overline{Q}\}=2\gamma^{\mu}P_{\mu}, ~~[Q,R]=\gamma_5Q.
\label{eq9p61}
\end{eqnarray}
To obtain the whole superconformal algebra. we only
need to calculate two new commutation relations, $[Q,K_\mu]$ and $[Q,D]$. 
The others can be determined from  the Jacobi identity. 
Thus, we have 
\begin{eqnarray}
[\overline{\zeta} Q,K_{\mu}]&=&[\overline{\zeta} Q,\int d^3xk_{\mu 0}]
=-i\int d^3x \delta k_{\mu 0}\nonumber\\
&=&-i\int d^3x (2x_{\mu}x^{\nu}
\delta \theta_{\nu 0}-x^2\delta \theta_{\mu 0})
=\overline{\zeta}\gamma_{\mu}S,
\label{eq9p63}\\[2mm]
[\overline{\zeta} Q,D]&=&[\overline{\zeta} Q,\int d^3xd_{0}]
=-i\int d^3x \delta d_{0}=-i\int d^3x x^{\nu}\delta\theta_{0\nu} \nonumber\\
&=&-\frac{1}{4}\overline{\zeta}\int d^3x \left[\eta^{i\nu}(\sigma_{0i}j_{\nu}
+\sigma_{\nu i}j_0-\sigma_{\nu 0}j_i)\right]\nonumber\\
&=&-\frac{1}{2}\overline{\zeta}\int d^3x\sigma_{0\nu}j^{\nu}
=-\frac{1}{2}\overline{\zeta}\int d^3x i\gamma_0\gamma_ij^i\nonumber\\
&=&\frac{i}{2}\overline{\zeta}\int d^3x j_0=\frac{1}{2}i\overline{\zeta}Q,
\label{eq9p63x}
\end{eqnarray}
where the condition $\gamma^{\mu}j_{\mu}=0$ was used. 
Eqs.\,(\ref{eq9p63}) and (\ref{eq9p63x}) yield new commutation relations,
\begin{eqnarray}
[Q,R]=\gamma_5Q, ~~[Q,K_{\mu}]=\gamma_{\mu}S,~~[Q,D]=\frac{i}{2}Q.
\label{eq9p64b}
\end{eqnarray}
In addition, since the $R$-symmetry is only related to super-coordinate
rotations, it actually belongs to the internal symmetries. Thus the generator
of $R$-symmetry must commute with the generators of the ordinary space-time
transformation,
\begin{eqnarray}  
[R,P_{\mu}]=[R,M_{\mu\nu}]=[R,D]=[R,K_{\mu}]=0.
\label{eq9p64}
\end{eqnarray}
With the algebraic relations listed in (\ref{eq9p61}),
(\ref{eq9p64b}) and (\ref{eq9p64}), the Jacobi identities
$(S,M,K)$, $(Q,P,K)$, $(Q,D,K)$, $(Q,K,K)$ and $(Q,K,R)$ yield the following
commutation relations, respectively,
\begin{eqnarray}
[S,M_{\mu\nu}]=\frac{1}{2}\sigma_{\mu\nu}S,~~[S,P_{\mu}]=\gamma_{\mu}Q,
~~[S,D]=-\frac{i}{2}S,~~[S,K_{\mu}]=0, ~~[S,R]=-\gamma_5S.
\end{eqnarray}
As an illustrative example, consider the Jacobi identity $(S,M,K)$. We have
\begin{eqnarray}
&&[[Q,M_{\mu\nu}],K_{\rho}]+[[K_{\rho},Q],M_{\mu\nu}]+[[M_{\mu\nu},K_{\rho}],Q]
=0,\nonumber\\
&&\frac{1}{2}\sigma_{\mu\nu}[Q,K_{\rho}]-\gamma_{\rho}[S,M_{\mu\nu}]-
[i(\eta_{\rho\mu}K_{\nu}-\eta_{\rho\nu}K_{\mu}), Q]=0,\nonumber\\
&& \frac{1}{2}\sigma_{\mu\nu}\gamma_{\rho}S-\gamma_{\rho}[S,M_{\mu\nu}]
+i\eta_{\rho\mu}\gamma_{\nu}S-i\eta_{\rho\nu}\gamma_{\mu}S=0,\nonumber\\
&&\frac{i}{4}\gamma_{\rho}[\gamma_{\mu},\gamma_{\nu}]S
-\gamma_{\rho}[S,M_{\mu\nu}]=0,\nonumber\\
&&[S,M_{\mu\nu}]=\frac{1}{2}\sigma_{\mu\nu}S,
\label{eq9p65}
\end{eqnarray}
where the $\gamma$-algebra operation $\gamma_{\mu}\gamma_{\nu}\gamma_{\rho}
=2\gamma_{\mu}{\eta}_{\nu\rho}-2{\eta}_{\mu\rho}
\gamma_{\nu}+\gamma_{\rho}\gamma_{\mu}\gamma_{\nu}$ was used.

The anticommutators $\{S,\overline{Q}\}$ (or equivalently $\{Q,\overline{S}\}$)
 and $\{S,\overline{S}\}$ need some special considerations. Since both
$S$ and $Q$ are fermionic generators, $\{S,\overline{Q}\}$
must be proportional to bosonic generators, 
so it must be of the following general form,
\begin{eqnarray}
\{S,\overline{Q}\}=a D+b\gamma^{\mu}P_{\mu}+c\sigma^{\mu\nu}M_{\mu\nu}
+d\gamma^{\mu}K_{\mu}+e R. 
\label{eq9p70}
\end{eqnarray}
The Jacobi identities $(\overline{Q},S,D)$; $(\overline{Q},S,P)$ and 
$(\overline{Q},S,\overline{Q})$ fix the indefinite coefficients
as $b=d=0$, $a=2i$, $d=1$ and $e=3{\gamma}_5$.
Hence we finally obtain  
\begin{eqnarray}
\{\overline{Q},S\}=2iD+\sigma^{\mu\nu}M_{\mu\nu}+3{\gamma}_5R.
\label{eq9p74}
\end{eqnarray}
Similarly one shows that  
\begin{eqnarray}
\{S, \overline{S}\}=2\gamma^{\mu}K_{\mu}.
\label{eq9p75}
\end{eqnarray}

The fermionic part of the whole 
$N=1$ superconformal algebra consists of the collection
of the above commutation relations,
\begin{eqnarray}
&&[Q, M_{\mu\nu}]=\frac{1}{2}\sigma_{\mu\nu}Q, ~~ [S, M_{\mu\nu}]
=\frac{1}{2}\sigma_{\mu\nu}S, \nonumber\\
&&[Q, D]=\frac{1}{2} i Q, ~~[S, D]=-\frac{1}{2} i S,\nonumber\\
&&[Q, P_{\mu}]=0,  ~~[S, P_{\mu}]=\gamma_{\mu}Q,\nonumber\\
&&[Q, K_{\mu}]=\gamma_{\mu}S, ~~[S, K_{\mu}]=0, \nonumber\\
&&[Q, R]=\gamma_5 Q,  ~~[S, R]=-\gamma_5 S,\nonumber\\
&& [R, M_{\mu\nu}]= [R,P_{\mu}]= [R,D]=[R,K_{\mu}]=0 ,\nonumber\\
&& \{Q,\overline{Q}\}=2\gamma^{\mu}P_{\mu}, ~~\{S,\overline{S}\}=2\gamma^{\mu}K_{\mu},
\nonumber\\
&&\{S,\overline{Q}\}=2iD+\sigma^{\mu\nu}M_{\mu\nu}+3\gamma_5  R.
\label{eq9p76}
\end{eqnarray}
The $N=1$ superconformal algebra is
isomorphic to $SU(2,2|1)$. 
For $N$-extended supersymmetry,
the supersymmetric algebra is isomorphic to $SU(2,2|N)$.
Furthermore, defining a conformal spinor,
\begin{eqnarray}
\Sigma {\equiv}\left(\begin{array}{c} Q_{\alpha} \\ 
\overline{S}^{\dot{\alpha}}\end{array}\right), ~~
\overline{\Sigma}=(S^{\alpha},\overline{Q}_{\dot{\alpha}}),
\label{eq9p79}
\end{eqnarray}
and
\begin{eqnarray}
&&\sigma_{ab}{\equiv}(\sigma_{\mu\nu}, \sigma_{\mu 5},\sigma_{\mu 6},
\sigma_{56}),~~a,b=0,{\cdots},3,5,6,\nonumber\\
&&\sigma_{\mu 5}=\gamma_{\mu}\gamma_5,~~\sigma_{\mu 6}=\gamma_{\mu},~~
\sigma_{56}=\gamma_5.
\label{eq9p80}
\end{eqnarray}
one can write (\ref{eq9p76}) in the form of the fermionic part of an $SU(2,2|1)$ algebra:
\begin{eqnarray}
&& [\Sigma, M_{ab}]=\frac{1}{2}\sigma_{ab}\Sigma;
 ~~ [\overline{\Sigma}, M_{ab}]=-\frac{1}{2}\overline{\Sigma}\sigma_{ab};\nonumber\\
&& [\Sigma, R]=\Sigma;~~ [\overline{\Sigma}, R]=-\overline{\Sigma};
~~[M_{ab},R]=0;\nonumber\\
&& \{\Sigma,\Sigma\}=\{\overline{\Sigma},\overline{\Sigma}\}=0;~~\{\Sigma,\overline{\Sigma}\}
=\sigma^{ab}M_{ab}-3R.
\label{eq9p81}
\end{eqnarray} 

\subsubsection{Representations of $N=1$ superconformal symmetry}
 \label{subsect8.4}

The method of finding a representation  
of the $N=1$ superconformal algebra on field operators
is the same as for the ordinary
conformal algebra, i.e. using Wigner's little group method. 
First, the commutation relation
$[Q,K_{\mu}]=\gamma_\mu S$
shows that the conformal spinor charge $S$ comes from the commutator
of the supercharge $Q$ with the special conformal transformation generators 
$K_{\mu}$. Thus it is possible to construct a superconformal multiplet
from a multiplet of Poincar\'{e} supersymmetry by working out the 
transformation of the fields under the special conformal transformations
generated by $K_\mu$. According to (\ref{eq2p1.31}), the transformation of 
the field $\phi$ under a special conformal transformation is  
\begin{eqnarray}
[K_{\mu},{\phi}(x)]
=i(2x_{\mu}x_{\nu}{\partial}^{\nu}-x^2{\partial}_{\mu}){\phi}(x)
+(k_{\mu}+2ix_{\mu}d+2x^{\nu}\Sigma_{\mu\nu}){\phi}(x).
\label{eq9p83}
\end{eqnarray} 
As shown in Sect.\,\ref{sect2}, 
in the field operator representation of the ordinary conformal 
algebra, there is no restriction on the little group representation
${\kappa}_{\mu}$, $d$ and $\Sigma_{\mu\nu}$ except that $d$ must be a number
due to $[D, M_{\mu\nu}]=0$. However, 
in the superconformal algebra, 
the situation is different: 
there is an important constraint for $\kappa_{\mu}$ coming from
the relation $[Q,K_{\mu}]=\gamma_\mu S$, thus a $\gamma_{\mu}$ should
 be ``separated out'' from
the representation of the commutator $[Q,K_{\mu}]$.
This constraint will restrict the possible little group representations 
on the components of a superconformal multiplet.

The little group of the superconformal algebra can still be found
by requiring that $x=0$ stays invariant. Then we see
that the little group is composed of 
not only the Lorentz group, scale transformations
and the special conformal transformations, but also of a $U_R(1)$ group
due to the algebraic relations 
$[R, M_{\mu\nu}]=[R,D]=[R,K_{\mu}]=0$.
Another difference between the representations of the superconformal
algebra and the ordinary conformal algebra is that owing to the 
supersymmetry, the representation of the superconformal algebra must be realized
on a supermultiplet. This is unlike the ordinary conformal algebra, where
only one type of field is enough. For the superconformal 
algebra, several kinds 
of fields such as (pseudo-)scalar fields, spinor fields and vector fields 
are required due to supersymmetry. Thus we consider the most general
supermultiplet, which can be constructed by starting from
a complex field $C(x)$ and performing the famous ``seven steps" 
introduced in Ref.\,\cite{so}. Since the concrete construction of this general
supermultiplet is very lengthy, 
we shall not repeat it explicitly. The basic method
of finding the chiral multiplet is illustrated detaily in Ref.\,\cite{so}.
Here, we only display this general multiplet and the 
supersymmetry transformations among 
its component fields \cite{so},
\begin{eqnarray}
G(x)&=&\left(C(x),{\chi}(x),M(x),N(x),A_{\mu}(x),{\lambda}(x),D(x)\right),
\nonumber\\
\delta G(x) &=& -i[G(x),\overline{\zeta} Q]=-i[G(x),\overline{Q}{\zeta}],\nonumber\\
\delta C(x) &= &-i\overline{\zeta}\gamma_5 {\chi}(x),\nonumber\\
\delta {\chi}(x) &= &(M(x)-i\gamma_5 N(x)){\zeta}
-i\gamma^{\mu}(A_{\mu}(x)-i\gamma_5 \partial_{\mu}C(x)){\zeta},\nonumber\\
\delta M(x) &= &\overline{\zeta}({\lambda}(x)-i\gamma^{\mu}\partial_{\mu}{\chi}(x)),
\nonumber\\
\delta N(x) &= &-i\overline{\zeta}\gamma_5 ({\lambda}(x)-i\gamma^{\mu}
\partial_{\mu}{\chi}(x)),\nonumber\\
\delta A_{\mu}(x)&=&i\overline{\zeta}\gamma_{\mu}{\lambda}(x)
+\overline{\zeta}\partial_{\mu}{\chi}(x),\nonumber\\
\delta {\lambda}(x)&=&-i \sigma^{\mu\nu}{\zeta}\partial_{\mu}A_{\nu}(x)
+i\gamma_5 {\zeta}D(x),\nonumber\\
\delta D(x)&=&-\overline{\zeta}\gamma^{\mu}\partial_{\mu}\gamma_5 {\lambda}(x).
\label{eq9p86}
\end{eqnarray}
Consequently, the supersymmetry transformations for the fields at 
the origin ($x=0$) should be the following:
\begin{eqnarray}
G(0)&=&(C(0),{\chi}(0),M(0),N(0),A_{\mu}(0),{\lambda}(0),D(0)),\nonumber\\
\delta G(0) &=& -i[G(0),\overline{\zeta} Q]=-i[G(0),\overline{Q}{\zeta}],\nonumber\\
\delta C(0)&=&-i\overline{\zeta}\gamma_5 {\chi}(0),\nonumber\\
\delta {\chi}(0)&=&(M(0)-i\gamma_5 N(0)){\zeta}
-i\gamma^{\mu}A_{\mu}(0){\zeta},\nonumber\\
\delta M(0)&=&\overline{\zeta}{\lambda}(0),
\nonumber\\
\delta N(0)&=&-i\overline{\zeta}\gamma_5 {\lambda}(0),\nonumber\\
\delta A_{\mu}(0)&=&i\overline{\zeta}\gamma_{\mu}{\lambda}(0),\nonumber\\
\delta {\lambda}(0)&=& i\gamma_5 {\zeta}D(0),\nonumber\\
\delta D(0)&=& 0.
\label{eq9p87}
\end{eqnarray}
We first find the representation of the little group on this 
supermultiplet. Since this supermultiplet is generated from $C(x)$ 
through a series of successive supersymmetry transformations,
as the first step, we define the action of the generators of
the little group on $C(0)$,
\begin{eqnarray}
[C(0),R] &=& n \,C(0),\nonumber\\[2mm]
[C(0),D] &=& i d \,C(0),\nonumber\\[2mm]
[C(0),M_{\mu\nu}] &=& {\Sigma}_{\mu\nu}C(0).
\label{eq9p88}
\end{eqnarray}
Then, translating $C(0)$ from the origin according to 
$C(x)=\mbox{exp}(-ix^{\mu}P_{\mu})C(0)\mbox{exp}(ix^{\mu}P_{\mu})$,
 we obtain 
\begin{eqnarray}
[C(x),R] &=& nC(x),\nonumber\\[2mm]
[C(x),D] &=& ix^{\mu}{\partial}_{\mu}C(x)+idC(x),\nonumber\\[2mm]
[C(x),M_{\mu\nu}] &=& i(x_{\mu}{\partial}_{\nu}-x_{\nu}{\partial}_{\mu})C(x)
+{\Sigma}_{\mu\nu}C(x).
\label{eq9p89}
\end{eqnarray}
The algebraic relations $[R,M_{\mu\nu}]=[D,M_{\mu\nu}]=0$ require
\begin{eqnarray}
[n,{\Sigma}_{\mu\nu}]=[d, {\Sigma}_{\mu\nu}]=0.
\label{eq9p91}
\end{eqnarray}
Since $\Sigma_{\mu\nu}$ is, by hypothesis, an irreducible representation,
according to Schur's lemma, $n$ and $d$ must be numbers.
They are the conformal dimension and the $R$-charge
of $C(x)$. The conformal dimensions and $R$-charges for 
other fields can be obtained from (\ref{eq9p87}), (\ref{eq9p88}),
the superconformal algebra (\ref{eq9p76}) and Jacobi identities. 
For example, using the Jacobi identities $(C,R,Q)$ and 
$({\chi},R,\overline{Q})$, we have
\begin{eqnarray}
&& [[C(0),R],Q]+[[Q,C(0)],R]+[[R,Q],C(0)]=0, \nonumber\\
&& [n C(0),Q]+[-{\gamma}_5{\chi}(0), R]+[-{\gamma}_5 Q,C(0)]=0, \nonumber\\
&& {\gamma}_5[{\chi}(0), R]=(n+{\gamma}_5)[C(0),Q]
=(n+{\gamma}_5){\gamma}_5{\chi}(0),\nonumber\\
&&[{\chi}(0), R]=(n+{\gamma}_5){\chi}(0);
\label{eq9p92}
\end{eqnarray}         
\begin{eqnarray}
&& [\{{\chi} (0) ,\overline{Q}\},R]+\{[R,{\chi}(0)],\overline{Q}\}
-\{[\overline{Q},R],{\chi}(0)\}=0, \nonumber\\
&& [iM(0)+{\gamma}_5 N(0)+\gamma^{\mu}A_{\mu}(0),R]
+\{-(n+{\gamma}_5){\chi}(0), \overline{Q}\}+\{-{\gamma}_5\overline{Q},{\chi}(0)\}
 =0, \nonumber\\
&& i[M(0),R]+{\gamma}_5[N(0), R]+\gamma^{\mu}[A_{\mu}(0),R]\nonumber\\
&&=(n+2{\gamma}_5)[iM(0)+{\gamma}_5 N(0)+\gamma^{\mu}A_{\mu}(0),R], \nonumber\\
&&[M(0),R]=(n+2{\gamma}_5)M(0);\nonumber\\
&&[N(0), R]=(n+2{\gamma}_5)N(0);\nonumber\\
&&[A_{\mu}(0),R]=(n+2{\gamma}_5)A_{\mu}(0).
\label{eq9p93}
\end{eqnarray}  
(\ref{eq9p92}) and (\ref{eq9p93}) 
show that the $R$-charges of $\chi$ and  $M$, $N$, $A_{\mu}$ 
are $(n+{\gamma}_5)$ and $(n+2{\gamma}_5)$, respectively. Similarly,
the Jacobi identities  $(M,R,Q)$, $(N,R,Q)$, 
$(A_{\mu},R,Q)$, $({\lambda},R,\overline{Q})$ and $(D,R,Q)$ yield the 
$R$-charges of ${\chi}(0)$, $M(0)$ etc. If we write $G(0)$
as a column vector,
\begin{eqnarray}
G(0)=\left(\begin{array}{c} C(0)\\{\chi}(0)\\M(0)\\N(0)\\A_{\mu}(0)
\\{\lambda}(0)\\D(0)\end{array}\right),
\label{eq9p94}
\end{eqnarray}      
the $R$-charges of the component field operators can be written as
a diagonal matrix,
\begin{eqnarray}
\widetilde{n}=n\, {\bf 1}+{\gamma}_5\,\mbox{diag}\left(0,1,2,2,2,3,4\right).
\label{eq9p95}
\end{eqnarray}
where ${\bf 1}$ denotes the $7{\times}7$ unit matrix.
The conformal dimensions of the component field operators can be worked out  
in the same way. For examples, from the Jacobi identities $(C,D,Q)$ 
and $(C,D,\overline{Q})$, we have
\begin{eqnarray}
&& [[C(0),D],Q]+[[Q,C(0)],D]+[[D,Q],C(0)]=0, \nonumber\\
&& [id C(0),Q]+[{\gamma}_5{\chi}(0), D]+[-\frac{i}{2}Q,C(0)]=0, 
\nonumber\\
&& {\gamma}_5[{\chi}(0), D]=i(d+\frac{1}{2})[C(0),Q]
=i(d+\frac{1}{2}){\gamma}_5{\chi}(0),\nonumber\\
&&[{\chi}(0), D]=i(d+\frac{1}{2}){\chi}(0);
\label{eq9p96}
\end{eqnarray}         
\begin{eqnarray}
&& [\{{\chi} (0) ,\overline{Q}\},D]+\{[D,{\chi}(0)],\overline{Q}\}
-\{[\overline{Q},D],{\chi}\}=0, \nonumber\\
&& [iM(0)+{\gamma}_5 N(0)+\gamma^{\mu}A_{\mu}(0),D]
+\{i(d+\frac{1}{2}){\chi}(0), \overline{Q}\}+\{i\frac{1}{2} \overline{Q},{\chi}(0)\}
 =0, \nonumber\\
&& i[M(0),D]+{\gamma}_5[N(0), D]+\gamma^{\mu}[A_{\mu}(0),D]\nonumber\\
&& =i(d+1)[iM(0)+{\gamma}_5 N(0)+\gamma^{\mu}A_{\mu}(0),R], \nonumber\\
&&[M(0),R]=i(d+1)M(0),~~[N(0), R]=i(d+1)N(0),\nonumber\\
&&[A_{\mu}(0),D]=i(d+1)A_{\mu}(0).
\label{eq9p97}
\end{eqnarray}  
Thus, the conformal dimensions of $\chi$ and $M$, $N$, $A_{\mu}$ 
are respectively $(d+1/2)$ and $(d+1)$. Furthermore,
the Jacobi identities  $(M,D,Q)$, $(N,D,Q)$, 
$(A_{\mu},D,Q)$, $({\lambda},D,\overline{Q})$, $(D,D,Q)$ give the 
conformal dimensions of other component fields such as
 ${\chi}(0)$, $M(0)$ etc. Therefore, the conformal dimension of
the supermultiplet $G(0)$ is
\begin{eqnarray}
\widetilde{d}=d\, {\bf 1}
+\mbox{diag}\left(0,\frac{1}{2},1,1,1,\frac{3}{2},2\right).
\label{eq9p98}
\end{eqnarray}

Working out the matrix representation 
${\kappa}_{\mu}$ of $K_{\mu}$ on the supermultiplet is more difficult.
The relation $[K_{\mu},D]=-iK_{\mu}$ requires that ${\kappa}_{\mu}$
should satisfy  
\begin{eqnarray}
[{\kappa}_{\mu},\widetilde{d}]=-{\kappa}_{\mu}.
\label{eq9p99}
\end{eqnarray}
In terms of matrix elements, since $\widetilde{d}$ is a diagonal matrix, 
the above equation becomes
\begin{eqnarray}
&&({\kappa}_{\mu})_{ik}\widetilde{d}_k{\delta}_{kj}
-\widetilde{d}_i{\delta}_{ik}({\kappa}_{\mu})_{kj}=-({\kappa}_{\mu})_{ij},
\nonumber\\
&&(\widetilde{d}_j-\widetilde{d}_i+1)({\kappa}_{\mu})_{ij}=0.
\label{eq9p99a}
\end{eqnarray}
This means that the matrix elements $({\kappa}_{\mu})_{ij}$ of 
${\kappa}_{\mu}$ will vanish unless $\widetilde{d}_i=\widetilde{d}_j+1$. According to
(\ref{eq9p98}), the matrix representation of ${\kappa}_{\mu}$ 
in the basis (\ref{eq9p94}) will be of following form (with $\ast$
representing a possible non-zero value), 
\begin{eqnarray}
&& ({\kappa}_{\mu})=\left(\begin{array}{ccccccc}
 0 & 0 & 0 & 0 & 0 & 0 & 0 \\
 0 & 0 & 0 & 0 & 0 & 0 & 0 \\
 {\ast} & 0 & 0 & 0 & 0 & 0 & 0 \\
 {\ast} & 0 & 0 & 0 & 0 & 0 & 0 \\
 {\ast} & 0 & 0 & 0 & 0 & 0 & 0 \\
 0 & {\ast} & 0 & 0 & 0 & 0 & 0  \\
 0 & 0 & {\ast} &{\ast}  & {\ast} & 0 & 0 \end{array}\right),\nonumber\\
&& ({\kappa}_{\mu})\left(\begin{array}{c} C(0)
\\{\chi}(0)\\M(0)\\N(0)\\A_{\mu}(0)\\{\lambda}(0)\\D(0)\end{array}\right)
=\left(\begin{array}{c} 0\\ 0\\{\ast} C(0)\\{\ast} C(0) \\{\ast} C(0)
\\{\ast}{\chi}(0) \\{\ast}M(0)+{\ast}N(0)+{\ast}A_{\mu}(0) \end{array}\right).
\label{eq9p100}
\end{eqnarray}
(\ref{eq9p100}) explicitly leads to  
\begin{eqnarray}
{\kappa}_{\mu} C(0)={\kappa}_{\mu}{\chi}(0)=0,
~~\mbox{i.e.}~~[C(0),K_{\mu}]=[{\chi}(0),K_{\mu}]=0.
\label{eq9p101}
\end{eqnarray}
It seems that ${\kappa}_{\mu}M(0){\neq}0$ and 
${\kappa}_{\mu}N(0){\neq}0$, but a careful analysis shows that this not
the case. The special conformal transformations and supersymmetry
transformations of $M$, $N$ and $C$ imply that ${\kappa}_{\mu}M$
and  ${\kappa}_{\mu}N$ have the same dimension as $C$, however,
${\kappa}_{\mu}M$ and  ${\kappa}_{\mu}N$ should be four vectors due to
Lorentz covariance. Since the transformation is linear, it is not allowed
to have a non-local operator such as ${\Box}^{-1}$ in the action
of $K_{\mu}$ on $M$ and $N$, hence there is no way of constructing
a vector with the same dimension as $C$. Thus, we must have  
\begin{eqnarray}
{\kappa}_{\mu}M(0)=0, ~{\kappa}_{\mu}N(0)=0, 
~~[M(0),K_{\mu}]=[N(0),K_{\mu}]=0.
\label{eq9p102}
\end{eqnarray}
The remaining components $A$, ${\lambda}$ and $D$ will not vanish under the 
action of $K_{\mu}$. Since the calculation is quite lengthy, we only list the main 
steps:

 First, we must know the matrix representation $\Sigma_{\mu\nu}$ of
$M_{\mu\nu}$ on the supermultiplet $G(0)$. This can be found by 
means of the Jacobi identities from the known representation 
(\ref{eq9p88}) on $C(0)$ For example,
using the Jacobi identity $(Q,C,M_{\mu\nu})$, we get
\begin{eqnarray}
&&[[Q,C(0)],M_{\mu\nu}]+[[M_{\mu\nu},Q],C(0)]+[[C(0),M_{\mu\nu}],Q]=0,
\nonumber\\
&&[\gamma_5 {\chi}(0),M_{\mu\nu}]+[-\frac{1}{2}{\sigma}_{\mu\nu} Q,C(0)]
+[{\Sigma}_{\mu\nu}C(0), Q]=0,\nonumber\\
&&[{\chi}(0), M_{\mu\nu}]
=(\frac{1}{2}{\sigma}_{\mu\nu}-{\Sigma}_{\mu\nu}){\chi}(0).
\label{eq9p102a}
\end{eqnarray}
Furthermore, the Jacobi identities such as $(\overline{Q},{\chi},M_{\mu\nu})$,
$({Q},M,M_{\mu\nu})$, $({Q},N,M_{\mu\nu})$,
 $({Q},A,M_{\mu\nu})$, $(\overline{Q},{\lambda},M_{\mu\nu})$ and
 $(\overline{Q},D,M_{\mu\nu})$ give
\begin{eqnarray}
&&[M(0), M_{\mu\nu}]=({\sigma}_{\mu\nu}-{\Sigma}_{\mu\nu})M(0),\nonumber\\
&&[N(0), M_{\mu\nu}]=({\sigma}_{\mu\nu}-{\Sigma}_{\mu\nu}N(0)),\nonumber\\
&&[A_{\rho}(0), M_{\mu\nu}]
=({\sigma}_{\mu\nu}-{\Sigma}_{\mu\nu})A_{\rho}(0),\nonumber\\
&&[{\lambda}(0), M_{\mu\nu}]
=(\frac{3}{2}{\sigma}_{\mu\nu}-{\Sigma}_{\mu\nu}){\lambda}(0) ,\nonumber\\
&&[D(0), M_{\mu\nu}]=({2}{\sigma}_{\mu\nu}-{\Sigma}_{\mu\nu})D(0).
\label{eq9p103}
\end{eqnarray}
Secondly, we must know the matrix representation of the special supersymmetry
charge $S$ on the supermultiplet. The Jacobi identities still play a role. 
For example, from the Jacobi identity $(Q,K,C)$ we have 
\begin{eqnarray}
&&[[Q,K_{\mu}],C(0)]+[[C(0),Q],K_{\mu}]+[[K_{\mu},C(0)],Q]=0,
\nonumber\\
&&[{\gamma}_{\mu}S,C(0)]+[{\gamma}_5{\chi}(0),K_{\mu}]=0 ,\nonumber\\
&&[{\gamma}_{\mu}S,C(0)]=0,~~[S,C(0)]=0,
\label{eq9p104}
\end{eqnarray}
where (\ref{eq9p101}) was used.
Further, using the Jacobi identity $(\overline{Q},S,C)$, we determine
$\{{\chi}(0), S\}$,
\begin{eqnarray}
&&[\{\overline{Q},S\},C(0)]+\{[C(0),\overline{Q}],S\}-\{[S,C(0)],\overline{Q}\}=0,\nonumber\\
&&[2i D+{\sigma}^{\mu\nu}M_{\mu\nu}+3{\gamma}_5 R,C(0)]
+\{{\gamma}_5{\chi}(0),S\}=0,\nonumber\\
&&\{ {\chi}(0),S\}={\gamma}_5(-2d+{\sigma}^{\mu\nu}{\Sigma}_{\mu\nu}
+3n{\gamma}_5)C(0).
\label{eq9p105}
\end{eqnarray}
The third step is to use the Jacobi identities $(\overline{Q},{\chi},K_{\mu})$,
 $({Q},{M},K_{\mu})$, $({Q},{N},K_{\mu})$ and $(\overline{Q},{\lambda},K_{\mu})$
for finding the matrix representation $\kappa_{\mu}$ of $K_{\mu}$ on $A$, 
$\lambda$ and $D$. For example, the Jacobi identity 
$(\overline{Q},{\chi},K_{\mu})$ gives
\begin{eqnarray}
&&[\{\overline{Q},{\chi}(0)\},K_{\mu}]+\{[K_{\mu},\overline{Q}],{\chi}(0)\}-
\{[{\chi}(0),K_{\mu}],\overline{Q}\}=0,\nonumber\\
&&\gamma^{\nu}[A_{\nu}(0),K_{\mu}]={\gamma}_{\mu}\{\overline{S},{\chi}(0)\}
={\gamma}_{\mu}3n C(0)+\gamma^{\nu}{\epsilon}_{\nu\mu\sigma\rho}
{\Sigma}^{\sigma\rho}C(0),\nonumber\\
&&[A_{\nu}(0),K_{\mu}]=3n C(0){\eta}_{\mu\nu}
-{\epsilon}_{\mu\nu\sigma\rho}{\Sigma}^{\sigma\rho}C(0).
\label{eq9p106}
\end{eqnarray}
Thus, we can work out 
\begin{eqnarray}
{\kappa}_{\mu}A_{\nu}(0)
&=&-{\epsilon}_{\mu\nu\sigma\rho}{\Sigma}^{\sigma\rho}C(0)
+3n C(0){\eta}_{\mu\nu},\nonumber\\
{\kappa}_{\mu}{\lambda}(0)&=&-\frac{1}{2}{\sigma}^{\nu\rho}{\gamma}_{\mu}
{\Sigma}_{\nu\rho}{\chi}(0)-d{\gamma}_{\mu}{\chi}(0)-
\frac{3}{2}n{\gamma}_{\mu}{\gamma}_5{\chi}(0),\nonumber\\
{\kappa}_{\mu} D(0)&=&-3n A_{\mu}(0)
+{\epsilon}_{\mu\nu\sigma\rho}{\Sigma}^{\sigma\rho}A^{\nu}(0). 
\label{eq9p107}
\end{eqnarray}
After translating (\ref{eq9p107}) from the origin, we finally obtain
\begin{eqnarray}
{\kappa}_{\mu}A_{\nu}(x)
&=&-{\epsilon}_{\mu\nu\sigma\rho}{\Sigma}^{\sigma\rho}C(x)
+3n C(x){\eta}_{\mu\nu},\nonumber\\
{\kappa}_{\mu}{\lambda}(x)&=&-\frac{1}{2}{\sigma}^{\nu\rho}{\gamma}_{\mu}
{\Sigma}_{\nu\rho}{\chi}(x)-d{\gamma}_{\mu}{\chi}(x)
-\frac{3}{2}n{\gamma}_{\mu}{\gamma}_5{\chi}(x) ,\nonumber\\
{\kappa}_{\mu} D(x)&=&-3n A_{\mu}(x)-2id{\partial}_{\mu}C(x)
-2{\Sigma}_{\mu\nu}{\partial}^{\nu}C(x)
+{\epsilon}_{\mu\nu\sigma\rho}{\Sigma}^{\sigma\rho}A^{\nu}(x).
\label{eq9p108}
\end{eqnarray}

Translating $[G(0),S\}$, combining this special supersymmetry transformation
with the Poincar\'{e} supersymmetry transformations in the following way
\begin{eqnarray}
\delta G(x)=-i[V(x),\overline{\zeta}Q+\overline{\epsilon}S],
\label{eq9p109}
\end{eqnarray}
 and defining the convenient combination
\begin{eqnarray} 
{\eta}\,\equiv \, {\zeta}-ix^{\mu}{\gamma}_{\mu}{\epsilon},~~
X^{\pm}\,{\equiv}\,d-\frac{3}{2}n{\gamma}_5{\pm}\frac{1}{2}
{\sigma}^{\mu\nu}{\Sigma}_{\mu\nu},
\label{eq9p110}
\end{eqnarray}
we finally get the superconformal transformation on the supermultiplet
expressed in the following compact form,
\begin{eqnarray} 
\delta C&=&-i\overline{\eta}{\gamma}_5 {\chi},\nonumber\\
\delta{\chi}&=&(M-i{\gamma}_5 N){\eta}-i{\gamma}^{\mu}(A_{\mu}
-i{\gamma}_5{\partial}_{\mu}C){\eta}-2iX^{+}{\gamma}_5{\epsilon}C,\nonumber\\
\delta M&=&\overline{\eta}({\lambda}-i\gamma^{\mu}{\partial}_{\mu}{\chi})+
\overline{\epsilon}X^{-}{\chi}-2\overline{\epsilon}{\chi},\nonumber\\
\delta N&=&-i\overline{\eta}{\gamma}_5
({\lambda}+i\gamma^{\mu}{\partial}_{\mu}{\chi})
-i\overline{\epsilon}{\gamma}_5X^{-}{\chi}-2i\overline{\epsilon}{\gamma}_5{\chi},
\nonumber\\
\delta A_{\mu}&=&i \overline{\eta}{\gamma}_{\mu}{\lambda}
 +{\partial}_{\mu}(\overline{\eta}{\chi})
+i\overline{\epsilon}X^{-}{\gamma}_{\mu}{\chi},\nonumber\\
\delta {\lambda}&=&-i{\sigma}^{\mu\nu}{\eta}{\partial}_{\mu}A_{\nu}
+i{\gamma}_5{\eta}D-X^{+}(M+i{\gamma}_5 N){\epsilon}+i{\gamma}^{\mu}X^{+}
(A_{\mu}-i{\gamma}_5{\partial}_{\mu} C){\epsilon},\nonumber\\
\delta D &=& -\overline{\eta}{\gamma}^{\mu}{\partial}_{\mu}{\gamma}_5{\lambda}
-2i\overline{\epsilon}{\gamma}_5X^{-}({\lambda}
-\frac{1}{2}i{\gamma}^{\mu}{\partial}_{\mu}{\chi}).
\label{eq9p111}
\end{eqnarray}
Eq.\,(\ref{eq9p111}) is the representation of the superconformal algebra
on a general supermultiplet. This supermultiplet is in general reducible.
One can impose reality and chirality conditions on this general multiplet
to reduce it to the representations on vector and chiral supermultiplets,
respectively.

First we consider the reduction to a vector supermultiplet.
If the lowest component field $C$ is real, then the whole supermultiplet 
will be real, $G=G^{\dagger}$, i.e. $G$ is a vector supermultiplet. 
Since $C$ is a real field, its $R$-charge $n$ must vanish  
\begin{eqnarray} 
n(C)=0,
\label{eq9p113}
\end{eqnarray}
as $U_R(1)$ acts nontrivially only on complex fields.
In particular, taking the Hermitian conjugate of
 the last relation of (\ref{eq9p89}), since $C=C^{\dagger}$ and 
$M_{\mu\nu}=M_{\mu\nu}^{\dagger}$, we obtain 
\begin{eqnarray} 
[C,M_{\mu\nu}]=i(x_{\mu}\partial_{\nu}-x_{\nu}\partial_{\mu})C
-{\Sigma}_{\mu\nu}^{\dagger}C.
\label{eq9p115}
\end{eqnarray}
This shows that $\Sigma_{\mu\nu}$ must be a real representation 
of the generators of the Lorentz group,
\begin{eqnarray} 
{\Sigma}_{\mu\nu}=-{\Sigma}_{\mu\nu}^{\dagger}.
\label{eq9p114}
\end{eqnarray} 

Next we turn to the reduction to a chiral supermultiplet through
imposing a chirality condition on the general supermultiplet $G$ . 
A chirality condition means choosing only the chiral (left- or right- handed)
part of Dirac spinor. This can be done by imposing the constraint,
\begin{eqnarray}
(1-{\gamma}_5){\chi}=0, ~~~((1+{\gamma}_5){\chi}=0).
\label{eq9p116}
\end{eqnarray}
This is equivalent to the definition of a chiral (anti-chiral) superfield
$\Phi$: $D_{\alpha}\Phi=0$ ($\overline{D}_{\dot{\alpha}}\Phi=0$),
with $C$ being the lowest component.
The chiral condition (\ref{eq9p116}) imposes a constraint on the
superconformal transformation of $\chi$ and hence selects
the chiral supermultiplet. According to the second equation 
in (\ref{eq9p111}), we should have 
\begin{eqnarray}
&&(1-{\gamma}_5){\delta}{\chi}=0,\nonumber\\
&&(1-{\gamma}_5)\left[(M-i{\gamma}_5 N){\eta}-i{\gamma}^{\mu}(A_{\mu}
-i{\gamma}_5{\partial}_{\mu}C){\eta}-2iX^{+}{\gamma}_5{\epsilon}C\right]
=0,\nonumber\\
&& (1-{\gamma}_5)(M+iN){\eta}-(1-{\gamma}_5){\gamma}^{\mu}(A_{\mu}-i
\partial_{\mu}C){\eta}-2i(1- {\gamma}_5)X^{+}{\gamma}_5{\epsilon}C=0.
\label{eq9p118}
\end{eqnarray}
Thus we obtain
\begin{eqnarray}
M+iN= A_{\mu}-i\partial_{\mu}C=0,
\label{eq9p119}
\end{eqnarray}
and 
\begin{eqnarray}
0&=&(1-{\gamma}_5)X^{+}=(1-{\gamma}_5)(d-\frac{3n}{2}{\gamma}_5+
\frac{1}{2}{\sigma}^{\mu\nu}{\Sigma}_{\mu\nu})\nonumber\\
&=&(1+{\gamma}_5)(d+\frac{3n}{2})+\frac{1}{2}(1-{\gamma}_5)
{\sigma}^{\mu\nu}{\Sigma}_{\mu\nu}.
\label{eq9p120}
\end{eqnarray}
Eq.\,(\ref{eq9p120}) gives
\begin{eqnarray}
n=-\frac{2d}{3},
\label{eq9p121}
\end{eqnarray}
and
\begin{eqnarray}
(1-{\gamma}_5){\sigma}^{\mu\nu}{\Sigma}_{\mu\nu}=0.
\label{eq9p122}
\end{eqnarray}
Using the self-dual property of $\gamma$-matrices \cite{ref27},
${\gamma}_5{\sigma}^{\mu\nu}={i}/{2}{\epsilon}^{\mu\nu\sigma\rho}
{\Sigma}_{\sigma\rho}$,
we can write (\ref{eq9p122}) as
${\sigma}_{\mu\nu}
({\Sigma}^{\mu\nu}-{i}/{2}{\epsilon}^{\mu\nu\sigma\rho}
{\Sigma}_{\sigma\rho})=0$,
and hence we get
\begin{eqnarray}
{\Sigma}_{\mu\nu}
=\frac{i}{2}{\epsilon}_{\mu\nu\sigma\rho}{\Sigma}^{\sigma\rho}.
\label{eq9p123}
\end{eqnarray}
This means that a chiral supermultiplet must be in a self-dual 
representation of the Lorentz group. Eq.\,(\ref{eq9p121}) shows that 
the $R$-charge of the chiral multiplet must be
$-2/3$ times its conformal dimension. These are two important properties
of the chiral conformal supermultiplet. Furthermore, the superconformal
transformations
\begin{eqnarray}
\delta (M+iN)=0,~~~\delta A_{\mu}=i{\partial}_{\mu} \delta C,
\label{eq9p124}
\end{eqnarray}
lead to
\begin{eqnarray}
{\lambda}(x)=D(x)=0.
\label{eq9p125}
\end{eqnarray}
Thus we are left with a chiral supermultiplet ${\Phi}
=(C, (1+{\gamma}_5){\chi}, M)$.

Finally, we reproduce the important results (\ref{eq9p121}) 
and (\ref{eq9p123}) concerning the chiral supermultiplet 
from the viewpoint of coset space \cite{so}. 
In Poincar\'{e} supersymmetry, the little group is the Lorentz group, 
 and Minkowski space is the coset space of the
super-Poincar\'{e} group modulo the Lorentz group. 
There exist chiral multiplets with arbitrary additional 
Lorentz indices, the chirality
condition $\overline{D}\Phi=0$ is covariant for arbitrary 
representations of the Lorentz group.
For example, there exists not only the scalar chiral supermultiplet
$\Phi =({\phi},{\psi}, F)$, but the supersymmetric 
Yang-Mills field strength $W=(F_{\mu\nu}, {\lambda},D)$ 
is also a chiral supermultiplet. However, this is not the case in conformal
supersymmetry.  Since a chiral superfield can be written as follows,
\begin{eqnarray}
\phi (x,\theta, \overline{\theta})=\phi (x+i{\theta}\sigma\overline{\theta},{\theta})=
\phi (y, {\theta}),
\label{eq9p129}
\end{eqnarray} 
a chiral superfield is independent of  the (anti-chiral) 
super-coordinate $\overline{\theta}$, hence one could from the 
beginning use a superspace $G/H$ with the little group 
$H$ generated by both $M_{\mu\nu}$ and $\overline{Q}_{\dot{\alpha}}$, 
$G$ being the super-Poincar\'{e} group. Obviously, this special 
superspace is parameterized only by $x^{\mu}$ and $\theta$ 
and hence this superspace is called a chiral superspace. 
For a general superfield defined in chiral superspace, the action of
$\overline{Q}_{\dot{\alpha}}$ would be as follows:
\begin{eqnarray}
[{\Phi}(x),\overline{Q}_{\dot{\alpha}}\}&=&2({\theta}{\sigma}^{\mu})_{\dot{\alpha}}
{\partial}_{\mu}{\Phi}(x)+\overline{q}_{\dot{\alpha}}{\Phi}(x),\nonumber\\[2mm]
[{\Phi}(0),\overline{Q}_{\dot{\alpha}}\}&=&\overline{q}_{\dot{\alpha}}{\Phi}(0).
\label{eq9p130}
\end{eqnarray} 
For a chiral superfield, the matrix representation $\overline{q}_{\dot{\alpha}}$ of
$\overline{Q}_{\dot{\alpha}}$ should vanish,
\begin{eqnarray}
\overline{q}_{\dot{\alpha}}=0.
\label{eq9p131}
\end{eqnarray}
This is the case of Poincar\'{e} supersymmetry. 
For conformal supersymmetry, as discussed in Sect.\,\ref{subsect21},
 the ordinary conformal group can be realized on Minkowski space,
and Minkowski space is the coset space of the conformal group 
modulo the Lorentz group, dilation and special conformal transformations.
Thus  the conformal supersymmetry can be realized on Minkowski 
superspace with a complicated $x$, $\theta$ and $\overline{\theta}$ 
dependence of the elements of the little
group. It is also possible to realize conformal 
supersymmetry on chiral superspace. The relevant little group will now
be generated by $M_{\mu\nu}$, $K_{\mu}$, $D$, $S^{\alpha}$, 
$\overline{S}^{\dot{\alpha}}$ and $\overline{Q}_{\dot{\alpha}}$
since the transformations generated by these generators 
keep $x={\theta}=0$ invariant. In this case the question whether there 
exists a chiral conformal supermultiplet is equivalent to whether the 
constraint (\ref{eq9p131}) is consistent. This is not naturally satisfied
since there is a non-vanishing anti-commutator from the superconformal algebra,
\begin{eqnarray}
\{\overline{S}^{\dot{\alpha}},\overline{Q}_{\dot{\beta}}\}
=(\overline{\sigma}^{\mu\nu})^{\dot{\alpha}}_{~\dot{\beta}}M_{\mu\nu}
-2{\delta}^{\dot{\alpha}}_{~\dot{\beta}}\left(\frac{3}{2}R-iD\right).
\label{eq9p133}
\end{eqnarray}
Requiring (\ref{eq9p130}) to be satisfied when 
$\{\overline{S}^{\dot{\alpha}},\overline{Q}_{\dot{\beta}}\}$ acts on a chiral
supermultiplet $\Phi$, we have
\begin{eqnarray}
0&=&[{\phi}(0),\{\overline{S}^{\dot{\alpha}},\overline{Q}_{\dot{\beta}}\}]\nonumber\\
&=&[{\phi}(0),(\overline{\sigma}^{\mu\nu})^{\dot{\alpha}}_{~\dot{\beta}}M_{\mu\nu}
-2{\delta}^{\dot{\alpha}}_{~\dot{\beta}}\left(\frac{3}{2}R-iD\right)]
\nonumber\\
&=&(\overline{\sigma}^{\mu\nu})^{\dot{\alpha}}_{~\dot{\beta}}
{\Sigma}_{\mu\nu}{\phi}(0)-2{\delta}^{\dot{\alpha}}_{~\dot{\beta}}
\left( \frac{3}{2}n+d \right){\phi}(0).
\label{eq9p134}
\end{eqnarray}
Furthermore, considering the anti-selfdual property 
of $\bar{\sigma}_{\mu\nu}$ \cite{so},
$\overline{\sigma}^{\mu\nu}
=-{i}/{2}{\epsilon}_{\mu\nu\sigma\rho}\overline{\sigma}^{\sigma\rho}$,
the relations (\ref{eq9p121}) and (\ref{eq9p123}) are reproduced.


\section{Non-perturbative dynamics in $N=1$ supersymmetric QCD}
\label{sect6}
\renewcommand{\thetable}{3.1.\arabic{equation}}
\setcounter{equation}{0}
\renewcommand{\theequation}{3.1.\arabic{equation}}
\setcounter{table}{0}

We shall next introduce the non-perturbative dynamical 
phenomena in supersymmetric gauge theory. This section is 
concentrating on supersymmetric QCD with gauge group
$SU(N_c)$ and $N_f$ flavours. The methods of analyzing the 
non-perturbative dynamics include the gauge and global 
symmetries; the holomorphic dependence of the non-perturbative 
dynamical superpotential not only on the chiral superfields 
but also on various parameters  such as the coupling and 
mass; instanton computations and the exact NSVZ beta 
function given in (\ref{eq1p1}) as well as the decoupling
relation of the heavy modes at low-energy. Furthermore,
the 't Hooft anomaly matching  between high and low energy
can provide a test of the non-perturbative result.
In addition to the global symmetries
$SU_L(N_f)\times SU_R(N_f)\times U_B(1)\times U_A(1)$
(some of them are broken or anomalous at the quantum level)
as in ordinary QCD, the $N=1$ supersymmetric QCD has another
$U(1)$ axial vector symmetry called the $R$-symmetry, which
becomes anomalous like the $U_A(1)$ at the quantum level.
However, this $R$-symmetry can be combined with the
$U_A(1)$ symmetry to an anomaly-free $U(1)$ symmetry.
The global and gauge symmetries together with the holomorphy
and instanton calculations uniquely fix the superpotential and
hence the moduli space in the range $N_f<N_c$. By considering the 
decoupling relation, the moduli spaces in the cases $N_f=N_c$ or 
$N_c+1$ can also be exactly determined. Consequently, a series
of non-perturbative physical phenomena such as dynamical 
supersymmetry breaking, chiral symmetry breaking and confinement
are exhibited. The 't Hooft anomaly matching confirms the
correctness of the physical pictures. 
The NSVZ beta function implies that in the range
$3N_c/2 < N_f< 3N_c$, the theory has a non-trivial IR fixed
point, at which the theory is a superconformal field
theory and the dual theory gives a completely equivalent 
description of the low-energy physics of the original theory
but with a weak coupling. This phenomenon is similar to
the electric-magnetic duality and is thus called
a non-Abelian electric-magnetic duality. It will be
discussed in detail in the next section.

\subsection{Introducing $N=1$ supersymmetric QCD}

\renewcommand{\theequation}{3.1.\arabic{equation}}
\setcounter{equation}{0}
\renewcommand{\thetable}{3.1.\arabic{table}}
\setcounter{table}{0}

\subsubsection{Classical action of $N=1$ supersymmetric QCD} 

Supersymmetric QCD (SQCD) is the generalization of ordinary 
QCD. As required by supersymmetry, corresponding to each particle,
there exists a superpartner. Corresponding to the gluon $A_{\mu}^a$,
the left-handed quark ${\psi}$ and the right-handed quark $\widetilde{\psi}$,
we have their superpartners: the gluino $\lambda^a$, the left-handed squark
$\phi$ and the right-handed squark $\widetilde{\phi}$. In superfield form,
the Lagrangian of (massive) SQCD is
\begin{eqnarray}
{\cal L}&=&\frac{1}{8\pi}\mbox{Im}
\left(\tau \mbox{Tr}W^{\alpha}W_{\alpha}|_{\theta^2}\right)+
\left[Q^{\dagger}e^{gV(N_c)}Q+\widetilde{Q}^{\dagger}
e^{gV(\overline{N}_c)}\widetilde{Q}\right]|_{{\theta}^2
\overline{\theta}^2}\nonumber\\[2mm]
&+&m(\widetilde{Q}{Q})|_{\theta^2}
+m^*({Q}^{\dagger}\widetilde{Q}^{\dagger})|_{\overline{\theta}^2}.
\end{eqnarray}
Here, $W_{\alpha}$ is the superfield strength which in the Wess-Zumino
gauge takes the form
\begin{eqnarray}
W_{\alpha}=T^a\left( -i{\lambda}_{\alpha}^a+{\theta}_{\alpha}D^a -
\frac{i}{2}({\sigma}^{\mu}\overline{\sigma}^{\nu}{\theta})_{\alpha}
F_{\mu\nu}^a+{\theta}^2{\sigma}^{\mu}_{\alpha\dot{\alpha}}
({\cal D}_{\mu}\overline{\lambda}^{\dot{\alpha}})^a
\right).
\end{eqnarray}
The quantity
\begin{eqnarray}
\tau = \frac{\theta}{2\pi} + i\frac{4\pi}{g^2}
\label{eq3p3}
\end{eqnarray}
combines the gauge coupling constant $g$ and the CP-violating parameter
$\theta $ into what can be effectively regarded as a constant chiral
superfield. $Q$ and $\widetilde{Q}$ are chiral left- and right-handed
quark superfields, respectively. $V$ is the vector superfield of
the gluon and the gluino. In Wess-Zumino gauge, they take the
following forms:
\begin{eqnarray}
Q_r(y)&=&{\phi}_r(y)+\sqrt{2}{\theta}^{\alpha}{\psi}_{\alpha r}(y)
+{\theta}^2F_r(y), ~~\widetilde{Q}_r(y)=\widetilde{\phi}_r(y)
+\sqrt{2}{\theta}^{\alpha}\widetilde{\psi}_{\alpha r}(y)
+{\theta}^2{\widetilde F}_r(y), \nonumber\\[2mm]
V^a(x,\theta,\overline{\theta})&=&-{\theta}^{\alpha}
({\sigma}^{\mu})_{\alpha\dot{\alpha}}
\overline{\theta}^{\dot{\alpha}}A_{\mu}^a(x)+
i{\theta}^2\overline{\theta}_{\dot{\alpha}}
\overline{\lambda}^{\dot{\alpha}a}(x)
-i\overline{\theta}^2{\theta}^{\alpha}{\lambda}^a_{\alpha}
+\frac{1}{2}{\theta}^2\overline{\theta}^2D^a(x).
\end{eqnarray}
The notation used is
\begin{eqnarray}
y^{\mu}&{\equiv}&x^{\mu}+i{\theta}^{\alpha}
{\sigma}^{\mu}_{\alpha\dot{\alpha}}
\overline{\theta}^{\dot{\alpha}},~
V(N_c){\equiv}V^aT^a(N_c), ~V(\overline{N}_c){\equiv}V^aT^a(\overline{N}_c),
\end{eqnarray}
where $T^a(N_c)\equiv T^a$ and $T(\overline{N}_c)$ are the generators of the
gauge group $SU(N_c)$ in the fundamental representation and its
conjugate representation, respectively.

Writing out the Lagrangian in terms of component fields, we have
\begin{eqnarray}
{\cal L}&=&-\frac{1}{4}F_{\mu\nu}^aF^{\mu\nu a}+i{\lambda}^{\alpha a}
({\sigma}^{\mu})_{\alpha\dot{\alpha}}({\cal D}_{\mu})^{ab}
\overline{\lambda}^{\dot{\alpha}b}+\frac{1}{2}D^aD^a
+i\overline{\psi}^{~r}_{\dot{\alpha}}
(\overline{\sigma}^{\mu})^{\dot{\alpha}\alpha}
(D_{\mu})_r^{~s}{\psi}_{\alpha s}\nonumber\\[2mm]
&+&i\widetilde{\psi}^{\alpha r}(\sigma^{\mu})_{\alpha\dot{\alpha}}
(D_{\mu})_r^{~s}\overline{\widetilde{\psi}}^{\dot{\alpha}}_{~s}
+(D^{\mu}\phi)^{*r}(D_{\mu}\phi)_r+
(\widetilde{D}^{\mu}\widetilde{\phi})^{r}(\widetilde{D}_{\mu}
\widetilde{\phi})^{*}_r\nonumber\\[2mm]
&+&i\sqrt{2}g\left[{\phi}^{*r}(T^a)_r^{~s}
{\lambda}^{a\alpha}{\psi}_{\alpha s}
-\overline{\lambda}^a_{~\dot{\alpha}}
\overline{\psi}^{\dot{\alpha}r}(T^a)_r^{~s}
{\phi}_s-\widetilde{\psi}^{\alpha s}
{\lambda}^a_{~\alpha}(T^a)_s^{~r}\widetilde{\phi}_r^*+
\widetilde{\phi}^s(T^a)_s^{~r}\overline{\widetilde{\psi}}_{\dot{\alpha} r}
\overline{\lambda}^{a\dot{\alpha}}\right]
\nonumber\\[2mm]
&+& gD^a\left[{\phi}^{*r}(T^a)_r^{~s}{\phi}_s-
\widetilde{\phi}^{r}(T^a)_r^{~s}\widetilde{\phi}^*_s\right]
-m\widetilde{\psi}^{\alpha r}{\psi}_{\alpha r}
-m^*\overline{\psi}^{r}_{~\dot{\alpha}}
\overline{\widetilde{\psi}}^{\dot{\alpha}}_{~r}+F^{*r}F_r
+\widetilde{F}^{*}_{r}\widetilde{F}^{r}\nonumber\\[2mm]
&+&m^*\widetilde{\phi}^{*}_{r}F^{*r}+m{\phi}_r\widetilde{F}^{r}
+m^*\phi^{*r}\widetilde{F}^{*}_{r}+m\widetilde{\phi}^{r}F_{r}
+\frac{i\theta}{32\pi^2}F_{\mu\nu}^a\widetilde{F}^{a\mu\nu},
\label{eq163}
\end{eqnarray}
where $D$, $F$ and $\widetilde{F}$ are auxiliary fields, 
$a,b=1,{\cdots}, N_c^2-1$, 
$r,s=1,{\cdots}, N_c$ and we suppress the flavour index.
The various covariant  derivatives 
are defined as follows:
\begin{eqnarray}
(D_{\mu}{\phi})_r&=&({\partial}_{\mu}{\delta}_r^{~s}+igA_{\mu}^a(T^a)_r^{~s})
{\phi}_s, ~~(D_{\mu}{\phi})^{* r}
={\partial}_{\mu}{\phi}^{*r}-ig{\phi}^{*s}(T^a)_s^{~r}A_{\mu}^a,
\nonumber\\[2mm]
(\widetilde{D}_{\mu}\widetilde{\phi})^r&=&{\partial}_{\mu}\widetilde{\phi}^r
-ig\widetilde{\phi}^s(T^a)^{~r}_sA_{\mu}^a, 
~~(\widetilde{D}_{\mu}\widetilde{\phi})^{*}_r
=({\partial}_{\mu}{\delta}_r^{~s}+igA_{\mu}^a(T^a)_r^{~s})
\widetilde{\phi}^{*}_{s},
\nonumber\\[2mm]
{\cal D}_{\mu}\overline{\lambda}^{a\dot{\alpha}}&=&{\partial}_{\mu}
\overline{\lambda}^{a\dot{\alpha}}+gf^{abc}A_{\mu}^b\overline{\lambda}^{c\dot{\alpha}}.
\end{eqnarray}
Eliminating the auxiliary fields $F$, $F^{*}$,
 $\widetilde{F}$, $\widetilde{F}^{*}$ and $D$ through
their equations of motion, 
\begin{eqnarray}
F_r=-m^{*}\widetilde{\phi}_r^{*}, ~~\widetilde{F}^r=-m^{*}{\phi}^{*r},~~
D^a=-g[{\phi}^{*r}(T^a)_r^{~s}{\phi}_s-
\widetilde{\phi}^{r}(T^a)_r^{~s}\widetilde{\phi}^*_s],
\end{eqnarray}
we can obtain the Lagrangian given in Ref.\,\cite{akmrv}.

\subsubsection{Global symmetries of massless supersymmetric QCD}
\label{subsect6.1}

\begin{figure}
\begin{center}
\leavevmode
{\epsfxsize=3.00truein \epsfbox{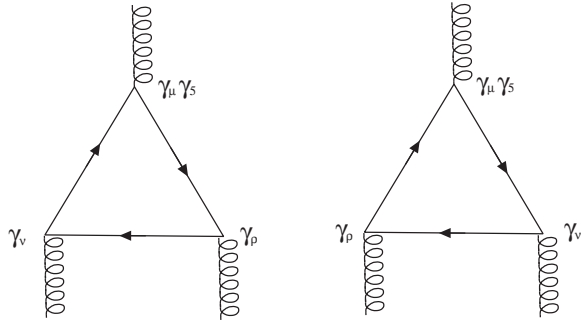}}
\caption{\protect\small Triangle diagrams 
$\widetilde{\Gamma}^{ab}_{\mu\nu\rho}(r,p,q)$.
\label{fig6.1}}
\end{center}
\end{figure}

Massless SQCD possesses the global
symmetries of ordinary QCD, i.e., $SU_L(N_f){\times}SU_R(N_f)$${\times}
U_B(1)$ ${\times}U_A(1)$. In addition, it has a new $U(1)$
symmetry called $R_0$-symmetry. In terms of superfields \cite{ref1p22}:
\begin{eqnarray}
W_{\beta}(\theta){\longrightarrow}e^{-i\alpha}W_{\beta}({\theta}e^{i\alpha}),
~{Q}({\theta}){\longrightarrow}{Q}(\theta e^{i\alpha}),
~\widetilde{Q}({\theta}){\longrightarrow}\widetilde{Q}(\theta e^{i\alpha}).
\end{eqnarray}
For the component fields, this means
\begin{eqnarray}
{\psi}_{r}{\longrightarrow}e^{-i{\alpha}}{\psi}_{r}, ~
\widetilde{\psi}^{r}{\longrightarrow}e^{-i{\alpha}}\widetilde{\psi}^{r},
~{\lambda}^a{\longrightarrow}e^{i{\alpha}}{\lambda}^a.
\label{eq170}
\end{eqnarray}
The total 
global symmetry of massless SQCD at the classical level is then 
\begin{eqnarray}
SU_L(N_f){\times}SU_R(N_f){\times}U_B(1){\times}U_A(1){\times}U_{R_0}(1).
\end{eqnarray}
However, like $U_A(1)$, the $R_0$-symmetry 
suffers from an anomaly at the quantum level. The current
corresponding to the $R_0$-symmetry (\ref{eq170}) is
\begin{eqnarray}
j^R_{\mu}=-\overline{\psi}_{\dot{\alpha}}
(\overline{\sigma}_{\mu})^{\dot{\alpha}\alpha}{\psi}_{\alpha}
+\widetilde{\psi}^{\alpha}({\sigma}_{\mu})_{\alpha\dot{\alpha}}
\overline{\widetilde{\psi}}^{\dot{\alpha}}+{\lambda}^{a\alpha}
({\sigma}_{\mu})_{\alpha\dot{\alpha}}
\overline{\lambda}^{a\dot{\alpha}},
\end{eqnarray}
or in four-component form
\begin{eqnarray}
j^{R_0}_{\mu}=j_{\mu}^5+\widetilde{j}_{\mu}^5=
\overline{\Psi}{\gamma}_{\mu}{\gamma}_5{\Psi}
+\frac{1}{2}\overline{\lambda}^a{\gamma}_{\mu}
{\gamma}_5{\lambda}^a,
\label{eq175}
\end{eqnarray}
$\Psi$ being the Dirac spinor of the quarks 
and $\Lambda$ the four component Majorana spinor of the gluino.
The operator anomaly equation for the $R$-current is
\begin{eqnarray}
{\partial}^{\mu}j_{\mu}^{R_0}
=(N_c-N_f)\frac{g^2}{32{\pi}^2}{\epsilon}^{\mu\nu\sigma\rho}
F_{\mu\nu}^aF_{\sigma\rho}^a.
\label{eq176}
\end{eqnarray}
This anomaly equation arises as follows. Eq.\,(\ref{eq175}) shows that 
$j^{R_0}_{\mu}$ is composed of two parts. The first part is  
the ordinary chiral current $j_{\mu}^5$. 
The triangle diagram 
${\langle}j_{\mu}^5(x)J_{\nu}^a(y)J_{\rho}^b(z){\rangle}$ gives
the familiar contribution $-g^2N_f/(16\pi^2)F_{\mu\nu}^a
\widetilde{F}^{\mu\nu a}$. 
The second part $\widetilde{j}_{\mu}^5$ is formed
by the gluino $\lambda$. The anomalous triangle diagram is (see
 Fig.\,\ref{fig6.1})
\begin{eqnarray}
\widetilde{\Gamma}^{ab}_{\mu\nu\rho}(z,x,y)
&=&{\langle}\widetilde{j}_{\mu}^5(z){\cal J}^a_{\nu}(x)
{\cal J}^b_{\rho}(y){\rangle},\nonumber\\
\widetilde{\Gamma}^{ab}_{\mu\nu\rho}(r,p,q)
(2{\pi})^4{\delta}^{(4)}(r+p+q)&=&{\int}
d^4xd^4yd^4ze^{i(p{\cdot}x+q{\cdot}y+r{\cdot}z)}
\widetilde{\Gamma}^{ab}_{\mu\nu\rho}(z,x,y). 
\end{eqnarray}


Since $\lambda$ is in the adjoint representation 
of $SU(N_c)$, ${\cal J}^a_{\nu}$
is the current corresponding to a global gauge transformation in
the adjoint representation,
\begin{eqnarray}
{\cal J}_{\mu}^a=if^{abc}\overline{\lambda}^b{\gamma}_{\mu}{\lambda}^c,
\end{eqnarray}
which couples with the gluon field $A_{\mu}^a$. Requiring
\begin{eqnarray}
{\partial}^{\mu}{\cal J}_{\mu}^a=0,~p^{\nu}\widetilde{\Gamma}^{ab}_{\mu\nu\rho}
(r,p,q)
=q^{\rho}\widetilde{\Gamma}^{ab}_{\mu\nu\rho}(r,p,q)=0,
\end{eqnarray}
we obtain  the anomalous Ward identity
\begin{eqnarray}
(p+q)^{\rho}\widetilde{\Gamma}^{ab}_{\rho\mu\nu}(p,q,r)
=2f^{adc}f^{bdc}\frac{1}{2\pi^2}p^{\alpha}q^{\beta}
{\epsilon}_{\mu\nu\alpha\beta} 
=2N_c\frac{1}{2\pi^2}p^{\alpha}q^{\beta}
{\epsilon}_{\mu\nu\alpha\beta}{\delta}^{ab}.
\label{eq6.18} 
\end{eqnarray}
Combining (\ref{eq6.18}) with the 
anomaly of $j_{\mu}^5$, we obtain (\ref{eq170}).

Since there are two anomalous $U(1)$ transformations, 
$U(1)_A$ and $U(1)_{R_0}$,
it is possible to combine them to get an anomaly-free $U(1)$
$R$-symmetry. Requiring that the linear combination
\begin{eqnarray}
j^{\mu}_R= mj^{\mu}_{R_0} + nj^{\mu}_A
\end{eqnarray}
has vanishing anomaly,
using Eq.\,(\ref{eq176}) and the corresponding 
equation for $j^{\mu}_A$ gives
\begin{eqnarray}
n=\frac{N_f-N_c}{N_f}m.
\label{nmeq}
\end{eqnarray}
The simplest choice in (\ref{nmeq}) is $m=1$. The charges
of the fields under this anomaly-free $U_R(1)$ symmetry are thus given
in terms of the $U_R(1)$ and $U_A(1)$ charges by
\begin{eqnarray}
R=R_0+\frac{N_f-N_c}{N_f}A.
\label{eq182}
\end{eqnarray}
The quantum numbers of every field are listed in Table \ref{ta6.1}.
 We thus have an anomaly-free global symmetry at the quantum level
\begin{eqnarray}
SU_L(N_f){\times}SU_R(N_f){\times}U_B(1){\times}U_{R}(1).
\end{eqnarray}

\begin{table}
\begin{center}
\begin{tabular}{|c|c|c|c|c|}
\hline
       & $U_B(1)$ & $U_A(1)$ & $U_{R_0}(1)$ & $U_R(1)$\\ \hline 
$\phi$ & $+1$ & $+1$ & $0$  &$(N_f-N_c)/N_f$ \\ \hline
$\psi$ & $+1$ & $+1$ & $-1$  &$ -N_c/N_f$ \\ \hline
$\widetilde{\phi}$ & $-1$ & $+1$ & $0$  &$(N_f-N_c)/N_f$ \\ \hline
$\widetilde{\psi}$ & $-1$ & $+1$ & $-1$  &$-N_c/N_f$ \\ \hline
${\lambda}$ & $0$ & $0$ & $+1$  & $+1$ \\ \hline
\end{tabular}
\end{center}
\caption{\protect\small Anomaly-free $R$-charges of fields. \label{ta6.1}}
\end{table}

A special consideration should be paid to the $N_c=2$ case. Since the
fundamental and anti-fundamental representations of $SU(2)$ are
equivalent, there is no difference between the left- and right-handed 
quarks. Thus, the theory has $2N_f$ quark chiral superfields 
$Q^i$, $i=1,{\cdots}, 2N_f$,
the anomaly-free global symmetry is
\begin{eqnarray}
SU(2N_f){\times}U_{R}(1)
\label{eq182b}
\end{eqnarray}
and the $U_{R}(1)$ charge of $Q$ is $(N_f-2)/N_f$.

\subsection{Holomorphy of supersymmetric QCD}
\label{subsect6.2}

\renewcommand{\theequation}{3.2.\arabic{equation}}
\setcounter{equation}{0}
\renewcommand{\thetable}{3.2.\arabic{table}}
\setcounter{table}{0}

Supersymmetric gauge theory possesses a powerful property:
the superpotential is a holomorphic (or anti-holomorphic) 
function of chiral superfields $Q$ (or anti-chiral 
superfields $\widetilde{Q}$);
Furthermore, the supersymmetric Ward identities 
determine that some Green functions have a holomorphic dependence  
on the mass parameter and coupling constant \cite{akmrv,ref1p2}, 
and so does the low energy effective Lagrangian. 
This property plays a key role in looking for the non-perturbative 
superpotential \cite{ref1p22}.

\subsubsection{Supersymmetric Ward identity}
\label{subsub6.2.1}

In supersymmetric theories, the matter fields are described 
by the chiral superfields. Chiral superfields have the 
important property that the product of two chiral superfields 
is still a chiral superfield. Thus, a chiral superfield
\begin{eqnarray}
{\Phi}(y)={\chi}(y)+\sqrt{2}
{\theta}^{\alpha}{\Psi}_{\alpha}(y)+{\theta}{\theta}F(y),
\end{eqnarray}
where $y=x+i{\theta}{\sigma}\overline{\theta}$, can be thought of as 
either a fundamental chiral superfield or a composite one, and the 
same goes for the component fields. Under a supersymmetry 
transformation the component fields of a chiral superfield transform 
as follows:
\begin{eqnarray}
&& a)~\left[\overline{Q}^{\dot{\alpha}},{\chi}\right]=0, ~
b)~\left\{\overline{Q}^{\dot{\alpha}},{\Psi}^{\alpha}(x)\right\}
=-i\sqrt{2}(\overline{\sigma}^{\mu})^{\dot{\alpha}\alpha}
{\partial}_{\mu}{\chi}(x),
~ c)~\left[\overline{Q}^{\dot{\alpha}},F(x)\right]
=i\sqrt{2}{\partial}^{\mu}{\Psi}_{\alpha}(x)
\overline{\sigma}_{\mu}^{\alpha\dot{\alpha}},
\nonumber\\[2mm]
&& d)~\left[Q_{\alpha},F\right]=0,~
e)~\left\{Q^{\alpha},\Psi^{\beta}(x)\right\}
=\sqrt{2}{\epsilon}^{\alpha\beta}F(x),~ f) 
~\left[Q_{\alpha},{\chi}(x)\right]=\sqrt{2}{\Psi}_{\alpha}.
\label{eq187}
\end{eqnarray}
Assuming that there is no spontaneous supersymmetry breaking,
\begin{eqnarray}
Q^{\alpha}|0{\rangle}=0,~~\overline{Q}^{\dot{\alpha}}|0{\rangle}=0,
\label{eq188}
\end{eqnarray}
we can derive strong constraints on Green functions from 
the Ward identities corresponding to the
above supersymmetry transformations.
First, we consider Green functions of the lowest components ${\chi}_i(x_i)$
of some chiral superfields ${\Phi}_i(x_i)$, here $i=1,{\cdots},n$,
\begin{eqnarray}
G(x_1,{\cdots}x_n){\equiv}{\langle}0|T\left[{\chi}_1(x_1){\cdots}
{\chi}_n(x_n)\right]|0{\rangle}.
\end{eqnarray}
From item $b)$ in (\ref{eq187}) and from (\ref{eq188}), we get
\begin{eqnarray}
0&=&\frac{i}{\sqrt{2}}{\langle}0|T\left[{\chi}_1(x_1){\cdots}
\{\overline{Q}^{\dot{\alpha}},{\Psi}_i^{\alpha}(x_i)\}{\cdots}
{\chi}_n(x_n)\right]|0{\rangle}\nonumber\\[2mm]
&=&{\langle}0|T\left[{\chi}_1(x_1){\cdots}
(\overline{\sigma}^{\mu})^{\dot{\alpha}\alpha}
\frac{\partial}{{\partial}x_i^{\mu}}
{\chi}_i(x_i){\cdots}{\chi}_n(x_n)\right]|0{\rangle}\nonumber\\[2mm]
&=&(\overline{\sigma}^{\mu})^{\dot{\alpha}\alpha}
\frac{\partial}{{\partial}x_i^{\mu}}
G(x_1,{\cdots},x_n).
\label{eq190}
\end{eqnarray}
Note that when we take the derivative outside the $T$-product 
in (\ref{eq190}), the equal time commutator terms arise but
all vanish. (\ref{eq190}) means that the Green function of the 
lowest component of chiral superfield operators
is space-time independent. Now we apply this result to supersymmetric 
QCD. 

\subsubsection{Holomorphic dependence of supersymmetric QCD}
\label{subsub6.2.2}

In supersymmetric QCD, the quark superfields $Q(x)$, $\widetilde{Q}(x)$ and
the gauge field strength superfield $W_{\alpha}$ are chiral superfields.
Since the product of two chiral superfields is still a chiral superfield,
$W_{\alpha}W^{\alpha}$ and $Q_{rj}\widetilde{Q}^{ri}$ are chiral superfields,
where $i$ and $j$ are flavour indices.
Their lowest component are, respectively,
\begin{eqnarray}
\frac{g^2}{32\pi^2}{\lambda}^{\alpha a}(x){\lambda}^a_{\alpha}(x){\equiv}
\frac{g^2}{32{\pi}^2}{\lambda}{\lambda}(x), 
~~\widetilde{\phi}^{ri}(x){\phi}_{rj}(x){\equiv}
\widetilde{\phi}^i{\phi}_j(x).
\end{eqnarray}
Note that here and the following $i,j=1,{\cdots},N_f$ denote flavour indices.
Considering the Green function of these operators,
\begin{eqnarray}
G^{(p,q)}&{\equiv}&G^{(p,q)i_1{\cdots}i_p}_{j_1{\cdots}j_p}
(x_1,{\cdots},x_p,x_{p+1},
{\cdots},x_{p+q})\nonumber\\[2mm]
&{\equiv}&{\langle}0|T\left[\widetilde{\phi}^{i_1}{\phi}_{j_1}(x_1)
{\cdots}\widetilde{\phi}^{i_p}{\phi}_{j_p}(x)
\frac{g^2}{32{\pi}^2}{\lambda}{\lambda}(x_{p+1})
{\cdots}\frac{g^2}{32{\pi}^2}{\lambda}{\lambda}(x_{p+q})\right]|0{\rangle},
\label{eq193}
\end{eqnarray}     
 we know from (\ref{eq190}) 
that $G^{(p,q)}$ is space-time independent. 
Furthermore, we  shall show that $G^{(p,q)}$ depends 
holomorphically on the mass parameters, 
that is, it depends only on  $m_i$, but not on $m^*_i$, $i=1,{\cdots},N_f$. 
The path integral representation of $G^{(p,q)}$ is
\begin{eqnarray}
G^{(p,q)}
= {\int}{\cal D}X \Pi_{k=1}^p\widetilde{\phi}^{i_k}{\phi}_{j_k}(x_k)
\Pi_{l=1}^q\frac{g^2}{32{\pi}^2}{\lambda}{\lambda}(x_{p+l})
e^{-i{\int}{\cal L}_{\rm eff}},
\label{eq194}
\end{eqnarray}
where ${\cal L}_{\rm eff}$ is the gauge-fixed effective Lagrangian of
supersymmetric QCD, $X$ is a shorthand for all fields integrated over,
including the ghost fields and their superpartners.
From Eq.\,(\ref{eq163}) we see that the coefficient $F^{*}_j$ of $m^{*}_j$
in the SQCD Lagrangian is the auxiliary field of the composite anti-chiral
superfield $Q^{\dagger j}\widetilde{Q}^{\dagger}_j$. Hence
\begin{eqnarray}
m_j^*\frac{\partial}{{\partial}m_j^*}G^{(p,q)}&=&
{\int}{\cal D}X\Pi_{k=1}^p
 \widetilde{\phi}^{i_k}{\phi}_{j_k}(x_k)\left[m_j^*{\int}d^4y F_j^*\right]
\Pi_{l=1}^q\frac{g^2}{32{\pi}^2}{\lambda}{\lambda}(x_{p+l})
e^{-i{\int}{\cal L}_{\rm eff}}\nonumber\\[2mm]
&=&{\langle}0|T\left[\Pi_{k=1}^p
\widetilde{\phi}^{i_k}{\phi}_{j_k}(x_k)
\Pi_{l=1}^q\frac{g^2}{32{\pi}^2}{\lambda}{\lambda}(x_{p+l})\right]
m_j^*{\int}d^4yF_j^*|0{\rangle}\\[2mm]
&=&\frac{m_j^*}{2\sqrt{2}}{\int}d^4y{\langle}0|
T\left[\Pi_{k=1}^p\widetilde{\phi}^{i_k}{\phi}_{j_k}(x_k)
\Pi_{l=1}^q\frac{g^2}{32{\pi}^2}{\lambda}{\lambda}(x_{p+l})\right]
\left\{\overline{Q}_{\dot{\alpha}},\overline{\Psi}^{\dot{\alpha}}_j(y)
\right\}|0{\rangle}=0,
\nonumber
\end{eqnarray} 
where $\overline{\Psi}_j$ is the spinor term of $Q^{\dagger j}
\widetilde{Q}^{\dagger}_j$ and we used item $a)$ and 
(the conjugate of) item $e)$ in (\ref{eq187}) and also (\ref{eq188}).
Thus $G^{(p,q)}$ is holomorphic with respect to the parameters $m_j$. 
Moreover, we can compute its explicit dependence
on them. Write the complex parameter $m_j$ as
$m_j=|m_j|e^{i{\alpha}_j}$. Then 
\begin{eqnarray}
m_j\frac{\partial}{{\partial}m_j}G^{(p,q)}&=&\left(
m_j\frac{\partial}{{\partial}m_j}-m_j^*\frac{\partial}{{\partial}m_j^*}\right)
G^{(p,q)} =-i\frac{\partial}{{\partial}{\alpha}_j}G^{(p,q)}.
\end{eqnarray}
The $m_j$-dependence of the Green function $G^{(p,q)}$ is thus given by the 
dependence on the phase angle of the quark mass of the $j$-th flavour. 
This dependence can be determined by defining a $U_A^{(j)}(1)$ transformation,
which is non-anomalous and is explicitly broken by the $j$-th quark mass $m_j$:
\begin{eqnarray}
{\lambda}{\rightarrow}e^{-i{\alpha}/(2N_c)}{\lambda},~~
(\widetilde{\psi}^l,{\psi}_l){\rightarrow}e^{i{\alpha}{\delta}_{lj}/2}
(\widetilde{\psi}^l,{\psi}_l), ~~
(\widetilde{\phi}^l,{\phi}_l)
{\rightarrow}e^{i{\alpha}({\delta}_{lj}-1/N_c)/2}
(\widetilde{\varphi}^l,{\varphi}_l).
\label{eq197}
\end{eqnarray}
From the process of combining the $U_A(1)$ and $U_R(1)$ symmetries 
to get the anomaly-free $U_R(1)$ in Subsect.\,\ref{subsect6.1},
we know that this is possible. It can easily be
checked that this $U_A^{(j)}(1)$ symmetry is indeed anomaly-free. 
The only terms in the Lagrangian which are not invariant under the
transformation (\ref{eq197}) are the quark mass term of the $j$-th flavour.
Performing the transformation of variables (\ref{eq197}) in the path
integral (\ref{eq194}) is the equivalent to changing the phase of
$m_j$: $m_j \longrightarrow m_je^{-i\alpha}$. Thus
\begin{eqnarray}
m^j\frac{{\partial}G^{(p,q)}}{{\partial}m^j}=q^{(j)}G^{(p,q)},
\label{mdGdm}
\end{eqnarray}
where
\begin{eqnarray}
q^{(j)}=\frac{p+q}{N_c}-\frac{1}{2}
\sum_{l=1}^p({\delta}_{i_l,j}+{\delta}_{j_l,j}).
\end{eqnarray}
Integrating the equations (\ref{mdGdm}) we find
\begin{eqnarray}
&&\frac{dG}{G^{(p,q)}}=\sum_{j=1}^{N_f}q^{(j)}\frac{dm_j}{m_j}, 
\nonumber\\
&&G^{(p,q)}{\equiv}G^{(p,q)i_1{\cdots}i_p}_{j_1{\cdots}j_p}  
=C^{i_1{\cdots}i_p}_{j_1{\cdots}j_p}
\Pi_{j=1}^{N_f}(m_j)^{(p+q)/N_c       
-\sum_{l=1}^p({\delta}_{i_l,j}+{\delta}_{j_l,j})/2}.
\label{eq202}
\end{eqnarray}
The last equation in (\ref{eq202}) can be expressed as follows,
\begin{eqnarray}
\Pi_{l=1}^p(m_{i_l}m_{j_l})^{1/2}G^{(p,q)i_1{\cdots}i_p}_{j_1{\cdots}j_p}  
=C^{i_1{\cdots}i_p}_{j_1{\cdots}j_p}(\mu, g)
\Pi_{j=1}^{N_f}(m_j)^{(p+q)/N_c},
\label{eq203}  
\end{eqnarray}
where after taking into account renormalization effects
we have written the integration constant $C$ as depending on
 the coupling constant $g$ and renormalization scale 
$\mu$ explicitly. We can use dimensional analysis to
determine the explicit dependence on $\mu$. Since $\mu$, $\lambda$ 
and $\phi$ ($\widetilde{\phi}$) have dimensions $1$, $3/2$ and $1$, 
respectively, so $G^{(p,q)}$ has dimension  $2p+3q$. 
Comparing both sides of (\ref{eq203}), we get
\begin{eqnarray}
C^{i_1{\cdots}i_p}_{j_1{\cdots}j_p}(\mu,g)=
C^{i_1{\cdots}i_p}_{j_1{\cdots}j_p}(g)\mu^{(p+q)(3-N_f/N_c)}.
\end{eqnarray}
Furthermore, since the left-hand side of (\ref{eq203}) is the 
vacuum expectation value of renormalization group invariant 
operators, its right-hand side  should also be expressed in terms
of the renormalization group invariant quantities:
\begin{eqnarray}
{\Lambda}={\mu}\mbox{exp}\left[-{\int}^g\frac{dg'}{\beta(g')}\right], ~~
m_{\rm inv}=m \mbox{exp}\left[-{\int}^g\frac{{\gamma}_m}{\beta(g')}dg'\right].
\end{eqnarray}
Hence (\ref{eq203}) can be rewritten as
\begin{eqnarray}
\Pi^p_{l=1}(m_{i_l}m_{j_l})^{1/2}
G^{(p,q)i_1{\cdots}i_p}_{j_1{\cdots}j_p}
=(\Lambda_{N_c,N_f})^{(p+q)(3-N_f/N_c)}
\left(\Pi_{l=1}^{N_f}m_{l~\rm inv}\right)^{(p+q)/N_c}
t^{(p,q)i_1{\cdots}i_p}_{j_1{\cdots}j_p},
\label{eq206}
\end{eqnarray}
where $t$ is a dimensionless constant tensor in flavour space, depending
only on $p$, $q$, the flavour number $N_f$ and colour number $N_c$.

So far $t$ is undetermined. Since $G^{(p,q)}$ is 
space-time independent, one can evaluate it
in the the limit $|x_i-x_j|{\longrightarrow}{\infty}$ for all $i{\neq}j$.
In this limit, if the vacuum is unique, the clustering condition
implies that the Green function factors into a product of the 
vacuum expectation values of each composite operator. 
Applying the clustering property to the Green function $G^{(p,q)}$, 
we obtain
\begin{eqnarray}
G^{(p,q)i_1{\cdots}i_p}_{j_1{\cdots}j_q}=
\left({\langle}\frac{g^2}{32{\pi}^2}{\lambda}{\lambda}{\rangle}\right)^q
\Pi_{l=1}^p{\langle}\widetilde{\phi}^{i_l}{\phi}_{j_l}{\rangle}.
\label{eq208}
\end{eqnarray}
Taking $p=0$, $q=1$ in (\ref{eq206}), we have
\begin{eqnarray}
{\langle}\frac{g^2}{32\pi^2}{\lambda}{\lambda}{\rangle}=
c_{\lambda}({\Lambda}_{N_c,N_f})^{3-N_f/N_c}
\left[\Pi_{l=1}^{N_f}m_{l\,{\rm inv}}\right]^{1/N_c},
\label{eq209}
\end{eqnarray}
while taking $q=0$, $p=1$ in (\ref{eq206}), we get 
\begin{eqnarray}
[(m_{i_l}m_{j_l})_{\rm inv}]^{1/2}
{\langle}\widetilde{\phi}^{i_l}{\phi}_{j_l}{\rangle}=
({\Lambda}_{N_c,N_f})^{3-N_f/N_c}
\left[\Pi_{l=1}^{N_f}m_{l\,{\rm inv}}\right]^{1/N_c}t^{i_l}_{~j_l}.
\label{eq210}
\end{eqnarray}
Inserting (\ref{eq208}) and (\ref{eq210}) into the left-hand side of 
(\ref{eq206}) and comparing with the right-hand side of (\ref{eq206}),
we see that the tensor $t^{(p,q)i_1,{\cdots},i_p}_{j_1,{\cdots},j_p}$ 
factorizes into a product of tensors $t^{i_l}_{~j_l}$,
\begin{eqnarray}
t^{(p,q)i_1{\cdots}i_p}_{j_1{\cdots}j_p}=\Pi_{l=1}^pt^{i_l}_{~j_l}.
\label{eq211}
\end{eqnarray}
Usually the vacuum is $\{U(1)\}^{N_f}$ invariant with $U(1)$ being
the rotation group in each flavour space, 
\begin{eqnarray}
Q_i|0{\rangle}=0, ~~i=1,{\cdots},N_f.
\end{eqnarray}
The operator ${\langle}\widetilde{\phi}^{i}{\phi}_{j}{\rangle}$ is also
$\{U(1)\}^{N_f}$ invariant and so is 
$t^i_{~j}$. Hence $t^{i_l}_{~j_l}$ should be
proportional to the identity matrix in flavour space,
\begin{eqnarray}
t^i_{~j}=c_{\phi}{\delta}^i_{~j}.
\label{eq213}
\end{eqnarray}
In (\ref{eq209}) and (\ref{eq213}), 
we have introduced two undetermined coefficients $c_{\lambda}$ and
$c_{\phi}$. From
the Konishi anomaly, which we introduce next, 
we can see they are in fact identical. First we 
have to explain this anomaly and then discuss its consequences.

\subsubsection{Konishi anomaly}
\label{subsub6.2.3}

The Konishi anomaly is another important characteristic
 of supersymmetric gauge theory. In operator form, the 
anomaly equation is \cite{ref1p20}, 
\begin{eqnarray}
\frac{1}{2\sqrt{2}}
\{\overline{Q}_{\dot{\alpha}},\overline{\psi}^{\dot{\alpha}i}(x){\phi}_j(x)\}
=-m_i\widetilde{\phi}^i{\phi}_j(x)+\frac{g^2}{32{\pi}^2}{\lambda}{\lambda}(x)
{\delta}^i_{~j}.
\label{eq214}
\end{eqnarray} 
A naive supersymmetric gauge transformation gives only the first 
term (see (\ref{eq187})). The second term on the right-hand 
side of the above equation is the
anomalous term. This anomaly equation can generate a series of
anomalous Ward identities when it is inserted into the operators
of various Green functions.

\begin{figure}
\begin{center}
\leavevmode
{\epsfxsize=3.00truein \epsfbox{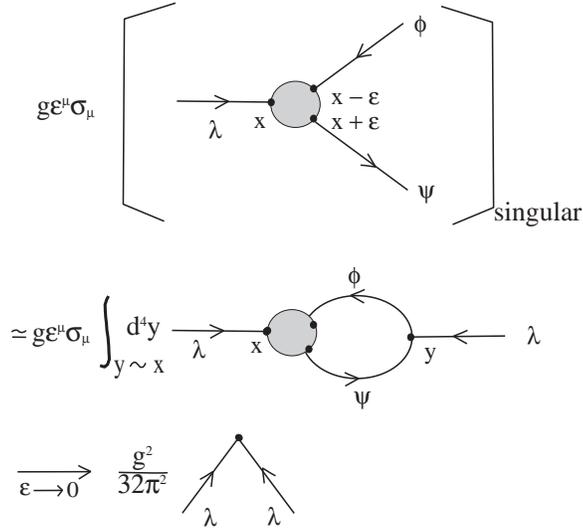}}
\caption{\protect\small Feynman diagram for Konishi anomaly.
\label{fig6.2}}
\end{center}
\end{figure}

To prove this anomaly equation, we first look at the composite operator
$m\widetilde{\phi}^i(x){\phi}_j(x)$. Similarly to regularizing
operator products in a gauge invariant way
by point-splitting \cite{wtt}, one can write it in a gauge-invariant 
non-local operator form. To keep supersymmetry manifest, it is better to 
work with superfields. Defining
\begin{eqnarray}
O^i_{~j}(x,u,\theta,\overline{\theta})&{\equiv}&m
\widetilde{Q}^{ir}(x,\theta,\overline{\theta})
U_r^{~s}(x,u,\theta,\overline{\theta})
Q_{js}(u,\theta,\overline{\theta}),\nonumber\\[2mm]
U_r^{~s}(x,u,\theta,\overline{\theta})&{\equiv}&
P\mbox{exp}\left(\frac{i}{4}
{\int}_u^xdz^{\mu}
\overline{\sigma}_{\mu}^{\dot{\beta}\alpha}\overline{D}_{\dot{\beta}}
e^{-V}D_{\alpha}e^V\right)_r^{~s},
\end{eqnarray}
where $P$ denotes path ordering. In a supersymmetric gauge
choice \cite{so}, the superspace component $A_{\dot{\alpha}}$ 
of the super-gauge potential vanishes, and 
$-1/4\overline{\sigma}_{\mu}^{\dot{\beta}\alpha}\overline{D}_{\dot{\beta}}
e^{-V}D_{\alpha}e^V$ is the usual Yang-Mills field $A_{\mu}$. One can easily
show that $O(x,u,\theta,\overline{\theta})$ is indeed gauge invariant under the
super-local gauge transformation,
\begin{eqnarray} 
{Q}{\rightarrow}e^{-i{\Lambda}}{Q},~~\widetilde{Q}^T{\rightarrow}
\widetilde{Q}^Te^{i{\Lambda}}, ~~e^V=e^{-i{\Lambda}^{\dagger}}e^V
e^{i{\Lambda}},
\end{eqnarray}
where $\Lambda (x,\theta,\overline{\theta})$ is an 
arbitrary chiral superfield. Correspondingly, 
$\overline{\sigma}_{\mu}^{\dot{\beta}\alpha}\overline{D}_{\dot{\beta}}
e^{-V}D_{\alpha}e^V$ transforms as follows:
\begin{eqnarray}
\overline{\sigma}_{\mu}^{\dot{\beta}\alpha}\overline{D}_{\dot{\beta}}
e^{-V}D_{\alpha}e^V {\rightarrow}e^{-i\Lambda}
\overline{\sigma}_{\mu}^{\dot{\beta}\alpha}(\overline{D}_{\dot{\beta}}
e^{-V}D_{\alpha}e^V+\overline{D}_{\dot{\beta}}D_{\alpha})e^{i{\Lambda}}.
\label{eq217}
\end{eqnarray}
Using the fact that
\begin{eqnarray}
\overline{\sigma}_{\mu}^{\dot{\beta}\alpha}
(\overline{D}_{\dot{\beta}}D_{\alpha}){\Lambda}
=\overline{\sigma}_{\mu}^{\dot{\beta}\alpha}
\{\overline{D}_{\dot{\beta}},D_{\alpha}\}{\Lambda} 
=-\overline{\sigma}_{\mu}^{\dot{\beta}\alpha}
(2i{\sigma}^{\nu}_{\alpha\dot{\beta}}{\partial}_{\nu}){\Lambda}
=-4i{\partial}_{\mu}{\Lambda},
\end{eqnarray}
where the definition of a chiral superfield, 
$\overline{D}{\Lambda}=0$, has been used,
one can discard the second term of (\ref{eq217}) in the integration, 
and thus $O(x,u,\theta,\overline{\theta})$ is gauge 
invariant. Now we concentrate
on the lowest component of the superfield operator 
$O(x,u,\theta,\overline{\theta})$.
In the Wess-Zumino gauge, 
\begin{eqnarray}
\overline{\sigma}_{\mu}^{\dot{\beta}\alpha}\overline{D}_{\dot{\beta}}
e^{-V}D_{\alpha}e^{V}
|_{\theta=\overline{\theta}=0}=-\overline{\sigma}_{\mu}^{\dot{\beta}\alpha}
{\sigma}^{\nu}_{\alpha\dot{\beta}}A_{\nu}=-2A_{\mu}(x),
\end{eqnarray}
so we have
\begin{eqnarray}
O^i_{~j}(x,u,\theta
=\overline{\theta}=0)|_{\rm WZ~gauge}&=&\widetilde{\phi}^{ir}(x)\left[
P\mbox{exp}\left(-\frac{i}{2}{\int}_u^xdz^{\mu}A_{\mu}(z)\right)
\right]_r^{~s}{\phi}_{js}(u)\nonumber\\[2mm]
&{\equiv}&\widetilde{\phi}^{ir}(x)U_r^{~s}(x,u){\phi}_{js}(u).
\label{eq220}
\end{eqnarray}
This is just the ordinary path-ordered integral for gauge invariant
non-local operators. 
We can now define the local product $\widetilde{\phi}^i(x){\phi}_j(x)$ as
\begin{eqnarray}
\widetilde{\phi}^i(x){\phi}_j(x){\equiv}\lim_{\epsilon{\rightarrow}0}
O^i_{~j}(x+{\epsilon},x-{\epsilon},\theta=\overline{\theta}=0)
|_{\rm WZ~gauge}.
\end{eqnarray}
Using the classical equation of motion $\widetilde{F}^{ir}(x)
=-m\widetilde{\phi}^{ir}(x)$, 
one finds 
\begin{eqnarray}
m\widetilde{\phi}^i(x){\phi}_j(x)&=&-\lim_{\epsilon{\rightarrow}0}F^{*ir}(x+
{\epsilon})U_r^{~s}(x+{\epsilon},x-{\epsilon}){\phi}_{js}
(x-{\epsilon})\nonumber\\   
&=&-\frac{1}{2\sqrt{2}}\lim_{\epsilon{\rightarrow}0}
\left\{\overline{Q},\overline{\psi}^{ir}(x+\epsilon)\right\}
U_r^{~s}(x+{\epsilon},x-{\epsilon}){\phi}_{js}(x-{\epsilon})   
\nonumber\\
&=&-\frac{1}{2\sqrt{2}}\lim_{\epsilon{\rightarrow}0}
\left[\{\overline{Q},\overline{\psi}^{ir}
(x+\epsilon)U_r^{~s}(x+{\epsilon},x-{\epsilon}){\phi}_{js}(x-{\epsilon})\}
\right.\nonumber\\[2mm]
&&-\left.{\epsilon}^{\mu}\overline{\psi}^{\dot{\alpha}ir}(x+\epsilon)
{\epsilon}_{\dot{\alpha}\dot{\beta}}\overline{\sigma}_{\mu}^{\dot{\beta}\alpha}
{\lambda}_{\alpha r}^{~~s}(x){\phi}_{js}(x-{\epsilon})\right],
\label{eq222}
\end{eqnarray}
where we have used the notation $\lambda_{\alpha r}^{~s}=\lambda^a_{\alpha}
\left(T^a\right)_r^{~~s}$ and the supersymmetry transformations
\begin{eqnarray}
[\overline{Q}_{\dot{\alpha}},A^a_{\mu}]=-i{\epsilon}_{\dot{\alpha}\dot{\beta}}
\overline{\sigma}_{\mu}^{\dot{\beta}\alpha}{\lambda}^a_{\alpha}(x), ~~
[\overline{Q}_{\dot{\alpha}},{\phi}]=0.
\end{eqnarray}
It is possible that the second term does not vanish in the limit ${\epsilon}
{\rightarrow}0$ since there is a Yukawa interaction vertex 
$i{\phi}^{\dagger}{\lambda}{\psi}/\sqrt{2}$ in the Lagrangian
(\ref{eq163}), so that  
$\overline{\psi}(x+{\epsilon}){\lambda}(x){\phi}(x-{\epsilon})$ contains
a linear singularity $\sim ({\epsilon}^{\nu}/{\epsilon}^2)$. One can see
this from simple dimensional analysis:
\begin{eqnarray}
\lambda (x){\int}d^4y{\langle}T[\psi (y)\overline{\psi}(x+{\epsilon})]{\rangle}
{\lambda}(y) {\langle}T[{\phi}(x-{\epsilon}){\phi}^{\dagger}(y)]{\rangle}
{\sim}{\lambda}^2(x){\epsilon}^4{\epsilon}^{-3}{\epsilon}^{-2}
{\sim}{\lambda}^2(x){\epsilon}^{-1}.
\end{eqnarray}
The exact form of the second term of (\ref{eq222}) can be obtained
from a simple Feynman diagram calculation in momentum 
space (see Fig.\,\ref{fig6.2}),
\begin{eqnarray}
-i\frac{1}{4}g^2{\epsilon}_{\dot{\alpha}\dot{\beta}}
\overline{\sigma}_{\mu}^{\dot{\beta}\alpha}{\lambda}_{\alpha r}^{~~s}(x)
{\lambda}^{\beta~~r}_{~s}(x)
{\int}\frac{d^4k}{(2\pi)^4}\frac{\partial}{{\partial}k^{\mu}}\left[
\frac{{\sigma}^{\nu}k_{\nu}}{(k^2-m^2)^2}\right]_{\beta}^{~\dot{\alpha}}
=\frac{g^2}{32{\pi}^2}{\lambda}^a(x){\lambda}^a(x).
\label{eq6.64}
\end{eqnarray}


Combining (\ref{eq6.64}) with (\ref{eq222}), one can see that
this gives (\ref{eq214}). It is worth mentioning that
the superfield form of (\ref{eq214}) is
\begin{eqnarray}
\frac{1}{4}\overline{D}^2({Q}^{\dagger i}e^{gV}{Q}_j)
=-m_i\widetilde{Q}^i{Q}_j+
\frac{g^2}{32{\pi}^2}W^{\alpha}W_{\alpha}{\delta}^i_{~j},
\end{eqnarray}
whose lowest component (${\theta}=\overline{\theta}=0$) 
is just the Konishi anomaly
equation. $\Sigma={Q}^{\dagger}e^{gV}{Q}$ is called
the Konishi supercurrent superfield, which plays an important role
in the operator product expansion in 4-dimensional superconformal
field theory \cite{ans,ans1}. If one expands the above superfield equation, 
one can see that the ${\theta}^2$ component is just 
the usual $U_A(1)$ anomalous equation. Hence the Konishi 
current (the lowest component of the Konishi supercurrent) is 
the superpartner of the $U_A(1)$ current. In this sense, the 
existence of the Konishi anomaly is not strange since the 
anomalies also form a supermultiplet. Later when we discuss 
the superconformal current multiplet, we shall return to this equation.  

Now we see the consequence of the Konishi anomaly equation.
Since the supersymmetry is not spontaneously broken for 
$m{\neq}0$ \cite{wit1,wit2}, the operator anomaly equation implies 
\begin{eqnarray}
m_i{\langle}\widetilde{\phi}^i{\phi}_i{\rangle}=
{\langle}\frac{g^2}{32{\pi}^2}{\lambda}{\lambda}{\rangle}, ~~
{\langle}\widetilde{\phi}^i{\phi}_j{\rangle}=0,~~i{\neq}j.
\end{eqnarray}
Therefore from (\ref{eq206}), (\ref{eq209}), (\ref{eq210}), (\ref{eq211}) 
and (\ref{eq213}), we have 
\begin{eqnarray}
c_{\lambda}=c_{\phi},
\end{eqnarray}
and hence
\begin{eqnarray}
(m_{i_l}m_{j_l})_{\rm inv}^{1/2} 
{\langle}\widetilde{\phi}^{i_l}{\phi}_{j_l}{\rangle}
=\frac{c_{\lambda}}{32{\pi}^2}({\Lambda}_{N_c,N_f})^{3-N_f/N_c}\left(
\Pi_{l=1}^{N_f}(m_l)_{\rm inv}\right)^{1/N_c}{\delta}^{i_l}_{~j_l}.
\end{eqnarray}
Thus the Green function $G^{(p,q)}$ can be specified by a single numerical
constant $c_{\lambda}$. Once  $G^{(p,q)}$ is determined, most of the
other Green functions can also be determined through 
supersymmetric Ward identities.

\subsubsection{Decoupling relation}
\label{subsub6.2.4}

Finally we briefly introduce the decoupling theorem in supersymmetric QCD
since it will be widely used in discussing the reasonableness of the 
non-Abelian electric-magnetic duality \cite{ref1p10}.
The decoupling theorem is a general result in field theory, it describes
the effects of the heavy particles in the low energy theory \cite{dgh}. 
In general this theorem states that {\it if, after integrating out the 
heavy particles, the remaining low energy
theory is renormalizable,  the effects of the heavy
particles appear either as a renormalization of the coupling constants
in the theory or are suppressed by powers of the heavy particle masses}. 
In electroweak model,
we have some obvious examples such as the decoupling of the heavy
$W^{\pm}$ and $Z$, their effects at low energy either renormalize the
 electric charge or are suppressed. Now we apply 
this decoupling theorem to supersymmetric QCD. We assume that one flavour, 
say the $N_f$-th flavour, becomes heavy, i.e. $m_{N_f}{\gg}{\Lambda}$. 
In this large-$m_{N_f}$ limit,
\begin{eqnarray}
{\langle}\frac{g^2}{32{\pi}^2}{\lambda}{\lambda}{\rangle}_{N_c,N_f}
=c_{\lambda}(N_c,N_f)({\Lambda}_{N_c,N_f})^{3-N_f/N_c}\left[\Pi_{l=1}^{N_f}
(m_l)_{\rm inv}\right]^{1/N_c}\nonumber\\[2mm]
\stackrel{m_{N_f}{\gg}{\Lambda}_{N_c,N_f}}{\longrightarrow}
c_{\lambda}(N_c,N_f-1)({\Lambda}_{N_c,N_f-1})^{3-(N_f-1)/N_c}
\left[\Pi_{l=1}^{N_f-1}(m_l)_{\rm inv}\right]^{1/N_c},
\label{eq230}
\end{eqnarray}
where the explicit flavour number $N_f$ and colour number $N_c$ dependence
of $c_{\lambda}$ is indicated in order to show the decoupling of the 
$N_f$-th heavy flavour. ${\Lambda}_{N_c,N_f-1}$ is the renormalization group 
invariant scale of supersymmetric $SU(N_c)$ QCD with $N_f-1$ flavours. 
We shall see that the  effect of the $N_f$th heavy
flavour is reflected in $\Lambda_{N_c,N_f-1}$, 
since the running coupling constant
depends on it. At the scale that the 
decoupling takes place, these two theories should  coincide. This means that
the coupling constants $g_{N_f}(q^2)$ and $g_{N_f-1}(q^2)$ should
be identical at the scale $q^2=m_{N_f}^2$,
\begin{eqnarray}
g^2_{N_f}(m_{N_f}^2)=g^2_{N_f-1}(m_{N_f}^2).
\end{eqnarray}
From the one-loop $\beta$-function of $N=1$ supersymmetric QCD,
\begin{eqnarray}
{\beta}_{N_c,N_f}&=&-\frac{g^2}{16\pi^2}(3N_c-N_f)
=-\frac{g^2}{16\pi^2}{\beta}_0,
\end{eqnarray}
and the running coupling constant
\begin{eqnarray}
\frac{4\pi}{g^2(q^2)}=\frac{\beta_0}{4\pi}\ln\frac{q^2}{\Lambda^2},
\end{eqnarray}
we obtain at $q^2=m^2_{N_f}$
\begin{eqnarray}
(3N_c-N_f)\ln\frac{m_{N_f}}{{\Lambda}_{N_c,N_f}}=
[3N_c-(N_f-1)]\ln\frac{m_{N_f}}{{\Lambda}_{N_c,N_f-1}}.
\end{eqnarray}
Thus
\begin{eqnarray}
{\Lambda}_{N_c,N_f-1}={\Lambda}_{N_c,N_f}
\left(\frac{m_{N_f}}{{\Lambda}_{N_c,N_f}}\right)^{1/(3N_c-N_f+1)}.
\end{eqnarray}
Later we shall see that this relation imposes a restrictive constraint on
 the form of the non-perturbative superpotential. This relation in fact
gives a link between the energy scales of theories with different number
of flavours. 

\subsection{Classical moduli space of supersymmetric QCD}

\renewcommand{\theequation}{3.3.\arabic{equation}}
\setcounter{equation}{0}
\renewcommand{\thetable}{3.3.\arabic{table}}
\setcounter{table}{0}

\label{subsect6.3}

\subsubsection{Classical moduli space}
\label{subsub6.3.1}

A field theory, be it classical, quantum or a low energy effective theory, 
is in general one of a whole family of theories, parametrized by a number
of parameters. Especially important for us are the vacuum expectation 
values of scalar fields, called {\it moduli}, which can range over a
moduli space.

 Lets us illustrate this concept by considering the classical moduli 
space of a simple theory, the Georgi-Glashow model with gauge group
$SO(3)$. The classical action is
\begin{eqnarray}
S={\int}d^4x\left[-\frac{1}{4}G^{a\mu\nu}G^a_{\mu\nu}+\frac{1}{2}
{\cal D}^{\mu}{\varphi}^{a}{\cal D}_{\mu}{\varphi}^a
-\frac{\lambda}{4}({\varphi}^2-a^2)^2\right],
\label{eqc1}
\end{eqnarray}
where $G_{\mu\nu}^a={\partial}_{\mu}W_{\mu}^a-{\partial}_{\nu}W_{\mu}^a
+g{\epsilon}^{abc}W_{\mu}^bW_{\mu}^c$, 
the Higgs field $\phi$ is in the adjoint (vector)
representation of gauge group of $SO(3)$, 
${\cal D}_{\mu}{\varphi}^a={\partial}_{\mu}
{\varphi}^a+g{\epsilon}^{abc}A_{\mu}^b{\phi}^c$ and
$\varphi^2=\phi^a\varphi^a$. The classical ground sates are given by
$W_\mu^a=0$ (up to a gauge transformation) and
\begin{eqnarray}
\varphi^2=a^2.
\label{eqc2}
\end{eqnarray}
The classical moduli space is thus the 2-sphere (\ref{eqc2}). Each point 
on this space defines a  semiclassical quantum field theory with the chosen
point being the expectation value of the Higgs field in the vacuum 
state of the theory. The (semi)classical dynamics of all these
theories are equivalent.

Quantum effects will, in general, change the dependence of the (effective)
theory on the moduli, and might even alter the topology of the moduli space.
For the Georgi-Glashow model, the structure of the quantum moduli space
is not known, but in supersymmetric theories we can often make definite 
statement about the quantum moduli space.

\subsubsection{Classical moduli space of supersymmetric QCD}
\label{subsub6.3.2}

We now turn to supersymmetric QCD. We first consider the classical 
moduli space and then turn to the quantum case.

Recall the Lagrangian (\ref{eq163}) of supersymmetric QCD. 
The scalar potential is\footnote{To emphasize that a field is a
component of the chiral superfield $Q$ ($\widetilde{Q}$), we add
an index $Q$ ($\widetilde{Q}$).}
\begin{eqnarray}
V&=&\frac{g^2}{2}D^aD^a,\nonumber\\[2mm]
D^a&=&{\phi}^{* r}_Q(T^a)_r^{~s}{\phi}_{Qs}-
\widetilde{\phi}^r_{\widetilde{Q}}(T^a)_r^{~s}
\widetilde{\phi}_{\widetilde{Q}s}^{*}.
\label{eq3419x}
\end{eqnarray}
Since the chiral superfields $Q$ and $\widetilde{Q}$ 
have both colour and flavour indices, one can arrange them 
in the form of an $N_c{\times}N_f$ matrix according to flavour 
and colour indices:
\begin{eqnarray}
(Q)=\left(\begin{array}{ccc}
{Q}_1^1 &{\cdots} &{Q}_1^{N_f}\\
{\vdots}   & {\ddots}   &{\vdots}   \\
{Q}_{N_f}^1 &{\cdots} &{Q}_{N_f}^{N_f}\\
{\vdots}   & {\ddots}   &{\vdots}   \\
{Q}_{N_c}^1 &{\cdots} &{Q}_{N_c}^{N_f}\end{array}\right), ~~
(\widetilde{Q})=\left(\begin{array}{ccc}
\widetilde{Q}_1^1 &{\cdots} &\widetilde{Q}_1^{N_f}\\
{\vdots}   & {\ddots}   &{\vdots}   \\
\widetilde{Q}_{N_f}^1 &{\cdots} &\widetilde{Q}_{N_f}^{N_f}\\
{\vdots}   & {\ddots}   &{\vdots}   \\
\widetilde{Q}_{N_c}^1 &{\cdots} &\widetilde{Q}_{N_c}^{N_f}\end{array}\right).
\end{eqnarray}
Both $(Q)$ and $(\widetilde{Q})$ can be globally rotated in the colour
and flavour spaces to make them diagonal since the Lagrangian
is globally $SU(N_c)$ and $SU(N_f)$ invariant.
The $D$-flatness condition $D^a=0$ does not lead to
$\phi_{Q}=\phi_{\widetilde{Q}}=0$. 
This is because, unlike ordinary QCD, supersymmetric 
QCD is very sensitive to the
relative number of flavours $N_f$ and colours $N_c$. 
Depending on the numbers of flavour and
colour, the restrictions on $Q$ and ${\widetilde{Q}}$ posed by 
the $D$-flatness conditions are different. In the following we
give a detailed analysis of the classical moduli spaces
for different ranges of $N_f$ and $N_c$.

\vspace{4mm}

\begin{flushleft}
{\it $N_f < N_c$}
\end{flushleft}

\vspace{4mm}

In this case, since the chiral superfield matrix $Q$ and $\widetilde{Q}$ can be
rotated in colour space and flavour space, 
we can always reduce the matrices $Q$ and $\widetilde{Q}$ 
to the following diagonal forms,
\begin{eqnarray}
(Q)=\left(\begin{array}{cccc}
a_1 & 0 &{\cdots} &0\\
0 & a_2 &{\cdots} & 0\\
{\vdots} &{\vdots} &{\ddots} & {\vdots}\\
0 & 0 & {\cdots} &a_{N_f}\\
0 & 0 & {\cdots} &0\\
{\vdots} &{\vdots} &{\ddots} &\vdots\\
0 & 0 & {\cdots} &0\end{array}\right) &,& ~~ 
(\widetilde{Q})=\left(\begin{array}{cccc}
\widetilde{a}_1 & 0 &{\cdots} &0\\
0 & \widetilde{a}_2 &{\cdots} & 0\\
{\vdots} &{\vdots} &{\ddots} & {\vdots}\\
0 & 0 & {\cdots} &\widetilde{a}_{N_f}\\
0 & 0 & {\cdots} &0\\
{\vdots} &{\vdots} &{\ddots} &\vdots\\
0 & 0 & {\cdots} &0\end{array}\right), 
\nonumber\\
a_1, {\cdots}, a_{N_f}{\neq}0 &,&
~~\widetilde{a}_1, {\cdots}, \widetilde{a}_{N_f}{\neq}0.
\label{eq6.91}
\end{eqnarray}
The $D$-flatness condition in this case requires 
\begin{eqnarray}
\phi_{Q}=\widetilde{\phi}_{\widetilde{Q}}^{\dagger}, 
\label{eq6.96}
\end{eqnarray}
whose superfield form is
\begin{eqnarray}
Q=\widetilde{Q}.
\label{eq6.97}
\end{eqnarray}
If $a_{i}=\widetilde{a}_i{\neq}0$,
the gauge symmetry ($SU(N_c)$ symmetry) will be spontaneously broken and the 
super-Higgs mechanism will occur: some of the chiral superfields will be
eaten up and the same number of gauge fields and their partners will 
become massive.

Now the problem is what kind of quantity 
describes this $D$-flat moduli space. From the viewpoint of 
dynamics, this means how the low energy dynamics is 
described. According to the idea of the effective 
field theory \cite{mon}, 
a general method to get the low energy effective action is
to integrate out the heavy modes (massive fields), the light modes (massless
fields) being the quantities describing the 
low energy dynamics. Therefore, for the case at hand, a 
$SU(N_c)$ global gauge invariant 
quantity would be
\begin{eqnarray}
M^i_{~j}{\equiv}\sum_{r=1}^{N_c}
\widetilde{Q}_{jr}Q^{ri}{\equiv}\widetilde{Q}_jQ^i,~~
i,j=1,{\cdots},N_f.
\end{eqnarray}
In the following
we give further arguments why this assumption is reasonable.                  

(\ref{eq6.91}) and (\ref{eq6.97}) give  
\begin{eqnarray}
M^i_{~j}=\sum_r a_i\delta_{ri}\widetilde{a}_j\delta_{rj}=
a_i\widetilde{a}_j{\delta}_{ij}=a_i^2{\delta}_{ij}.
\label{eq6.92}
\end{eqnarray}
(\ref{eq6.92}) can be written in the explicit matrix form
\begin{eqnarray}
(M^i_{~j})=\left(\begin{array}{cccc}
a^2_1 & 0 &{\cdots} &0\\
0 & a^2_2 &{\cdots} & 0\\
{\vdots} &{\vdots} &{\ddots} &\vdots \\
0 & 0 & {\cdots} &a^2_{N_f}\end{array}\right).
\end{eqnarray}
Furthermore, (\ref{eq6.91}) 
and (\ref{eq6.96}) imply that
 the gauge symmetry $SU(N_c)$
is broken to $SU(N_c-N_f)$. The original gauge group has $N_c^2-1$ generators 
and the remaining gauge group has $(N_c-N_f)^2-1$ generators, the number of 
eaten chiral superfields is thus $[N_c^2-1]-[(N_c-N_f)^2-1]$$=2N_cN_f-N_f^2$.
This number is also the number of particles becoming massive. After these
massive particles have been integrated out, only the massless particles are
left. The number of the original matter fields 
is $2N_fN_c$ ($Q_{ir}$, $\widetilde{Q}_{ir}$, $i=1,{\cdots},N_f$, 
$r=1,{\cdots},N_c$), so there are $2N_fN_c-(2N_cN_f-N_f^2)$
$=N_f^2$ massless particles left. This is exactly the number of degrees of
freedom of $M_{~j}^{i}$, thus we can use $M_{~j}^{i}$ to describe the 
moduli space. Later we shall return to the dynamics.

\vspace{4mm}

\begin{flushleft}
{\it $N_f{\geq}N_c$}
\end{flushleft}

\vspace{4mm}

In this case, after an appropriate rotation in flavour
space and colour space, the $(Q)$ and $(\widetilde{Q})$ matrix 
takes the following diagonal form,
\begin{eqnarray}
&& (Q)=\left(\begin{array}{ccccccc}
a_1 & 0 &{\cdots} &0 &0 & {\cdots}& 0\\
0 & a_2 &{\cdots} & 0 & 0 &{\cdots} & 0\\
{\vdots} &{\vdots} &{\ddots} &{\vdots} &{\vdots} &{\ddots}& {\vdots}  \\
0 & 0 & {\cdots} &a_{N_c} & 0 & {\cdots} &0\\
\end{array}\right), ~~
(\widetilde{Q})=\left(\begin{array}{ccccccc}
\widetilde{a}_1 & 0 &{\cdots} &0 &0 & {\cdots}& 0\\
0 & \widetilde{a}_2 &{\cdots} & 0 & 0 &{\cdots} & 0\\
{\vdots} &{\vdots} &{\ddots} &{\vdots} &{\vdots} &{\ddots}&{\vdots}  \\
0 & 0 & {\cdots} &\widetilde{a}_{N_c} & 0 & {\cdots} &0\\
\end{array}\right),\nonumber\\[2mm]
&&a_1, {\cdots}, a_{N_c}{\neq}0, 
~~\widetilde{a}_1, {\cdots}, \widetilde{a}_{N_c}{\neq}0.
\label{eq6.93}
\end{eqnarray}
Let us analyze the $D$-flatness condition,
\begin{eqnarray}
D^a&=&\sum_{i=1}^{N_f}\sum_{r,s=1}^{N_c}[\phi_{Qi}^{*r}
(T^a)_r^{~s}\phi_{Qis}-\widetilde{\phi}_{\widetilde{Q}i}^r
(T^a)_r^{~s}\phi_{\widetilde{Q}is}^*]\nonumber\\
&=&\sum_{i=1}^{N_c}\sum_{r,s=1}^{N_c}[\phi^*_{ai}{\delta}_{i}^{~r}
(T^{'a})_r^{~s}\phi_{ai}
{\delta}_{is}-\widetilde{\phi}_{\widetilde{a}i}
{\delta}_{i}^{~r}(T^{'a})_r^{~s}\widetilde{\phi}^*_{\widetilde{a}i}
{\delta}_{is}]\nonumber\\
&=& \sum_{i=1}^{N_c}[|\phi_{ai}|^2(T^{'a})_i^{~i}
-|\widetilde{\phi}_{\widetilde{a}i}|^2(T^{'a})_i^{~i}]  
 =\sum_{i=1}^{N_c}(|\phi_{ai}|^2-|
\widetilde{\phi}_{\widetilde{ai}}|^2)(T^{'a})_i^{~i},
\end{eqnarray}
where $\phi_{ai}$ ($\widetilde{\phi}_{\widetilde{ai}}$) 
is the scalar component of chiral superfield $a_i$,
 $T^{'a}$ is the  rotated $T^a$ in colour space, 
$T^{'a}=U^{\dagger}T^aU$ with $U$ being certain unitary 
matrix. So the the condition defining
a $D$-flat configuration is
\begin{eqnarray}
&&|\phi_{ai}|^2-|\widetilde{\phi}_{\widetilde{ai}}|^2=C~~ (\mbox{constant}),
\nonumber\\[2mm]
&&D^a=C\mbox{Tr}T^{'a}=C\mbox{Tr}T^{a}=0,
\label{eq6.103x}
\end{eqnarray}
where we have used the fact that $T^a$ is the generator  of gauge
group $SU(N_c)$. Since $N_f{\geq}N_c$, the possible gauge invariant chiral
superfield operators parameterizing the moduli space are not only 
the meson-type chiral superfield operators, but
also the baryon-type chiral field operators:
\begin{eqnarray} 
M_{~j}^{i}&=&\widetilde{Q}_j{\cdot}Q^i, \nonumber\\[2mm]
B^{i_1{\cdots}i_{N_c}}&=&\frac{1}{N_c!} 
{\epsilon}^{r_1{\cdots}r_{N_c}}Q^{i_1}_{r_1}{\cdots}Q^{i_{N_c}}_{r_{N_c}}
=Q^{[i_1}{\cdots}Q^{i_{N_c}]},
\nonumber\\[2mm] 
\widetilde{B}^{j_1{\cdots}j_{N_c}}&=&\frac{1}{N_c!} 
{\epsilon}^{s_1{\cdots}s_{N_c}}\widetilde{Q}^{j_1}_{s_1}{\cdots}
\widetilde{Q}^{j_{N_c}}_{s_{N_c}}
=\widetilde{Q}^{[j_1}{\cdots}\widetilde{Q}^{j_{N_c}]},
\label{eq259}
\end{eqnarray}
where $i_k,j_k=1,{\cdots},N_f$ are flavour indices and $r_p, s_p=1,{\cdots},
N_c$ are colour indices. Since the flavour indices in the baryons are
antisymmetric, the number of $B^{i_1{\cdots}i_{N_c}}$ 
(or $\widetilde{B}^{i_1{\cdots}i_{N_c}}$) fields is
\begin{eqnarray}
C^{N_c}_{N_f}=\frac{N_f!}{N_c!(N_f-N_c)!}.
\end{eqnarray}
We shall see that not all of $M$, $B$ and $\widetilde{B}$
can be independently used to label the moduli space, there are
additional constraints imposed on them.
We first consider two simple cases.

\vspace{2mm}

\begin{flushleft}
a. {\it $N_f=N_c$}
\end{flushleft}

\vspace{2mm}

Since now the $SU(N_c)$ 
gauge symmetry is completely broken, the number of the 
eaten particles (also the number of particles becoming massive)
is $N_c^2-1$$=N_f^2-1$. The original number of chiral superfields 
is $2N_fN_c=2N_f^2$, so the dimension of moduli space is
$2N_f^2-(N_f^2-1)$ $=N_f^2+1$ $=N_c^2+1$. However, from (\ref{eq259}), 
the number of mesons and baryons is $N_f^2+2$, so they are not independent
in parameterizing the moduli space. From the definitions of these mesons
and baryons, one can easily see that they satisfy the constraint
\begin{eqnarray}
{\det}M=B\widetilde{B}.
\label{eq6.99}
\end{eqnarray}
This equation is obvious from the the diagonal form (\ref{eq6.93}),
 since $N_f=N_c$, $B$ ($\widetilde{B}$) has only one nonvanishing
component,
\begin{eqnarray}
M&=&\widetilde{q}^Tq,~~ 
M^i_{~j}=\sum_r\widetilde{a}_i{\delta}_{ir}a_j{\delta}_{rj}
=\widetilde{a}_ia_j{\delta}_{ij},\nonumber\\
\det M&=&\Pi_{i=1}^{N_f}\widetilde{a}_ia_i
=\widetilde{a}_1a_1{\cdots}\widetilde{a}_{N_c}a_{N_c},~~
B=a_1{\cdots}a_{N_c},~~
\widetilde{B}=\widetilde{a}_1{\cdots}\widetilde{a}_{N_c}.
\end{eqnarray}
We shall see that at the quantum level, the constraint (\ref{eq6.99}) 
will be modified owing to the non-perturbative quantum 
correction coming from instantons.

\vspace{2mm}

\begin{flushleft}
b. {\it $N_f=N_c+1$}
\end{flushleft}

\vspace{2mm}

In this case, like in $N_f=N_c$, the $SU(N_c)$ gauge symmetry
is also completely broken. From (\ref{eq6.93}), 
the number of the eaten superfields is $N_c^2-1$$=(N_f-1)^2-1$,
the number of the original chiral superfields
is still $2N_fN_c=2N_f(N_f-1)$, so the number of massless particles is
$2N_f(N_f-1)-[(N_f-1)^2-1]=N_f^2$. However, the number 
of parameters describing  the  moduli space
is $N_f^2+2C_{N_f}^{N_f-1}$$=N_f^2+2N_f$, so  $2N_f$ chiral
variables in $M$, $B$ and $\widetilde{B}$ are redundant. However,
one can exactly find $2N_f$ constraints to remove them.
First we write the baryon operators in (\ref{eq259}) in 
their Hodge dual form
\begin{eqnarray}
\overline{B}_{i_{N_c+1}i_{N_c+2}{\cdots}i_{N_f}}&{\equiv}&
\frac{1}{(N_f-N_c)!}
{\epsilon}_{i_1{\cdots}i_{N_c}i_{N_c+1}{\cdots}i_{N_f}}
{B}^{i_{N_1}{\cdots}i_{N_c}}, \nonumber\\
\overline{\widetilde{B}}^{i_{N_c+1}i_{N_c+2}{\cdots}i_{N_f}}&{\equiv}&
\frac{1}{(N_f-N_c)!}{\epsilon}^{i_1{\cdots}i_{N_c}i_{N_c+1}{\cdots}
i_{N_f}}B_{i_{N_1}{\cdots}i_{N_c}}.
\label{eq263} 
\end{eqnarray}
For the case at hand $N_f=N_c+1$, the baryon operators are $B_i$
($\widetilde{B}_i$). With the definition (\ref{eq263}), one can 
easily find that they satisfy the constraints
\begin{eqnarray}
\overline{B}_iM^i_{~j}=0, ~~M^i_{~j}\overline{\widetilde{B}}^j=0,
~~~\widetilde{M}^i_{~j}=B_j\widetilde{B}^i,
\label{eq265}
\end{eqnarray} 
where $\widetilde{M}$ is the matrix whose elements 
are defined as $(-1)^{i+j}{\times}$ the determinant of the matrix obtained
from $M$ by deleting the $i$-th row and the $j$-th column.
(\ref{eq265}) can be written in the following explicit form:
\begin{eqnarray}
\frac{1}{N_c!}{\epsilon}^{i_1{\cdots}i_{N_c}i}
{\epsilon}_{j_1{\cdots}j_{N_c}j}M_{~i_1}^{j_1}{\cdots}
M_{~i_{N_c}}^{j_{N_c}}=B_j\widetilde{B}^i.
\label{eq266}
\end{eqnarray}
 The constraint equations can be checked easily,
\begin{eqnarray}
\overline{B}_iM^i_{~j}
&=&\frac{1}{N_c!}({\epsilon}_{ii_1{\cdots}i_{N_c}}
{\epsilon}^{r_1{\cdots}r_{N_c}}Q^{i_1}_{r_1}{\cdots}Q^{i_{N_c}}_{r_{N_c}}
Q^i_r)\widetilde{Q}^r_j=0.
\end{eqnarray}
Using the fact that
\begin{eqnarray}
M^i_{~k}\widetilde{M}^{k}_{~j}= \det M{\delta}^i_{~j}, ~~
\widetilde{M}^i_{~j}= \det M (M^{-1})^i_{~j}, 
\end{eqnarray}
we can formally write (\ref{eq265}) in the following form:
\begin{eqnarray}
\det M (M^{-1})^i_{~j} =B^i\widetilde{B}_j.
\label{eq6.113}
\end{eqnarray}
Note that since in this case $\det M=0$, (\ref{eq6.113}) is only a formal
expression and the true meaning is given by (\ref{eq265}).

Finally, we consider the special case, $N_c=2$. In the classical
direction, due to the general relation (\ref{eq6.103x}), 
the matrix from of the quark superfield is
\begin{eqnarray}
&& (Q)=\left(\begin{array}{cccccc}
a & 0 &{\cdots} &0 & {\cdots}& 0\\
0 & a &{\cdots} & 0 &{\cdots} & 0\\
\end{array}\right).
\label{eq6.113a}
\end{eqnarray}
 The classical moduli space is parameterized by
the gauge invariant
\begin{eqnarray}
V^{ij}=\epsilon^{rs}Q^i_{~r}Q^{~j}_s=Q^i{\cdot}Q^j, ~~~V^{ij}=-V^{ji}
\label{eq6.113b}
\end{eqnarray}
but subject to the constraint
\begin{eqnarray}
\epsilon_{i_1{\cdots}i_{N_f}}V^{i_1i_2}V^{i_3i_4}=0,
\label{eq6.113c}
\end{eqnarray}
since in the flat directions only $V^{12}=-V^{21}=a^2{\neq}0$.
For non-zero $V$, the gauge symmetry is completely
broken. Furthermore, since $V^{ij}$ are relevant to
the quark mass terms of the fundamental theory  and 
$V^{12}=-V^{21}\,{\neq}0$ implies that two flavours get equal mass, 
the global symmetry (\ref{eq182b}) is broken to 
\begin{eqnarray}
SU(2){\times}SU(2N_f-2){\times}U_R(1). 
\label{eq6.113d}
\end{eqnarray}

\subsection{Quantum moduli space and low energy non-perturbative dynamics}
\label{subsub6.3.4}

\renewcommand{\theequation}{3.4.\arabic{equation}}
\setcounter{equation}{0}
\renewcommand{\thetable}{3.4.\arabic{table}}
\setcounter{table}{0}

In the previous section we have discussed various classical moduli spaces 
parametrized by classical composite fields --- mesons and  baryons. 
In some cases these  composite field should satisfy some
constraints. At the quantum level, the moduli
space will be parametrized by the vacuum expectation values
of the corresponding composite 
operators. The constraints may be changed due to possible 
non-perturbative quantum corrections. The quantum moduli space may
differ from the classical one and hence the 
corresponding physical consequences may
also change. To see the quantum effects on the
moduli space, it is necessary to investigate 
the non-perturbative low energy dynamics. 
Since the moduli spaces vary with  the relative
numbers of colours and flavours, we shall discuss them according to
different ranges of the numbers of flavour and colour.

\subsubsection{$N_f < N_c$: Erasing of vacua}
\label{subsub6.3.5}

Consider the non-perturbative dynamics for this case. Recall 
that the global symmetry of supersymmetric QCD at the quantum level
is $SU_L(N_f){\times}SU_R(N_f){\times}U_B(1){\times}U_R(1)$. 
The dynamically generated superpotential should respect this symmetry.
Under the chiral symmetry $SU_L(N_f){\times}SU_R(N_f)$, 
$\widetilde{Q}{\sim}(0,\overline{N}_f)$ and
$Q{\sim}(N_f,0)$, $M^i_{~j}{\sim}(N_f,\overline{N}_f)$, and the simplest
invariant under $SU_L(N_f){\times}SU_R(N_f)$ is the determinant $\det M$.
Let us determine the quantum numbers of $\det M$ 
under the $U(1)$ transformations. Since the $U(1)$ quantum numbers
are additive, Table \ref{ta6.1} gives
\begin{eqnarray}
&&B(M^i_{~j})=B(Q)+B(\widetilde{Q})=0, 
~~R(M^i_{~j})=R(Q)+R(\widetilde{Q})=
\frac{N_f-N_c}{N_f}, \nonumber\\
&& B(\det M)=0, ~~R(\det M)=2N_f \frac{N_f-N_c}{N_f}=2(N_f-N_c). 
\end{eqnarray}
To construct the superpotential, we need another quantity with 
mass dimension. The natural choice is 
the non-perturbative dynamical scale $\Lambda$.
It is $SU_L(N_f){\times}SU_R(N_f)$ invariant, and its transformation
under the $U(1)$ symmetries is related to the vacuum angle
$\theta$. We can argue this from the perturbative viewpoint despite
the fact that $\Lambda$ is a non-perturbative energy scale.
The running gauge coupling in  perturbative QCD is
\begin{eqnarray}
g^2(q^2)&=&\frac{g^2(q^2_0)}{1+g^2/(16\pi^2){\beta}_0\ln(q^2/q_0^2)},~~ 
\nonumber\\
\frac{4\pi}{g^2(q^2)}&=&\frac{4\pi}{g^2(q_0^2)}\left[
1+\frac{g^2(q_0^2)}{16\pi^2}{\beta}_0\ln\frac{q_0^2}{q^2}\right],
\nonumber\\[2mm]
&=&\frac{4\pi}{g^2(q_0^2)}+\frac{\beta_0}{2\pi}\ln\frac{q_0}{q}
{\equiv}\frac{\beta_0}{2\pi}\ln\frac{q}{\Lambda_{\rm pert}},
\label{eq272}
\end{eqnarray}
where ${\beta}_0$ is the coefficient of 
one-loop $\beta$-function and $\Lambda_{\rm pert}$
is the perturbative energy scale. As a result,
\begin{eqnarray}
\frac{8\pi^2}{g^2(q^2)}=\ln\left(\frac{q}{\Lambda_{\rm pert}}\right)^{\beta_0},
~~~{\Lambda_{\rm pert}}^{\beta_0}=q^{\beta_0}e^{-8\pi^2/[g^2(q^2)]}.
\label{eq6.112x}
\end{eqnarray}
For the non-perturbative scale $\Lambda$, the $\theta$ parameter 
will arise since it is associated with the complex coupling 
constant (\ref{eq3p3})
\begin{eqnarray}
{\Lambda}^{\beta_0}=q^{\beta_0}e^{-8\pi^2/[g^2(q^2)]+i{\theta}}
=q^{\beta_0}e^{2\pi i[(4\pi i/g^2)+\theta/(2\pi)]}
{\equiv}q^{\beta_0}e^{2\pi i \tau}.
\label{eq6.112y}
\end{eqnarray}
Thus ${\Lambda}^{\beta_0}$ should transform as $e^{i\theta}$ 
under the $U(1)$ transformations. Note that the non-perturbative
scale $\Lambda$ is a complex parameter. For clarity 
the quantum numbers of $\det M$ and ${\Lambda}^{\beta_0}$ are listed 
in Table \ref{ta6.2}.

\begin{table}
\begin{center}
\begin{tabular}{|c|c|c|c|c|}\hline
       & $U_B(1)$ & $U_A(1)$ & $U_{R_0}(1)$ & $U_R(1)$ \\ \hline
$\det M$ & $0$ & $2N_f$ & $0$ & $2(N_f-N_c)$ \\ \hline
${\Lambda}^{\beta_0}$ & $0$ & $2N_f$ & $2(N_c-N_f)$ & 0\\ \hline
\end{tabular}
\end{center}
\caption{\protect\small Quantum numbers of $\det M$ and ${\Lambda}^{\beta_0}$.
\label{ta6.2}}
\end{table}

Since the low energy superpotential must be
a holomorphic function of $Q$ and $\widetilde{Q}$ (i.e. $\det M$, not
$\overline{Q}$ or $\overline{\widetilde{Q}}$) 
and the scale parameter $\Lambda$, it 
should also be $SU_L(N_f){\times}SU_R(N_f){\times}U_B(1)$ invariant.
Its $R$ charge is $2$ since  the superpotential is the $F$-term of
the effective action
\begin{eqnarray}
\overline{W}_{\rm eff}{\sim}{\int}d^4x{\int}d^2{\theta}W_{\rm eff}.
\label{eq6.112}
\end{eqnarray}
Thus this superpotential must be composed only of
$\det M$ and $\Lambda$.  In addition, from  (\ref{eq6.112}) 
the mass dimension of $W_{\rm eff}$ should be $3$ owing to the
mass dimensions - $[\theta^{-1}]=[d\theta]=1/2$. Since the dimension
of $\Lambda$ is $[\Lambda]=1$, the coefficient of one-loop beta function
in $N=1$ supersymmetric QCD is ${\beta}_0=3N_c-N_f$, $[M^{i}_{~j}]=2$, 
$[\det M]=2N_f$, the $R$-charge-$R(\theta)=R(d\theta)=-1$,
the only possible combination with $R=2$ and mass dimension 3 should be
\begin{eqnarray}
W_{\rm eff}=C(N_c,N_f)\left(\frac{\Lambda^{\beta_0}}
{\det M}\right)^{1/(N_c-N_f)},
\label{eq276}
\end{eqnarray}
where $C(N_c,N_f)$ is a dimensionless constant depending only on
$N_f$ and $N_c$, and the factor $1/(N_c-N_f)$ 
 comes purely from dimensional analysis.  The superpotential
(\ref{eq276}) is called the Affleck-Dine-Seiberg (ADS) 
superpotential \cite{ref1p22}. Owing to the
famous nonrenormalization theorem in supersymmetric theory,
this superpotential cannot be generated from 
perturbative quantum corrections.
However, the non-renormalization theorem only concerns at perturbation
theory and imposes no restrictions on non-perturbative quantum 
corrections. Thus this superpotential can only possibly come from
instanton contributions and the constant $C$
can be determined from a one-instanton calculation.

The superpotential can be further determined by 
considering various limiting cases. First,
we consider the case that the expectation value of the $N_f$-th
flavour ${\langle}Q_{N_f}{\rangle}$$=$ ${\langle}\widetilde{Q}_{N_f}{\rangle}$ 
$=a_f$ becomes very large. Using the diagonal form (\ref{eq6.91})
and the fact that $\det M$ is a rotational invariant, as well as
the $D$-flatness condition (\ref{eq6.97}), we have 
\begin{eqnarray}
\widetilde{Q}_iQ^j&=& \sum_{r=1}^{N_c}\widetilde{Q}_{ir}Q^{rj}=
\sum_{r=1}^{N_c}\widetilde{a}_i{\delta}_{ir}a_{j}{\delta}^{rj}
=\sum_{r=1}^{N_f}\widetilde{a}_i{\delta}_{ir}a_{j}{\delta}^{rj}
=a_i^2{\delta}_{i}^{~j},
\nonumber\\
\det M&=&\det \widetilde{Q}{\cdot}Q =a_1^2{\cdots}a_{N_f-1}^2a_{N_f}^2
=\det\widetilde{Q}'{\cdot}Q'a_{N_f}^2,
\label{eq277}
\end{eqnarray}
where $Q'$ or ($\widetilde{Q}'$) only contains $N_f-1$ flavours. At an
energy scale less than $a_{N_f}$, the $N_f-1$ flavours 
can be thought as the light flavours since $a_{N_f}$ is very 
big. Compared with $a_{N_f}$, $a_i$, $i=1,{\cdots},N_f-1$, can 
be regarded as approximately zero on this energy scale.
Due to the super-Higgs mechanism, the $SU(N_c)$ gauge theory
with $N_f$ flavours is broken to $SU(N_c-1)$ supersymmetric
QCD with $N_f-1$ flavours. 
According to (\ref{eq272}), at the energy $q>a_{N_f}$ 
the running coupling  is
\begin{eqnarray}
\frac{4\pi}{g^2(q^2)}=\frac{3N_c-N_f}{2\pi}\ln\frac{q}{\Lambda},
\label{eq279}
\end{eqnarray}
while at the energy scale $q<a_{N_f}$, the running coupling is
\begin{eqnarray}
\frac{4\pi}{g^2(q^2)}=\frac{3(N_c-1)-(N_f-1)}{2\pi}
\ln\frac{q}{\Lambda_{L}},
\end{eqnarray}
since now the theory becomes supersymmetric QCD with gauge 
group $SU(N_c-1)$ and $N_f-1$ flavours, the 
$N_f$th heavy flavour having been integrated out. 
At the energy $q^2=a^2_{N_f}$, the running coupling 
constants should match,
\begin{eqnarray}
\frac{4\pi}{g^2(a_{N_f}^2)}=\frac{3N_c-N_f}{2\pi}\ln\frac{a_{N_f}}{\Lambda}
=\frac{3(N_c-1)-(N_f-1)}{2\pi}\ln\frac{a_{N_f}}{\Lambda_{L}}.
\label{eq281}
\end{eqnarray}
Thus one can obtain the relation between the energy scales,
\begin{eqnarray}
{\Lambda}_L^{3(N_c-1)-(N_f-1)}=\frac{{\Lambda}^{3N_c-N_f}}{a_{N_f}^2}.
\label{eq282}
\end{eqnarray}
Requiring that the ADS potentials should coincide at $q=a_f$, we have
from (\ref{eq276}), (\ref{eq277}) and (\ref{eq282}), 
\begin{eqnarray}
C(N_c,N_f)\left(\frac{\Lambda^{3N_c-N_f}}{\det\widetilde{Q}'{\cdot}Q'a_{N_f}^2}
\right)^{1/(N_c-N_f)}
&=&C(N_c-1,N_f-1)\nonumber\\
&{\times}&\left(\frac{\Lambda_L^{3(N_c-1)-(N_f-1)}}
{\det\widetilde{Q}'{\cdot}Q'}\right)^{1/(N_c-N_f)}.
\end{eqnarray}
This implies
\begin{eqnarray}
C(N_c,N_f)=C(N_c-1,N_f-1)=C(N_c-N_f).
\end{eqnarray}
i.e. $C(N_c,N_f)$ should only be a function of $N_c-N_f$.

Further, the explicit form of $C(N_c-N_f)$ can be determined from 
another limit:
giving $Q_{N_f}$ and $\widetilde{Q}_{N_f}$ a large mass by adding
a mass term (only the holomorphic part) to the 
superpotential at tree level,
\begin{eqnarray}
W_{\rm tree}=mM_{~N_f}^{N_f}=mQ^{N_f}{\cdot}\widetilde{Q}_{N_f}.
\end{eqnarray}
Similarly to the previous case, consider the energy 
scale $m$. When the energy $q>m$, the theory is a 
$SU(N_c)$ supersymmetric QCD with $N_f$ flavours and the running
coupling constant is (\ref{eq279}).
When the energy $q<m$, the theory is $SU(N_c-1)$ supersymmetric QCD
with $N_f$ flavours. The running coupling constant is now
\begin{eqnarray}
\frac{4\pi}{g^2(q^2)}=\frac{3(N_c-1)-N_f}{2\pi}\ln\frac{q}{\Lambda_{L}}. 
\end{eqnarray}
Matching the coupling constants at $q=m$ gives
\begin{eqnarray} 
\frac{4\pi}{g^2(m^2)}&=&\frac{3N_c-N_f}{2\pi}\ln\frac{m}{\Lambda}
=\frac{3(N_c-1)-N_f}{2\pi}\ln\frac{m}{\widetilde{\Lambda}_{L}}.
\label{eq287} 
\end{eqnarray}
Hence we obtain
\begin{eqnarray}
\widetilde{\Lambda}_L^{3N_c-(N_f-1)}=m{\Lambda}^{3N_c-N_f}.
\label{eq288}
\end{eqnarray}
Now with the mass term $mM^{N_f}_{~N_f}$ for the $N_f$-th flavour, 
at the energy $q>m$, the superpotential is
\begin{eqnarray}
W_{\rm eff}
=C(N_c-N_f)\left(\frac{\Lambda^{\beta_0}}{\det M}\right)^{1/(N_c-N_f)}+
mM^{N_f}_{~N_f}.
\label{eq6.126}
\end{eqnarray}
As we know, the $F$-term associated with the superpotential 
is given by
\begin{eqnarray}
F=\frac{\partial W}{\partial {\phi_O}},
\end{eqnarray}
where $\phi_{O}$ is the lowest component of 
an elementary or composite chiral superfield $O$. Since the form of the 
superpotential in terms of chiral superfield is the same as that of
its lowest component, in the following we shall discuss 
the $F$-flatness condition of the corresponding chiral
superfield to manifest supersymmetry.
Unbroken supersymmetry requires that the $F$-term must vanish (i.e.
$F$-flatness). The $F$-flatness conditions for $M_{N_fi}$ and $M_{iN_f}$,
${\partial W_{\rm eff}}/{{\partial}M_{N_f i}}=0$ and
${\partial W_{\rm eff}}/{{\partial}M_{i N_f}}=0$,
lead to 
\begin{eqnarray}
M_{N_f i}=0, ~~~M_{i N_f}=0,
\end{eqnarray}
where we have used that for a matrix $M$
\footnote{This formula can be derived as follows:
\begin{eqnarray}{\delta}\ln \det M &=&\ln\det(M+{\delta}M)-\ln\det M
=\ln\displaystyle\frac{\det(M+{\delta}M)}{\det M}\nonumber\\
&=&\ln\det(1+M^{-1}{\delta}M){\sim}\ln(1+\mbox{Tr}M^{-1}{\delta}M) 
{\sim}\mbox{Tr}M^{-1}{\delta}M.\nonumber \end{eqnarray}}
\begin{eqnarray}
\mbox{Tr}M^{-1}{\delta}M =
\delta\ln\det M=\frac{1}{\det M}{\delta}\det M,
~~\frac{{\partial}\det M}{{\partial}M^i_{~j}}=\det M\, M^{~j\,-1}_{i}.
\label{eq6.129}
\end{eqnarray}
The meson operator hence takes the following form:
\begin{eqnarray}
M=\left(\begin{array}{cc}\widetilde{M} & 0\\ 0 & M_{~N_f}^{N_f}
\end{array}\right),
~~~\det M=\det \widetilde{M} M_{~N_f}^{N_f}.
\end{eqnarray}
As a consequence, the superpotential (\ref{eq6.126}) becomes
\begin{eqnarray}
W_{\rm eff}
&=&C(N_c-N_f)\left(\frac{\Lambda^{\beta_0}}{\det M}\right)^{1/(N_c-N_f)}+
mM^{N_f}_{~N_f}\nonumber\\
&=&C(N_c-N_f)\left(\frac{\Lambda^{\beta_0}}{\det \widetilde{M}}
\right)^{1/(N_c-N_f)}(M_{~N_f}^{N_f})^{-1/(N_c-N_f)}+mM^{N_f}_{~N_f}.
\label{eq3na} 
\end{eqnarray}
The $F$-flatness condition for $M_{~N_f}^{N_f}$,
\begin{eqnarray}
\frac{\partial W}{\partial M_{~N_f}^{N_f}}=\frac{C(N_c-N_f)}{N_c-N_f}
\left(\frac{\Lambda^{\beta_0}}{\det \widetilde{M}}
\right)^{1/(N_c-N_f)}(M_{N_f}^{~N_f})^{-[1+1/(N_c-N_f)]}+m=0, 
\end{eqnarray}
gives 
\begin{eqnarray}
M_{~N_f}^{N_f}=\left[\frac{m(N_c-N_f)}{C(N_c-N_f)}
\right]^{-(N_c-N_f)/(1+N_c-N_f)}\left[\left(\frac{\Lambda^{\beta_0}}
{\det \widetilde{M}}\right)^{1/(1+N_c-N_f)}\right].
\label{eq6.133}
\end{eqnarray}  
Inserting this expectation value into the superpotential (\ref{eq3na}), 
we get
\begin{eqnarray}
W_{\rm eff}
&=& C(N_c-N_f) \left[\frac{m(N_c-N_f)}{C(N_c-N_f)}
\right]^{1/(1+N_c-N_f)}\left[
\frac{\Lambda^{\beta_0}}{\det \widetilde{M}}\right]^{1/(1+N_c-N_f)} 
\nonumber\\
&+&m \left[\frac{m(N_c-N_f)}{C(N_c-N_f)}
\right]^{-(N_c-N_f)/(1+N_c-N_f)}
\left[\left(\frac{\Lambda^{\beta_0}}{\det \widetilde{M}}\right)^{1/(1+N_c-N_f)}
\right] \nonumber\\
&=&\left[\frac{m\Lambda^{\beta_0}}{\det \widetilde{M}}\right]^{1/(1+N_c-N_f)} 
\left[\frac{C(N_c-N_f)}{N_c-N_f}\right]^{(N_c-N_f)/(1+N_c-N_f)} 
\left(1+N_c-N_f\right).                      
\end{eqnarray}
Using (\ref{eq288}), we can write the superpotential as follows,
\begin{eqnarray}
W_{\rm eff}=\left(\frac{\Lambda_L^{\widetilde{\beta}_0}}
{\det\widetilde{M}}\right)^{1/[N_c-(N_f-1)]}
\left[\frac{C(N_c-N_f)}{N_c-N_f}\right]^{(N_c-N_f)/[N_c-(N_f-1)]}
\left[N_c-(N_f-1)\right],
\label{eq298}
\end{eqnarray}
where $\widetilde{\beta}_0=3N_c-(N_f-1)$ is the one-loop 
beta function coefficient of $SU(N_c-1)$ supersymmetric QCD 
with $N_f$ flavours. Recalling that when $m$ becomes big, 
the theory will become an $SU(N_c)$ theory
with $N_f-1$ flavours. Requiring that (\ref{eq298}) leads to the 
correct superpotential in the low energy theory ($q<m$), we must have 
\begin{eqnarray}
C(N_c-N_f)=(N_c-N_f)C^{1/(N_c-N_f)},
\end{eqnarray}
with $C$ being a universal constant. Hence  we get a more
transparent form of the superpotential
\begin{eqnarray}
W_{\rm eff}=(N_c-N_f)C^{1/(N_c-N_f)}
\left(\frac{\Lambda^{3N_c-N_f}}
{\det\widetilde{Q}{\cdot}Q}\right)^{1/(N_c-N_f)}.
\label{eq300}
\end{eqnarray}
The universal constant $C$ can be determined by a 
concrete instanton calculation. 
Here we only cite the result of Ref.\,\cite{arv}. 
For $N_f=N_c-1$, the superpotential (\ref{eq300}) is 
proportional to the one-instanton action and thus the 
constant $C$ can be exactly computed in a one-instanton
background. In particular, in this case the gauge group $SU(N_c)$
is completely broken since $\det M{\neq}0$. There is no 
infrared divergence and the instanton calculation is reliable. 
Ref.\,\cite{arv} has presented a detailed calculation in the 
dimensional regularization method and in the modified 
minimal subtraction scheme, 
and the result shows that $C=1$. Thus, for the case
of $N_f<N_c$ we finally obtain the exact ADS superpotential, 
\begin{eqnarray}
W_{\rm eff}=(N_c-N_f)
\left(\frac{\Lambda^{3N_c-N_f}}{\det\widetilde{Q}{\cdot}Q}
\right)^{1/(N_c-N_f)}.
\label{eq6.138}
\end{eqnarray} 
Note that this superpotential is the Wilsonian effective potential
due to the scale $\Lambda$ \cite{alw}. 
For the case $N_f<N_c-1$, this superpotential is associated with the
the gaugino condensate of the unbroken $SU(N_c-N_f)$ gauge group.

For $N_c=2$, the meson matrix $V$ is a $2N_f{\times}2N_f$ 
antisymmetric matrix, $\det V$ is not the simplest gauge and global 
invariant to constitute the superpotential, since it can be written as
the square of a simpler invariant, the Pfaffian of $V$, 
\begin{eqnarray}
\mbox{Pf}V=\sqrt{\det V}=\frac{1}{2^{N_f}N_f!}\sum_P\epsilon_P
V_{i_1i_2} V_{i_3i_4} {\cdots} V_{i_{2N_f-1}i_{2N_f}}, 
\label{eq6.138a}
\end{eqnarray} 
where $P$ denotes the permutation $\{i_1,{\cdots},i_{2N_f}\}$
and $\epsilon_P$ the signature of $P$.
A similar procedure to the derivation of (\ref{eq6.138})
yields the dynamical superpotential of the $SU(2)$ case,
 \begin{eqnarray}
W_{\rm eff}=(2-N_f)
\left(\frac{\Lambda^{6-N_f}}{\mbox{Pf}V}\right)^{1/(2-N_f)}.
\label{eq6.138b}
\end{eqnarray}  

Let us see what are the physical consequences 
of the dynamical superpotential (\ref{eq6.138}).
As we know, the relation between the usual potential 
and the superpotential is
\begin{eqnarray}
V&=&|F_{Qir}|^2=|\frac{\partial W_{\rm eff}}
{\partial \phi_{Qir}}|^2,\nonumber\\
F_{ir}&=&\frac{\partial W_{\rm eff}}{\partial Q_{ir}}
=(N_c-N_f)\left(\frac{\Lambda^{3N_c-N_f}}
{\det\widetilde{Q}{\cdot}Q}\right)^{1/(N_c-N_f)-1}
\det\widetilde{Q}{\cdot}Q\mbox{Tr}\left[(\widetilde{Q}{\cdot}Q)^{-1}
\frac{\partial}{\partial Q_{ir}}\widetilde{Q}{\cdot}Q\right]\nonumber\\
&=&(N_c-N_f)\left(\Lambda^{3N_c-N_f}\right)^{1/(N_c-N_f)-1}
\left(\frac{1}{\det\widetilde{Q}{\cdot}Q}\right)^{1/(N_c-N_f)}
Q^{-1}_{ir}\nonumber\\
&{\sim}&\frac{1}{Q}\left(\frac{1}{\det \widetilde{Q}{\cdot}Q}
\right)^{1/(N_c-N_f)},
\label{eq302}
\end{eqnarray} 
where (\ref{eq6.129}) was employed again. 
(\ref{eq302}) shows that the dynamically generated 
superpotential leads to
a squark potential, which tends to zero only when
 $\det M{\rightarrow}{\infty}$.
Therefore, the quantum theory does not have a stable ground state. 
In classical theory we have a vacuum configuration, but at the 
quantum level no vacuum state exists!

Finally, we consider the massive case.  
The mass term is one part of the tree-level superpotential, 
\begin{eqnarray}
W_{\rm tree}=\mbox{Tr}m{\cdot}M=m_{~i}^{j}M_{~j}^i.
\end{eqnarray}
In the weak coupling and small mass limit, the full superpotential is
\begin{eqnarray}
W_{\rm full}=W_{\rm eff}+W_{\rm tree}=
(N_c-N_f) \left(\frac{\Lambda^{3N_c-N_f}}
{\det\widetilde{Q}{\cdot}Q}\right)^{1/(N_c-N_f)}+
m_{~i}^jM_{~j}^i.
\label{eq305}
\end{eqnarray}
The vacua are still labeled by $M{\equiv}{\langle}M{\rangle}$, which is
determined by the $F$-flatness condition
\begin{eqnarray}
\frac{\partial W_{\rm full}}{{\partial} M^i_{~j}}|_M&=&
\left(\frac{\Lambda^{3N_c-N_f}}{\det M}\right)^{1/(N_c-N_f)-1}
\Lambda^{3N_c-N_f}\frac{\partial}{{\partial} M^i_{~j}}\frac{1}{\det M}
+m_i^{~j}\nonumber\\    
&=&-\left(\frac{\Lambda^{3N_c-N_f}}
{\det M}\right)^{1/(N_c-N_f)}(M^{-1})_i^{~j} +m_i^{~j}=0.
\label{eq306}
\end{eqnarray}
This gives 
\begin{eqnarray}
m^i_{~j}&=&\left(\frac{\Lambda^{3N_c-N_f}}{\det M}\right)^{1/(N_c-N_f)}
(M^{-1})^i_{~j}, \nonumber\\[2mm]
\det m&=&\left(\frac{\Lambda^{3N_c-N_f}}{\det M}\right)^{N_f/(N_c-N_f)}
\frac{1}{\det M}\nonumber\\[2mm]
&=& (\Lambda^{3N_c-N_f})^{N_f/(N_c-N_f)}
\left(\frac{1}{\det M}\right)^{N_c/(N_c-N_f)},
\nonumber\\[2mm]
\frac{1}{\det M}&=&(\det m)^{(N_c-N_f)/N_c}(\Lambda^{3N_c-N_f})^{-N_f/N_c}.
\label{eq307}
\end{eqnarray}
Taking $1/\det M$ back into (\ref{eq306}), we get 
\begin{eqnarray}
m^i_{~j}
&=&\left({\Lambda}^{3N_c-N_f}\det M \right)^{1/N_c}(M^{-1})^i_{~j},
\nonumber\\[2mm]  
M^i_{~j}&=&\left[\left(\det m\right) {\Lambda}^{3N_c-N_f}\right]^{1/N_c}
(m^{-1})^i_{~j}.
\label{eq308}
\end{eqnarray}
When $N_c=2$, the quark mass term is $w_{\rm tree}=
m_{ij}V^{ji}$, and a similar calculation gives
\begin{eqnarray}
V^{ij}=\Lambda^{(6-N_f)/2}\left(\mbox{Pf}\,m\right)^{1/2}
\left(m^{-1}\right)^{ij}.
\label{eq308a}
\end{eqnarray}

Now we consider the case $q<m^i_{~j}$. This means that
the matter fields get very big masses and hence will decouple. 
The theory will become an $SU(N_c)$ Yang-Mills theory. 
(\ref{eq305}) and (\ref{eq307}) give the full superpotential 
of this case,
\begin{eqnarray}
W(m)_{\rm eff}
&=&(N_c-N_f)\left[{\Lambda}^{3N_c-N_f}\det m\right]^{1/N_c}
+N_f\left[\left(\det m\right){\Lambda}^{3N_c-N_f}\right]^{1/N_c}\nonumber\\[2mm]
&=& N_c \left[\left(\det m\right){\Lambda}^{3N_c-N_f}\right]^{1/N_c}=N_c{\Lambda}^3_L,
\label{eq309}
\end{eqnarray}
where ${\Lambda}_L^3=(\det m{\Lambda}^{3N_c-N_f})^{1/N_c}$ is the energy 
scale of the low energy $SU(N_c)$ Yang-Mills theory, a many-flavour 
generalization of (\ref{eq288}). Comparing with (\ref{eq230}), 
one can see that the superpotential is generated by gaugino 
condensation in the low energy $SU(N_c)$ Yang-Mills theory. 
Thus in the case that $N_f<N_c-1$, the superpotential
is associated with gaugino condensation, while when $N_f=N_c-1$, the
superpotential arises from instanton contributions. 

Furthermore, we can show that in the massive case, 
the Wilsonian effective superpotential (\ref{eq6.138})
 is the same as the 1PI effective superpotential. 
Let us first explain the definition of 1PI effective superpotential 
in a general supersymmetric theory.

Consider a supersymmetric theory with the tree-level superpotential
\begin{eqnarray}
W_{\rm tree}=\sum_iJ_iX^i.
\label{eq311}
\end{eqnarray}
$X^i$ can be fundamental or composite superfields or their
gauge invariant polynomials. (\ref{eq311}) is similar to the source terms
in the usual functional integration with $J_i$ being the background
external sources. The generating functional is
\begin{eqnarray}
Z[J_i]&=&e^{iG [J_i]}
={\int}{\cal D}[f(X_i)]\mbox{exp}\left(iS+i{\int}d^4x
{\int}d^2{\theta}\sum_i J_iX^i\right),
\nonumber\\[2mm]
G[J]&=&{\cdots}+{\int}d^4x {\int}d^2{\theta}W(J){\equiv} {\cdots}+\overline{W}[J].
\label{eq312}
\end{eqnarray}
$G[J]$ is the generating functional of connected Green functions. 
Here we only write out its part related with the
 quantum superpotential. Correspondingly, $W(J)$ is 
the connected superpotential. Using the expectation value
calculated from $\overline{W}(J)$
\begin{eqnarray}
\frac{{\delta}\overline{W}(J)}{{\delta}J_i}={\langle}X^i{\rangle}{\equiv}\widetilde{X}_i,
\label{eq313}
\end{eqnarray}
If the omitted part $(\cdots)$ is independent of $J_i$, $\widetilde{X}_i$
is the usual vacuum expectation value in the presence of 
the external sources $J_i$,
\begin{eqnarray}
\widetilde{X}_i=\frac{{\delta}G(J)}{{\delta}J_i}
={\int}{\cal D}[f(X_i)]X_i\,\mbox{exp}\left(iS+i{\int}d^4x
{\int}d^2{\theta}\sum_i J_iX^i\right).
\label{eq313x}
\end{eqnarray}
Performing a Legendre transformation, we can get the
1PI effective action for $\widetilde{X}_i$:
\begin{eqnarray}
{\Gamma}_{\rm dyn}(\widetilde{X}^i)&=&\left[G(J)-{\int}d^4x{\int}d^2\theta
 \sum_iJ_i\widetilde{X}^i\right]_{J_i}\nonumber\\
&=&\left[{\cdots}+{\int}d^4x {\int}d^2{\theta}
\left(W(J)-\sum_iJ_i\widetilde{X}^i\right)\right]_{J_i}, 
\label{eq3131} 
\end{eqnarray}
where $J_i$ are solutions to (\ref{eq313}). Correspondingly, 
the dynamical superpotential part is 
\begin{eqnarray}
\overline{W}_{\rm dyn}(\widetilde{X})
=\left[{\int}d^4x {\int}d^2{\theta}
\left(W(J)-\sum_iJ_i\widetilde{X}^i\right)\right]_{J_i},
\label{eq3132}
\end{eqnarray}
and obviously, $W(J_i)$ can be obtained from $W_{\rm dyn}(X^i)$ by the 
inverse Legendre transformation,
\begin{eqnarray}
\overline{W}(J_i)=\overline{W}_{\rm dyn}(\widetilde{X}^i)
+\int d^4x{\int} d^2\theta \sum_iJ_i\widetilde{X}^i,
\label{eq3133}
\end{eqnarray}
where $\widetilde{X}^i$ is the expectation value of the operator
$X^i$ satisfying the following equation,
\begin{eqnarray}
\frac{\partial \overline{W}_{\rm dyn}}{{\partial}\widetilde{X}_i}+J_i=0.
\label{eq3134}
\end{eqnarray}
The 1PI effective potential is defined as
\begin{eqnarray}
\overline{W}_{\rm eff}(\widetilde{X},J)=\overline{W}_{\rm dyn}(\widetilde{X}^i)
+{\int}d^4x{\int}d^2{\theta}\sum_i J_i\widetilde{X}^i.
\label{eq3135}
\end{eqnarray}
This procedure is similar to the calculation of effective
potential \cite{ref27,ref217}.
For the case at hand, $X=M^i_{~j}$, $J=m^{j}_{~i}$. 
From (\ref{eq308}) and (\ref{eq309}), we obtain
\begin{eqnarray}
\frac{\partial W_{\rm eff}(m)}{\partial m^i_{~j}}&=& 
\left[\left(\det m\right){\Lambda}^{3N_c-N_f}\right]^{1/N_c-1}{\Lambda}^{3N_c-N_f}
\frac{\partial (\det m)}{\partial m^i_{~j}}\nonumber\\
&=& \left[\left(\det m\right){\Lambda}^{3N_c-N_f}\right]^{1/N_c}(m^{-1})_i^{~j}
={\langle}M{\rangle}^j_{~i}.
\label{eq3136}
\end{eqnarray}
With the definition (\ref{eq3132}), using (\ref{eq3136}),
(\ref{eq309}) and the second equation in (\ref{eq307}), we have
\begin{eqnarray}
W_{\rm dyn}(M)&=&N_c{\Lambda}_L^3-M^i_{~j}m^{j}_{~i}
=N_c{\Lambda}_L^3-(\det m{\Lambda}^{3N_c-N_f})^{1/N_c}(m^{-1})_{~j}^im^{j}_{~i}
\nonumber\\
&=&(N_c-N_f)\left[\left(\det m\right){\Lambda}^{3N_c-N_f}\right]^{1/N_c}\nonumber\\
&=&(N_c-N_f)\left[
\left(\frac{\Lambda^{3N_c-N_f}}{\det M}\right)^{N_f/(N_c-N_f)}
\frac{\Lambda^{3N_c-N_f}}{\det M}\right]^{1/N_c}\nonumber\\
&=&(N_c-N_f)\left(\frac{\Lambda^{3N_c-N_f}}{\det M}\right)^{1/(N_c-N_f)}.
\label{eq3137}
\end{eqnarray}
Comparing (\ref{eq3137}) with the ADS superpotential (\ref{eq6.138}),
which is a Wilsonian effective superpotential,
we can see they are identical. Therefore, in the massive case,
the 1PI effective superpotential is the same as the Wilsonian 
effective superpotential.
 
\subsubsection{$N_f=N_c$: Confinement with chiral symmetry breaking or baryon
number violation}
\label{subsub6.3.6}

We have seen that in the case $N_f<N_c$, the non-perturbative
superpotential lifts the vacuum degeneracy. All the 
classical vacua disappear. What the situation for $N_f{\geq}N_c$?
We shall see that in this case no non-perturbative superpotential can be
generated dynamically and hence the vacuum degeneracy remains.
The reasons are as follows:

For the case $N_f=N_c$, Table \ref{ta6.1} shows that the $R$-charges
of the chiral superfield $Q$($\widetilde{Q}$) 
and of ${\Lambda}^{\beta_0}={\Lambda}^{2N_f}$ both vanish.
However, the superpotential should have $R$-charge $2$, 
since it is an $F$-term. Thus in this case it is 
not possible to construct a superpotential.

For the case $N_f>N_c$, considering only the $R$-charge
and dimensionality, we could have a dynamically generated 
superpotential 
$$W{\propto}\left(\frac{{\Lambda}^{3N_c-N_f}}{\det M}\right)^{1/(N_c-N_f)}.$$
However, since $N_c-N_f<0$, ${\Lambda}^{3N_c-N_f}$ will be in the
denominator of the superpotential, and this can not match the expression
generated by instantons. It is known that
the contribution from instantons is
proportional to ${\Lambda}^{3N_c-N_f}$ \cite{arv}; 
it is not possible to have a superpotential of the form
${\Lambda}^{-(3N_c-N_f)}$. In particular, when $N_f>N_c$, from
the previous diagonal form, $\det M=0$, no non-perturbative superpotential
can be generated and the $D$-flatness still remains.

However, for $N_f{\geq}N_c$, some more interesting 
phenomena will arise. First, we see that in the case $N_f=N_c$,
although the vacuum degeneracy can not be lifted, owing to
non-perturbative quantum effects, the quantum moduli space 
will be different from the classical one. This is reflected in 
the change of the constraint,
\begin{eqnarray}
&& \det M-\widetilde{B}B={\Lambda}^{2 N_c};\nonumber \\
&& \mbox{Pf}V=\Lambda^4 ~~~~~\mbox{for}~N_c=2.
\label{eq318}
\end{eqnarray}
The above constraints must be manifested in the
low energy effective Lagrangian. One natural way is to introduce a
Lagrange multiplier field $X$ to add  the following superpotentials
to the effective Lagrangian,
\begin{eqnarray}
W_{\rm eff}&=&X \left(\det M-\widetilde{B}B-{\Lambda}^{2 N_c}\right);
\nonumber\\
W_{\rm eff}&=&X \left(\mbox{Pf}V-\Lambda^4\right)~~~~ \mbox{for}~N_c=2.
\label{eq318a} 
\end{eqnarray}

The reasonableness of the modified constraint (\ref{eq318}) 
can be argued as follows. 
We first consider a superpotential at tree level by adding a large 
mass term for the $N_f$th flavour, 
\begin{eqnarray}
W_{\rm tree}=mM^{N_f}_{~N_f}.
\label{eq319}
\end{eqnarray}
Since for $N_f=N_c$, no dynamical superpotential is generated, this 
tree level superpotential should be the full superpotential.
At the energy $q<m$, after integrating out the $N_f$-th  flavour, 
the theory is an $SU(N_f)$ gauge theory with $N_f-1$ flavours.
A dynamical effective superpotential (\ref{eq6.138}) is
generated by instanton contributions,
\begin{eqnarray}
W_{\rm eff}=\frac{\Lambda_L^{3N_c-(N_f-1)}}
{\det\widetilde{M}}=\frac{m\Lambda^{2N_c}}{\det\widetilde{M}},
\label{eq320}
\end{eqnarray}
where we have used (\ref{eq288}) and $\widetilde{M}$ gets contributions 
from the $N_f-1$ light flavours. 
The $F$-flatness conditions  $\partial W_{\rm eff}/\partial M_{iN_f}=$ 
$\partial W_{\rm eff}/\partial \widetilde{M}_{iN_f}=0$ ($i<N_f$) lead to
\begin{eqnarray}
M_i^{~N_f}=M^i_{~N_f}=0, ~~~
M=\left(\begin{array}{cc} \widetilde{M} & 0 \\
0 & M_{~N_f}^{N_f}\end{array}\right). 
\end{eqnarray}
This gives
\begin{eqnarray}
\det M=\det \widetilde{M}M_{~N_f}^{N_f}, 
~~~~M_{~N_f}^{N_f}=\frac{\det M}{\det\widetilde{M}}.
\label{eq322}
\end{eqnarray}
Inserting (\ref{eq322}) back into (\ref{eq319}), we obtain
\begin{eqnarray}
W_{\rm eff}=\frac{m\det M}{\det\widetilde{M}}.
\label{eq323}
\end{eqnarray}                                               
Comparing (\ref{eq323}) with (\ref{eq320}),
 one can see that only by choosing $\det M={\Lambda}^{2N_c}$, can one
get the low energy effective superpotential. This is the case for
$\langle\widetilde{B}B\rangle=0$. In the case that 
$\langle\widetilde{B}B\rangle {\neq}0$, it can also be proved
that (\ref{eq318}) is satisfied \cite{in}. Since the right-hand side
of (\ref{eq318}) is proportional to the one-instanton action \cite{ref1p22},
the quantum modification of the classical constraint must 
arise from the one-instanton contribution.

The quantum constraint (\ref{eq318}) has important physical consequences. 
The singular point $B=\widetilde{B}=M=0$ ($M=0$ means that the 
eigenvalues of $M$ vanish) has been eliminated by quantum effects 
through the generation of a mass gap since the point 
$B=\widetilde{B}=M=0$ does not satisfy the constraint. 
A vivid explanation was given by Intriligator and Seiberg by considering a 
two-dimensional surface defined by $XY=\mu$ in 
three-dimensional space \cite{ref1p11}. If $XY=0$, then 
either $X=0$ or $Y=0$, and the surface is $X$-plane or $Y$-plane. 
If $\mu{\neq}0$, these two cones are smoothed out to a hyperboloid,  
and the origin is expelled from the surface. So the only 
massless particles are the moduli, the quantum 
fluctuations of $M$, $B$, $\widetilde{B}$ satisfying the 
constraint. In geometric language, they are the tangent vectors 
of the surface determined by the constraint (\ref{eq318}) in (composite)
chiral superfield space. In the region of large $M$, $\widetilde{B}$ and $B$, 
the gauge symmetry is spontaneously broken and the the theory is in 
the Higgs phase. In the region of small $M$, $\widetilde{B}$ and $B$ 
(near the origin), the theory is in the confinement phase due to the 
quantum constraint (\ref{eq318}). In particular, 
the anomaly-free global symmetry
$SU_L(N_f){\times}SU_R(N_f){\times}U_B(1){\times}U_R(1)$ is broken 
since now the origin $B=\widetilde{B}=M=0$ is not on the quantum 
moduli space. Different points on the quantum moduli space exhibit 
different dynamical pictures. In the following we shall consider two 
typical points in the quantum moduli space:

\begin{flushleft}
{1. $M^i_{~j}={\Lambda}^2{\delta}_{~j}^i$, ~~$B=\widetilde{B}=0$
\footnote{Strictly speaking, one should write 
${\langle}M^{i}_{~j}{\rangle}={\Lambda}^2{\delta}_{~j}^{i}$, 
${\langle}B{\rangle}={\langle}\widetilde{B}{\rangle}=0$.}:
{\it Confinement and chiral symmetry breaking} }
\end{flushleft}

Obviously, this point lies in the quantum moduli space. In this case  
a quark condensation occurs since 
$M^{i}_{~j}={\Lambda}^2{\delta}_{~j}^{i}{\neq}0$, so the chiral symmetry 
$SU_L(N_f){\times}SU_R(N_f)$ is spontaneously broken to the diagonal 
$SU_V(N_f)$. However, the $U_B(1){\times}U_{R}(1)$
symmetry still remains. Thus the breaking pattern is
\begin{eqnarray}
SU_L(N_f){\times}SU_R(N_f){\times}U_B(1){\times}U_{R}(1){\rightarrow}
SU_V(N_f){\times}U_B(1){\times}U_{R}(1).
\end{eqnarray}
Let us analyze the transformation behaviours of $M$, $B$ and $\widetilde{B}$
under $SU(N_f)_V{\times}U_B(1){\times}U_{R}(1)$. From
$M^i_{~j}=\widetilde{Q}^i{\cdot}Q_j$ one may naively think 
that the number of the mesons is $N_f^2$. However, 
since we are considering quantum fluctuations of the 
moduli fields around the expectation values  
\begin{eqnarray}
{\langle}M^{i}_{~j}{\rangle}={\Lambda}^2{\delta}_{~j}^{i},
 ~~~{\langle}B{\rangle}
={\langle}\widetilde{B}{\rangle}=0,
\end{eqnarray}
the fluctuation matrix $M^i_{~j}-{\Lambda}^2{\delta}_{~j}^i$ 
should be traceless, 
\begin{eqnarray} 
\mbox{Tr}\left(M^i_{~j}-{\Lambda}^2{\delta}_{~j}^{i}\right)=0. 
\end{eqnarray}
Hence there are actually $N_f^2-1$ (super-)mesons;
the fluctuations $M^i_{~j}$ span the adjoint 
representation of the vector group $SU_V(N_f)$. 
The $U_B(1)$ quantum numbers of these fluctuations 
are $0$ since they are meson operators and their $U_R(1)$ quantum numbers 
should also be $0$ from the Table \ref{ta6.2}. Consequently, the fermionic 
component $\psi_M$ is also in the adjoint representation of $SU(N_f)$ and its 
 baryon number is $0$. In particular, the $R$ quantum number of $\psi_M$ 
is $-1$, since $M^i_{~j}=\widetilde{Q}^iQ_j$ is a chiral superfield 
\begin{eqnarray}
M^i_{~j}={\phi}^{i}_{M}+{\theta}{\psi}^{i}_{Mj}+{\theta}^2F^{i}_{Mj},
\end{eqnarray}
and $R(\theta)=1$ ($\theta$ is the super-coordinate).

The quantum fluctuations of $B$ and $\widetilde{B}$ do not carry any 
flavour index and hence are in the trivial representation of $SU_V(N_f)$.
Their baryon numbers are respectively $N_f$ and $-N_f$ 
due to the additivity of the  $U(1)$ quantum number. 
For clarity, we list the quantum numbers of the various fields
in Table \ref{ta6.3}.

\begin{table}
\begin{center}
\begin{tabular}{|c|c|c|c|}\hline
                      &$SU_V(N_f)$    & $U_B(1)$    & $U_R(1)$ \\ \hline
  $Q$                 & $N_f$         & $1$         &  $0$\\ \hline 
  $\widetilde{Q}$    & $\overline{N}_f$         & $-1$        &  $0$\\ \hline
  $\psi_Q$            & $N_f$         & $1$         &  $-1$\\ \hline
  $\psi_{\widetilde{Q}}$ & $\overline{N}_f$    & $-1$        &  $-1$\\ \hline 
  $\lambda$           & $1$           & $0$         &  $+1$\\ \hline
  $M$                 & $N_f^2-1$     & $0$         &  $0$\\  \hline
  $B$                 & $1$           & $N_f$       &  $0$\\  \hline
  $\widetilde{B}$         & $1$     & $-N_f$       &  $0$\\ \hline
  $\psi_{M}$          & $N_f^2-1$     & $0$         &  $-1$\\ \hline
  $\psi_{B}$          & $1$           & $N_f$       &  $-1$\\  \hline
  $\psi_{\widetilde{B}}$    & $1$           & $-N_f$       &  $-1$\\ \hline
\end{tabular}
\end{center}  
\caption{\protect\small $SU_V(N_f){\times}U_B(1){\times}U_R(1)$ 
quantum numbers of elementary and composite fields.
\label{ta6.3}}
\end{table}

A strong support to the dynamical pattern comes from 't Hooft anomaly
matching. As introduced in Sect.\ref{subsub243}, 
in a theory with confinement, 
the anomalies contributed by massless composite fermions at the
macroscopic level and those from the elementary massless confined 
fermions should match. 
Now we give a detail check whether the anomalies match. 
Corresponding to the global symmetry 
$SU_V(N_f){\times}U_B(1){\times}U_{R}(1)$ and the quantum numbers
listed in table \ref{ta6.3}, 
the currents for elementary fermions and massless 
composite fermions are collected in Table \ref{ta6.4}.
In addition, the fermionic part of the energy-momentum tensor
is in Table \ref{ta6.5} to allow for a discussion of
a possible axial gravitational anomaly.

\begin{table}
\begin{center}
\begin{tabular}{|c|c|c|c|} \hline
                      &$SU_V(N_f)$    & $U_B(1)$    & $U_R(1)$ \\ \hline
  $\psi_Q$ & $j^A_{\mu}=\overline{\psi}_{Q}{\sigma}_{\mu}t^A{\psi}_{Q}$       
           & $j^{(B)}_{\mu}=\overline{\psi}_{Q}{\sigma}_{\mu}{\psi}_{Q}$       
           & $j^{(R)}_{\mu}=-\overline{\psi}_{Q}{\sigma}_{\mu}{\psi}_{Q}$\\
 \hline
  $\psi_{\widetilde{Q}}$ & $\widetilde{j}^A_{\mu}
=\overline{\psi}_{\widetilde{Q}}
                         {\sigma}_{\mu}\overline{t}^A{\psi}_{\widetilde{Q}}$    
               & $\widetilde{j}^{(B)}_{\mu}=-\overline{\psi}_{\widetilde{Q}}
{\sigma}_{\mu}{\psi}_{\widetilde{Q}}$       
             & $ \widetilde{j}^{(R)}_{\mu}=-\overline{\psi}_{\widetilde{Q}}
{\sigma}_{\mu}{\psi}_{\widetilde{Q}}$\\ \hline 
  $\lambda$           & $0$  
                      & $0$         
     & $j^{(R)}_{\mu}(\lambda)
=\overline{\lambda}^a{\sigma}_{\mu}{\lambda}^a $\\ \hline
$\psi_{M}$ 
& $j^A_{\mu}(M)=f^{ABC}\overline{\psi}^B_{M}{\sigma}_{\mu}{\psi}^C_{M}$ 
    & $0$       
 &  $j^{(R)}_{\mu}=-\overline{\psi}_{M}^A{\sigma}_{\mu}{\psi}_{M}^A$\\ \hline
  $\psi_{B}$          & $0$           
    & $j^{(B)}_{\mu}=N_f\overline{\psi}_{B}{\sigma}_{\mu}{\psi}_{B}$       
    & $j^{(R)}_{\mu}=-\overline{\psi}_{B}{\sigma}_{\mu}{\psi}_{B}$\\ \hline
  $\psi_{\widetilde{B}}$    & $0$           
    & $\widetilde{j}^{(B)}_{\mu}=-N_f\overline{\psi}_{\widetilde{B}}
                        {\sigma}_{\mu}{\psi}_{\widetilde{B}} $       
   &  $\widetilde{j}^{(R)}_{\mu}=-\overline{\psi}_{\widetilde{B}}
                  {\sigma}_{\mu}{\psi}_{\widetilde{B}}$ \\ \hline
\end{tabular}
\end{center}  
\caption{\protect\small Currents corresponding to global symmetry
$SU_V(N_f){\times}U_B(1){\times}U_{R}(1)$.
\label{ta6.4} }
\end{table}  

\begin{table}
\begin{center}
\begin{tabular}{|c|c|} \hline
           & $T_{\mu\nu}$ \\ \hline
$\psi_Q$ & $ i/4\left[\left(\overline{\psi}_Q{\sigma}_{\mu}{\nabla}_{\nu}\psi_Q
-{\nabla}_{\nu}\overline{\psi}_Q{\sigma}_{\mu}\psi_Q\right)
+\left(\mu\longleftrightarrow\nu\right)\right]-g_{\mu\nu}{\cal L}[\psi_Q]$\\ \hline
$\psi_{\widetilde{Q}}$ & 
$\psi_{Q}{\longrightarrow}\psi_{\widetilde{Q}}$\\ \hline
$\lambda$    & $i/4\left[\left(\overline{\lambda}^a{\sigma}_{\mu}{\nabla}_{\nu}{\lambda}^a
-{\nabla}_{\nu}\overline{\lambda}^a{\sigma}_{\mu}\lambda^a\right)
+\left(\mu\longleftrightarrow\nu\right) \right]
-g_{\mu\nu}{\cal L}[\lambda]$\\ \hline
$\psi_{M}$   & 
$ {i}/{4}\left[\left(\overline{\psi}_M^A{\sigma}_{\mu}{\nabla}_{\nu}\psi_M^A
-{\nabla}_{\nu}\overline{\psi}_M^A{\sigma}_{\mu}\psi_M^A\right)
+\left(\mu\longleftrightarrow\nu\right)\right]-g_{\mu\nu}{\cal L}[\psi_M] $\\ \hline
$\psi_{B}$   & $ i/4\left[\left(\overline{\psi}_B{\sigma}_{\mu}{\nabla}_{\nu}\psi_B
-{\nabla}_{\nu}\overline{\psi}_B{\sigma}_{\mu}\psi_B\right)+
\left(\mu\longleftrightarrow\nu\right)\right]-g_{\mu\nu}{\cal L}[\psi_B] $\\ \hline
$\psi_{\widetilde{B}}$  & $\psi_{B}{\longrightarrow}\psi_{\widetilde{B}} $
\\ \hline
\end{tabular}
\end{center}  
\caption{\protect\small Energy-momentum tensor composed of the fermionic
components of chiral superfields; ${\cal L}[\psi]=i/2(\overline{\psi}\sigma^\mu\nabla_\mu\psi
-\nabla_\mu\overline{\psi}\sigma^\mu\psi)$,
$\nabla_\mu=\partial_\mu
-\omega_{KL\mu}\sigma^{KL}/2$, $\sigma^{KL}=1/4[\sigma^K,\sigma^L]$,
$\sigma^K=e^K_{~\mu}\sigma^\mu$. \label{ta6.5}}
\end{table}

The above tables give the currents corresponding to the global symmetry 
$SU_V(N_f){\times}U_B(1)$ ${\times}U_{R}(1)$ at both fundamental
and composite levels:

\begin{itemize}
\item For the elementary fermions:

$SU_V(N_f)$ current:
\begin{eqnarray}
J_{\mu}^A&{\equiv}&j^A_{\mu}(Q)+\widetilde{j}^A_{\mu}(\widetilde{Q})
=\overline{\psi}_{Qir}{\sigma}_{\mu}t^A_{ij}{\psi}_{Qjr}
+\overline{\psi}_{\widetilde{Q}ir}
{\sigma}_{\mu}\overline{t}^A_{ij}{\psi}_{\widetilde{Q}jr},
\nonumber\\
A&=&1,{\cdots},N_f^2-1, ~~i,j=1,{\cdots},N_f,~~r=1,{\cdots}, N_c.
\end{eqnarray}
$U_B(1)$ current:
\begin{eqnarray}
J^{(B)}_{\mu}&{\equiv} 
&j^{(B)}_{\mu}(Q)+\widetilde{j}^{(B)}_{\mu}(\widetilde{Q})
=\overline{\psi}_{Qir}{\sigma}_{\mu}{\psi}_{Qir}
-\overline{\psi}_{\widetilde{Q}ir}{\sigma}_{\mu}{\psi}_{\widetilde{Q}ir}.
\end{eqnarray}
$U_R(1)$ current:
\begin{eqnarray}
J^{(R)}_{\mu}&{\equiv}&
j^{(R)}_{\mu}(Q)+\widetilde{j}^{(R)}_{\mu}(\widetilde{Q})
+j^{(R)}_{\mu}(\lambda)\nonumber\\
&=&-\overline{\psi}_{Qir}{\sigma}_{\mu}{\psi}_{Qir}
-\overline{\psi}_{\widetilde{Q}ir}{\sigma}_{\mu}{\psi}_{\widetilde{Q}ir}
+\overline{\lambda}^a{\sigma}_{\mu}{\lambda}^a;~~
{a}=1,{\cdots}, N_c^2-1.
\end{eqnarray}

\item For the composite fermions:

$SU_V(N_f)$ current:
\begin{eqnarray}
J_{\mu}^A{\equiv} j^A_{\mu}(M)
=f^{ABC}\overline{\psi}^B_{M}{\sigma}_{\mu}{\psi}^C_{M}.
\end{eqnarray}
$U_B(1)$ current:
\begin{eqnarray}
J^{(B)}_{\mu}&{\equiv}&
j^{(B)}_{\mu}(B)+\widetilde{j}^{(B)}_{\mu}(\widetilde{B})
=N_f\overline{\psi}_{B}{\sigma}_{\mu}{\psi}_{B}
-N_f\overline{\psi}_{\widetilde{B}}{\sigma}_{\mu}{\psi}_{\widetilde{B}}.
\end{eqnarray}
$U_R(1)$ current:
\begin{eqnarray}
J^{(R)}_{\mu}&{\equiv}& j^{(R)}_{\mu}(M)
+j^{(R)}_{\mu}+\widetilde{j}^{(R)}_{\mu}
=-\overline{\psi}_{M}^A{\sigma}_{\mu}{\psi}_{M}^A
-\overline{\psi}_{B}{\sigma}_{\mu}{\psi}_{B}
-\overline{\psi}_{\widetilde{B}}{\sigma}_{\mu}{\psi}_{\widetilde{B}}.
\end{eqnarray}
\end{itemize}
One can directly calculate the anomaly coefficients 
of various triangle diagrams composed
of the above currents. In order to calculate the anomaly coefficient, 
one must introduce the gauge fields associated with 
$SU_V(N_f)$, $U_B(1)$ and $U_R(1)$, that is, turn these global groups 
into local ones. In general, 
these new gauge fields are not physical ones, 
except in some special cases such as electroweak theory, 
where the gauge symmetry is the local flavour symmetry. Therefore,
 't Hooft called these assumed gauge fields 
``spectator gauge fields'' \cite{ref218}. 
Also the gauge coupling constants associated to
these ``spectator gauge fields'' should be very small so that the 
dynamics of the  real physical strong coupling gauge 
interaction cannot be affected. 
As 't Hooft pointed out, one may either think of 
these ``spectator gauge fields'' as completely quantized fields 
or simply as artificial background fields with 
non-trivial topology. 

The calculation of the possible anomalous
triangle diagrams shows that the anomaly coefficients really are
identical. They are listed in Table \ref{ta6.6}. Note that in 
Table \ref{ta6.6} (and in what follows) we use the groups 
to represent the triangle diagrams 
composed of the corresponding currents, for example, 
$SU_V(N_f)^2U_R(1)$ represents
$\langle J^{(R)}_{\mu}J_{\nu}^AJ_{\rho}^B \rangle$ and $U_R(1)$
means the axial gravitational anomalous triangle diagram
$\langle J^{(R)}_{\mu}T_{\nu\rho}T_{\alpha\beta}\rangle$ etc. 
In principle, there are many possible triangle combinations of 
the currents, but only the anomaly coefficients listed in 
Table \ref{ta6.6} do not vanish.

The calculation of the anomaly coefficients 
listed in Table \ref{ta6.6} is straightforward.
For example, considering the $U_R(1)^3$ triangle digram. For elementary
fermions, the amplitude is 
\begin{eqnarray}
{\langle}J^{(R)}_{\mu}J^{(R)}_{\nu}J^{(R)}_{\rho}\rangle
&=& {\langle}j^{(R)}_{\mu}(Q)j^{(R)}_{\nu}(Q)j^{(R)}_{\rho}(Q)\rangle+
{\langle}j^{(R)}_{\mu}(\widetilde{Q})j^{(R)}_{\nu}(\widetilde{Q})
j^{(R)}_{\rho}(\widetilde{Q}) \rangle\nonumber\\[2mm] 
&&+{\langle}j^{(R)}_{\mu}(\lambda)j^{(R)}_{\nu}(\lambda)
j^{(R)}_{\rho}(\lambda)\rangle ,
\end{eqnarray}
and the anomaly coefficient is
\begin{eqnarray}  
R(Q)R(Q)R(Q)\mbox{Tr}({\bf 1})
&+&R(\widetilde{Q})R(\widetilde{Q})R(\widetilde{Q})\mbox{Tr}({\bf 1})+
R(\lambda)R(\lambda)R(\lambda)\mbox{Tr}({\bf 1})_{\rm adj}  \nonumber\\
&=&2{\delta}_{ij}{\delta}_{jk}{\delta}_{ki}{\delta}_{rs}
{\delta}_{st}{\delta}_{tr}(-1)^3
+{\delta}^{ab}{\delta}^{bc}{\delta}^{ca}\nonumber\\
&=&-2N_f^2+N_f^2-1=-N_f^2-1.
\end{eqnarray}
For composite fermions, the corresponding triangle diagram
amplitude is
\begin{eqnarray}
{\langle}J^{(R)}_{\mu}J^{(R)}_{\nu}J^{(R)}_{\rho}\rangle
&=& {\langle}j^{(B)}_{\mu}(B)j^{(B)}_{\nu}(B)j^{(R)}_{\rho}(B)\rangle+
{\langle}j^{(R)}_{\mu}(\widetilde{B})j^{(R)}_{\nu}(\widetilde{B})
j^{(R)}_{\rho}(\widetilde{B}) \rangle\nonumber\\[2mm] 
&+&{\langle}j^{(R)}_{\mu}(M)j^{(R)}_{\nu}(M)j^{(R)}_{\rho}(M)\rangle.
\end{eqnarray}
The anomaly coefficient is
\begin{eqnarray}  
R(B)R(B)R(B)&+&R(\widetilde{B})R(\widetilde{B})R(\widetilde{B})+
R(M)R(M)R(M)\mbox{Tr}({\bf 1})_{\rm adj}   \nonumber\\[2mm]
&=&2 (-1)^3+(-1)^3\mbox{Tr}({\bf 1} )_{\rm adj}\nonumber\\
&=&-2+N_f^2+1=-N_f^2-1.
\end{eqnarray}
Thus the anomaly coefficients at both elementary and composite
levels are equal.

As another illustrative example, take the axial gravitational anomaly
$U_R(1)$. At fundamental level, the amplitude is 
$\langle J_\mu^{(R)}T_{\nu\rho}T_{\sigma\delta}\rangle$, and the anomaly 
coefficient is
\begin{eqnarray}
-\mbox{Tr}(1)_{\rm c.f.}\mbox{Tr}(1)_{\rm f.f.}
-\mbox{Tr}(1)_{\rm c.f.}\mbox{Tr}(1)_{\rm f.f.} +
\mbox{Tr}(1)_{\rm adj}=-2N_f^2+N_f^2-1=-N_f^2-1,
\end{eqnarray}
where the subscripts ``c.f." and ``´f.f." denote
the fundamental representations of colour gauge group
$SU(N_c)$ and flavour group $SU(N_f)$, respectively.
At composite level, the anomaly coefficient is
\begin{eqnarray} 
-1-1-{\delta}^{AC}{\delta}^{CB}{\delta}^{BA}
=-2-(N_f^2-1)=-(N_f^2+1).
\end{eqnarray}
The anomaly coefficients again match exactly.

\begin{table}
\begin{center}

\begin{tabular}{|c|c|c|} \hline 
Triangle diagram   & Elementary anomaly coefficient  
& Composite anomaly coefficient \\ \hline 
$(U_R(1))^3$    &    $-N_f^2-1$    &    $-N_f^2-1$ \\ \hline
$(U_R(1))^2U_R(1)$  & $-2N_f$ & $-2N_f$ \\ \hline
$(SU_V(N_f))^2U_R(1)$  & $-N_f\mbox{Tr}(t^At^B)$ & 
$-N_f\mbox{Tr}(t^At^B)$\\ \hline
$U_R(1)$ & $-N_f^2-1 $ & $-N_f^2-1 $ \\ \hline 
\end{tabular}
\caption{\protect\small $SU_V(N_f){\times}U_R(1)$ 
't Hooft anomaly coefficients. \label{ta6.6}.}
\end{center}
\end{table}

\begin{flushleft}
{2. $M^i_{~j}=0$, ~~$B=-\widetilde{B}={\Lambda}^{N_f}$:
{\it Confinement and baryon number violation}}
\end{flushleft}

In this case, the symmetry breaking pattern is
\begin{eqnarray}
SU_L(N_f){\times}SU_R(N_f){\times}U_B(1){\times}U_R(1){\longrightarrow}
SU_L(N_f){\times}SU_R(N_f){\times}U_R(1).
\end{eqnarray}
The chiral symmetry does not break since $M^i_{~j}=0$. The 
$R$ charges  $R(M)=R(B)=R(\widetilde{B})=R({\Lambda})=0$ due to $N_f=N_c$ , 
so $R$ symmetry still remains. Obviously the baryon number symmetry is broken, 
usually $B(B)=N_f$, $B(\widetilde{B})=-N_f$, $B(\Lambda)=0$, and
the equation $B=-\widetilde{B}={\Lambda}^{N_f}$ 
does not satisfy the baryon number conservation law. 
Note that a breaking of baryon number conservation 
is not possible in ordinary QCD, as
Vafa and Witten proposed a strict theorem stating that the spontaneous
breaking of vector symmetries is forbidden \cite{vw}.
However, their proof of this theorem is based on the vector nature
of the quark-quark-gluon vertex, while in supersymmetric QCD, 
there exist scalar quarks,
and the quark-squark-gluino interaction vertex, which is an axial 
vector vertex, so this spoils the initial assumption of 
the theorem and hence the baryon number symmetry can be 
spontaneously broken.

Let us check the 't Hooft anomaly matching conditions. In this case
the quantum number for the elementary and composite fields are obvious
since all the quantum numbers remain intact. We list the quantum numbers
and the currents for the elementary and composite fields 
in the Tables \ref{ta6.9}, \ref{ta6.10}, \ref{ta6.11} and \ref{ta6.12}.

\begin{itemize}

\item At elementary level:

\begin{table}

\begin{center}

\begin{tabular}{|c|c|c|c|} \hline 
    & $SU_L(N_f)$ &$ SU_R(N_f)$ &$ U_R(1)$ \\ \hline
$ Q$   & $ N_f$ & $1$ &$ 0$ \\ \hline
$ \widetilde{Q}$ & $1$ &$\overline{N}_f$ &$ 0$\\ \hline
$ \psi_Q$ &$ N_f$ &$ 1$ & $-1$ \\ \hline
$ \widetilde{\psi}_Q$ &$ 1$ &$ \overline{N}_f$ &$-1$\\ \hline
$ \lambda$ &$ 1$ &$ 1$ &$ 1$ \\ \hline 
\end{tabular}
\caption{\protect\small $SU_L(N_f)\times SU_R(N_f)\times U_R(1)$
 quantum numbers for elementary fields. \label{ta6.9} }
\end{center}

\end{table}

\begin{table}
\begin{center}
\begin{tabular}{|c|c|c|c|} \hline 
    &$ SU_L(N_f)$ &$ SU_R(N_f)$ &$ U_R(1)$ \\ \hline
 $\psi_Q$ & $j^A_{L\mu}(Q)=\overline{\psi}_Q{\sigma}_{\mu}t^A{\psi}_Q$ 
 &$ 0$ &$j^{(R)}_{\mu}(Q)=-\overline{\psi}_Q{\sigma}_{\mu}{\psi}_Q$ \\ \hline
$\widetilde{\psi}_Q$ &$0$ 
&$\widetilde{j}^a_{R\mu}(\widetilde{Q})
=\overline{\psi}_{\widetilde{Q}}{\sigma}_{\mu}\overline{t}^A
{\psi}_{\widetilde{Q}}$  
&$ \widetilde{j}^{(R)}_{\mu}(\widetilde{Q})
=-\overline{\psi}_{\widetilde{Q}}{\sigma}_{\mu}{\psi}_{\widetilde{Q}}$
 \\ \hline
$\lambda$ & $0$ & $0$ &
 $j^{(R)}_{\mu}(\lambda)=\overline{\lambda}^a{\sigma}_{\mu}
{\lambda}^a$ \\ \hline 
\end{tabular}
\end{center}
\caption{\protect\small Currents composed of the fermionic components
of elementary chiral superfields. \label{ta6.10}}
\end{table}

The currents corresponding to the global symmetries
$SU_L(N_f){\times}SU_R(N_f){\times}U_R(1)$ are as follows:
\begin{eqnarray}
J^A_{L\mu}&{\equiv}&j^A_{L\mu}(Q)
=\overline{\psi}_{Qir}{\sigma}_{\mu}t^A_{ij}{\psi}_{Qjr}; \nonumber \\
J^A_{R\mu}&{\equiv} &\widetilde{j}^A_{R\mu}(\widetilde{Q})
=\overline{\psi}_{\widetilde{Q}ir}{\sigma}_{\mu}\overline{t}^A_{ij}
{\psi}_{\widetilde{Q}jr};\\ 
\nonumber
J^{(R)}_{\mu}&{\equiv}&
j^{(R)}_{\mu}(Q)+\widetilde{j}^{(R)}_{\mu}(\widetilde{Q})
+j^{(R)}_{\mu}(\lambda) \nonumber \\
&=&-\overline{\psi}_{Qir}{\sigma}_{\mu}{\psi}_{Qir}
-\overline{\psi}_{\widetilde{Q}ir}{\sigma}_{\mu}{\psi}_{\widetilde{Q}ir}
+\overline{\lambda}^{a}{\sigma}_{\mu}{\lambda}^{a}.
\end{eqnarray}

\item At composite level:

\begin{table}
\begin{center}
\begin{tabular}{|c|c|c|c|} \hline 
    &$ SU_L(N_f)$ &$ SU_R(N_f)$ &$ U_R(1)$ \\ \hline
$M$   & $ N_f$ & $\overline{N}_f$ &$ 0$ \\ \hline
$B$ & $1$ &$ 1$ &$ 0$\\ \hline
$\widetilde{B}$ &$1$ & $1$ & $0$\\ \hline
$\psi_M$ &$ N_f$& $ \overline{N}_f$ & $-1$\\ \hline
$\psi_B$ & $1$ &$ 1$ &$ -1$\\ \hline
$\widetilde{\psi}_B$ &$1$ &$ 1$ &$-1$ \\ \hline 
\end{tabular}
\caption{\protect\small Representation quantum numbers for composite
chiral superfields. \label{ta6.11}}
\end{center}
\end{table}

\begin{table}
\begin{center}
\begin{tabular}{|c|c|c|c|} \hline 
    & $SU_L(N_f)$ &$ SU_R(N_f)$ &$ U_R(1)$ \\ \hline
$\psi_M$ &$ j^A_{L\mu}(M)=\overline{\psi}_M{\sigma}_{\mu}t^A
{\psi}_M$ 
&$j^A_{R\mu}(M)=\overline{\psi}_{M}{\sigma}_{\mu}\overline{t}^A{\psi}_M $ 
 &$j^{(R)}_{\mu}(M)=-\overline{\psi}_M{\sigma}_{\mu}{\psi}_M$  \\ \hline
$\psi_B$ &$ 0$ & $0$ & $j^{(R)}_{\mu}(B)
=-\overline{\psi}_B{\sigma}_{\mu}{\psi}_B$ 
  \\ \hline
$\widetilde{\psi}_B$ &$0$ & $0$ &$\widetilde{j}^{(R)}_{\mu}(\widetilde{B})
=-\overline{\psi}_{\widetilde{B}}{\sigma}_{\mu}{\psi}_{\widetilde{B}}$ 
\\ \hline 
\end{tabular}
\caption{\protect\small Currents composed of the fermionic components
of composite chiral superfields. \label{ta6.12}}
\end{center}
\end{table}

The currents corresponding to 
$SU_L(N_f){\times}SU_R(N_f){\times}U_R(1)$ are as follows:
\begin{eqnarray}
J^A_{L\mu}&{\equiv}&j^A_{L\mu}(M)
=\overline{\psi}_{M}^{~i}{\sigma}_{\mu}t^A_{ij}{\psi}_{M}^j;
 \nonumber \\
J^A_{R\mu}&{\equiv} &j^a_{R\mu}(M)
=\overline{\psi}_{Mi}{\sigma}_{\mu}\overline{t}^A_{ij}{\psi}_{Mj};
 \nonumber \\
J^{(R)}_{\mu}&{\equiv} &j^{(R)}_{\mu}(M)+j^{(R)}_{\mu}(B)
+\widetilde{j}^{(R)}_{\mu}(\widetilde{B})\nonumber \\
&=&-\overline{\psi}_{Mj}^{i}{\sigma}_{\mu}t^A_{ij}{\psi}_{Mi}^{j}
-\overline{\psi}_B{\sigma}_{\mu}{\psi}_B
-\overline{\psi}_{\widetilde{B}}{\sigma}_{\mu}{\psi}_{\widetilde{B}}. 
\end{eqnarray}

\end{itemize}
One can easily check that for both elementary and composite fermions only
the triangle diagrams $SU_{L(R)}(N_f)^3$, $SU_{L(R)}(N_f)^2U_R(1)$, 
$U_R(1)^3$ and the axial gravitational anomaly $U_R(1)$ do not vanish.
 The corresponding anomaly coefficients can be calculated in the same way
and the results are listed in Table \ref{ta6.12a}.  
As expected, the anomaly coefficients match exactly.

\begin{table}
\begin{center}

\begin{tabular}{|c|c|c|}
 \hline 
Triangle diagram   & Elementary anomaly coefficient  
& Composite anomaly coefficient \\ \hline 
$SU_{L(R)}(N_f)^3$    & $+(-) d^{ABC}N_f$ 
 &    $+(-) d^{ABC}N_f$  \\ \hline
$SU_{L(R)}(N_f)^2U_R(1)$  & $-N_f\mbox{Tr}(t^At^B)$ & $-N_f\mbox{Tr}(t^At^B)$ \\ \hline
$U_R(1)^3$  & $-(N_f^2+1)$ & $-(N_f^2+1)$\\ \hline
$U_R(1)$ & $-N_f^2-1 $ & $-N_f^2-1 $ \\ \hline 
\end{tabular}
\caption{\protect\small $SU_{L(R)}(N_f){\times}U_R(1)$ 't Hooft anomaly 
coefficients, $d^{ABC}{\equiv}\mbox{Tr}(t^A\{t^B,t^C\})$. \label{ta6.12a} }
\end{center}
\end{table}

Note that  we have used the fact 
that the fluctuations of $B$ and $\widetilde{B}$
are not independent due to the constraint 
\begin{eqnarray}
B-{\lambda}^{N_f}=\widetilde{B}+{\lambda}^{N_f}.
\end{eqnarray}
Thus in calculating the anomaly coefficient of $U_R(1)^3$, only
the contribution from $\psi_B$ is considered. 
It should be emphasized  that in the above section (and in what
follows) we have assumed (will assume) that 
the global symmetries are realized linearly 
on $M$, $B$ and $\widetilde{B}$.

For $N_c=2$, the point 
\begin{eqnarray}
V=\Lambda^2\left(\begin{array}{cc}-i\sigma^2 & 0\\
                                   0       & -i\sigma^2 \end{array}\right)
=\Lambda^2\left(\begin{array}{cccc} 0 & -1 & 0 & 0\\
                                    1 &  0 & 0 & 0\\
                                    0 &  0 & 0 & -1\\
                                    0 &  0 & 1 & 0 \end{array}\right) 
\label{eqa101}
\end{eqnarray}
is obviously in the moduli space. Consequently, the global flavour
symmetry $SU(4)$ breaks to $Sp(4)$, while $R$-symmetry is preserved.
The 't Hooft anomaly matching associated with the global symmetry
$Sp(4){\times}U_R(1)$ can be checked. The fundamental massless fermions
are quarks and gauginos, their quantum numbers with respect to 
$Sp(4){\times}U_R(1)$ are $4_{-1}$ and $1_1$, respectively. 
The low energy fields are the quantum fluctuations of $V$ around the
the expectation value (\ref{eqa101}) subject to the constraint
(\ref{eq318}). Their fermionic component transform as $5_{-1}$ under 
$Sp(4){\times}U_R(1)$. With these quantum numbers, the various
conserved currents can be easily constructed. Considering
the relation between the quadratic Casimir operators of
the 4- and 5-dimensional representations of $Sp(4)$,
$2 \mbox{Tr}(t^At^B)_{4}=\mbox{Tr}(t^At^B)_{5}$, we see that
the anomaly coefficients at low energy and high energy levels
are indeed equal, as shown in Table \ref{ta6.8a}.

\begin{table}  
\begin{center}
\begin{tabular}{|c|c|c|}
 \hline 
Triangle diagram   & Elementary anomaly coefficient  
& Composite anomaly coefficient \\ \hline 
$Sp(4)^2U_R(1)$    &    $-2 \mbox{Tr}(t^At^B)_{4}$ &  
$-\mbox{Tr}(t^At^B)_{5}$ \\ \hline
$U_R(1)^3$  & $2{\times}4{\times}(-1)^3+3$ & $5{\times}(-1)^3$\\ \hline
$U_R(1)$ & $2{\times}4{\times}(-1)+3$ & $5{\times}(-1)$ \\ \hline 
\end{tabular}
\caption{\protect\small $Sp(4){\times}U_R(1)$ 
't Hooft anomaly coefficients. \label{ta6.8a} }
\end{center}
\end{table}

\subsubsection{$N_f=N_c+1$: Confinement without chiral symmetry breaking}
\label{subsub6.3.7}

At the classical level, the gauge invariant chiral superfields 
describing the moduli space in this case are  $M^{i}_{~j}$ and 
the baryon superfields  
\begin{eqnarray}
B_i&=&{\epsilon}_{ij_1{\cdots}j_{N_c}}{\epsilon}^{r_1{\cdots}r_{N_c}}
Q_{~r_1}^{j_1}{\cdots}Q_{~r_{N_c}}^{j_{N_c}},\nonumber\\
\widetilde{B}_i&=&
{\epsilon}_{ij_1{\cdots}j_{N_c}}{\epsilon}^{r_1{\cdots}r_{N_c}}
\widetilde{Q}_{~r_1}^{j_1}{\cdots}\widetilde{Q}_{~r_{N_c}}^{j_{N_c}}.
\label{eq365}
\end{eqnarray}
Under the $SU_L(N_f){\times}SU_R(N_f)$, the composite superfields 
$M$, $B$ and $\widetilde{B}$ transform as follows,
\begin{eqnarray}
M: (N_f, \overline{N}_f), ~~~~B: (\overline{N}_f,1), 
~~~~\widetilde{B}: (1,N_f).
\end{eqnarray}
Let us first analyze what the quantum moduli space is in this case.
The $SU(N_c)$ gauge symmetry and the global symmetry
$SU_L(N_f){\times}SU_R(N_f){\times}U_B(1){\times}U_R(1)$ as well as the 
mass dimension 3 determine that the 
effective superpotential $W_{\rm eff}$ must be of the following form,
\begin{eqnarray} 
W_{\rm eff}{\equiv}\frac{1}{\Lambda^{\beta_0}}\left(a \det M
+b B^iM_i^{~j}\widetilde{B}_j\right),
\end{eqnarray}
where $a$ and $b$ are non-vanishing constants which need to be determined. 
Note that this superpotential is not the dynamically generated superpotential,
it is rather an artificial one for describing the moduli space.
Owing to the holomorphic decoupling, if we give one
flavour, say, the $N_f$-th flavour, a large mass $m$, then at the energy
scale $q<m$, the low energy theory should be an $SU(N_c)$ theory with
$N_f$ flavours. Thus, after integrating out this heavy flavour, 
we should get the nontrivial constraints (\ref{eq318}) of the 
$N_f=N_c$ case. After giving the last flavour a mass, the superpotential 
becomes 
\begin{eqnarray} 
W_{\rm eff}(m)
=\frac{1}{\Lambda^{\beta_0}}(a \det M+b B^iM_i^{~j}\widetilde{B}_j)
-mM_{N_f}^{~N_f}.
\end{eqnarray}
The $F$-flatness conditions (due to the unbroken supersymmetry) 
for $M_{N_f}^{~i}$ and $M_{i}^{~N_f}$ 
with $i<N_f$ reduce $M$ to the following form, 
\begin{eqnarray}
M=\left(\begin{array}{cc}\widetilde{M} & 0 
\\ 0 & M_{N_f}^{~N_f} \end{array}\right).
\end{eqnarray}
With $M_{N_f}^{~i}=0$ and $M_{i}^{~N_f}=0$, the  $F$-flatness conditions
for $B_i$ and $\widetilde{B}_i$ yield
\begin{eqnarray}
B=\left(\begin{array}{c} 0 \\ B_{N_f}\end{array}\right), ~~
\widetilde{B}=\left(\begin{array}{c} 0 \\ \widetilde{B}_{N_f}
\end{array}\right).
\end{eqnarray}
So the effective superpotential takes the following form:
\begin{eqnarray} 
W_{\rm eff} &=&\frac{1}{\Lambda^{\beta_0}}(a \det\widetilde{M} M_{N_f}^{~N_f}
+b B^{N_f}M_{N_f}^{~N_f}\widetilde{B}_{N_f})-mM_{N_f}^{~N_f} \nonumber \\[2mm]
&{\equiv}&\frac{1}{\Lambda^{\beta_0}}(a \det\widetilde{M} M_{N_f}^{~N_f}
+bM_{N_f}^{~N_f} B\widetilde{B})-mM_{N_f}^{~N_f},
\end{eqnarray}
where we denoted $B_{N_f}{\equiv}B$ and $\widetilde{B}_{N_f}{\equiv}
 \widetilde{B} $. 
The $F$-flatness condition for $M_{N_f}^{~N_f}$ gives
\begin{eqnarray} 
a\det\widetilde{M} +b B\widetilde{B}=m\Lambda^{\beta_0}.
\label{eq372}
\end{eqnarray}
At the energy $q=m$, the running coupling constant
of the $SU(N_c)$ theory with $N_f=N_c+1$ flavours should match with
that of the $SU(N_c)$ theory  with $N_f(=N_c)$ flavours:
\begin{eqnarray}
\frac{4\pi}{g^2 (m^2)}=\frac{3(N_c+1)-N_c}{2\pi}\ln\frac{\Lambda}{m}
=\frac{3N_c-N_c}{2\pi}\ln\frac{\Lambda_L}{m}.
\end{eqnarray}
This gives
\begin{eqnarray}
{\Lambda}^{2N_c}_L=m{\Lambda}^{2N_c-1},
\end{eqnarray}
i.e.
\begin{eqnarray}
{\Lambda}^{\widetilde{\beta}_0}_L=m{\Lambda}^{\beta_0}.
\end{eqnarray}
If we choose the undetermined parameters 
$a=1$, $b=-1$, (\ref{eq372}) will lead to the
constraint (\ref{eq318}) in the case $N_f=N_c$.
Therefore, the effective superpotential 
in the case $N_f=N_c+1$ should be of the form
\begin{eqnarray} 
W_{\rm eff}=\frac{1}{\Lambda^{\beta_0}}(\det M- B^iM_i^{~j}\widetilde{B}_j).
\label{eq6.225}
\end{eqnarray}
The moduli space of vacuum states is described by the $F$-flatness conditions,
 $\partial W_{\rm eff}/\partial B^i=$
$\partial W_{\rm eff}/\partial \widetilde{B}_i=$
$\partial W_{\rm eff}/\partial M_i^{~j}=0$,
\begin{eqnarray} 
M{\cdot}B=\widetilde{B}{\cdot}M=0, 
~~~\det M\, (M^{-1})^{ij}=B^i\widetilde{B}^j.
\end{eqnarray} 
Obviously, the origin $M=B=\widetilde{B}=0$ is on the moduli space
 since it satisfies the above conditions. Thus the whole global symmetry
$SU_L(N_f){\times}SU_R(N_f){\times}U_B(1){\times}U_R(1)$
is preserved in the origin of the moduli space. 
If the above dynamical picture is correct, 't Hooft's anomaly matching
with respect to this global symmetry must be satisfied. 

According to Table \ref{ta6.1}, 
we list the quantum numbers for
the elementary fermions and composite fermions in 
Tables \ref{ta6.13} and \ref{ta6.14}, respectively.
The corresponding
conserved currents corresponding to the global symmetry 
$SU_L(N_f){\times}SU_R(N_f){\times}U_B(1){\times}U_R(1)$ are
collected in Table \ref{ta6.15}.
One can easily find that the non-vanishing anomaly coefficients 
at both the elementary and composite fermion levels
are identical. They are listed in Table \ref{ta6.16}. 

\begin{table}
\begin{center}

\begin{tabular}{|c|c|c|c|c|} \hline
   & $ SU_L(N_f)$ & $SU_R(N_f)$ & $ U_B(1)$ & $U_R(1)$\\ \hline
 $ Q$ &$ N_f$   &$ 1$  &$ 0$ & $1/N_f$\\ \hline
$\widetilde{Q}$ &$ 1$ &$ \overline{N}_f$ &$ 0$  &$ 1/N_f$\\ \hline
$\psi_{Q}$ & $ N_f$   &$ 1$  & $0$ & $1/N_f-1$\\ \hline
$\psi_{\widetilde{Q}}$ &$ 1$ &$ \overline{N}_f$ &$ 0$  & $1/N_f-1$\\ \hline
$\lambda$ & $1$ & $1$ & $0$ & $1$ \\ \hline
\end{tabular}
\end{center}
\caption{\protect\small Representation quantum numbers for elementary fields. \label{ta6.13}}
\end{table}

\begin{table}
\begin{center}
\begin{tabular}{|c|c|c|c|c|} \hline
   &$ SU_L(N_f)$ &$SU_R(N_f)$ &$ U_B(1)$ &$U_R(1)$\\ \hline
 $ M$ &$ N_f$   &$\overline{N}_f$  &$ 0$ &$ 2/N_f$\\ \hline
 $ B$ &$ \overline{N}_f$ &$ 1$ & $ N_c(=N_f-1)$ & 
$N_c/N_f=(N_f-1)/N_f$ \\ \hline
$\widetilde{B}$ &$ 1$ &$ N_f$ & $N_c(=N_f-1)$  &$N_c/N_f=(N_f-1)/N_f$ \\ \hline
$\psi_{M}$ & $ N_f$   &$ \overline{N}_f$  & $0$ & $2/N_f-1$\\ \hline
$\psi_{B}$ &  $\overline{N}_f$ &$ 1$ &$ N_c(=N_f-1)$ &$-1/N_f$ \\ \hline
$\psi_{\widetilde{B}}$ & $1$ &$ N_f$ & $0$  & $-1/N_f$\\ \hline 
\end{tabular}
\end{center}
\caption{\protect\small Representation quantum numbers for composite fields. \label{ta6.14}}
\end{table}

\begin{table}
\begin{center}

\begin{tabular}{|c|c|c|} \hline
           & Elementary fermions  & Composite fermions \\ \hline
$SU_L(N_f)$ & $J_{L\mu}^A=\overline{\psi}_Q{\sigma}_{\mu}t^A{\psi}_Q$ &
$J_{L\mu}^A=\overline{\psi}_M{\sigma}_{\mu}t^A{\psi}_M
+\overline{\psi}_B{\sigma}_{\mu}\overline{t}^A{\psi}_B   $\\ \hline
$SU_R(N_f)$ &$J_{R\mu}^A=\overline{\psi}_{\widetilde{Q}}
{\sigma}_{\mu}\overline{t}^A{\psi}_{\widetilde{Q}}$
 &$J_{R\mu}^A=\overline{\psi}_M{\sigma}_{\mu}\overline{t}^A{\psi}_M
+\overline{\psi}_{\widetilde{B}}{\sigma}_{\mu}t^A{\psi}_{\widetilde{B}}$
\\ \hline
$U_B(1)$  &$J^{(B)}_{\mu}=\overline{\psi}_Q{\sigma}_{\mu}{\psi}_Q
-\overline{\psi}_{\widetilde{Q}}{\sigma}_{\mu}
{\psi}_{\widetilde{Q}}$ &$ J^{(B)}_{\mu}=
N_c\overline{\psi}_B{\sigma}_{\mu}{\psi}_B
+N_c\overline{\psi}_{\widetilde{B}}
{\sigma}_{\mu}{\psi}_{\widetilde{B}}$\\ \hline
$U_R(1)$  &${\begin{array}{rcl}J^{(R)}_{\mu}
&=&-N_c/N_f\overline{\psi}_Q{\sigma}_{\mu}{\psi}_Q\\
&-&N_c/N_f\overline{\psi}_{\widetilde{Q}}{\sigma}_{\mu}{\psi}_{\widetilde{Q}}\\
&+&\overline{\lambda}^a{\sigma}_{\mu}\lambda^a \end{array}} $ &
 ${\begin{array}{rcl}J^{(R)}_{\mu}&=&
(-1+2/N_f)\overline{\psi}_M{\sigma}_{\mu}{\psi}_M\\
&-&1/N_f\overline{\psi}_B{\sigma}_{\mu}{\psi}_B\\
&-&1/N_f\overline{\psi}_{\widetilde{B}}{\sigma}_{\mu}
{\psi}_{\widetilde{B}}\end{array}} $
 \\ \hline
\end{tabular}

\end{center}
\caption{\protect\small Currents corresponding to the global
symmetry $SU_L(N_f){\times}SU_R(N_f){\times}U_B(1){\times}U_R(1)$.
\label{ta6.15}}
\end{table}

\begin{table}
\begin{center}
\begin{tabular}{|c|c|c|} \hline
         & Elementary level & Composite level \\ \hline
$(SU_{L(R)}(N_f))^3 $ & $+(-)\mbox{Tr}(t^A\{t^B,t^C\}) N_c$ 
& $+(-)\mbox{Tr}(t^A\{t^B,t^C\}) N_c$ \\ \hline
$(SU_{L(R)}(N_f))^2U_B(1) $ & $N_c\mbox{Tr}(t^At^B)$ 
& $N_c\mbox{Tr}(t^At^B)$\\ \hline
$(SU_{L(R)}(N_f))^2U_R(1) $ & $ -N_c^2/N_f\mbox{Tr}(t^At^B)$ &  
$-N_c^2/N_f\mbox{Tr}(t^At^B) $\\ \hline
$(U_B(1))^2U_R(1)$ & $ -2N_c^2$ &  $-2N_c^2 $\\ \hline
$(U_R(1))^3$ & $ -N_f^2+6N_f-12+8/N_f-2/N_f^2$ & 
$-N_f^2+6N_f-12+8/N_f-2/N_f^2 $\\
\hline 
$(U_R(1))^2U_B(1)$ & $0$ & $0$ \\ \hline
$U_R(1)$ & $-N_f^2+2N_f-2$ & $-N_f^2+2N_f-2$ \\
\hline
\end{tabular}
\end{center}
\caption{\protect\small 't Hooft anomaly coefficients.\label{ta6.16}}
\end{table}

In the case $N_c=2$, the superpotential in the low energy effective 
Lagrangian is
\begin{eqnarray}
W_{\rm eff}=-\frac{1}{\Lambda^3}\mbox{Pf}\,V.
\end{eqnarray} 
The quarks and the gaugino transform as $6_{1/3}$ and 
$1_1$, respectively, and the quantum fluctuation
of the composite field $V$ as $15_{2/3}$ under the global
symmetry group $SU(6){\times}U_R(1)$. The 't Hooft anomaly conditions 
are  satisfied as can be seen from Table \ref{ta6.16a}.

\begin{table}
\begin{center}
\begin{tabular}{|c|c|c|} \hline
         & Elementary level & Composite level \\ \hline
$(SU(6))^3 $ & $2\mbox{Tr}(t^A\{t^B,t^C\})_{6} $ & 
$\mbox{Tr}(t^A\{t^B,t^C\})_{15}$ \\ \hline
$(SU(6))^2U_R(1) $ & $ 2{\times} (-2/3)\mbox{Tr}(t^At^B)_6 $ 
& $-1/3\mbox{Tr}(t^At^B)_{15}$ \\ \hline
$U_R(1)^3$ & $ 12{\times} (-2/3)^3+3 $ &  $15{\times} (-1/3)^3 $\\ \hline
$U_R(1)$ & $12{\times}(-2/3)+3 $ & $ 15{\times} (-1/3)$\\ 
\hline
\end{tabular}

\end{center}
\caption{\protect\small 't Hooft anomaly coefficients. \label{ta6.16a}}
\end{table}

\subsubsection{Supersymmetric QCD for $N_f>N_c+1$}
\label{subsub6.3.8}

Now we continue to add the flavours and the more interesting phenomena
will arise. In the following we shall concentrate on several typical
ranges of the flavour and colours.

\vspace{2mm}

\begin{flushleft}
{\bf $3N_c/2<N_f<3N_c$: {\it Non-Abelian Coulomb phase, 
conformal window and electric-magnetic duality}}
\end{flushleft}

\vspace{2mm}

We now specialize to $N=1$ $SU(N_c)$ supersymmetric QCD with 
$N_f$ flavours. The NSVZ beta function (\ref{eq1p1}) and 
anomalous dimension of the quark and squark fields are
\begin{eqnarray}
\beta (g) &=& -\frac{g^3}{16\pi^2}\frac{3N_c-N_f+N_f \gamma (g^2)}
{1-N_cg^2/(8\pi^2)},\nonumber \\
\gamma (g^2) &=& -\frac{g^2}{8\pi^2}\frac{N^2_c-1}{N_c}+{\cal O}(g^4).
\label{eq6.224}
\end{eqnarray}
Thus we see that $\beta_0 (g) <0 $, and the theory in this range is
asymptotically free.
However, the anomalous dimensions of the quark and squark
fields imply $\beta_1 (g) >0$, i.e. the
one-loop beta function is negative and the two-loop beta function is positive.
This fact will make the beta function have non-trivial zero points, i.e.
non-trivial fixed points. One explicit nontrivial fixed point can be 
observed in the following way:
taking the limit $N_f{\rightarrow}{\infty}$ and $N_c{\rightarrow}{\infty}$
but keeping $N_cg^2$ and $N_f/N_c=3-\epsilon$  
with $\epsilon {\ll}1 $ fixed, we have
\begin{eqnarray}
\gamma (g^2) &\simeq & -\frac{N_cg^2}{8\pi^2},\nonumber \\
\beta (g) &=& -\frac{g^3}{16\pi^2}\frac{3N_c-N_f+N_f \gamma (g^2)}
{1-N_cg^2/(8\pi^2)}\simeq -\frac{g^3}{16\pi^2}
\frac{3-N_f/N_c+N_f/N_c\gamma (g^2)}{-g^2/(8\pi^2)}\nonumber \\
 &=&\frac{g}{2}\left[\epsilon +(3-\epsilon)\frac{N_cg^2}{8\pi^2}\right].
\label{eq6.224x}
\end{eqnarray}
Accordingly, the beta function has a second zero at
\begin{eqnarray}
N_cg^2=-\frac{8\pi^2\epsilon}{3}+{\cal O}(\epsilon^2).
\end{eqnarray}
So, at least for large $N_c$, $N_f$ and $\epsilon
=3-\frac{N_f}{N_c}{\ll}1$, the theory has a nontrivial IR fixed point. At
this non-trivial IR fixed point a four-dimensional superconformal
field theory will arise due to the relation between the trace
of the energy-momentum tensor and the $\beta$-function. Therefore, in the
range $3N_c/2<N_f< 3N_c$, the infrared region of $N=1$ supersymmetric
QCD is described by a 
four-dimensional superconformal field theory and  this range
is called Seiberg's conformal window. The quarks and gluons
are not confined but appear as interacting (effective) massless particles.
The effective non-relativistic interaction potential between 
two static electric charged quarks takes the form of the Coulomb potential:
$V(r){\sim}{1}/{r}$.
Consequently, the theory is now in the non-Abelian Coulomb phase.

\vspace{2mm}

\begin{flushleft}
{\bf $N_f>3N_c$: {\it Non-Abelian free electric phase}}
\end{flushleft}

\vspace{2mm}

In this range, since the one-loop beta function ${\beta}_0>0$, the
theory is not asymptotically free. At large distance (low energy) 
the coupling constant becomes smaller, and  
the particle spectrum of the theory 
consists of elementary quarks and gluons. 
Hence in this range of  $N_f$, the theory is in a so-called
``non-Abelian free electric phase''.
As mentioned in Sect.\ref{subsect27},  
a theory in a free electric phase is not well defined 
due to the Landau singularity.
This can be easily seen from the definition of the beta function
(in the modified minimal subtraction scheme)
\begin{eqnarray}
{\beta}_0=\frac{(N_f-3N_c)g^3}{16\pi^2}
=\mu\frac{\partial g}{\partial \mu}.
\end{eqnarray}
Integrating this equation, we obtain
\begin{eqnarray}
g^2=\frac{g^2_0}{1-g_0^2(N_f-3N_c)/(16 \pi^2)\ln(\mu/\mu_0)}.
\end{eqnarray}
The coupling constant increases with $\mu$, and theory 
breaks down at the Landau pole \cite{ref27}
\begin{eqnarray}
\mu=\mu_0 \exp\frac{8\pi^2}{g_0^2(N_f-3N_c)}.
\end{eqnarray}
However, it can be regarded as the low energy limit of another theory.

To conclude this section, we explain why one can discuss non-trivial
four-dimensional superconformal theory in the range $3/2N_c<N_f<3N_c$.
$N_f<3N_c$ ensures that the theory is asymptotically free.
As for the lower bound $3/2N_c<N_f$, 
this requires some knowledge about the representations of the
four-dimensional superconformal algebra on the chiral supermultiplet 
introduced in Sect.\,\ref{sect8}. As we know, 
for the representation of the superconformal algebra on the chiral 
supermultiplet, there is a simple relation between the $R$-charge and the 
conformal dimension of a gauge invariant chiral superfield 
operator ${\cal O}$ (see (\ref{eq9p121})),
\begin{eqnarray}
R({\cal O})=-\frac{2}{3}d({\cal O}),~~\mbox{or}~~ d({\cal O})
=\frac{3}{2}|R({\cal O})|.
\end{eqnarray}
However, from the unitary representations listed in Table \ref{ta2p2},
we see that except for the trivial representation, 
which is not interesting, the conformal dimension $d$ of a physical operator 
should be larger than $1$. Otherwise
the highest weight representation
of the conformal algebra will have a negative norm state, and 
this is not allowed
in a unitary representation. The simplest gauge invariant
chiral superfield is the meson $M$. 
The $R$-charges listed in Table \ref{ta6.1} give
\begin{eqnarray}
d(M)=\frac{3}{2} R(M)=3\frac{N_f-N_c}{N_f}>1,
\end{eqnarray}
and hence one gets $N_f>3/2N_c$. This is the reason why  
a superconformal field theory arises in 
the range $3/2N_c<N_f<3N_c$.

\setcounter{section}{3}
\section{ $N=1$ supersymmetric dual QCD and 
 non-Abelian electric-magnetic duality }
\label{sect7n}

The global symmetry
and the 't Hooft anomaly matching require that for $N_f>N_c+1$
the low-energy phenomena of supersymmetric $SU(N_c)$ QCD with $N_f$ 
flavours need a distinct description through the introduction
of a dual theory, which takes the same form as the fundamental
supersymmetric gauge theory except for a gauge singlet
field and an additional superpotential. In this section, we shall 
concentrate on Seiberg's conformal window $3N_c/2<N_f<3N_c$ and 
show how the non-Abelian electric-magnetic duality arises at the 
IR fixed point. We shall discuss why the two dual theories have 
inverse couplings and how the theory behaves when some of the heavy modes
decouple.
A generalization of supersymmetric QCD with a matter
field in the adjoint representation of the gauge group
and a non-trival superpotential and its duality
will also be discussed.

\subsection{ Dual supersymmetric QCD}
\label{subsect7.1}

\renewcommand{\theequation}{4.1.\arabic{equation}}
\setcounter{equation}{0}
\renewcommand{\thetable}{4.1.\arabic{table}}
\setcounter{table}{0}

In general, when the flavour $N_f> N_c+1$, the gauge invariant
chiral superfields parameterizing moduli space are the meson superfields
$M^i_{~j}$ and the baryon superfields $B_{ij{\cdots}k}$ and
$\widetilde{B}_{\widetilde{i}\widetilde{j}{\cdots}\widetilde{k}}$. 
Naively one may think that the effective superpotential should have
the following form:
\begin{eqnarray}
W_{\rm eff}
{\sim}\left(\det M-B_{ij{\cdots}k}M^{i\widetilde{i}}M^{j\widetilde{j}}
{\cdots}M^{k\widetilde{k}}\widetilde{B}_{\widetilde{i}\widetilde{j}{\cdots}\widetilde{k}}\right).
\label{eq7.1}
\end{eqnarray}
Although this superpotential is $SU_L(N_f){\times}SU_R(N_f)$ and gauge
invariant, it cannot be regarded as an effective superpotential
since its $R$-charge is not equal to $2$. In addition, one can
easily find that the 't Hooft anomaly matching condition for
the fermionic components of $(Q,\widetilde{Q}, \lambda)$ 
and $(M, B,\widetilde{B})$ cannot be satisfied
if we adopt the effective superpotential (\ref{eq7.1}).
A clever way out has been found by Seiberg and this has led to 
the invention of dual supersymmetric QCD. 
From the Hodge dual form of the baryon superfields (\ref{eq263}),
\begin{eqnarray}  
\overline{B}_{i_{N_c+1}i_{N_c+2} {\cdots}i_{N_f}}&{\equiv}&
\frac{1}{(N_f-N_c)!}{\epsilon}_{i_1{\cdots}i_{N_c}i_{N_c+1}{\cdots}i_{N_f}}
B^{i_{N_1}{\cdots}i_{N_c}}, \nonumber\\[2mm]
\overline{\widetilde{B}}^{i_{N_c+1}i_{N_c+2}, {\cdots}i_{N_f}}&{\equiv}&
\frac{1}{(N_f-N_c)!}{\epsilon}^{i_1{\cdots}i_{N_c}i_{N_c+1}{\cdots}i_{N_f}}
B_{i_{N_1}{\cdots}i_{N_c}},
\label{eq387}
\end{eqnarray}
one can see that $\overline{B}$ and $\overline{\widetilde{B}}$ have
\begin{eqnarray}
\widetilde{N_c}=N_f-N_c
\end{eqnarray}
indices. Thus one can assume that these baryon superfields are bound
states of $\widetilde{N_c}$ chiral superfields $q$ and $\widetilde{q}$,
\begin{eqnarray}
\overline{B}_{i_{N_c+1}i_{N_c+2} {\cdots}i_{N_f}}&{\equiv}&
 B_{i_1{\cdots}i_{\widetilde{N}_c}}
{\equiv}\frac{1}{\widetilde{N}_c}{\epsilon}_{\widetilde{r}_1{\cdots}
\widetilde{r}_{\widetilde{N}_c}}
q^{\widetilde{r}_1}_{~i_1}q^{\widetilde{r}_2}_{~i_2}{\cdots}
q^{\widetilde{r}_{\widetilde{N}_c}}_{~i_{\widetilde{N}_c}},
\nonumber\\
\overline{\widetilde{B}}^{i_{N_c+1}i_{N_c+2} {\cdots}i_{N_f}}&{\equiv}&
\widetilde{B}_{j_1{\cdots}j_{\widetilde{N}_c}}
{\equiv}\frac{1}{\widetilde{N}_c}{\epsilon}_{\widetilde{s}_1{\cdots}
\widetilde{s}_{\widetilde{N}_c}}
q^{\widetilde{s}_1}_{~j_1}q^{\widetilde{s}_2}_{~j_2}{\cdots}
q^{\widetilde{s}_{\widetilde{N}_c}}_{~j_{\widetilde{N}_c}},
\nonumber\\
i,j&=&1,{\cdots}, N_f, ~~~\widetilde{r},\widetilde{s}
=1,{\cdots}, \widetilde{N}_c=N_f-N_c.
\label{eq389}
\end{eqnarray}
Obviously $q^i_{~\widetilde{r}}$ and $\widetilde{q}^i_{~\widetilde{ r}}$ 
belongs to the fundamental 
representation of $SU(N_f){\times}SU(\widetilde{N}_c)$.
To bind these elementary constituents into gauge 
invariant baryon superfields, we must construct a dynamical Yang-Mills theory 
with gauge group $SU(\widetilde{N_c})$, which provides the dynamics.

Seiberg proposed that the low energy 
supersymmetric $SU(N_c)$ QCD with $N_f>N_c+1$
can be described by a supersymmetric Yang-Mills theory with gauge group 
$SU(\widetilde{N_c})$ coupled to the chiral superfields  
$q_{~i}^r$ and $\widetilde{q}_{~j}^r$
as well as a new colour singlet chiral superfield ${\cal M}_i^{~j}$
together with an additional gauge invariant effective superpotential 
\begin{eqnarray} 
W_{\rm eff}=q{\cdot}{\cal M}{\cdot}\widetilde{q}=q^{\widetilde{ r}}_{~i}{\cal M}^{ij}
{\delta}_{\widetilde{ r}\widetilde{s}}\widetilde{q}^{\widetilde{s}}_{~j}.
\label{eq390}
\end{eqnarray}
Note that this new colour singlet superfield ${\cal M}$ cannot be 
directly constructed from $q$ and $\widetilde{q}$ like
that the meson superfield $M$ is constructed from $Q$ 
and $\widetilde{Q}$. Its quantum numbers can be determined from
the superpotential (\ref{eq390}) once we have determined the quantum
numbers for $q$ and $\widetilde{q}$.

\begin{table}
\begin{center}

\begin{tabular}{|c|c|c|c|c|} \hline
      &$ SU_L(N_f)$ &$ SU_R(N_f)$ &$ U_B(1)$ &$ U_R(1)$\\ \hline
 $ Q$   & $N_f$ & $1$  & $1$ & $1-N_c/N_f$ \\ \hline
$\widetilde{ Q}$ & $1$ &$ \overline{N}_f$ & $-1$ &$ 1-N_c/N_f$ \\ \hline
$\psi_Q$ &$ N_f$ & $1$ &$ 1$ &$ -N_c/N_f$ \\ \hline
$\psi_{\widetilde{Q}}$ &$ 1$ &$ \overline{N}_f$ &$ 1$ &$ -N_c/N_f$ \\ \hline
$\lambda$ &$ 1$ &$ 1$ & $0$ &$ 1$ \\ \hline
 $M^{ij}$ &$ N_f$ &$ \overline{N}_f$ &$ 0$ & $2-2N_c/N_f$\\ \hline
$\psi_M$ &$ N_f$ &$ \overline{N}_f$ &$ 0$ &$ 1-2N_c/N_f$\\ \hline
$B_{i_1{\cdots}i_{\widetilde{N}_c}}$ &$ \overline{N}_f
\widetilde{N}_c$ & $1$ &$ N_c$ 
&$ N_c (1-N_c/N_f)$ \\ \hline
$\widetilde{B}_{j_1{\cdots}j_{\widetilde{N}_c}}$ &$1$ & 
$ N_f\widetilde{N}_c$ & $N_c$ & $ N_c (1-N_c/N_f)$ \\ \hline
\end{tabular}
\caption{\protect\small Representation quantum numbers of the chiral superfields
in the original supersymmetric QCD. \label{ta7.1} } 
\end{center}

\end{table}

\begin{table}
\begin{center}

\begin{tabular}{|c|c|c|c|c|} \hline
      &$ SU_L(N_f)$ &$ SU_R(N_f)$ & $U_B(1)$ &$ U_R(1)$\\ \hline
 $ q$   &$ \overline{N}_f$ &$ 1$  &$ N_c/\widetilde{N}_c$ &$ N_c/N_f$ \\ \hline
$\widetilde{ q}$ &$ 1$ &$ N_f$ &$ -N_c/\widetilde{N}_c$ &$ N_c/N_f$ \\ \hline
$\psi_q$ &$\overline{N}_f$ & $1$ & $1$ &$ N_c/N_f-1$ \\ \hline
$\psi_{\widetilde{q}}$ & $1$ &$ N_f$ & $-N_c/\widetilde{N}_c$ &$ N_c/N_f-1$ \\
\hline
$\widetilde{\lambda}$ & $1$ & $1$ & $0$ & $1$ \\ \hline
\end{tabular}
\caption{\protect\small Representation quantum numbers of the chiral superfields and their
fermionic components in dual supersymmetric QCD. \label{ta7.2}}
\end{center}
\end{table}

From (\ref{eq389}) and the quantum numbers of the original 
quarks of the meson and baryon superfields listed in Table \ref{ta7.1}, 
we can determine the quantum numbers of the 
elementary dual superfields and their
fermionic components listed in Table \ref{ta7.2}.
Obviously, the superpotential is $SU(\widetilde{N_c})$ gauge
invariant and globally $SU_L(N_f){\times}SU_R(N_f)$ invariant. 
Further, the superpotential should be $U_B(1){\times}U_R(1)$ 
invariant, i.e.
the superpotential should have baryon number $0$ and $R$-charge 2. 
This requires that 
the quantum numbers of the new meson superfield ${\cal M}$ under
$SU_L(N_f){\times}$$SU_R(N_f)$${\times}$$U_B(1)$~${\times}U_R(1)$
should be
\begin{eqnarray}
{\cal M}: 
\left(N_f, \overline{N}_f, 0, 2-2\frac{N_c}{N_f}\right).
\label{eq7.6}
\end{eqnarray}

To summarize, the low energy supersymmetric $SU(N_c)$ QCD with 
$N_f>N_c+1$ flavours can be described by a dual theory. The dynamical 
variables in the dual theory are the gauge singlet operator ${\cal M}$, 
the $SU(\widetilde{N_c})$ gluons and the dual quarks
$q$ and $\widetilde{q}$. In addition to the standard form
of the supersymmetric QCD Lagrangian,
the dual Lagrangian includes a kinetic energy term of the
colour singlet ${\cal M}$ and the effective superpotential
(\ref{eq390}), which is $SU_L(N_f){\times}SU_R(N_f){\times}U_B(1)
{\times}U_R(1)$ invariant as the original supersymmetric QCD. 
Seiberg further found that these two theories are dual in the sense
of electric-magnetic duality in the conformal window, 
i.e. as the flavour number $N_f$ decreases, 
 the original supersymmetric $SU(N_c)$ QCD becomes more strongly
coupled, but the dual supersymmetric $SU(\widetilde{N}_c)$ QCD
becomes more weakly coupled. 
We shall give a detailed explanation in the next subsection.
This relation between the  
$SU(\widetilde{N}_c)$ gauge theory and the original $SU(N_c)$ is called
non-Abelian electric-magnetic duality. The original theory is usually
called the electric theory and the dual theory is called the magnetic theory.
In addition, the $U(1)$ quantum numbers (i.e. the baryon
numbers and $R$-charges) listed in  
Tables (\ref{ta7.1}) and (\ref{ta7.2}) show that there 
is a very complicated relation between 
the baryon  numbers and $R$-charges of the quarks in the electric and magnetic
theories. Since the $U(1)$ quantum numbers such as baryon number and $R$-charge
are additive quantum numbers, this relations implies that the quarks 
$q$ and $\widetilde{q}$ cannot be simply expressed as  polynomials
of the quarks in the electric theory. This connection between ``electric''
and ``magnetic'' quarks is highly non-local and complicated.
Only in some special cases can the explicit connection between them be 
worked out \cite{kss}. The ``magnetic'' quarks $q$ and $\widetilde{q}$ and 
the $SU(N_f-N_c)$ gluons can be interpreted as solitons of 
the electric theory, i.e. as non-Abelian magnetic monopoles.

Now one question naturally arises: does this $SU(\widetilde{N}_c)$ theory 
describe the real physical dynamics? From the above statements, it seems as if
this $SU(\widetilde{N}_c)$  gauge theory is only a formal device
to introduce dual quark superfields $q$ and $\widetilde{q}$ 
supposedly bound together to form the baryon superfield. However,
in two-dimensional space time, there exist examples where
the gauge field parametrizing a constraints becomes dynamical. The most
famous example is the two-dimensional $CP^N$ model \cite{zak}. 
Thus we can assume that 
this dual $SU(\widetilde{N}_c)$ supersymmetric QCD is
a dynamical theory in which there are gauge bosons and their superpartners
--- gauginos. A strong support for this picture is still the
't Hooft anomaly matching: the anomaly coefficients of supersymmetric QCD 
and of the dual supersymmetric QCD
are identical, so the dual supersymmetric QCD has indeed provided
a dynamical description of the composite superfields of the electric theory.
 In the following we shall explicitly show this.

The quantum numbers for dual quarks superfields listed 
in Table \ref{ta7.2} mean that the currents composed of the 
dual quarks, the mesons and 
the dual gauginos $\widetilde{\lambda}$ corresponding to the 
global symmetry $SU_L(N_f){\times}SU_R(N_f){\times}U_B(1){\times}U_R(1)$
are the following: 
 
\begin{itemize}

\item $SU_{L}(N_f)$
\begin{eqnarray}
\widetilde{J}_{L\mu}^A&{\equiv}&
\overline{\psi}_{qi\widetilde{r}}{\sigma}_{\mu}\overline{t}^A_{ij}
{\psi}_{qj\widetilde{ r}}+
\overline{\psi}_{{\cal M}i}{\sigma}_{\mu}t^A_{ij}{\psi}_{{\cal M}j};
\end{eqnarray}
\item $SU_{R}(N_f)$
\begin{eqnarray}
\widetilde{J}_{R\mu}^A&{\equiv}&
\overline{\psi}_{\widetilde{q}i\widetilde{ r}}{\sigma}_{\mu}
t^A_{ij}{\psi}_{\widetilde{q}j\widetilde{ r}}+
\overline{\psi}_{{\cal M}}{\sigma}_{\mu}\overline{t}^A_{ij}{\psi}_{{\cal M}j};
\end{eqnarray}
\item $U_B(1)$
\begin{eqnarray}
\widetilde{J}^{(B)}_{\mu}
&{\equiv}&
\frac{N_c}{N_f-N_c}\overline{\psi}_{qi\widetilde{ r}}
{\sigma}_{\mu}{\psi}_{qi\widetilde{ r}}
-\frac{N_c}{N_f-N_c}\overline{\psi}_{\widetilde{q}i\widetilde{ r}}{\sigma}_{\mu}
{\psi}_{\widetilde{q}i\widetilde{ r}};
\end{eqnarray}
\item $U_R(1)$
\begin{eqnarray}
\widetilde{J}^{(R)}_{\mu}
&{\equiv}&
\left(-1+\frac{N_c}{N_f}\right)
\overline{\psi}_{qi\widetilde{r}}{\sigma}_{\mu}{\psi}_{qi\widetilde{ r}}
+\left(-1+\frac{N_c}{N_f}\right) 
\overline{\psi}_{\widetilde{q}i\widetilde{ r}}
{\sigma}_{\mu}{\psi}_{\widetilde{q}i\widetilde{ r}}\nonumber\\
&&+\overline{\widetilde{\lambda}}^{\widetilde{a}}{\sigma}_{\mu}
\widetilde{\lambda}^{\widetilde{a}}
+\left(1-\frac{2N_c}{N_f}\right)\overline{\psi}_{{\cal M}i}
\sigma_{\mu}{\psi}_{{\cal M}i}.
\end{eqnarray}
\end{itemize}
Note that in the above currents the range for the 
flavour indices is $i,j=1,{\cdots},N_f$, for the colour 
indices $\widetilde{r}=1,{\cdots},N_f-N_c$, for the magnetic gauge group indices
$\widetilde{a}=1,{\cdots}, (N_f-N_c)^2-1$ and for the flavour group
indices $A=1,{\cdots},N_f^2-1$.

The $SU_L(N_f){\times}SU_R(N_f){\times}U_B(1){\times}U_R(1)$ currents 
composed of the electric quarks and gauginos
are listed below:  

\begin{itemize}
\item $SU_{L}(N_f)$
\begin{eqnarray}
J_{L\mu}^A=\overline{\psi}_Q{\sigma}_{\mu}t^A{\psi}_Q;
\end{eqnarray}
\item $SU_{R}(N_f)$
\begin{eqnarray}
J_{R\mu}^A=\overline{\psi}_{\widetilde{Q}}
{\sigma}_{\mu}\overline{t}^A{\psi}_{\widetilde{Q}};
\end{eqnarray}
\item $U_B(1)$
\begin{eqnarray}
J^{(B)}_{\mu}
=\overline{\psi}_Q{\sigma}_{\mu}{\psi}_Q
+\overline{\psi}_{\widetilde{Q}}{\sigma}_{\mu}{\psi}_{\widetilde{Q}};
\end{eqnarray}
\item $U_R(1)$
\begin{eqnarray}
J^{(R)}_{\mu}
=-\frac{N_c}{N_f}\overline{\psi}_Q{\sigma}_{\mu}{\psi}_Q
-\frac{N_c}{N_f} \overline{\psi}_{\widetilde{Q}}
{\sigma}_{\mu}{\psi}_{\widetilde{Q}}
+\overline{\lambda}^a{\sigma}_{\mu}{\lambda}^a.
\end{eqnarray}
\end{itemize}
It can be easily checked that the non-vanishing 
anomaly coefficients are those collected in Table \ref{ta7.3}.

\begin{table}
\begin{center}

\begin{tabular}{|c|c|c|} \hline
      &$ SU(N_c)$ (or elementary fermions) & $ SU(\widetilde{N}_c) $
(or composite fermions) \\ \hline
$(SU_{L(R)}(N_f))^3$ & $ +(-)d^{ABC}N_c$ & $d^{ABC}N_c$ \\ \hline
$(SU_{L(R)}(N_f))^2U_B(1)$ & $ N_c \mbox{Tr}(t^At^B)$
  & $ N_c \mbox{Tr}(t^At^B)$\\ \hline
$(SU_{L(R)}(N_f))^2U_R(1)$  &  $-N_c^2/N_f\mbox{Tr}(t^At^B)$ 
& $ -N_c^2/N_f\mbox{Tr}(t^At^B)$  \\ \hline
$(U_B(1)(1))^2U_R(1)$ & $ -2N_c^2$ & $-2N_c^2$   \\ \hline
$(U_R(1) )^3$ & $ N_c^2-1-2N_c^4/N_f^2$ & $ N_c^2-1-2N_c^4/N_f^2$  \\ \hline
\end{tabular}

\caption{\protect\small Anomaly coefficients of the original and dual 
supersymmetric QCD, $d^{ABC}=\mbox{Tr}(t^A\left\{t^B,T^C\right\})$.
\label{ta7.3}}
\end{center}

\end{table}

\subsection{Non-Abelian electric-magnetic duality}
\label{subsect7.2}

\renewcommand{\theequation}{4.2.\arabic{equation}}
\setcounter{equation}{0}
\renewcommand{\thetable}{4.2.\arabic{table}}
\setcounter{table}{0}

Subsect.\,\ref{subsect7.1}  
gives a dual description of low energy supersymmetric QCD. 
However, only at the IR fixed point of the range $3N_c/2<N_f<3N_c$, can
the ``electric'' theory and ``magnetic'' theory describe the 
same physics. In the following we give a detailed analysis of 
this non-Abelian electric-magnetic duality. 

First, the range $3N_c/2<N_f<3N_c$ implies the inequality
\begin{eqnarray}
\frac{3}{2}\widetilde{N}_c<N_f<3\widetilde{N}_c,
\end{eqnarray}
where $\widetilde{N}_c=N_f-N_c$ is the number of colours in the
dual theory.
Hence the range $3N_c/2<N_f<3N_c$ for the ``electric'' 
theory, for which a non-trivial IR fixed point exists,  implies the range
$3/2\widetilde{N}_c<N_f<3\widetilde{N}_c$ for the magnetic theory. 
Thus the fixed point of the magnetic theory is identical to that 
of the original electric theory, i.e. they have the 
same non-Abelian Coulomb phase! Note that these two theories have different 
gauge groups and different numbers of interacting particles, but they have 
the same fixed point, and at the fixed point they describe the 
same physics. In the non-relativistic case, both theories have  
a $1/r$ potential in these two ranges,  and there is
no experimental way to tell whether the electric or magnetic gauge bosons
are mediating the interaction.

Secondly, we look at the dynamics. Eq.\,(\ref{eq390}) gives an additional
superpotential of the magnetic theory.
It can be argued that at the IR fixed point there exists a simple relation
between an ``electric'' meson and a colour singlet of the ``magnetic" theory
\begin{eqnarray} 
M^i_{~j}={\mu}{\cal M}^i_{~j}.
\label{eq7.18}
\end{eqnarray}
Since both $SU(N_c)$ and $SU(\widetilde{N}_c)$ QCD are 
asymptotically free in the range $3N_c/2<N_f<3N_c$, 
both of them have an UV fixed point $g=0$. (\ref{eq6.224}) shows  
that the anomalous dimensions of $M$
in the electric theory and of ${\cal M}$ in the magnetic theory vanish
at the UV fixed point. Thus, their dimensions at the UV fixed point
are the canonical dimensions. From its definition (\ref{eq259})
the canonical dimension of $M$ is 2 at the IR fixed point, while from
the beta function (\ref{eq6.224}), the vanishing of the beta 
function gives the anomalous dimension
\begin{eqnarray}
\gamma =-3\frac{N_c}{N_f}+1.
\end{eqnarray}
Thus, the full dimension of $M$ at the IR fixed point is
\begin{eqnarray}
D(M)_{\rm IR~fixed~point}=2+\gamma=3\frac{N_f-N_c}{N_f}.
\end{eqnarray}
This can also obtained from the relation between the conformal dimension
and $R$-charge of a chiral meson superfield operator in a superconformal field
theory (see (\ref{eq9p121})),
\begin{eqnarray}
D(\widetilde{Q}Q)=\frac{3}{2}R(\widetilde{Q}Q)
=\frac{3}{2}[R(\widetilde{Q})+R(Q)]=3\frac{N_f-N_c}{N_f}.
\end{eqnarray}
In the magnetic theory, the discussion in Subsect.\,\ref{subsect7.1}
shows that ${\cal M}$ cannot be constructed from
the elementary magnetic quarks $q$ and $\widetilde{q}$. It should be
regarded as an elementary chiral superfield, thus its canonical dimension
is $1$ and so its dimension at the UV fixed point is $1$. 
At the IR fixed point,
${\cal M}$ and $M$ should become identical under the renormalization group 
flow and ${\cal M}$ should have the same dimension as 
$M$, i.e. $3(N_f-N_c)/N_f$. This can be
seen from its $R$-charge in (\ref{eq7.6}). 
In order to relate ${\cal M}$ to $M$ at 
the UV fixed point, we must introduce a scale $\mu$ and hence (\ref{eq7.18})
arises. In the following, we shall replace ${\cal M}$ by $M/{\mu}$, and  
then the additional superpotential (\ref{eq390}) 
in the magnetic theory can be written as follows:
\begin{eqnarray}
W=\frac{1}{\mu}q_iM^i_{~j}\widetilde{q}^j.
\label{eq7.22}
\end{eqnarray}
We shall see that the introduction of this parameter with mass dimension 
will make the 
non-Abelian electric-magnetic duality explicit. From dimensional
analysis, the scale $\widetilde{\Lambda}$ of the 
magnetic theory and ${\Lambda}$
of the electric theory should satisfy
\begin{eqnarray} 
{\Lambda}^{3N_c-N_f}\widetilde{\Lambda}^{3(N_f-N_c)-N_f}{\propto}{\mu}^{N_f}.
\end{eqnarray}
The explicit relation involves a phase factor $(-1)^{N_f-N_c}$,
\begin{eqnarray} 
{\Lambda}^{3N_c-N_f}\widetilde{\Lambda}^{3(N_f-N_c)-N_f}
=(-1)^{N_f-N_c}{\mu}^{N_f}.
\label{eq7.24}
\end{eqnarray}
The necessity of introducing this phase factor will 
be explained in the following. Let us first see 
the consequences of the relation (\ref{eq7.24}).

First, (\ref{eq7.24}) implies that 
the gauge coupling of the electric theory becomes stronger while
the coupling of the magnetic theory will become weaker 
and vice versa. This can be seen
from the running coupling constant (\ref{eq6.112}),
\begin{eqnarray} 
{\Lambda}^{3N_c-N_f}=q^{3N_c-N_f}e^{-8 \pi^2/[g^2(q^2)]},~~
\widetilde{\Lambda}^{3(N_f-N_c)-N_f}
=q^{3(N_f-N_c)-N_f}e^{-8 \pi^2/[\widetilde{g}^2(q^2)]}.
\end{eqnarray}
The relation (\ref{eq7.24}) leads to
\begin{eqnarray} 
(-1)^{N_f-N_c}{\mu}^{N_f}=q^{N_f}e^{-8 \pi^2/[g^2(q^2)]}
e^{-8 \pi^2/[\widetilde{g}^2(q^2)]}.
\end{eqnarray}
At a certain fixed scale $q=\Lambda$, we have
\begin{eqnarray}
(-1)^{N_f-N_c}\left(\frac{\mu}{\Lambda}\right)^{N_f}e^{8 \pi^2/[g^2(q^2)]}
=e^{-8 \pi^2/[\widetilde{g}^2(q^2)]}.
\end{eqnarray}
This can be thought of as the analogue of the usual electric-magnetic
duality $g{\longrightarrow}1/g$ in an asymptotically free theories.

Secondly, (\ref{eq7.24}) gives the connection between 
gluino condensations of the electric
theory and the magnetic theory. (\ref{eq7.24}) gives
\begin{eqnarray}
&& \ln {\Lambda}^{3N_c-N_f}+\ln \widetilde{\Lambda}^{3(N_f-N_c)-N_f}
=\ln (-1)^{N_f-N_c}{\mu}^{N_f},\nonumber\\
&& d\ln {\Lambda}^{3N_c-N_f}=-d \ln \widetilde{\Lambda}^{3(N_f-N_c)-N_f}.
\end{eqnarray}
As it is well known, the quantum one-loop effective action
of supersymmetric QCD can be expressed in the following form:
\begin{eqnarray}
\Gamma&=&\frac{1}{4}\frac{1}{g^2_{\rm eff}}{\int}d^4x{\int}d^2{\theta}
\mbox{Tr}(W^{\alpha}W_{\alpha})+\mbox{h.c.}\nonumber\\
&=&\frac{1}{8\pi^2}\ln\left(\frac{q}{\Lambda}\right)^{\beta_0}
\int d^4x\int d^2\theta\mbox{Tr}(W^{\alpha}W_{\alpha})+\mbox{h.c.}\, ,
\end{eqnarray}
where $\beta_0=3N_c-N_f$ for the electric theory 
and $\beta_0=3(N_f-N_c)-N_f$ for the
magnetic theory. Differentiating the effective action with respect to
$\ln {\Lambda}^{\beta_0}$, we obtain 
\begin{eqnarray}
\langle W_{\alpha}^2 \rangle =-\langle \widetilde{W}_{\alpha}^2 \rangle,
\label{eq7.30}
\end{eqnarray}
whose lowest component shows that the gluino condensates of the 
electric and magnetic theories are related by
\begin{eqnarray}
\langle \lambda \lambda \rangle =-\langle \widetilde{\lambda}
\widetilde{\lambda}\rangle.
\end{eqnarray}
(\ref{eq7.30}) is very 
similar to the ordinary electric-magnetic duality
if we consider its $\theta^2$ component, which contains the term
$F_{\mu\nu} F^{\mu\nu}$ and is in electric-magnetic components
\begin{eqnarray}
E^2-B^2=-(\widetilde{E}^2-\widetilde{B}^2).
\end{eqnarray}

Now let us see the effect of the phase factor $(-1)^{N_f-N_c}$. A
 duality transformation of the relation (\ref{eq7.24}) 
(i.e. $N_c{\longrightarrow} N_f-N_c$) gives
\begin{eqnarray}  
\Lambda^{3(N_f-N_c)-N_f}\widetilde{\Lambda}^{3N_c-N_f}
=(-1)^{N_c}\widetilde{{\mu}}^{N_f}=(-1)^{N_c-N_f}{\mu}^{N_f}.
\label{eq7.33}
\end{eqnarray}
Hence
\begin{eqnarray}  
\widetilde{\mu}=-\mu.
\label{eq7.34}
\end{eqnarray}
(\ref{eq7.33}) and (\ref{eq7.34}) 
imply that the dual of the magnetic theory (i.e. the dual of the dual) 
is an $SU(N_c)$ theory with scale $\Lambda$. 
We denote its quark fields 
as $d^{ir}$ and $\widetilde{d}_{ir}$. Now there are two 
independent colour singlets, one is the original $M^i_{~j}$ and another
is constructed from the magnetic quarks
\begin{eqnarray}  
N^j_{~i}=q_i\cdot\widetilde{q}^j.
\end{eqnarray}
The mass dimension $3$ and $R$-charge $2$ of 
the superpotential imply that the possible
gauge invariant and $SU_L(N_f){\times}SU_R(N_f){\times}U_B(1){\times}U_R(1)$ 
invariant superpotential for the dual description of the 
magnetic theory must be of the following form:
\begin{eqnarray} 
W=\frac{1}{\widetilde{\mu}}N^j_{~i}d^i
\widetilde{d}_j+\frac{1}{\mu}M^i_{~j}N^j_{~i}
=\frac{1}{\mu}N^j_{~i}(-d^i\widetilde{d}_j+M^i_{~j}).
\end{eqnarray}
$M$ and $N$ are massive chiral superfields and they can be integrated out by
their equations of motion:
\begin{eqnarray} 
M^i_{~j}=d^i\widetilde{d}_j,~~N=0.
\end{eqnarray}
This shows that the duals of the magnetic quarks can be identified with
the original electric quarks $Q$ and $\widetilde{Q}$. Therefore, the dual
of the magnetic theory coincides with the original electric theory. 
The phase factor $(-1)^{N_f-N_c}$ plays an important role in revealing this.

Finally, we stress the correspondence between 
the operators in the electric and
magnetic theories. As mentioned above, the explicit connection
between the electric quarks and magnetic quarks is very difficult to find.
However, the relation between gauge invariant composite operators 
in both theories is obvious. 
The meson operator 
$M^i_{~j}=Q^i\widetilde{Q}_j$ is identical to the colour singlet
operator ${\cal M}^i_{~j}$ at the infrared fixed point. 
As for the baryon operators, (\ref{eq387}) shows that 
at the IR fixed point the baryon operators are related as follows:
\begin{eqnarray}
B^{i_1{\cdots}i_{N_c}}&=&Q^{i_1}{\cdots}Q^{i_{N_c}}=
C{\epsilon}^{i_1{\cdots}i_{N_c}j_1{\cdots}j_{\widetilde{N}_c}i}
B_{j_1{\cdots}j_{\widetilde{N}_c}}
=C{\epsilon}^{i_1{\cdots}i_{N_c}j_1{\cdots}j_{\widetilde{N}_c}}
q_{j_1}{\cdots}q_{j_{\widetilde{N}_c}},\nonumber\\[2mm]
\widetilde{B}^{i_1{\cdots}{i}_{N_c}}&=&\widetilde{Q}^{i_1}{\cdots}\widetilde{Q}^{i_{N_c}}
=C{\epsilon}^{i_1{\cdots}i_{N_c}j_1{\cdots}j_{\widetilde{N}_c}}
B_{j_1{\cdots}j_{\widetilde{N}_c}}
=C{\epsilon}^{i_1{\cdots}i_{N_c}j_1{\cdots}j_{\widetilde{N}_c}} 
\widetilde{q}_{j_1}{\cdots}\widetilde{q}_{j_{\widetilde{N}_c}},
\label{eq7.38}
\end{eqnarray}
where the normalization constant $C$ is
\begin{eqnarray}
C=\sqrt{-(-\mu)^{N_c-N_f}{\Lambda}^{3N_c-N_f}},
\end{eqnarray}
which is required by the various limits to 
be discussed in following subsection. 
Note that the dual relations (\ref{eq7.24}) and (\ref{eq7.38}) respect 
the idenpotency of the duality transformation.

\subsubsection{Various deformations of non-Abelian electric-magnetic duality}
\label{subsub7.2.1}

By deformation we mean the various limiting cases mentioned in 
Subsect.\,\ref{subsub6.3.5}.
We shall show how in these limits the electric-magnetic duality
exchanges strong with weak coupling and the Higgs phase with
the confinement phase in the dual theories.

First, we consider the limit of a large mass of the $N_f$-th
flavour in the electric theory ($SU(N_c)$ supersymmetric 
QCD with $N_f$ flavours) 
by introducing a superpotential $W_{\rm tree}=mM^{N_f}_{~N_f}$.
After integrating out this heavy mode, the low energy theory will be
 an $SU(N_c)$ supersymmetric QCD with $N_f-1$ light flavours. As stated in 
Subsect.\,\ref{subsub6.3.5}, the matching of coupling constants 
at the energy scale $q=m$ gives a connection between the scale $\Lambda$ 
of the high energy theory and the scale $\Lambda_L$ of the 
low energy theory (see (\ref{eq288})),
$\Lambda_L^{3N_c-(N_f-1)}=m\Lambda^{3N_c-N_f}$. 
Since supersymmetric QCD in the range $3N_c/2<N_f<3N_c$
is an asymptotically free theory, the low energy electric theory will have
the stronger coupling. Let us see how the dual magnetic theory 
($SU(\widetilde{N}_c)$ supersymmetric QCD with $N_f$ flavours) behaves. 
With the added tree-level superpotential,  
the full superpotential is
\begin{eqnarray}
W=\frac{1}{\mu}q_iM^i_{~j}\widetilde{q}^j+mM^{N_f}_{~\widetilde{N}_f},
~~i,j=1,{\cdots},N_f.
\end{eqnarray}
The $F$-flatness conditions for 
$M^{N_f}_{~\widetilde{N}_f}$, $M^{i}_{~\widetilde{N}_f}$ and 
$M^{N_f}_{~\widetilde{i}}$ yield
\begin{eqnarray}
q_{N_f}\widetilde{q}^{N_f}=q_{N_f r}\widetilde{q}^{N_f}_r=
-{\mu} m,~~q_i\widetilde{q}^{N_f}=q_{N_f}\widetilde{q}^{i}=0.
\end{eqnarray}
So the lowest components of $q_{N_f}$ and
$\widetilde{q}_{N_f}$ will break 
the magnetic gauge group $SU(N_f-N_c)$ 
to $SU(N_f-N_c-1)$ with $N_f-1$ light magnetic quarks.  
The equations of motion of the massive quarks 
$q_{N_f}$, $\widetilde{q}^{N_f}$ and the composite
quantity $q_{N_f}\widetilde{q}^{N_f}$ lead to
\begin{eqnarray}
M^{N_f}_{~\widetilde{N}_f}=M^{i}_{~\widetilde{N}_f}
=M^{N_f}_{~\widetilde{i}}=0.
\end{eqnarray}    
Thus the low energy superpotential is composed of the light fields
$\widehat{M}$, $\widehat{q}$ and $\widehat{\widetilde{q}}$,
\begin{eqnarray} 
W=\frac{1}{\mu}\widehat{q}_i\widehat{M}^i_{~j}\widehat{\widetilde{q}}^j
=\frac{1}{\mu}\widehat{q}_{is}\widehat{M}^i_{~j}
\widehat{\widetilde{q}}^{j}_s,~~
i,j=1,{\cdots},N_f-1,~~s=1,{\cdots},N_f-N_c-1.
\end{eqnarray}
Similarly to the electric theory, the coupling constants of the 
high energy theory ($SU(\widetilde{N}_c)$ 
with $N_f$ flavours) and the low energy magnetic theory 
($SU(\widetilde{N}_c-1)$ with $N_f-1$ flavours) should match at 
the energy scale $q^2=\langle q_{N_f} \widetilde{q}^{N_f} \rangle $,
\begin{eqnarray} 
\frac{4\pi}{g^2(\langle q_{N_f} \widetilde{q}^{N_f} \rangle )}&=&
\frac{3\widetilde{N}_c-N_f}{2\pi}
\ln\frac{\sqrt{\langle q_{N_f}\widetilde{q}^{N_f} 
\rangle}} {\widetilde{\Lambda}}\nonumber\\
&=& \frac{3(\widetilde{N}_c-1)-(N_f-1)}{4\pi}
\ln\frac{\sqrt{\langle q_{N_f} \widetilde{q}^{N_f} \rangle}}
{\widetilde{\Lambda}_L},\nonumber\\
\widetilde{\Lambda}_L^{3(\widetilde{N}_c-1)-(N_f-1)}
&=&\frac{\widetilde{\Lambda}^{3\widetilde{N}_c-N_f}}
{\langle q_{N_f}\widetilde{q}^{N_f} \rangle}.
\end{eqnarray}
Using the above relation, one can easily find 
\begin{eqnarray} 
&&\Lambda_L^{3N_c-(N_f-1)}\widetilde{\Lambda}_L^{3(\widetilde{N}_c-1)-(N_f-1)}
=m\Lambda^{3N_c-N_f}\frac{\widetilde{\Lambda}^{3\widetilde{N}_c-N_f}}
{\langle q_{N_f}\widetilde{q}^{N_f} \rangle} \nonumber\\
&=&\frac{ m(-1)^{N_f-N_c}{\mu}^{N_f}}{-\mu m}
=(-1)^{N_f-N_c-1}{\mu}^{N_f-1}.
\label{eq7.46}
\end{eqnarray}
Thus the relation (\ref{eq7.24}) is preserved in the large mass limit. 
Further it can be easily verified that (\ref{eq7.38}) is also preserved.
Therefore, under the mass deformation duality is preserved. From the
running gauge coupling, one can see that the duality makes a more 
strongly coupled electric theory equivalent to a more weakly coupled
magnetic theory at the IR fixed point.

However, for the case $N_f=N_c+2$, the magnetic theory is an $SU(2)$
gauge theory and has only two colours. In this case the above discussion 
of the mass deformation is incomplete. If we introduce a large mass 
term for the $N_f$($=N_c+2$)-th flavour, the $SU(2)$ magnetic
gauge symmetry will be completely broken. The low 
energy electric theory contains the mesons $\widehat{M}^{i}_{~j}$,
$i,j=1,{\cdots}, N_c+1$, while in the low energy magnetic theory,
after integrating out the massive quarks, 
there are only massless, colourless magnetic 
quarks $\widehat{q}_i$ and $\widehat{\widetilde{q}}_i$
left. The low energy superpotential is still of the form (\ref{eq7.22}) but  
with no summation over colour indices. From the relation (\ref{eq7.38}), 
which gives the connection
between the baryons $B_i$ and $\widetilde{B}^i$ of the low energy 
electric theory and the colour singlet magnetic quarks of low energy 
magnetic theory, we have 
\begin{eqnarray}
B^{i_1{\cdots}i_{N_c}}=Q^{i_1}{\cdots}Q^{i_{N_c}}=
\widehat{C}{\epsilon}^{i_1{\cdots}i_{N_c}i}\widehat{q}_i,\nonumber\\
\widetilde{B}^{i_1{\cdots}i_{N_c}}=\widetilde{Q}^{i_1}
{\cdots}\widetilde{Q}^{i_{N_c}}=
\widehat{C}{\epsilon}^{i_1{\cdots}i_{N_c}i} \widehat{\widetilde{q}}_i,
\end{eqnarray}  
where $\widehat{C}=\sqrt{\Lambda^{2N_c-1}/\mu}$.
If we adopt the dual form (\ref{eq387}) of the baryon, 
one can see 
\begin{eqnarray} 
\overline{B}_i&=&\sqrt{\frac{\Lambda_L^{2N_c-1}}{\mu}}\frac{1}{N_f!} 
{\epsilon}_{i_1{\cdots}i_{N_c}i}B^{i_1{\cdots}i_{N_c}}
= \sqrt{\frac{\Lambda_L^{2N_c-1}}{\mu}}\widehat{q}_i,\nonumber\\
\overline{\widetilde{B}}_i&
=&\sqrt{\frac{\Lambda_L^{2N_c-1}}{\mu}}\frac{1}{N_f!} 
{\epsilon}_{i_1{\cdots}i_{N_c}i}\widetilde{B}^{i_1{\cdots}i_{N_c}}
=\sqrt{\frac{\Lambda_L^{2N_c-1}}{\mu}}\widehat{\widetilde{q}}_i.
\label{eq7.48}
\end{eqnarray}  
This means that the gauge singlet magnetic quarks are in fact the baryons of the
low energy electric theory, i.e. the baryons are magnetic monopoles  
of elementary quarks and gluons. This idea was actually proposed 
many years ago by Skyrme. Later Witten further showed that at least
in the large $N_c$ case, the baryons can be regarded as 
solitons of the low energy effective 
Lagrangian of ordinary QCD \cite{wit3}. In supersymmetric gauge theory,
this idea comes out naturally in the context of electric-magnetic duality.

Furthermore, with $N_f=N_c+2$, we can obtain 
the superpotential (\ref{eq6.225}) of the case of $N_f=N_c+1$
light flavours
from the duality between the low energy electric and magnetic theory.
From (\ref{eq7.46}) and (\ref{eq7.48}), the superpotential in 
low energy magnetic theory is
\begin{eqnarray} 
W=\frac{1}{\mu}\widehat{q}_i\widehat{M}^i_{~j}\widehat{\widetilde{q}}^j
=\frac{1}{\Lambda_L^{2N_c-1}}\overline{B}_i\widehat{M}^i_{~j}\overline{B}^j,
\label{eq7.49}
\end{eqnarray} 
where $\Lambda_L$ is the scale of the low energy electric theory with
$N_f(=N_c+1)$ light flavours. However, since the gauge 
symmetry in the magnetic theory is completely broken, 
the low energy effective action should include
the contribution from instantons for the broken magnetic group, 
since in this case the magnetic theory is completely Higgsed 
and the instanton calculation
is reliable. According to (\ref{eq300}), the instanton contribution 
in the magnetic theory is
\begin{eqnarray} 
W_{\rm inst}=\frac{\widetilde{\Lambda}_L^{6-(N_c+2)}\det(\mu^{-1}\widehat{M})}
{q^{N_f+2}\widetilde{q}^{N_f+2}}=-\frac{\det\widehat{M}}
{\Lambda_L^{3N_c-(N_c+1)}} ,
\label{eq7.50}
\end{eqnarray} 
where (\ref{eq7.46}) and (\ref{eq322}) were used. 
Putting (\ref{eq7.49}) and 
(\ref{eq7.50}) together, we obtain 
the superpotential of the low energy theory with $N_f=N_c+1$ 
light flavours (the index $L$ is omitted)
\begin{eqnarray}
W=\frac{1}{\Lambda^{2N_c-1}}(B_i M^i_{~j}\widetilde{B}^j-\det M).
\end{eqnarray} 
This is exactly the superpotential (\ref{eq6.225}), 
where we got it from the strongly coupled
electric theory, while here we rederived it from the weakly 
coupled magnetic theory. 

In a similar way one can consider the general mass deformation by
introducing a mass term $m_{~i}^{j}M^i_{~j}$ for all the (electric) quarks. 
The full superpotential $W=q_iM^i_{~j}\widetilde{q}^j+m_{~i}^{j}M^i_{~j}$
implies that the dual magnetic quarks get the mass (matrix) 
\begin{eqnarray}
m^i_{{\rm mag}~j}=\frac{M^i_{~j}}{\mu},
\label{eq7.52}
\end{eqnarray}
so the low energy magnetic theory is a pure $SU(N_f-N_c)$ Yang-Mills theory.
From (\ref{eq309}) we have 
\begin{eqnarray}
\widetilde{\Lambda}^3_L
&=&\left[\det(\mu^{-1}M)\Lambda^{3\widetilde{N}_c-N_f}
\right]^{1/\widetilde{N}_c},\nonumber\\
\widetilde{\Lambda}^{3(N_c-N_f)}_L&=&\mu^{-N_f}\widetilde{\Lambda}^{3(N_f-N_c)-N_f}
\det M.
\end{eqnarray}
Correspondingly, there exists an effective superpotential produced 
from gluino condensation,
\begin{eqnarray}
W_{\rm eff}&=&\widetilde{N}_c\widetilde{\Lambda}^3_L
=(N_f-N_c)\left[\mu^{-N_f}\widetilde{\Lambda}^{3(N_f-N_c)-N_f}
\det M\right]^{1/(N_f-N_c)}\nonumber\\
&=&(N_c-N_f)\left(\frac{\Lambda^{3N_c-N_f}}{\det M}\right)^{1/(N_c-N_f)},
\label{eq7.54}
\end{eqnarray}
where we have used (\ref{eq7.24}).  
One can see that (\ref{eq7.54}) is the same as (\ref{eq300}), 
but (\ref{eq7.54}) is obtained from the dual magnetic theory. 
This implies that the superpotential (\ref{eq300}) and the expectation
 values (\ref{eq308}) are reproduced correctly when mass terms are 
added to the magnetic theory. It should be emphasized that
the factor $(-1)^{N_f-N_c}$ plays a crucial role in getting (\ref{eq7.54}), 
otherwise the sign will be opposite.

There is another possible deformation, which is realized by making
the quark superfields have big expectation values along the $D$-flat
direction (\ref{eq6.103x}) 
in the electric theory. For simplicity, we only choose
one flavour, say, the $N_f$-th flavour, to have a large expectation value,
 i.e. $\langle Q^{N_f} \rangle =
\langle \widetilde{Q}^{N_f} \rangle$  is large. 
The lowest components of $Q^{N_f}$
and $\widetilde{Q}^{N_f}$ will break the electric $SU(N_c)$ theory with 
$N_f$ flavours to $SU(N_c-1)$ with $N_f-1$ flavours. From (\ref{eq282}),
the matching of running coupling constant at 
 $\langle Q^{N_f} \rangle =
\langle \widetilde{Q}^{N_f} \rangle $ leads to 
\begin{eqnarray}
\Lambda_L^{3(N_c-1)-(N_f-1)}=\frac{\Lambda_L^{3N_c-N_f}}
{\langle Q^{N_f}\widetilde{Q}^{N_f} \rangle}. 
\end{eqnarray}
In the magnetic theory, similarly to (\ref{eq7.52}), a large
$\langle M^{N_f}_{~N_f} \rangle $ yields a large mass 
$\mu^{-1}\langle M^{N_f}_{~N_f} \rangle $ for the $N_f$-th magnetic
quarks, $q^{N_f}$ and $\widetilde{q}^{N_f}$. The low energy magnetic 
theory is $SU(N_f-N_c)$ with $N_f-1$ light flavours and 
the low energy scale is
\begin{eqnarray}
\widetilde{\Lambda}_L^{3(N_f-N_c)-(N_f-1)}
=\mu^{-1}\langle M^{N_f}_{~N_f} \rangle
\widetilde{\Lambda}^{3(N_c-N_f)-N_f}.
\end{eqnarray}
One can easily check that the relation (\ref{eq7.24}) 
is satisfied and hence that duality
is preserved. Similar discussions for $\langle B\rangle {\neq}0$
show that the duality is also preserved in this case \cite{aha}. 

\subsubsection{Non-Abelian free magnetic phase: $N_c+2{\leq}N_f{\leq}3/2 N_c$}
\label{subsub7.2.2}

Now we consider a special range of colour and flavour numbers,
$N_c+2{\leq}N_f{\leq}3/2 N_c$. 
From the NSVZ beta function of the dual theory 
\begin{eqnarray}
\beta (g)&=&-\frac{g^3}{16\pi^2}\frac{3(N_f-N_c)-N_f
+N_f\gamma (g^2)}{1-(N_f-N_c)g^2/(8\pi^2)},\nonumber\\
\gamma (g^2)&=&-\frac{g^2}{8\pi^2}
\frac{(N_f-N_c)^2-1}{(N_f-N_c)}+{\cal O}(g^4),
\end{eqnarray}
one can see that when $N_f{\leq}3/2 N_c$, $3(N_f-N_c){\leq}N_f$, the 
beta function is positive, so the magnetic $SU(N_f-N_c)$ theory is not 
asymptotically free and is weakly coupled at low energy. Thus at low energy
the magnetic quarks are not confined  and the particle spectrum consists
of the singlet $M$, and the magnetic quarks $q$ and 
$\widetilde{q}$. The relations (\ref{eq7.38}) show that the 
massless magnetic particles are composites of the 
elementary electric degrees of freedom. Comparing with
the case $N_f{\geq}3N_c$ in the electric theory, which is in
the non-Abelian free electric phase, we say that the theory is
in a non-Abelian free magnetic phase since there are massless 
``magnetic'' charged fields.

\setcounter{section}{4}
\setcounter{subsection}{2}
\subsection{Duality in Kutasov-Schwimmer model}
\label{subsect43}
\renewcommand{\theequation}{4.3.\arabic{equation}}
\setcounter{equation}{0}
\renewcommand{\thetable}{4.3.\arabic{table}}
\setcounter{table}{0}

\subsubsection{Kutasov's observation}

The Kutasov-Schwimmer model is the usual $N=1$ supersymmetric 
$SU(N_c)$ QCD but with an additional matter field $X$ in the adjoint
representation of the gauge group and an associated superpotential   
of the general form \cite{ref431,ref432,kss}
\begin{eqnarray}
W=\sum_{i=0}^k\frac{s_i}{k+1-i}\mbox{Tr}X^{k+1-i}.
\label{eq4p31}
\end{eqnarray}
It seems that the presence of these non-renormalizable 
interactions described by this superpotential is irrelevant 
for the short-distance behaviour of the theory, but 
they may have strong effects on the infrared dynamics. Thus this
model has a rich electric-magnetic duality structure. In the 
following we give a brief introduction to the duality present in 
this model.

 We start from the original observation made by Kutasov \cite{ref431}.
According to the $NSVZ$ $\beta$-function of $N=1$ supersymmetric Yang-Mills
theory with matter fields $\Phi_i$ in representations $r_i$ of the gauge
group $G$ given by (\ref{eq1p1}) 
\begin{eqnarray}  
\beta (\alpha )=-\frac{\alpha^2}{2\pi}\frac{3T(G)
-\sum_i T(r_i)[1-\gamma_i(\alpha)]}{1-T(G)\alpha/(2\pi)},
\label{eq4p32}
\end{eqnarray}
where $i$ labels the species of fields, $\alpha=g^2/(4\pi)$
is the fine structure constant, and $\gamma_i(\alpha)$ are the anomalous
dimensions of $\Phi_i$, 
\begin{eqnarray}  
\gamma_i(\alpha)=-C_2(r_i)\frac{\alpha}{\pi}+{\cal O}(\alpha^2)
\label{eq4p33}
\end{eqnarray}  
with 
\begin{eqnarray}
\mbox{Tr}(T^aT^b)&=&T(R)\delta^{ab},~~~T^aT^a=C_2(R){\bf 1},
~~~T(G)\,{\equiv}\,T(R=\mbox{adjoint}).
\label{eq4p34}
\end{eqnarray} 
Non-trivial fixed points arise when 
\begin{eqnarray}
3T(G)-\sum_i T(r_i)[1-\gamma_i(\alpha)]=0.
\label{eq4p35}
\end{eqnarray} 
Note that now these fixed points are not necessarily 
the infrared fixed points.
At a fixed point the physics is described by a superconformal 
field theory, and the scaling dimension of $\Phi_i$ should be
the sum of the canonical and anomalous dimensions
\begin{eqnarray}
d_i=1+\frac{\gamma_i}{2}.
\label{eq4p36}
\end{eqnarray}
On the other hand, for a superconformal theory there exists a simple 
relation between the scaling dimension and the anomaly-free 
$R$-charge
\begin{eqnarray}
R_i=\frac{2}{3}d_i.
\label{eq4p37}
\end{eqnarray}
With (\ref{eq4p36}) and (\ref{eq4p37}), (\ref{eq4p35}) can be written as
\begin{eqnarray}
T(G)+\sum_i T(r_i)[R_i-1]=0.
\label{eq4p38}
\end{eqnarray} 
(\ref{eq4p38}) actually provides a condition for an $R$-symmetry to be 
anomaly-free, under which the supercoordinates $\theta_{\alpha}$ 
have the $R$-charge $1$ and the fields $\Phi_i$ have $R$-charges $R_i$.
Thus (\ref{eq4p38}) is also called the anomaly cancellation condition. 
In general, there may be many $U_R(1)$ symmetries whose $R$-charges satisfy
(\ref{eq4p38}) and thus they are anomaly-free. 
However, since we are only interested 
in infra-red (IR) fixed points, it is necessary to know
 which of the $R$-symmetries
 is the right one, i.e. which $R$-symmetry becomes a part of the IR 
superconformal algebra. There are some cases in which the answer is known,
for instance, if all the matter fields $\Phi_i$ belong to the same 
representation $r_i=r$, $i=1,{\cdots}, M$, then we have
\begin{eqnarray}
R_i=R=1-\frac{T(G)}{MT(R)}.
\label{eq4p39}
\end{eqnarray} 
The $SU(N_c)$ supersymmetric QCD with $N_f$ quark and anti-quark 
superfields, $Q_i$ and $\widetilde{Q}_i$, is just this case,
where we have
\begin{eqnarray}
T(SU(N_c))=N_c, ~~~T(N_c)=1/2, ~~~M=2N_f.
\label{eq4p310}
\end{eqnarray}
Thus anomaly-free $R$ charges for quark and anti-quark superfields are
$R=1-{N_c}/{N_f}$,
which naturally agrees with the $R$-charges listed in Table \ref{ta6.1}.
However, there is no general method of 
determining $R_i$ in (\ref{eq4p38}) to
pick the right $R$-charge so that it becomes a generator of the IR 
superconformal algebra. It is not even clear in general whether 
the theory ends up in the infrared region  of a non-Abelian Coulomb phase.
Among the phases introduced in Sect.\ref{subsect27}, 
we know that only the Coulomb phase allows
a self-dual description.

  Based on the above observation, Kutasov considered a straightforward
generalization of supersymmetric QCD \cite{ref431}, i.e. supersymmetric 
$SU(N_c)$ Yang-Mills theory with a matter superfield $X$ in the adjoint 
representation of the gauge group, $N_f$ multiplets $Q^i$ in the
$N_c$ and $N_f$ supermultiplets $\widetilde{Q}_i$ 
in the $\overline{N}_c$ representations; $i=1,{\cdots},N_f$.
Due to the presence of the matter field in the adjoint representation,
the one-loop $\beta$ function coefficients for this model becomes
\begin{eqnarray}
\beta_0=2N_c-N_f.
\label{eq4p312}
\end{eqnarray} 
Thus the theory is asymptotically free only for $N_f<2N_c$. It is natural
to assign the same $R$ charge $R_f$ to all the fundamental multiplets
$Q^i$, $\widetilde{Q}_i$ and a different $R$ charge $R_a$ to the 
adjoint one, $X$. From (\ref{eq4p310}) 
and $T(\mbox{adjoint})=N_c$, the anomaly 
cancellation condition (\ref{eq4p38}) takes the form
\begin{eqnarray}
&& N_c+2N_f\, \frac{1}{2}(R_f-1)+N_c(R_a-1)=0,\nonumber\\
&& N_fR_f+N_cR_a=N_f.
\label{eq4p313}
\end{eqnarray}            
There are many possible assignments for $R_f$ and $R_a$ satisfying
(\ref{eq4p313}), the problem is which $R$-symmetry becomes part of the IR 
superconformal algebra. 

 Kutasov used the following technique to solve this problem \cite{ref431}: 
first formally finding the dual description, then making use of the
consistency of the theory to work out the correct dynamics so that
the right anomaly-free $R$-symmetry can be distinguished. The method of
searching for the dual magnetic description is the same as in the 
supersymmetric QCD case discussed in Sect.\ref{subsect7.1}. 
Since the baryons can reveal the form of
the duality transformation, one first defines the baryon-like operators 
in the above model,
\begin{eqnarray}
B^{[i_1{\cdots}i_k][i_{k+1}{\cdots}i_{N_c}]} 
=\epsilon^{\alpha_1{\cdots}\alpha_{N_c}}X^{~\beta_1}_{\alpha_1}
X^{~\beta_2}_{\alpha_2}{\cdots}X^{~\beta_k}_{\alpha_k}
Q_{\beta_1}^{~i_1}{\cdots}Q_{\beta_k}^{~i_k}Q_{\alpha_{k+1}}^{~i_{k+1}}
{\cdots}Q_{\alpha_{N_c}}^{~i_{N_c}},
\label{eq4p314}  
\end{eqnarray}
where $\alpha_i, \beta_j=1,{\cdots},N_c$ are colour indices.
For a given $k$, there are $\left(\begin{array}{c}N_f\\k\end{array}\right)
\left(\begin{array}{c}N_f\\N_c-k\end{array}\right)$ baryons
$B^{[i_1{\cdots}i_k][i_{k+1}{\cdots}i_{N_c}]}$, so the total
number of the baryon operators is
\begin{eqnarray}
\sum_{k=0}^{N_c}\left(\begin{array}{c}N_f\\k\end{array}\right)
\left(\begin{array}{c}N_f\\N_c-k\end{array}\right)=\left(\begin{array}{c}2N_f\\N_c
\end{array}\right).
\label{eq4p315}
\end{eqnarray}
(\ref{eq4p314}) and (\ref{eq4p315}) show that 
$k{\leq}N_f$, $N_c-k{\leq}N_f$, and
hence the baryon operators (\ref{eq4p314}) exist only for $N_f{\geq}N_c/2$.

(\ref{eq4p315}) reveals that the spectrum of baryons has a symmetry under
$N_c{\longleftrightarrow}2N_f-N_c$ (with the flavour number $N_f$ fixed)
since $\left(\begin{array}{c}2N_f\\N_c\end{array}\right)
=\left(\begin{array}{c}2N_f\\2N_f-N_c\end{array}\right)$. Thus the
corresponding dual (``magnetic") theory should be an  
$SU(\widetilde{N}_c=2N_f-N_c)$ gauge theory with the fundamental
supermultiplets $q_i$, $\widetilde{q}^i$ and adjoint supermultiplet $Y$
as well as some other colour singlets.   
Similarly to (\ref{eq7.38}), an identification of 
the baryon operators in electric and magnetic theories should be 
possible:
\begin{eqnarray}
B^{[i_1{\cdots}i_k][i_{k+1}{\cdots}	i_{N_c}]}_{\rm el}
=B^{[j_1{\cdots}j_p][j_{p+1}{\cdots}j_{2N_f-N_c}]}_{\rm mag},
\label{eq4p316}
\end{eqnarray}
with $p=N_f-N_c+k$. Assume that the magnetic quark superfields
$q_i$ and $\widetilde{q}^i$ have the $R$-charges $\widetilde{R}_f$
and that $Y$ has the same $R$-charges $R_a$ as $X$. Then 
the anomaly cancellation
condition (\ref{eq4p313}) in both electric and magnetic theories and
the identification (\ref{eq4p316}) yield
\begin{eqnarray}
&& N_fR_f+N_cR_a=N_f, \nonumber\\
&& N_f\widetilde{R}_f+(2N_f-N_c)R_a=N_f, \nonumber\\
&& kR_a+N_cR_f=(N_f-N_c+k)R_a+(2N_f-N_c)\widetilde{R}_f.
\label{eq4p317}
\end{eqnarray}
Hence we get
\begin{eqnarray}
R_a=\frac{2}{3},~~~R_f=1-\frac{2}{3}\frac{N_c}{N_f},~~~
\widetilde{R}_f=1-\frac{2}{3}\frac{2N_f-N_c}{N_f}.
\label{eq4p318}
\end{eqnarray}
The $R$-charge of the adjoint supermultiplets in (\ref{eq4p318})
seems to present a puzzle to us: the theory we are considering 
is asymptotically free,
so $g=0$ is the UV fixed point, at which the $R$-charge of $X$ is
$2/3$, whereas (\ref{eq4p318}) shows that at the 
IR fixed point the $R$-charge of $X$
is also $2/3$. Usually this not possible since  (\ref{eq4p32}) 
and (\ref{eq4p33}) show that
even for a perturbative fixed point the conformal dimension and hence
the $R$-charge of $X$ can receive a contribution at least
at first order in $\alpha$.

It was realized by Kutasov that the reason for this lies in the ambiguous
action of $U_R(1)$ on the chiral supermutiplets in the adjoint and
fundamental representations, and only an interaction superpotential
relevant to the chiral supermutiplet in the adjoint representation
can fix this ambiguity. Thus
Kutasov's resolution to this puzzle was to assume that the model considered 
should be subject to a Wess-Zumino superpotential  composed of
the adjoint matter \cite{ref431}
\begin{eqnarray}
W_{\rm el}(X){\propto}\mbox{Tr}X^3.
\label{eq4p319}
\end{eqnarray}
A perturbative calculation in the Wess-Zumino model shows that the interaction
provided by this superpotential contributes to the anomalous dimensions
of $X$ and hence to the $\beta$ function of the gauge coupling. Thus with the 
superpotential (\ref{eq4p319}) there is 
a possible additional fixed point at which
$X$ has the canonical conformal dimension $1$ and hence the $R$-charge
$2/3$. In particular, the superpotential 
(\ref{eq4p319}) respects the $R$-symmetry given by (\ref{eq4p318}),
since as a superpotential it should have $R$-charge $2$. Therefore,
 the introduction of this superpotential also chooses the right $R$-symmetry
so that it becomes a part of the IR superconformal symmetry. Of course,
the $R$-charge (\ref{eq4p318}) at the IR fixed point is 
actually found by requiring the existence of a duality, which 
is independent of the discussion of the superpotential (\ref{eq4p319}).

 Next we briefly review the duality mentioned above. 
First, the electric theory has an anomaly-free global symmetry
$SU_L(N_f){\times}SU_R(N_f){\times}U_B(1){\times}U_R(1)$
at the IR fixed point under which the quantum numbers of the 
matter fields are listed in Table (\ref{ta4p31}).

\begin{table}
\begin{center}
\begin{tabular}{|c|c|c|c|c|} \hline
     & $SU_L(N_f)$  & $SU_R(N_f)$ &  $U_B(1)$  & $U_R(1)$ \\ \hline
 $Q$ & $N_f$        & $1$         &    $1$     &  $1-2N_c/(3N_f)$ \\ \hline
 $\widetilde{Q}$ & $1$ & $\overline{N_f}$ & $-1$ & $1-2N_c/(3N_f)$ \\ \hline
 $X$ &  $1$ & $1$ & $0$ & $2/3$\\ \hline 
\end{tabular}
\caption{\protect\small Representation quantum numbers of the matter fields
of the electric theory under 
$SU_L(N_f){\times}SU_R(N_f){\times}U_B(1){\times}U_R(1)$. \label{ta4p31} }
 \end{center}

\end{table}

The dual description is the $SU(2N_f-N_c)$ gauge theory. The 
matter fields include not only the dual quarks $q_i$, $\widetilde{q}^i$
and the adjoint field $Y$, but also two gauge singlet chiral superfields
$M_{~j}^i$ and $N^i_{~j}$. At the IR fixed point, these gauge singlets
are the meson and meson-like operators of the original electric theory
\begin{eqnarray}
 M_{~j}^i=Q^i_{~r}\widetilde{Q}^{r}_{~j}=
Q^i{\cdot}\widetilde{Q}_j, 
~~~N_{~j}^i=Q^i_{~r}X^r_{~s}\widetilde{Q}^s_{~j}=
Q^i{\cdot}X{\cdot}\widetilde{Q}_j.
\label{eq4p321}
\end{eqnarray}  
The quantum numbers of all the matter fields under the 
global symmetry $SU_L(N_f){\times}SU_R(N_f)$ ${\times}U_B(1){\times}U_R(1)$ 
are listed in Table (\ref{ta4p32}).

\begin{table}
\begin{center}
\begin{tabular}{|c|c|c|c|c|} \hline
     & $SU_L(N_f)$  & $SU_R(N_f)$ &  $U_B(1)$  & $U_R(1)$ \\ \hline
 $q$ & $\overline{N}_f$        & $1$         &    $N_c/(2N_f-N_c)$     &  
$1-2(N_f-N_c)/(3N_f)$ \\ \hline
 $\widetilde{q}$ & $1$ & $N_f$ & $-N_c/(2N_f-N_c)$ & $1-2(N_f-N_c)/(3N_f)$ 
\\ \hline
 $Y$ &  $1$ & $1$ & $0$ & $2/3$\\ \hline
 $M$ &  $N_f$ & $\overline{N}_f$ & $0$ & $2-4N_c/(3N_f)$ \\ \hline
 $N$ &  $N_f$ & $\overline{N}_f$ & $0$ & $8/3-4N_c/(3N_f)$ \\ \hline
\end{tabular}
\caption{\protect\small Representation quantum numbers of the matter fields
of the magnetic theory under 
$SU_L(N_f){\times}SU_R(N_f){\times}U_B(1){\times}U_R(1)$. \label{ta4p32} }
 \end{center}

\end{table}

Strong support to this duality pattern comes from the 
't Hooft anomaly matching for the above global symmetry group . 
Using the quantum numbers listed in Tables (\ref{ta4p31}), (\ref{ta4p32}) 
and the quantum numbers of the electric and magnetic gauginos,
\begin{eqnarray}
\lambda: \, \, (1,1,0,1); ~~~~\widetilde{\lambda}:\, \, (1,1,0,1),
\label{eq4p322}
\end{eqnarray}
one can easily write down the conserved currents and the energy-momentum 
tensor composed of the massless fermionic components and calculate the
anomaly coefficient. The explicit calculation 
shows that the coefficients are indeed equal in both 
theories and they are listed in Table (\ref{ta4p33}). Note
that now there are more non-vanishing triangle diagrams in comparison
with the usual $N=1$ supersymmetric QCD. 

\begin{table}
\begin{center}
\begin{tabular}{|c|c|} \hline
  Triangle diagrams    &  Anomaly coefficients\\ \hline
 $SU_{L(R)}(N_f)^3$ & $+(-)N_c\mbox{Tr}(t^A\{t^B,t^C\})$ \\ \hline
 $SU_{L(R)}(N_f)^2U_R(1)$ & $-2N_c^2/(3N_f)\mbox{Tr}(t^At^B)$  \\ \hline
 $SU_{L(R)}(N_f)^2U_B(1)$ &  $N_c\mbox{Tr}(t^At^B)$\\ \hline
 $U_R(1)$ &  $-2(N_c^2+1)/3$ \\ \hline
 $U_R(1)^3$ & $26(N_c^2-1)/27-16N_c^4/(27N_f^2)$   \\ \hline
 $U_B(1)^2 U_R(1)$ & $-4N_c^2/3$\\ \hline
\end{tabular}
\caption{\protect\small  't Hooft anomaly coefficients.\label{ta4p33} }
 \end{center}

\end{table}

As in the usual supersymmetric QCD case, the global 
symmetry and holomorphy determine that the superpotential 
of the dual theory should be of the form
\begin{eqnarray}
W_{\rm mag}=M^i_{~j}q_i{\cdot}Y{\cdot}\widetilde{q}^j+
N^i_{~j}q_i{\cdot}\widetilde{q}^j+\mbox{Tr}Y^3.
\label{eq4p323}
\end{eqnarray}
Note that the first term of the above superpotential is
an operator with UV dimensions $4$. However, as argued
in the following, $M^i_{~j}q_i{\cdot}Y{\cdot}\widetilde{q}^j$ is actually
not always irrelevant, i.e. its dimension can become less than $4$.

The dynamical structure of the theory can be qualitatively analyzed
through the scaling dimension of the meson fields $M^i_{~j}$ at the
IR fixed point,
\begin{eqnarray}
d(M)=\frac{3}{2}R=3-\frac{2N_c}{N_f}.
\label{eq4p324}
\end{eqnarray}
The one-loop beta function coefficient (\ref{eq4p312}) shows that when
$N_f>2N_c$, the electric theory is not asymptotically free
and thus in a free electric phase. At $N_f{\sim}2N_c$, $d(M){\sim}2$,
(\ref{eq4p323}) shows that the electric theory 
should be weak coupled. As $N_f$
decreases, the running coupling in the IR electric theory increases and
$d(M)$ decreases. At $N_f=N_c$, $d(M)$ becomes $1$ and $M$ behaves
as a free scalar field. It is then natural to expect that
the dimension of $M$ in the IR region remains $1$ for $N_f<N_c$ as well.

 In the dual magnetic description, the dynamical behaviour of $M$ is also
remarkable. From the one-loop beta function coefficient of 
the magnetic theory,
\begin{eqnarray}
\widetilde{\beta}=2(2N_f-N_c)-N_f=3N_f-2N_c,
\label{eq4p325}
 \end{eqnarray} 
one can see that for  
\begin{eqnarray}
N_f{\simeq}\frac{2}{3}N_c
\label{eq4p326}
\end{eqnarray}
the coupling is weak,
and at low energy $M$ will become a free field coupled to the 
gauge sector through the first term of the nonrenormalizable 
superpotential (\ref{eq4p323}). If $N_f>2N_c/3$, the 
coupling of $MqY\widetilde{q}$
will decrease at large distances. For $N_f$ not much larger than 
$2N_c/3$, perturbation theory works.  One can see that $M$ becomes
a free field with dimension $1$. However, as $N_f$ increases, the coupling
in the magnetic theory becomes stronger, the anomalous dimension
of $MqY\widetilde{q}$ may become more and more negative until, if
the coupling is strong enough, this irrelevant operator actually becomes 
relevant, i.e.  the full (canonical + anomalous) dimension may become
less than $4$. Hence it will have effects on the IR dynamics of $M$ and the
strongly interacting magnetic gauge degrees of freedom. Obviously, if
this phenomenon occurs, it is completely due to non-perturbative effects
in the magnetic theory, since usually the IR magnetic {\it gauge}
coupling should be larger than the critical coupling 
so that $MqY\widetilde{q}$ becomes relevant. Note that the interaction
of $M$ with the gauge sector is a flavour interaction and the gauge
interaction occurs through the colour degrees of freedom. 
Usually at low energy the colour interaction is much stronger 
than the flavour one.

 One can perform a similar discussion for the operators $N^i_{¨j}$ and the
related $Nq\widetilde{q}$ interaction. From Table 
(\ref{ta4p32}) the operators $N^i_{~j}$ have dimension
\begin{eqnarray}
d(N)=\frac{3}{2}R(N)= 4-\frac{2N_c}{N_f},
\label{eq4p327}
\end{eqnarray} 
which will go to $1$ at $N_f=2N_c/3$. For $N_f<2N_c/3$, 
(\ref{eq4p325}) shows that
the magnetic theory is not asymptotically free, and hence the full
theory is free in the infrared region. Consequently,
$N^i_{~j}$ become free field operators when $N_f<2N_c/3$. 
Otherwise they will participate in the interaction non-trivially.

 In summary, the general dynamical structure of this set of theories
can be stated as follows: the theory is free in the ranges 
$N_f >2N_c$ and $N_f<2N_c/3$. The former corresponds to a free electric
phase, with the fields $M$ and $N$ corresponding to quark composites 
with dimensions $2$ and $3$, respectively, due to (\ref{eq4p324}) 
and (\ref{eq4p327}); the latter corresponds to 
a free magnetic phase with $M$ and $N$ being
free fields with dimension $1$. The theory is in an interacting 
non-Abelian Coulomb phase for the range $2N_c/3 <N_f<2N_c$, in which
there is a dual magnetic description, i.e. supersymmetric $SU(2N_c-N_f)$ QCD
with two colour singlets $M$ and $N$. Obviously, the model with $N_f=N_c$
is self-dual under the duality $N_c{\leftrightarrow}2N_f-N_c$, which exchanges 
the range $N_c<N_f<2N_c$ with $2N_c/3<N_f<N_c$. (\ref{eq4p324}) 
shows that as $N_f<N_c$ 
the field operators $M^i_{~j}$ are free field operators 
since they have dimension $1$,
but the full theory is not.

 A discussion similar to the $SU(N_c)$ QCD case can show that the duality
is preserved under a mass deformation and in the flat directions. Note that
the flat directions of the magnetic theory are composed of 
the $D$-flat directions of the gauge sector and the 
$F$-flat directions obtained through the superpotential 
(\ref{eq4p323}) \cite{ref431}.

\subsubsection{Duality in the Kutasov-Schwimmer model}

\vspace{2mm}

\begin{flushleft}
{\it Kutasov-Schwimmer model}
\end{flushleft}

The Kutasov-Schwimmer model is a generalization of the model introduced
in last section. The field content is the same, only the 
superpotential (\ref{eq4p319}) is replaced by a general one,
\begin{eqnarray}
W= g_k\mbox{Tr}X^{k+1}.
\label{eq4p328}
\end{eqnarray}
For simplicity, only the case $k<N_c$ will be considered. 
(\ref{eq4p328}) shows that for $k=1$ this superpotential 
is the mass term for $X$, so
one can integrate out the adjoint superfield in the IR region 
and return back to the usual supersymmetric QCD. The $k=2$ case is
just the one discussed in the last section. 

 We first have a look at the moduli space of this model. 
As in the usual supersymmetric QCD, the moduli space is labeled by two kinds
of gauge invariant operators: the meson operators
\begin{eqnarray}
(M_l)^i_{~j}=Q^i_{~\alpha}(X^{l-1})^\alpha_{~\beta}\widetilde{Q}_{~j}^{\beta},
l=1,{\cdots},k,
\label{eq4p329}
\end{eqnarray}
with
\begin{eqnarray}
(X^{l-1})^\alpha_{~\beta}{\equiv}X^\alpha_{~\alpha_1}X^{\alpha_1}_{~\alpha_2}
{\cdots}X^{\alpha_{l-1}}_{~\beta},
\label{eq4p330}
\end{eqnarray}
and the baryon-like operators
\begin{eqnarray}
B^{(n_1n_2{\cdots}n_k)}=Q^{n_1}_{(1)}{\cdots}Q^{n_k}_{(k)};~~~~~
\sum^k_{l=1}n_l=N_c,
\label{eq4p331}
\end{eqnarray}
where $Q_{(l)}$ are  the ``dressed'' quarks 
\begin{eqnarray}
Q_{(l)\alpha}^i{\equiv}\left(X^{l-1}Q^i\right)_\alpha
=\left(X^{l-1}\right)_\alpha^{~\beta} Q^i_{~\beta}, ~~~~l=1,{\cdots},k,
\label{eq4p332}
\end{eqnarray}
and consequently
\begin{eqnarray}
Q^{n_l}_{(l)}=\frac{1}{l!}\epsilon^{\alpha_1{\cdots}\alpha_l}
Q^{i_1}_{(l)\alpha_1}{\cdots}Q^{i_{n_l}}_{(l)\alpha_l}.
\label{eq4p333}
\end{eqnarray}
( \ref{eq4p331}) and (\ref{eq4p333}) show that the total number 
of the baryon operators is
\begin{eqnarray}
\sum_{\{n_l\}}\left(\begin{array}{c}N_f\\n_1\end{array}\right){\cdots}
\left(\begin{array}{c}N_f\\n_k\end{array}\right)
=\left(\begin{array}{c}kN_f\\N_c\end{array}\right).
\label{eq4p334}
\end{eqnarray}

Before discussing the duality, we  find the anomaly-free
global symmetry. Classically,  the action of $N=1$ 
the theory has the global symmetry
\begin{eqnarray}
SU_L(N_f){\times}SU_R(N_f){\times}U_B(1){\times}U_A(1){\times}U_{R_0}(1).
\label{eq4p335}
\end{eqnarray} 
The standard form of the classical action of the $N=1$ supersymmetric 
gauge theory with quark superfields in the fundamental 
representation and a matter field in 
the adjoint representation and the superpotential (\ref{eq4p328}) 
imply that the quantum numbers of matter fields under (\ref{eq4p335}) 
should be as listed in Table (\ref{ta4p34}).

\begin{table}
\begin{center}
\begin{tabular}{|c|c|c|c|c|c|} \hline
     & $SU_L(N_f)$  & $SU_R(N_f)$ &  $U_B(1)$  & $U_A(1)$ & $U_R(1)$ \\ \hline
 $Q$ & $N_f$        & $1$         &    $1$     & $1$    &  $1-2N_c/(3N_f)$ 
\\ \hline
 $\widetilde{Q}$ & $1$ & $\overline{N_f}$ & $-1$ & $1$ & $1-2N_c/(3N_f)$ \\ \hline
 $X$ &  $1$ & $1$ & $0$ & $0$ & $2/(k+1)$\\ \hline 
\end{tabular}
\caption{\protect\small Representation quantum numbers of the matter fields
of the electric theory
under $SU_L(N_f){\times}SU_R(N_f){\times}U_B(1){\times}U_R(1)$. \label{ta4p34} }
 \end{center}

\end{table}

At the quantum level, the $U_A(1)$ and $U_{R_0}(1)$ will become anomalous.
 Like in the $SU(N_c)$ QCD case, there exists an 
anomaly-free $U_{R_0}(1)$ symmetry
coming from a combination of $U_{R_0}(1)$ and $U_A(1)$ transformations. 
The method of searching for such an anomaly-free $U_R(1)$ 
is the same as the one discussed in Subsect.\,\ref{subsect6.1}.  
The classical conserved currents
corresponding to the classical $U_{R_0}(1)$ and $U_A(1)$ transformations
are, respectively,
\begin{eqnarray}
&& j_{(R_0)\mu}^{5}=\overline{\Psi}_Q\gamma_\mu\gamma_5\Psi_Q
                 -\frac{k-1}{k+1}\overline{\Psi}_X\gamma_\mu\gamma_5\Psi_X
                 -\overline{\lambda}^a\gamma_\mu\gamma_5\lambda^a ;\nonumber\\
&& j_{(A)\mu}^{5}=\overline{\Psi}_Q\gamma_\mu\gamma_5\Psi_Q.
\label{eq4p336}
\end{eqnarray}
Their operator anomaly equations read 
\begin{eqnarray}
\partial_{\mu}j_{(R_0)}^{5\mu}
&=&2\left(\frac{2N_c}{k+1}-N_f\right)\frac{g^2}{32\pi^2}
F_{\mu\nu}^a\widetilde{F}^{\mu\nu a};
\nonumber\\
\partial_{\mu}j_{(A)}^{5\mu}&=&2N_f\frac{g^2}{32\pi^2}
F_{\mu\nu}^a\widetilde{F}^{\mu\nu a}.
\label{eq337}
\end{eqnarray} 
According to (\ref{eq6.112y}) 
the anomaly-free $U_R(1)$ charge should be the 
following combination 
 \begin{eqnarray}
R=R_0+\left(1-\frac{2}{k+1}\frac{N_c}{N_f}\right)A.
\label{eq4p338}
\end{eqnarray}
Therefore, the full quantum symmetry is
$SU_L(N_f){\times}SU_R(N_f){\times}U_B(1){\times}U_R(1)$,
with the anomaly-free $R$-charges listed in table (\ref{ta4p35}).

\begin{table}
\begin{center}
\begin{tabular}{|c|c|c|c|c|} \hline
               & $U_B(1)$  & $U_A(1)$ & $U_R(1)$ \\ \hline
 $Q$           & $1$     & $1$    &  $1-2N_c/(3N_f)$ \\ \hline
 $\widetilde{Q}$ & $-1$ & $1$ & $1-2N_c/(3N_f)$ \\ \hline
 $X$           &  $0$ & $0$ & $2/(k+1)$\\ \hline 
\end{tabular}
\caption{\protect\small Anomaly-free $U_R(1)$ combination of 
quantum numbers of the matter fields in the electric theory. 
\label{ta4p35}}
 \end{center}

\end{table}

\begin{flushleft}
{\it Duality}
\end{flushleft}

The discussion in the last section and (\ref{eq4p334}) show that the dual
magnetic  description should
be an $SU(kN_f-N_c)$ gauge theory with the following matter content:
$N_f$ flavours of dual quarks $q_i$, $\widetilde{q}^i$, an adjoint 
field $Y$ and gauge singlets $M_j$, which are 
identical to (\ref{eq4p329}) at the
IR fixed point. The quantum numbers of these matter fields under
the global symmetry are listed in Table (\ref{ta4p36}). 
They are determined by (\ref{eq4p329}) and the identification of the 
baryon-like operators in both the electric and magnetic theories,
\begin{eqnarray}
B^{(n_1n_2{\cdots}n_k)}_{\rm elec}
{\sim}B^{(m_1m_2{\cdots}m_k)}_{\rm mag}
\label{eq4p340}
\end{eqnarray}
with 
\begin{eqnarray}
B^{(m_1m_2{\cdots}m_k)}_{\rm mag}=q^{m_1}_{(1)}{\cdots}q^{m_k}_{(k)};
~~~~\sum_l m_l=kN_f-N_c, ~~~l=1,2,{\cdots},k.
\label{eq4p341}
\end{eqnarray}    
(\ref{eq4p340}) and (\ref{eq4p341}) suggest that
\begin{eqnarray}
m_l=N_f-n_{k+1-l}, ~~~l=1,2,{\cdots},k.
\label{eq4p342}
\end{eqnarray}

\begin{table}
\begin{center}
\begin{tabular}{|c|c|c|c|c|} \hline
               & $SU_L(N_f)$  & $SU_R(N_f)$ & $U_B(1)$ & $U_R(1)$ \\ \hline
 $q$           & $\overline{N}_f$ & $1$    &  $\frac{N_c}{kN_f-N_c}$ &
 $1-\frac{2}{k+1}\frac{kN_f-N_c}{N_f}$\\ \hline
 $\widetilde{q}$ & $1$ & $N_f$ & $-\frac{N_c}{(kN_f-N_c)}$ &
 $1-\frac{2}{k+1}\frac{(kN_f-N_c)}{N_f}$\\ \hline
 $Y$   &  $1$ & $1$ & $0$ & $\frac{2}{N+1}$\\ \hline
 $M_j$ &  $N_f$ & $\overline{N}_f$ & $0$ & $2-\frac{4}{N+1}\frac{N_c}{N_f}+
\frac{2}{k+1}(j-1)$\\ \hline
\end{tabular}
\caption{\protect\small Representation quantum numbers of the matter fields
in the magnetic theory under the 
global symmetry $SU_L(N_f){\times}SU_R(N_f){\times}U_B(1){\times}U_R(1)$.
\label{ta4p36} }
 \end{center}

\end{table}

As in the usual supersymmetric QCD, this dual picture  is supported by
the 't Hooft anomaly matching. Considering the various currents
corresponding to the global symmetry 
$SU_L(N_f){\times}$ $SU_R(N_f){\times}U_B(1){\times}U_R(1)$ 
and the energy-momenta composed of the fermionic components 
of the above matter fields and gauginos
in both the electric and magnetic theories and calculating the non-vanishing
triangle diagrams, one can easily find that the anomaly coefficient
are indeed identical. They are listed in table (\ref{ta4p37}).

\begin{table}
\begin{center}
\begin{tabular}{|c|c|} \hline
     Triangle diagrams    &  Anomaly coefficients\\ \hline
 $SU(N_f)^3$           & $N_c\mbox{Tr}(t^A\{t^B,t^C\})$ \\ \hline
 $SU(N_f)^2U_R(1)$ & $-\frac{2}{k+1}\frac{N_c^2}{N_f}\mbox{Tr}(t^At^B)$\\
 \hline
 $SU(N_f)^2U_B(1)$   &  $N_c\mbox{Tr}(t^At^B)$\\ \hline
 $U_R(1)^3$  & $\left[\left(\frac{2}{k+1}-1\right)^3+1\right]
\left(N_c^2-1\right)-\frac{16}{(k+1)^3}\frac{N_c^4}{N_f^2}$\\ \hline
$U_B(1)^2U_R(1)$ & $-\frac{4}{k+1}N_c^2$ \\ \hline
 $U_R(1)$ & $-\frac{2}{k+1}(N_c^2+1)$ \\ \hline
\end{tabular}
\caption{\protect\small 't Hooft anomaly coefficients. \label{ta4p37} }
 \end{center}

\end{table}

The quantum numbers given in Table (\ref{ta4p36})) 
and the holomorphy determine that the superpotential of 
the dual, magnetic  theory should be of the following form
\begin{eqnarray}
W_{\rm mag}=\mbox{Tr}Y^{k+1}+\sum_{j=1}^kM_j\widetilde{q}Y^{k-j}q,
\label{eq4p343}
\end{eqnarray} 
where the normalization coefficients 
are chosen equal to 1. In principle
these coefficients can be calculated and they are relevant for a 
more clear understanding of duality. Especially, the case $k=1$ corresponds to 
the Seiberg duality discussed in Sect.\,\ref{subsect7.2},
 since now the superpotential 
(\ref{eq4p328}) and (\ref{eq4p343}) show that $X$, $Y$ are massive and 
can be integrated out. The case $k=2$ is just the Kutasov model 
discussed in the previous section.

\begin{flushleft}
{\it Deformation}
\end{flushleft}

There are many interesting  deformations, which can provide a non-trivial
check on the above duality. Two cases will be 
discussed \cite{ref432}. The first one is the mass deformation, 
which is implemented by giving a mass to
the $N_f$-th electric quarks; thus, the whole superpotential of the electric
theory at tree level becomes
\begin{eqnarray}
W_{\rm el}=g_k\mbox{Tr}X^{k+1}+m\widetilde{Q}_{N_f}{Q}_{N_f}.
\label{eq4p344}
\end{eqnarray}
Consequently, the low energy theory becomes an $SU(N_c)$ gauge theory 
with $N_f-1$ flavours since the heavy flavour should be integrated out.
On the other hand, in the magnetic theory 
the introduction of the mass term for the electric 
quarks gives a new term to the magnetic superpotential 
(\ref{eq4p343}) due to the duality:
\begin{eqnarray}
W_{\rm mag}=g_k\mbox{Tr}Y^{k+1}+\sum_{j=1}^kM_j\widetilde{q}Y^{k-j}q+
m(M_1)^{N_f}_{~N_f}.
\label{eq4p345}
\end{eqnarray}
The equations of motion for $(M_1)^{N_f}_{~N_f}$ show that the vacuum
should satisfy
\begin{eqnarray}
q_{N_f}Y^{l-1}\widetilde{q}^{N_f}=-\delta_{lk}m, ~~~~l=1,{\cdots},k,
\label{eq4p346}
\end{eqnarray}
which together with the following conditions determine the 
expectation values:
\begin{eqnarray}
\widetilde{q}^{N_f}_\alpha &=&\delta_{\alpha,1}; ~~~q^{\alpha}_{N_f}
=\delta^{\alpha,k};\nonumber\\
Y^{\alpha}_{~\beta}&=&\left\{\begin{array}{ll}\delta^{\alpha}_{~\beta+1},
 &\beta=1,{\cdots},k-1 \\
0 & \mbox{otherwise} \end{array} \right.
\label{eq4p347}
\end{eqnarray}
(\ref{eq4p346}) and (\ref{eq4p347}) imply that the Higgs mechanism 
occurs. Consequently, the low energy magnetic theory is
the $SU(kN_f-N_c-k)$ theory with $N_f-1$ flavours. It 
corresponds exactly to the low energy electric theory. Thus the duality
is preserved under this mass deformation. 

 Another deformation is achieved by perturbating the electric 
superpotential (\ref{eq4p328}) by the following term
\begin{eqnarray}
W(X)=\mbox{Tr}\left(X^3+\frac{m}{2}X^2+\lambda X\right),
\label{eq4p348}
\end{eqnarray}
where the last term $\lambda\mbox{Tr}X$ is a Lagrange multiplier term 
since the $SU(N_c)$
adjoint field $X$ is traceless.
Due to  the relation between  the scalar potential and the superpotential,
$V(\phi)=|\partial W/\partial X|^2_{X\rightarrow \phi_X}$, the vacuum 
solutions are given by diagonal matrices $X=(x_r\delta^r_{~s})$
with eigenvalues $x_r$ satisfying the quadratic equations:
\begin{eqnarray}
3x^2+mx+\lambda =0.
\label{eq4p349}
\end{eqnarray}
Thus there are two solutions,
\begin{eqnarray}
x_{\pm}=\frac{-m\pm\sqrt{m^2-12\lambda}}{6},
\label{eq4p350}
\end{eqnarray}
which correspond to the two minima of the scalar potential.
Since the supersymmetry is not broken, the Witten index implies 
that there should exist $N_c+1$ possible vacua labelled
by $k=0,1,{\cdots},N_c$. Without loss of generality, we assume that 
there are $k$ eigenvalues $x_i$ equal to $x_+$ and $N_c-k$ ones equal to 
$x_-$. Consequently, the gauge group is broken,
\begin{eqnarray}
SU(N_c){\longrightarrow} SU(k){\times} SU(N_c-k){\times}U(1).
\label{eq4p351}
\end{eqnarray}
In each vacuum the theory is reduced to the usual 
supersymmetric QCD since $X$ becomes massive and 
can be integrated out.

In the dual magnetic theory, the gauge group is $SU(2N_f-N_c)$
and hence there seems to exist $2N_f-N_c+1$ supersymmetric vacua. A similar 
analysis shows that the introduction of the superpotential (\ref{eq4p348}) 
breaks the symmetry of the magnetic theory as follows:
\begin{eqnarray}
SU(2N_f-N_c){\longrightarrow} SU(l){\times} SU(2N_f-N_c-l){\times}U(1).
\label{eq4p352}
\end{eqnarray}
However, as indicated by ADS superpotential in Ref.\,\cite{ref1p22}, 
a supersymmetric vacuum is not stable if the the number of flavours 
is smaller than the number of colours in supersymmetric QCD. Therefore,
the $l$-th vacuum is stable if and 
only if $l{\leq}N_f$ and $2N_f-N_c-l{\leq}N_f$. Thus the true vacua are
labelled by $l=N_f-N_c,{\cdots},N_f$ and there are only $N_c+1$ vacua.
This coincides exactly with the requirement of duality. One can naturally 
think that the explicit correspondence between the vacua given 
in (\ref{eq4p351}) and (\ref{eq4p352}) is $l=N_f-k$.

 Special consideration should be paid to two particular 
cases, $k=0$ or $N_c$. Now the
$SU(N_c)$ gauge group remains unbroken. Consequently, the trivial
vacuum $\langle X\rangle =0$ in the electric theory corresponds to a
magnetic vacuum with $\langle Y\rangle {\neq}0$, in which the breaking
pattern of the magnetic gauge group is
\begin{eqnarray}
SU(2N_f-N_c){\longrightarrow}SU(N_f){\times}SU(N_f-N_c){\times}U(1).
\label{eq4p353}
\end{eqnarray}
However, we can still get a consistent duality mapping.  Eq.\,(\ref{eq390})
shows that in the magnetic theory there exists a superpotential
$W=M^{ij}\widetilde{q}_{i}^rq_{jr}$. The equations of motion for $M$ will
set $\widetilde{q}{\cdot}q=0$. Furthermore,  according 
to Eq.\,(\ref{eq318}), the quantum  moduli space is given by 
$\det(\widetilde{q}\cdot q)-B\widetilde{B}=\Lambda^{2N_f}$, which relates
$\widetilde{q}$ and $q$ to the baryons $\widetilde{B}$, $B$. Thus 
in the quantum moduli space of vacua,
$\widetilde{q}{\cdot}q=0$ means $B\widetilde{B}=-\Lambda^{2N_f}$.
This non-vanishing expectation value of $B$ will break the $U(1)$ symmetry
of (\ref{eq4p353}). Therefore, Seiberg's duality pattern arises:
\begin{eqnarray}
SU(N_f){\times}SU(N_c){\longleftrightarrow}SU(N_f){\times}SU(N_f-N_c).
\label{eq4p354}
\end{eqnarray}

In the case $k>2$, there are many possible choices for the 
superpotential $W$ leading to deformations. The situation
becomes more complicated, but the analysis is conceptually similar to
the case $k=2$. By Eq.\,(\ref{eq4p328}), the vacua are determined
by the equation $W^{\prime}(x)=0$. Since $W^{\prime}(x)$ is a polynomial
of degree $k>2$, there exist $k$ solutions, and in general
these solutions are different. The ground states are labelled by the number 
$i_l$ ($l=1,{\cdots},k$) of eigenvalues $x$ residing in the
$l$-th minimum of the scalar potential (
$i_1{\leq}i_2{\leq}{\cdots}i_k$ and $\sum_{l=1}^ki_l=N_c$). 
Correspondingly, the electric gauge group $SU(N_c)$ is broken by
the non-vanishing expectation values as follows:
\begin{eqnarray}
SU(N_c){\longrightarrow}SU(i_1){\times}SU(i_2){\times}{\cdots}{\times}
SU(i_k){\times}U(1)^{k-1}.
\label{eq4p356}
\end{eqnarray}
Note that each of the $SU(i_l)$ factors describes a supersymmetric
QCD model since $X$ becomes massive and can be removed. In the dual
magnetic theory, a similar breaking pattern occurs. Assume that
there are $j_l$ eigenvalues in the $l$-th minimum
($\sum_l j_l=kN_f-N_c$) 
and the breaking of the magnetic gauge group is
\begin{eqnarray}
SU(kN_f-N_c){\longrightarrow}SU(j_1){\times}SU(j_2){\times}{\cdots}
{\times}SU(j_k){\times}U(1)^{k-1}.
\label{eq4p357}
\end{eqnarray} 
The duality mapping is just given by $j_l=N_f-i_l$.


\setcounter{section}{4}
\section{Non-perturbative phenomena in $N=1$ supersymmetric 
gauge theories with gauge group $SO(N_c)$ and $Sp(N_c=2n_c)$}
\label{sect5}
\renewcommand{\thetable}{5.\arabic{table}}
\setcounter{equation}{0}

In this section we shall review the non-perturbative dynamics of
$N=1$ supersymmetric gauge theories with 
gauge groups $SO(N_c)$ and $Sp(N_c=2n_c)$. We shall see that
in these supersymmetric gauge theories, some new
dynamical phenomena arise such as inequivalent branches,
oblique confinement and electric-magnetic-dyonic triality \cite{ref51}.
In addition, there emerges an explicit phase transition from the Higgs 
phase to the confining phase \cite{ref51} in supersymmetric $SO(N_c)$ 
gauge theory with quarks in the fundamental representation 
(i.e. the vector representation). This is different from the 
supersymmetric $SU(N_c)$ gauge theory, where the 
transition from the Higgs phase
to the confining phase is smooth. The reason behind this is that
in the $SO(N_c)$ supersymmetric gauge theory, the dynamical quarks
cannot screen and therefore, at large distances 
confinement occurs suddenly. 

In the following, the non-perturbative phenomena in
the $SO(N_c)$ supersymmetric gauge theory will be discussed
in detail. The new dynamical phenomena different from the $SU(N_c)$ case will
be emphasized. The non-perturbative dynamics of the $Sp(2n_c)$
supersymmetric gauge theory is similar to the $SU(N_c)$ and $SO(N_c)$ cases.
In fact, it is proposed that by using the trick
of extrapolating the colour parameter $N_c$ of $SO(N_c)$ to 
a negative value, the dynamics
of the $Sp(N_c=2n_c)$ supersymmetric gauge theory can be read
out  from that of the $SO(N_c)$ theory. Thus the $Sp(2n_c)$ case 
will be only briefly introduced.  

\subsection{$N=1$ supersymmetric $SO(N_c)$ gauge theory with $N_f$ flavours}
\label{subsect51}
\renewcommand{\theequation}{5.1.\arabic{equation}}
\renewcommand{\thetable}{5.1.\arabic{table}}
\setcounter{equation}{0}
\setcounter{table}{0}
\subsubsection{Global symmetries of $N=1$ supersymmetric $SO(N_c)$ QCD}
\label{subsub511}

The fundamental representation of the $SO(N_c)$ group is its
vector representation and hence it is 
always real. Thus in $SO(N_c)$ gauge theory there is no difference 
between left- and right-handed quarks. This is different from the 
$SU(N_c)$ case, where the right-handed quark is the left-handed 
anti-quark. The classical Lagrangian of $N=1$ supersymmetric 
$SO(N_c)$ QCD is
\begin{eqnarray}
{\cal L}=\frac{1}{4}\mbox{Tr}\left(W^{\alpha}W_{\alpha}|_{\theta^2}+
\overline{W}^{\alpha}\overline{W}_{\alpha}|_{\overline{\theta}^2}\right)
+Q^{\dagger}e^{gV(N_c)}Q|_{\theta^2\overline{\theta}^2}.
\label{eq5p11}
\end{eqnarray}
In the Wess-Zumino gauge, the (four-) component field 
form of above Lagrangian reads
\begin{eqnarray}
\hspace{-4mm}{\cal L}&=&-\frac{1}{4}F_{\mu\nu}^aF^{a\mu\nu}+
\frac{1}{2}i\overline{\lambda}^a\gamma^{\mu}{\cal D}_{\mu}^{ab}\lambda^b
+i\overline{\Psi}^{ir}\gamma^{\mu}\left(D_{\mu}\Psi\right)_{ir}+
\left(D^{\mu}\Phi\right)^{* ir}\left(D_{\mu}\Phi\right)_{ir} \nonumber\\
\hspace{-4mm}&-&i\sqrt{2}g\left[\Phi_{ir}(T^a)^r_{~s}
\overline{\Psi}^{is}\lambda^a-
\Phi^{*ir}\left(T^a\right)^{~s}_r\overline{\lambda}^a\Psi_{is}\right]
-\frac{g^2}{2}[\Phi^{\dagger},\Phi]^2+\frac{i\theta}{32\pi^2}
\epsilon^{\mu\nu\lambda\rho}F^a_{\mu\nu}F^a_{\lambda\rho},
\label{eq5p12}
\end{eqnarray}
where the group indices $a=1,\cdots , N_c(N_c-1)/2$; the colour indices
$r,s=1,\cdots, N_c$ and the flavour indices $i=1,\cdots, N_f$. 
${\cal D}_{\mu}$ is the covariant derivative in the adjoint 
representation and $D_{\mu}$ in the vector representation,   
${\cal D}_{\mu}^{ab}=\partial_{\mu}\delta^{ab}-gf^{abc}A_{\mu}^c$; 
$D^{\mu}_{rs}=\partial^{\mu}\delta_{rs}-igA^{\mu a}T^a_{rs}$.
The classical Lagrangian (\ref{eq5p12}) possesses the global symmetry 
$SU(N_f)\times U_A(1)\times U_{R_0}(1)$ corresponding to 
the following transformations:
\begin{itemize}
\item $SU(N_f)$:  
\begin{eqnarray}
&&\Psi_{ir}\rightarrow\left(e^{i\alpha^A t^A}\right)_i^{~j}\Psi_{jr},~ 
\overline{\Psi}^{ir}\rightarrow\overline{\Psi}^{jr}
\left(e^{-i\alpha^A t^A}\right)_j^{~i};~
\nonumber\\
&& \Phi_{ir}\rightarrow\left(e^{i\alpha^At^A}\right)_i^{~j}\Phi_{jr},~ 
\Phi^{*ir}\rightarrow\Phi^{*jr}\left(e^{-i\alpha^A t^A}\right)_j^{~i}.
\label{eq5p14}
\end{eqnarray}
\item $U_A(1)$:
\begin{eqnarray} 
\Psi\rightarrow e^{i\gamma_5\alpha}\Psi,~\overline{\Psi}\rightarrow
 \overline{\Psi}e^{i\gamma_5\alpha}.
\label{eq5p15}
\end{eqnarray}
\item $U_{R_0}(1)$:
\begin{eqnarray} 
\Psi\rightarrow e^{i\gamma_5\alpha}\Psi, ~\overline{\Psi}\rightarrow
 \overline{\Psi}e^{i\gamma_5\alpha};~
\lambda^a\rightarrow e^{-i\gamma_5\alpha}\lambda^a,~
\overline{\lambda}^a\rightarrow\overline{\lambda}^a e^{-i\gamma_5\alpha}.
\label{eq5p15m}
\end{eqnarray}
\end{itemize}

Like in the $SU(N_c)$ case, the $U_A(1)$ and $U_{R_0}(1)$ symmetries suffer
from an Adler-Bell-Jackiw (ABJ) anomaly, and they can be combined into 
an anomaly-free 
$U_R(1)$ symmetry. To do this, let us first calculate the fermionic 
triangle anomaly diagrams of the relevant currents and get 
the coefficients in the anomaly operator equations for $j^{\mu}_{(A)}$ 
and $j^{\mu}_{(R_0)}$, and then try to construct
an anomaly-free $U_R(1)$ symmetry. The Noether currents corresponding to
the above $U_A(1)$ and $U_{R_0}(1)$ transformations are
 \begin{eqnarray}
j_{(A)\mu}&=&\overline{\Psi}^{ir}\gamma_{\mu}\gamma_5\Psi_{ir}
{\equiv}j_{\mu}^5,\nonumber\\
j_{(R_0)\mu}&=&\overline{\Psi}^{ir}\gamma_{\mu}\gamma_5\Psi_{ir}
-\frac{1}{2}\overline{\lambda}^a\gamma_{\mu}\gamma_5\lambda^a
{\equiv}j_{\mu}^5-{\cal J}_{\mu}^5.
 \label{eq5p17}
 \end{eqnarray}       
Considering the triangle diagrams 
$\langle j_{\mu}^5 J_{\mu}^aJ_{\nu}^b\rangle$, 
$\langle {\cal J}_{\mu}^5 {\cal J}_{\mu}^a{\cal J}_{\nu}^b\rangle$
with $J_{\mu}^a$ and ${\cal J}_{\mu}^a$ being the dynamical currents
corresponding to  the global gauge transformations, which are composed 
of the quarks and the gaugino, respectively,
\begin{eqnarray}
J_{\mu}^a =\overline{\Psi}^{ir}\gamma_{\mu}\left(T^a\right)_r^{~s}\Psi_{is};~~~
{\cal J}_{\mu}^a =f^{abc}\overline{\lambda}^b\gamma_\mu{\lambda}^c,
\label{eq5p18}
\end{eqnarray} 
we find the operator equation for the anomalies of
$j_{(A)\mu}$ and $j_{(R_0)\mu}$:
\begin{eqnarray}
\partial^{\mu}j_{(A)\mu}&=&
2 N_f \frac{1}{32\pi^2}\epsilon^{\mu\nu\lambda\rho}
F_{\mu\nu}^aF^{a}_{\lambda\rho};\label{eq5p19i}\\
\partial^{\mu}j_{(R_0)\mu}&=&\left[2 N_f-2(N_c-2)\right]
 \frac{1}{32\pi^2}\epsilon^{\mu\nu\lambda\rho}
F_{\mu\nu}^aF^{a}_{\lambda\rho}.
\label{eq5p19}
\end{eqnarray} 
Using the same arguments as in the $SU(N_c)$ case, namely find the 
above anomalies lead to a shift of the vacuum angle $\theta$ in the 
Lagrangian (\ref{eq5p11}) and hence the absence of anomalies is  
reflected in the vanishing of the shift, one can easily 
obtain the non-anomalous $U_R(1)$ symmetry with the charge 
being the linear combination
\begin{eqnarray}
R=R_0+\frac{N_f-N_c+2}{N_f}A.
\label{eq5p110}
\end{eqnarray} 
The $R_0$-charges and the $U_A(1)$ charges of all the fields
(and even of some parameters) should be combined in such ways so that
an anomaly-free $U_R(1)$ symmetry is achieved. Therefore,
at the quantum level the anomaly-free global symmetry is 
$SU(N_f){\times}U_R(1)$.
The anomaly-free $R$-charges of some fundamental fields are
listed in Table (\ref{ta5pon}) and 
the transformation properties of every fundamental field 
corresponding to the  global symmetry
are given in Table (\ref{ta5ptw}).

\begin{table}
\begin{center}
\begin{tabular}{|c|c|c|c|} \hline
  & $U_A(1)$ & $U_{R_0}(1)$ & $U_R(1)$ \\ \hline
 $\phi_Q$ & $+1$ & $0$ & $(N_f+2-N_c)/N_f$ \\ \hline
 $\psi_Q$ & $+1$ & $-1$ & $-(N_c-2)/N_f$  \\ \hline
$ Q$ & $+1$ & $0$ & $(N_f+2-N_c)/N_f$   \\ \hline
$\lambda$ & $0$ & $+1$ & $+1$  \\ \hline
\end{tabular}
\caption{\protect\small Combination of anomaly-free $R$-charges for
 fundamental fields; $\phi_Q$ and $\psi_Q$ denote the lowest component
 and the (two-) component spinor of the quark chiral superfield.\label{ta5pon}}
 \end{center}
\end{table}

\begin{table}
\begin{center}
\begin{tabular}{|c|c|c|} \hline
  & $SU(N_f)$ &  $R$ \\ \hline
 $\phi_Q$ & $N_f$ &  $(N_f+2-N_c)/N_f$ \\ \hline
 $\psi_Q$ & $N_f$ &  $-(N_c-2)/N_f$  \\ \hline
$ Q$ & $N_f$ &  $(N_f+2-N_c)/N_f$   \\ \hline
$\lambda$ & $0$ & $+1$  \\ \hline
\end{tabular}
\caption{\protect\small Representation quantum numbers of fundamental fields
under global symmetries $SU(N_f){\times}U_B(1){\times}U_R(1)$.
\label{ta5ptw} }
 \end{center}
\end{table}

In addition, the theory has an explicit $Z_2$ charge conjugation
symmetry ${\cal C}$ \cite{ref51}.
In particular, 
the $U_A(1)$ transformation (\ref{eq5p15}) and the operator 
anomaly equation (\ref{eq5p19i}),
show that at the quantum level the theory is also invariant 
under a discrete $Z_{2N_f}$ symmetry,
\begin{eqnarray}
Q\longrightarrow e^{in2\pi/(2N_f)}Q, ~~~n=1,\cdots, 2N_f.
\label{eq5p114}
\end{eqnarray}
(For $N_c=3$, the symmetry is enhanced to $Z_{4N_f}$, which will 
be discussed in detail in Sect.\,\ref{subsub551}. 
In fact, the anomaly equation
(\ref{eq5p19i}) means that under the $U(1)_A$ transformation (\ref{eq5p15})
the vacuum angle $\theta$ in the Lagrangian (\ref{eq5p12}) is shifted to
\begin{eqnarray}
\theta {\longrightarrow}\theta +2N_f \alpha,
\label{eq5p115i}
\end{eqnarray}
and hence we have the discrete symmetry (\ref{eq5p114}).
However, since the
even elements of $Z_{2N_f}$, i.e. those with $n=2,4,\cdots,2N_f$, also
belong to $SU(N_f)$, this global flavour symmetry should be
$SU(N_f){\times}Z_{2N_f}/Z_{N_f}$.
 
\subsubsection{Classical moduli space}
\label{subsub512}

 If written in two-component field form, 
the scalar potential of the supersymmetric
$SO(N_c)$ QCD has the same form as in the $SU(N_c)$ case, (\ref{eq3419x}).  
The classical moduli space is still determined by the
$D$-flatness condition,  $D^a=0$. 
As in the $SU(N_c)$ theory, the $D$-flatness condition does not
require that the expectation values of the squarks vanish, only that
they equal some constant values. 
To make the supersymmetry manifest
we work in the superfield
form. Like in the $SU(N_c)$ case, we write the quark
superfield in an $N_f{\times}N_c$ matrix form 
in terms of the flavour and colour
indices. Since the scalar potential is $SO(N_c)$ gauge and 
$SU(N_f)$ global transformation invariant, in the D-flat
directions we can use these gauge and global rotations to make
the quark superfield matrix diagonal. In the following we shall 
discuss the classical moduli space for different relative
colour and flavour numbers.

\vspace{4mm}
{\noindent \it $N_f < N_c$}
\vspace{4mm} 

 The quark superfield matrix can in this case be diagonalized
as follows:
\begin{eqnarray}
Q=\left(\begin{array}{ccccccc}
          a_1 & 0   & \cdots &   0    & \cdots  & 0\\
           0  & a_2 & \cdots &   0    & \cdots & 0 \\
          \vdots &\vdots & \ddots &\vdots &\ddots & 0 \\
           0  & 0 & \cdots & a_{N_f}& \cdots & 0\end{array}\right).
\label{eq5p115}
\end{eqnarray}
If all the $a_i{\neq}0$, their scalar components will break the 
gauge group $SO(N_c)$ to $SO(N_c-N_f)$
for $N_f <N_c-2$, and completely break $SU(N_c)$ for $N_f>N_c-2$.  
The moduli space will still be described by the expectation value
of the ``meson" superfields
\begin{eqnarray}
M^{ij}=Q^i_{~r}Q^{~rj}{\equiv}Q^i{\cdot}Q^j,
\label{eq5p116}
\end{eqnarray}  
which is explicitly gauge invariant. We shall show that these 
$N_f(N_f+1)/2$ mesons are the appropriate low energy dynamical degrees of 
freedom since they correspond precisely to the matter superfields
left massless by the Higgs mechanism. Due to the spontaneous
symmetry breaking $SO(N_c)$$\rightarrow$$SO(N_c-N_f)$, among the
$N_fN_c$ quark superfields $Q_i^{~r}$ there are 
\begin{eqnarray}
\frac{1}{2}N_c(N_c-1)-\frac{1}{2}(N_c-N_f)(N_c-N_f-1)=N_cN_f
-\frac{1}{2}N_f(N_f+1)
\label{eq5p117}
\end{eqnarray}  
fields which become massive through the Higgs mechanism.
$N_fN_c-[N_fN_c-N_f(N_f+1)/2]$$=
N_f(N_f+1)/2$ fields remain massless. 

\vspace{4mm}
{\noindent \it  $N_f {\geq} N_c$}
\vspace{4mm} 

In this case the $D$-flat directions are given by
\begin{eqnarray}
Q=\left(\begin{array}{cccc}
          a_1 &  0   & \cdots & 0\\
           0  & a_2  & \cdots & 0 \\
      \vdots  &\vdots& \ddots & \vdots \\
          0   &  0   & \cdots & a_{N_c}\\
          0   &  0   & \cdots & 0  \\
       \vdots &\vdots & \ddots &\vdots  \\
         0 & 0 & \cdots & 0
                          \end{array}\right).
\label{eq5p119}
\end{eqnarray}
The low energy gauge invariant degrees of freedom that can be constructed
to describe the moduli space are not only the meson superfields $M^{ij}=
Q^i{\cdot}Q^j$, but also the baryon superfields
\begin{eqnarray}
B^{i_1{\cdots}i_{N_c}}=\frac{1}{N_c!}\epsilon^{r_1{\cdots}r_{N_c}}
Q^{i_1}_{~r_1}Q^{i_2}_{~r_2}{\cdots}Q^{i_{N_c}}_{~r_{N_c}}.
\label{eq5p1191}
\end{eqnarray}
Along the flat directions given by $(\ref{eq5p119})$, we have
\begin{eqnarray}
M&=&\left(\begin{array}{ccccc}
        a_1^2 & 0      & \cdots  & 0 \\
          0   & a_2^2  & \cdots  & 0 \\
      \vdots  & \vdots & \ddots  & \vdots \\
          0   & 0      & \cdots  & a_{N_c}^2\\
          0   & 0      & \cdots  & 0  \\
       \vdots & \vdots & \ddots  & \vdots\\
         0    & 0      & \cdots  & 0   
                          \end{array}\right);\nonumber\\
      \nonumber\\
B^{1\cdots N_c}&=&a_1a_2{\cdots}a_{N_c}
\label{eq5p121}
\end{eqnarray}
with all other components of $M$ and $B$ vanishing. Thus one can see that
the rank of $M$ is at most $N_c$. If the rank of $M$ is less than $N_c$, 
i.e. one $a_i$ or several $a_i$s vanish, then the baryon fields $B=0$.
If the rank of $M$ is equal to $N_c$, then $B$ has rank 1 and its 
non-zero component is the square root of the product of non-zero
diagonal values of $M$, up to a sign. So the baryon field should not
be regarded as an independent variable to parameterize the moduli 
space. Therefore, for $N_f{\geq}N_c$, the
classical moduli space is described by a set of $M$ of rank at most 
$N_c$ with an additional sign coming from taking the square root 
\begin{eqnarray}
B=\pm\sqrt{{\det}^{\prime}M}
\label{eq5p122}
\end{eqnarray}   
for $M$ with rank $N_c$, where the prime denotes that only the non-zero
diagonal values of $M$ are considered.

\subsection{Dynamically generated superpotential and decoupling relation}
\label{subsect52}
\renewcommand{\thetable}{5.2.\arabic{table}}
\setcounter{table}{0}
\renewcommand{\theequation}{5.2.\arabic{equation}}
\setcounter{equation}{0}


The quantum moduli space and the corresponding non-perturbative
dynamics will become more complicated than in the $SU(N_c)$ case due 
to the peculiarities of the $SO(N_c)$ group. Some new dynamical phenomena 
will arise. The quantum theory is also more sensitive
to the relative numbers of colours and flavours than the $SU(N_c)$ 
gauge theory. In each case ($N_f<N_c$ or $N_f{\geq}N_c$) we must give
a detailed classification of the relative numbers of colours and flavours
and discuss the corresponding non-perturbative phenomena. Before going into
details we first present the general form of the dynamically generated
superpotential in the case $N_f <N_c$ and the connections between 
different energy scales related by the decoupling limits in various ways. 

To construct the dynamically generated superpotential, we need a dynamical
scale associated with the running coupling. For $N_c>4$, the
coefficient of one-loop beta function of the $SO(N_c)$ gauge theory with
$N_f$ quarks in the vector representation of the gauge group is
\begin{eqnarray}
\beta_0=3(N_c-2)-N_f.
\label{eq5p21}
\end{eqnarray}
With Eq.\,(\ref{eq6.112y}), the running of the gauge coupling is
\begin{eqnarray}
{\Lambda}_{N_c,N_f}^{3(N_c-2)-N_f}
=q^{3(N_c-2)-N_f}e^{-8{\pi}/g^2(q^2)+i\theta }.
\label{eq5p22}
\end{eqnarray}    
(\ref{eq5p22}) implies that  
the dynamically generated scale $\Lambda$ becomes a complex number
due to the presence of the vacuum angle,  
and further it can be formally thought of as the 
scalar component  of a (space-time independent) chiral superfield. 
The $U_A(1)$ charge and the
$R_0$-charge of this superfield should be the same as those of 
$\theta$, i.e. $2N_f$ and $2(N_f-N_c+2)$, respectively, since the
anomalies of $U_A(1)$ and $U_{R_0}(1)$ give corresponding shifts
in $\theta$.

In the cases $2< N_c{\leq}4$, there are some peculiarities.
When $N_c=4$, since 
\begin{eqnarray}
SO(4){\simeq}SU(2)_X{\times}SU(2)_Y,
\label{eq5p23}
\end{eqnarray}
with the subscripts $X$ and $Y$ labelling two $SU(2)$ branches,
the theory must be decomposed into two independent gauge theories with
gauge group $SU(2)$. The quark superfields, which before were 
in the vector representation
of $SO(4)$ before, should now constitute the fundamental
(spinorial) representation $(2,2)$ of $SU(2)_X{\times}SU(2)_Y$,
\begin{eqnarray}
Q^i_{~r}=Q^i_{~\alpha,\beta}, ~~~r=1,{\cdots},4;~~\alpha, \beta =1,2,
\label{eq5p24}
\end{eqnarray}
i.e. under a gauge transformation,
\begin{eqnarray}
Q^i_{~\alpha,\beta}{\longrightarrow}A_{\alpha}^{(X)~\gamma}
A_{\beta}^{(Y)~\delta} Q^i_{~\gamma,\delta}, ~~~~
A_{\alpha}^{(X)~\gamma}{\in}SU(2)_X; ~~A_{\beta}^{(Y)~\delta}{\in}SU(2)_Y.
 \label{eq5p25} 
\end{eqnarray}
There are two independent gauge couplings, one for each $SU(2)_s$, $s=X,Y$.    
Since the one-loop beta function coefficient in an $SU(2)$
gauge theory with $N_f$ quarks in the fundamental representation
is $6-N_f$, we have
\begin{eqnarray}
e^{-8\pi^2/g^2_s(q^2)+i\theta}
=\left(\frac{\Lambda^{(s)}}{q}\right)^{6-N_f}.
\label{eq5p26}
\end{eqnarray}
(\ref{eq5p26}) shows that the 
running of the gauge coupling in each $SU(2)$ branch
accidentally coincides with the general form of the $SO(N_c)$ theory 
with $N_c=4$.
 
When $N_c=3$, the vector representation of $SO(3)$ coincides with its
adjoint representation. In this case the one-loop $\beta$-function
coefficient is $\beta_0 =6-2N_f$, and  the running of the gauge 
coupling is
\begin{eqnarray}
e^{-8\pi^2/g^2_s(q^2)+i\theta}
=\left(\frac{\Lambda^{(s)}}{q}\right)^{6-2N_f}.
\label{eq5p27}
\end{eqnarray}
One can see that the running coupling in this case does not
coincide with the general form for $SO(N_c)$, and thus it needs a 
special consideration.

A gauge invariant quantity composed of the dynamically
generated superpotential should be a function of the low energy
dynamical degree of freedom $M$. The global symmetry $SU(N_f){\times}U_R(1)$
restricts it to be a function of $\det M$. We list
the various quantum numbers of $\Lambda^{\beta_0}$ and $\det M$ in
Table (\ref{ta5pth}) for $N_c{\neq}3$. With the 
requirement that the superpotential 
should have $R$-charge 2 and dimension 3, and should be holomorphic, 
the only possible form is
\begin{eqnarray}
 W 
=C_{N_c,N_f} \left(\frac{{\Lambda}_{N_c,N_f}^{3N_c-6-N_f}}
{\det M}\right)^{1/(N_c-2-N_f)},
\label{eq5p28}
\end{eqnarray}
The coefficient $C_{N_c,N_f}$ can be determined, like in the $SU(N_c)$
case, through an explicit calculation.
Since the scale $\Lambda$ is a complex quantity, the superpotential
can pick up a ${\bf Z}_{N_c-N_f-2}$ phase factor 
due to the power $1/(N_c-N_f-2)$,
\begin{eqnarray}
 W
&=&C_{N_c,N_f} e^{2in\pi/(N_c-2-N_f)}
\left(\frac{{\Lambda}_{N_c,N_f}^{3N_c-6-N_f}}
{\det M}\right)^{1/(N_c-2-N_f)}\nonumber\\
&{\equiv}& C_{N_c,N_f}\epsilon_{(N_c-2-N_f)}
\left(\frac{{\Lambda}_{N_c,N_f}^{3N_c-6-N_f}}
{\det M}\right)^{1/(N_c-2-N_f)},\nonumber\\ 
&n=&1,2,\cdots, N_c-2-N_f.
\label{eq5p28i}
\end{eqnarray}
The phase factor in (\ref{eq5p28i}) labels different but 
physically equivalent vacua of the theory coming from the 
spontaneous breaking of a discrete symmetry
induced by gaugino condensation in the low energy $SO(N_c-N_f)$ 
Yang-Mills theory.
 
\begin{table}
\begin{center}
\begin{tabular}{|c|c|c|c|} \hline
        & $U_A(1)$ & $U_{R_0}(1)$ & $U_R(1)$ \\ \hline
 $\Lambda^{\beta_0}$ & $2N_f$ & $-2(N_f+2-N_c)$& 0\\ \hline
 $\det M$ & $2N_f$ & $0$ & $2(N_f+2-N_c)$  \\ \hline
\end{tabular}
\caption{\protect\small  $U(1)$ quantum numbers of the quantities
 composed of the dynamical superpotential ($N_c>3$).  \label{ta5pth} }
 \end{center}
\end{table}

 In the following sections, we shall see that a superpotential of the
form (\ref{eq5p28}) can indeed be generated by gaugino condensation,
like in the $SU(N_c)$ case with $N_f<N_c-1$. For $N_f=N_c-2$, the above
superpotential does not makes sense. For
$N_c-2<N_f{\leq}N_c$, a superpotential (\ref{eq5p28}) cannot be
generated, since it would lead to non-physical dynamical behaviour.
For $N_f>N_c$, $\det M=0$, and the superpotential
(\ref{eq5p28}) does not exist. Overall, there will be no dynamically 
generated superpotential for $N_f{\geq}N_c-2$. For the special
case $N_c=3$, we shall see that there is also no dynamically
generated superpotential for any $N_f$. Consequently, these theories, with 
no dynamically generated superpotential, will have a quantum moduli 
space of exactly degenerate but physically inequivalent vacua, and
they will present interesting non-perturbative dynamical phenomena different 
from the $SU(N_c)$ case. 

For later use, we give the relations between different 
energy scales connected by the decoupling of heavy modes. Due to 
the peculiarity of the $SO(N_c)$ group when $2<N_c{\leq}4$, we must 
give special consideration to the decoupling in  
$SO(3)$ and $SO(4)$ theories.

 Giving the $N_f$-th quarks a large mass
$W_{\rm tree}={1}/{2}mM^{N_fN_f}={1}/{2}mQ^{N_f}_{r}Q^{rN_f}$,
and making this heavy mode decouple, the theory with $N_f$ quarks will yield 
a low energy theory with $N_f-1$ quarks. The running couplings 
should match at the scale $m$. When $N_c>4$, we have
\begin{eqnarray}
\frac{4\pi}{g^2(m)}=\frac{3(N_c-2)-N_f}{2\pi}\ln\frac{m}{\Lambda_{N_c,N_f}}&=&
\frac{3(N_c-2)-(N_f-1)}{2\pi}\ln\frac{m}{\Lambda_{N_c,N_f-1}}; \nonumber\\
\Lambda_{N_c,N_f}^{3(N_c-2)-N_f}m &=&\Lambda_{N_c,N_f-1}^{3(N_c-2)-(N_f-1)}.
\label{eq5p210}
\end{eqnarray}
When $N_c=4$, due to Eq.\,(\ref{eq5p23}), in each $SU(2)$ branch the 
coupling constants should match,
\begin{eqnarray}
\frac{6-N_f}{2\pi}\ln\frac{m}{\Lambda_{s,N_f}}&=&
\frac{6-(N_f-1)}{2\pi}\ln\frac{m}{\Lambda_{s,N_f-1}}; \nonumber\\
\Lambda_{s,N_f}^{6-N_f}m &=&\Lambda_{s,N_f-1}^{6-(N_f-1)}, ~~~s=X,Y.
\label{eq5p211}
\end{eqnarray}
In the case $N_c=3$, the quarks are in the adjoint representation 
of the gauge group and the one-loop beta function coefficient changes to 
$\beta_0=6-2N_f$, so we have
\begin{eqnarray}
\frac{6-2N_f}{2\pi}\ln\frac{m}{\Lambda_{3,N_f}}&=&
\frac{6-2(N_f-1)}{2\pi}\ln\frac{m}{\Lambda_{3,N_f-1}}; \nonumber\\
\Lambda_{3,N_f}^{6-N_f}m^2&=&\Lambda_{3,N_f-1}^{6-2(N_f-1)}.
\label{eq5p212}
\end{eqnarray} 

Another way of decoupling is through the Higgs mechanism with
a large expectation value $a_{N_f}$ in (\ref{eq5p115}). The $SO(N_c)$ theory
with $N_f$ quarks will decouple into an $SO(N_c-1)$ theory with $N_f-1$
quarks. Under the requirement that the running couplings should match
at the energy $a_{N_f}$, we  get a relation between the high
energy scale $\Lambda_{N_c,N_f}$ and the low energy scale 
$\Lambda_{N_c-1,N_f-1}$. In the case $N_c>5$, we have
\begin{eqnarray}
\frac{4\pi}{g^2(a_{N_f})}=\frac{3(N_c-2)-N_f}{2\pi}
\ln\frac{a_{N_f}}{\Lambda_{N_c,N_f}}&=&
\frac{3[(N_c-1)-2]-(N_f-1)}{2\pi}\ln\frac{a_{N_f}}
{\Lambda_{N_c-1,N_f-1}}; \nonumber\\
\Lambda_{N_c,N_f}^{3(N_c-2)-N_f}a_{N_f}^{-2}
&=&\Lambda_{N_c-1,N_f-1}^{3(N_c-2)-N_f-2}.
\label{eq5p213}
\end{eqnarray}
Since in the moduli space, $Q^i_{~r}=a_i\delta^i_{r}$ and hence 
$M^{N_fN_f}=a_{N_f}^2$, (\ref{eq5p213}) can be written as
\begin{eqnarray}
\Lambda_{N_c,N_f}^{3(N_c-2)-N_f}(M^{N_fN_f})^{-1}
=\Lambda_{N_c-1,N_f-1}^{3(N_c-2)-N_f-2}.
\label{eq5p214}
\end{eqnarray}
When $N_c=5$, a decoupling $SO(5){\rightarrow}SU(2)_X{\times}SU(2)_Y$
occurs, and the high energy running coupling should match the low energy
ones in each $SU(2)$ branch,
\begin{eqnarray}
\frac{9-N_f}{2\pi}\ln\frac{a_{N_f}}{\Lambda_{5,N_f}}&=&
\frac{6-(N_f-1)}{2\pi}\ln\frac{a_{N_f}}
{\Lambda_{s,N_f-1}}; \nonumber\\
\Lambda_{5,N_f}^{9-N_f}a_{N_f}^{-2}&=&\Lambda_{5,N_f}^{9-N_f}(M^{N_fN_f})^{-1}
=\Lambda_{s,N_f-1}^{6-(N_f-1)}.
\label{eq5p215}
\end{eqnarray}
For the case $N_c=4$, the decoupling pattern is 
$SU(2)_X{\times}SU(2)_Y{\rightarrow}SO(3)$. This needs a special 
consideration. From (\ref{eq5p24}) and 
\begin{eqnarray}
M^{N_fN_f}=Q^{N_f}{\cdot}Q^{N_f}=Q^{N_f}_{\alpha_X,\alpha_Y}
Q^{N_f}_{\beta_X,\beta_Y}\epsilon^{\alpha_X\beta_X}
\epsilon^{\alpha_Y\beta_Y},
\label{eq5p216}
\end{eqnarray}
it follows that the sum of the two running couplings of
each $SU(2)$ branch should match the coupling of the low energy $SO(3)$ theory,
\begin{eqnarray}
&&\frac{6-N_f}{2\pi}\left(\ln\frac{a_{N_f}}{\Lambda_{X,N_f}}+
\ln\frac{a_{N_f}}{\Lambda_{Y,N_f}}\right)=\frac{6-2(N_f-1)}{2\pi}
\ln\frac{a_{N_f}}{\Lambda_{3,N_f-1}};\nonumber\\
&&\Lambda_{X,N_f}^{6-N_f}\Lambda_{Y,N_f}^{6-N_f}(a_{N_f})^{-4}=
4\Lambda_{X,N_f}^{6-N_f}\Lambda_{Y,N_f}^{6-N_f}(M^{N_fN_f})^{-2}=
\Lambda_{3,N_f-1}^{6-2(N_f-1)}.
\label{eq5p217}
\end{eqnarray}
The numerical factor $4$ is due to the fact that
\begin{eqnarray}
M^{N_fN_f}=Q^{N_f}_{\alpha_X,\alpha_Y}Q^{N_f}_{\beta_X,\beta_Y}
\epsilon^{\alpha_X\beta_X}\epsilon^{\alpha_Y\beta_Y}
=2(Q^{N_f}_{1,1}Q^{N_f}_{2,2}-Q^{N_f}_{1,2}Q^{N_f}_{2,1}),
\label{eq5p218}
\end{eqnarray}
and hence in the moduli space,
\begin{eqnarray}
M^{N_fN_f}=2a_{N_f}^2.
\label{eq5p219}
\end{eqnarray}
In general, $\Lambda_X{\neq}\Lambda_Y$ since the dynamics of each
$SU(2)$ branch is independent, but for convenience, we shall
 limit the discussions on $SO(4)$ to the case $\Lambda_X=\Lambda_Y$.

\subsection{Non-perturbative dynamical phenomena when $N_c{\geq}4$, 
$N_f{\leq}N_c-2$}
\label{subsect53}
\renewcommand{\thetable}{5.3.\arabic{table}}
\setcounter{table}{0}
\renewcommand{\theequation}{5.3.\arabic{equation}}
\setcounter{equation}{0}

\subsubsection{$N_f{\leq}N_c-5$: dynamically 
generated superpotential by gaugino condensation}
\label{subsub531}

This range is similar to the $SU(N_c)$ case when $N_f<N_c-1$ \cite{ref1p22}.
A superpotential arises generated by gaugino condensation in
the $SO(N_c-N_f)$ supersymmetric Yang-Mills theory, which is
the remainder of the $SO(N_C)$ QCD broken
by the scalar component of $\langle Q\rangle$.
According to (\ref{eq5p28}), we have \cite{ref51}
\begin{eqnarray}
W=(N_c-N_f-2)\langle\lambda\lambda\rangle 
 =\frac{1}{2}(N_c-N_f-2)\epsilon_{(N_c-N_f-2)}
\left(\frac{16\Lambda^{3N_c-N_f-6}_{N_c,N_f}}{\det M}\right)^{1/(N_c-N_f-2)},
\label{eq5p31}
\end{eqnarray}
where the phase factor $\epsilon_{(N)}{\equiv}\exp (i2n\pi/N)$.
Like in the $SU(N_c)$ case when $N_f<N_c-1$, this quantum effective 
superpotential will lift the classical vacuum degeneracy and make
the theory have no vacuum. This can be explicitly seen from the $F$-term
relevant to this dynamical superpotential:
\begin{eqnarray}
F_{ir}&=&\frac{\partial W}{\partial Q^{ir}}
=-\frac{1}{2}\epsilon_{(N_c-N_f-2)}
\left(\frac{16\Lambda^{3N_c-N_f-6}_{N_c,N_f}}{\det M}\right)^{1/(N_c-N_f-2)}
Q^{-1}_{ir}\nonumber\\
&\sim &\frac{1}{Q}\left(\frac{1}{\det M}\right)^{1/(N_c-N_f-2)}{\neq}0.
\label{eq5p32}
\end{eqnarray}
Thus all the supersymmetry vacua disappear and
this is another typical example of dynamical supersymmetry 
breaking \cite{wit1,wit2}.

If we consider a mass term for the matter fields,
$w_{\rm tree}=\mbox{Tr}(mM)/2$, this situation will change
greatly. The full superpotential with this mass term is
\begin{eqnarray}
W_{\rm full}=\frac{1}{2}(N_c-N_f-2)\epsilon_{(N_c-N_f-2)}
\left(\frac{16\Lambda^{3N_c-N_f-6}_{N_c,N_f}}{\det M}\right)^{1/(N_c-N_f-2)}
+\frac{1}{2}m^j_{~i}M^i_{~j}.
\label{eq5p33}
\end{eqnarray}
The moduli space is still given by the following $F$-flat direction 
labelled by $\langle M \rangle{\equiv}M$,
\begin{eqnarray}
F^{j}_{~i}=\frac{\partial W_{\rm full}}{\partial M^i_{~j}}
          =-\frac{1}{2}\epsilon_{(N_c-N_f-2)}
\left(\frac{16\Lambda^{3N_c-N_f-6}_{N_c,N_f}}
{\det M}\right)^{1/(N_c-N_f-2)}(M^{-1})^{j}_{~i}
+\frac{1}{2}m^j_{~i}=0,
\label{eq5p34}
\end{eqnarray}
which gives
\begin{eqnarray}
m^j_{~i}&=&\epsilon_{(N_c-N_f-2)}
\left(\frac{16\Lambda^{3N_c-N_f-6}_{N_c,N_f}}{\det M}\right)^{1/(N_c-N_f-2)}
(M^{-1})^j_{~i},\nonumber\\
\det m &=&\left[\epsilon_{(N_c-N_f-2)}\right]^{N_f}
\frac{(16\Lambda^{3N_c-N_f-6}_{N_c,N_f})^{N_f/(N_c-N_f-2)}}
{(\det M)^{(N_c-2)/(N_c-N_f-2)}}
\label{eq5p35}
\end{eqnarray}
and 
\begin{eqnarray}
\frac{1}{\det M}&=&\left(\epsilon_{(N_c-2)}\right)^{-N_f}
\frac{(\det m)^{(N_c-N_f-2)/(N_c-2)}}
{(16\Lambda^{3N_c-N_f-6}_{N_c,N_f})^{N_f/(N_c-2)}}.
\label{eq5p36}
\end{eqnarray}
Inserting Eq.\,(\ref{eq5p36}) into (\ref{eq5p34}), we have
\begin{eqnarray}
M^i_{~j} =\epsilon_{(N_c-2)}
\left[16\left(\det m\right)\Lambda^{3N_c-N_f-6}_{N_c,N_f}\right]^{1/(N_c-2)}
\left(m^{-1}\right)^i_{~j}.
\label{eq5p37}
\end{eqnarray}
Therefore, with this mass term we obtain a theory with $N_c-2$ 
supersymmetric vacua labelled by $\langle M \rangle$. This result
can also be derived by calculating the Witten index \cite{wit2}. 

Further, we can easily check that the superpotential is indeed
generated by gaugino condensation in the $SO(N_c-N_f)$ Yang-Mills 
theory. Integrating all massive modes out by inserting (\ref{eq5p35})
and (\ref{eq5p36}) into (\ref{eq5p33}), we can see that
\begin{eqnarray}
W_{\rm full}&=&\frac{1}{2}(N_c-2)\epsilon_{(N_c-2)}
\left[16\left(\det m\right)\Lambda^{3N_c-N_f-6}_{N_c,N_f}\right]^{1/(N_c-2)}
\nonumber\\
&=&\frac{1}{2}(N_c-2)\epsilon_{(N_c-2)} \Lambda^3_{N_c-N_f,0},
\label{eq5p38}
\end{eqnarray}
where $\Lambda^3_{N_c-N_f,0}{\equiv}
\left[16\left(\det m\right)\Lambda^{3N_c-N_f-6}_{N_c,N_f}\right]^{1/(N_c-2)}$
is the low energy scale for the $SO(N_c-N_f)$ Yang-Mills theory, the
many-flavour generalization of (\ref{eq5p210}).

If not all of the matter fields are massive, we can integrate out 
the massive quarks and get the effective superpotential at low energy
for the massless ones. It has the same 
form as (\ref{eq5p31}) but with the scale
replaced by the low energy one. For instance, if we only add a mass term
for the $N_f$-th quark, $W_{\rm tree}=m^{N_f}_{~N_f}M^{N_f}_{~N_f}/2$,
decoupling this heavy quark, we obtain the low energy effective 
superpotential:
\begin{eqnarray}
W_L=\frac{1}{2}\left[N_c-(N_f-1)-2\right]\epsilon_{(N_c-(N_f-1)-2)}
\left(\frac{16\Lambda^{3N_c-(N_f-1)-6}_{N_c,N_f-1}}
{\det M}\right)^{1/[N_c-(N_f-1)-2]},
\label{eq5p39}
\end{eqnarray}
where $\Lambda_{N_c,N_f-1}$ is given by Eqs.\,(\ref{eq5p210}), (\ref{eq5p211}) 
and (\ref{eq5p212}), respectively, depending on the concrete case.

\subsubsection{$N_f=N_c-4$: Two inequivalent branches and novel dynamics}
\label{subsub532}

 In this range, 
Eq.\,(\ref{eq5p115}) indicates that $SO(N_c)$ is broken to
$SO(4){\simeq}SU(2)_X{\times}SU(2)_Y$ by the scalar component of 
$\langle Q \rangle$. The phase factor for each $SU(2)$ branch is
\begin{eqnarray}
\epsilon_s=e^{i2n\pi/2}=e^{in\pi}=\pm 1, ~~s=X, Y; ~~n=1,2.
\label{eq5p310}
\end{eqnarray}
The scale for the low energy $SU(2)_s$ Yang-Mills theory 
(i.e. all the matter fields integrated out) of each branch is
\begin{eqnarray}
\Lambda^6_{X,0}=\Lambda^6_{Y,0}=
\frac{\Lambda^{3N_c-2)-(N_c-4)}_{N_c,N_c-4}}{\det M}=
\frac{\Lambda^{2(N_c-1)}_{N_c,N_c-4}}{\det M},
\label{eq5p311}
\end{eqnarray}
which is the many-flavour generalization of Eq.\,(\ref{eq5p216}). 
The number $6$ is
the one-loop $\beta$-function coefficient of $SU(2)$ Yang-Mills theory.
Note that we have used $\Lambda_X=\Lambda_Y$ as discussed above. 
The dynamical superpotential is generated by gaugino condensation 
in the unbroken $SU(2)_X{\times}SU(2)_Y$ Yang-Mills theory. 
According to Eqs.\,(\ref{eq5p31}) and (\ref{eq5p311}) the 
dynamical superpotential is
\begin{eqnarray}
W&=&W_X+W_Y=\frac{1}{2}(4-2)\epsilon_X\Lambda^3_{X,0}+
\frac{1}{2}(4-2)\epsilon_Y\Lambda^3_{Y,0}\nonumber\\
&=&2 \langle \lambda\lambda \rangle_X+
2 \langle \lambda\lambda \rangle_Y\nonumber\\
&=&\frac{1}{2}\left(\epsilon_X+\epsilon_Y\right)
\left(\frac{16\Lambda^{2(N_c-1)}_{N_c,N_c-4}}{\det M}\right)^{1/2}.
\label{eq5p312}
\end{eqnarray}
Since the phase factor $\epsilon$ labels different vacua, 
there are four ground states,
which are labelled by the four possible combinations of 
$(\epsilon_X,\epsilon_Y)$, i.e.
 \begin{eqnarray}
1. ~(1,1);~~~2.~(-1,-1);~~~3.~(1,-1);~~~4.~(-1,1).
\label{eq5p313}
\end{eqnarray}
The first two ground states, characterized by $\epsilon_X=\epsilon_Y$,
are physically equivalent, since they are related by a discrete $R$-symmetry 
given in Eq.\,(\ref{eq5p114}). Similarly, the last two ground states with
$\epsilon_X=-\epsilon_Y$ are also physically equivalent. Therefore, the sign
of $\epsilon_X\epsilon_Y$ labels two physically inequivalent 
branches of the low energy effective theory. The non-perturbative dynamics
in these two branches is greatly different, as shown in the following.

 The dynamics in the branch with $\epsilon_X\epsilon_Y=1$ is the same as 
in the case $N_f=N_c-4$. The dynamical superpotential 
$W=(16\Lambda^{2(N_c-1)}_{N_c,N_c-4}/\det M)^{1/2}$ lifts all the vacuum 
degeneracy and the quantum theory has no vacuum.   

 The two ground states with $\epsilon_X\epsilon_Y=1$ present a completely 
different physical pattern \cite{ref51}. The superpotential 
(\ref{eq5p312}) is zero, and hence the vacua
in the quantum moduli space are still degenerate but physically inequivalent.
These vacua are parameterized by $\langle M \rangle$. The two different
values $\pm 1$ of $\epsilon_X(=-\epsilon_Y)$ in this branch mean that
for every $\langle M \rangle$ there are two ground states. However, in the 
origin of the moduli space, $\langle M\rangle =0$, 
there is only one vacuum. Classically, in this vacuum
the gauge symmetry  $SO(4)$ is enhanced to $SO(N_c)$, i.e. at the 
origin of the moduli space, the original $SO(N_c)$ symmetry does not
break at all and the low energy effective theory will have a singularity,
corresponding to the $SO(N_c)/SO(4)$ vector bosons which become massless. 
This singularity can show up
in the kinetic term $K_{\rm clas.}(M,M^{\dagger})$, the classical 
K\"{a}hler potential. In quantum theory, the situation will change:
such a singularity is either smoothed out or it is associated with
some fields which become massless. In the theory we are considering, 
Intriligator and Seiberg conjectured that the classical singularity
at the origin is simply smoothed out \cite{ref51}. This means that 
the massless particle spectrum at the origin is the same as it is 
at other generic points, consisting only of the fields $M$. 
Similar phenomena have happened in low energy $N=2$ supersymmetric
Yang-Mills theory \cite{ref1p1,ref1p1a} and in a toy model proposed 
in Ref.\,\cite{ref57}. 

This conjecture can be subjected to several independent and 
nontrivial tests. The first is the 't Hooft anomaly matching. 
Since at both microscopic and macroscopic
levels the theory has a global $SU(N_f){\times}U_R(1)$ symmetry which is
unbroken at the origin, one can check the massless (fundamental 
and composite) particle spectrum by looking whether the 't Hooft anomaly at
the fundamental level matches with that at the composite level. The quantum 
numbers of the fundamental massless fermions (quark and gluino) under
the global symmetry $SO(N_c){\times}SU(N_f){\times}U_R(1)$ and  
the currents corresponding to $SU(N_f){\times}U_R(1)$
as well as the energy-momentum tensor are listed in Tables (\ref{ta5p3on}), 
(\ref{ta5p3tw}) and (\ref{ta5p3th}), respectively. The 't Hooft 
anomaly coefficients contributed by the massless elementary 
fermions can be easily calculated as in the
$SU(N_c)$ case and they are collected in Table (\ref{ta5p3fo}). 

\begin{table}
\begin{center}
\begin{tabular}{|c|c|c|c|} \hline
        & $SO(N_c)$ & $SU(N_f)$ &  $U_R(1)$ \\ \hline
 $\psi_Q$ & $N_c$     & $N_f$ &  $-(N_c-2)/N_f$  \\ \hline
$ Q$   & $N_c$  & $N_f$ &  $(N_f+2-N_c)/N_f$   \\ \hline
$\lambda$ & $N_c(N_c-1)/2$  & $1$ & $+1$  \\ \hline
\end{tabular}
\caption{\protect\small Representation quantum numbers of fundamental fields
under the global symmetry $SO(N_c){\times}SU(N_f){\times}U_R(1)$.
\label{ta5p3on}}
 \end{center}
\end{table}

\begin{table}
\begin{center}
\begin{tabular}{|c|c|c|c|} \hline
            & $SU(N_f)$ &  $U_R(1)$ \\ \hline
 $\psi_Q$  & $j_{\mu}^A(Q)=\overline{\psi}_Qt^A\sigma_{\mu}\psi_Q$ 
& $j_{\mu}(Q)=(2-N_c)/N_f\overline{\psi}_Q\sigma_{\mu}\psi_Q$
  \\ \hline
$\lambda$   & $0$       & $j_{\mu}(\lambda)
=\overline{\lambda}^a\sigma_{\mu}{\lambda}^a$  
\\ \hline
\end{tabular}
\caption{\protect\small Currents composed of  fundamental 
fermionic fields corresponding to the global symmetry 
$SU(N_f){\times}U_R(1)$. \label{ta5p3tw}}
 \end{center}
\end{table}

\begin{table}
\begin{center}
\begin{tabular}{|c|c|} \hline
            & $T_{\mu\nu}$\\ \hline
 $\psi$  & $i/4\left[\left(\overline{\psi}_Q\sigma_\mu\nabla_\nu\psi_Q
-\nabla_\nu\overline{\psi}_Q\sigma_\nu\psi_Q\right)
+\left(\mu\longleftrightarrow\nu\right)
\right]-g_{\mu\nu}{\cal L}[\psi_Q] $       \\ \hline
$\lambda$ &  $i/4\left[\left(\overline{\lambda}^a\sigma_\mu\nabla_\nu\lambda^a
-\nabla_\nu\overline{\lambda}^a\sigma_\mu\lambda^a\right)
+\left(\mu\longleftrightarrow\nu\right)\right]-g_{\mu\nu}{\cal L}[\lambda] $   \\ \hline
\end{tabular}
\caption{\protect\small Contribution of the fundamental fermionic fields
 to the energy-momentum tensor; 
${\cal L}[\psi]=i/2(\overline{\psi}\sigma^\mu\nabla_\mu\psi
-\nabla_\mu\overline{\psi}\sigma^\mu\psi)$,
$\nabla_\mu=\partial_\mu-\omega_{KL\mu}\sigma^{KL}/2$, 
$\sigma^{KL}=i/4[\sigma^K,\overline{\sigma}^L]$
and $\gamma^K=e^K_{~\mu}\sigma^{\mu}$.
\label{ta5p3th} }
 \end{center}
\end{table}

\begin{table}
\begin{center}
\begin{tabular}{|c|c|} \hline
 Triangle diagrams and      & 't Hooft anomaly \\ 
 gravitational anomaly      &    coefficients \\ \hline
 $ U_R(1)^3 $  & $N_c(N_c-1)/2+N_c/N_f^2(2-N_c)^3 $       \\ \hline
 $SU(N_f)^3 $ &  $N_c\mbox{Tr}(t^A\{t^B,t^C\})$            \\ \hline
$SU(N_f)^2U_R(1) $ &  $ (2-N_c){N_c}/{N_f}\mbox{Tr}(t^At^B)$\\ \hline
$U_R(1)$ &  $-N_c(N_c-3)/2$            \\ \hline
\end{tabular}
\caption{\protect\small  't Hooft anomaly coefficients from elementary 
massless fermions. \label{ta5p3fo}}
 \end{center}
\end{table}

At the macroscopic level, the only massless fermion is the
fermionic component $\psi_M$ of $M^{ij}$, which belongs to the
$N_f(N_f+1)/2$-dimensional representation of the $SU(N_f)$ group.
Its $R$-charge, from Table (\ref{ta5pon}), is 
\begin{eqnarray}
2\frac{N_f+2-N_c}{N_f}-1=\frac{N_f-2N_c+4}{N_f}.
\label{eq5p314}
\end{eqnarray} 
Thus the $SU(N_f){\times}U_R(1)$ Noether currents are, respectively:
\begin{eqnarray}
&& j^{A}_{\mu}(M)=\overline{\psi}_M^p\gamma_\mu t^A_{pq}{\psi}_M^q,~~
j^{(R)}_\mu (M)=\frac{N_f-2N_c+4}{N_f}\overline{\psi}_M^p\gamma_\mu
{\psi}_M^p, \nonumber\\
&& p,q=1,2,\cdots, N_f(N_f+1)/2.
\label{eq5p315}
\end{eqnarray}
The corresponding 't Hooft anomaly coefficients 
are collected in Table (\ref{ta5p3fi}),
and one can easily see that when $N_f=N_c-4$, the anomalies of the 
macroscopic theory exactly match those of the microscopic theory. 
Note that in calculating the anomaly 
coefficients of Table (\ref{ta5p3fi}) we have used
the relation between the quadratic and cubic $SU(N_f)$ Casimirs 
in the $N_f(N_f+1)/2$ dimensional representation and 
in the fundamental ($N_f$-dimensional) representation:
\begin{eqnarray}
\mbox{Tr}(t^A\{t^B,t^C\})_{N_f(N_f+1)/2}&=&(N_f+4)
\mbox{Tr}(t^A\{t^B,t^C\})_{N_f},\nonumber\\
\mbox{Tr}(t^At^B)_{N_f(N_f+1)/2}&=&(N_f+2)
\mbox{Tr}(t^At^B)_{N_f}.
\label{eq315i}
\end{eqnarray}
Thus the 't Hooft anomaly matching supports the conjecture: the classical
singularity of the K\"{a}hler potential near the origin is smoothed out by
quantum effects and hence \cite{ref1p1,ref57} 
\begin{eqnarray}
K(M^{\dagger},M)\stackrel{M{\rightarrow}0}
{\sim}\frac{\mbox{Tr}M^{\dagger}M}{|\Lambda|^2},
\label{eq5p315m}
\end{eqnarray}
where the dynamical scale $\Lambda$ is introduced from dimensional
considerations.
 
\begin{table}
\begin{center}
\begin{tabular}{|c|c|} \hline
 Triangle diagrams and      & 't Hooft anomaly \\ 
 gravitational anomaly      &    coefficients \\ \hline
 $ U_R(1)^3 $  & $(N_f+1)(N_f-2N_c+4)^3/(2N_f^2) $       \\ \hline
 $SU(N_f)^3 $ &  $(N_f+4)\mbox{Tr}(t^A\{t^B,t^C\})$            \\ \hline
$SU(N_f)^2U_R(1) $ &  $ (N_f+2)(N_f-2N_c+4)/N_f\mbox{Tr}(t^At^B)$\\ \hline
$U_R(1)$ &  $(N_f+1)(N_f-2N_c+4)/2$            \\ \hline
\end{tabular}
\caption{\protect\small  't Hooft anomaly coefficients for 
the composite fermions. \label{ta5p3fi} }
 \end{center}
\end{table}

Another test of the above conjecture is the decoupling of a heavy mode. Giving
$Q_{N_f}$ a large mass and integrating it out, the resulting low energy 
effective theory should agree with the $N_f=N_c-4$ case discussed
in the last section. We first look at the branch with $\epsilon_X\epsilon_Y
=1$. Adding the mass term $W_{\rm tree}=mM^{N_f}_{~N_f}/2$ to 
the dynamical superpotential (\ref{eq5p312}), we have
\begin{eqnarray}
W_{\rm full}=\left(\frac{16\Lambda^{2(N_c-1)}_{N_c,N_c-4}}{\det M}
\right)^{1/2}+\frac{1}{2}mM^{N_f}_{~N_f}.
\label{eq5p317}
\end{eqnarray}
The $F$-flatness condition for $M^{N_f}_{~N_f}$ gives
\begin{eqnarray}
M^{N_f}_{~N_f}=\left(\frac{16\Lambda^{2(N_c-1)}_{N_c,N_c-4}}{\det M}
\right)^{1/2}\frac{1}{m}.
\label{eq5p318}
\end{eqnarray}
Thus, the low energy superpotential is 
\begin{eqnarray}
W=\frac{3}{2}mM^{N_f}_{~N_f}=\frac{3}{2}
\left(\frac{16\Lambda^{2(N_c-1)-1}_{N_c-1,N_c-5}}{\det M'}\right)^{1/2},
\label{eq5p319}
\end{eqnarray}
where we have used the decoupling relation (\ref{eq5p210}) 
and $\det M=\det M' M^{N_f}_{~N_f}$.
(\ref{eq5p319}) coincides exactly with the superpotential 
(\ref{eq5p31}) with $N_f=N_c-5$.

In the branch with $\epsilon_X\epsilon_Y=-1$ the dynamical
generated superpotential is zero. Adding a mass term, the full
superpotential is $W_{\rm full}=W_{\rm tree}=mM^{N_f}_{~N_f}/2$. It is
not possible to return back to the theory with $N_f=N_c-5$ using this 
superpotential and thus this branch should be eliminated from the
low energy effective theory with $N_f{\leq}N_c-5$. This supports
the conjecture, since if one requires that the K\"{a}hler potential
is smooth everywhere in $M$, this branch will have no supersymmetric ground 
state due to the fact that the scalar potential will be proportional to 
$1/K(M^{\dagger},M)$. This means that the supersymmetry is dynamically
broken. Further, if we suppose that in the branch
with $\epsilon_X\epsilon_Y=-1$ there are new massless states somewhere,
then the addition of this $Q^{N_f}$ mass term will result in additional
ground states. Since all the ground states for $N_f<N_c-4$  are already 
exhausted by (\ref{eq5p37}), there will be no such extra 
ground states. Therefore, one can conclude that the manifold 
of quantum vacua (i.e. the quantum moduli space) must be smooth 
everywhere and without any new massless fields except $M$.  
  
 The physical phenomena at the origin in $M$ are very 
interesting. The above discussion shows that at the classical 
level there are massless quarks and gluons and their superpartners, 
while in the quantum theory, only the $M$ quanta are massless. 
Since $M$ are colour singlets, 
this clearly shows that the elementary degrees of freedom are confined. 
However, the global chiral symmetry $SU(N_f){\times}U_R(1)$ is not
broken at the origin. Therefore, we have a novel physical phenomenon that
there is confinement but without chiral symmetry breaking. The same 
phenomenon has been observed in the $SU(N_c)$ case with $N_f=N_c+1$.

\subsubsection{$N_f=N_c-3$: Two dynamical branches and 
massless composite particles (glueball and exotic states)}
\label{subsub533}

For this value it will be shown that the ground state of the theory still
has two branches, but with different dynamics.

 From (\ref{eq5p115}), we know that the expectation value 
of the scalar component of $Q$ breaks $SO(N_c)$ to $SO(3)$, but 
the dynamically generated superpotential at low energy is not a simple 
continuation of (\ref{eq5p28}), since in breaking $SO(N_c)$ 
to $SO(3)$ by the Higgs mechanism, there is a special case 
$SO(4){\simeq}SU(2)_X{\times}SU(2)_Y$ between $SO(N_c)$ and $SO(3)$.
The dynamically generated superpotential is not only contributed
by the gaugino condensation of supersymmetric $SO(3)$ Yang-Mills 
theory, but it also receives contributions from the instanton in each 
$SU(2)_s$ branch. A concrete method to find the dynamically
generated superpotential is as follows. First by choosing  
$N_f-1$ eigenvalues of $\langle Q \rangle$ to be large 
we break the $SO(N_c)$ theory to $SU(2)_X{\times}SU(2)_Y$ 
with one quark superfield $Q^{N_f}$ left massless. 
Matching the running gauge couplings at the scales of the Higgs
mechanism gives us the relation between the low energy and high energy
scales:
\begin{eqnarray}
\Lambda^5_{X,1}=\Lambda^5_{Y,1}=\frac{\Lambda^{2N_c-3}_{N_c,N_c-3}}
{\det M'}, ~~~~\det M'=\frac{\det M}{M^{N_f}_{~N_f}}.
\label{eq5p320}
\end{eqnarray}  
(This relation is actually the many-flavour generalization 
of (\ref{eq5p215}).) Then the
expectation value $\langle Q^{N_f}\rangle$ breaks the 
$SU(2)_X{\times}SU(2)_Y$ gauge group of this intermediate theory to a 
diagonally embedded $SO(3)_d$. According to (\ref{eq5p217}), 
the  low energy scale is
\begin{eqnarray}
\Lambda^6_{(d)3,0}=4\Lambda^5_{X,1}\Lambda^5_{Y,1}(M^{N_fN_f})^{-2}.
\label{eq5p321}
\end{eqnarray} 
The gaugino condensation in the unbroken $SO(3)$ will generate a 
superpotential
\begin{eqnarray}
W_d=2\langle \lambda\lambda\rangle=2(\Lambda_{(d)3,0}^6)^{1/2}
=4\epsilon \frac{\Lambda^5_{X(Y),1}}{M^{N_f}_{~N_f}}=
4\epsilon \frac{\Lambda^{2N_c-3}}{\det M}, ~~\epsilon=\pm 1.
\label{eq5p322}
\end{eqnarray}
In addition instantons in the broken $SU(2)_X$ generate
another superpotential
\begin{eqnarray}
W_X=2 \frac{\Lambda^5_{X,1}}{M^{N_f}_{~N_f}}=
2\frac{\Lambda^{2N_c-3}_{N_c,N_c-3}}{\det M},
\label{eq5p323}
\end{eqnarray}
and instantons in $SU(2)_Y$ give
\begin{eqnarray}
W_Y=2 \frac{\Lambda^5_{Y,1}}{M^{N_f}_{~N_f}}=
2\frac{\Lambda^{2N_c-3}_{N_c,N_c-3}}{\det M}.
\label{eq5p324}
\end{eqnarray}
Adding these three contributions together, we obtain the superpotential
for $SO(N_c)$ with $N_f=N_c-3$,
\begin{eqnarray}
W=W_d+W_X+W_Y=4(1+{\epsilon})\frac{\Lambda^{2N_c-3}_{N_c,N_c-3}}{\det M}.
\label{eq5p325}
\end{eqnarray}

Here the generation of the superpotential from the broken $SU(2)_s$
needs some delicate explanation \cite{ref51}. 
The instanton contributions in $SU(2)_s$
contain those from the instantons in the broken part.
Usually when a gauge group $G$ is broken
to a non-Abelian subgroup $H$ along a flat direction, there is no need
to consider instantons in the broken $G/H$ part. This is because 
an instanton in the broken $G/H$ is not well-defined,
and can be rotated to become an instanton of the $H$ 
gauge theory. But the dynamics described by the $H$ gauge 
theory will be stronger than these instanton effects.
However, when the instantons in the broken part like 
$(SU(2)_X\times SU(2)_Y)/SO(3)$ are well defined, their effect must 
be taken into account when one integrates out the massive gauge fields. This 
situation occurs when $G$ (or one of its factors if $G$ is fully
reducible) is completely broken or broken to an Abelian subgroup, or 
when the index of the embedding of $H$ in $G$ is large than 1\footnote{The 
index of a group is defined as the expansion coefficient 
of the leading term when expanding the trace of the product of
any number of its generators in some representation 
in terms of  the fundamental symmetric invariant tensors of 
the group \cite{ref58}. For the trace of $n$ generators,
the index is called $n$th index of this representation.}. In our case, the 
the second index of $SO(3)$ in the adjoint representation is $2$ and 
thus one should consider the contribution from these instantons.

 Now let us see what physics the superpotential (\ref{eq5p325}) describes.
First, the low energy theory again has two physically inequivalent branches
classified by $\epsilon$. The branch with $\epsilon=1$ is
the continuation of (\ref{eq5p31}) to $N_f=N_c-3$. Thus at the quantum 
level the classical degeneracy will be lifted and there is no vacuum. 
The branch with $\epsilon=-1$ has vanishing superpotential, and thus 
there exists a quantum moduli space of degenerate vacua. However, the 
dynamics in this case is greatly different from the 
$\epsilon_X\epsilon_Y=-1$ branch of the $N_f=N_c-4$ case. This 
can be observed from the decoupling of the heavy mode.

 The decoupling is done in the standard way. 
Adding a mass term  $W_{\rm tree}=mM^{N_f}_{~N_f}/2$ 
and then integrating out $Q^{N_f}$ in the same
way as above, the branch with $\epsilon=1$ will give the two ground states
of the $\epsilon_X\epsilon_Y=1$ branch of the case $N_f=N_c-4$. This is exactly
what we expect. If we add the mass term $W_{\rm tree}=mM^{N_f}_{~N_f}/2$
to the mass term in the $\epsilon=-1$ branch, we should get the two ground 
states of the $\epsilon_X\epsilon_Y=1$ branch of the $N_f=N_c-4$ case.
However, since the dynamical superpotential vanishes, a similar argument
as in the $N_f=N_c-4$ case shows that this branch has no decoupling
limit and must be eliminated upon adding $W_{\rm tree}$. In oder to 
avoid this disease, Intriligator and Seiberg conjecture that there
must be additional massless particles at the origin of the moduli 
space \cite{ref51}.
Since these massless fields should not appear at generic points of
the moduli space, there must be a superpotential responsible for their
masses away from the origin $\langle M\rangle=0$. 
The simplest way (perhaps the only 
possible way) to implement this conjecture is to introduce some 
(chiral super-) fields $k_i$ with $i$ being the flavour indices, and
make them couple to $M^{ij}$. We will see that
these fields indeed have a natural physical interpretation.
Near the origin, the corresponding
superpotential for the mass term of these new particles 
should have the asymptotic form:  
\begin{eqnarray}
W\sim \frac{1}{2\mu}M^{ij}k_ik_j, ~~~\mbox{when}~M\sim 0,
\label{eq5p326}
\end{eqnarray}
where $\mu$ is a scale with mass dimension. It is necessary to
introduce this scale since the field $k$, as a scalar superfield,
should have dimension $1$, $M$ has
dimension $2$ and the superpotential has dimension $2$. 
Requiring this superpotential to respect the global $SU(N_f){\times}U_R(1)$
symmetry, we see that the $k_i$ should belong 
to the $N_f$-dimensional conjugate
representation of $SU(N_f)$. Since the superpotential should have
$R$ charge $2$ and $M$ has $R$-charge $2(N_f+2-N_c)/N_f$,  the fields
$k_i$ should have $R$-charge 
\begin{eqnarray}
\frac{1}{2}\left[2-2\frac{N_f+2-N_c}{N_f}\right]
=\frac{N_c-2}{N_f}=1+\frac{1}{N_f}.
\label{eq5p327}
\end{eqnarray}
Now making the heavy mode $M^{N_f}_{~N_f}$ decouple near the origin 
by adding the mass term $W_{\rm tree}=mM^{N_f}_{~N_f}/2$ to 
(\ref{eq5p326}),
\begin{eqnarray}
W_{\rm full}=\frac{1}{2}mM^{N_f}_{~N_f}+\frac{1}{2\mu}M^{ij}k_ik_j,
\label{eq5p328}
\end{eqnarray}
and then integrating out the $M^{N_f}_{~N_f}$, we immediately obtain
\begin{eqnarray}
\langle k_{N_f}\rangle=\pm i\sqrt{m\mu}.
\label{eq5p329}
\end{eqnarray}
The two sign choices in $\langle k_{N_f}\rangle$ can be interpreted as
two physically equivalent ground states with $W=0$, which exactly 
correspond to the two choices in the $\epsilon_X\epsilon_Y=-1$ branch 
of the low energy $N_f=N_c-4$ theory.

Eq.\,(\ref{eq5p326}) is the approximate form of the superpotential near
$M=0$. The most general superpotential respecting the 
$SU(N_f){\times}U_R(1)$ symmetry and having the right mass dimension 
is then
\begin{eqnarray}
W=\frac{1}{2\mu}f\left[t=\frac{(\det M)
(M^{ij}k_ik_j)}{\Lambda^{2N_c-2}_{N_c,N_c-3}}\right]M^{ij}k_ik_j.
\label{eq5p330}
\end{eqnarray}
In order for this general superpotential to yield the ground states
(\ref{eq5p329}), $f(t)$ must be a holomorphic function in the neighborhood
of $t=0$. $f(0)$ can be set to $1$ by rescaling $q_i{\rightarrow}q_i/f(0)$.

 How can we check the reasonableness of the conjecture that
in addition to the massless fields $M$ there still exist other 
massless particles? We again resort to the 't Hooft anomaly matching.  
It will be a highly non-trivial verification of the above 
conjecture if the anomalies contributed by the 
massless spectrum consisting of $M^{ij}$ and $k_i$ match  those of
the fundamental massless particle spectrum with $N_f=N_c-3$. The
conserved $SU(N_f){\times}U_R(1)$ currents composed of the fermionic 
component of $k_i$ are:
\begin{eqnarray}
j^A_{\mu}(\psi_k)=\overline{\psi}_k{t}^A\gamma_{\mu}\psi_k;
~~~j_{\mu}(\psi_k) =\left(1+\frac{1}{N_f}\right)\overline{\psi}_k
\gamma_{\mu}\psi_k.
\label{eq5p331}
\end{eqnarray}
The relevant fermionic part of the energy-momentum tensor for 
the 't Hooft axial gravitational anomaly formally reads:
\begin{eqnarray}
T_{\mu\nu}=\frac{i}{4}\left(\overline{\psi}_k\gamma_{\mu}\nabla_{\nu}\psi_k
-\nabla_{\nu}\overline{\psi}_k\gamma_{\mu}\psi_k\right)\psi
-g_{\mu\nu}{\cal L}[\psi_k].
\label{eq5p332}
\end{eqnarray} 
The 't Hooft anomaly coefficients corresponding to the triangle
diagrams composed of these currents are listed 
in Table (\ref{ta5p3si}). Adding the anomaly 
coefficients to those contributed by the field $M$ listed 
in Table (\ref{ta5p3fi}) and comparing them 
with the anomaly coefficients from the 
massless elementary particles listed in Table (\ref{ta5p3fo}), one can see 
that they are precisely equal for $N_f=N_c-3$.  

\begin{table}
\begin{center}
\begin{tabular}{|c|c|} \hline
 Triangle diagrams and      & 't Hooft anomaly \\ 
 gravitational anomaly      &    coefficients \\ \hline
 $ U_R(1)^3 $  & $1/N_f^2 $       \\ \hline
 $SU(N_f)^3 $ &  $-\mbox{Tr}(t^A\{t^B,t^C\})$            \\ \hline
$SU(N_f)^2U_R(1) $ &  $ 1/N_f\mbox{Tr}(t^At^B)$\\ \hline
 $U_R(1)$ &  $1$            \\ \hline
\end{tabular}
\caption{\protect\small 't Hooft anomaly coefficients for 
composite fermions. \label{ta5p3si} }
 \end{center}
\end{table}

Finally let us see what physical objects these fields $k_i$ can be 
interpreted as. $k_i$ should be constructed from the fundamental chiral 
superfields since the product of any chiral superfields is still a
chiral superfield. We know that the fundamental chiral superfields in the
theory are the matter fields $Q^i_{~r}$ and the gauge superfield strength
$W_\alpha^a$. From the mass dimensions and the quantum numbers under
$SU(N_f){\times}U_R(1)$, one can immediately identify $k_i$ as
\begin{eqnarray}
k_i
=\Lambda^{2-N_c}_{N_c,N_c-3}b_i,
\label{eq5p333}
\end{eqnarray}
where
\begin{eqnarray}
b_i&=&(Q)^{N_c-4}_i\mbox{Tr}(W_{\alpha}W^{\alpha})\nonumber\\
&{\equiv} &\frac{1}{(N_c-4)!}\epsilon_{ii_1i_2{\cdots}i_{N_c-4}}
\epsilon^{r_1r_2{\cdots}r_{N_c-4}}Q^{i_1}_{~r_1}Q^{i_2}_{~r_2}{\cdots}
Q^{i_{N_c-4}}_{~r_{N_c-4}}(W_\alpha^aW^{a\alpha}).
\label{eq5p334}
\end{eqnarray}
Obviously, $b_i$ has mass dimension $N_c-1$. The 
superpotential (\ref{eq5p330}) can now be rewritten 
in terms of $b_i$, 
\begin{eqnarray}
W=\frac{1}{2\Lambda^{2N-c-3}}
f\left[t=\frac{(\det M)(M^{ij}b_ib_j)}
{\Lambda^{4N-c-6}_{N_c,N_c-3}}\right]M^{ij}b_ib_j,
\label{eq5p335}
\end{eqnarray}
where the scale $\mu$ appearing in (\ref{eq5p330}) has 
been absorbed into the definition of $f$.  

(\ref{eq5p333}) shows that the $k_i$ fields describe exotic particles. One 
can intuitively think of such exotics as being some heavy bound states. 
They become massless at the origin of the quantum moduli space.
These exotic particles are similar to the massless mesons and baryons
in $SU(N_c)$ supersymmetric QCD with $N_f=N_c+1$, as discussed in
Sect.\,\ref{subsub6.3.7}. Since they are colour singlets and respect 
the chiral symmetry $SU(N_f){\times}U_R(1)$, this is again 
a phase in which there exists confinement but without chiral symmetry 
breaking.

\subsubsection{$N_f=N_c-2$: Coulomb phase with 
massless monopoles and dyons and confinement and 
oblique confinement}
\label{subsub534}

For this number of flavours, the dynamics is more 
complicated than in the cases discussed above. 
From Table (\ref{ta5ptw}), the $R$-charge of $M^{ij}$ 
vanishes, thus no superpotential of the form (\ref{eq5p28i}) 
can be dynamically generated  from gaugino condensation.
Consequently, the quantum theory has a moduli space of physically 
inequivalent vacua, which should still be parametrized by the 
expectation values $\langle M^{ij}\rangle $. Classically, according 
to (\ref{eq5p115})  in the moduli space determined by the $D$-flat 
directions, the $SO(N_c)$ gauge group breaks
to $SO(2){\cong}U(1)$. Thus the low energy theory is in the 
Coulomb phase with the gauge vector superfield being a massless photon 
supermultiplet. Like in the various cases discussed above, 
there exists a singularity at the origin $\langle M\rangle =0$ 
(or equivalently, $\det M=0$) of the classical moduli space, 
which is associated with an unbroken gauge symmetry. In the quantum theory, 
we will see that a distinct kind of singularity arises 
at $M=0$, which is related to massless monopoles rather 
than massless vector bosons. In fact, from the discussion
on the Coulomb phase in Subsect.\,\ref{subsect27}, 
we can imagine this situation arising since the Coulomb phase has a 
natural electric-magnetic duality, and hence monopoles should emerge.

 How can we explore the non-perturbative dynamics in the Coulomb
phase? The Coulomb phase looks simple, but actually its non-perturbative
dynamics, as a consequence of electric-magnetic duality, is very 
complicated. In the cases considered above, the dynamical 
superpotential and the decoupling limit give almost all of the 
non-perturbative phenomena, at least qualitatively. However, the
situation in the Coulomb phase is completely different. 
There is no dynamical superpotential to depend on. 
Fortunately, the investigation of $N=2$
supersymmetric gauge theories by Seiberg and Witten provided 
some clues \cite{ref1p1,ref1p1a}. The Coulomb phase of the theory 
can be investigated by determining the effective gauge coupling
$\tau =\frac{\theta}{\pi}+i\frac{8\pi}{g^2}$
of the massless photon supermultiplet on the moduli space of 
vacua. This is because the general form of the low energy effective action 
in the Coulomb phase is
\begin{eqnarray} 
{\cal L}\sim \mbox{Im}\int d^2\theta \tau W_{\alpha}W^{\alpha}.
\label{eq5p337}  
\end{eqnarray} 
The non-perturbative phenomena for the value $N_f=N_c-2$ have much in common 
with the Coulomb phase of the $N=2$ supersymmetric gauge theory. As 
will be discussed  in the following, there
are massless monopoles and dyons at some points of the moduli space,
and their condensation will cause confinement and oblique confinement.
It is then not so strange that both cases exhibit similar 
non-perturbative phenomena since their  low-energy theories
are both in the Coulomb phase. 

 How can we determine $\tau$? In general, in the Coulomb phase $\tau$
receives two contributions. When $\langle M^{ij}\rangle$ is very large,
the microscopic theory is weakly coupled and perturbation theory works.
Thus, at the quantum level, one part of $\tau$ in the Coulomb phase
comes from the gauge running coupling evaluated at the energy scale
characterized by $\langle M^{ij}\rangle$. The explicit form
of this perturbative contribution is given 
by the one-loop beta function  of the microscopic theory, i.e.
\begin{eqnarray}
\beta_0=3(N_c-2)-N_f=2N_c-4=2N_f.
\label{eq5p338}
\end{eqnarray}
From the requirement of holomorphicity, the 
effective coupling constant $\tau$ in the Coulomb 
phase should be a holomorphic function of the parameters 
$\langle M^{ij}\rangle$. Because of the global flavour 
symmetry $SU(N_f)$, it should depend on 
$SU(N_f)$ invariant combinations of $M^{ij}$. The natural choice is
\begin{eqnarray} 
U{\equiv}\det M^{ij}.
 \label{eq5p339}  
\end{eqnarray}
As shown in Eq.\,(\ref{eq6.112y}), the  
perturbative one-loop exact $\tau$ at the energy
scale characterized by $\langle M^{ij}\rangle$ is
\begin{eqnarray} 
e^{i2\pi \tau}{\det M}=\Lambda^{\beta_0};
~~~~\tau
= -\frac{i}{2\pi}\ln\left(\frac{\Lambda^{\beta_0}}{U}\right). 
\label{eq5p340}  
\end{eqnarray} 
The general form of the one-loop running gauge coupling 
and dimensional analysis imply that the perturbative contribution 
to $\tau$ must be of this form.  

However, for small $\langle M^{ij}\rangle$ (or equivalently small $U$), 
another non-perturbative contribution arises from the 
instantons, and hence 
the determination of $\tau$  will become complicated.
There are two ways to determine the explicit functional form of $\tau$. 
The first one is using a straightforward 
instanton calculation \cite{ref1p1,ref1p2,ref1p25},
which gives an $F$-term of the form: 
\begin{eqnarray} 
\int d^2\theta \left[\left(W^{\alpha}W_{\alpha}\right)
\left(\frac{\Lambda^{2N_c-4}}{U}\right)\right],
\label{eq5p341}  
\end{eqnarray} 
where $W^{\alpha}$ is the low energy $U(1)$ photon field strength 
supermultiplet. The $n$-instanton contribution to the low energy effective
action in the dilute gas approximation is  
\begin{eqnarray} 
\int d^2\theta
\left[W^{\alpha}W_{\alpha}\left(\frac{\Lambda^{2N_c-4}}{U}\right)^n\right].
\label{eq5p342}  
\end{eqnarray} 
Since the effective Lagrangian can always be written in
the following general form,
\begin{eqnarray}
{\cal L}_{\rm eff}= \frac{1}{16\pi}\mbox{Im}\int d^2\theta 
\tau_{\rm eff}W_{\alpha}W^{\alpha}+\cdots ,
\label{eq5p343}  
\end{eqnarray} 
the $F$-term gives the $n$-instanton correction to
$\tau$ of the form:
\begin{eqnarray}
\left(\frac{\Lambda^{2N_c-4}}{U}\right)^n.
\label{eq5p344}  
\end{eqnarray}
To get the full non-perturbative $\tau$, one should sum 
all the instanton contributions (\ref{eq5p342}): 
\begin{eqnarray}
\sum_{n=0}^{\infty} a_n \left(\frac{\Lambda^{2N_c-4}}{U}\right)^n 
W^\alpha W_\alpha,
\label{eq5p345}  
\end{eqnarray} 
where the $a_n$ are some constant coefficients, which are determined
by explicit instanton calculations. Despite the fact that this method is 
very physical and can be carried out up to three instantons,  
it is actually not possible to perform the above summation. Thus 
this method can only be used as a check of the non-perturbative 
result \cite{ref512}.  

 A beautiful and powerful method to determine $\tau$ 
was worked out by Seiberg and Witten \cite{ref1p1} based on 
\begin{itemize} 
\item electric-magnetic duality conjecture in the Coulomb phase;
\item global geometric structure of the quantum moduli space;
\item holomorphicity; 
\item gauge invariance and various global symmetries; 
\item correct decoupling limit. 
\end{itemize}
A crucial observation of Seiberg
and Witten is that in the $SU(2)$ case, the quantum moduli space
of $N=2$ supersymmetric Yang-Mills theory can be determined exactly:
the Riemann surface of the 
quantum moduli space is a torus and $\tau$ is equivalent to 
the modular parameter of the torus, i.e. the ratio
of the periods of the torus. Then one can use the global geometric
property of the torus and the electric-magnetic duality to determine
$\tau$. Before going into the details of determining $\tau$, we first
see how the $SO(N_c)$ symmetry spontaneously breaks to 
$SO(2){\cong}U(1)$.

The spontaneous breaking from $SO(N_c)$ to $SO(2)$ is not 
straightforward, since there are the two special groups 
$SO(4)$ and $SO(3)$ between them. The  
breaking pattern should be as follows: first
the breaking  $SO(N_c)\longrightarrow SO(4)
{\cong}SU(2)_X{\times}SU(2)_Y$ occurs; then $SU(2)_X{\times}SU(2)_Y$ 
${\longrightarrow}$ $SU(2){\cong}SO(3)$; finally $SU(2)$ breaks to $U(1)$.

The first step can be taken by considering the region of the moduli 
space where $N_c-4$ eigenvalues of $\langle M^{ij}\rangle$ become large,
breaking the $SO(N_c)$ theory to a low energy 
$SO(4){\cong}$$SU(2)_X$ ${\times}SU(2)_Y$ theory with two light flavours 
($N_f=2$) in the fundamental representation $(2,2)$ of 
the low energy gauge group. Matching the running gauge couplings at
the scale $\langle M^{ij}\rangle$ gives the relation between the
high energy dynamical scale and the low energy scale,
\begin{eqnarray}
\Lambda^4_{X,2}=\Lambda^4_{Y,2}=\frac{\Lambda^{2N_c-4}_{N_c,N_c-2}}
{\det^{\prime} M_{2N_c-4}}{\equiv}\frac{\Lambda^{2N_c-4}}{U_H}.
\label{eq5p346}  
\end{eqnarray}
Eq.\,(\ref{eq5p346}) is the $N_f-4$ flavour generalization 
of (\ref{eq5p215}) and (\ref{eq5p217}). 
$U_H{\equiv}\det^{\prime} M_{2N_c-4}$ 
means the determinant for $N_f-4$ heavy flavours, or equivalently, 
the product of the $N_c-4$ large eigenvalues of $M^{ij}$. 
Explicitly, the $SU(N_f=2)$ invariant combination of two light 
flavours in the low energy $SU(2)_X{\times}SU(2)_Y$ theory is
\begin{eqnarray}
\widehat{U}=\frac{\det M}{\det^{\prime} M_{2N_c-4}}
=\frac{U}{U_H}=\det M_{fg},
\label{eq5p347}  
\end{eqnarray}        
where $M_{fg}$ are the $SU(2)_X{\times}SU(2)_Y$ singlets,
\begin{eqnarray}
M_{fg}=Q_f{\cdot}Q_g{\equiv}\frac{1}{2}Q_{f,\alpha_1\alpha_2}
Q_{g,\beta_1\beta_2}\epsilon^{\alpha_1\alpha_2}\epsilon^{\beta_1\beta_2},
~~~f,g=1,2;~~\alpha_i=\beta_i=1,2.
\label{eq5p348}  
\end{eqnarray}  

The second step $SU(2)_X{\times}SU(2)_Y
{\longrightarrow}SU(2)$ can be performed 
by considering the limit of large $\widehat{U}$ and taking one eigenvalue, 
say $M_{11}$, large. Then the gauge symmetry is broken to $SU(2)_d$, with
the light flavour $Q_{2,\alpha\beta}$ decomposing into an $SU(2)_d$
singlet $S$ (i.e. in the trivial representation) and a triplet 
$\phi_d$ (i.e. in the adjoint representation) \cite{ref1p9},
\begin{eqnarray}
Q_2=\phi_d\oplus S.
\label{eq5p349}  
\end{eqnarray} 
(\ref{eq5p349}) and the definition 
$M_{22}{\equiv}Q_2\cdot Q_2$ mean that
\begin{eqnarray}
\phi_d^2=\frac{U}{M_{11}}{\equiv}\widetilde{U}.
\label{eq5p350}  
\end{eqnarray}
The subscript $d$ indicates that this $SU(2)$ is a diagonally 
embedded subgroup of $SU(2)_X{\times}SU(2)_Y$. 
This index will be omitted later. Integrating out the heavy flavour, we
obtain the low energy $SU(2)_d$ theory. According to (\ref{eq5p217}), 
the relation between the low energy dynamical 
scale $\Lambda_d$ and the intermediate dynamical 
scales $\Lambda_s$, $s=X,Y$ is
\begin{eqnarray}
\Lambda^4=\frac{16\Lambda_X^4\Lambda_Y^4}{M^2_{11}}.
\label{eq5p351}
\end{eqnarray} 

Now we have an $SU(2)$ theory with a triplet $\phi$. It is very
similar to the $N=2$ Seiberg-Witten model except for two extra gauge 
singlets ($M_{11}$ and another one from $Q_2$). The last 
step $SU(2){\longrightarrow}U(1)$ is induced by the scalar potential
of $\phi^a$, $a=1,2,3$. This is a standard Higgs mechanism. 
Therefore, we can use the 
Seiberg-Witten method to determine the low energy effective coupling in
the Coulomb phase. There already exist several excellent reviews on the
Seiberg-Witten solution \cite{ref514}, here we only repeat 
the main points.

\vspace{3mm}
\begin{flushleft}
{\it Seiberg-Witten algebraic curve solution}
\end{flushleft}
\vspace{3mm}

The scalar potential of $N=2$ supersymmetric $SU(2)$ 
Yang-Mills theory is 
\begin{eqnarray}
V(\phi)=-\frac{g^2}{2}(D^a)^2,~~~~D^a=[\phi^{\dagger}, \phi]^a.
\label{eq5p352}
\end{eqnarray}
Classically, like in supersymmetric QCD, there is a $U_R(1)$ symmetry,
under which the gaugino $\lambda$, the triplet $\phi$ and its superpartner
 $\psi$ transform as
\begin{eqnarray}
\lambda{\longrightarrow}e^{i\alpha}\lambda;~~~ 
\phi{\longrightarrow}e^{i2\alpha}\phi;~~~
\psi{\longrightarrow}e^{i\alpha}\psi.
\label{eq5p353}
\end{eqnarray}
At the quantum level, this symmetry is broken by a gauge anomaly, 
which is equivalent to a shift of the vacuum angle $\theta$,
\begin{eqnarray}
\theta{\longrightarrow}\theta - 2N_c\alpha-2N_c\alpha=\theta - 8\alpha .
\label{eq5p354}
\end{eqnarray}
The shift of $\theta$ by $2\pi$ is still a symmetry of the theory since
the generating functional is invariant.
Thus at the quantum level the $U_R(1)$ symmetry reduces to a 
discrete symmetry
\begin{eqnarray}
\lambda{\longrightarrow}e^{i\pi/4}\lambda;~~~ 
\phi{\longrightarrow}e^{i\pi/2}\phi;~~~
\psi{\longrightarrow}e^{i\pi/4}\psi.
\label{eq5p355}
\end{eqnarray}     
In the moduli space of the vacua, the scalar potential vanishes, but
$\langle \phi^a\rangle$ may be non-vanishing. 
This will break $SU(2)$ to $U(1)$, 
and as a consequence, the theory will be in the Coulomb phase. 
One can choose 
\begin{eqnarray}
\langle \phi^b\rangle =a\delta^{b3}
\label{eq5p356}
\end{eqnarray} 
by a gauge rotation with $a$ being a complex number. The parameter
labelling the moduli space will be the gauge invariant chiral superfield
\begin{eqnarray}
u=a^2=\mbox{Tr}\langle\phi^2\rangle 
 \label{eq5p358} 
\end{eqnarray}
rather than $a$. The discrete symmetry (\ref{eq5p355}) 
is realized on $u$ as a $Z_2$ symmetry,
\begin{eqnarray}
u{\longrightarrow}e^{i\pi}u=-u.
\label{eq5p359}
\end{eqnarray}
In the BPS limit all the particles including
the gauge bosons, monopoles and dyons have a universal
mass formula:
\begin{eqnarray}
m=\sqrt{2}|Z|, ~~~Z=a n_e+a_Dn_m,
 \label{eq5p360}
\end{eqnarray} 
where $n_e$ and $n_m$ are the electric and magnetic charges
carried by the particle. $n_e{\neq}0$, $n_m=0$ corresponds to
the usual electrically charged particles;
$n_m{\neq}0$, $n_e=0$ corresponds to the magnetic monopoles and 
$n_m{\neq}0$, $n_e{\neq}0$ to the dyons. In the weak-coupling region,
\begin{eqnarray}
a_D=\tau a,
\label{eq5p361e}
\end{eqnarray}
the subscript $D$ indicating the dual variable. 
Since there exists an  electric-magnetic duality in the Coulomb
phase, the duality transformation maps
\begin{eqnarray}
a \longleftrightarrow a_D,~~~~\tau (a)
\longleftrightarrow -\frac{1}{\tau(a)}
{\equiv}\tau_D,
\label{eq5p361}
\end{eqnarray}
which is a typical feature of electric-magnetic (S-)duality, i.e.
the strong and the weak couplings are exchanged, while 
the mass spectrum of the particles remains the same. Notice that
there exists another invariance of $\tau$,
\begin{eqnarray}
\tau\longrightarrow\tau +1,
\label{eq5p362}
\end{eqnarray}  
since this means, from Eq.\,(\ref{eq3p3}),  
the vacuum angle is shifted by $2\pi$ and hence
the generating functional remains unchanged.

The transformations (\ref{eq5p361}) and (\ref{eq5p362}) 
of $\tau$  can be implemented on $(a,a_D)^T$ by the matrices
\begin{eqnarray}
S=\left(\begin{array}{cc} 0 & 1 \\ -1 & 0 \end{array}\right),~~~
T=\left(\begin{array}{cc} 1 & 1 \\ 0 & 1 \end{array}\right).
\label{eq5p363}
\end{eqnarray}
These matrices generate the group $SL(2,Z)$, 
under which $\tau$ is transformed as
\begin{eqnarray}
\tau \longrightarrow 
\left(\begin{array}{cc} a & b \\  c & d \end{array}\right)
\tau =\frac{a\tau+b}{c\tau +d},~~~a,b,c,d\in Z;~~~
ad-bc{\neq}0.
\label{eq5p364}
\end{eqnarray}
One immediately recognizes that the invariance 
of $\tau$ is the same as the modular invariance of
one-loop amplitudes of strings \cite{ref515}. This fact gives 
a hint that $\tau$ can be interpreted as the modular parameter
of a certain torus. Seiberg and Witten proposed the following 
formula \cite{ref1p1, ref514}
\begin{eqnarray}
\tau=\frac{da_D}{da}.
\label{eq5p365}
\end{eqnarray} 
We will see that $a_D$ and $a$ can indeed be regarded as 
the two periods of a torus.

(\ref{eq5p340}) implies that when $a$ is large, 
the $SU(2)$ theory is weakly coupled. Hence 
$\tau$ can be determined perturbatively. Since the one-loop 
beta function coefficient $\beta_0=4$
for $N=1$ $SU(2)$ gauge theory with a triplet matter field, 
one can according to (\ref{eq5p340}) write down 
the holomorphic relation between $\tau$ and $u$ as 
\begin{eqnarray}
\tau (u)=\frac{i}{\pi}\ln\frac{u}{\Lambda^2}.
\label{eq5p366}
\end{eqnarray}
In the region of the moduli space corresponding to strong coupling, 
there is not only the one-loop perturbative contribution, 
but also the non-perturbative corrections from instantons. This makes
the $u$-dependence of $\tau$ quite complicated. Fortunately, as Seiberg
and Witten did \cite{ref1p1,ref1p1a}, one can use the geometric 
structure of the quantum moduli space to determine $\tau (u)$. 

At large $u$ $\tau (u)$ is a multi-valued function, 
\begin{eqnarray}
\tau (u)=\frac{i}{\pi}\ln\frac{u}{\Lambda^2}+i2k\pi, 
~~~~k=0,\pm 1,\pm 2, \cdots,
\label{eq5p367}
\end{eqnarray}
i.e. for each point $u$ in the moduli space, there exists an 
infinite number of $\tau (u)$. From complex analysis we 
know that to make the correspondence between $\tau (u)$ and $u$
one-to-one, we must cut the $u$-plane 
along a line connecting two branch points.
For example, both $u=0$ and $u=\infty$
are branch points of $\ln (u/\Lambda^2)$, we cut
the $u$-plane along the positive real axis. 
The moduli space divides into branches
and $\tau (u)$ will be a single-valued function on each branch. 
These branches can be glued together by identifying the lower 
lip of the previous branch with the upper lip of the next branch 
and the lower lip of the last branch with the upper lip of the 
first branch. The complex surface obtained in 
this way is called the Riemann surface of the moduli space, 
and $\tau (u)$ will be a single valued function on this Riemann surface. 

Seiberg and Witten found the exact solution of $\tau (u)$ by determining
the singularity structure of the above Riemann surface, and we shall
follow their reasoning. Later Flume et al showed that their solution
is unique assuming only that supersymmetry is unbroken and that the
number of singularities in $u$ is finite \cite{ref515a}. They started 
from the solution (\ref{eq5p366}) for $\tau (u)$ at large $u$. 
(\ref{eq5p365}) and (\ref{eq5p366}) imply that in the weak 
coupling limit, 
\begin{eqnarray}
a_D=\frac{2i}{\pi}a\ln\frac{a}{\Lambda}-\frac{2i}{\pi}a.
 \label{eq5p368}
\end{eqnarray}     
Since $u=\infty$ is a singular point of $\tau (u)$, moving along 
a closed path around
$u=\infty$\footnote{ It is a large loop around $u=0$, but
we consider the region outside the loop.} shifts $\ln u$ by
\begin{eqnarray}
\ln u\longrightarrow\ln u+2i\pi
\label{eq5p369}
\end{eqnarray}
and hence 
\begin{eqnarray}
\ln a\longrightarrow \ln a+i\pi
\label{eq5p370}
\end{eqnarray} 
due to $u=a^2$. Such a transformation of a complex function around 
a singularity is called a monodromy. From (\ref{eq5p368}) 
and (\ref{eq5p370}), we have
\begin{eqnarray}
a_D \longrightarrow -a_D+2a,~~~
a \longrightarrow -a.
\label{eq5p371}
\end{eqnarray} 
Therefore, there exists a non-trivial monodromy at infinity 
on the Riemann surface,
\begin{eqnarray}
M_{\infty}=\left(\begin{array}{cc} -1 & 2 \\ 0 & -1\end{array}\right)=
PT^{-2},
\label{eq5p372}
\end{eqnarray}
where $T$ is the generator of the $SL(2,Z)$ group given in (\ref{eq5p363})
and $P$ is the element $-1$ of $SL(2,Z)$. 

This non-trivial monodromy at $u=\infty$ implies that there must exist
other monodromies on the $u$-plane. Equivalently speaking,
$\tau (u)$ has a branch point at large $u$, 
requiring further branch point or points in the interior 
of the $u$-plane, corresponding to strong coupling of the theory, since now $u$ is 
finite and non-perturbative effects have already become noticeable.  

Now let us analyze these extra singularities. Due to the discrete symmetry
$u{\longrightarrow}-u$, singularities of the moduli space must
emerge in pairs, i.e. for each singularity at $u=u_0$, there must exist
another one at $u=-u_0$. If there are only two singularities, they must
be $u=\infty$ and $u=0$ since they are the only fixed points of the $Z_2$
symmetry. We have already seen that $u=\infty$ is a singularity 
and that it corresponds to
the weak coupling limit of theory. Since the monodromy around $0$ is the 
same as the monodromy around $\infty$, $M_0=M_\infty$, 
$u=a^2$ is not affected by any monodromy and hence
$u$ would be a global coordinate of the moduli space. Consequently, 
$\tau (u)$ would be an analytic function on the moduli space, and its
imaginary part $\mbox{Im}\tau (u)\sim 1/g^2_{\rm eff}$ would be a harmonic 
function due to the Cauchy-Riemann equation 
\begin{eqnarray}
\partial_z\partial_{\overline{z}}\mbox{Im}\tau (u)=0.
\label{eq5p373}
\end{eqnarray} 
Since the Laplacian $\partial_z\partial_{\overline{z}}$
 is a positive definite operator, (\ref{eq5p373}) means
that $\mbox{Im}\tau (u)$ cannot be positive definite 
everywhere on the moduli space.
Therefore, there must exist regions in the moduli space
where the low energy effective gauge coupling 
$g_{\rm eff}\sim 1/\sqrt{\mbox{Im}\tau}$ becomes imaginary. To avoid this
unphysical conclusion, there have to exist at least two singularities in the 
interior of the moduli space \cite{ref516}. 
So we can consider three singularities, i.e. $\infty$, $u_0$ and $-u_0$ 
for some $u_0{\neq}0$. $u=0$ will not be a singular point, although its
existence respects the $Z_2$ symmetry.
       
  What is the physical interpretation of the singularities at $u=\pm u_0$?
Classically, $u=0$ is a singular point since classically this means $a=0$,
or that the full gauge symmetry $SU(2)$ is restored and no 
Higgs mechanism occurs.
The massive gauge bosons and their 
superpartners become massless. However, there is no singularity
at $u=0$ in the quantum moduli space. The singularities at $u=\pm u_0$
do not imply that the gauge bosons become massless. This is because 
the theory at $u=\pm u_0$ is in the strong coupling region. The existence
of massless gauge bosons would imply an asymptotically conformal 
invariant theory in the infrared limit, but conformal invariance
implies that the dimensional parameter 
$u=\langle\mbox{Tr}\phi^2\rangle =0$. 
This is obviously inconsistent.
Therefore, the singularities at $u=\pm u_0$ do not 
correspond to massless gauge 
bosons. Seiberg and Witten found that the singularities at $u=\pm u_0$
in fact correspond to massless magnetic monopoles and dyons. According to
the BPS mass formula (\ref{eq5p360}) \cite{ref1p1}, the mass of the 
magnetic monopole is
\begin{eqnarray}
m^2=2|a_D|^2.
\label{eq5p374}
\end{eqnarray}
Thus a massless magnetic monopole will give $a_D=0$. If we choose $a_D$ 
to vanish at $u_0$, i.e. that the magnetic monopoles become massless there,
we will see that at the singularity $u=-u_0$ 
the dyon becomes massless.
From the Montonen-Olive duality conjecture, the magnetic monopole 
hypermultiplet couples to the dual fields
of the original $N=2$ photon supermultiplet, and the dynamics 
is exactly the same as that of the $N=2$ supersymmetric QED with
massless electrons. The magnetic coupling is weak due to the 
electric-magnetic duality and hence the perturbative method can be adopted.
From the one-loop beta function of the $N=2$ supersymmetric Abelian gauge 
theory with a massless hypermultiplet, the magnetic coupling should run 
according to
\begin{eqnarray}
\mu \frac{d}{d\mu}g_D=\frac{g_D^3}{8\pi}.
\label{eq5p375}
\end{eqnarray}       
Since the scale $\mu$ is proportional to 
$a_D$ and $4\pi i/[g^2_D(a_D)]$ is
$\tau_D$ when $\theta_D=0$\footnote{Note that supersymmetric QED, 
unlike supersymmetric QCD, does not allow a non-vanishing vacuum 
angle, except if it is embedded into a larger gauge group.}, 
then near $u=u_0$ (or near $a_D=0$) we have
\begin{eqnarray}
a_D\frac{d}{da_D}\tau_D = -\frac{i}{\pi};
~~~\tau_D&=&-\frac{i}{\pi}\ln a_D.
\label{eq5p376}
\end{eqnarray}       
(\ref{eq5p361}) and (\ref{eq5p365}) lead to
\begin{eqnarray}
\tau_D=-\frac{da}{da_D},
\label{eq5p377}
\end{eqnarray}
which can be integrated to give $a$ near $u=u_0$,
\begin{eqnarray}
a=a_0+\frac{i}{\pi}a_D\ln a_D-\frac{i}{\pi}a_D.
\label{eq5p378}
\end{eqnarray} 
Since $a_D(u_0)$ vanishes, in the vicinity of $u_0$, $a_D$ should be a 
good coordinate like $u$ and depend linearly on $u$, so we obtain
\begin{eqnarray}
&&a_D=c_0 (u-u_0);\nonumber\\
&&a=a_0+\frac{i}{\pi}c_0 (u-u_0)\ln (u-u_0)-\frac{i}{\pi}c_0\ln (u-u_0).
\label{eq5p379}
\end{eqnarray}   
From these expressions, we immediately read off the monodromy matrix as 
$u$ turns around $u_0$ counterclockwise, $u-u_0{\longrightarrow}
e^{i2\pi}(u-u_0)$,
\begin{eqnarray}
&&\left(\begin{array}{c} a_D \\a\end{array} \right)
\longrightarrow \left(\begin{array}{c} a_D \\a-2a_D\end{array} \right)
=M_{u_0}\left(\begin{array}{c} a_D \\a\end{array} \right),\nonumber\\
&& M_{u_0}=\left(\begin{array}{cc} 1 & 0\\-2 & 1\end{array} \right)
=ST^2S^{-1}.
\label{eq5p380}
\end{eqnarray}   
The monodromy matrix at $u=-u_0$ is easy to find. Since there
are only three singularities, the contour around $u=\infty$ can 
be deformed
into a contour encircling $u_0$ and a contour encircling $-u_0$,
both being counterclockwise and having same base point. 
The factorization of the monodromy matrices gives
\begin{eqnarray}
M_{\infty}=M_{u_0}M_{-u_0}.
\label{eq5p381}
\end{eqnarray} 
and hence 
\begin{eqnarray}
M_{-u_0}=\left(\begin{array}{cc} -1 & 2\\-2 & 3\end{array} \right)
=(TS)T^2(TS)^{-1}.
\label{eq5p382}
\end{eqnarray}
The physical interpretation of the singularity at $u=-u_0$ can be found
from the way the BPS mass formula (\ref{eq5p360}) transforms
under the action of the monodromy matrix. We write
the central charge $Z$ as
\begin{eqnarray}
Z=(n_m, n_e)\left(\begin{array}{c} a_D\\ a\end{array} \right).
\label{eq5p383}
\end{eqnarray}
The monodromy transformation 
\begin{eqnarray}
\left(\begin{array}{c}a_D \\ a\end{array} \right ){\longrightarrow}
M\left(\begin{array}{c}a_D \\ a\end{array} \right )
\label{eq5p384}
\end{eqnarray}
can be equivalently thought of as changing the magnetic and electric 
quantum numbers as
\begin{eqnarray}
(n_m,n_e){\longrightarrow}(n_m,n_e)M.
\label{eq5p385}
\end{eqnarray}
The state with vanishing mass should be invariant under the monodromy 
and hence should be a left eigenvector of $M$ with unit eigenvalue.
For the singularity $u_0$, we have assumed that it corresponds to
massless monopoles, so the monopole $(1,0)$ should be a left
eigenvector of $M_{u_0}$ with unit eigenvalue. It is easy to check
that the left eigenvector of $M_{-u_0}$ with unit eigenvalue is
$(n_m,n_e)=(1,-1)$. This is a dyon since both the electric and 
magnetic charges do not vanish. Thus the singularity at $-u_0$ can be
interpreted as being due to a $(1,-1)$ dyon becoming massless.  
The general monodromy matrix corresponding to  
a massless dyon $(n_m,n_e)$ is 
\begin{eqnarray}
M(n_m,n_e)=\left(\begin{array}{cc} 1+2n_m n_e & 2n_e^2\\-2n_m^2 &
1-2n_m n_e  \end{array} \right).
\label{eq5p386}
\end{eqnarray}

 One special point should be emphasized. Since 
$M_{u_0}M_{-u_0}{\neq}M_{-u_0}M_{u_0}$ the monodromy 
relation (\ref{eq5p381})
seems not to be invariant under $u_0{\longrightarrow}-u_0$. This
does not contradict the $Z_2$ symmetry, since we have not
indicated the base point in defining the composition of two 
monodromies. Assume the composition $M_{u_0}M_{-u_0}$ happens in 
the base point $u=P$, then another choice of base point 
$u=-P$ will lead to
\begin{eqnarray}
M_{\infty}=M_{-u_0}M_{u_0}.
\label{eq5p388}
\end{eqnarray}  
Then from (\ref{eq5p372}), (\ref{eq5p380}) and 
(\ref{eq5p388}) we get the monodromy
\begin{eqnarray}
M_{-u_0}=\left(\begin{array}{cc} 3 & 2\\-2 & -1\end{array} \right),
\label{eq5p389}
\end{eqnarray}
one of whose left eigenvectors with unit eigenvalue is the dyon (1,1).
The $Z_2$ symmetry on the quantum moduli space not only exchanges
the singularity, but also exchanges the base point $P$, hence
\begin{eqnarray}
M_{u_0}M_{-u_0}{\longleftrightarrow}M_{-u_0}M_{u_0}.
\label{eq5p390}
\end{eqnarray}
At the same time the $(1,-1)$ dyon is exchanged with the $(1,1)$ 
dyon. 

Using the monodromy corresponding to the fundamental monopole 
or dyon, one can construct the monodromy corresponding to the 
composite dyon \cite{ref1p1,ref1p1a}.
For example, turning first $n_e$ times around $\infty$, then around
$-u_0$, and then $n_e$ times around $\infty$ in the opposite direction,
we obtain the monodromy
\begin{eqnarray}
M_{\infty}^{-n_e}M_{-u_0}M_{\infty}^{n_e}
&=&T^{-2n_e}(TS)T^2(TS)^{-1}T^{2n_e} \nonumber\\
&=& \left(\begin{array}{cc} -1-4n_e & 2+8n_e+8n_e^2 \\
                         -2 & 3+4n_e  \end{array}\right)
=M(1,-1-2n_e),
\label{eq5p391}
\end{eqnarray}    
which corresponds to a massless dyon with $n_m=1$ and any $n_e\in Z$.
Similarly,  we have
\begin{eqnarray}
M_{\infty}^{-n_e}M_{u_0}M_{\infty}^{n_e}
&=&T^{-2n_e}ST^2S^{-1}T^{2n_e}\nonumber\\
&=& \left(\begin{array}{cc} 1-4n_e & 8n_e^2 \\
                         -2 & 1+4n_e  \end{array}\right)
=M (1,-1-2n_e).
\label{eq5p392}
\end{eqnarray}   

We know that $a_D(u)$ and $a(u)$ are multiple valued functions
on the moduli space, and they are single-valued function 
on each branch with the branch points being at 
$\infty$, $u_0$ and $-u_0$, or equivalently,
single valued on the Riemann surface of the moduli space.
Since the value of $u_0$ depends on the choice of renormalization scheme,
one can always choose $u_0=1$. There are two approaches to
determining $a(u)$ and $a_D(u)$ and hence $\tau$ 
exactly by (\ref{eq5p365}). The first one 
is to look for a differential equation satisfied by $a (u)$ and $a_D(u)$,
since meromorphic functions with such singularities should satisfy
the characteristic equations of the hypergeometric function, which is
called the Picard-Fuchs equations \cite{ref517}. 
This method is straightforward and one can easily write
the explicit forms of $a (u)$ and $a_D(u)$ in terms of the integral 
representation of the hypergeometric function \cite{ref1p1},
\begin{eqnarray}
&& a_D(u)=\frac{\sqrt{2}}{\pi}\int^u_1 dx\frac{\sqrt{x-u}}{\sqrt{x^2-1}};
\nonumber\\
&& a(u)=\frac{\sqrt{2}}{\pi}\int^1_{-1} dx\frac{\sqrt{x-u}}{\sqrt{x^2-1}}.
\label{eq5p393}
\end{eqnarray}
Consequently, the low energy effective coupling is given by
\begin{eqnarray}
\tau (u)=\frac{da_D(u)/du}{da(u)/du}.
\label{eq5p394}
\end{eqnarray} 

However, Seiberg and Witten gave an indirect but 
theoretically more beautiful expression for above solutions
 --- the algebraic curve of the Riemann surface of the moduli 
space \cite{ref1p1,ref1p1a}. 
They made use of the
geometric structure of the Riemann surface of 
the moduli space in constructing the solutions. 
Here we shall only use the explicit expression (\ref{eq5p393}) for 
the solution to understand the Seiberg-Witten construction.
 From (\ref{eq5p393}), we see that
the integrand has square-root branch cuts 
with branch points at $+1$, $-1$,
$u$ and $\infty$. The two branch cuts can be chosen 
from $-1$ to $+1$ and
from $u$ to $\infty$. The Riemann surface of the integrand
is composed of two sheets glued along the cuts. If 
one adds the point 
at infinity to each of the two sheets, the topology of 
the Riemann surface 
is that of two spheres connected by two tubes, this is just a torus. 
How can we describe a torus, or more generally,
 an arbitrary Riemann surface quantitatively?
We know that the torus is formed by the 
compactification of a complex plane, identifying
\begin{eqnarray}
z\longrightarrow z+\omega_1, ~~~z\longrightarrow z+\omega_2,
\label{eq5p395}
\end{eqnarray}
with $\omega_2/\omega_1=\tau$ and $\mbox{Im}\tau >0$. 
$\omega_1$ and $\omega_2$ 
are called the periods of the torus. $\tau$ is 
the modular parameter of the torus, which
determines the geometric structure of the torus. 
Two tori with modular parameters
 $\tau^{\prime}$ and $\tau$ related by a $SL(2,Z)$ transformation
\begin{eqnarray}
\tau^{\prime}=\frac{a\tau  +b}{c\tau +d}, ~~~ad-bc=1, ~~a,b,c,d{\in}Z.
\label{eq5p396}
\end{eqnarray}
have the same geometric shape (i.e. are conformally 
equivalent) \cite{ref517}. (\ref{eq5p395}) shows that the torus
is in fact the coset space $C/G$ of the complex plane $C$, with 
the group $G$ consisting of the action on $C$
\begin{eqnarray}
z\longrightarrow z+n+m\tau, ~~~\tau\in C,~~~\mbox{Im}\tau >0.
\label{eq5p397}
\end{eqnarray}
A natural choice to describe a Riemann surface  
is to use the meromorphic functions defined on it. 
For a torus, a meromorphic function should be
elliptic (doubly periodic) functions with periods $1$ and $\tau$
due to (\ref{eq5p397}).
A typical example is the Weierstrass elliptic function with 
periods $1$ and $\tau$ \cite{ref518,ref519}:
\begin{eqnarray}
\xi (z)=\frac{1}{z^2}+\sum_{n,m}\left[\frac{1}{(z-n-m\tau)^2}
-\frac{1}{(n+m\tau)^2}\right], ~~~n,m \in Z, ~~n, m{\neq}0.
 \label{eq5p398}
\end{eqnarray}
The function $\xi$ satisfies the following differential equation
\begin{eqnarray}
\xi^{\prime 2}(z)=4(\xi -e_1)(\xi -e_2)(\xi -e_3),
\label{eq5p399}
\end{eqnarray}
with
\begin{eqnarray}
e_1=\xi\left(\frac{1}{2}\right), ~~~e_2=\xi\left(\frac{\tau}{2}\right),
~~~e_3=\xi\left(\frac{1+\tau}{2}\right).
\label{eq5p3100}
\end{eqnarray}
$\xi^{\prime}(z)$ is again an elliptic function and hence is 
another meromorphic function defined on the torus. 
If we define $\xi^{\prime}(z){\equiv}y$,
$x{\equiv}\xi (z)$, the differential equation (\ref{eq5p399}) 
can be written as
\begin{eqnarray}
y^2=4(x-e_1)(x-e_2) (x-e_3).
\label{eq5p3101}
\end{eqnarray}
This is a plane cubic curve with the plane 
coordinates $(x,y)$ being
meromorphic functions \cite{ref519}. 
In fact, every Riemann surface can be equivalently 
represented by such an algebraic curve. The  domain of definition
of an algebraic curve is a Riemann surface. Conversely, given
an algebraic curve, one can immediately know what 
the Riemann surface is. The
zero points of $y$ are just the singular points of the Riemann surface.
Therefore, a Riemann surface and its algebraic
curve are in one-to-one correspondence. For example, 
the plane cubic curve (\ref{eq5p3101}) shows that $y$ has 
three zero points $x=e_i$, $i=1,2,3$,  
in the $x$-space. From (\ref{eq5p3101})
\begin{eqnarray}
y=\pm \sqrt{4(x-e_1)(x-e_2) (x-e_3)},
\label{eq5p3102}
\end{eqnarray}
This shows that $y$ is a double-valued function on $x$-space. The 
branch points are obviously $x=e_i, \infty$. 
We choose one branch cut from $e_1$ to $e_2$ and 
another one from $e_3$ to $\infty$.
$y$ is then a single valued function on the Riemann 
surface defined by joining the two sheets of the $x$-plane along the cuts. 
It is now easy to see that the Riemann surface is a torus. 
A torus has two independent non-trivial closed paths called
cycles. The loop on the two-sheeted covering of the $x$-plane 
that goes around one of the two cuts
corresponds to one of the cycles 
and the loop which intersects both cuts, 
i.e. the loop goes around $e_2$ and $e_3$ pairing into
the second sheet for half of the way, 
corresponds to the other cycle. 

 For the theory we are discussing,  
the singularities are at $1$, $-1$ and $\infty$
in the $u$-plane, and the corresponding Riemann surface is a torus,
which, according to (\ref{eq5p3101}), should 
be parametrized by a family of curves with the parameter $u$,
\begin{eqnarray}
y^2=(x-1)(x+1) (x-u).
\label{eq5p3103}
\end{eqnarray}
The periods $a_D(u)$ and $a(u)$ of the torus represented by this 
algebraic curve and hence $\tau (u)$, as argued by Seiberg 
and Witten, can be found as follows \cite{ref1p1,ref514}. 
First, a torus is represented 
by two cycles, which we denote as $\gamma_1$ and 
$\gamma_2$. These cycles form a local canonical basis for 
the first homology group $H_{1,0}(T,C)$ of the  torus, or equivalently, 
the first homology group of the curve, $T$ denoting the torus or the
algebraic curve family. According to Stokes' theorem, one can 
construct a homotopy invariant by pairing an element in the 
first homology group with an element in its dual, the first 
cohomology group. Since $a_D(u)$ and $a(u)$ should be homotopic 
invariant, we define 
\begin{eqnarray}
a_D=\oint_{\gamma_1}\lambda,~~~a=\oint_{\gamma_2}\lambda,
\label{eq5p3104}
\end{eqnarray}
where $\lambda$ is an element of the first cohomology group 
$H^{1,0}(T,C)$. As $H^{1,0}(T,C)$ is two-dimensional, its 
basis must be provided by any two linearly independent 
elements. One typical choice is
\begin{eqnarray}
\lambda_1=\frac{dx}{y}, ~~~~\lambda_2=\frac{xdx}{y}.
\label{eq5p3105}
\end{eqnarray}
In general, $\lambda$ should be a linear combination of
the above basis elements with $u$-dependent coefficients,
\begin{eqnarray}
\lambda =a_1 (u)\lambda_1+a_2 (u)\lambda_2.
\label{eq5p3106}
\end{eqnarray}
In addition, $\lambda$ must be a form  with vanishing residue 
so that, on encircling a singularity, $a_D$ and $a$ transform 
in the way that $\gamma_1$ and 
$\gamma_2$ transform under a subgroup of $SL(2,Z)$. Especially, 
the physical requirement $\mbox{Im}\tau >0$ must be satisfied.
(\ref{eq5p394}) and (\ref{eq5p3104}) give
\begin{eqnarray}
\tau =\frac{da_D/du}{da/du}, ~~~\frac{da_D}{du}=\oint_{\gamma_1}
\frac{d\lambda}{du}, ~~~\frac{da}{du}=\oint_{\gamma_2}
\frac{d\lambda}{du}.
\label{eq5p3107}
\end{eqnarray}  
On the other hand, on a torus defined by the above
curve there exists a natural definition for the periods,
\begin{eqnarray}
b_i=\oint_{\gamma_i}\lambda_1, ~~~i=1,2,\label{eq5p3108}
\end{eqnarray} 
and the modulus parameter
\begin{eqnarray}
\tau_u=\frac{b_1}{b_2}
\label{eq5p3109}
\end{eqnarray} 
should possess the fundamental property $\mbox{Im}\tau_u >0$. 
Therefore, one can choose
\begin{eqnarray}
\frac{d\lambda}{du}=f(u)\lambda_1=f(u)\frac{dx}{y}
\label{eq5p3110}
\end{eqnarray}
with $f(u)$ a function of $u$ only. Then we get
\begin{eqnarray}
\tau=\frac{da_D/du}{da/du}=\frac{b_1}{b_2}=\tau_u=
\frac{\oint_{\gamma_1}dx/y}{\oint_{\gamma_2}dx/y},
\label{eq5p3111}
\end{eqnarray}
which naturally satisfies $\mbox{Im}\tau >0$. The reverse is also
true, i.e. if $\mbox{Im}\tau >0$ everywhere, then $\tau =\tau_u$
and $\lambda$ satisfy (\ref{eq5p3110})
and (\ref{eq5p3111}). The function $f(u)$ can be determined by 
demanding the asymptotic behaviours (\ref{eq5p368}), (\ref{eq5p379}) 
of $a_D$ and $a$ near $u=1$, $-1$, $\infty$,
\begin{eqnarray}
f(u)=-\frac{\sqrt{2}}{4\pi}.
\label{eq5p3112}
\end{eqnarray}
Therefore, with (\ref{eq5p3103}) and (\ref{eq5p3112}), 
one can integrate (\ref{eq5p3110}) over $u$ and get
\begin{eqnarray}
\lambda =\frac{\sqrt{2}}{2\pi}\frac{\sqrt{x-u}}{\sqrt{x^2-1}}dx=
\frac{\sqrt{2}}{2\pi}\frac{y}{x^2-1}dx.
\label{eq5p3113}
\end{eqnarray}  
We finally have from (\ref{eq5p3107}) 
\begin{eqnarray}
a_D=\oint_{\gamma_1}\frac{\sqrt{2}}{2\pi}
\frac{\sqrt{x-u}}{\sqrt{x^2-1}}dx,
~~~a=\oint_{\gamma_2}\frac{\sqrt{2}}{2\pi}
\frac{\sqrt{x-u}}{\sqrt{x^2-1}}dx.
\label{eq5p3114}
\end{eqnarray}  
Deforming the cycles $\gamma_1$ and $\gamma_2$ continuously 
into the branch cuts, we immediately get (\ref{eq5p393}), 
the same result as obtained from the differential equation approach. 

Since the Riemann surface and the algebraic curve are exactly
in one-to-one correspondence, and given a Riemann surface, its
modulus $\tau$ is fixed up to an $SL(2,Z)$ transformation, one can 
conveniently use the algebraic curve to represent the solution
of the low energy effective theory in the Coulomb phase.

The Seiberg-Witten method can be naturally generalized to the
case with matter fields, i.e. $N=2$ supersymmetric QCD \cite{ref1p1a}.
The matter fields belong to $N=2$ hypermultiplets, which are
pairs of $N=1$ chiral supermultiplets ($Q_i$, $\widetilde{Q}_i$)
in the fundamental representation of the gauge group and its conjugate 
($i=1,\cdots, N_f$ is the flavour index) \cite{so,ref521}. 
In $N=1$ language, the classical Lagrangian 
consists of the standard coupling of the $N=2$ gauge 
supermultiplet to $Q_i$, $\tilde{Q}_i$,
plus the following superpotential
\begin{eqnarray}
W=\int d^2\theta \left[\sqrt{2}\sum_{i=1}^{N_f}\widetilde{Q}^i\Phi^aT^aQ_i
+\sum_{i=1}^{N_f}m_i\widetilde{Q}^iQ_i\right],
\label{eq5p3115}
\end{eqnarray}
$\Phi$ is the $N=1$ chiral supermultiplet part of the $N=2$ gauge 
supermultiplet, since an $N=2$ gauge supermultiplets 
is composed of an $N=1$ gauge supermultiplet
(gauge fields and gaugino) and $N=1$ chiral supermultiplet in the 
adjoint representation. The classical moduli space of
the vacua will be determined by both $D$-flatness and $F$-flatness
conditions. There are two types of possible solutions 
to these conditions \cite{ref1p1a}.
The first type leads to a Coulomb branch of the moduli space
with the expectation values of 
the scalar components of $\Phi$, $Q$ and $\tilde{Q}$ satisfying 
\begin{eqnarray}
\langle \Phi \rangle {\neq}0,~~~\langle Q_i \rangle 
=\langle \tilde{Q}_i^{\dagger}  \rangle  =0.
\label{eq5p3116}
\end{eqnarray}
The second one gives a Higgs branch in
the moduli space in the massless case ($m_i=0$) with
\footnote{For $N_f=0,1$, the moduli space has no Higgs
branch. When $N_f=2$, there are two Higgs branches in the moduli space,
which coincide with the Coulomb branch at the origin
of the moduli space. These two branches are exchanged by a
parity-like symmetry. For $N_f{\geq}3$, there is only
one Higgs branch, which meets the Coulomb branch at the
origin of the moduli space \cite{ref1p1a}.}  
\begin{eqnarray}
\langle \Phi \rangle =0,~~\langle Q_i \rangle 
=\langle \tilde{Q}_i^{\dagger}\rangle{\neq}0, ~~i=1,{\cdots},k{\leq}N_f,
~N_f{\geq}2.
\label{eq5p3117}
\end{eqnarray}
We are only interested in the Coulomb branch. 
To ensure that $(n_m,n_e)$ are both integers even in the presence
of matter fields, one should rescale the electric charge 
$n_e$ by a factor $2$ and simultaneously divide $a$ by $2$ so that the
BPS mass spectrum $m^2=2|a_Dn_m+an_e|^2$ remains the same.
This rescaling can be thought of as a transformation
\begin{eqnarray}
\left(\begin{array}{c} a_D\\ a\end{array}\right)
\longrightarrow \left(\begin{array}{c} a_D\\ a/2\end{array}\right)
=\left(\begin{array}{cc} 1 & 0\\ 0 & 1/2 \end{array}\right)
\left(\begin{array}{c} a_D\\ a\end{array}\right)
{\equiv}S\left(\begin{array}{c} a_D\\ a\end{array}\right).
\label{eq5p3118}
\end{eqnarray}
Consequently, the monodromy matrix will transform as
\begin{eqnarray}
\left(\begin{array}{cc} m & n\\ p & q \end{array}\right)
\longrightarrow S\left(\begin{array}{cc} m & n\\ p & q \end{array}\right)
S^{-1}&=&\left(\begin{array}{cc} 1 & 0\\ 0 & 1/2 \end{array}\right)
\left(\begin{array}{cc} m & n\\ p & q \end{array}\right)
\left(\begin{array}{cc} 1 & 0\\ 0 & 2 \end{array}\right)\nonumber\\
&=&\left(\begin{array}{cc} m & 2n\\ p/2 & q \end{array}\right).
\label{eq5p3119}
\end{eqnarray}
Therefore, in the new convention the monodromy matrices 
(\ref{eq5p372}), (\ref{eq5p380}) and (\ref{eq5p382}) with 
$N_f=0$ are
\begin{eqnarray}
M_{u_0}=\left(\begin{array}{cc} 1 & 0\\ -1 & 1 \end{array}\right);
~~M_{-u_0}=\left(\begin{array}{cc} -1 & 4\\ -1 & 3 \end{array}\right);
~~M_{\infty}=\left(\begin{array}{cc} -1 & 4\\ 0 & -1 \end{array}\right).
\label{eq5p3120}
\end{eqnarray}
With these new monodromy matrices
the algebraic curve corresponding to (\ref{eq5p3103})
becomes \cite{ref1p1a,ref1p9}
\begin{eqnarray}
y^2=x^3-ux^2+\frac{1}{4}\Lambda^4x,
\label{eq5p3121}
\end{eqnarray}  
where $u_0=\Lambda^2$ was chosen. 
The curve (\ref{eq5p3121}) describes the same physics as 
the curve (\ref{eq5p3103}). However
this curve is more general since it can be applied to
the case of matter fields. Thus, one usually
adopts the curve solution of the form (\ref{eq5p3121}) even in 
the case $N_f=0$. The branch points of the moduli space 
are the zeros of the plane cubic curve (\ref{eq5p3121})
\begin{eqnarray}
x=0, ~~~x=x_{\pm}=\frac{1}{2}\left(u-\sqrt{u^2-\Lambda^4}\right),
  \label{eq5p3122}
\end{eqnarray}  
together with the point at infinity. 
The natural choice for the two branch cuts
is $(x_-,x_+)$ and $(0,\infty)$, and the Riemann surface of course
remains a torus. When two branch points coincide, one cycle of the torus
will vanish and hence the torus will become singular. This 
can be related to the singular points in the $u$-plane, where 
there will exist massless particles. 
Two solutions of a cubic equation can coincide 
only when the discriminant $\Delta$ vanishes. 
For a general cubic equation
\begin{eqnarray}
x^3+Bx^2+Cx+D=0,
\label{eq5p3123}
\end{eqnarray}  
the discriminant is
\begin{eqnarray}
\Delta 
=-\frac{1}{108}\left(B^2C^2-4C^3-4B^4D+18BCD-27D^2\right)
\label{eq5p3124}
\end{eqnarray}
The discriminant for the algebraic curve (\ref{eq5p3121}) is hence 
\begin{eqnarray}
\Delta (u)=\frac{1}{16} (u^2-\Lambda^4)\Lambda^8.
\label{eq5p3125}
\end{eqnarray}  
Therefore, the torus will become singular at $u=\pm \Lambda^2$,
the zeros of $\Delta (u)$. Massless monopoles and dyons will exist 
in these two singular points.
 
In the Coulomb phase Seiberg and Witten worked out
the explicit algebraic curve solutions for $N_f=1,2,3$ using
a similar method as in the $N_f=0$ case \cite{ref1p1a}. 

 In summary, in the Coulomb phase, we can get insight into the dynamics 
by determining the effective gauge coupling $\tau (u)$ with
$u=\langle\mbox{Tr}\phi^2\rangle$ being the coordinate of 
the moduli space, while
$\tau (u)$ is indirectly given by a family of elliptic curves 
parametrized by $u$. From the algebraic curve, one can 
determine the Riemann surface of the moduli space
and its periods $a_D(u)$ and $a (u)$, and hence the $\tau (u)$
through (\ref{eq5p394}). In particular, 
the singular points in the moduli space,
which will lead to a singular Riemann surface,  
correspond to the zeros of the  discriminant of the vanishing 
elliptic curve. There exist massless particles in these singular points
of the moduli space.

\vspace{4mm}

 Now we go back to the theory at hand. In the Coulomb phase 
the effective gauge coupling at large $\widehat{U}$ is given 
by the curve (\ref{eq5p3121}),
\begin{eqnarray}
y^2=x^3-x^2\widetilde{U}+\frac{1}{4}\Lambda^4_d x,
\label{eq5p3126}
\end{eqnarray}
in which $\widetilde{U}$ is the light field
\begin{eqnarray}
\widetilde{U}=\langle\mbox{Tr}\phi^2\rangle =\frac{2\widehat{U}}{M_{11}},
\label{eq5p3127}
\end{eqnarray}
where we have used (\ref{eq5p350}), and the factor $2$ comes 
from taking the trace. Considering the relation (\ref{eq5p3127}) 
and rescaling
$y{\longrightarrow}\left({M_{11}}/{2}\right)^{3/2}y$,
$x {\longrightarrow}\left(M_{11}/2\right)x$,
we write the curve at large $U$ as
\begin{eqnarray}
y^2=x^3-x^2\widehat{U}+x \Lambda_X^4\Lambda_Y^4.
\label{eq5p3129}
\end{eqnarray}
Using this asymptotic solution, we can find the exact curve solution. 
We first assume that the exact solution has the general 
form of a plane cubic curve,
\begin{eqnarray}
y^2=x^3+\alpha x^2+\beta x+\gamma.
\label{eq5p3130}
\end{eqnarray}
The coefficients $\alpha$, $\beta$ and $\gamma$ are functions of 
$U$, the gauge coupling and the scales $\Lambda_X$, $\Lambda_Y$. 
There exist some important constraints on 
them \cite{ref1p1a,ref1p9}:
\begin{enumerate}
\item In the weak coupling limit $\Lambda=0$, 
the curve should give a singular Riemann surface for 
every $U$, i.e. the curve should vanish.
So generally one can assume that the curve 
should be of the form  
\begin{eqnarray}
y^2_0=x^2 (x-U), \mbox{when} ~\Lambda=0.
\label{eq5p3131}
\end{eqnarray}  
\item $\alpha$, $\beta$ and $\gamma$ should be holomorphic functions 
in $U$ and the various coupling constants,  since this 
guarantees that $\tau$ is also holomorphic in them.
\item The solution expressed by the curve (\ref{eq5p3129}) 
should be compatible with all the global symmetries of the 
theory including the discrete symmetry
and those explicitly broken by the anomaly.
\item In various limits we should
get the curves of the models obtained in the corresponding limit.
\item The curve should have the correct monodromies around the 
singular points.
\end{enumerate}   
 
 We use these constraints to determine the coefficients. The global
flavour symmetry $SU(2)_f$ and the discrete symmetry $Z_2$ as well 
as the large $U$ limit (\ref{eq5p3129}) show that the coefficient 
$\alpha$ must be \cite{ref1p9}
\begin{eqnarray}
\alpha=-\widehat{U}+\delta (\Lambda_X^4+\Lambda_Y^4)
\label{eq5p3132}
\end{eqnarray}  
with $\delta$ being some constant. In addition, the asymptotic form
at large $U$ gives
\begin{eqnarray}
\beta=\Lambda_X^4 \Lambda_Y^4, ~~~~\gamma=0.
\label{eq5p3133}
\end{eqnarray}  
This choice also ensures that the curve is singular when either 
$\Lambda_X$ or $\Lambda_Y$ vanishes.
So the exact curve solution should be
\begin{eqnarray}
y^2=x^3+\left[-U+\delta (\Lambda_X^4+\Lambda_Y^4)\right]x^2
+\Lambda_X^4 \Lambda_Y^4 x.
\label{eq5p3134}
\end{eqnarray} 
To determine the parameter $\delta$, we consider the limit
$\Lambda_Y{\gg}\Lambda_X$. In this limit the theory is
approximately an $SU(2)_X$ gauge theory with three singlets 
$M_{fg}$, $f,g=1,2$ and a triplet field $\widetilde{\phi}$. 
This model is just the $N_f=N_c=2$ case of the low energy  
supersymmetric $SU(N_c)$ QCD with $N_f$ flavours. From the 
discussion in Sect.\,\ref{subsub6.3.6}, the quantum moduli 
space is parametrized by $V$
with the constraint (\ref{eq318}).
So here we  have
\begin{eqnarray}
\det M_{fg}+\frac{1}{2}\mu^2 \mbox{Tr}(\widetilde{\phi}^2)
=\widehat{U}+\mu^2\widetilde{U}=\Lambda^4_Y,
\label{eq5p3136}
\end{eqnarray}  
where $\mu$ is a dimensional normalization necessary for 
making the dimensions right. The low energy Coulomb phase 
of $SU(2)_X$ with a triplet $\widetilde{\phi}$
is just the Seiberg-Witten model, whose
exact solution is given by the curve (\ref{eq5p3126}). 
Its branch points (\ref{eq5p3122}) and the discriminant
(\ref{eq5p3125}) show that the solution is singular 
at $\widetilde{U}=\pm \Lambda_X^2$ since two branch 
points coincide. Therefore, for $SU(2)_X$ with the matter fields
containing not only $\widetilde{\phi}$, but 
also $M_{fg}$, and considering (\ref{eq5p3126}),
the $\tau$ of this $SU(2)_X$ theory should be singular at
\begin{eqnarray}
\widehat{U}{\approx}\Lambda_Y^4\pm \mu^2\Lambda_X^2
\label{eq5p3137}
\end{eqnarray} 
in the $\Lambda_Y{\gg}\Lambda_X$ limit. On the other hand, the
 discriminant of the curve (\ref{eq5p3134}) is
\begin{eqnarray}
\Delta =-\frac{1}{108}(\Lambda_X\Lambda_Y)^8\left\{\left[-\widehat{U}+\delta
\left(\Lambda_X^4+\Lambda_Y^4\right)\right]^2-4
\Lambda_X^4\Lambda_Y^4\right\},
\label{eq5p3138}
\end{eqnarray}
and hence the curve is singular at 
\begin{eqnarray}
\widehat{U}=\delta\left(\Lambda_X^4+\Lambda_Y^4\right)
\pm 2\Lambda_X^2\Lambda_Y^2.
\label{eq5p3139}
\end{eqnarray}
Comparing (\ref{eq5p3137}) with (\ref{eq5p3138}), we find that 
$\delta=1$. So finally
we obtain the solution in the large $U$ limit,
\begin{eqnarray}
y^2=x^3+\left(-\widehat{U}+\Lambda_X^4
+\Lambda_Y^4\right)x^2+\Lambda_X^4\Lambda_Y^4x
\label{eq5p3140}
\end{eqnarray}   
Usually, for convenience of discussion, we rescale
$\Lambda_s\rightarrow\sqrt{2}\Lambda_s$, $s=X,Y$. Then the
curve (\ref{eq5p3140}) is rewritten as
\begin{eqnarray}
y^2=x^3+ \left(-\widehat{U}+4\Lambda_X^4+4\Lambda_Y^4\right) x^2
+16\Lambda_X^4\Lambda_Y^4x
\label{eq5p3141}
\end{eqnarray}
Expressing (\ref{eq5p3141}) in terms of the original 
high energy scale (\ref{eq5p346}), using (\ref{eq5p347}) and rescaling
\begin{eqnarray}
y{\longrightarrow}U_H^{3/2}y,~~~x{\longrightarrow}U_H x,
\label{eq5p3142}
\end{eqnarray}
we have
\begin{eqnarray}
y^2=x^3+\left(-U+8\Lambda^{2N_c-4}_{N_c,N_c-2}\right)x^2+
16\Lambda^{4N_c-8}_{N_c,N_c-2}x,
\label{eq5p3143}
\end{eqnarray}
where the relation (\ref{eq5p346}) was used.
It should be stressed that (\ref{eq5p3143}) 
is the curve where $U$ is large enough
to compare with the scale $\Lambda^{2N_c-4}_{N_c,N_c-2}$. The exact
curve solution should reproduce it in the 
large $U$ limit. As Seiberg and Witten
did \cite{ref1p1,ref1p1a}, assuming that the quantum corrections 
to (\ref{eq5p3143}) are polynomials
in the instanton factor $\Lambda^{2N_c-4}_{N_c,N_c-2}$, as implied
by (\ref{eq5p344}), the holomorphy and the various global 
symmetries prohibit any corrections to (\ref{eq5p3143}). Therefore, 
the curve (\ref{eq5p3143}) is an exact solution \cite{ref51}.

We can now discuss the physical consequences. First the branch points 
given by the curve are $x=0$, $\infty$ and 
\begin{eqnarray}
x_{\pm}=\frac{1}{2}\left[U-8\Lambda^{2N_c-4}_{N_c,N_c-2}
\pm\sqrt{U(U-16\Lambda^{2N_c-4}_{N_c,N_c-2})}
\right].
\label{eq5p3144}
\end{eqnarray}  
According to (\ref{eq5p3110}) and (\ref{eq5p3111}), 
the periods of the torus are
\begin{eqnarray}
a_D (U) &\sim& \int_{x_-}^{x^+}dx\left[x^3
+x^2\left(-U+8\Lambda^{2N_c-4}_{N_c,N_c-2}\right)
+16\Lambda^{4N_c-8}_{N_c,N_c-2}x\right]^{1/2}\nonumber\\
&=&\int_{x_-}^{x^+}dx\left[x (x-x_+) (x-x_-)\right]^{1/2},\nonumber\\
a (U)&\sim& \int_{0}^{x^-}dx
\left[x (x-x_+) (x-x_-)\right]^{1/2}
\label{eq5p3145}
\end{eqnarray}
and the effective coupling is given by the ratio
\begin{eqnarray}
\tau (U)=\frac{\int_{x_-}^{x^+}dx/y(x)}{\int_{0}^{x^-}dx/y(x)}.
\label{eq5p3146}
\end{eqnarray}  
The singularities of the effective
gauge coupling $\tau (U)$ can be inferred
from the vanishing of the discriminant
\begin{eqnarray}
\Delta =-\frac{1}{108}\left(\Lambda^{2N_c-4}_{N_c,N_c-2}\right)^2\left[
\left(-U+8\Lambda^{2N_c-4}_{N_c,N_c-2}\right)^2
-64\Lambda^{4N_c-8}_{N_c,N_c-2}\right]=0,
\label{eq5p3147}
\end{eqnarray}
which gives singularities at $U=0$ and 
$U=16\Lambda^{2N_c-4}_{N_c,N_c-2}{\equiv}U_1$. The monodromy matrices
around $U=0$ and $U=U_1$ can be immediately obtained by observing  
the change in the asymptotic
expansion form of $a_D(U)$ and $a(U)$ near $U=0$ and $U=U_1$ 
when taking $U\longrightarrow e^{2i\pi}U$.
They are, respectively, \cite{ref1p9}
\begin{eqnarray}
M_0=S^{-1}TS=\left(\begin{array}{cc} 1 & 0\\
                                    -1 & 1 \end{array}\right),
~~~M_1=(ST^{-2})^{-1}T(ST^{-2})=\left(\begin{array}{cc} -1 & 4\\
                                    -1 & 3 \end{array}\right),
\label{eq5p3148}
\end{eqnarray} 
up to an overall conjugation by $T^2$. 
According to (\ref{eq5p386}) and (\ref{eq5p3148}) we can see 
that $(1,0)$ is the left eigenvector
of $M_0$ and $(1,-1)$  the left eigenvector of $M_1$. This reveals
that there must exist massless monopoles or dyons in two subspaces
$\langle M^{ij}\rangle =M^*$ of the moduli space of vacua determined by
$\det M^*=0$ or $\det M^*=U_1$. Note that the spaces of these singular
vacua $M^*$ are non-compact. Ignoring the overall $T^2$
conjugation, which from (\ref{eq5p392}) only 
shifts the electric charges of the monopole or dyon, we can 
consider that magnetic monopoles exist in the singular vacuum
$M^*$ with $\det M^*=0$ and dyons in the singular vacuum $M^*$
with $\det M^*=U_1$.    
 
 The number of massless monopoles or dyons existing in 
the singular vacua $M^*$ with $\det M^*=0$ or $\det M^*=U_1$ 
can be detected from the  monodromy of $\tau$ upon 
taking $M$ around $M^*$. 
We will see that these two cases are different.
Let us first consider the vacua with $U=U_1$. Since 
in this case $U=f(M)=\det M$ is a single-valued
function in the $M$ complex plane, moving $M$ around  
$M^*$, $(M-M^*){\rightarrow} 
e^{2i\pi}(M-M^*)$ leads to $(U-U_1){\rightarrow} e^{2i\pi}(U-U_1)$
and gives the monodromy $M_1$. From the above discussion, we know that 
this monodromy is associated with a single pair of dyons $E^{\pm}$ 
with magnetic charge $\pm 1$. At $U=U_1$ the dyons are 
massless and away from $U=U_1$ the dyons become massive. Thus
the global symmetries, the holomorphy and mass dimension determine
that the superpotential near $U_1$ should be
\begin{eqnarray}
W=(U-U_1)\left[1+{\cal O}
\left(\frac{U-U_1}{\Lambda^{2(N_c-2)}_{N_c,N_c-2}}
\right)\right]E^+E^-.
\label{eq5p3149}
\end{eqnarray}   
   
The situation for the singular vacuum $M^*$ with $\det M^*=0$ is more
interesting. $\det M^*=0$ means that $r < N_f$ 
with $r$ being the rank of $M^*$, so $M^*$ has $N_f-r$ zero
eigenvalues. Thus when taking $M$ around a vacuum $M^*$,
$(M-M^*){\rightarrow}e^{2i\pi}(M-M^*)$, the complex function
$U=\det M$ will behave as $U{\rightarrow}e^{2i\pi (N_f-r)}U$. Since 
the transformation $U{\rightarrow}e^{2i\pi }U$ leads to the monodromy
$M_0$, the above transformation should yield the monodromy $M_0^{N_f-r}$.
Therefore, there must exist $N_f-r$ pairs of massless monopoles in the
vacuum parameterized by $\langle M\rangle =M^*$ with $M^*$ having 
rank $r$. This requires the superpotential for $N_f$ pairs of monopoles 
$q_i^+$ and $q_i^-$ with magnetic charge $\pm 1$ to be of the form
 \begin{eqnarray}
W=\frac{1}{2\mu}f\left(t=\frac{\det M}
{\Lambda^{2(N_c-2)}_{N_c,N_c-2}}\right)M^{ij}q^+_iq^-_j,
\label{eq5p3151}
\end{eqnarray} 
where $f(t)$ should be holomorphic around $t=0$ and normalized
as $f(0)=1$. The scale $\mu$ is introduced to ensure the
correct dimension $3$ of the superpotential because $M$ has dimension
$2$ and $q^{\pm}_i$ has dimension $1$. The superpotential automatically
makes $N_f-r$ monopoles massless at $M^*$ since it has rank $r$. In 
addition, to make the above superpotential respect the 
global flavour symmetry
$SU(N_f)$ and $R$-symmetry, the monopole $q^{\pm}_i$ must belong to
the conjugate fundamental representation $\overline{N}_f$ of $SU(N_f)$ and
have $R$-charge $1$. 

  The magnetic monopole and dyon are the left 
eigenvectors of the monodromy matrices at the singularities 
obtained from the curve solution (\ref{eq5p3143}). 
Considering the various representation quantum numbers of
of the electric quark $Q$ and the magnetic quark $q$
under the global symmetry $SU(N_f){\times}U_R(1)$ and their
electric and magnetic charges,
one can regard the dyon $E$ as the bound state of the 
electric quark $Q$ and magnetic quark $q$ \cite{ref51},
\begin{eqnarray}
E^{\pm}\sim q_i^{\pm}Q^i.
\label{eq5p3152}
\end{eqnarray} 
This construction can be checked from the left 
eigenvectors of $M_0$ and $M_1$. 

In addition, there exists a massless supermultiplet left by the breaking
$SO(N_c){\longrightarrow}SO(2){\cong}U(1)$ in the Coulomb phase,
i.e. an $N=1$ low energy (effective) photon field, whose field 
strength is a chiral superfield ${\cal W}_{\alpha}$
and which can be given a gauge invariant construction on the moduli space of 
vacua in terms of the fundamental fields,
\begin{eqnarray}
{\cal W}_{\alpha}{\sim}\epsilon_{i_1i_2{\cdots}i_{N_c-2}}
\epsilon^{rsr_1r_2{\cdots}r_{N_c-2}}(W_{\alpha})_{rs}
Q^{i_1}_{r_1}Q^{i_2}_{r_2}
{\cdots}Q^{i_{N_c-2}}_{r_{N_c-2}}{\equiv}W_{\alpha}(Q)^{N_c-2}.
\label{eq5p3153}
\end{eqnarray} 
Later we will see that a similar relation also exists in 
the non-Abelian Coulomb phase.

 The reasonableness of the above massless particle spectrum
is supported by two non-trivial consistency checks. 
The first one is still 't Hooft anomaly matching. At the origin
$\langle M \rangle =0$ of the moduli space of vacua, 
the global symmetry $SU(N_f){\times}U_R(1)$ is unbroken like in
the $SO(N_c)$ theory. The massless particle spectrum consists of
the  meson field $M^{ij}$, the photon supermultiplet 
with the field strength (\ref{eq5p3153}) and the monopole 
pair $q_i^{\pm}$ 
associated with the singularity $U=\det M=0$. Let us check whether
the 't Hooft anomalies contributed from this low energy 
massless particle
spectrum match those contributed by the massless fundamental 
quarks as listed in Table (\ref{ta5p3fo}). 
The contributions from massless $M^{ij}$ have already
been collected in Table (\ref{ta5p3fi}), 
so we need only consider the contributions
from the fermionic components of the $q_i^{\pm}$ and the photino.
(\ref{eq5p3153}) shows that the $R$-charge of the photino $\lambda_{\cal W}$
is $1+(N_f+2-N_c) (N_c-2)/N_f=1$ for $N_f=N_c-2$.  
The relevant currents and energy-momentum tensors are listed in
Tables (\ref{ta5p3se}), (\ref{ta5p3ei}) 
and the corresponding anomalies in Table (\ref{ta5p3ni}).

\begin{table}
\begin{center}
\begin{tabular}{|c|c|c|} \hline
            & $SU(N_f)$ &  $U_R(1)$ \\ \hline
 $\psi_{qi}^{\pm}$ & $j_{\mu}^A=\overline{\psi}_{qi}^+
\overline{t}^A_{ij}\sigma_{\mu}\psi^-_{qj}
+\overline{\psi}_{qi}^-\overline{t}^A_{ij} 
\sigma_{\mu}\psi^+_{qj}$ &
 $0$    \\ \hline
$\lambda_{\cal W} $   & $1$       & $j_{\mu}(\lambda)
=\overline{\lambda}^a_{\cal W}\sigma_{\mu}{\lambda}^a_{\cal W}$
\\ \hline
\end{tabular}
\caption{\protect\small Currents composed of the fermionic
components of monopole and photino corresponding to the global symmetry
$SU(N_f){\times}U_R(1)$.    \label{ta5p3se} }
 \end{center}
\end{table}

\begin{table}
\begin{center}
\begin{tabular}{|c|c|} \hline
            & $T_{\mu\nu}$\\ \hline
 $\psi_{qi}^{\pm}$  & 
$i/4\left[(\overline{\psi}_{qi}^+\sigma_\mu\nabla_\nu\psi_{qi}^-
-\nabla_\nu\overline{\psi}_{qi}^+\sigma_\mu\psi_{qi}^-)
+(\mu\longleftrightarrow\nu)\right]-g_{\mu\nu}{\cal L}[\psi_q]$       \\ \hline
$\lambda_{\cal W} $ &  $i/4\left[\left(\overline{\lambda}_{\cal W}
\sigma_\mu\nabla_\nu\lambda_{\cal W}-\nabla_\nu\overline{\lambda}_\mu{\cal W}
\sigma_\mu\lambda_{\cal W}\right)
+\left(\mu\longleftrightarrow\nu\right)\right]-g_{\mu\nu}{\cal L}
[\lambda_W]$
            \\ \hline
\end{tabular}
\caption{\protect\small  Energy-momentum tensor contributed by 
the fermionic components of $q^{\pm}_i$ and 
the photino; ${\cal L}[\psi]=i/2(\overline{\psi}\sigma^\mu\nabla_\mu\psi
-\nabla_\mu\overline{\psi}\sigma^\mu\psi)$,
$\nabla_\mu=\partial_\mu-\omega_{KL\mu}\sigma^{KL}/2$, 
$\sigma^{KL}=i/4[\sigma^K,\overline{\sigma}^L]$
and $\gamma^K=e^K_{~\mu}\sigma^{\mu}$.  \label{ta5p3ei}}
 \end{center}
\end{table}

\begin{table}
\begin{center}
\begin{tabular}{|c|c|} \hline
 Triangle diagrams and      & 't Hooft anomaly \\
 gravitational anomaly      &    coefficients \\ \hline
 $ U_R(1)^3 $  & $1$       \\ \hline
 $SU(N_f)^3 $ &  $-2\mbox{Tr}(t^A\{t^B,t^C\})$            \\ \hline
$SU(N_f)^2U_R(1) $ &  $0$\\ \hline
 $U_R(1)$ &  $1$            \\ \hline
\end{tabular}
\caption{\protect\small 't Hooft anomaly coefficients. \label{ta5p3ni}}
 \end{center}
\end{table}

Adding the contributions from the low energy particles 
to the contributions from the field $M$ given in Table (\ref{ta5p3fi}), 
one can see the anomalies indeed match the high energy 
anomalies given in Table (\ref{ta5p3fo}) for $N_f=N_c-2$.

Another check is to verify that the decoupling of a heavy flavour will
yield the description of the $N_f=N_c-3$ case discussed 
in Subsec.\,\ref{subsub533}. Without losing generality, we 
choose the $N_f$-th flavour to be heavy by adding a
large mass term $W_{\rm tree}=mM^{N_f}_{~N_f}/2$. There are 
two branches in the moduli space, a branch with $\det M^*=U_1$ 
and a branch with $\det M^*=0$.  We first discuss the branch 
with $\det M^*=U_1$. According to (\ref{eq5p3149}), 
the full superpotential near $U=U_1$ with the above large 
mass term is
\begin{eqnarray}
W=(U-U_1)E^+E^-+\frac{1}{2}mM^{N_f}_{~N_f}.
\label{eq5p3154}
\end{eqnarray} 
Integrating out $M^{N_f}_{~N_f}$ by its equation 
of motion
\begin{eqnarray}
\frac{\partial W}{\partial M^{N_f}_{~N_f}}=
\frac{\det M}{M^{N_f}_{~N_f}} E^+E^-+\frac{1}{2}m=0
\label{eq5p3155}
\end{eqnarray} 
gives
\begin{eqnarray}
\langle E^+ E^-\rangle=-\frac{m}{2\det \widehat{M}},
\label{eq5p3156}
\end{eqnarray} 
where $\widehat{M}$ denotes the mesons for the 
remaining $N_f-1=N_c-3$ flavours.
Obviously, the non-vanishing expectation value (\ref{eq5p3156}) of 
$\langle E^+ E^-\rangle$ has 
made the electric charge confined.    
Since $E^{\pm}$ are a pair of dyons, according to the discussions in
Sect.\,\ref{subsect27}, this 
phenomenon is just the oblique confinement proposed by 
't Hooft. The low energy superpotential at $U=U_1$ is
\begin{eqnarray}
W=\frac{1}{2}mM^{N_f}_{~N_f}=\frac{m\det M^*}{2\det \widehat{M}}
=8\frac{m\Lambda^{2N_c-4}_{N_c,N_c-2}}{\det \widehat{M}}.
\label{eq5p3157}
\end{eqnarray} 
Using the relation (\ref{eq5p210}) between the high energy scale 
$\Lambda_{N_c,N_c-2}$ and the low energy scale  $\Lambda_{N_c,N_c-3}$,
(\ref{eq5p3157}) is just the superpotential of the $\epsilon=1$ branch 
of (\ref{eq5p325}).

 As for the branch with $\det M^*=0$, we add a large mass term for
the $N_f$-th flavour to (\ref{eq5p3151}). The classical equation of 
motion of $M^{N_f}_{~N_f}$, 
\begin{eqnarray}
\frac{\partial W}{\partial M^{N_f}_{~N_f}}=
\frac{1}{2\mu}f(t)q_{N_f}^+q_{N_f}+\frac{1}{2\mu}\frac{df}{dt}
\frac{1}{\Lambda^{2N_c-2}_{N_c,N_c-2}}
\frac{\det M}{M^{N_f}_{~N_f}}+\frac{m}{2}=0,
\label{eq5p3158}
\end{eqnarray}
shows that $\langle q^{\pm}_{N_f}\rangle {\neq}0$
and hence that the magnetic $U(1)$ group is Higgsed. From the discussion
in Sect.\,\ref{subsect27}, the dual Meissner effects occurs 
and the original electric variables are confined. Due to the 
non-trivial function $f(t)$ in (\ref{eq5p3151}) and the constraint 
from the magnetic $U(1)$ $D$-term, 
$q_i^+e^{gV_D}q_i^-|_{\theta^2\overline{\theta}^2}$, there is 
a difficulty in explicitly integrating out the massive 
modes. However, from the classical
equations for $M^{N_f}_{~N_f}$, $q_{N_f}^+$ and $q_{N_f}^-$, 
one can see that the low energy massless modes are only 
$\widehat{M}^{\widehat{i}}_{~\widehat{j}}$, $q^+_{\widehat{i}}$
and $q^-_{\widehat{i}}$, $\widehat{i},\widehat{j}=1,\cdots, N_f-1$. 
Usually, for convenience of discussion, one defines the 
gauge invariant interpolating fields 
\begin{eqnarray}
q_{\widehat{i}}=\frac{1}{2\sqrt{m\mu}}\left(q_{\widehat{i}}^+q_{N_f}^--
q_{\widehat{i}}^-q_{N_f}^+\right)
\label{eq5p3159}
\end{eqnarray}
of $q_{\widehat{i}}^{\pm}$ to replace $q_{\widehat{i}}^{\pm}$
as massless modes. Considering the contribution from the magnetic $U(1)$
$D$-term, one can get the low energy effective superpotential
\begin{eqnarray}
W&=&\frac{1}{2\mu}\widehat{f}\left(\widehat{t}=\frac{(\det \widehat{M})
(\widehat{M}^{ij}q_{\widehat{i}}q_{\widehat{j}})}
{m\Lambda^{2(N_c-2)}_{N_c,N_c-2}}\right)
\widehat{M}^{ij}q_{\widehat{i}}q_{\widehat{j}}\nonumber\\
&=& \frac{1}{2\mu}\widehat{f}\left(\widehat{t}=\frac{(\det \widehat{M})
(\widehat{M}^{ij}q_{\widehat{i}}q_{\widehat{j}})}
{\Lambda^{2(N_c-2)}_{N_c,N_c-3}}\right)
\widehat{M}^{ij}q_{\widehat{i}}q_{\widehat{j}}.
\label{eq5p3160}
\end{eqnarray}
This is just the superpotential of the $\epsilon =-1$ branch of the 
low energy $N_f=N_c-3$ theory given by (\ref{eq5p335}). 
Consequently, according to (\ref{eq5p334}) we have
\begin{eqnarray}
q_{\widehat{i}}=\Lambda^{2-N_c}_{N_c,N_c-3}b_{\widehat{i}}
=\Lambda^{2-N_c}_{N_c,N_c-3}(Q)^{N_c-4}(W^{\alpha}W_{\alpha}),
\label{eq5p3161}
\end{eqnarray}
i.e. the remaining massless monopoles can be identified as massless
exotics (glueballs). Note that in (\ref{eq5p3160})
$\widehat{f}(\widehat{t})$ depends on $f(t)$ in 
(\ref{eq5p3151}), and the condition from the $U(1)$ $D$-term 
is important in showing that a non-trivial
$f(t)$ leads to $\widehat{f}(\widehat{t})$. 

In summary, the physical phenomena in the branch with $\det M^*=0$
are very interesting. Upon giving a heavy mass to $Q^{N_f}$, some of
the massless magnetic monopoles $q^{\pm}_i$ condense and this 
condensation leads to the confinement of $Q_i$. Especially, the 
remaining massless magnetic monopoles can be identified 
as massless exotics (glueballs). According to the discussion 
in Sect.\,\ref{subsect7.2}, this phenomenon was also observed
in the $SU(N_c)$ theories where massless magnetic quarks become
massless baryons. So one can conclude that the following non-perturbative 
physical phenomenon  is generic: some of gauge invariant 
composite operators such as baryons, glueballs and other 
exotics can be thought of as (Abelian or non-Abelian) magnetic 
objects.

\subsection{$N_c{\geq}4$, $N_f{\geq}N_c-1$: Dual magnetic $SO(N_f-N_c+4)$ description}
\label{subsect54}
\renewcommand{\thetable}{5.4.\arabic{table}}
\setcounter{table}{0}
\renewcommand{\theequation}{5.4.\arabic{equation}}
\setcounter{equation}{0}
\subsubsection{General introduction to magnetic description}
\label{subsub541}

 In this range, the infrared behaviour of these theories can be 
equivalently described by a dual magnetic theory. 
The reason we resort to a magnetic description
is the same as in the $SU(N_c)$ case: the ``electric" mesons 
(\ref{eq5p116}) and baryons (\ref{eq5p1191}) are not the appropriate 
variables to describe the moduli space of vacua
and we cannot use them to construct a consistent dynamical 
superpotential. One can easily check that the 't Hooft 
$SU(N_f){\times}U_R(1)$ anomalies 
cannot match for the fermionic components 
of $(Q,\lambda )$ in the microscopic
theory with those of $(M,B)$.  The matter field variables 
in the magnetic theory are the original ``electric" variables 
$M^{ij}$ and magnetic quarks $q_i^{\widetilde{r}}$ with 
the superpotential
\begin{eqnarray}
W=\frac{1}{2\mu}M^{ij}q_i{\cdot}q_j=\frac{1}{2\mu}M^{ij}
q_{i\widetilde{r}}q_j^{\widetilde{r}}.
\label{eq5p41}
\end{eqnarray} 
To ensure that $W$ is $SU(N_f){\times}U_R(1)$ invariant, $q_i$ must
be in the conjugate fundamental representation $\overline{N_f}$ of $SU(N_f)$ and
have $R$-charge $(N_c-2)/N_f$ since the $R$-charge of $M$ is
$2(N_f-N_c+2)/N_f$. The subscript $\widetilde{r}$ 
is the magnetic colour index.
Note that the magnetic quarks  cannot introduced 
from the baryons (\ref{eq5p1191}) as in the $SU(N_c)$ case. 
Before we explain the role
played by the scale $\mu$, we first explain what the dual gauge group
$SO(\widetilde{N}_c)$ is, which is not as easily 
found as in the $SU(N_c)$ case. This gauge group is restricted by the
requirement that is $U_R(1)$ anomaly-free in the 
magnetic theory. Since the $U_R(1)$ charge of the magnetic 
gaugino $\widetilde{\lambda}$ should be $1$, so from the above 
assignments of the $U_R(1)$ charge to $M$ and the magnetic 
quarks $q_{i\widetilde{r}}$, the anomaly-free $U_R(1)$ current in terms
of four-component is
\begin{eqnarray}
j_{\mu}^{(R)}=\left(\frac{N_c-2}{N_f}-1\right)\overline{\Psi}^i_{~q\widetilde{r}}
\gamma_{\mu}\gamma_5{\Psi}_{qi\widetilde{r}}
+\overline{\widetilde{\lambda}}^{\widetilde{a}}\gamma_{\mu}\gamma_5
\widetilde{\lambda}^{\widetilde{a}}+\frac{2(N_f-N_c+1)}{N_f}
\overline{\psi}_M\gamma_{\mu}\gamma_5{\psi}_M.
\label{eq5p42}
\end{eqnarray} 
The dynamical vector current for this magnetic $SO(\widetilde{N}_c)$ 
gauge theory, i.e. the Noether current corresponding to global
$SO(\widetilde{N}_c)$ gauge transformations, is
\begin{eqnarray}
J_{\mu}^{\widetilde{a}}=\overline{\Psi}^i_{~q\widetilde{r}}\gamma_{\mu}
T^{\widetilde{a}}_{\widetilde{r}\widetilde{s}}{\Psi}_{qi\widetilde{s}}
+f^{\widetilde{a}\widetilde{b}\widetilde{c}}
\overline{\widetilde{\lambda}}^{\widetilde{b}}\gamma_{\mu}
\widetilde{\lambda}^{\widetilde{c}}.
\label{eq5p43}
\end{eqnarray} 
The triangle diagram 
$\langle j_{\mu}^{(R)} J_{\mu}^{\widetilde{a}}J_{\mu}^{\widetilde{b}}\rangle$
gives the $U_R(1)$ anomaly
\begin{eqnarray}
\partial^{\mu}j_{\mu}^{(R)}&=&\left[
2 N_f\left(\frac{N_c-2}{N_f}-1\right)
+2f^{\widetilde{a}\widetilde{b}\widetilde{c}}
f^{\widetilde{a}\widetilde{b}\widetilde{c}}\right]\frac{1}{32\pi^2}
\epsilon^{\mu\nu\lambda\rho}F_{\mu\nu}^{\widetilde{a}}F_{\lambda\rho}^{\widetilde{a}}
\nonumber\\
&=&\left[2 \left(N_c-2-N_f\right)+2(\widetilde{N_c}-2)\right]
\frac{1}{32\pi^2}
\epsilon^{\mu\nu\lambda\rho}F_{\mu\nu}^{\widetilde{a}}F_{\lambda\rho}^{\widetilde{a}}.
\label{eq5p44}
\end{eqnarray}
That $U_R(1)$ is anomaly-free means that the above anomaly 
coefficient should vanish and hence we have 
$\widetilde{N_c}=N_f-N_c+4$.
Thus the magnetic gauge group should be 
$SO(N_f-N_c+4)$ \cite{ref51}. 
This theory is asymptotically free since the 
one-loop $\beta$ function coefficient is
\begin{eqnarray}
\widetilde{\beta}_0&=&3(\tilde{N_c}-2)-N_f=3(N_f-N_c+4-2)-N_f
=2N_f-3N_c+6 >0 
\label{eq5p46}
\end{eqnarray}  
because $N_c{\geq}4$ and $N_f{\geq}N_c-1$. 
The scale $\mu$ is introduced for the following reason. 
In the electric description the meson field
$M^{ij}=Q^i{\cdot}Q^j$ has dimension $2$ at 
the UV fixed point $g=0$ and generally
acquires some anomalous dimension at the IR fixed point. In the magnetic
description, since now $M$ is thought of as an elementary matter field,
its dimension should be its canonical dimension $1$ at the UV
fixed point $\widetilde{g}=0$. We denote $M$ in the magnetic 
description by $M_m$. In order to relate $M_m$ to $M$ of the 
electric description, a scale $\mu$ must be introduced with
the relation 
$M=\mu  M_m$. 
Later all the expressions will be written in terms of $M$ and $\mu$
rather than in terms of $M_m$.

  Dimensional considerations show that for generic $N_c$ and $N_f$
the scale $\widetilde{\Lambda}$ of the magnetic theory should be related
to $\Lambda$, the scale of the electric theory by
\begin{eqnarray}
\Lambda^{\beta_0}\widetilde{\Lambda}^{\widetilde{\beta}_0}=
\Lambda^{3(N_c-2)-N_f}\widetilde{\Lambda}^{3(N_f-N_c+2)-N_f}=
C(-1)^{N_f-N_c}\mu^{N_f},
 \label{eq5p48}
\end{eqnarray}
where $C$ is a dimensionless constant which will be determined 
below. Like in the $SU(N_c)$ gauge theory, the scale relation (\ref{eq5p48}) 
has several consequences:

\begin{enumerate}
\item It is preserved under mass deformation
and  along the flat directions, in which spontaneous symmetry
breaking occurs. The phase $(-1)^{N_f-N_c}$ is necessary to ensure that 
this relation is preserved. 
\item It reveals that as the electric theory becomes stronger the
magnetic theory becomes weaker and vice versa, 
\begin{eqnarray}
q^{N_f}e^{-8\pi^2/[g^2(q^2)]}e^{-8\pi^2/[\widetilde{g}^2(q^2)]}
{\propto}(-1)^{N_f-N_c}.
\label{eq5p410}
\end{eqnarray}
\item It implies that the dual theory of the magnetic description 
is identical to the original electric theory;  
\item Combined with the quantum effective action, it yields a relation between
the gaugino condensations in both the electric and magnetic theories,
$\lambda\lambda=-\widetilde{\lambda}\widetilde{\lambda}$.
\end{enumerate}

Now let us look at how the discrete symmetry behaves 
in the dual magnetic theory. The discussion in 
Sect.\,\ref{subsub511} shows that there exist discrete 
symmetries $Z_{2N_f}$ and $Z_2$ 
generated  by the transformation $Q{\longrightarrow}e^{i2\pi/(2N_f)}Q$ and
the charge conjugation ${\cal C}$, respectively. In the magnetic theory,
due to the kinetic term of the gauge singlet $M/\mu$ and 
the superpotential (\ref{eq5p41}) in the classical 
action, the corresponding discrete symmetries is generated by 
$q{\longrightarrow}e^{-i2\pi/(2N_f)}{\cal C}q$ and ${\cal C}$,respectively.
Note that the $Z_{2N_f}$ symmetry commutes with the electric gauge 
group $SO(N_c)$ but does not commute with the magnetic gauge group
$SO(N_f-N_c+4)$ \cite{ref51}. 

 Finally, it is worth seeing what the correspondences are 
in the magnetic theory for the gauge invariant chiral operators 
such as the meson, baryon and exotics used in the electric theory. 
The gauge invariant chiral operators appearing in the previous 
subsections, which can also be formally defined for $N_f{\geq}N_c-1$
and $N_c{\geq}4$, are collected in the following:
\begin{eqnarray}
M^{ij}&=& \frac{1}{2}Q^i{\cdot}Q^j,\nonumber\\
B^{i_1{\cdots}i_{N_c}}&=&\frac{1}{N_c!}\epsilon^{r_1{\cdots}r_{N_c}}
Q^{i_1}_{~r_1}{\cdots}Q^{i_{N_c}}_{~r_{N_c}}
{\equiv}B^{[i_1{\cdots}i_{N_c}]} ,\nonumber\\
b^{i_1{\cdots}i_{N_c-4}}&=&\frac{1}{(N_c-4)!}
\left(W^{\alpha}W_{\alpha}\right)
\epsilon^{r_1{\cdots}r_{N_c-4}}Q^{i_1}_{~r_1}{\cdots}
Q^{i_{N_c-4}}_{~r_{N_c-4}}{\equiv}b^{[i_1{\cdots}i_{N_c-4}]}
 ,\nonumber\\
{\cal W}_{\alpha}^{i_1{\cdots}i_{N_c-2}}&=&
\frac{1}{(N_c-2)!}W_{\alpha}
\epsilon^{r_1{\cdots}r_{N_c-2}}Q^{i_1}_{~r_1}{\cdots}
Q^{i_{N_c}-2}_{~r_{N_c-2}}{\equiv}{\cal W}_{\alpha}^{[i_1{\cdots}i_{N_c-2}]}.
\label{eq5p420}
\end{eqnarray} 
Based on $SU(N_f){\times}U_R(1)$ symmetry, these operators 
should be mapped to the following gauge invariant operators of 
the magnetic theory
\begin{eqnarray}
M^{ij} &{\longrightarrow}& M^{ij},\nonumber\\
B^{[i_1{\cdots}i_{N_c}]} &{\longrightarrow}&\epsilon^{i_1{\cdots}i_{N_f}}
\widetilde{b}_{[i_{N_c+1}{\cdots}i_{N_f}]},\nonumber\\
b^{[i_1{\cdots}i_{N_c-4}]} &{\longrightarrow}&\epsilon^{i_1{\cdots}i_{N_f}}
\tilde{B}_{[i_{N_c+1}{\cdots}i_{N_f}]},\nonumber\\
{\cal W}_{\alpha}^{[i_1{\cdots}i_{N_c-2}]}&{\longrightarrow}&
\epsilon^{i_1{\cdots}i_{N_f}}
\left(\widetilde{W}_{\alpha}\right)_{[i_{N_c-1}{\cdots}i_{N_f}]},
\label{eq5p421}
\end{eqnarray} 
where
\begin{eqnarray}
\widetilde{b}_{[i_{N_c+1}{\cdots}i_{N_f}]}&{\equiv}&\frac{1}{(N_f-N_c)!}
\epsilon_{\widetilde{r}_1{\cdots}\widetilde{r}_{N_f-N_c}}
q_{i_{N_c+1}}^{~\widetilde{r}_1}{\cdots}
q_{i_{N_f}}^{~\widetilde{r}_{N_f-N_c}},~~~ \mbox{for}~~N_f>N_c,\nonumber\\
 \widetilde{B}_{[i_{N_c-3}{\cdots}i_{N_f}]}&{\equiv}&\frac{1}{(N_f-N_c+4)!}
\epsilon_{\widetilde{r}_1{\cdots}\widetilde{ r}_{N_f-N_c+4}}
q_{i_{N_c-3}}^{~\widetilde{r}_1}{\cdots}
q_{i_{N_f}}^{~\widetilde{r}_{N_f-N_c+4}},\nonumber\\
\left(\widetilde{W}_{\alpha}\right)_{[i_{N_c-1}{\cdots}i_{N_f}]}
&{\equiv}&\frac{1}{(N_f-N_c+2)!}\widetilde{W}_{\alpha}
\epsilon_{\widetilde{r}_1{\cdots}\widetilde{r}_{N_f-N_c+2}}
q_{i_{N_c-1}}^{~\widetilde{r}_1}{\cdots}
q_{i_{N_f}}^{~\widetilde{r}_{N_f-N_c+2}}.
\label{eq5p422}
\end{eqnarray} 
The $R$-charge of each ``electric" operator in (\ref{eq5p421}) is 
equal to that of its ``magnetic" image, for example, 
the $R$-charge of the baryon operator $B^{[i_1{\cdots}i_{N_c}]}$ 
of the electric theory is $N_c(N_f-N_c+2)/N_f$, while the $R$-charge of 
$\epsilon^{i_1{\cdots}i_{N_f}}\widetilde{b}_{[i_{N_c+1}{\cdots}i_{N_f}]}$
is $2+(N_f-N_c)(N_c-2)/N_f=N_c(N_f-N_c+2)/N_f$. 
The mappings (\ref{eq5p421}) show that the baryons of 
the electric theory are dual to the exotics of the magnetic 
description, and the electric photon supermultiplet is dual to 
the magnetic photon supermultiplet.  

 The above is a general introduction to the dual magnetic 
description of supersymmetric $SO(N_c)$ gauge theory in the range of 
colours and flavours, $N_c {\geq}4$ and $N_f{\geq}N_c-1$. It is shown
that the magnetic theory is a supersymmetric $SO(N_f-N_c+4)$ gauge
theory with colour singlets $M^{ij}$ and $N_f$ magnetic quarks in the 
conjugate fundamental representation of $SU(N_f)$. 
Since the gauge groups $SO(3)$ and $SO(4)$ are special, 
we shall in the following two sections give a special discussion
of the magnetic $SO(3)$ and $SO(4)$ gauge theories, and 
then return to the general case.  

\subsubsection{$N_f=N_c-1$: Dual magnetic $SO(3)$ gauge theory}
\label{subsub542}

$N_f=N_c-1$ means that the dual magnetic description is an $SO(3)$ 
gauge theory. The matter fields, as discussed in the general
introduction, are the $SO(3)$ singlets $M^{ij}$ and the magnetic quarks
$q_i^{~\widetilde{r}}$. Note that now the magnetic quarks belong to the adjoint
representation of $SO(3)$ due to the coincidence of the adjoint 
representation and vector representation of $SO(3)$. We first 
introduce the dynamics of this case. 
Since from Table (\ref{ta5pon}) the $R$-charge of $M^{ij}$ is
$2/N_f$  for $N_f=N_c-1$, the gauge 
invariant and $SU(N_f){\times}U_R(1)$
invariant superpotential consists of not only (\ref{eq5p41}), 
but also of a term proportional to $\det M$. Taking into account the 
mass dimension, the full superpotential should be of the form
\begin{eqnarray}
W=\frac{1}{2\mu}M^{ij}q_i{\cdot}q_j
-\frac{1}{2^6\Lambda^{2N_c-5}_{N_c,N_c-1}}\det M,
\label{eq5p423}
\end{eqnarray}
where $\Lambda^{N_c,N_f=N_c-1}$ is the dynamically generated scale
and the normalization factor $2^6$ is chosen to make 
various deformations of theory consistent, as shown in detail
later. The scale relation here is also different from the general case,
since the one-loop $\beta$ function coefficient 
of the magnetic $SO(3)$ theory is $\widetilde{\beta}_0=6-2N_f$, while
for $N_f=N_c-1$ the factor $\widetilde{\Lambda}^{3(N_f-N_c+2)-N_f}$
in the general scale relation (\ref{eq5p48}) 
is just $\widetilde{\Lambda}^{3-N_f}$;
thus here the scale relation should be something like the square
of the general scale relation for $N_f=N_c-1$,
\begin{eqnarray}
2^{14}\left(\Lambda^{2N_c-5}_{N_c,N_c-1}\right)^2
\widetilde{\Lambda}^{6-2(N_c-1)}_{3,N_c-1}=\mu^{2N_c-1},
\label{eq5p424}
\end{eqnarray}  
where the normalization factor $2^{14}$ is determined in the same way
 as the factor $2^6$ and will be explained later. 

 At the origin of the moduli space $\langle M\rangle=0$, the 
matter fields $M^{ij}$ and $q_i^{~\widetilde{r}}$ and the $SO(3)$ vector 
bosons of the magnetic theory are massless.
Both the electric and the magnetic theories have the global 
symmetry $SU(N_f){\times}U_R(1)$.
One can verify the reasonableness of this magnetic magnetic 
theory by checking whether its 't Hooft $SU(N_f){\times}U_R(1)$ anomalies
match those of the electric theory given in 
Table (\ref{ta5p3fo}). The corresponding
currents composed of the fermionic components and the energy-momentum 
tensors for the gravitational anomaly are listed in Tables (\ref{ta5p4on}) 
and (\ref{ta5p4tw}), respectively,
and the 't Hooft anomaly coefficients are collected in 
Table (\ref{ta5p4th}). It is shown that by explicit calculations
that for $N_f=N_c-1$ the 't Hooft anomalies in  the electric and magnetic 
theories indeed match. 
  
\begin{table}
\begin{center}
\begin{tabular}{|c|c|c|} \hline
                    & $SU(N_f)$ &  $U_R(1)$ \\ \hline
 $\psi_{M}^{ij}$    & ${\psi}_M^{ij}t^A_{ij,kl}\sigma_{\mu}\psi^{kl}_M$ &
 $(N_f-2N_c+4)/N_f\,\overline{\psi}_M^{ij}\sigma_{\mu}\psi^{ij}_M $    
\\ \hline
$\psi_{q\widetilde{r}}^i$ & $\epsilon^{\widetilde{r}\widetilde{s}}
 \overline{\psi}_{q\widetilde{r}}^i\sigma_{\mu}
\overline{t}^A_{ij}{\psi}_{q\widetilde{s}}^j$ 
& $(N_c-2)/N_f\,\epsilon^{\widetilde{r}\widetilde{s}} 
\overline{\psi}_{q\widetilde{r}}^i\sigma_{\mu}
{\psi}_{q\widetilde{s}}^j$ \\ \hline
$\widetilde{\lambda} $   & $0$       & $
\overline{\widetilde{\lambda}}^{\widetilde{a}}
\sigma_{\mu}{\widetilde{\lambda}}^{\widetilde{a}}$
\\ \hline
\end{tabular}
\caption{\protect\small Currents composed of the fermionic
components of the singlets $M$, magnetic quarks and the magnetic 
$SO(3)$ gluino corresponding to the global symmetry
$SU(N_f){\times}U_R(1)$.  \label{ta5p4on}}
 \end{center}
\end{table}

\begin{table}
\begin{center}
\begin{tabular}{|c|c|} \hline
            & $T_{\mu\nu}$\\ \hline
 $\psi_M$  & $i/4\left[\left(\overline{\psi}_M^{ij}
\sigma_\mu\nabla_\nu\psi_M^{ij}-\nabla_\nu\overline{\psi}_M^{ij}
\sigma_\mu\psi_M^{ij}\right)+\left(\mu\longleftrightarrow\nu\right)\right]
-g_{\mu\nu}{\cal L}[\psi_M]$       \\ \hline
$ \psi_q$  & $i/4\left[\epsilon^{\widetilde{r}\widetilde{s}}
 \left(\overline{\psi}_{q\widetilde{r}}^i
\sigma_\mu\nabla_\nu\psi_{q\widetilde{s}}^i-\nabla_\nu\overline{\psi}_{q\widetilde{r}}^i
\sigma_\mu\psi_{q\widetilde{s}}^i\right)
+\left(\mu\longleftrightarrow\nu\right)\right]-g_{\mu\nu}{\cal L}[\psi_q] $  \\ \hline
$\widetilde{\lambda} $ &  $i/4\left[\left(
(\overline{\widetilde{\lambda}}^{\widetilde{a}}\sigma_\mu\nabla_\nu
 \widetilde{\lambda}^{\widetilde{a}}
-\nabla_\nu\overline{\widetilde{\lambda}}^{\widetilde{a}} \sigma_\mu
\widetilde{\lambda}^{\widetilde{a}}\right)
+\left(\mu\longleftrightarrow\nu\right)
\right]-g_{\mu\nu}{\cal L}[\lambda] $
            \\ \hline
\end{tabular}
\caption{\protect\small 
Energy-momentum tensor contributed from the fermionic component 
of $M^{ij}$, $q^i_r$ and the magnetic $SO(3)$ gluino;
${\cal L}[\psi]=i/2 \epsilon^{\widetilde{r}\widetilde{s}} 
(\overline{\psi}_{\widetilde{r}} \sigma^\mu\nabla_\mu\psi_{\widetilde{s}} 
-\nabla_\mu\overline{\psi}_{\widetilde{r}} \sigma^\mu\psi_{\widetilde{s}} )$,
$\nabla_\mu=\partial_\mu-\omega_{KL\mu}\sigma^{KL}/2$, $\sigma^{KL}
=i/4[\sigma^K,\overline{\sigma}^L]$
and $\sigma^K=e^K_{~\mu}\sigma^{\mu}$.
\label{ta5p4tw}}
 \end{center}
\end{table}
\begin{table}
\begin{center}
\begin{tabular}{|c|c|} \hline
 Triangle diagrams and      & 't Hooft anomaly \\
 gravitational anomaly      &    coefficients \\ \hline
 $ U_R(1)^3 $  & $(N_f+1)(N_f-2N_c+4)^3/(2N_f^2)+N_c(N_c-2)^3/N_f^2$\\
               & $+(N_f-N_c+4)(N_f-N_c+3)/2$ 
  \\ \hline
 $SU(N_f)^3 $ &  $[(N_f+4)-(N_f-N_c+4)]\mbox{Tr}(t^A\{t^B,t^C\})$\\ \hline
$SU(N_f)^2U_R(1) $ &  $2\left [(N_f+2)(N_f-2N_c+4)/N_f
+(N_c-2)/N_f\right]\mbox{Tr}(t^At^B)$\\ \hline
 $U_R(1)$ &  $(N_f-2N_c+4)(N_f+1)+N_c(N_c-2)$ \\ \hline
\end{tabular}
\caption{\protect\small 't Hooft anomaly coefficients. \label{ta5p4th}}
 \end{center}
\end{table}

A special point for the magnetic $SO(3)$ theory should be mentioned. The
one-loop $\beta$-function coefficient, $\widetilde{\beta}_0=6-2N_f$ implies that
when $N_f{\geq} 3$ this magnetic $SO(3)$ gauge 
theory is not asymptotically free. Consequently, the value $\widetilde{g}=0$
of the magnetic coupling is the infrared fixed point
and thus the theory is free in the infrared.
The fields $q_i$ and $M$ behave as free scalar fields and should 
have dimension $1$. In this case the superpotential (\ref{eq5p423})
does not exist and hence the infrared theory has a large accidental 
symmetry. When $N_f <3$, the global symmetries (including the
discrete symmetry $Z_{2N_f}$) and the holomorphy around
$M=q=0$ determine that the superpotential (\ref{eq5p423}) 
of the magnetic theory is unique. Unlike (\ref{eq5p330}) 
and (\ref{eq5p3151}), (\ref{eq5p423}) cannot 
be modified by a non-trivial function of the 
global and gauge invariants. In the following we shall see 
that the $\det M$ term of (\ref{eq5p423}) is very important in 
properly describing the theory when it is
perturbed by a large mass term or along the flat directions.  
In particular, without the $\det M$ term, the term 
$M^{ij}q_i{\cdot}q_j$ would possess a $Z_{4N_f}$ symmetry 
in contrast to the $Z_{2N_f}$ symmetry (\ref{eq5p114}) 
of the electric theory, while as a dual theory the magnetic theory
should have a $Z_{2N_f}$ symmetry. 

In the following we shall show that  
the moduli space of vacua of the magnetic description
agrees with that of the original electric theory.
We shall see that some interesting phenomena will arise.

\vspace{2mm}
\begin{flushleft}
{\it Flat directions} 
\end{flushleft}
\vspace{2mm}

The moduli space of the magnetic theory is given by 
the $F$-flat directions of the superpotential (\ref{eq5p423}) 
and the $D$-flat directions of the gauge theory part. Here we only 
consider the $F$-flat directions. (\ref{eq5p423}) shows that for 
$M{\neq}0$ the magnetic quarks will have a mass matrix $\mu^{-1}M$.
At low energy, the heavy quarks will decouple and there are only
$k=N_f-\mbox{rank}(M)$ massless dual quarks $q$ left. We first consider
the case $\mbox{rank}(M)=N_f$. All the dual quarks 
become massive and hence decouple, so the low energy theory is 
a pure $SO(3)$ Yang-Mills theory with massless
singlets $M$. According to (\ref{eq5p212}) 
this low energy magnetic theory has a scale
\begin{eqnarray}
\widetilde{\Lambda}^6_{3,0}=\widetilde{\Lambda}^{6-2(N_c-1)}_{3,N_f}
\det (\mu^{-1}M)^2.
\label{eq5p426}
\end{eqnarray}
Gluino condensation generates a dynamical superpotential 
\begin{eqnarray}
W_{\rm dyn}=2\langle \widetilde{\lambda}\widetilde{\lambda}\rangle
=2\epsilon \widetilde{\Lambda}^3_{3,0}=2\epsilon 
\widetilde{\Lambda}^{3-(N_c-1)}_{3,N_f}\mu^{N_c-1}\det M,
\label{eq5p427}
\end{eqnarray}    
where $\epsilon =\pm 1$ means that the gaugino condensation
leads to two vacua. Considering the term proportional to $\det M$
in (\ref{eq5p423}) and adding it to (\ref{eq5p427}), we have 
the full low energy superpotential
\begin{eqnarray}
W_{\rm full}=2\epsilon 
\widetilde{\Lambda}^{3-(N_c-1)}_{3,N_f}\mu^{N_c-1}\det M
-\frac{1}{2^6\Lambda^{2N_c-5}_{N_c,N_c-1}}\det M
=(\epsilon-1)\frac{1}{2^6\Lambda^{2N_c-5}_{N_c,N_c-1}}\det M.
\label{eq5p428}
\end{eqnarray} 
Therefore, the $\epsilon=1$ branch reproduces the 
moduli space of supersymmetric ground sates represented by generic 
$\langle M\rangle$. The $\epsilon =-1$ branch will be discussed later.
(\ref{eq5p428}) also shows why the normalization 
factors $2^{14}$ and $2^6$ were chosen
in (\ref{eq5p423}) and (\ref{eq5p424}), respectively.
 
 When $\mbox{rank}(M)=N_f-1$, the 
corresponding low energy theory is a magnetic $SO(3)$ with one massless
flavour, which can be taken to be the $N_f$-th flavour, $q_{N_f}$.
This is just the magnetic version of the $N=2$ Seiberg-Witten 
model \cite{ref1p1}, whose exact solution is given by the algebraic 
curve (\ref{eq5p3134}). From the discussions in Subsect.\,\ref{subsub534} 
we know that it has a massless photon and massless monopoles at 
the singularity
\begin{eqnarray}
{\langle u \rangle}{\equiv}u_1 =\langle q^2_{N_f}\rangle 
=\sqrt{16\widetilde{\Lambda}^4_{3,1}}
=4\epsilon \widetilde{\Lambda}^2_{3,1},
~~~\epsilon=\pm 1,
 \label{eq5p429}
\end{eqnarray}   
where, according to (\ref{eq5p424}) and (\ref{eq5p426}), 
the low energy scale $\widetilde{\Lambda}_{3,1}$ is given by
\begin{eqnarray}
\widetilde{\Lambda}^4_{3,1}=\frac{\mu^2}{2^{14}
(\Lambda^{2N_c-5}\det\widehat{M})^2}
\label{eq5p430}
\end{eqnarray}
with $\det\widehat{M}=\det M/M^{N_fN_f}$, the product of the
$N_f-1$ non-zero eigenvalues of $M$. Considering the contribution
of the massless magnetic monopoles to the low energy superpotential, 
the full low energy superpotential near the massless monopole points 
$u{\approx}4{\epsilon}\widetilde{\Lambda}^2_{3,1}$ should be the sum of
the low energy version of (\ref{eq5p423}) and 
the monopole contribution (\ref{eq5p3149}),
\begin{eqnarray}
W&=&\frac{1}{2\mu}M^{N_f}_{~N_f}(q_{N_f})^2
-\frac{1}{64\Lambda^{2N_c-5}_{N_c,N_c-1}}\det \widehat{M}-
\frac{1}{2\mu}\left(u-u_1\right)\widetilde{E}^+_{(\epsilon)}
\widetilde{E}^-_{(\epsilon)}\nonumber\\
&=&\frac{1}{2\mu}M^{N_f}_{~N_f}\left( u
-\frac{1}{32\Lambda^{2N_c-5}_{N_c,N_c-1}}\mu\det \widehat{M}\right)
-\frac{1}{2\mu}\left(u-4{\epsilon}\widetilde{\Lambda}^2_{3,1}\right)
\widetilde{E}^+_{(\epsilon)}
\widetilde{E}^-_{(\epsilon)},
\label{eq5p431}
\end{eqnarray}
where the choice of the normalization factor 
$f(u_1)=-1$ of the $\widetilde{E}^+_{(\epsilon)}
\widetilde{E}^-_{(\epsilon)}$ term is purely for convenience.
From this low energy superpotential, the equation for $M^{N_fN_f}$ gives
\begin{eqnarray}
\langle u \rangle =\frac{\mu\det \widehat{M}}
{2^5\Lambda^{2N_c-5}_{N_c,N_c-1}}
=4 \widetilde{\Lambda}^2_{3,1}.
\label{eq5p432}
\end{eqnarray}
This has fixed the low energy magnetic theory with $\epsilon=1$ 
in which there are a massless photon supermultiplet and a pair 
of massless monopoles $\widetilde{E}^{\pm}_{(+)}$. 
The $u$ equation of motion further yields
\begin{eqnarray}
M^{N_fN_f}=\widetilde{E}^{+}_{(+)}\widetilde{E}^{-}_{(+)}.
\label{eq5p433}
\end{eqnarray}
This relation can be understood from 
electric-magnetic duality: the monopoles $ \widetilde{E}^{\pm}_{(+)}$
are magnetic relative to the magnetic variables; they will be electric
in terms of the original electric variables. Now we consider the 
low energy electric theory. $\mbox{rank}(M)=N_f-1=N_c-2$
means that the $SO(N_c)$ gauge group breaks to $SO(2){\cong}U(1)$
in the flat direction but with one of the elementary quarks 
which is charged under $U(1)$ remaining massless . According to 
the duality relation, it should be a massless collective excitation
in the magnetic description --- the monopole. (\ref{eq5p433}) is 
just a reflection of this correspondence.  

 For the case $\mbox{rank}(M){\leq}N_f-2$, the low energy magnetic theory
is an $SO(3)$ gauge theory with $k=N_f-\mbox{rank}(M){\geq}2$ light flavours.
When $k=2$, $\beta_0=0$, the low energy theory is at a non-trivial
fixed point of the beta function and hence it is described by a
four dimensional superconformal field theory. If $k>2$, the low energy
theory is in a free magnetic phase, it is not well defined due to
the Landau pole. All the results in the magnetic theory of these
cases are dual to those of
the original electric description.

\vspace{2mm}
\begin{flushleft}
{\it Mass deformations} 
\end{flushleft}
\vspace{2mm}

Let us see whether the magnetic $SO(3)$ theory leads to the correct
description of the $N_f=N_c-2$ electric theory 
if one heavy flavour decouples. As usual, adding a large $Q^{N_f}$ mass 
term $W_{\rm tree}=mM^{N_fN_f}/2$ to the superpotential (\ref{eq5p423}) 
of the magnetic theory, we have the full superpotential
\begin{eqnarray}
W_{\rm full}=\frac{1}{2\mu}M^{ij}q_i{\cdot}q_j
-\frac{1}{2^6\Lambda^{2N_c-5}_{N_c,N_c-1}}\det M+\frac{1}{2}mM^{N_fN_f}.
\label{eq5p434}
\end{eqnarray} 
The $M^{N_fN_f}$ equation of motion 
$\partial W_{\rm full}/\partial M^{N_fN_f}=0$ gives
\begin{eqnarray}
\langle q^2_{N_f}\rangle 
=\frac{\det \widehat{M}}{2^5 \Lambda^{2N_c-5}_{N_c,N_c-1}}-\mu m
\label{eq5p435}
\end{eqnarray} 
with $\det \widehat{M}=\det M/M^{N_fN_f}$. This non-vanishing 
expectation value
breaks the magnetic $SO(3)$ group to $SO(2)$. After integrating out 
the massive fields, the low energy magnetic $SO(2)$ gauge theory has 
matter fields consisting of neutral fields 
$M^{\widehat{i}\widehat{j}}$ and 
fields $q^{\pm}_{\widehat{i}}$ of $SO(2)$ charge $\pm 1$ and the 
corresponding tree-level superpotential is 
$W_{\rm tree}=M^{\widehat{i}\widehat{j}}q_{\widehat{i}}q_{\widehat{j}}/2$. 
However, according to the discussion 
in Subsect.\,\ref{subsub534}, the contribution 
to the superpotential from the instantons 
in the broken magnetic $SO(3)$ should also be 
included since there are well
defined instantons in the broken part of the 
$SO(3)$ group. Therefore, the superpotential is modified to
\begin{eqnarray}
W=\frac{1}{2\mu}f\left[\frac{\det\widehat{M}}
{\Lambda^{2(N_c-2)}_{N_c,N_c-2}}\right]M^{\widehat{i}
\widehat{j}}q^+_{\widehat{i}}q^-_{\widehat{j}}.
\label{eq5p436}
\end{eqnarray}
This is just the superpotential (\ref{eq5p3151}) and 
the $q^{\pm}_{\hat{i}}$ can be identified as the 
monopoles of the electric theory with $N_f=N_c-2$.

 Now we consider another special point in the moduli space
of vacua. After we choose a small mass $m$ for the $N_f$th 
flavour but large $\det\widehat{M}$, the first
$N_f-1$ flavours $q_i$ will 
become very heavy and must be integrated out.
The low energy theory is the magnetic $SO(3)$ theory with the 
quarks $q_{N_f}$ in the adjoint representation and the scale 
given by (\ref{eq5p430}). This is just the 
Seiberg-Witten model discussed in Sect.\,\ref{subsub533} \cite{ref1p1}. 
From its exact algebraic curve solution we know
there exist massless monopoles or dyons 
 $\widetilde{E}^{\pm}_{(\epsilon)}$at 
$u_1=4\epsilon\widetilde{\Lambda}^2_{3,1}$. Thus near
one of these two vacua, the low energy superpotential 
should include contributions from these solitons,
\begin{eqnarray}
W&{\approx}&\frac{1}{2\mu}M^{N_fN_f}(q_{N_f})^2-
\frac{1}{2^6\Lambda^{2N_c-5}_{N_c,N_c-1}} \det\widehat{M}+
\frac{1}{2}mM^{N_fN_f}-\frac{1}{2\mu}
\left((q_{N_f})^2-4\epsilon\widetilde{\Lambda}^2_{3,1}\right)
\widehat{E}^+_{(-)} \widehat{E}^-_{(-)}\nonumber\\
&=& \frac{1}{2\mu}M^{N_fN_f}
(u-\frac{1}{2^5\Lambda^{2N_c-5}_{N_c,N_c-1}}\mu \det\widehat{M}+\mu m)-
\frac{1}{2\mu}\left(u-4\epsilon\widetilde{\Lambda}^2_{3,1}\right)
\widehat{E}^+_{(-)} \widehat{E}^-_{(-)},
\label{eq5p437}
\end{eqnarray}
where we denoted $u{\equiv}q^2_{N_f}$. The $M^{N_fN_f}$ and $u$ equations
of motion, respectively, give
\begin{eqnarray}
u=\frac{1}{2^5\Lambda^{2N_c-5}_{N_c,N_c-1}}\mu \det\widehat{M}-\mu m;
~~~M^{N_fN_f}=\widehat{E}^+_{(-)} \widehat{E}^-_{(-)}.
\label{eq5p438}
\end{eqnarray}                         
Inserting (\ref{eq5p438}) into (\ref{eq5p437}) we get 
the low energy superpotential
as $m$ becomes large
\begin{eqnarray}
W_L=\frac{1}{2}\left[m-(1-\epsilon)
\frac{1}{2^5\Lambda^{2N_c-5}_{N_c,N_c-1}}\det\widehat{M}\right].
\label{eq5p439i}
\end{eqnarray}    
This $W_L$ shows that for $\epsilon =1$ there is no 
supersymmetric vacuum since the $F$-term does not 
vanish. The low energy superpotential
for the $\epsilon=-1$ vacuum is
\begin{eqnarray}
W_L
&=& \frac{1}{2}\left(m
-\frac{1}{2^4\Lambda^{2(N_c-2)-1}_{N_c,N_c-1}}\det\widehat{M}\right)
\widehat{E}^+_{(-)} \widehat{E}^-_{(-)}.
\label{eq5p439}
\end{eqnarray}    
The theory can be identified as the one described by (\ref{eq5p3149}) 
with $\widehat{E}^{\pm}_{(-)}$ as the dyons, which become massless at   
$\det\widehat{M}=16 m \Lambda^{2(N_c-2)-1}_{N_c,N_c-1}
=16\Lambda^{2N_c-4}_{N_c,N_c-2}$.
This conclusion is the same as that obtained from low energy electric
theory discussed in Subsect.\,\ref{subsub534}.

 If we assign large masses to more flavours, the number of massless flavours
in the low energy theory is correspondingly reduced. The non-perturbative 
phenomena observed in the $N_f=N_c-3$ and $N_f=N_c-4$ electric theory will
be produced: the monopoles or dyons condense and lead to confinement 
or oblique confinement. If there are $N_f <N_c-4$ massless flavours
in the low energy theory, the moduli space of vacua will not exist 
and the confining branch disappears \cite{ref1p1}. However, when all 
the flavours are given large masses, there exist an oblique 
confinement branch with the superpotential
\begin{eqnarray}
W_{\rm obl}=-\frac{1}{32\Lambda^{2N_c-5}_{N_c,N_c-1}}\det M.
\label{eq5p441}
\end{eqnarray}
This superpotential can be formally obtained by setting $m=0$ in
(\ref{eq5p439}) and using the $u$ equation of motion since all 
the flavours should be integrated out. 

Overall, the monopoles $q_i^{\pm}$ of the electric theory at the origin
(u=0) of the low energy $N_f=N_c-2$ electric theory have a weakly 
coupled magnetic description in terms of the magnetic quarks 
$q_i^{\pm}$ of dual theory. The massless dyons $E^{\pm}$ of 
the $N_f=N_c-2$ electric theory at 
$u=u_1=\det \widehat{M}=16\Lambda^{2N_c-4}_{N_c,N_c-2}$ 
are described by strongly coupled dyons.

\subsubsection{$N_f=N_c$: Dual magnetic $SO(4)$ gauge theory}
\label{subsub543}

 The discussion in Subsect.\,\ref{subsub512} shows that
the classical moduli space 
of the electric $SO(N_c)$ theory with $N_f=N_c$ can be parametrized by
the mesons $M^{ij}=Q^i{\cdot}Q^j$ and the baryon $B=\det Q$ but  with
the constraint $B=\pm \sqrt{\det M}$. The quantum moduli space in 
this case should be described in the context of the dual 
magnetic description, which is an 
$SO(4){\cong}SU(2)_X{\times}SU(2)_Y$ gauge theory with $N_f$ flavours
of quarks in the $(2,2)$ representation of the gauge group
and the $SO(4)$ singlets $M^{ij}$. The symmetries and the 
holomorphy around $M=q=0$ determine the superpotential
as given by (\ref{eq5p41}) and it cannot be modified by any 
non-trivial gauge invariant function. The one-loop beta function 
coefficients of the $SU(2)$ theory and the $SO(N_c)$ gauge theory 
with $N_f=N_c$ flavours together with dimensional
analysis give the scale relation \cite{ref1p1}
\begin{eqnarray}
2^8\widetilde{\Lambda}^{\widetilde{\beta}_0}_{s,N_c}
{\Lambda}^{\beta_0}_{N_c,N_c}=2^8\widetilde{\Lambda}^{6-N_f}_{s,N_c}
{\Lambda}^{2N_c-6}_{N_c,N_c}=\mu^{N_c}, ~~~~s=X,Y. 
\label{eq5p443}
\end{eqnarray}  
The normalization factor $2^8$ is chosen for consistency under deformation
and symmetry breaking along the flat directions.
The  one-loop beta function coefficient $\tilde{\beta}_0=6-N_f$ 
of $SU(2)$ shows that the magnetic $SU(2)_X{\times}SU(2)_Y$ theory
is not asymptotically free for $N_f{\geq}6$. Thus the  magnetic theory
in the case of $N_f{\geq}6$ is free in the infrared region. For
$N_c=N_f=4,5$, the theory is asymptotically free and has a 
non-trivial infrared fixed point, 
at which the magnetic theory is in an interacting non-Abelian
Coulomb phase. Thus the magnetic theory should be dual to the
original $SO(N_c)$ electric theory with $N_f=N_c$ quarks $Q^i$ in a 
non-Abelian Coulomb phase. 

The magnetic theory has an anomaly-free 
$SU(N_f){\times}U_R(1)$ global symmetry, under which the 
magnetic quarks $q_i$ are in the conjugate fundamental representation
$\overline{N}_f$ and have the $R$-charge $(N_c-2)/N_c$, the 
singlets $M^{ij}$ are in the $N_f(N_f+1)/2$ dimensional 
representation and have $R$-charge $4/N_c$. The global 
$SU(N_f){\times}U_R(1)$ symmetry is unbroken at the origin 
of the moduli space of both the electric and the magnetic theories and
$q_i$ and $M^{ij}$ are massless. We can verify this massless
particle spectrum by checking the 't Hooft anomaly matching between 
the electric and magnetic theories. The currents and the energy-momentum
tensors are listed in Tables \ref{ta5p4fo} and \ref{ta5p4fi}
and the 't Hooft anomalies from the massless
fermions of magnetic theory are collected in Table \ref{ta5p4si}. 
The anomaly coefficients are indeed equal to those listed 
in Table \ref{ta5p3fo} for $N_f=N_c$. 

\begin{table}
\begin{center}
\begin{tabular}{|c|c|c|} \hline
                    & $SU(N_f)$ &  $U_R(1)$ \\ \hline
 $\psi_{M}^{ij}$    & $\overline{\psi}_M^{ij}t^A_{ij,kl}
\sigma_{\mu}\psi^{kl}_M$ &
 $\overline{\psi}_M^{ij}\sigma_{\mu}\psi^{ij}_M $    
\\ \hline
$\psi_{q \widetilde{r}_X\widetilde{r}_Y}^i$ & 
$\epsilon^{\widetilde{r}_X\widetilde{s}_X}\epsilon^{\widetilde{r}_Y\widetilde{s}_Y}
\overline{\psi}_{q\widetilde{r}_X\widetilde{r}_Y}^i\sigma_{\mu}\overline{t}^A_{ij}
{\psi}_{q\widetilde{s}_X\widetilde{s}_Y}^j$ & $-2/N_f
\epsilon^{\widetilde{r}_X\widetilde{s}_X}\epsilon^{\widetilde{r}_Y\widetilde{s}_Y} 
\overline{\psi}_{q\widetilde{r}_X\widetilde{r}_Y}^i
\sigma_{\mu}{\psi}_{q\widetilde{s}_X\widetilde{s}_Y}^i$ \\ \hline
$\widetilde{\lambda} $   & $0$       & $
\overline{\widetilde{\lambda}}^{\widetilde{a}}
\sigma_{\mu}{\widetilde{\lambda}}^{\widetilde{a}}$
\\ \hline
\end{tabular}
\caption{\protect\small Currents composed of the fermionic
components of the singlets $M$, magnetic quarks and the magnetic 
$SO(3)$ gluino corresponding to the global symmetry
$SU(N_f){\times}U_R(1)$. \label{ta5p4fo}}
 \end{center}
\end{table}

\begin{table}
\begin{center}
\begin{tabular}{|c|c|} \hline
            & $T_{\mu\nu}$\\ \hline
 $\psi_M$  & $i/4\left[\left(\overline{\psi}_M^{ij}\sigma_\mu\nabla_\nu
\psi_M^{ij}-\nabla_\nu\overline{\psi}_M^{ij}\sigma_\mu
\psi_M^{ij}\right)+\left(\mu\longleftrightarrow\nu\right)\right]
-g_{\mu\nu}{\cal L}[\psi_M] $   \\ \hline
$ \psi_q$  & $i/4\epsilon^{\widetilde{r}_X\widetilde{s}_X}
\epsilon^{\widetilde{r}_Y\widetilde{s}_Y} \left[\left(
\overline{\psi}_{q\widetilde{r}_X\widetilde{r}_Y}^i\sigma_\mu\nabla_\nu
\psi_{q\widetilde{s}_X\widetilde{s}_Y}^i-
\nabla_\nu\overline{\psi}_{q\widetilde{r}_X\widetilde{r}_Y}^i
\sigma_\mu\psi_{q\widetilde{s}_X\widetilde{s}_Y}^i\right)
+\left(\mu\longleftrightarrow\nu\right)\right]-g_{\mu\nu}{\cal L}[\psi_q] $       \\ \hline
$\widetilde{\lambda}_s $ &  $i/4\left[
\left(\overline{\widetilde{\lambda}}^{\widetilde{a}}_s\sigma_\mu\nabla_\nu
 \widetilde{\lambda}^{\widetilde{a}}_s
-\nabla_\nu\overline{\widetilde{\lambda}}^{\widetilde{a}}_s \sigma_\mu
\widetilde{\lambda}^{\widetilde{a}}_s\right)
+\left(\mu\longleftrightarrow\nu\right)\right]
-g_{\mu\nu}{\cal L}[\widetilde{\lambda}] $
            \\ \hline
\end{tabular}
\caption{\protect\small Energy-momentum tensor composed of 
the fermionic components 
of $M^{ij}$, $q^i_r$ and the magnetic $SO(3)_s$ gluino, $s=X,Y$; 
${\cal L}[\psi]=i/2 \epsilon^{\widetilde{r}_X\widetilde{s}_X}
\epsilon^{\widetilde{r}_Y\widetilde{s}_Y}
(\overline{\psi}_{\widetilde{r}_X\widetilde{r}_Y} \sigma^\mu\nabla_\mu
\psi_{\widetilde{s}_X\widetilde{s}_Y}
-\nabla_\mu\overline{\psi}_{\widetilde{r}_X\widetilde{r}_Y} 
\sigma^\mu\psi_{\widetilde{s}_X\widetilde{s}_Y} )$,
$\nabla_\mu=\partial_\mu-\omega_{KL\mu}\sigma^{KL}/2$, 
$\sigma^{KL}=i/4[\sigma^K,\overline{\sigma}^L]$
and $\sigma^K=e^K_{~\mu}\sigma^{\mu}$.
\label{ta5p4fi}}
 \end{center}
\end{table}

\begin{table}
\begin{center}
\begin{tabular}{|c|c|} \hline
 Triangle diagrams and      & 't Hooft anomaly \\
 gravitational anomaly      &    coefficients \\ \hline
 $ U_R(1)^3 $  & $N_f(N_f-1)/2+1/N_f(2-N_f)^3 $ \\ \hline
 $SU(N_f)^3 $ &  $N_f\mbox{Tr}(t^A\{t^B,t^C\})$ 
\\ \hline
$SU(N_f)^2U_R(1) $ &  $(2-N_f)\mbox{Tr}(t^At^B)$\\ \hline
$U_R(1)$ &  $-N_f(N_f-3)/2$  \\ \hline
\end{tabular}
\caption{\protect\small 't Hooft anomaly coefficients.
\label{ta5p4si}}
 \end{center}
\end{table}

\vspace{2mm}
\begin{flushleft}
{\it Flat directions} 
\end{flushleft}
\vspace{2mm}

The flat directions are given by the $F$-term of (\ref{eq5p41}) 
and the $D$-term of the magnetic $SO(4)$ gauge theory. 
The $M^{ij}$ equations of motion give $q_i{\cdot}q_j=0$. Furthermore, 
the vanishing of $D$-terms 
gives the solution of $\langle q_i\rangle =0$. Thus the low energy theory around such 
a point $\langle M\rangle$  should be a magnetic $SO(4)$ gauge theory with
$k=N_f-\mbox{rank}(M)$ dual quarks $q_i$ since the equations of 
motion for $M$ and the vanishing of the $D$-term do not require them to vanish. In 
the following we consider several typical cases.

 When $\mbox{rank}(M)=N_f$, there are no massless dual quarks. The 
low energy magnetic theory is a pure $SU(2)_X{\times}SU(2)_Y$ Yang-Mills
theory with the scale
\begin{eqnarray}
\widetilde{\Lambda}^6_{s,0}{\equiv}\widetilde{\Lambda}^6=
\frac{{\mu}^{N_c}}{2^8\Lambda^{2N_c-6}_{N_c,N_c}/\det (\mu^{-1}M)}=
\frac{\det M}{2^8\Lambda^{2N_c-6}_{N_c,N_c}}, ~~~s=X,Y,
\label{eq5p444}
\end{eqnarray}
where the scale relations (\ref{eq5p210}) and (\ref{eq5p443}) 
are used. The discussion in Subsect.\,\ref{subsub532} shows that
the low energy supersymmetric $SO(4){\cong}SU(2)_X{\times}SU(2)_Y$ 
gauge theory have four vacua labelled by 
$\epsilon_X\epsilon_Y={\pm}1$. The gaugino condensation
in each $SU(2)_s$ generates the superpotential
\begin{eqnarray}
W=2\left(\langle\widetilde{\lambda}\widetilde{\lambda}{\rangle}_X
+\langle\widetilde{\lambda}\widetilde{\lambda}{\rangle}_Y\right)
=2({\epsilon}_X+{\epsilon}_Y)\widetilde{\Lambda}^3=
2({\epsilon}_X+{\epsilon}_Y)\frac{(\det M)^{1/2}}
{2^4\Lambda^{N_c-3}_{N_c,N_c}}.
\label{eq5p445}
\end{eqnarray}
(\ref{eq5p445}) shows that 
these two vacua with $\epsilon_X\epsilon_Y=1$ have the superpotential
\begin{eqnarray}
W{\approx}\pm\frac{1}{4}\frac{(\det M)^{1/2}}{\Lambda^{N_c-3}_{N_c,N_c}}
\label{eq5p446}
\end{eqnarray}
and hence are not supersymmetric vacua since 
$F_M=\partial W/\partial M$ does not vanish 
unless $M=0$. The other two vacua with 
$\epsilon_X\epsilon_Y=-1$ lead to
$W=0$
and hence lead to two supersymmetric ground states. Therefore,
there exist two vacua for $\mbox{rank}(M)=N_f$ corresponding to
the sign $\pm$ of $\pm\langle (\widetilde{W}_{\alpha})^2_X
-\widetilde{W}_{\alpha})^2_Y \rangle$, the superfield form of the
gaugino condensation $\langle\widetilde{\lambda}\widetilde{\lambda}\rangle_X
+\langle\widetilde{\lambda}\widetilde{\lambda}\rangle_Y$. By the identification
\begin{eqnarray}
B{\sim}(\widetilde{W}_{\alpha})^2_X - (\widetilde{W}_{\alpha})^2_Y,
\label{eq5p448}
\end{eqnarray}
one can see that these two vacua for $\langle M\rangle$ of rank $N_f=N_c$
correspond to the sign of $B=\pm \sqrt{\det M}$ and hence are 
identical to the classical moduli space of vacua discussed in 
Subsect.\,\ref{subsub532}.

For the case $\mbox{rank}(M)=N_f-1$, the low energy theory is a magnetic
$SO(4)$ theory with one flavour, say, $q_{N_f}$.  This low energy theory 
was discussed in Subsect.\,\ref{subsub533},
 where the low energy theory has 
no massless gauge fields but $q_i{\sim}b_i$. For the present case,
the massless magnetic composite should be something like a glueball
\begin{eqnarray}
\widetilde{q} {\sim}(\widetilde{W}_{\alpha})^2_X
- (\widetilde{W}_{\alpha})^2_Y.
\label{eq5p449}
\end{eqnarray}
According to (\ref{eq5p41}) one can 
construct a low energy effective superpotential
\begin{eqnarray}
W=\frac{1}{2\mu}M^{N_fN_f}q_{N_f}{\cdot}q_{N_f}
-\frac{1}{2\mu}q_{N_f}{\cdot}q_{N_f}\widetilde{q}^2
{\equiv}\frac{1}{2\mu}N_{N_fN_f}(M^{N_fN_f}-\widetilde{q}^2).
\label{eq5p450}
\end{eqnarray}   
Integrating out $N_{N_fN_f}=q_{N_f}{\cdot}q_{N_f}$ 
from the superpotential, we obtain 
\begin{eqnarray}
M^{N_fN_f}=\widetilde{q}^2,
\label{eq5p451}
\end{eqnarray}
so the composite field $\widetilde{q}$ of the magnetic theory is
a semi-classical construction in the electric theory. On the other hand,
(\ref{eq5p121}) implies that for 
$\mbox{rank}(M)=N_f-1=N_c-1$, the $SO(N_c)$ gauge 
symmetry is broken to $U(1)$ in the moduli space and $N_f-1$ of 
the $N_f$ quarks become massive. So the electric 
theory is completely Higgsed but one of the quarks, $q^{N_f}$,
 remains massless. Consequently, the baryon field is
\begin{eqnarray}
B=\pm\sqrt{\det M}=
\pm\sqrt{ M^{N_fN_f}},
\label{eq5p452}
\end{eqnarray} 
and thus from (\ref{eq5p449}) and (\ref{eq5p451}) we have
\begin{eqnarray}
B=\pm \widetilde{q}{\sim}(\widetilde{W}_{\alpha})^2_X 
- (\widetilde{W}_{\alpha})^2_Y.
\label{eq5p453}
\end{eqnarray}  
This means that the baryon field of the electric theory is indeed
mapped to a massless glueball 
$\widetilde{q}{\sim}(\widetilde{W}_{\alpha})^2_X 
-(\widetilde{W}_{\alpha})^2_Y$ of the magnetic theory
under the duality, as expected.

 For $\mbox{rank}(M)=N_f-2$, the low energy theory is a magnetic
$SO(4){\cong}SU(2)_X{\times}SU(2)_Y$ with two flavours 
$q_i$, $i=N_f-1,N_f$ in the representation $(2,2)$. This theory was also
discussed in Subsect.\,\ref{subsub534}. It is in the Coulomb 
phase and its exact solution
is given by the curve (\ref{eq5p3140}),
$y^2=x^3+(-U+\Lambda_X^4+\Lambda_Y^4)x^2+\Lambda_X^4\Lambda_Y^4x$.
Here for the magnetic theory  $U=\det N_{ij}$ and 
$N_{ij}=\epsilon^{rs}q_{ir}q_{js}$, $i,j=N_f-1,N_f$. 
The discriminant
\begin{eqnarray}
\Delta =[U-(\Lambda_X^2+\Lambda_Y^2)^2] [U-(\Lambda_X^2-\Lambda_Y^2)^2],
\label{eq5p455}
\end{eqnarray} 
shows that there exist massless monopoles or dyons at the points of
the moduli space $U=0$ and $U=U_1=(\Lambda_X^2+\Lambda_Y^2)^2{\equiv}
4\Lambda^4$. We only consider the massless
monopoles near $U=U_1$. According to (\ref{eq5p3149}) and 
(\ref{eq5p41}), the superpotential near $U=U_1$ should be
\begin{eqnarray}
W=\frac{1}{2\mu}\sum^{N_f}_{i,j=N_f-1}M^{ij}N_{ij}
-\frac{1}{2\mu}\left[\det (N_{ij})-U_1\right]\widetilde{E}^+\widetilde{E}^-.
\label{eq5p456}
\end{eqnarray} 
At $U=U_1$, the $N^{ij}$ equations of motion give 
\begin{eqnarray}
\langle \widetilde{E}^+\widetilde{E}^-\rangle =
\frac{M^{ij}}{U_1}=\frac{M^{ij}}{4\Lambda^4}{\sim}M^{ij}.
\label{eq5p457}
\end{eqnarray}
 This shows that the monopoles or dyons in the magnetic theory 
are identified as some of the components of the elementary quarks of
the electric theory.

When $\mbox{rank}(M)=N_f-3$, the low energy theory is a magnetic 
$SO(4){\cong}SU(2)_X{\times}SU(2)_Y$ with three flavours $q_i$. This 
is just the case discussed in the last subsection since $SU(2)_s{\cong}SO(3)$.
Due to the vanishing of the one-loop beta function coefficient: 
$\widetilde{\beta}_0=6-2N_f=0$, the low energy theory is in a free 
non-Abelian magnetic phase with the gauge group $SO(3)$ and 
three flavours of magnetic quarks. These
magnetic quarks can be identified as 
the quarks $Q^i$, $i=N_f-2$, $N_f-1$, $N_f$, of the $SO(N_c)$ 
electric theory, since along the flat directions with
$\mbox{rank}(M)=N_f-3$, the theory is Higgsed to an electric $SO(3)$ gauge 
theory. It is very interesting that these elementary
quarks and gluons emerge out of strong coupling dynamics in the dual
magnetic theory. 

For the case $\mbox{rank}(M){\leq}N_f-4$, the low energy theory is a 
magnetic theory $SO(4){\cong}SU(2)_X{\times}$ $SU(2)_Y$ with more than
three flavours. Since the one-loop beta function coefficient
$\widetilde{\beta}_0=6-2N_f <0$, the theory is either not asymptotically free 
or a free theory in the infrared region.

 Overall, above discussions shows that for $\mbox{rank}(M)=N_f$
there are two vacua, the same as in the classical case, while for
$\mbox{rank}(M) <N_f$ there is a unique ground state which can be
interpreted either as the one of the electric or as the one of 
the magnetic theory.

\vspace{2mm}
\begin{flushleft}
{\it Mass deformation}
\end{flushleft}
\vspace{2mm}

 We add as usual a large mass term to the $N_f$-th electric quark
by introducing a  tree level superpotential $W_{\rm tree}=mM^{N_fN_f}/2$.
With (\ref{eq5p41}), the full superpotential of the magnetic theory is thus
$W_{\rm full}=\frac{1}{2\mu}M^{ij}q_i{\cdot}q_j+\frac{1}{2}mM^{N_fN_f}$.
The $M^{N_fN_f}$ equation of motion gives
\begin{eqnarray}
\langle q^2_{N_f}\rangle =-\mu m.
\label{eq5p459}
\end{eqnarray}
This non-vanishing expectation value breaks the gauge symmetry
$SU(2)_X{\times}SU(2)_Y$ to the diagonal subgroup $SU(2)_d$. Consequently,
the $N_f-1$ quarks $q_{\widehat{i}r_Xr_Y}$ will be decomposed into $SU(2)_d$
triplets $\widehat{q}_{\widehat{i}}$ and singlets $S_{\widehat{i}}$ as
in (\ref{eq5p349}). The $q_{N_f}$ equations of motion give 
$M^{iN_f}q_i=0$ for any $q_i$ and hence lead to 
\begin{eqnarray}
M^{iN_f}=0.
\label{eq5p460}
\end{eqnarray}
Further, the $M^{\widehat{i}N_f}$
equations of motion give 
\begin{eqnarray}
q_{\widehat{i}}{\cdot}q_{N_f}=0,~~~\widehat{i}=1,{\cdots},N_f-1. 
\label{eq5p461}
\end{eqnarray}
With (\ref{eq5p459}), (\ref{eq5p460}) and (\ref{eq5p461}), 
the remaining low energy theory is the diagonal 
$SO(3){\cong}$ $SU(2)_d$ gauge theory with
$N_f-1$ $SO(3)$ triplets $\widehat{q}_{\widehat{i}}$ and 
the singlets $M^{\widehat{i}\widehat{j}}$, 
$\widehat{i},\widehat{j}=1,{\cdots}, N_f-1$. 
The dynamics of these fields is described by the low energy 
superpotential inherited from (\ref{eq5p41}),
\begin{eqnarray}
\widehat{W}=\frac{1}{2\mu}M^{\widehat{i}\widehat{j}}
\widehat{q}_{\widehat{i}a}\widehat{q}_{\widehat{j}a}, ~~~~a=1,2,3.
\label{eq5p462}
\end{eqnarray}
The instantons in the broken magnetic $SU(2)_X{\times}SU(2)_Y$ 
give an additional contribution to the superpotential
since they are well defined. The superpotential generated by the instantons
can be analyzed as follows. For $\det \widehat{M}{\neq}0$, the superpotential
(\ref{eq5p462}) gives masses $\mu^{-1}\widehat{M}$ to the first $N_f-1$ dual 
quarks $q_{\widehat{i}}$, hence the low energy theory has one 
dual quark $q_{N_c}$ and the $SU(2)_s$ scale, which, according to 
(\ref{eq5p210}) and (\ref{eq5p443}), is 
\begin{eqnarray}   
\widetilde{\Lambda}^5_{s,1}
=\frac{\mu^{N_c}}{2^8\Lambda^{2N_c-6}_{N_c,N_c}/\det (\mu^{-1}\widehat{M}) }=
\frac{\mu\det\widehat{M}}{2^8\Lambda^{2N_c-6}_{N_c,N_c}}.
\label{eq5p463}
\end{eqnarray}
(\ref{eq5p323}) and (\ref{eq5p324}) imply that 
the superpotential generated by the instantons in the
broken magnetic $SU(2)_s$ has the form
\begin{eqnarray}
W_{\rm inst}=2\frac{\widetilde{\Lambda}^5_{X,1}+\widetilde{\Lambda}^5_{Y,1}}
{q_{N_f}{\cdot}q_{N_f}}=4\frac{\widetilde{\Lambda}^5}{q_{N_f}{\cdot}q_{N_f}}
=-\frac{\det\widehat{M}}{2^6m\Lambda^{2N_c-6}_{N_c,N_c}},
\label{eq5p464}
\end{eqnarray}
where we have used (\ref{eq5p463}), (\ref{eq5p443}) and the decoupling 
relation (\ref{eq5p210}) for $N_f=N_c$,
\begin{eqnarray}
\Lambda^{2N_c-6}_{N_c,N_c}m=\Lambda^{2N_c-5}_{N_c,N_c-1}.
\label{eq5p467}
\end{eqnarray}  
Combining $W_{\rm inst}$ with the superpotential (\ref{eq5p462}), 
one can see that the low energy dynamics properly reproduces the 
magnetic $SO(3)$ theory with $N_f-1$ flavours with the superpotential 
(\ref{eq5p423}). Moreover, the scale relation (\ref{eq5p443}) 
reproduces the scale relation 
(\ref{eq5p424}) for the low energy electric and magnetic theories.  
First, the square of the scale relation (\ref{eq5p443}) now becomes 
\begin{eqnarray}
2^{16}\left(\widetilde{\Lambda}^{6-N_c}_{s,N_c}\right)^2
\left(\Lambda^{2N_c-6}_{N_c,N_c}\right)^2=\mu^{2N_c}.
\label{eq5p466}
\end{eqnarray} 
Then in the magnetic theory, the non-vanishing expectation value
(\ref{eq5p459}), according to (\ref{eq5p217}), leads to
\begin{eqnarray}
&&4\left(\widetilde{\Lambda}^{6-N_c}_{s,N_c}\right)^2
\left(\langle q_{N_f}{\cdot}q_{N_f}\rangle\right)^{-2}
=\widetilde{\Lambda}^{6-2(N_c-1)}_{3,N_c-1}, 
~~~\left(\widetilde{\Lambda}^{6-N_c}_{s,N_c}\right)^2
=\widetilde{\Lambda}^{6-2(N_c-1)}_{3,N_c-1}\frac{\mu^2 m^2}{2^2}.
\label{eq5p468}
\end{eqnarray}  
Inserting (\ref{eq5p467}) and (\ref{eq5p467}) into 
(\ref{eq5p466}), we immediately get (\ref{eq5p424}).

Having discussed these two specific cases, we shall
review the general $N_f>N_c$ case for which the dual theory
is an $SO(N_f-N_c+4)$ gauge theory. 

\subsubsection{$N_f >N_c$: General dual magnetic 
$SO(N_f-N_c+4)$ gauge theory}
\label{subsub544}

The superpotential in this range 
is given by (\ref{eq5p41}), which, like before, is uniquely 
determined by the symmetries and holomorphy 
around $M=q=0$. The scale relation (\ref{eq5p48})
between the electric and magnetic theories now becomes
\begin{eqnarray}
2^8 \Lambda^{3(N_c-2)-N_f}_{N_c,N_f}
\widetilde{\Lambda}^{3(N_f-N_c+2)-N_f}_{N_f-N_c+4,N_f}
=(-1)^{N_f-N_c}\mu^{N_f},
\label{eq5p469}
\end{eqnarray}  
where the normalization factor $C=1/2^8$ is chosen to get the consistent
low energy behaviour under large mass deformation and along 
flat directions.

We first have a look at the dynamical behaviour of the magnetic
$SO(N_f-N_c+4)$ gauge theory with $N_f$ flavours. Its one-loop
beta function coefficient $\widetilde{\beta}_0=3(N_f-N_c+2)-N_f$ reveals
that for $N_f{\leq}3(N_c-2)/2$, $\widetilde{\beta}_0 {\leq}0$ and hence the theory
is not asymptotically free. Consequently, in the infrared region the
theory will provide a weakly coupled magnetic description
to the strongly coupled electric theory. When $3(N_c-2)/2 <N_f <3(N_c-2)$
both the magnetic $SO(N_f-N_c+4)$ and the electric $SO(N_c)$ theories
are asymptotically free and have the same interacting infrared 
fixed point, at which both the electric and the magnetic 
descriptions are in a non-Abelian Coulomb 
phase and are physically equivalent. Due to the scale relation
(\ref{eq5p469}), the magnetic description is 
at strong coupling as the number of flavours
$N_f$ increases while the electric description is at weak coupling
and vice versa. For $N_f >3(N_c-2)$, the one-loop electric beta
function coefficient $\widetilde{\beta}_0 <0$, so the electric description is 
a free theory in the infrared region whereas the magnetic coupling
is strongly coupled. Therefore, the high energy magnetic theory
has provided a dual low energy description of the
electric theory and vice versa.

At the origin of the moduli space, the fields
$M^{ij}$, the magnetic quarks $q_i$ and the $SO(N_f-N_c+4)$ gauge
field multiplets are all massless, and the global symmetry
$SU(N_f){\times}U_R(1)$ remains unbroken. One can check that
the 't Hooft anomalies of this massless spectrum of the magnetic
theory indeed match the anomalies listed in Table \ref{ta5p3fo} 
which receives contributions from the fundamental
particles of the electric theory. The relevant particulars such as
the currents, the energy-momentum tensor parts and the anomaly
coefficients are listed in Tables \ref{ta5p4se}, 
\ref{ta5p4ei} and \ref{ta5p4ni}, respectively. 
 
 In the following, we further discuss the decoupling behaviour of the 
magnetic theory under large mass deformation and along the flat
directions and show that these phenomena indeed coincide with
the original electric description.

\begin{table}
\begin{center}
\begin{tabular}{|c|c|c|} \hline
                    & $SU(N_f)$ &  $U_R(1)$ \\ \hline
 $\psi_{M}^{ij}$    & $\overline{\psi}_M^{ij}t^A_{ij,kl}
\sigma_{\mu}\psi^{kl}_M$ &
 $(3N_f-2N_c)/N_f\overline{\psi}_M^{ij}\sigma_{\mu}\psi^{ij}_M $    
\\ \hline
$\psi_{q\widetilde{r}}^i$ & 
$ \overline{\psi}_{q\widetilde{r}}^i\sigma_{\mu}\overline{t}^A_{ij}
{\psi}_{q\widetilde{r}}^j$ & $(N_c-N_f-2)/N_f 
\overline{\psi}_{q\widetilde{r}}^i\sigma_{\mu}
{\psi}_{q\widetilde{r}}^i$ \\ \hline
$\widetilde{\lambda}^a $   & $0$       & $
\overline{\widetilde{\lambda}}^{\widetilde{a}}
\sigma_{\mu}{\widetilde{\lambda}}^{\widetilde{a}}$
\\ \hline
\end{tabular}
\caption{\protect\small Currents composed of the fermionic
components of the singlet $M$, magnetic quarks and the magnetic 
$SO(N_f-N_c+4)$ gluino corresponding to the global symmetries
$SU(N_f){\times}U_R(1)$. \label{ta5p4se} }
 \end{center}
\end{table}

\begin{table}
\begin{center}
\begin{tabular}{|c|c|} \hline
            & $T_{\mu\nu}$\\ \hline
 $\psi_M$  & $i/4\left[\left(\overline{\psi}_M^{ij}\sigma_\mu\nabla_\nu\psi_M^{ij}-
\nabla_\nu\overline{\psi}_M^{ij}\sigma_\mu\psi_M^{ij}\right)
+\left(\mu\longleftrightarrow\nu\right)\right]-g_{\mu\nu}{\cal L}[\psi_M]$  \\ \hline
$ \psi_q$  & $i/4 \left[\left(\overline{\psi}_{q\widetilde{r}}^i
\sigma_\mu\nabla_\nu\psi_{q\widetilde{r}}^i-
\nabla_\nu\overline{\psi}_{q\widetilde{r}}^i\sigma_\mu \psi_{q\widetilde{r}}^i\right)
+\left(\mu\longleftrightarrow\nu\right)\right]-g_{\mu\nu}{\cal L}[\psi_q] 
$  \\ \hline
$\widetilde{\lambda} $ &  $i/4\left[\left(\overline{\widetilde{\lambda}}^{\widetilde{a}}
\sigma_\mu\nabla_\nu \widetilde{\lambda}^{\widetilde{a}}
-\nabla_\nu\overline{\widetilde{\lambda}}^{\widetilde{a}} \sigma_\mu
\widetilde{\lambda}^{\widetilde{a}}\right)+\left(\mu\longleftrightarrow\nu\right)\right]
-g_{\mu\nu}{\cal L}[\widetilde{\lambda}] $
            \\ \hline
\end{tabular}
\caption{\protect\small Energy-momentum tensor composed 
of fermionic components 
of $M^{ij}$, $q^i_r$ and the magnetic $SO(N_f-N_c+4)$ gluino; 
${\cal L}[\psi]=i/2(\overline{\psi}_{\widetilde{r}}\sigma^\mu\nabla_\mu\psi_{\widetilde{r}}
-\nabla_\mu\overline{\psi}_{\widetilde{r}} \sigma^\mu\psi_{\widetilde{r}} )$,
$\Delta_\mu=\partial_\mu-\omega_{KL\mu}\sigma^{KL}/2$, 
$\sigma^{KL}=i/4[\sigma^K,\overline{\sigma}^L]$
and $\sigma^K=e^K_{~\mu}\sigma^{\mu}$. \label{ta5p4ei} }
 \end{center}
\end{table}

\begin{table}
\begin{center}
\begin{tabular}{|c|c|} \hline
 Triangle diagrams and      & 't Hooft anomaly \\
 gravitational anomaly      &    coefficients \\ \hline
 $ U_R(1)^3 $  & $ N_c(N_c-1)/2+N_c(2-N_c)^3N_f^2$ 
  \\ \hline
 $SU(N_f)^3 $ &  $N_c\mbox{Tr}(t^A\{t^B,t^C\})$ \\ \hline
$SU(N_f)^2U_R(1) $ &  $N_c(2-N_c)/N_f\,\mbox{Tr}(t^At^B)$\\ \hline
 $U_R(1)$ &  $-N_c(N_c-3)/2$            \\ \hline
\end{tabular}
\caption{\protect\small 't Hooft anomaly coefficients. \label{ta5p4ni}}
 \end{center}
\end{table}

\vspace{2mm}
\begin{flushleft}
{\it Flat directions} 
\end{flushleft}
\vspace{2mm}

In the flat directions parametrized by $\langle M\rangle$,
the superpotential (\ref{eq5p41}) shows 
that there are $k=N_f-\mbox{rank}(M)$ massless magnetic quarks $q_i$.
The $F$ term, i.e. the $M$ equation of motion from (\ref{eq5p461}), 
and the vanishing of the $D$-term from the gauge coupling part of 
the magnetic theory require
$\langle q_i\rangle=0$.
So along the flat directions the gauge symmetry $SO(N_f-N_c+4)$ 
remains unbroken but there are only
$k=N_f-\mbox{rank}(M)$ flavours in the low energy magnetic theory. 
In the following we discuss the dynamics of the 
low energy magnetic theory along the flat directions according to 
the rank of $M$.

 For $\mbox{rank}(M)>N_c$, more than $N_c$ quarks are massive,   
so the flavour number $k$ in the low energy magnetic theory satisfies 
$k=N_f-\mbox{rank}(M)<N_f-N_c$, and thus 
\begin{eqnarray}
k {\leq}(N_f-N_c)-1=(N_f-N_c+4)-5.
\label{eq5p471}
\end{eqnarray}
Therefore, this low energy magnetic $SO(N_f-N_c+4)$ theory with 
$k$ flavours is similar to the electric $SO(N_c)$ theory with  
$N_f{\leq}N_c-5$ flavours discussed in Subsect.\,\ref{subsub531}. Thus 
a superpotential like (\ref{eq5p31})
\begin{eqnarray}
W
=\frac{1}{2}(N_f-N_c-k+2)\epsilon_{N_f-N_c-k+2}
\left(\frac{16\Lambda^{3(N_f-N_c+2)-k}_{N_f-N_c+4,k}}
{\det M}\right)^{1/(N_f-N_c-k+2)}.
\label{eq5p472}
\end{eqnarray}
will ne generated. Therefore, there exists no supersymmetric ground state 
at $\langle q_i\rangle=0$ due to the above dynamical superpotential.

For $\mbox{rank}(M)=N_c$, the low energy magnetic $SO(N_f-N_c+4)$ gauge 
theory has
\begin{eqnarray}
k=N_f-\mbox{rank}(M)=N_f-N_c=(N_f-N_c+4)-4
\label{eq5p473}
\end{eqnarray}
flavours. Thus it is analogous to the electric $SO(N_c)$ theory with 
$N_f=N_c-4$ flavours considered in in Subsect.\,\ref{subsub532}. 
Similarly, a superpotential
\begin{eqnarray}
W=\frac{1}{2}(\epsilon_X+\epsilon_Y)
\left(\frac{16\Lambda^{2(N_f-N_c+4)}_{N_f-N_c+4, k}}{\det M}\right)^{1/2}.
\label{eq5p474}
\end{eqnarray}
arises. Consequently, two supersymmetric ground states exist at the origin 
$\langle q_i\rangle=0$, corresponding to the two sign choices for 
$\epsilon_X$ (or $\epsilon_Y$) in $\epsilon_X\epsilon_Y=-1$. There 
is no supersymmetric ground state for $\epsilon_X\epsilon_Y=1$.
The same is also true in the underlying electric theory.

For $\mbox{rank}(M)=N_c-1$, the low energy magnetic 
theory is $SO(N_f-N_c+4)$ with
\begin{eqnarray}
k=N_f-N_c+1=(N_f-N_c+4)-3
\label{eq5p475}
\end{eqnarray}  
flavours. So it is analogous to the theory considered in 
Subsect.\,\ref{subsub533} . 
Thus the low energy magnetic theory 
has no massless gauge fields but has massless composites, and they
can be interpreted  as some of the components of the elementary electric 
quarks as in (\ref{eq5p334}).

For $\mbox{rank}(M)=N_c-2$, the low energy magnetic theory is 
$SO(N_f-N_c+4)$ with $k=(N_f-N_c+4)-2$ flavours. It is analogous 
to the theory discussed in Subsect.\,\ref{subsub534}. 
A similar analysis shows 
that this magnetic theory has a massless photon which is confined
because of the existence of magnetic monopoles at the origin 
$\langle q_i\rangle=0$. This is just the dual description of the 
the corresponding low energy electric theory when 
$\mbox{rank}(M)=N_c-2$, in which  there is a massless photon 
with massless elementary quarks.

 For $3N_c/2-N_f/2-3{\leq}\mbox{rank}(M)<N_c-2$, the number
of massless flavours number $k$ lies in the range
\begin{eqnarray}
(N_f-N_c+4)-2 <k=N_f-\mbox{rank}(M){\leq}\frac{3}{2}[(N_f-N_c+4)-2].
\label{eq5p476}
\end{eqnarray} 
The low energy magnetic theory is still strongly coupled 
since the one-loop beta function coefficient 
$\widetilde{\beta}_0=3(N_f-N_c+2)-k>0$.
If we dualize this magnetic theory, 
the  electric theory will be a free $SO(N_c-\mbox{rank}(M))$ gauge theory
with $N_f-\mbox{rank}(M)$ massless quarks due to the fact that
$\beta_0 =3(N_c-\mbox{rank}(M)-2)-(N_f-\mbox{rank}(M)) <0$.
This result is obvious in the original electric description. 

 For $\mbox{rank}(M)<3N_c/2-N_f/2-3$, the $\widetilde{\beta}_0 <0$ and hence
the variables in the low energy magnetic theory are the same as the free ones.

 In summary, when $N_f>N_c$, the moduli space of supersymmetric 
vacua is described by $\langle M\rangle$, whose rank is 
at most $N_c$ along with
an additional sign when $\mbox{rank}(M)=N_c$. Thus one has obtained
a consistent description of the classical moduli space of the electric 
theory as discussed in Subsect.\,\ref{subsub512} in 
terms of the strong coupling effects 
of the magnetic theory. 

\vspace{2mm}
\begin{flushleft}
{\it Mass deformation} 
\end{flushleft}
\vspace{2mm}

Like in the above two special cases, we consider a tree-level superpotential 
$W_{\rm tree}=mM^{N_fN_f}/2$. In the electric theory this term
gives a mass to the $N_f$-th quark $Q^{N_f}$ and the corresponding 
low energy theory is an $SO(N_c)$ gauge theory with $N_f-1$ quarks.
In the magnetic theory, combining this term with the 
superpotential (\ref{eq5p41}), we have the full superpotential
$W_{\rm full}=\frac{1}{2\mu}M^{ij}q_i{\cdot}q_j+\frac{1}{2}mM^{N_fN_f}$.
The $M^{N_fN_f}$ equations of motion lead to
$\langle q_{N_f}^2\rangle =-\mu m$.
This non-vanishing expectation value further breaks
the magnetic $SO(N_f-N_c+4)$ gauge theory with $N_f$ quarks to an
 $SO(N_f-N_c+3)$ gauge theory with $N_f-1$ quarks. The $q_{N_f}$
equation of motion gives
$M^{\widehat{i}N_f}=0$, $\widehat{i}=1,{\cdots},N_f-1$,
and the $M^{\widehat{i}N_f}$ equations of motion yield
$q_{\widehat{i}}{\cdot}q_{N_f}=0$.
Therefore, the remaining low energy magnetic theory is a magnetic
$SO(N_f-N_c+3)$ gauge theory with $N_f-1$ flavours and the 
superpotential
$W_L={1}/{2\mu}M^{\widehat{i}\hat{j}}q_{\widehat{i}}{\cdot}q_{\widehat{j}}$.
This low energy magnetic theory is dual to the low energy $SO(N_c)$
gauge theory with $N_f-1$ massless quarks. This can be seen from following
two aspects. First, the scale relation (\ref{eq5p469}) for the high 
energy theories can be precisely reduced to the one that relates the 
low energy electric and magnetic theories mentioned above. (\ref{eq5p210}) 
gives the scale of the low energy electric $SO(N_c)$ 
theory with $N_f-1$ massless quarks, 
\begin{eqnarray}
\Lambda^{3(N_c-2)-(N_f-1)}_{N_c,N_f-1}=m\Lambda^{3(N_c-2)-N_f}_{N_c,N_f},
\label{eq5p483}
\end{eqnarray} 
while (\ref{eq5p214}) gives the scale of the low energy magnetic theory 
\begin{eqnarray}
\widetilde{\Lambda}^{3[(N_f-N_c+3)-2]-(N_f-1)}_{N_f-N_c+3,N_f-1}
=\Lambda^{3[(N_f-N_c+4)-2]-N_f}_{N_f-N_c+4,N_f}(-\mu m)^{-1}.
\label{eq5p484}
\end{eqnarray} 
Inserting (\ref{eq5p483}) and (\ref{eq5p484}) into 
(\ref{eq5p469}) we indeed get the scale relation
that relates the $SO(N_c)$ electric theory with $N_f-1$
flavours to the $SO(N_f-N_c+3)$ magnetic theory with $N_f-1$ flavours,
\begin{eqnarray}
2^8\Lambda^{3(N_c-2)-(N_f-1)}\widetilde{\Lambda}^{3(N_f-N_c+1)-(N_f-1)}
=(-1)^{N_f-(N_c-1)}\mu^{N_f-1}.
\label{eq5p485}
\end{eqnarray}  
Secondly, if we consider a concrete case, 
$N_f=N_c+1$, the corresponding magnetic description
is an $SO(5)$ theory with $N_f+1$ flavours. The mass 
term $mM^{N_fN_f}/2$ breaks the $SO(N_c)$ electric 
theory with $N_f=N_c+1$ flavours to the low energy
$SO(N_c)$ electric theory with $N_f=N_c$ flavours, while at the same time,
the non-vanishing expectation value $\langle q_{N_f}\rangle$ 
breaks $SO(5)$ with $N_f=N_c+1$ flavours to the low energy 
magnetic $SO(4){\cong}SU(2)_X{\times}SU(2)_Y$ with $N_f=N_c$, which 
was discussed in the last subsection. This explicit
low energy pattern coincides with the general consideration \cite{ref51}.

\subsection{Electric-magnetic-dyonic triality in supersymmetric $SO(3)$
 gauge theory}
\label{subsect55}
\renewcommand{\thetable}{5.5.\arabic{table}}
\setcounter{table}{0}
\renewcommand{\theequation}{5.5.\arabic{equation}}
\setcounter{equation}{0}

\subsubsection{Peculiarities of supersymmetric $SO(3)$ gauge theory} 
\label{subsub551}

Supersymmetric $SO(3)$ gauge theory is somehow exceptional compared to the
general cases introduced above. New non-perturbative phenomena 
arise. The most remarkable of them is the occurrence of two
theories dual to the original one. If the original 
$SO(3)$ gauge theory is ``electric", one dual theory is 
``magnetic" and the other one is called ``dyonic".
One refers to this dynamical pattern as 
electric-magnetic-dyonic triality. Moreover,
a discrete symmetry ($Z_{4N_f}$) which is explicit in the 
original $SO(3)$ theory can be realized in the dual magnetic and dyonic 
theories. However, such symmetries can not be explicitly observed in the 
dual Lagrangians because they are implemented by non-local transformations
on the fields and hence are called ``quantum" symmetries. In the following
we first give a general introduction to the special points of 
supersymmetric $SO(3)$ gauge theory and then, following the route
of Ref.\,\cite{ref1p1}, introduce some concrete cases with the definite matter 
contents.

  Let us first have a look at the discrete symmetries 
of the $SO(3)$ gauge theory with $N_f$ quarks. 
It is invariant under the $Z_2$ charge conjugation transformation 
${\cal C}$ and an enhanced $Z_{4N_f}$ symmetry,
\begin{eqnarray}
Q{\longrightarrow}e^{i2n\pi/{4N_f}}Q.
\label{eq5p51}
\end{eqnarray} 
This is because the quarks $Q^i$ are in the adjoint representation
of $SO(3)$. With the normalization of $SO(3)$ generators
$\mbox{Tr}(T^{\widetilde{a}}T^{\widetilde{b}})
=\delta^{\widetilde{a}\widetilde{b}}$, 
$\widetilde{a},\widetilde{b}=1,2,3$, the $U_A(1)$ 
operator anomaly equation is
\begin{eqnarray}
\partial^{\mu}j_{\mu}^{(A)}=4N_f\frac{1}{32\pi^2}
\epsilon^{\mu\nu\lambda\rho}F_{\mu\nu}^{\widetilde{a}}F_{\lambda\rho}^{\widetilde{a}}. 
\label{eq5p52}  
\end{eqnarray} 
This $Z_{4N_f}$ symmetry will be realized non-locally 
in the dual theories. We can roughly understand this as follows 
in the dual magnetic theory, which is an $SO(N_f+1)$ theory
with $N_f$ dual quarks $q_i$ and has the discrete symmetry
$Z_{2N_f}$ and charge conjugation $Z_2$ generated by the operation
\begin{eqnarray}
q{\longrightarrow}e^{-2i\pi/(2N_f)}{\cal C}q.
\label{eq5p53}
\end{eqnarray}
The full $Z_{4N_f}$ symmetry in this dual magnetic theory should be
generated by the `` square root " of the operation (\ref{eq5p53}). 
There are many possible choices in the ``square root" of the 
charge conjugation ${\cal C}$, but Intriligator and Seiberg, using 
the concrete examples, found that it should be a special element 
of the $SL(2,Z)$ electric-magnetic duality modular 
transformation \cite{ref51},
\begin{eqnarray}
A=TST^2S.
\label{eq5p54}
\end{eqnarray}   
The discrete $Z_{4N_f}$ is thus a non-local ``quantum symmetry".

 Another special point is that in the dual description of $SO(3)$
gauge theory a new superpotential term proportional to
\begin{eqnarray}
\det (q_i{\cdot}q_j)
\label{eq5p55}
\end{eqnarray}
arises with $q_i$ being the magnetic quarks. The necessity of adding this
superpotential term stems from ensuring that the dual of the dual 
of the $SO(3)$ gauge theory should be $SO(3)$ itself. The discussion in 
Subsect.\,\ref{subsub542} shows that 
the dual description of $SO(N_f+1)$ with $N_f$ flavours is
$SO(3)$ with $N_f$ flavours and an extra interaction term proportional 
to $\det M$. By analogy, we can see that only with the inclusion of 
(\ref{eq5p55}) the dual of the dual of the $SO(3)$ superpotential 
(\ref{eq5p423}) is identical to that of the original theory.
For $N_f{\geq}3$, the full superpotential of $SO(N_f+1)$ should be of 
the form (\ref{eq5p423}) with the replacements
\begin{eqnarray}
 M^{ij}&{\longleftrightarrow}&q^i{\cdot}q^j;\nonumber\\
\Lambda^{2N_c-5}_{N_c,N_c-1}=\Lambda^{2(N_f-1)-1}_{N_f+1,N_f}
&{\longleftrightarrow}& -\widetilde{\Lambda}^{2(N_f-1)-1}_{N_f+1,N_f},
\label{eq5p56}
\end{eqnarray}
that is,
\begin{eqnarray}
W=\frac{1}{2\mu}M^{ij}q_i{\cdot}q_j
+\frac{1}{2^6\widetilde{\Lambda}^{2(N_f-1)-1}_{N_f+1,N_f}}
\det (q_i{\cdot}q_j).
\label{eq5p57}
\end{eqnarray}
The scale relation (\ref{eq5p424}) 
should remain the same with
the exchange $\widetilde{\Lambda}{\leftrightarrow}{\Lambda}$
since now $SO(3)$ is the original theory, and thus we get 
the scale relation
\begin{eqnarray}
2^{14}(\widetilde{\Lambda}^{2(N_f-1)-1}_{N_f+1,N_f})^2
{\Lambda}^{6-2N_f}_{3,N_f}=\mu^{2N_f},
 \label{eq5p58}
\end{eqnarray}  
and its square root
\begin{eqnarray}
2^{7}\epsilon \widetilde{\Lambda}^{2(N_f-1)-1}_{N_f+1,N_f}
{\Lambda}^{3-N_f}_{3,N_f}=(-1)^{3-N_f}\mu^{N_f},
 \label{eq5p59}
\end{eqnarray} 
where $\epsilon=\pm 1$ comes from taking the square root
of the instanton factor ${\Lambda}^{3-N_f}_{3,N_f}$ of the
electric $SO(3)$ theory and the phase $(-1)^{3-N_f}$ preserves
the relation (\ref{eq5p59}) along the flat directions 
and under mass deformation. 

 The term (\ref{eq5p55}) in the superpotential 
(\ref{eq5p57}) brings some new phenomena
into the dual $SO(N_f+1)$ theory. Despite the invariance 
of the term (\ref{eq5p55}) under the global symmetry 
$SU(N_f){\times}U_R(1)$, it breaks some of 
the discrete symmetries, since under the transformation
$q_i{\longrightarrow}e^{i2n\pi/(4N_f)}$
\begin{eqnarray}
\det (q_i{\cdot}q_j){\longrightarrow}e^{in\pi}
\det (q_i{\cdot}q_j).
 \label{eq5p511}
\end{eqnarray}
This shows that only the $Z_{2N_f}$ subgroup of the $Z_{4N_f}$
symmetry and the charge conjugation ${\cal C}$ remain unbroken. Due to 
the anomaly (\ref{eq5p19}) in the $SO(N_f+1)$ gauge theory, except for
$N_f=1,2$, the transformations $q_i{\longrightarrow}e^{i2\pi/(4N_f)}q_i$ 
shift the vacuum angle $\theta$:
\begin{eqnarray}
\theta{\longrightarrow}\theta+2N_f \frac{2\pi}{4N_f}=\theta +\pi.
\label{eq5p512}
\end{eqnarray} 
Therefore, the symmetry transformation 
$q_i{\longrightarrow}e^{i2\pi/(4N_f)}q_i$ in the original electric
$SO(3)$ theory, which should also exist in the dual $SO(N_f+1)$ theory,
changes the sign of (\ref{eq5p55}) (i.e. the $n=1$ case of (\ref{eq5p511})) 
and shifts the vacuum angle by $\pi$ (see (\ref{eq5p512})) for 
$N_f{\neq}1,2$. Intriligator and Seiberg interpreted
this phenomenon as follows \cite{ref51}.  
The original electric $SO(3)$ theory 
has, in fact, two dual descriptions corresponding to the two signs
of this term for $N_f{\neq}1,2$. One of them is
``magnetic", which is the electric $SO(N_f+1)$ theory of 
discussed in Subsect.\,\ref{subsub542}; The 
other dual theory is ``dyonic", which will be reviewed later. These
two dual theories are related to another by a duality 
transformation. We will see that these triality transformations
are the extension of the $Z_2$ group of $N=1$ duality transformations
to the modular transformation group $SL(2,Z)$. Precisely speaking, 
it is the subgroup of $SL(2,Z)$,
\begin{eqnarray}  
S_3{\cong}SL(2,Z)/\Gamma (2),
\label{eq5p513}
\end{eqnarray}
which permutes these three theories,
where $\Gamma(2)$ is the subgroup of $SL(2,Z)$ generated by the 
monodromies (\ref{eq5p3120}). In one word, the full $Z_{4N_f}$ 
includes the moduli transformation which exchanges the magnetic
and dyonic theories and which appears as a quantum symmetry in the dual
descriptions. 

 Since the theories with $N_f=1,2$ are exceptional cases, we first
discuss how the quantum symmetry is realized in these two cases
and then turn to the cases with more flavours.

\subsubsection{One-flavour case: Abelian Coulomb phase 
and quantum symmetries}
\label{subsub552}

Since the quark now is in the adjoint representation of the $SO(3)$ group,
the model is just the $N=2$ theory discussed by Seiberg
and Witten \cite{ref1p1}. The theory has a quantum moduli space labeled by the 
expectation value $u=\langle M\rangle/2=Q^2/2$. The $SO(3)$
gauge symmetry is spontaneously broken to $SO(2){\cong}U(1)$ on this
moduli space. Consequently, the low energy theory has a Coulomb phase
with a massless photon. As discussed in Subsect.\,\ref{subsub534}, 
the low energy gauge 
coupling is given by the algebraic curve solution (\ref{eq5p3121}),
\begin{eqnarray}
y^2=x^3-\frac{1}{2}Mx^2+\frac{1}{4}\Lambda^4_{3,1}x.
\label{eq5p514}
\end{eqnarray} 
One usually chooses for convenience
the normalization $\Lambda_{3,1}{\rightarrow}2\Lambda_{3,1}$ and
$M{\rightarrow}2M$. With this convention the curve solution changes
to
\begin{eqnarray}
y^2=x^3-Mx^2+4\Lambda^4_{3,1}x.
\label{eq5p515}
\end{eqnarray}
As discussed in Subsect.\,\ref{subsub534}, the singularities 
of the quantum moduli space is given by the zeroes 
$M_{\pm}{\equiv}\pm 4\Lambda^2_{3,1}$
of the discriminant
$\Delta (M)=M^2-16\Lambda^4_{3,1}$.
There exists a pair of massless magnetic monopoles $q^{\pm}_{(+)}$ 
at $M_+=4\Lambda^2_{3,1}$ and a pair of massless dyons $q^{\pm}_{(-)}$
at $M_-=-4\Lambda^2_{3,1}$. Correspondingly, the effective superpotentials,
for the monopoles and dyons, according to (\ref{eq5p3151}), are, 
respectively
\begin{eqnarray}
W_+=f_+\left(\frac{M}{\Lambda^2_{3,1}}\right)q^+_{(+)}q^-_{(+)};
~~~W_-=f_-\left(\frac{M}{\Lambda^2_{3,1}}\right)q^+_{(-)}q^-_{(-)}.
\label{eq5p518}
\end{eqnarray}
Near the singularities $M=M_{\pm}$, from (\ref{eq5p3149}), the 
superpotentials are, respectively,
\begin{eqnarray}
W_+&{\approx}&\frac{1}{2\mu}\left(M-4\Lambda^2_{3,1}\right)
q^+_{(+)}q^-_{(+)};
~~~W_-{\approx}\frac{1}{2\mu}\left(M+4\Lambda^2_{3,1}\right)
q^+_{(-)}q^-_{(-)}.
\label{eq5p519}
\end{eqnarray}
Comparing (\ref{eq5p519}) with (\ref{eq5p57}), one can see 
that the first term is the $Mq^+q^-$
term and the second one is (\ref{eq5p55}).

 Let us see what the quantum symmetry in this model is. Classically,
the $N=2$ $SU(2)$ gauge theory has a global $R$-symmetry 
$SU_R(2){\times}U_R(1)$ \cite{ref1p1a,so}. 
At the quantum level, the $U_R(1)$ symmetry
is broken by an anomaly. Since all the fermionic fields, which include
the gaugino and quarks, are in the adjoint
representation and their transformations under $U_R(1)$ have the same
form, the operator anomaly equation for this $R$-current
in the adjoint representation is 
\begin{eqnarray}
\partial^{\mu}{\cal J}_{\mu}^{(R)}=4N_c\frac{1}{32\pi^2}
\epsilon^{\mu\nu\lambda\rho}F_{\mu\nu}^aF_{\lambda\rho}^a=
8\frac{1}{32\pi^2}
\epsilon^{\mu\nu\lambda\rho}F_{\mu\nu}^aF_{\lambda\rho}^a, ~~~a=1,2,3.
\label{eq5p520}
\end{eqnarray}
This anomaly shifts the vacuum angle:
$\theta{\longleftrightarrow}\theta+8\alpha$,
and hence the $U_R(1)$ is broken to $Z_8^R$ under which the field
$Q$ and its fermionic component transform as
\begin{eqnarray}
Q{\longrightarrow}e^{2in\pi/4}Q, 
~~~\psi_Q{\longrightarrow}e^{2in\pi/8}\psi_Q,
~~~ n=1,2,{\cdots},8.
\label{eq5p522}
\end{eqnarray}
The index $R$ in $Z_8^R$ indicates that it is the remanent from
the $U_R(1)$. 
Since the center elements of $SU_R(2)$ are contained in $Z_8^R$, the full 
global symmetry including the charge conjugation ${\cal C}$ is
\begin{eqnarray}
\left(\left(SU_R(2){\times}Z_8^R\right)/Z_2\right){\times}{\cal C}.
\label{eq5p523}  
\end{eqnarray}
 
 Now let us observe how the discrete symmetries of (\ref{eq5p523}) 
are realized in a given vacuum. From (\ref{eq5p522}), the $Z_8^R$ generator 
$e^{2i\pi/8}$ acts on the scalar component of $M$ as the 
operation $R$,
\begin{eqnarray}
R:~M{\longrightarrow}e^{i\pi}M=-M.
\label{eq5p524}
\end{eqnarray}
Hence the $Z_8^R$ symmetry is spontaneously broken 
to $Z_4^R$ for $M{\neq}0$ since
the elements $e^{2in\pi/8}$ of $Z_8^R$ with $n$ even still
leave $M$ invariant. At the origin $M=0$ of the moduli space 
the full $Z_8^R$ symmetry is restored.
It should be emphasized that for a given vacuum, i.e. 
a given point on the moduli space, only a $Z_4^R$ symmetry is left of
the $U_R(1)$ symmetry, while on the whole moduli space the full $Z_8^R$
symmetry is still preserved. $R^2$ acts as charge conjugation
on the scalar component of $Q$, so does on $Q$,
\begin{eqnarray}
R^2:~Q{\longrightarrow}e^{i\pi}Q=-Q.
\label{eq5p525}
\end{eqnarray}
Thus for $M{\neq}0$ this remaining $Z_4^R$ should be generated 
by $R^2{\cal C}$. 
Intriligator and Seiberg further observed that the generators 
of $Z_8^R$ includes an $SL(2,Z)$ modular
transformation \cite{ref51}
\begin{eqnarray}
w=RA
\label{eq5p526}
\end{eqnarray}
with
\begin{eqnarray}
A=(TS)^{-1}S(TS)=\left(\begin{array}{cc} -1 & 1 \\ -2 & 1\end{array}\right),
\label{eq5p527}
\end{eqnarray}     
where $T$ and $S$ are the $SL(2,Z)$ generators 
given in (\ref{eq5p363}). Since that $A^2=-1={\cal C}$, 
${\cal C}$ being the charge conjugation generators. Thus, $w$ is
the square root of the $Z_4^R$ generator,  $R^2{\cal C}$, i.e.
a generator of $Z_8^R$. 

The following consideration shows why 
the moduli transformation $A$ necessarily appears 
in $w$. Consider the central charge $Z=a n_e+a_Dn_m$ of the $N=2$ 
superalgebra at $M=0$. The action
of $U_R(1)$ on the supercharge (two-component form), ${\cal Q}$, is
${\cal Q}{\longrightarrow}e^{i\alpha}{\cal Q}$.
After $U_R(1)$ is broken to $Z_8^R$, the above transformation is 
carried out by the elements of $Z_8^R$ and is generated 
by the transformation 
$R:{\cal Q}{\longrightarrow}e^{2i\pi/8}{\cal Q}$.
Due to the $N=2$ superalgebra, 
$Z{\sim}\{{\cal Q},{\cal Q}\}$,
and $Z$ must transform under $w$ as
$Z{\longrightarrow}e^{i\pi/2}Z=iZ$.
On the other hand, from the integral expressions 
(\ref{eq5p3114}) for $a(M)$ and $a_D(M)$,
\begin{eqnarray}
a_D &=& \frac{\sqrt{2}}{\pi}\int^{M/(4\Lambda^2_{3,1})}_1
\frac{dx \sqrt{x-M/(4\Lambda^2_{3,1})}}{\sqrt{x^2-1}},\nonumber\\
a &=& \frac{\sqrt{2}}{\pi}\int^{+1}_{-1}
\frac{dx \sqrt{x-M/(4\Lambda^2_{3,1})}}{\sqrt{x^2-1}},
 \label{eq5p532}
\end{eqnarray}
it is easily seen that near $M=0$ \cite{ref51}
\begin{eqnarray}
a^{\prime}n_e^{\prime}+a_D^{\prime}n_e^{\prime}=i(an_e+a_Dn_m),
\label{eq5p533}
\end{eqnarray} 
if $n_e^{\prime}$ and $n_m^{\prime}$ are related to $n_e$ and $n_m$ by
the modular transformation $A=(TS)^{-1}S(TS)$. Since the modular 
transformation $A$ can be rewritten as ${\cal C}T(S^{-1}T^2S)$ whereas
according to (\ref{eq5p380}) $S^{-1}T^2S$ is the 
monodromy around $M_+=4\Lambda^2_{3,1}$, thus $A$ is actually 
congruent to ${\cal C}T$ modulo the multiplication by 
the monodromy \cite{ref51}.

  If $M{\neq}0$, the $Z_8^R$ symmetry is broken to 
$Z_4^R$. The broken $Z_8^R$ generator $w$ maps the massless monopole at 
the singular point $M=4\Lambda^2_{3,1}$ to the massless dyon at 
$M=-4\Lambda^2_{3,1}$. At the origin the $Z_8^R$ symmetry
is restored and these massless soliton states are degenerate and are 
mapped from one to another by the $Z_8^R$ symmetry. Since the monopoles
and dyons are collective excitations, the fields representing them
are not local, so it is not possible to give $w$ 
a local realization. This fact further confirms that
$A$ should be a modular transformation. In particular, $A$ cannot be
diagonalized by an $SL(2,Z)$ transformation, i.e. we cannot find
a $2{\times}2$ matrix $X$ with integer elements to make $A$ diagonal
by the operation $X^{-1}AX$. This means that there exists no photon field
multiplet which is invariant under the action of $A$, since the photon
supermultiplet is the only local object in the low energy theory. Therefore, even 
$A$ cannot be realized locally in the low energy effective 
theory \cite{ref51}.

  Now we consider how the above non-local symmetry is reflected in
the dynamics (\ref{eq5p518}) of monopoles and dyons. 
The interpretation is as follows. 
The electric $SO(3)$ theory has two dual theories,
one of them, which is called  the magnetic dual, describes the 
physics around $M=4\Lambda^2_{3,1}$ with the superpotential
$W_+$ in (\ref{eq5p518}). The other dual theory, which can be called the
dyonic dual, gives the physics near $M=-4\Lambda^2_{3,1}$
with the superpotential $W_-$ in (\ref{eq5p518}). 
The magnetic dual is related to
the electric theory by the modular transformation $S$ of $SL(2,Z)$ 
modulo $\Gamma(2)$, while the dyonic dual is related to the electric
description by the $SL(2,Z)$ transformation $ST$ modulo $\Gamma (2)$.

\subsubsection{Two-flavour case: non-Abelian Coulomb phase and quantum
symmetries} 
\label{subsub553}
 
In this case the classical moduli space is parametrized by the
expectation value of the colour singlets $M^{ij}=Q^i{\cdot}Q^j$,
$i,j=1,2$, which transform as the adjoint representation of the global 
$SU(2)_f$ flavour symmetry. For $M^{ij}{\neq}0$ the $SU(2)$ gauge symmetry
is completely broken and the theory is in the Higgs phase. On the
submanifold of the moduli space where $\det M=0$ , the rank of $M$ should be
$1$, and the $SU(2)$ gauge symmetry will break to $U(1)$.
There now only exists a photon supermultiplet in the low energy theory, 
and the theory is in a Coulomb phase \cite{ref51}. 

There are not only these two phases in the low energy theory, a
confining phase can also arise. To see this, we add a 
tree level superpotential $W_{\rm tree}=M^{ij}m_{ij}/2$. 
For $\det m{\neq}0$, both $Q_1$ and $Q_2$ get a mass, 
and after integrating out the massive matter fields, the low energy
theory is an $N=1$ pure $SU(2)$ gauge theory, which is known to be in the
confining phase. For $\det m =0$ but $m{\neq}0$, only one of 
the matter fields gets a mass, the low energy theory is the 
$N=2$ supersymmetric Yang-Mills theory discussed by Seiberg 
and Witten. Since now the matter field is in the adjoint representation, 
the theory is in the Coulomb phase as shown in Ref.\,\cite{ref1p1}. 
Thus we see that the theory can be in three different phases, the 
Coulomb, confining and the Higgs phases. Note that since the theory 
has no field in the fundamental representation of gauge group, as 
discussed in Subsect.\,\ref{subsect27}, the 
confining phase and the Higgs phases are distinct. The 
order parameter, the Wilson loop, obeys an area law in the 
confining phase, but a perimeter law in the Higgs phase.
    
 In the following we discuss the dynamics of each phase. Let us first
consider the confining phase. The low energy
effective Lagrangian for $N=1$ supersymmetric $SU(N_c)$ 
gauge theory was constructed in the 1980s from the $U_R(1)$ 
anomaly \cite{akmrv,veya}. Since locally $SO(3){\cong}SU(2)$, the low energy
effective action for the present $N=1$ pure $SO(3)$ Yang-Mills gauge 
theory is the same as in the pure $SU(2)$ case \cite{ref525,in}:
\begin{eqnarray}
W_{\rm eff}=S\left[\ln\left(\frac{\Lambda_{3,0}^6}{S^2}\right)+2\right]
=S\left[\ln\left(\frac{\Lambda_{3,2}^2(\det m)^2}{S^2}\right)+2\right],
\label{eq5p534}
\end{eqnarray}
where $S=W^{\alpha}W_{\alpha}$ is the composite (glueball) field
and we have used the decoupling relation (\ref{eq5p212}) for the
$SO(3)$ theory,
$\Lambda^{2}_{3,2}(\det m)^2=\Lambda^{6}_{3,0}$.
According to Intriligator's  ``integrating in" technique \cite{in}, 
the low energy superpotential for the $SO(3)$ gauge
theory with matter fields can be obtained from 
\begin{eqnarray}
W=W_{\rm eff}+\frac{1}{2}M^{ij}m_{ij}
=S\left[\ln\left(\frac{\Lambda_{3,2}^2(\det m)^2}{S^2}\right)+2\right]
-\frac{1}{2}M^{ij}m_{ij}
\label{eq5p536}
\end{eqnarray} 
by integrating out $m_{ij}$. $\partial W/\partial m_{ij}=0$ gives 
\begin{eqnarray}
m_{ij}=4SM_{ij}^{-1}
\label{eq5p537}
\end{eqnarray}
and hence 
\begin{eqnarray}
\det m= \frac{16S^2}{\det M}.
 \label{eq5p538}
\end{eqnarray}
Inserting (\ref{eq5p537}) and (\ref{eq5p538}) into (\ref{eq5p536})
we obtain the low energy superpotential
for the $SO(3)$ gauge theory with two flavours,
\begin{eqnarray}
W=S\left[\ln\left(\frac{16^2\Lambda_{3,2}^2S^2}{(\det M)^2}\right)
-2\right].
\label{eq5p539}
\end{eqnarray}
With the inclusion of the mass term for the two flavours, the full
low energy superpotential is
\begin{eqnarray}
W_{\rm full}=S\left[\ln\left(\frac{16^2\Lambda_{3,2}^2S^2}{(\det M)^2}\right)
-2\right]+\frac{1}{2}M^{ij}m_{ij}.
 \label{eq5p540}
\end{eqnarray}
Because $S$, as a glueball field, is always massive, it should be 
integrated out. The equation of motion for $S$, 
$\partial W_{\rm full}/\partial S=0$, gives
\begin{eqnarray}
S=\pm\frac{\det M}{16\Lambda_{3,2}}.
\label{eq5p541}
\end{eqnarray} 
Thus after integrating out $S$ the full superpotential for the confining 
phase is
\begin{eqnarray}
W_{\rm full}=\mp \frac{\det M}{8\Lambda_{3,2}}+\frac{1}{2}\mbox{Tr}mM.
 \label{eq5p542}
\end{eqnarray} 

The dynamics of the Coulomb and the Higgs phase can be discussed
as follows. Integrating out $M$ from (\ref{eq5p542})
we obtain 
\begin{eqnarray}
\langle M_{ij}\rangle =\pm 4\left(\Lambda_{3,2}\det m\right)m^{-1}_{ij},
~~\langle S\rangle =\pm \Lambda_{3,2}\det m.
 \label{eq5p543}
\end{eqnarray}
Taking
\begin{eqnarray}
m=\left(\begin{array}{cc} 0 & 0 \\ 0 & m_{22}\end{array}\right),
 \label{eq5p544}
\end{eqnarray}
i.e. choosing only the second flavour $Q_2$ to 
be massive. After integrating out $Q_2$,
the low energy theory is the Seiberg-Witten model and hence is in the 
Coulomb phase \cite{ref1p1}. Note that now $\det M=0$ since 
from (\ref{eq5p543}) $\det m=0$. Nevertheless, (\ref{eq5p543}) 
and (\ref{eq5p544}) give 
\begin{eqnarray}
\mbox{Tr}(Q_1)^2=\mbox{Tr}M_{11}=\pm 4\Lambda_{3,2}m_{22}
=\pm 2\Lambda_{3,1}^2,
 \label{eq5p545}
\end{eqnarray}
where we have used the scale relation (\ref{eq5p212}) between 
$\Lambda_{3,2}$ and $\Lambda_{3,1}$.
The exact solution is given by the curve (\ref{eq5p515}),
$y^2=x^3-M_{11}x^2+4\Lambda_{3,2}m_{22}x$.
There exist massless monopoles and dyons at the points 
$M_{11}=\pm 4m_{22}\Lambda_{3,2}$.
The Higgs phase is represented by the generic points in the moduli 
space with mass matrix $m=0$ (i.e. $m_{ij}=0$). 

 In summary, the above discussion has shown that 
all the three phases can be described by the superpotential
\begin{eqnarray}
W=e\frac{\det M}{8\Lambda_{3,2}}+\frac{1}{2}\mbox{Tr}m M,
 \label{eq5p547}
\end{eqnarray}
where $e=0,\pm 1$ labels the three branches. The branch with
$e=0$ describes the Higgs or Coulomb phases of the theory.
Both of these phases are obtained for $\det m=0$. For 
$m=0$, the generic point in the moduli space is in the Higgs 
phase. When only $m_{22}{\neq}0$, the low energy theory is in
the Coulomb phase, and it has a massless monopole
at the point $M_{11}=4m_{22}\Lambda_{3,2}$ and a massless
dyon at the point  $M_{11}=-4m_{22}\Lambda_{3,2}$. When
$\det m{\neq}0$, the monopole (the dyon) will condense and lead
to confinement (oblique confinement). $e=-1$ represents
the confining branch and  $e=1$ the oblique confinement
branch \cite{ref51}. 

 Now let us turn to the dual description of the electric $SO(3)$ theory
with $N_f=2$. It has two dual theories and they are $SO(3)$
gauge theories with two flavours. The holomorphy, dimensional analysis,
the global and gauge symmetries determine the superpotentials of
the dual theories, which should take following form:
\begin{eqnarray}
W=\frac{2}{3\mu}\mbox{Tr}\left(Mq{\cdot}q\right)+
\epsilon\left(\frac{8\widetilde{\Lambda}_{3,2}}{3\mu^2}\det M+
\frac{1}{24 \widetilde{\Lambda}_{3,2}}\det (q {\cdot}q)\right),
 \label{eq5p548}
\end{eqnarray}
where $\epsilon =\pm 1$ labels the two dual theories and the scales
of the theories are related by the square root of (\ref{eq5p424}),
\begin{eqnarray}
64{\Lambda}_{3,2} \widetilde{\Lambda}_{3,2}=\mu^2.
 \label{eq5p549}
\end{eqnarray}
The sign ambiguity introduced by taking the square root is reflected 
in (\ref{eq5p548}) by the sign $\epsilon$. The first term 
in (\ref{eq5p548}) is the standard one (\ref{eq5p41}) of the dual 
theory. The $\det M$ term, from (\ref{eq5p423}), should appear for
$N_f=N_c-1$ and the $\det (q{\cdot}q)$ term is as in (\ref{eq5p55}). 
The coefficients in (\ref{eq5p548}) and (\ref{eq5p549}) are chosen 
to guarantee the duality; later we shall explain in detail why they 
are chosen as those given in (\ref{eq5p548}).   

 The dual magnetic theory should also have three branches. In 
comparison with (\ref{eq5p547}), the superpotential is not only 
of the form (\ref{eq5p548}), there should also emerge another term 
characterizing the branches. Thus the full superpotential of the dual 
magnetic theory should be
\begin{eqnarray}
W_{\widetilde{e}}=\frac{2}{3\mu}\mbox{Tr}(MN)+{\epsilon}\left(
\frac{8\widetilde{\Lambda}_{3,2}}{3\mu^2}\det M
+\frac{1}{24 \widetilde{\Lambda}_{3,2}}\det N\right)+
\widetilde{e}\frac{\det M}{8 \widetilde{\Lambda}_{3,2}},
 \label{eq5p550}
\end{eqnarray}
where $N_{ij}=q_i{\cdot}q_j$ and $\widetilde{e}=0,{\pm}1$ labels the 
three branches.
Integrating out $N_{ij}$ by the equations of motion for $N_{ij}$
and adding a $Q^i$ mass term, $W_{\rm tree}=\mbox{Tr}(m M)/2$, 
we immediately obtain
\begin{eqnarray}
W_{\widetilde{e}}&=&\frac{8 \widetilde{\Lambda}_{3,2}}{\mu^2}
\left(\frac{\widetilde{e}-\epsilon}{1+3\widetilde{e}\epsilon}\right)
+\frac{1}{2}\mbox{Tr}(m M).
 \label{eq5p551}
\end{eqnarray}
This superpotential is the same as (\ref{eq5p547}) with the identification
\begin{eqnarray}
e=\frac{\widetilde{e}-\epsilon}{1+3\widetilde{e}\epsilon}.
 \label{eq5p552}
\end{eqnarray}
Eq.\,(\ref{eq5p551}) shows that the coefficients in (\ref{eq5p548}) are chosen
 to guarantee the identification of this dual potential with (\ref{eq5p547}).

The $\widetilde{e}=0$ branch describes the 
the weakly coupled Higgs phase of the dual theory.
On the other hand, when $\widetilde{e}=0$ (\ref{eq5p552}) gives 
$e=-\epsilon =\mp 1$. From the above discussion, we know that 
in the electric theory this corresponds to the two strongly coupled branches. 
In particular, when $\epsilon=1$, the dual theory is in the 
Higgs phase, and $e=-1$ states that the corresponding
electric theory is in the confining phase. The Higgs branch
of the $\epsilon=-1$ dual theory describes the oblique confinement
branch of the electric theory since now $e=1$. Therefore, the $\epsilon=1$
branch should be regarded as the magnetic dual and  the $\epsilon=-1$
branch as the dyonic dual. These are  exactly the duality patterns 
introduced in Subsect.\,\ref{subsub551}. 

The two other branches of the dual theories are strongly coupled. 
The branches with $\widetilde{e}=\epsilon$, which describe
the oblique confinement of the magnetic theory and confinement 
of the dyonic theory, give the $e=0$ branch and hence 
correspond to the Higgs branch of the electric theory. The 
branches with $\widetilde{e}=-\epsilon$, which correspond to
the confinement phase of the magnetic theory and the oblique 
confinement phase of the dyonic theory, also yield ${e}=\epsilon$. 
and hence give another description of the strongly coupled branches of
the electric theory \cite{ref51}. 

 Overall, in this two-flavour case, there are three equivalent 
theories: electric, magnetic and dyonic. The moduli space of 
each of them has three branches: Higgs, confinement and 
oblique confinement. The map between
the branches of the different theories is the $S_3$ permutation
given by (\ref{eq5p552}). Let us argue this from the 
discrete symmetry, along the line of the discussion 
in Subsect.\,\ref{subsub552}. The electric theory 
has a $Z_8^R$ symmetry generated by $Q{\longrightarrow}e^{2i\pi/8}Q$ 
and charge conjugation ${\cal C}$. In the magnetic theory 
the $Z_8^R$ symmetry takes the form: $M{\longrightarrow}e^{2i\pi/4}M$ and 
$q{\longrightarrow}e^{-2i\pi/8}AQ$, where $A$ is the non-local moduli
transformation and satisfies that $A^2={\cal C}$. Thus the $Z_8^R$ 
symmetry is a ``quantum symmetry''.
  
 In the following we explicitly verify the above duality patterns by working
out one example, the Coulomb phase of the electric theory. 
This phase is obtained by adding
$W_{\rm tree}=mM^{22}/2$
to (\ref{eq5p550}) with $\widetilde{e}=\epsilon$ 
and integrating out the massive 
fields. The equations of motion for $M^{22}$, $N_{22}$, $M^{12}$ and 
$N_{12}$ from the superpotential 
\begin{eqnarray}
W_{\widetilde{e}=\epsilon}+W_{\rm tree}=
\frac{2}{3\mu}\mbox{Tr}(MN)+\epsilon \left(
\frac{8\widetilde{\Lambda}_{3,2}}{3\mu^2}\det M
+\frac{1}{16 \widetilde{\Lambda}_{3,2}}\det N\right)+\frac{1}{2}mM^{22}
\label{eq5p554}
\end{eqnarray} 
give, respectively,
\begin{eqnarray}
 \frac{2}{3\mu}(q_2{\cdot}q_2)
+\frac{8\epsilon\widetilde{\Lambda}_{3,2}}{3\mu^2}M^{11}+\frac{1}{2}m=0,
~~ M^{22}=-\frac{\epsilon}{16\widetilde{\Lambda}_{3,2}}
\frac{q_1{\cdot}q_1}{\mu},
~~ q_1{\cdot}q_2=0, ~~M^{12}=0.
\label{eq5p555}
\end{eqnarray}
(\ref{eq5p555}) shows that for 
$M^{11}+3{\epsilon}m\mu^2/(16\widetilde{\Lambda}_{32})$${\neq}0$, 
the expectation value of $q_2$ will break the gauge group to $SO(2)$. 
(\ref{eq5p554}) and (\ref{eq5p555}) lead to the 
low energy superpotential
for the remaining massless fields $q_1$, denoted as $q_1^+$ and $q_1^-$
according to their $SO(2)$ charges,
\begin{eqnarray}
W_L&=&\frac{1}{2\mu}\left(M^{11}
-m\frac{\epsilon\mu^2}{16\widetilde{\Lambda}_{3,2}}\right)q_1^+q_1^-
=\frac{1}{2\mu}\left(M^{11}-4\epsilon m {\Lambda}_{3,2}\right)q_1^+q_1^-,
\label{eq5p556}
\end{eqnarray}
where the relation (\ref{eq5p549}) was used. 
This low energy superpotential
will be modified by quantum corrections from instantons in the broken
magnetic $SO(3)$ part. For large $m$, the instanton contribution is
small and can be ignored. The superpotential shows that
the low energy theory has massless fields $q_1^{\pm}$ at 
$M^{11}=4\epsilon m\Lambda_{3,2}=4\epsilon m\Lambda^2_{3,1}$. These
massless fields can be interpreted as the monopoles for $\epsilon =1$
and the dyons for $\epsilon =-1$ of the $N_f=1$ case. This is
precisely the dual interpretation according to which the $\epsilon=1$ branch
is the magnetic description and the $\epsilon=-1$ the dyonic one. 

In addition to the above two monopoles in the Coulomb phase,
one can still find other monopoles in 
the $N_f=1$ theory arising from the strong coupling 
dynamics of the dual theories. 
The superpotential (\ref{eq5p554}) shows that
$q_1$ gets an effective mass 
\begin{eqnarray}
\widetilde{m}=\frac{4M^{11}}{3\mu}
+\frac{\epsilon}{12\widetilde{\Lambda}_{3,2}}u
\label{eq5p557}
\end{eqnarray}
for 
$u=q_2^2{\neq}0$.
Thus $q_1$ should be integrated out first and an $N_f=1$ theory is 
obtained. (\ref{eq5p557}) and (\ref{eq5p212}) give 
the scale of the low energy magnetic theory 
\begin{eqnarray}
\widetilde{\Lambda}_{3,1}^4=\left(\frac{4}{3\mu}
\widetilde{\Lambda}_{3,2}+\frac{\epsilon}{12}u\right)^2.
\label{eq5p559}
\end{eqnarray}
There should exist massless monopoles at
\begin{eqnarray} 
u&=&\pm 4\widetilde{\Lambda}_{3,1}^2=\mp 4\left(\frac{4}{3\mu}
\widetilde{\Lambda}_{3,2}+\frac{\epsilon}{12}u\right)
=\frac{16 \widetilde{\Lambda}_{3,2}M^{11}}{\mu (\pm 3-\epsilon)}.
\label{eq5p560}
\end{eqnarray}
The $M^{22}$ equation of motion in (\ref{eq5p555}) gives
\begin{eqnarray} 
\mu^{-1} u=-\frac{4\epsilon\widetilde{\Lambda}_{3,2}}{\mu^2}M^{11}
-\frac{3}{4}m=-\frac{\epsilon}{16 {\Lambda}_{3,2}}M^{11}-\frac{3}{4}m.
 \label{eq5p561}
\end{eqnarray}
From (\ref{eq5p560}) and (\ref{eq5p561}), we obtain
\begin{eqnarray} 
M^{11}=4m{\Lambda}_{3,2}\frac{\mp 3+\epsilon}{1\pm \epsilon}.
\label{eq5p562}
\end{eqnarray} 
For $m{\neq}0$, since $\epsilon =\pm 1$, the above equation gives only
one solution:
\begin{eqnarray} 
M^{11}= -4{\epsilon}m{\Lambda}_{3,2}=-4{\epsilon}{\Lambda}_{3,1}^2.
\label{eq5p563}
\end{eqnarray}  
Therefore, another monopole of the $N_f=1$ theory has been found,
whose existence is a consequence of the strong coupling dynamics 
of the dual theory.

 In the $m=0$ case, a similar analysis shows that
there exists a strongly coupled state in the dual theories along the 
flat directions with $\det M=0$. This state can be interpreted as the 
massless quark of the electric theory. 
 
 Finally, we consider the dual of the dual description (\ref{eq5p548}) 
and see how the triality behaves \cite{ref51}. From the scale relation 
(\ref{eq5p549}), the dual of the dual theories (\ref{eq5p548}) should 
be an $SO(3)$ theory with $N_f=2$ quarks, say $d^i$, gauge singlet 
fields $M$ and $N$ and the scale
\begin{eqnarray} 
\widetilde{\widetilde{\Lambda}}_{3,2}={\Lambda}_{3,2}.
 \label{eq5p564}
\end{eqnarray}
By analogy with (\ref{eq5p548}), the superpotential should be  
\begin{eqnarray} 
W&=&\frac{2}{3\mu}\mbox{Tr}(MN)
+\epsilon\left(\frac{8\widetilde{\Lambda}_{3,2}\det M}{3\mu^2}+
\frac{\det N}{24\widetilde{\Lambda}_{3,2}}\right)
-\frac{2}{3\mu}\mbox{Tr}[N(d{\cdot}d)]\nonumber\\
&& +\eta\left(
\frac{8\widetilde{\widetilde{\Lambda}}_{3,2}\det N}{3\mu^2}
+\frac{\det(d{\cdot}d)}{24\widetilde{\Lambda}_{3,2}}\right)\nonumber\\
&=&\frac{2}{3\mu}\mbox{Tr}(MN)+\epsilon\left(\frac{\det M}{24{\Lambda}_{3,2}}
+\frac{\det N}{24\widetilde{\Lambda}_{3,2}}\right)
-\frac{2}{3\mu}\mbox{Tr}[N(d{\cdot}d)]\nonumber\\
&&+\eta\left(\frac{\det N}{24\widetilde{\Lambda}_{3,2}}
+\frac{\det(d{\cdot}d)}{24\Lambda_{3,2}}\right),
\label{eq5p565}
\end{eqnarray}
where $\epsilon=\pm 1$ and $\eta=\pm 1$ label different duals. There are
two possible choices for $\epsilon$ and $\eta$. The first one is
$\epsilon=-\eta$. Then the superpotential shows that $N$ is a Lagrangian
multiplier implementing the constraint
$M=d{\cdot}d$
and consequently the superpotential is 
$W=0$.
Thus with this choice the dual of dual theory is just the original
electric theory with $d^i=Q^i$. The other choice is $\epsilon=\eta$, and 
the superpotential (\ref{eq5p565}) becomes
\begin{eqnarray} 
W=\frac{2}{3\mu}\mbox{Tr}(MN)-\frac{2}{3\mu}\mbox{Tr}[N(d{\cdot}d)]
+\epsilon\left(\frac{\det M}{24\Lambda_{3,2}}
+\frac{\det N}{12\widetilde{\Lambda}_{3,2}}
+\frac{\det(d{\cdot}d)}{24\Lambda_{3,2}}\right).
 \label{eq5p568}
\end{eqnarray}
Now we integrate out $N_{ij}$. The equations of motion 
for $N_{ij}$ give
\begin{eqnarray}
\det N N_{ij}^{-1}&=&\frac{8\widetilde{\Lambda}_{3,2}}{\mu}
(d_i{\cdot}d_j-M_{ij}),\nonumber\\
N_{ij}&=&{\Lambda}_{3,2}^2\left[\det (d_i{\cdot}d_j-M_{ij})\right]
(d_i{\cdot}d_j-M_{ij})^{-1}.
\label{eq5p569}
\end{eqnarray}
Inserting (\ref{eq5p569}) into (\ref{eq5p568}) we get the superpotential
\begin{eqnarray}
W=\frac{\epsilon}{12 {\Lambda}_{3,2}}M^{ij}{\epsilon}_{ik}{\epsilon}_{jl}
(d^k{\cdot}d^l)-\frac{\epsilon}{24 \Lambda_{3,2}}
\left[\det M+(d{\cdot}d)\right].
\label{eq5p570}
\end{eqnarray}
(\ref{eq5p570}) does not describe new dual theories. Defining
\begin{eqnarray}
q_i{\equiv}{\epsilon}_{ij}\sqrt{\epsilon}
\left(\frac{\Lambda_{3,2}}{\widetilde{\Lambda}_{3,2}}\right)^{1/4}d^j
\label{eq5p571}
\end{eqnarray}
and using the scale relation (\ref{eq5p549}), we rewrite 
(\ref{eq5p570}) in terms of the new variables,
\begin{eqnarray}
W&=&\frac{1}{12\Lambda_{3,2}}
\left(\frac{\Lambda_{3,2}}{\widetilde{\Lambda}_{3,2}}\right)^{1/2}M^{ij}
(q^i{\cdot}q^j)-\frac{\epsilon}{24 {\Lambda}_{3,2}}\left[\det M+
\frac{\Lambda_{3,2}}{\widetilde{\Lambda}_{3,2}}\det (q{\cdot}q)\right]
\nonumber\\
&=&\frac{2}{3\mu}\mbox{Tr}[M(q{\cdot}q)]+{\epsilon}\left[
\frac{8\widetilde{\Lambda}_{3,2}}{3\mu^2}\det M+
\frac{24 \widetilde{\Lambda}_{3,2}}\det(q{\cdot}q)\right].
\label{eq5p572}
\end{eqnarray}
This is precisely the magnetic dual superpotential 
(\ref{eq5p548}). Thus it does not
give a new dual description.  

In summary, the supersymmetric $SO(3)$ 
gauge theory with $N_f=2$ has three equivalent descriptions: the original
electric theory and the two magnetic dual ones. The above discussions show
that taking the duals of the dual theories permutes these three
descriptions.

\subsubsection{$N_f=N_c=3$ case: Identification of $N=1$ duality
  with $N=4$ Montonen-Olive-Osborn duality} 
\label{subsub554}

 This case has several special points \cite{ref51}. First, 
the one-loop beta function vanishes,
and hence the bare coupling constant $\tau_0=\theta_0/(2\pi)+4i\pi/(g_0^2)$
is not renormalized at one-loop. The two-loop beta function is negative,
so the theory is not asymptotically free and the theory is free in
the infrared region; Secondly, the (magnetic and dyonic) dual descriptions 
are $SO(4){\cong}SU(2)_X{\times}SU(2)_Y$ gauge theories, thus 
we should discuss the duality for each $SU(2)_s$ branch. In particular,
since the electric quark superfields $Q$ can be written as a $3{\times}3$
square matrix in terms of their flavour and colour indices, the theory
allows a cubic superpotential $\sim\det Q$ at tree level. With this 
cubic superpotential,  the $N=1$ duality is in fact
the duality proposed by Montonen and Olive \cite{ref1p6} 
and found by Osborn \cite{ref1p7} in $N=4$ supersymmetric 
Yang-Mills theory. 

 Like in the discussions in previous sections, 
the various symmetries and holomorphy as well as
the mass dimension determine the superpotential of the magnetic and
dyonic theories to be
\begin{eqnarray}
W=\frac{1}{2\mu}\mbox{Tr}[M(q{\cdot}q)]
+\frac{1}{64\widetilde{\Lambda}_{s,3}}\det(q{\cdot}q), ~~~~~s=X,Y, 
\label{eq5p573}
\end{eqnarray}
where $q{\cdot}q={\epsilon}^{\widetilde{r}_X\widetilde{s}_X}
{\epsilon}^{\widetilde{r}_Y\widetilde{s}_Y}q_{\widetilde{r}_X\widetilde{r}_Y}
q_{\widetilde{s}_X\widetilde{s}_Y}$
since the magnetic quarks $q$ belong to the fundamental representation of
each $SU(2)_s$. The scales $\widetilde{\Lambda}_{s,3}$ of the magnetic $SU(2)_s$
are chosen to be equal and are given by the relation
\begin{eqnarray}
\epsilon 2^7e^{i\pi\tau_0}\widetilde{\Lambda}_{s,3}^3=\mu^3,
\label{eq5p574}
\end{eqnarray}
where $\tau_0$ is the bare gauge coupling mentioned above, and 
$\epsilon=\pm 1$ shows that the term $e^{i\pi\tau_0}$ is the square root
of the $SO(3)$ instanton factor. The sign of $\epsilon$ determines
whether the dual theory is magnetic or dyonic. Note that in the two-flavour
case $\epsilon$ only appears in the superpotential 
(\ref{eq5p548}) and not in
the the scale relation (\ref{eq5p549}). 
However, here $\epsilon$ appears in the scale relation
so that it can relate the instanton factor for the $SU(2)_s$ to the 
square root of the instanton factor for the electric $SO(3)$ theory.

 As usual, we analyze the duality by looking at the flat directions
and mass deformation. The analysis in the flat directions is similar to
the general $N_f=N_c$ case and the $\det (q{\cdot}q)$ term in the 
superpotential does not modify the analysis essentially. Thus, in 
the following we only discuss the mass deformation.

Introducing a large mass term $W_{\rm tree}=mM^{33}/2$ for 
the third flavour, according to (\ref{eq5p212}) the decoupling 
of this heavy flavour in the electric theory will lead to a low 
energy theory with two flavours and the scale
 \begin{eqnarray}
 \Lambda^2_{3,2}=m^2e^{2i\pi\tau_0}.
\label{eq5p575}
\end{eqnarray}
In the dual theory, adding 
the above mass term to the superpotential (\ref{eq5p573}), we have
\begin{eqnarray}
W_{\rm full}=\frac{1}{2\mu}\mbox{Tr}[M(q{\cdot}q)]
+\frac{1}{64\widetilde{\Lambda}^3_{s,3}}\det(q{\cdot}q)
+\frac{1}{2}mM^{33}.
\label{eq5p576} 
\end{eqnarray}
The equation of motion for $M^{33}$ from (\ref{eq5p576}) gives
the expectation value,
$\langle q_3^2\rangle =-\mu m$,
which breaks the dual $SO(4){\cong}SU(2)_X{\times}SU(2)_Y$ gauge group
to a diagonal subgroup $SO(3)_d$. (\ref{eq5p217}) gives the decoupling 
relation of the dual theory
\begin{eqnarray}
4(\widetilde{\Lambda}^3_{s,3})^2(\mu m)^{-2}=\widetilde{\Lambda}_{3,2}^2.
\label{eq5p578}
\end{eqnarray}  
(\ref{eq5p574}), (\ref{eq5p575}) and (\ref{eq5p578}) 
immediately yield the relation (\ref{eq5p549}) between
the scale $\widetilde{\Lambda}_{3,2}$ of the low energy magnetic theory
and the scale ${\Lambda}_{3,2}$ of low energy electric theory. This
is another argument for the correctness of the relation (\ref{eq5p574}). 
Integrating out the massive field $M^{33}$ using the equations of motion 
for $N_{33}=q_3{\cdot}q_3$, we have
\begin{eqnarray}
M^{33}=-\frac{\mu}{32 \widetilde{\Lambda}^3_{s,3}}\det (q{\cdot}q) 
(q^3{\cdot}q^3)^{-1}.
\label{eq5p579}
\end{eqnarray} 
Inserting (\ref{eq5p579}) into (\ref{eq5p576}), we obtain 
\begin{eqnarray}
W=\frac{1}{2\mu}\mbox{Tr}[\widehat{M}(\widehat{q}{\cdot}\widehat{q})]+
\frac{1}{2}mM^{33}
=\frac{1}{2\mu}\mbox{Tr}[\widehat{M}(\widehat{q}{\cdot}\widehat{q})]+
\frac{\epsilon}{32 \widetilde{\Lambda}_{3,2}}
\det (\widehat{q}{\cdot}\widehat{q})
\label{eq5p580}
\end{eqnarray}
where we have taken into account (\ref{eq5p574}) and (\ref{eq5p578}). 
$\widehat{q}$ denotes the
two light flavours. In addition, the contribution generated
by instantons in the broken part of the $SU(2)_X{\times}SU(2)_Y$
gauge group should be included. To get the instanton contribution,
we introduce the Lagrangian multiplier field $L$ and rewrite
the second term as $\mbox{Tr}[L(\widehat{q}{\cdot}\widehat{q})]$
$-32\epsilon\widetilde{\Lambda}_{3,2}\det L$. The original term can 
be recovered upon integrating out $L$. Consequently, the superpotential 
(\ref{eq5p580}) is rewritten as
\begin{eqnarray}
W=\frac{1}{2\mu}\mbox{Tr}[\widehat{M}(\widehat{q}{\cdot}\widehat{q})]
+\mbox{Tr}[L(\widehat{q}{\cdot}\widehat{q})]
-32\epsilon\widetilde{\Lambda}_{3,2}\det L.
\label{eq5p581}
\end{eqnarray}
(\ref{eq5p581}) implies that $\widehat{q}^i_{r_Xr_Y}$ get the effective 
mass $2L+\widehat{M}/\mu$ and hence that
they should be integrated out. The low energy theory is a pure $SO(3)_d$
gauge theory with the scale given by (\ref{eq5p212}),
\begin{eqnarray}
\widetilde{\Lambda}^6_{3,0}=\widetilde{\Lambda}^2_{3,2}
\left[\det\left(\frac{\widehat{M}}{\mu}
+2L\right)\right]^2=\widetilde{\Lambda}^2_{3,2}
\frac{[\det(\widehat{M}+2{\mu} L)]^2}{\mu^4}.
\label{eq5p582}
\end{eqnarray}
A superpotential contribution by instantons is
\begin{eqnarray}
W_{\rm ins}=2\epsilon \widetilde{\Lambda}^3_{3,0}
=\frac{2\epsilon \widetilde{\Lambda}_{3,2}}{\mu^2}
\det(\widehat{M}+2{\mu} L).
\label{eq5p583}
\end{eqnarray}
The whole low energy superpotential is the combination of this 
instanton generated one and (\ref{eq5p583}),
\begin{eqnarray}
W=\frac{1}{2\mu}\mbox{Tr}[(\widehat{M}+2\mu L)(\widehat{q}{\cdot}\widehat{q})]
-32\epsilon\widetilde{\Lambda}_{3,2}\det L
+\frac{2\epsilon \widetilde{\Lambda}_{3,2}}{\mu^2}
\det(\widehat{M}+2{\mu} L).
\label{eq5p584}
\end{eqnarray}
Integrating out $L$, ( \ref{eq5p584}) gives
\begin{eqnarray}
W=\frac{2}{3\mu}\mbox{Tr}\left[\widehat{M}(\widehat{q}{\cdot}\widehat{q})
\right]+\epsilon \left[\frac{8 \widetilde{\Lambda}_{3,2}}{3\mu^2}
\det\widehat{M}+\frac{1}{24 \widetilde{\Lambda}_{3,2}}
\det(\widehat{q}{\cdot}\widehat{q})\right],
\label{eq5p585}
\end{eqnarray}
which is just the superpotential of the dual $SO(3)$ gauge theory of
the $N_f=2$ case given by (\ref{eq5p548}).

  Next we add a perturbation to  the electric theory in the form of a 
cubic superpotential \cite{ref51} 
\begin{eqnarray}
W_{\rm tree}=\beta \det Q,
\label{eq5p586}
\end{eqnarray}
where  $\beta$ corresponds to the Yukawa coupling. 
This kind of superpotential is special for 
$N_f=3$ since only in this case $Q$ can be written as a square matrix
in terms of its flavour and colour indices. If we choose
$\beta=\sqrt{2}$,
the gauge coupling will be identical to the Yukawa coupling,
and the theory is very similar to the $N=4$ $SO(3)$ supersymmetric 
Yang-Mills theory since all of these three flavours are in the 
adjoint representation and the one-loop beta function is zero. 
However, in general this theory is not identical to the $N=4$ 
supersymmetric $SO(3)$ 
Yang-Mills theory since the two-loop beta function is negative,
and thus only in the infrared region, the theory with the cubic 
superpotential agrees with the $N=4$ supersymmetric Yang-Mills theory.   
The choice $\beta=\sqrt{2}$ means that we rescale $Q$ as
\begin{eqnarray}
Q{\longrightarrow}\left(\frac{\sqrt{2}}{\beta}\right)^{1/3} Q.
\label{eq5p588}
\end{eqnarray}
Due to the non-vanishing two-loop beta function, there is
a conformal anomaly connected to the scale transformation (\ref{eq5p588})
in the infrared region \cite{ref1p15}, which is proportional to 
\begin{eqnarray}
4\ln\left(\frac{\sqrt{2}}{\beta}\right)
\int d^2\theta (W^{\alpha}W_{\alpha})+\mbox{h.c.}\,.
\label{eq5p589}
\end{eqnarray}
This leads to a relation between $\tau_E$, which is the effective gauge 
coupling in the infrared region, and $\tau_0=\theta/(2\pi)+4i\pi/g_0^2$,
the bare gauge coupling: 
\begin{eqnarray}
e^{2i\pi\tau_E}=e^{2i\pi\tau_0}\left(\frac{\beta}{\sqrt{2}}\right)^4=
\frac{1}{4}e^{2i\pi\tau_0}\beta^4,
\label{eq5p590}
\end{eqnarray} 
since the classical Lagrangian takes the form 
${1}/{g_0^2}\int d^2\theta (W^{\alpha}W_{\alpha})+\mbox{h.c.}$\,.

 What are the effects of the cubic superpotential in magnetic theory?
(\ref{eq5p420}) and the second relation in (\ref{eq5p421}) mean that 
the electric operator $\det Q$ is mapped to $(\widetilde{W}_\alpha)^2_X-
(\widetilde{W}_\alpha)^2_Y$. Therefore, the addition of the above cubic 
superpotential to the electric theory makes the magnetic 
$SU(2)_X{\times}SU(2)_Y$ theory have 
$\widetilde{\Lambda}^3_{X,3}{\neq} \widetilde{\Lambda}^3_{Y,3}$. 
Consequently, the various symmetries determine that the superpotential
(\ref{eq5p573}) should be modified to
\begin{eqnarray}
W=\frac{1}{2\mu}\mbox{Tr}[M(q{\cdot}q)]+\frac{2\epsilon e^{i\pi\tau_0}}{\mu^3}
f(\tau_E,\epsilon)\det (q{\cdot}q).
 \label{eq5p591}
\end{eqnarray}  
Similarly, the scale relation (\ref{eq5p574}) is modified to
\begin{eqnarray}
\epsilon 2^7e^{i\pi\tau_0}\widetilde{\Lambda}^3_{s,3}g_s(\tau_E,\epsilon)=
\mu^3,
\label{eq5p592}
\end{eqnarray} 
where $f$ and $g_s$ are functions of $\tau_E$ and 
$\epsilon$ whose explicit forms are not known.
When $\beta=0$, the scales should coincide: $\widetilde{\Lambda}^3_{X,3}=
\widetilde{\Lambda}^3_{Y,3}$, and thus from (\ref{eq5p590}),
(\ref{eq5p591}) and (\ref{eq5p592}) we should have
\begin{eqnarray}
\tau_E=i\infty, ~~~~f(\tau_E,\epsilon)|_{\tau_E=\infty}
=g_s(\tau_E,\epsilon)|_{\tau_E=\infty}=1.
\label{eq5p593}
\end{eqnarray} 

 In the infrared region the magnetic $SO(4){\cong}SU(2)_X{\times}SU(2)_Y$ 
theory with $\widetilde{\Lambda}_X{\neq} \widetilde{\Lambda}_Y$ also flows 
to the $N=4$ supersymmetric gauge theory. This can be observed from the 
$\widetilde{\Lambda}_X{\gg}\widetilde{\Lambda}_Y$ limit. 
We denote $\tau_E$ in this limit
by $\tau_*$. Since supersymmetric $SU(2)$ gauge theory with $N_f=3$
is free in the infrared region, for $\widetilde{\Lambda}_Y=0$
the $SU(2)_Y$ gauge symmetry has become a global symmetry and hence
is not a dynamical symmetry any more. Consequently, the magnetic quarks
$q_{r_Xr_Y}^i$ can be written as $q^A_{r_X}$, $A=1,{\cdots},6$. 
Thus the magnetic theory is an $SU(2)_X$ theory with six doublets coupled 
through the superpotential (\ref{eq5p591}). 
This superpotential breaks the global
$SU(6)$ to $SU(3)_f{\times}SU(2)_Y$ under which the $SU(2)_X$ doublets
$q_{r_X}$ are in the representation $(\overline{3},2)$. The strong
$SU(2)_X$ dynamics confines them to be the meson fields
$N_{ij}{\equiv}q_i{\cdot}q_j$  in the representation $(\overline{6},1)$, and
$\phi^i$ in the representation $(3,3)$ of $SU(3)_f{\times}SU(2)_X$.
This can be seen from the decompositions of the fundamental
representations of  $SU(3){\times}SU(3)$
and $SU(2){\times}SU(2)$: $3{\times}3=3{\oplus}6$, $2{\times}2=1{\oplus}3$.
The above global and $SU(2)_X$ gauge symmetries and the holomorphy 
determine that the interaction of these fields should be 
given by the superpotential
\begin{eqnarray}
W_{\rm int}=-\frac{1}{2}\frac{N_{ij}}{\widetilde{\Lambda}_X}
(\phi^i{\cdot}\phi^j)+\frac{1}{8}\frac{\det N}{\widetilde{\Lambda}_X^3}
+2\det \phi,
\label{eq5p594}
\end{eqnarray}
where the $\phi^i$ have been rescaled to have 
mass dimension $1$ instead of $2$. 
Combining (\ref{eq5p594}) with
(\ref{eq5p591}) and adding a mass term $\mbox{Tr}(mM)/2$, we have the full 
superpotential
\begin{eqnarray}
W_{\rm full}&=&\frac{1}{2\mu}M^{ij}N_{ij}
+\frac{2{\epsilon}e^{i\pi\tau_0}}{\mu^3}
\left[f(\tau_*,\epsilon)+2^3 g_X(\tau_*,\epsilon)\right]\det N\nonumber\\
&&-\frac{1}{2}\frac{N_{ij}(\phi^i{\cdot}\phi^j)}{\widetilde{\Lambda}_X}
+2\det \phi+\frac{1}{8}\frac{\det N}{\widetilde{\Lambda}_X^3}
+\frac{1}{2}\mbox{Tr}(mM).
\label{eq5p595}
\end{eqnarray}  

 Now we gauge the group $SU(2)_Y$, 
i.e. localize this group and 
introduce new degrees of freedom, the $SU(2)_Y$ gauge fields.
If the energy is higher than $\widetilde{\Lambda}_X$,
the coupling of the $SU(2)_Y$ gauge theory is very weak and 
runs with the scale $\widetilde{\Lambda}_Y$. If the energy 
is much lower than $\widetilde{\Lambda}_X$, 
the $SU(2)_Y$ gauge fields will couple to the three triplets
$\phi^i$. The coupling $\tau_Y$ will become very strong 
and will not run. Thus it should satisfy 
\begin{eqnarray}
e^{2i\pi\tau_Y}{\sim}\frac{\widetilde{\Lambda}_Y^3}
{\widetilde{\Lambda}_X^3}.
\label{eq5p596}
\end{eqnarray}  
This means that for $\widetilde{\Lambda}_Y{\ll}\widetilde{\Lambda}_X$, 
$\tau_Y{\simeq}i\infty$. 

The fields $M$ and $N$ in (\ref{eq5p595}) are massive and should 
be integrated out. The $M_{ij}$ equations of motion set
\begin{eqnarray}
 N_{ij}=-\mu m_{ij},
\label{eq5p597}
\end{eqnarray}  
and the $N_{ij}$ equations of motion give
\begin{eqnarray}
 M_{ij}=\frac{\phi^i{\cdot}\phi^j}{\widetilde{\Lambda}_L}
+\frac{4\epsilon e^{i\pi\tau_0}}{\mu^2}
\left[f(\tau_*,\epsilon)+2^3 g_L(\tau_*,\epsilon)\right]m_{ij}^{-1}.
\label{eq5p598}
\end{eqnarray}  
After integrating out $M$ and $N$ the superpotential 
(\ref{eq5p595}) becomes
\begin{eqnarray}
W=2\det\phi +\frac{1}{2}\frac{\mu}{\widetilde{\Lambda}_X}m_{ij}
(\phi^i{\cdot}\phi^j)-2\epsilon e^{i\pi\tau_0}
\left[f(\tau_*,\epsilon)+2^3 g_X(\tau_*,\epsilon)\right]\det m.
\label{eq5p599}
\end{eqnarray}  
For $m=0$, this superpotential is proportional to the Yukawa potential
$\det \phi$. Thus in the infrared region the magnetic theory can be
identified with an $N=4$ $SU(2)$ supersymmetric gauge theory with weak 
coupling $\tau_Y$.

 The above discussion on the infrared magnetic theory was based on the 
limit $\widetilde{\Lambda}_X{\gg}\widetilde{\Lambda}_Y$.  
Away from this limit, in the infrared magnetic theory 
also flows to an $N=4$ supersymmetric gauge theory with the coupling 
$\tau_Y$ being a function of $\tau_E$. 

It should be emphasized that the $N=4$ supersymmetric theory with $\tau_Y$,
as the infrared limit of the magnetic theory, is the not the same
as the $N=4$ supersymmetric gauge theory with $\tau_E$ given by 
(\ref{eq5p590}), i.e. the infrared limit of the original electric theory. 
The original electric $N=4$ supersymmetric gauge theory, 
with coupling $\tau_E$, is weakly coupled for $\beta{\ll}1$; this 
can be seen from (\ref{eq5p590}):
\begin{eqnarray}
\tau_E{\sim}\frac{2}{i\pi}\ln \beta\sim\infty.
\label{eq5p5100}
\end{eqnarray} 
(Note that $\tau=\theta/(2\pi)+4i\pi/g^2$.) The magnetic $N=4$ 
theory, with coupling $\tau_Y$, is strongly coupled for $\beta{\ll}1$. 
This is because when  $\beta{\ll}1$, the mapping from the electric 
operator to the magnetic one leads to
\begin{eqnarray}
\beta\det Q{\longrightarrow}\beta 
\left[(\widetilde{W}_{\alpha})^2_X
-\widetilde{W}_{\alpha})^2_Y\right]{\simeq}0.
\label{eq5p5101}
\end{eqnarray}  
Consequently
\begin{eqnarray}
\widetilde{\Lambda}_X{\approx}\widetilde{\Lambda}_Y,
\label{eq5p5102}
\end{eqnarray}    
and hence according to (\ref{eq5p596}), 
$\tau_Y{\approx}0$. Conversely, the magnetic 
$N=4$ gauge theory is weakly coupled when 
$\widetilde{\Lambda}_X{\gg}\widetilde{\Lambda}_Y$. This limit occurs
for $\beta\sim 1$, where the original electric $N=4$ theory is weakly 
coupled. Based on this fact, Intriligator and Seiberg assumed that 
\cite{ref51}
\begin{eqnarray}
\tau_E=-\frac{1}{\tau_R}.
\label{eq5p5103}
\end{eqnarray} 
In this sense the $N=1$ duality can be interpreted as
a generalization of the $N=4$ duality proposed by Osborn based 
on the Montonen-Olive conjecture. 

 In this duality, the meson operator $M^{ij}$ of the 
electric theory can be related to the corresponding operator 
of the magnetic theory in the $\tau_Y{\rightarrow}i\infty$ limit
by differentiating the superpotential (\ref{eq5p599}) 
with respect to $m_{ij}$,
\begin{eqnarray}
M_{ij}=\frac{\mu}{\widetilde{\Lambda}_X}\left(\phi^i{\cdot}\phi^j\right)
-4 \epsilon e^{i\pi\tau_0}\left[f(\tau_*)+2^3g_X(\tau_*)\right]
\det (m)m^{-1}_{ij}.
 \label{eq5p5104}
\end{eqnarray} 
One notices that $M^{ij}$ has not the simple form of
$\left(\phi^i{\cdot}\phi^j\right)$ but  is shifted. A similar shift
was observed by Seiberg and Witten in discussing the flow from
the $N=4$ to the $N=2$ theory when $m$ has one vanishing
eigenvalue and the two other eigenvalues are equal. This gives a strong support
to the above assumption that $N=1$ duality is related to $N=4$ 
duality \cite{ref51}.

\subsection{Dyonic dual of $SO(N_c)$ theory with $N_f=N_c-1$ flavours}
\label{subsect56}
\renewcommand{\thetable}{5.6.\arabic{table}}
\setcounter{table}{0}
\renewcommand{\theequation}{5.6.\arabic{equation}}
\setcounter{equation}{0}

 It was shown in the last section that there exists a dyonic dual 
description in the $SO(3)$ gauge theory, which is a new duality 
phenomenon. In this section we shall explore some aspects of this 
dual theory in more detail such as its flat directions and mass 
deformation etc.

 We start from the electric $SO(N_c)$ ($N_c>4$) theory 
with $N_f=N_c-1$ flavours discussed in Subsect.\,\ref{subsub542}. 
Its dual magnetic description is an $SO(3)$ gauge theory 
with $N_f$ quarks and the superpotential (\ref{eq5p423}). 
Now we consider the dual of this magnetic theory \cite{ref51}. 
The discussion in Subsect.\,\ref{subsub542} shows that the
$SO(3)$ theory has both magnetic and dyonic dual descriptions.
Both of them are $SO(N_c)$ gauge theories with $N_f$
matter fields $d^i$, gauge singlet fields $M^{ij}$ and $N_{ij}$ and 
the superpotential
\begin{eqnarray}
W=\frac{1}{2\mu}\mbox{Tr}\left[N(M-d{\cdot}d)\right]-
\frac{1}{2^6\Lambda^{2(N_c-2)-1}_{N_c,N_c-1}}\left[\det M
-\epsilon \det (d{\cdot}d)\right]
\label{eq5p61}
\end{eqnarray}
according to  (\ref{eq5p423}), 
where $\epsilon =\pm 1$ and the scales satisfy
\begin{eqnarray} 
\widetilde{\widetilde{\Lambda}}^{2(N_c-2)-1}_{N_c,N_c-1}=\epsilon
 {\Lambda}^{2(N_c-2)-1}_{N_c,N_c-1}
\label{eq5p62}
\end{eqnarray}
due to the scale relations (\ref{eq5p424}) and (\ref{eq5p59}). 
The equation of motion
from (\ref{eq5p61}) yields $M^{ij}=d^i{\cdot}d^j$. One can easily 
see that the theory (\ref{eq5p61}) with $\epsilon =1$ gives $W=0$ 
and (\ref{eq5p62}) shows that
\begin{eqnarray}
\widetilde{\widetilde{\Lambda}}^{2(N_c-2)-1}_{N_c,N_c-1}=
{\Lambda}^{2(N_c-2)-1}_{N_c,N_c-1}.
\label{eq5p63}
\end{eqnarray}
 Thus this is just the  original electric theory with the matter fields
identified with the electric quarks $Q^i$. On the other hand,
the theory (\ref{eq5p61}) with $\epsilon=-1$ has the superpotential
\begin{eqnarray}
W=-\frac{1}{32 {\Lambda}^{2(N_c-2)-1}_{N_c,N_c-1}}\det (d{\cdot}d)=
\frac{1}{32 \widetilde{\widetilde{\Lambda}}^{2(N_c-2)-1}_{N_c,N_c-1}}
\det (d{\cdot}d)
 \label{eq5p64}
\end{eqnarray}   
and the scale
\begin{eqnarray}
\widetilde{\widetilde{\Lambda}}^{2(N_c-2)-1}_{N_c,N_c-1}=
-{\Lambda}^{2(N_c-2)-1}_{N_c,N_c-1}
=e^{i\pi}{\Lambda}^{2(N_c-2)-1}_{N_c,N_c-1}.
\label{eq5p65}
\end{eqnarray}
According to (\ref{eq3p3}),  we have
\begin{eqnarray}
\tau =\frac{\theta}{2\pi}+\frac{4\pi}{g^2}i{\sim}
\frac{1}{2i\pi}\ln\Lambda^{\beta_0}_{N_c,N_f}
=\frac{1}{2i\pi} \ln {\Lambda}^{3(N_c-2)-N_f}_{N_c,N_f} 
=\frac{1}{2i\pi} \ln {\Lambda}^{2(N_c-2)-1}_{N_c,N_c-1},
\label{eq5p66}
\end{eqnarray}
so the scale difference (\ref{eq5p65}) with the original electric theory
means a shift of the vacuum angle $\theta$ by $\pi$,
\begin{eqnarray}
\theta{\longrightarrow}\theta +\pi.
\label{eq5p67}
\end{eqnarray}  
Therefore, the theory with the superpotential (\ref{eq5p64}) is 
called the dyonic description of the original electric theory. Let 
us next consider the physics in the flat directions of the dyonic
dual theory and the effects of mass deformation.

\vspace{2mm}
\begin{flushleft}
{\it Flat directions}
\end{flushleft}
\vspace{2mm}

The flat directions of this theory are very subtle. The natural 
variables parametrizing the classical moduli space are the 
singlet fields $M^{ij}=d^i{\cdot}d^j$. 
The $M$ equation of motion from (\ref{eq5p64}) 
gives $\det M=0$. Since $M$ is an $N_f{\times}N_f$
matrix, one might conclude that the classical moduli space of vacua
is given by all the values of $M^{ij}$ subject to the constraints
$\det M=0$, i.e. $\mbox{rank}(M){\leq}N_f-1$,  and
the gauge symmetry at most breaks to $SO(2){\cong}U(1)$. 
However, this conclusion is not correct since from (\ref{eq5p64}) 
this theory is scale $\Lambda$ dependent. This can be made more
clear from the following arguments \cite{ref51}. Consider the flat
directions where $M$ is diagonal and has $N_f-1$ non-zero
equal eigenvalues $a$. In the case of the vacua far away 
from the origin of moduli space, i.e. $a{\gg}\Lambda_{N_c,N_c}^2$,
the $SO(N_c)$ gauge symmetry will break to $SO(2){\cong}U(1)$ due to
the non-vanishing $\langle M\rangle$. From the superpotential 
(\ref{eq5p64}) and the Higgs mechanism,
some quarks will acquire masses of order $a^{N_f-1}/\Lambda^{2N_f-3}$
while the massive gauge bosons are much lighter, their masses are of 
the order $\sqrt{a}$. In the case that the energy of the theory lies 
between these two values, $\Lambda^2_{N_c,N_f}<q^2<a$, the gauge 
symmetry is neither broken nor are the quarks in the vector representation
of $SO(N_c)$. This occurs because the interaction described by the 
superpotential (\ref{eq5p64}) is not renormalizable. Therefore, the above
symmetry breaking pattern cannot be applied to the large $a$ case. 
On the other hand, if the the expectation values of $d^i$ are very large,
the gauge symmetry is broken at a high energy scale and the $SO(N_c)$ 
gauge interaction is weak. This is because its one-loop beta function 
is positive. However, in this case the superpotential (\ref{eq5p64}) 
leads to strong coupling for the massive fields so that they cannot be
easily integrated out and hence the classical analysis gives the wrong 
conclusion.
   
  What will happen if we consider the origin of the moduli 
space? Near the origin, the expectation value
 $\langle M_{ij}\rangle {\ll} \Lambda^2_{N_c,N_f-1}$, and 
thus one can analyze the flat direction by first putting the 
superpotential (\ref{eq5p63}) aside. Then the dyonic dual 
description is similar to the electric theory. From the discussion 
on the electric theory in Subsect.\,\ref{subsub552}, 
we know that this dyonic dual theory should have several 
branches as does the electric theory. We only consider
its oblique confining branch, which should still be described by 
the superpotential (\ref{eq5p441}), only with the scale replaced by 
$\widetilde{\widetilde{\Lambda}}$. This $W_{\rm obl}$ will differ 
from (\ref{eq5p441}) by a sign due to the relation (\ref{eq5p62})
with $\epsilon =-1$. Now considering the superpotential (\ref{eq5p64}) on 
this branch and adding to it $W_{\rm obl}$ will give the full 
superpotential $W_{\rm full}=0$. Thus on this branch with oblique confinement 
of the dyonic description, one finds that the flat directions given
by the space of $\langle M^{ij}\rangle$ are identical with the ones 
in the original electric
theory, except that this theory is strongly coupled.

\vspace{2mm}
\begin{flushleft}
{\it Mass deformation}
\end{flushleft}
\vspace{2mm}

To discuss the mass deformation, we again add a large mass term for
the $N_f$-th flavour, $W_{\rm tree}=mM^{N_fN_f}/2$ to the
superpotential (\ref{eq5p64}),
\begin{eqnarray}
W_{\rm full}=-\frac{1}{32\Lambda^{2N_c-5}_{N_c,N_c-1}}\det M+
\frac{1}{2}mM^{N_fN_f}.
\label{eq5p68}
\end{eqnarray}
The above discussions shows that in flat directions, near 
the origin of moduli space, the dynamics is strongly coupled, 
and the theory has a confinement branch and hence 
there exist monopoles. Away from the origin, in the 
case $m{\ll}\Lambda$, the massive fields can be integrated 
out. From (\ref{eq5p68}) the equations of motion of these 
massive fields lead to:
\begin{eqnarray}
d^{N_f}{\cdot}d^{N_f}=d^{\widehat{i}}{\cdot}d^{N_f}=0, 
~~~~\widehat{i}=1,{\cdots}, N_f-1(=N_f-2).
\label{eq5p69}
\end{eqnarray} 
(\ref{eq5p69}) implies that $d^{N_f}=0$ while  $d^{\widehat{i}}$ 
may not vanish. If $d^{\widehat{i}}{\neq}0$, the $SO(N_c)$ 
gauge symmetry will break to $SO(2){\cong}U(1)$, 
the massless fields being 
$\widehat{M}^{\widehat{i}\widehat{j}} =d^{\widehat{i}}{\cdot}d^{\widehat{j}}$.
However, rewriting the superpotential
\begin{eqnarray}
W_{\rm full}&=&-\frac{1}{32\Lambda^{2N_c-5}_{N_c,N_c-1}}\det\widehat{M}
M_{N_fN_f}+\frac{1}{2}mM^{N_fN_f}\nonumber\\
&=&\frac{1}{2}m\left(1-\frac{\det\widehat{M}}{16m\Lambda^{2N_c-5}_{N_c,N_c-1}}
\right)d^{N_f}{\cdot}d^{N_f}
= \frac{1}{2}m\left(1-\frac{\det\widehat{M}}
{16\Lambda^{2(N_c-2)}_{N_c,N_c-2}}
\right)d^{+}{\cdot}d^{-},
\label{eq5p610}
\end{eqnarray}   
where we have used the decoupling relation (\ref{eq5p210}), we 
see that in the region 
$\det\widehat{M}=16\Lambda^{2N_c-2}_{N_c,N_c-2}$, there are also light
fields coming from $d^{N_f}$, denoted as $d^{\pm}$ according to their
$U(1)$ charges. The fields $d^{\pm}$ can be interpreted as the dyons
$E^{\pm}$  of the low energy $N_f=N_c-2$ theory. Recall that these
dyons were found in Subsect.\,\ref{subsub534} 
by means of a strong coupling analysis of 
the electric theory and in Subsect.\,\ref{subsub542} 
by a strong coupling analysis of the magnetic 
theory, while here we recognize them in a weak coupling analysis of the 
dyonic theory. This means that in the dyonic dual theory, the dyon
is a fundamental particle. As a natural consequence, an oblique 
confining superpotential like (\ref{eq5p441}) should be present in
the tree level Lagrangian of the dyonic theory (\ref{eq5p64}).

 Finally, to clearly show the triality between electric, magnetic
and dyonic theories, let us see what the magnetic dual and dyonic dual
descriptions of this dyonic theory look like \cite{ref51}. 
First we consider the magnetic dual
of the dyonic theory with superpotential (\ref{eq5p64}). 
It is an $SO(3)$ gauge theory with $N_f$ quarks $q_i$ and singlet 
fields $M^{ij}$. Its superpotential should be composed 
of the superpotential (\ref{eq5p423}) of the magnetic $SO(3)$ gauge theory
and the tree level superpotential (\ref{eq5p64}) of the dyonic theory,
\begin{eqnarray}
W=\frac{1}{2\mu}M^{ij}\left(q_i{\cdot}q_j\right)
-\frac{1}{64\widetilde{\widetilde{\Lambda}}^{2N_c-5}_{N_c,N_c-1}}\det M
+\frac{1}{32\widetilde{\widetilde{\Lambda}}^{2N_c-5}_{N_c,N_c-1}}\det M.
\label{eq5p611}
\end{eqnarray}     
According to (\ref{eq5p62}) and the square root of (\ref{eq5p424}), 
$\widetilde{\widetilde{\Lambda}}^{2N_c-5}_{N_c,N_c-1}
=-{\Lambda}^{2N_c-5}_{N_c,N_c-1}$, we see that (\ref{eq5p611}) is
the same as (\ref{eq5p423}), so the magnetic dual of the 
dyonic dual theory is exactly the
magnetic dual of the original electric theory. If we take the dyonic dual
of the dyonic theory with superpotential (\ref{eq5p64}), 
according to (\ref{eq5p66}) and (\ref{eq5p67}), the vacuum theta angle
will be shifted by $\pi$ again and this will lead to a superpotential
which cancels (\ref{eq5p64}). Thus the dyonic dual of the dyonic 
dual theory is the original electric theory.

 To summarize, the $SO(N_c)$ theory with $N_f=N_c-1$ flavours has three
equivalent descriptions: the original electric $SO(3)$ theory discussed in 
Subsect.\,\ref{subsub534}, the magnetic $SO(3)$ theory described 
in Subsect.\,\ref{subsub542} and the dyonic $SO(N_c)$ 
theory considered here. Taking the dual of the dual theory permutes 
these three descriptions.

 \subsection{A brief introduction to $Sp(N_c=2n_c)$ gauge theory with 
$N_f=2n_f$ quarks }
\label{subsect57}
\renewcommand{\thetable}{5.7.\arabic{table}}
\setcounter{table}{0}
\renewcommand{\theequation}{5.7.\arabic{equation}}
\setcounter{equation}{0}

 In this section we shall give a brief introduction to the 
non-perturbative phenomena, especially the electric-magnetic duality, 
of $N=1$ supersymmetric $Sp(N_c)$ gauge theory with $N_f$ flavours of 
matter in the fundamental representation. These phenomena
such as the generation of a dynamical superpotential, the erasing
of the classical vacuum degeneracy by non-perturbative quantum
effects, the appearance of a conformal window and the relevant
duality are qualitatively  similar to those found in the $SU(N_c)$ gauge 
theory  with matter in the fundamental representation and
in the $SO(N_c)$ gauge theory with matter in the vector 
representation \cite{ref530}.
In fact, it was shown that
the dynamics behaviours of the $Sp(N_c)$ gauge theory are 
parallel to that of the $SO(N_c)$ theory since by formally extrapolating the 
parameter $N_c$ in the $SO(N_c)$ to negative value, the result obtained
in the $SO(N_c)$ theory can be easily adapted to the $Sp(N_c)$ theory
 \cite{ref531}.
Thus in the following we shall only state the main results. 

\subsubsection{Some aspects of $Sp(N_c=2n_c)$ gauge theory}
\label{subsub571}

\vspace{2mm}
\begin{flushleft}
{\it $Sp(N_c)$ gauge theory}
\end{flushleft}
\vspace{2mm}

First we briefly introduce the unitary symplectic group $Sp(N_c)$. It is 
composed of the transformations that preserve the antisymmetric 
inner product $\eta_AJ^{AB}\xi_B$ \cite{ref532},
with 
\begin{eqnarray}
(J^{AB})=\left(\begin{array}{cc} 0 & {\bf 1} \\ -{\bf 1} & 0
                                        \end{array}\right)
={\bf 1}_{N_c/2{\times}N_c/2}{\otimes}i\sigma_2,
\label{eq5p72} 
\end{eqnarray}
where the element ${\bf 1}$ denotes $N_c/2{\times}N_c/2$ unit matrix.
Thus the number of colours $N_c$ should be even, $N_c=2n_c$. The 
dimension of this group is $N_c(N_c+1)/2=n_c(2n_c+1)$.
In particular, the number $N_f$ of flavours must be 
even since for an odd number of (chiral) fermions 
there exists a discrete global anomaly \cite{ref533}, which 
will make the theory inconsistent at the quantum level. 
So, we write  $N_f=2n_f$. It should be emphasized that the 
fundamental representation of $Sp(N_c)$ is always pseudo-real. 

The classical Lagrangian of supersymmetric $Sp(2n_c)$ gauge theory
has the same form as (\ref{eq5p11}) and (\ref{eq5p12}). The classical 
global flavour symmetry is $SU(2n_f){\times}U_A(1){\times}U_{R_0}(1)$
and the explicit transformations of the fields are similar to those 
listed in (\ref{eq5p14}), (\ref{eq5p15}) and (\ref{eq5p15m}).
At the quantum level, the $U_A(1)$ and $U_{R_0}(1)$ symmetries 
will suffer from the ABJ chiral anomaly. The corresponding operator 
anomaly equations are 
\begin{eqnarray}
\partial_{\mu}j_A^\mu &=&4n_f\frac{1}{32\pi^2}
\epsilon^{\mu\nu\lambda\rho}F_{\mu\nu}^aF_{\lambda\rho}^a;\nonumber\\
\partial_{\mu}j_{R_0}^\mu&=&\left[4n_f-4(n_c+1)\frac{1}{32\pi^2}\right]
\epsilon^{\mu\nu\lambda\rho}F_{\mu\nu}^aF_{\lambda\rho}^a, ~~
a=1,2,{\cdots}, n_c(2n_c+1),
\label{eq5p73} 
\end{eqnarray}
respectively. In the same way as in the $SU(N_c)$ and $SO(N_c)$ 
cases, one can combine $U_A(1)$ and $U_{R_0}(1)$ to get an 
anomaly-free $R$-symmetry with the $R$-charge \cite{ref1p26}
\begin{eqnarray}
R=R_0+\frac{n_f-n_c-1}{n_f}A.
\label{eq5p74} 
\end{eqnarray}
Thus the quantum theory has anomaly-free global symmetries
$SU(2n_f){\times}U_R(1)$. According to (\ref{eq5p74}), the 
various $U(1)$ quantum numbers of the quark superfield
and gaugino and their representation dimension under $SU(2n_f)$ are 
listed in Table \ref{ta5p7on}. The perturbative theory gives the 
one-loop beta function coefficient \cite{ref1p26}:
\begin{eqnarray}
\beta_0=3(N_c+2)-2N_f=3(2n_c+2)-2n_f.
\label{eq5p76} 
\end{eqnarray}

\begin{table}
\begin{center}
\begin{tabular}{|c|c|c|c|c|} \hline
            & $SU(2n_f)$  & $U_A(1)$    &  $U_{R_0}(1)$  & $U_R(1)$\\ \hline
 $Q^i$      & $2n_f$      & $+1$   &  $0$  & $(n_f-1-n_c)/n_f$ \\ \hline
 $\lambda$  & $0$         & $0$    &  $+1$ & $+1$ \\ \hline
\end{tabular}
\caption{\protect\small Representation quantum numbers of fundamental fields.
\label{ta5p7on}}
 \end{center}
\end{table}

\vspace{2mm}
\begin{flushleft}
{\it Classical moduli space}
\end{flushleft}
\vspace{2mm}

The classical moduli space is described by the $D$-flatness directions.
$Q$ may have non-vanishing expectation values in a $D$-flat direction. 
Up to gauge and global rotations, these expectation values 
$\langle Q^i_{~r}\rangle$ can be written in the following matrix 
forms \cite{ref530}
\begin{eqnarray}
Q=\left(\begin{array}{ccccccc} a_1 &  0   &\cdots &   0    & 0  & \cdots & 0\\
                               0  & a_2  &\cdots &   0    & 0  & \cdots & 0   \\
                          \vdots  &\vdots&\ddots & \vdots & \vdots &\ddots &0 \\
                               0  &  0   &\cdots &a_{n_f} & 0 & \cdots & 0 
\end{array}\right){\otimes}{\bf 1}_{2{\times}2}
\label{eq5p77} 
\end{eqnarray}
for $n_f<n_c$ and
\begin{eqnarray}
Q=\left(\begin{array}{cccc} a_1 & 0    & \cdots & 0\\
                             0  & a_2  & \cdots & 0 \\
                         \vdots &\vdots& \ddots & \vdots \\
                             0  &  0   & \cdots &a_{n_c} \\
                             0  &  0   & \cdots & 0        \\
                         \vdots &\vdots& \ddots & 0 \\
                             0  &  0   &\cdots  & 0
\end{array}\right){\otimes}{\bf 1}_{2{\times}2}
\label{eq5p78} 
\end{eqnarray}
for $n_f{\geq}n_c$. (\ref{eq5p77}) shows that for generic $a_i$ these 
expectation values break $Sp(2n_c)$ to $Sp(2n_c-2n_f)$ by the Higgs 
mechanism for $n_f<n_c$ and completely break $Sp(n_c)$ for $n_f{\geq}n_c$. 
Thus the moduli space of vacua is described by the expectation values of 
the ``meson" superfields
\begin{eqnarray}
M_{ij}=Q_i{\cdot}Q_j={\epsilon}^{rs}Q_{ir}{\cdot}Q_{js}=-M_{ji}.
\label{eq5p79} 
\end{eqnarray} 
Note that the fundamental representation of $Sp(2n_c)$ 
is always pseudo-real, so the metric in the colour space 
is antisymmetric. The number of fields $M_{ij}$ is $n_f(2n_f-1)$. 
For $n_f<n_c$, this is just the number of the matter fields 
left massless after the Higgs mechanism. For $n_f>n_c$, 
since from (\ref{eq5p78}) and (\ref{eq5p79}) 
$\mbox{rank}(\langle M\rangle){\leq}2n_c$, the $M_{ij}$ should 
be subjected to the classical constraints \cite{ref530}
\begin{eqnarray}
\epsilon^{i_1{\cdots}i_{2n_f}}M_{i_1i_2}M_{i_3i_4}{\cdots}
M_{i_{2n_c+1}i_{2n_c+2}}=0.
\label{eq5p710} 
\end{eqnarray}
In particular, for $n_f=n_c+1$, the above constraint can be written as
\begin{eqnarray}
\mbox{Pf}\,M=0,
\label{eq5p711} 
\end{eqnarray}
where $\mbox{Pf}\,M$ is the Pfaffian of the antisymmetric matrix $M$.
Note that the moduli space of 
$\langle M\rangle$ subject to the constraint (\ref{eq5p710}) is 
singular on submanifolds with $\mbox{rank}(\langle M\rangle){\leq}2(n_c-1)$
since in this case some of the $a_i$ are zero and
$Sp(2n_c)$ is not broken to $Sp(2n_c-2n_f)$. Consequently, there 
exist additional massless bosons on these singular manifolds.

The number of the meson superfields $M^{ij}$ 
subject to the constraints (\ref{eq5p710}) is $4n_fn_c-n_c(2n_c+1)$.
It is precisely the number of the matter fields $Q^i_r$ 
remaining massless after the Higgs mechanism. Note that there 
are no baryons in distinction to 
the $SU(N_c)$ and $SO(N_c)$ cases since 
the invariant tensor $\epsilon^{r_1{\cdots}r_{2n_c}}$ of $Sp(2n_c)$
is always reducible, i.e. it can always be broken up into sums
of products of the second rank anti-symmetric tensor 
$J^{ij}$. Therefore, the baryons always decompose into mesons.

\subsubsection{Quantum moduli space and non-perturbative dynamics }
\label{subsub572}

The non-perturbative quantum effects will modify the classical moduli 
space. Like in the $SU(N_c)$ and $SO(N_c)$ cases, the dynamics is very 
sensitive to the relative numbers of colours and flavours. Different 
ranges of the colour and flavour numbers will present distinct 
physical pictures  \cite{ref530}.

\vspace{2mm}
\begin{flushleft}
{\it $n_f{\leq}n_c$: dynamically generated superpotential and erasing 
of classical vacuum}
\end{flushleft}
\vspace{2mm}

This range is very similar to the $SU(N_c)$ case. In the low energy
theory there is a dynamically generated superpotential and it
lifts all the classical vacuum degeneracy. The explicit form
of this dynamically generated superpotential is determined by the 
holomorphicity and the anomaly-free global $SU(2n_f){\times}U_R(1)$ symmetry
as well as the mass dimension $3$. The representation quantum numbers 
of the quantities entering the superpotential 
are listed in Table \ref{ta5p7tw}. 
Similarly to the dynamical superpotential (\ref{eq6.138b}) 
of the $SU(2)$ case, and since $M$ is an antisymmetric matrix, 
the dynamical superpotential should be 
\begin{eqnarray}
W=A(n_c,n_f)\left(\frac{\Lambda^{\beta_0/2}}{\mbox{Pf}M}
\right)^{1/(n_c+1-n_f)}
=A(n_c,n_f)\left(\frac{\Lambda^{3(n_c+1)-n_f}_{n_c,n_f}}{\mbox{Pf}M}
\right)^{1/(n_c+1-n_f)}.
\label{eq5p713} 
\end{eqnarray}

\begin{table}
\begin{center}
\begin{tabular}{|c|c|c|c|c|} \hline
            & $SU(2n_f)$     & $U_A(1)$    &  $U_{R_0}(1)$  & $U_R (1)$
\\ \hline
$M^{ij}$    & $n_f(2n_f-1)$  & $+2$   &  $0$  & $2(n_f-1-n_c)/n_f$ 
\\ \hline
$\det M$    & $0$            & $4n_f$ &  $0$  & $4(n_f-1-n_c)$ \\ \hline
 $\Lambda^{\beta_0}$ & $0$   & $4n_f$ & $-4(n_f-1-n_c)$   & $0$\\ \hline
\end{tabular}
\caption{\protect\small Representation quantum numbers of the quantities
composing of the dynamical superpotential, $\Lambda$ being 
the dynamical scale.  \label{ta5p7tw}}
 \end{center}
\end{table}

$A(n_c,n_f)$ can be determined from the low energy limit
and the explicit instanton calculation.
First, the one-loop
beta function coefficient (\ref{eq5p76}) and (\ref{eq6.112y}) 
give the running coupling
\begin{eqnarray}
e^{2i\pi\tau}=e^{-8\pi^2g^{-2}(q)+i\theta}
=\left(\frac{\Lambda}{q}\right)^{3(n_c+1)-n_f}.
\label{eq5p714} 
\end{eqnarray}
Considering the large $a_{N_f}$ limit, (\ref{eq5p77})
shows that the $Sp(2n_c)$ theory with $2n_f$ flavours 
becomes the low energy $Sp(2n_c-2)$ theory 
with $2n_f-2$ flavours. In the low energy theory  
\begin{eqnarray}
\mbox{Pf}\,\widehat{M}=a_{n_f}^{-2}\,\mbox{Pf}M,
\label{eq5p715} 
\end{eqnarray}
where $\widehat{M}$ denotes the mesons corresponding to the
light flavours. With (\ref{eq5p714}) the identification of 
the running coupling at the energy $q=a_{n_f}$ gives the 
relation between the high energy and low energy scales 
\begin{eqnarray}
\Lambda^{3n_c-(n_f-1)}_{n_c-1,n_f-1}
=\frac{2\Lambda^{3n_c-n_f}_{n_c,n_f}}{a_{n_f}^2}.
\label{eq5p716} 
\end{eqnarray}
As in the $SU(N_c)$ case discussed in Sect.\,\ref{subsub6.3.5}, 
the requirement that the superpotential (\ref{eq5p713}) 
properly reproduces the superpotential of the low energy theory,  
restricts the coefficients $A(n_c,n_f)$ to the form:
\begin{eqnarray}
A(n_c,n_f)=2^{n_f/(n_c+1-n_f)}A(n_c-n_f, 0)=2^{n_f/(n_c+1-n_f)}A(n_c-n_f).
\label{eq5p717} 
\end{eqnarray}
Further, giving the $(2n_f-1)$-th and $2n_f$-th flavours a large mass 
by introducing a tree level superpotential,
$W_{\rm tree}=mM_{2n_f-1,2n_f}$,
the low energy theory is the $Sp(2n_c)$ theory with $2n_f-2$ flavours with
the scale given by the matching of the running couplings of the high and low
energy theories at the energy $q=m$:
\begin{eqnarray}
\Lambda^{3(n_c+1)-(n_f-1)}_{n_c,n_f-1}=m\Lambda^{3(n_c+1)-n_f}_{n_c,n_f}.
\label{eq5p719} 
\end{eqnarray}  
After integrating out the two heavy flavours, the low energy superpotential
coincides with the general form of (\ref{eq5p713}) only 
when the $A(n_c,n_f)$ satisfy
\begin{eqnarray}
\left(\frac{A(n_c,n_f)}{n_c+1-n_f}\right)^{n_c+1-n_f}=
\left(\frac{A(n_c,0)}{n_c+1}\right)^{n_c+1}.
\label{eq5p720} 
\end{eqnarray}    
To ensure (\ref{eq5p717}) and (\ref{eq5p720}) 
$A(n_c,n_f)$ must be equal to
\begin{eqnarray}
&& A(n_c,n_f)=(n_c+1-n_f)e^{2in\pi/(n_c+1-n_f)}
\left(2^{n_c-1}A\right)^{1/(n_c+1-n_f)}, \nonumber\\
&& n=1,2,{\cdots}, n_c+1-n_f.
\label{eq5p721} 
\end{eqnarray}
The constant $A$ can be determined from the instanton contribution. 
An explicit calculation in the modified dimensional regularization 
scheme shows that $A=1$ \cite{ref534}. The dynamically 
generated superpotential is now finally fixed:
\begin{eqnarray}
&& W=(n_c+1-n_f)e^{2in\pi/(n_c+1-n_f)}
\left(\frac{2^{n_c-1}\Lambda^{3(n_c+1)-n_f}_{n_c,n_f}}{\mbox{Pf}M}
\right)^{1/(n_c+1-n_f)},\nonumber\\ 
&& n=1,2,{\cdots}, n_c+1-n_f.
\label{eq5p722}
\end{eqnarray}

As mentioned in the $SU(N_c)$ case, the concrete dynamical
mechanisms generating the superpotential (\ref{eq5p722}) for 
$n_f<n_c$ and $n_f=n_c$ are different. For $n_f=n_c$, the gauge 
group $Sp(2n_c)$ is completely broken, and (\ref{eq5p722}) is 
generated by an instanton in the broken $Sp(2n_c)$. A similar 
calculation as in the $SU(N_c)$ case can explicitly verify 
this \cite{ref1p22}. For $n_f<n_c$, (\ref{eq5p722}) is associated 
with gaugino condensation in the $Sp(2n_c-2n_f)$ supersymmetric 
Yang-Mills theory,
\begin{eqnarray}
W=(n_c+1-n_f)\left(-\frac{1}{32\pi^2}W^{\alpha}W_{\alpha}
\right)|_{Sp(2n_c-2n_f)}.
\label{eq5p723}
\end{eqnarray}

Like in the $N_f<N_c$ case of the $SU(N_c)$ theory, the dynamically 
generated superpotential (\ref{eq5p723}) has lifted all the classical
vacua since it leads to non-vanishing $F$-term, 
$F=\partial W/\partial Q{\neq}0$.
As the $SU(N_c)$ and $SO(N_c)$ cases, this is another typical example 
of dynamical supersymmetry breaking. However, if we add a mass term
$W_{\rm tree}=m_{ij}M^{ij}/2$ to (\ref{eq5p722}) and integrate the massive
fields, the low energy $Sp(2n_c-2n_f)$ Yang-Mills theory has $n_c+1$
supersymmetric vacua represented by the expectation values of $M_{ij}$,
\begin{eqnarray}
\langle M_{ij}\rangle =
e^{2in\pi/(n_c+1)}\left(2^{n_c-1}\mbox{Pf}\,m\,\Lambda^{3(n_c+1)-n_f}_{n_c,n_f}
\right)^{1/(n_c+1)}(m^{-1})_{ij}.
\label{eq5p724}
\end{eqnarray}
This conclusion can also be obtained from calculating the
Witten index \cite{wit2}. Note that now the mass matrix of 
the $Sp(2n_c)$ quarks is antisymmetric.

\vspace{2mm}
\begin{flushleft}
{\it $n_f=n_c+1$:  Smoothing of classical singular moduli space and chiral 
symmetry breaking}   
\end{flushleft}
\vspace{2mm}

The superpotential (\ref{eq5p722}) does not make sense for 
$n_f{\geq}n_c+1$, so the dynamically generated 
superpotential $W=0$. Consequently,
the vacuum degeneracy is not lifted for $n_f{\geq}n_c+1$ and there exists
a continuous quantum moduli space parametrized by the expectation 
values of $M_{ij}$. Giving masses to all of these
$2n_f$ flavours, their expectation values should 
still be given by (\ref{eq5p724}).
This leads to a constraint to the quantum moduli space expressed in 
terms of the Pfaffian
\begin{eqnarray}
\mbox{Pf}\,M=2^{n_c-1}\Lambda^{2n_c+1}_{n_c,n_c+1}.
\label{eq5p725}
\end{eqnarray}
This is a quantum deformation of the classical constraint 
(\ref{eq5p710}) as for the $N_f=N_c$ case of the $SU(N_c)$ theory. 
This quantum correction is due to the contribution from instantons.  
Because of the quantum deformation, there are no longer any classical 
singularities on the quantum moduli space. Thus
the quantum moduli space of vacua is smooth and there are no other 
additional massless fields. This conclusion
can be verified by the 't Hooft anomaly matching. 

It is easy to see from (\ref{eq5p78}) and 
(\ref{eq5p79}) that the non-vanishing 
$\langle M_{ij}\rangle$ satisfying (\ref{eq5p725}) break the 
global $SU(2n_f){\times}U_R(1)$ chiral symmetry 
to $Sp(2n_f)$${\times}$$U_R(1)$
since $M_{ij}$ is antisymmetric. We can check the 't Hooft matching
conditions for this unbroken global symmetry. 
The fundamental massless fermions are the gaugino $\lambda^a$ and 
the quarks $Q^i_{~r}$. Their representation dimensions and 
anomaly-free $R$-charges under the
gauge group $Sp(2n_c)$ and the global symmetry $Sp(2n_f){\times}U_R(1)$ 
are $(n_c(2n_c+1),1)_{-1}$ and $(2n_c, 2n_f)_{-1}$, 
respectively. The massless composite fermions are 
the fermionic components of the  fluctuations of $M$ 
around $\langle M\rangle$ satisfying (\ref{eq5p725}) 
and their representation quantum numbers  
are $(1,n_f(2n_f-1)-1)_{-1}$. We can write down the 
$Sp(2n_f){\times}U_R(1)$ Noether currents and the energy-momentum 
tensors composed of the fundamental massless fermions and 
the fermionic components of quantum moduli
as in the $SU(N_c)$ and $SO(N_c)$ cases. It can be easily calculated
that for $n_f=n_c+1$, the anomalies do match. The explicit
non-vanishing anomaly coefficients are listed in Table \ref{ta5p7th} 

\begin{table}
\begin{center}
\begin{tabular}{|c|c|} \hline
   triangle diagrams       &        anomaly coefficients \\
   and gravitational anomaly & \\ \hline
 $Sp(2n_f)^2U_R(1)$    & $-2n_c\mbox{Tr}(\widetilde{t}^A\widetilde{t}^B)$\\ 
\hline
 $U_R(1)^3$ & $-n_c(2n_c+3)$ \\ \hline
 $U_R(1)$ & $-n_c(2n_c+3)$ \\ \hline
\end{tabular}
\caption{\protect\small 't Hooft anomaly coefficients, $\widetilde{t}^A$
here denoting the generators of $Sp(2n_f)$.  \label{ta5p7th}}
\end{center}
\end{table}

 We make two flavours heavy by adding a mass term 
$W_{\rm tree}=mM_{2n_f-1,2n_f}$. From
the high energy and low energy scale relation (\ref{eq5p719}) 
and the constraint (\ref{eq5p725}), after integrating out the 
two heavy flavours, we immediately get the superpotential
(\ref{eq5p722}) for the low energy $n_f=n_c$ theory.

\vspace{2mm}
\begin{flushleft}
{\it $n_f=n_c+2$:  Confinement without chiral symmetry breaking }  
\end{flushleft}
\vspace{2mm}

Before discussing the quantum moduli space let us first have 
a look at the classical moduli space given by 
the constraints (\ref{eq5p710}), 
which now become
\begin{eqnarray}
\epsilon^{i_1{\cdots}i_{2n_c}i_{2n_c+1}i_{2n_c+2}i_{2n_c+3}i_{2n_c+4}}
M_{i_1i_2}{\cdots}M_{i_{2n_c+1}i_{2n_c+2}}=0.
\label{eq5p726}
\end{eqnarray}
(\ref{eq5p726}) shows that the number of constraints is 
$\left(\begin{array}{c}4\\2\end{array}\right)=6$.
Thus the classical low energy theory has $n_f(2n_f-1)-6=$$(n_c+2)(2n_c+3)-6
=$$n_c(2n_c+7)$ light fields $M$. In the quantum theory, the light
fields in the low energy theory are $n_f(2n_f-1)=$$(n_c+2)(2n_c+3)
=$$n_c(2n_c+7)+6$ antisymmetric fields $M$ but subject to the
dynamics given by the superpotential
\begin{eqnarray}
W=-\frac{\mbox{Pf}\,M}{2^{n_c-1}\Lambda^{2n_c+1}_{n_c,n_c+2}}.
\label{eq5p727}
\end{eqnarray}
The classical constraints can arise as the equations of motion
from this superpotential. This can be easily seen for 
$\mbox{rank}\left(\langle M\rangle\right)=2n_c$. 
The superpotential gives masses to
$\left(\begin{array}{c}2n_f-2n_c\\2\end{array}\right)=6$ 
components of $M$, hence only $n_c(2n_c+7)$ fields remain 
massless. This coincides with the classical case as it should be. 

 A new phenomenon is that there will be singular submanifolds in the 
quantum moduli space when $\mbox{rank}\left(\langle M\rangle\right)
{\leq}2(n_c-1)$. In distinction from the classical physical interpretation,
here the singularity implies that some of the quantum components of $M$
become massless on these submanifolds. At the origin of the moduli 
space, $\mbox{rank}\left(\langle M\rangle\right)=0$, all the 
$(n_c+2) (2n_c+3)$ components of $M$ are massless and 
the full global $SU(2n_f){\times}U_R(1)$ chiral symmetry 
is unbroken. Thus we can check whether the 't Hooft anomalies
match for the massless fermions at both fundamental and composite levels. 
The massless fundamental fermions are the gaugino and quarks,
and their quantum numbers under $SU(2n_f){\times}U_R(1)$ are given 
in Table \ref{ta5p7on}. The massless composite 
fermions are the fermionic components of the quantum moduli 
around the origin $\langle M\rangle=0$ and their 
$SU(2n_f){\times}U_R(1)$ quantum numbers are $(n_f(2n_f-1))_{-1}$. One can 
easily calculate the 't Hooft triangle anomaly diagrams and the $U_R(1)$
axial gravitational anomaly. The non-vanishing anomaly coefficients 
are collected in Table \ref{ta5p7fo} and they indeed match. Note that 
at $M_{ij}=0$ the chiral symmetry does not break but the colour degrees 
of freedom are confined. A similar phenomenon was also observed in the 
$N_f=N_c+1$ case of the $SU(N_c)$ theory.

\begin{table}
\begin{center}
\begin{tabular}{|c|c|} \hline
   Triangle diagrams       &        Anomaly coefficients \\
   and gravitational anomaly & \\ \hline
$SU(2n_f)^3$  & $2n_c\mbox{Tr}(t^A\{t^B,t^C\})$ \\ \hline
$SU(2n_f)^2U_R(1)$    & $-2n_c(n_c+1)/(n_c+2)\mbox{Tr}(t^At^B)$\\ \hline
 $U_R(1)^3$ & $-n_c^3(2n_c+3)/(n_c+2)^2$ \\ \hline
 $U_R(1)$ & $-n_c(2n_c+3)$ \\ \hline
\end{tabular}
\caption{\protect\small 't Hooft anomaly coefficients for both high and 
low-energy theories, $t^A$ being the generators of $SU(2n_f)$ \label{ta5p7fo}}
 \end{center}
\end{table}

 Making two flavours heavy by adding 
$W_{\rm tree}=mM_{2n_f-1,2n_f}$ to the superpotential 
(\ref{eq5p727}) and integrating
out these two massive fields, we can get the constraint 
(\ref{eq5p725}) in the low energy $n_f=n_c+1$ theory as 
an equation of motion.

\vspace{2mm}
\begin{flushleft}
{\it $n_f>n_c+2$:  Conformal window and duality}   
\end{flushleft}
\vspace{2mm}
  
Increasing the number of flavours, we now reach the theories
with $n_f>n_c+2$. (\ref{eq5p76}) shows that for $n_f{\geq}3(n_c+1)$ 
the one-loop beta function coefficient $\beta_0{\leq}0$, the theory is not
asymptotically free. It is free in the infrared region. So
this range is not interesting. In the range $3(n_c+1)/2<n_f<3(n_c+1)$,
there is a non-trivial infrared fixed point of the renormalization group 
flow at which the theory is in an interacting non-Abelian Coulomb phase.
From the discussions on the phases of gauge theory in 
Subsect.\,\ref{subsect27}, 
the theory in this phase
can have a self-dual description. It was found \cite{ref530}
that the dual description is
an $Sp(2n_f-2n_c-4)$ gauge theory with $2n_f$ matter
fields $q^i$ in the fundamental conjugate representation of $SU(2n_f)$,
gauge singlets $M_{ij}=-M_{ji}$ and a superpotential
\begin{eqnarray}
W=\frac{1}{4\mu}M_{ij}q^i_{~r}q^j_{~s}J^{rs},
\label{eq5p728}
\end{eqnarray} 
where $\mu$ is a parameter with mass dimension.
The one-loop beta function coefficient of the gauge coupling 
of the dual theory is, according to (\ref{eq5p76}),   
\begin{eqnarray}
\widetilde{\beta}_0=3\left[2(n_f-n_c-2)+2\right]-2n_f=
6(n_f-n_c-1)-2n_f.
 \label{eq5p729}
\end{eqnarray} 
(\ref{eq5p729}) shows the reason to require 
$n_f>3(n_c+1)/2$: this choice makes
$\widetilde{\beta}_0>0$ and the dual theory has an
identical fixed point of the renormalization group flow as the 
electric theory, at which the electric and magnetic theories give 
physically equivalent description. Similarly to the $SU(N_c)$ and $SO(N_c)$ 
cases, the relation between the dynamical scale $\Lambda$ of 
the electric theory and the scale $\widetilde{\Lambda}$ of magnetic 
theory is
\begin{eqnarray}
\Lambda^{3(n_c+1)-n_f}_{n_c,n_f}
\widetilde{\Lambda}^{3(n_f-n_c-1)-n_f}_{n_f-n_c-2,n_f}=C
(-1)^{n_f-n_c-1}\mu^{n_f}.
\label{eq5p730}
\end{eqnarray}
This scale relation will lead to typical electric-magnetic 
duality features as described for the $SO(N_c)$ and $SU(N_c)$ 
theories. The low energy electric theory is equivalent to 
high energy magnetic theory and vice versa.  

 Taking the singlet $M$ to transform as the meson fields 
$M_{ij}=Q_i{\cdot}Q_j$ of the electric theory, the magnetic theory 
theory with the superpotential (\ref{eq5p728}) has a global anomaly-free
$SU(2n_f){\times}U_R(1)$ flavour symmetry as the electric´ theory,
under which $M$ is in the representation $(n_f(2n_f-1))_{2(1-(n_c+1)/n_f)}$ 
and the magnetic quark superfields $q_i$ in $(2\overline{n}_f)_{(n_c+1)/n_f}$.
At $\langle M\rangle =0$, the global $SU(2n_f){\times}U_R(1)$ is unbroken
in both the electric and magnetic theories. One can easily check
that the 't Hooft anomalies contributed by the massless fermions 
in the electric $Sp(2n_c)$ theory, gaugino and quarks, match those 
contributed from the massless fermions in the magnetic theory: magnetic 
gaugino, magnetic quarks and the fermionic components of the singlet 
$M$. Both sets of massless fermions give identical 't Hooft anomaly 
coefficients as listed in Table \ref{ta5p7fi}.

\begin{table}
\begin{center}
\begin{tabular}{|c|c|} \hline
   Triangle diagrams       &        Anomaly coefficients \\
   and gravitational anomaly & \\ \hline
$SU(2n_f)^3$  & $2n_c\mbox{Tr}(t^A\{t^B,t^C\})$ \\ \hline
$SU(2n_f)^2U_R(1)$    & $-2n_c(n_c+1)/n_f\mbox{Tr}(t^At^B)$\\ \hline
 $U_R(1)^3$ & $-n_c(2n_c+3)$ \\ \hline
 $U_R(1)$ & $n_c(2n_c+1)-4n_c(n_c+1)^3/n_f^2$ \\ \hline
\end{tabular}
\caption{\protect\small 't Hooft anomaly coefficients
contributed by the massless fermions of the electric 
and magnetic theories, $t^A$ being  the generators 
of $SU(2n_f)$. \label{ta5p7fi}}
 \end{center}
\end{table}

Along the lines of discussion of the $SU(N_c)$ and $SO(N_c)$ theories,
we shall discuss the dynamical behaviour in the
flat directions and under the mass deformation of 
the magnetic theory. Let us first find the flat directions 
in the magnetic theory. The equations of motion for $M$ from the 
superpotential (\ref{eq5p728}) and the $D$-terms of the 
$Sp(2n_c-2n_f-4)$ gauge theory give the flat directions parametrized 
by $\langle q^i\rangle =0$ and arbitrary $\langle M_{ij}\rangle$. Now 
giving $M_{2n_f-1,2n_f}$ a large 
expectation value, the quark superfields $Q_{2n_f-1}$ and
$Q_{2n_f}$ of the electric theory will get large expectation values
and they will break the electric $Sp(2n_c)$ theory with $2n_f$ flavours
to a low energy  $Sp(2n_c-2)$ theory with $2n_f-2$ flavours
and the scale 
\begin{eqnarray}
\Lambda^{3n_c-(n_f-1)}_{n_c-1,n_f-1}
=\frac{2\Lambda^{3(n_c+1)-n_f}}{\langle M_{2n_f-1,2n_f}\rangle},
\label{eq5p731}
\end{eqnarray}
which is obtained from matching the running couplings
at the energy $\langle M_{2n_f-1,2n_f}\rangle$. On the other hand,
in the magnetic theory, a large $\langle M_{2n_f-1,2n_f}\rangle$
will give a large mass $(2\mu)^{-1}\langle M_{2n_f-1,2n_f}\rangle$
to $q^{2n_f-1}$ and $q^{2n_f}$. The low energy
magnetic theory is an $Sp(2n_f-2n_c-4)$ theory with $2n_f-2$ flavours,
a superpotential of the form (\ref{eq5p728}) 
and a scale
\begin{eqnarray}
\widetilde{\Lambda}^{3(n_f-n_c-1)-(n_f-1)}_{n_f-n_c-2,n_f-1}
=(2\mu)^{-1}M_{2n_f-1,2n_f}\widetilde{\Lambda}^{3(n_c+1)-n_f}.
\label{eq5p732}
\end{eqnarray}    
(\ref{eq5p728}) and (\ref{eq5p732}) show that 
the scale relation is preserved for 
the low energy electric and magnetic theories and hence the duality
relation remains.
 
Another interesting example is giving large values to 
$\mbox{rank}(\langle M\rangle)$ eigenvalues of $M$. Then
$\mbox{rank}(\langle M\rangle)$ magnetic quarks get heavy masses,
and the low energy magnetic theory is the $Sp(2n_f-2n_c-4)$ gauge
theory with $2n_f-\mbox{rank}(\langle M\rangle)$ flavours. From the 
above discussions, we know that the electric $Sp(2n_c)$ gauge theory 
with $n_f{\geq}n_c+2$ has a vacuum at the origin, 
$\langle Q\rangle =0$, while the $Sp(2n_c)$ theory 
with $n_f{\leq}n_c+1$ does not: when $n_f{\leq}n_c$, the dynamical 
superpotential has erased all the vacua and the classical
vacuum at the origin cannot escape from this fate; when $n_f=n_c+1$,
the instanton correction smoothes out the singularity at the origin. 
In the magnetic theory, for 
$2n_f-\mbox{rank}(\langle M\rangle){\leq}2n_f-2n_c-2=2(n_f-n_c-2)+2$,
i.e. $\mbox{rank}(\langle M\rangle){\geq}n_c+1$
there should exist no vacuum at $\langle q\rangle=0$ due to strong 
coupling effects. Since the equation of motion of $M$ requires the vacuum
to be at $\langle q\rangle=0$, there is no supersymmetric vacuum for
$\mbox{rank}(\langle M\rangle){\geq}n_c+1$. This is an obvious classical
constraint in the electric theory, while in the magnetic theory it is
recovered from the strong coupling dynamics and the equations of motion
of $M$.     

 In the following we consider mass deformation. Giving a 
mass to the $(2n_f-1)$-th and $2n_f$-th flavours by adding the 
superpotential $W_{\rm tree}=mM_{2n_f-1,2n_f}$,
the low energy electric theory is an $Sp(2n_c)$ gauge theory 
with $2n_f-2$ flavours and the scale 
\begin{eqnarray}
\Lambda^{3(n_c+1)-(n_f-1)}_{n_c,n_f-1}
=m\Lambda^{3(n_c+1)-n_f}_{n_c,n_f}.
 \label{eq5p733}
\end{eqnarray}
On the other hand, the equations of motion for $M_{2n_f-1,2n_f}$ obtained 
from $W_{\rm tree}$ and the superpotential (\ref{eq5p728}) yield
\begin{eqnarray}
\langle q^{2n_f-1}{\cdot}q^{2n_f} \rangle =-2\mu m.
\label{eq5p734}
\end{eqnarray} 
This expectation value breaks  the dual $Sp(2n_f-2n_c-4)$ theory to
$Sp(2n_f-2n_c-6)$. Furthermore, the equations of motion of 
$M_{\widehat{i}2n_f-1}$ and $M_{\widehat{i}2n_f}$, 
$\widehat{i}=1,{\cdots}, 2(n_f-1)$  yield
\begin{eqnarray} 
\langle q^{\widehat{i}}{\cdot}q^{2n_f-1}\rangle 
=\langle q^{\widehat{i}}{\cdot}q^{2n_f}\rangle =0.
\label{eq5p735}
\end{eqnarray} 
This means that
\begin{eqnarray}
M_{\widehat{i},2n_f-1}=M_{\widehat{i},2n_f}=0.
\label{eq5p736}
\end{eqnarray}
(\ref{eq5p734}) and (\ref{eq5p736}) imply that the light 
fields in the low energy $Sp(n_f-n_c-3)$ theory are the 
singlets $M_{\widehat{i}\widehat{j}}$ and the $2n_f-2$
magnetic quarks $q^{\widehat{i}}$ with a superpotential of the form 
(\ref{eq5p728}). The matching of the running couplings at the 
energy $\langle q^{2n_f-1}{\cdot}q^{2n_f} \rangle$
gives the scale relation between the low energy
and high energy magnetic theories,    
\begin{eqnarray}
\widetilde{\Lambda}^{3(n_f-n_c-2)-(n_f-1)}_{n_f-n_c-3,n_f-1}=
-(\mu m)^{-1}\widetilde{\Lambda}^{3(n_f-n_c-1)-n_f}_{n_f-n_c-1,n_f}.
\label{eq5p737}
\end{eqnarray}
(\ref{eq5p733}) and (\ref{eq5p737}) show that the dual 
scale relation (\ref{eq5p730}) is still satisfied
and hence the duality is preserved in the low energy theory.   

 Another kind of mass deformation consists of assigning 
masses to some of the flavours by adding a 
mass term $W_{\rm tree}=m^{ij}M_{ij}/2$ with 
$\mbox{rank}(m^{ij})=2r$. The low energy electric theory is an
$Sp(2n_f-2r-2n_c-4)$ gauge theory with $2n_f-r$ flavours. If 
one chooses the mass matrix $m$ to satisfy $n_f-r-n_c-2=0$, the magnetic
gauge group will be completely broken and there will arise a contribution
to the superpotential from the instanton in 
the broken magnetic gauge group. On the
other hand, this also occurs in the low energy electric theory when
$n_f=n_c+2$. Turning on the mass term $W_{\rm tree}=mM_{2n_f-1,2n_f}$
in the electric $Sp(2n_c)$ theory with $n_f=n_c+3$
and integrating out the $(2n_f-1)$-th and $2n_f$-th flavours, 
we get the low energy $Sp(2n_c)$ gauge 
theory with $n_f=n_c+2$. In the magnetic theory, with the addition 
of $W_{\rm tree}$, the equation of motion 
for $M_{2n_f-1,2n_f}$ yields 
$\langle q^{2n_f-1}{\cdot}q^{2n_f}\rangle =-2\mu m$.
The other magnetic quarks $q^{\widehat{i}}$, 
$\widehat{i}=1,{\cdots}, 2n_f-2=2(n_c+2)$ get masses
$\langle \widehat{M}^{\widehat{i}\widehat{j}}\rangle/(2\mu)$.
This is implied by the superpotential (\ref{eq5p728}). 
Thus $Sp(2n_c)$ breaks to $Sp(2)$ and the instanton in this
magnetic $Sp(2)$ gauge group yields a low energy superpotential  
\begin{eqnarray}
W&=&\frac{\widetilde{\Lambda}^{6-(n_c+3)}_{1,n_c+3}\mbox{Pf}
\left(\widehat{M}/(2\mu)\right)}{q^{2n_f-1}q^{2n_f}}=
-\frac{\widetilde{\Lambda}^{6-(n_c+3)}_{1,n_c+3}\mbox{Pf}
\,\widehat{M}}{(2\mu)^{n_c+3}m}\nonumber\\
&=&-\frac{C\,\mbox{Pf}\,\widehat{M}}
{2^{n_c+3}(m{\Lambda}^{2n_c}_{n_c,n_c+3}}
=-\frac{C\,\mbox{Pf}\,\widehat{M}}
{2^{n_c+1}{\Lambda}^{2n_c+1}_{n_c,n_c+2}},
\label{eq5p739}
\end{eqnarray} 
where we have used the scale relation (\ref{eq5p730})
$\Lambda^{2n_c}_{n_c,n_c+3}\widetilde{\Lambda}^{6-(n_c+3)}_{1,n_c+3}=
C\mu^{n_c+3}$, 
and the decoupling relation 
$\Lambda^{2n_c+1}_{n_c,n_c+2}=m\Lambda^{2n_c}_{n_c,n_c+3}$.
(\ref{eq5p739}) shows that if the constant appearing 
in (\ref{eq5p730}) is $C=16$, 
the superpotential (\ref{eq5p739}) is precisely 
the superpotential (\ref{eq5p727}) for
the $n_f=n_c+2$ case of the electric theory. Therefore, all the results
for $n_f{\leq}n_c+2$ in the electric theory can be obtained from
the dual magnetic theory with $n_f{\leq}n_c+3$
by flowing down through introducing  mass terms. 

 To summarize, supersymmetric $Sp(2n_c)$ gauge theory with
matter fields in the fundamental representation is another typical
example that exhibits duality and other interesting non-perturbative 
dynamical phenomena. Therefore, all the $N=1$ supersymmetric gauge 
theories have conformal windows and dual descriptions. This is 
in fact a generalization of Montonen-Olive-Osborn duality of 
the $N=4$ supersymmetric theory and Seiberg-Witten duality
of the low energy $N=2$ supersymmetric theory. However, there 
are some differences between $N=1$ duality and $N=2,4$ duality. 
In the $N=1$ duality the original electric
theory and the dual magnetic description have different gauge groups
and matter field contents. Especially, the $N=4$ and $N=2$ dualities are
thought to be exact, while $N=1$ duality only arises in the infrared
region. In spite of this limitation, the potential application of $N=1$
duality in exploring the non-perturbative dynamics should not be 
underestimated since $N=1$ supersymmetry can be easily broken to 
get a non-supersymmetric theory. In addition,
the associated conformal windows also have great significance 
since we get a large number of non-trivial interacting four-dimensional
conformal quantum field theories \cite{luc}.


\section{New features of $N=1$ four-dimensional superconformal field theory 
and some of its relevant aspects}
\label{sect7}
\renewcommand{\theequation}{6.\arabic{equation}}
\setcounter{equation}{0}

The discussion in the previous  sections shows 
that non-Abelian electro-magnetic duality emerges 
in the IR fixed point of $N=1$ supersymmetric
gauge theory, where the theory is described by 
an interacting superconformal invariant field theory. 
This is a new conformal invariant quantum 
field theory in four-dimensions \cite{luc}.
The only known non-trivial superconformal field 
theories known before in four dimensions 
were the $N=4$ supersymmetric Yang-Mills theory \cite{ref819} and
the special $N=2$ theories with vanishing $\beta$-function
\cite{hsw}. In fact, conformal invariance and duality depend 
on each other: superconformal symmetry, manifested by the vanishing 
of the NSVZ $\beta$-function,
is the necessary environment for the survival of 
electric-magnetic duality, otherwise
the running of the coupling will ruin the 
electric-magnetic duality.
On the other hand, duality is very useful in understanding the
dynamical structure of superconformal field theory. The combination
of electric-magnetic duality and superconformal symmetry 
has revealed a large number of non-perturbative dynamical phenomena.
Furthermore, some non-perturbative information beyond the IR fixed
point can be acquired by investigating  
the renormalization group flow. Note that the realistic
colour number $N_c=3$ and flavour number $N_f=6$ satisfy
the conformal window condition, $3N_c/2<N_f<3N_c$.
In this section we shall introduce some of the new features of these 
non-trivial four-dimensional superconformal field theories 
and some related aspects including the critical behaviour of 
the various anomalous currents, anomaly matching in the 
presence of higher order quantum corrections, the
universality of the operator product expansion and the evidences
for the existence of a four-dimensional $c$-theorem provided by 
duality and superconformal symmetry. 

\subsection{Critical behaviour of anomalous currents and anomaly matching}
\label{subsect71}
\renewcommand{\theequation}{6.1.\arabic{equation}}
\setcounter{equation}{0}

As shown in (\ref{eq9px}),  
in a superconformal field theory, the supersymmetry algebra 
determines that the energy-momentum tensor, the supersymmetry 
current and the chiral $R$-current all lie in a single supermultiplet. 
Consequently, the axial anomaly of the $R$-current,
the $\gamma$-trace anomaly of the supersymmetry supercurrent 
and the trace anomaly of the energy-momentum tensor also 
belong to the same supermultiplet.
However, the trace anomaly of the energy-momentum 
tensor is proportional to the $\beta$-function and thus cannot be 
saturated by the one-loop quantum correction, while it was for 
a long time believed that the axial anomaly only receives
contributions from the one-loop quantum correction \cite{ref1}. This 
used to be the famous anomaly puzzle  
in supersymmetric gauge theory. A series of investigations
have concluded that this paradox is actually due to the difference between
the operator form and the matrix element form of the chiral anomaly 
equation \cite{ref2}-\cite{ref8}: the operator form
of the anomaly equation is one-loop only, while
the matrix element form can receive
multi-loop contributions. It was pointed out in Ref.\,\cite{ref8}
that this difference actually occurs in any gauge theory, since the
gauge invariance of the regularization schemes adopted in calculating
the anomaly can only unambiguously fix the form of the renormalized 
matrix elements, while the form of the operator equation is conditional.
Only in supersymmetric gauge theory, has this anomaly paradox been 
exposed because of the anomaly supermultiplet.

In previous sections we only considered the one-loop 
't Hooft anomaly matching as a check of the duality and other
non-perturbative dynamical phenomena. Since $N=1$ duality only 
exists in the IR fixed point of the theory, where the
theory is in a strong coupling region, 
the effects of higher order quantum corrections 
cannot be neglected. Therefore, we must now check 't Hooft's 
anomaly matching including higher order quantum corrections.

Among the various anomalous currents participating in 
't Hooft anomaly matching, only the singlet chiral $R$-current is 
affected by higher order quantum corrections \cite{ksv}. 
Eq.\,(\ref{eq182}) shows that in supersymmetric $SU(N_c)$ QCD with
$N_f$ flavours the anomaly-free $R$-current  
at the one-loop level is a combination of two classically
conserved but quantum mechanically anomalous 
axial vector currents. One is the current $R_0$ lying in the
same supermultiplet with the energy-momentum and the supersymmetry 
supercurrent. It was given by Eq.\,(\ref{eq175})  
and can be rewritten in the following 
two-component field form \cite{ksv}
\begin{eqnarray}
R_{0\alpha\dot{\alpha}}&=&\frac{2}{g^2}\mbox{Tr}
(\lambda^{\dagger}_{\dot{\alpha}}\lambda_{\alpha})-
\sum_{i=1}^{N_f}\left(\psi_{\dot{\alpha}}^{i\dagger}\psi_{\alpha}^i
+\widetilde{\psi}_{\dot{\alpha}}^{i\dagger}
\widetilde{\psi}_{\alpha}^i\right)
\label{eq711}
\end{eqnarray}
which is the fermionic part of the 
lowest ($\theta=\overline{\theta}=0$) component 
of the supercurrent superfield \cite{ref2,ksv},
\begin{eqnarray}
J_{0\alpha\dot{\alpha}}&=& -\frac{2}{g^2}\mbox{Tr}\left(W_{\alpha}e^V
 W^{\dagger}_{\dot{\alpha}}e^{-V}\right)
+\frac{Z}{4}\left[\left(\left\{\left[D_{\alpha}\left(
e^{-V}Q\right)\right]e^V\overline{D}_{\dot{\alpha}}
\left(e^{-V}Q^{\dagger}\right)
\right.\right. \right.\nonumber \\
&+& \left.\left.Qe^{-V}D_{\alpha}\left[e^V\overline{D}_{\dot{\alpha}}
\left(e^{-V}Q^{\dagger}\right)\right]+Q\overline{D}_{\dot{\alpha}}
\left(e^{-V}D_{\alpha}Q^{\dagger}\right)\right\}
-\left\{Q\rightarrow Q^{\dagger},V\rightarrow -V\right\}\right)
\nonumber\\
&+&\left.\left(Q{\rightarrow} \widetilde{Q}, V{\rightarrow}-V \right)\right],
\label{eq712}
\end{eqnarray}
where $Z$ is the wave function renormalization constant
of the quark chiral superfields.
Note that here and in what follows, for the convenience of discussion,
we have rescaled the vector supermultiplet $V{\longrightarrow}V/g$.
Another axial current is the flavour singlet axial vector current $K_\mu$ 
composed only of the matter fields, and
its two-component field form is 
$K_{\alpha\dot{\alpha}}=\sum_{i=1}^{N_f}\left(\psi_{\dot{\alpha}}^{i\dagger}
\psi_{\alpha}^i+\widetilde{\psi}_{\dot{\alpha}}^{i\dagger}
\widetilde{\psi}_{\alpha}^i\right)$.
This chiral current is usually called the Konishi current, it is 
the fermionic part of the $\overline{\theta}\theta$ component of 
the following superfield  
\cite{ref1p20}, 
\begin{eqnarray}
\widetilde{K}_{\alpha\dot{\alpha}}
&=&-\frac{Z}{4}\left(\left\{\left[D_\alpha\left(e^{-V}Q\right)
\right]e^V\overline{D}_{\dot{\alpha}}\left(e^{-V}Q^\dagger\right)
-\frac{1}{2}Qe^{-V}D_\alpha\left[e^V
\overline{D}_{\dot{\alpha}}\left(e^{-V}Q^\dagger\right)\right]
\right.\right.\nonumber\\
&-&\left.\left.Q\overline{D}_{\dot{\alpha}}
\left(e^{-V}D_\alpha Q^\dagger\right)
-\left(Q\rightarrow Q^{\dagger}, V\rightarrow -V\right)\right\}
+\left\{Q\rightarrow \widetilde{Q}, V\rightarrow -V\right\}\right)\nonumber\\
&=&-\frac{Z}{2}\left[D_\alpha, D_{\dot{\alpha}}\right]
\left(\widetilde{Q}e^VQ\right)+
\left(Q\rightarrow Q^{\dagger}, V\rightarrow -V\right),
\end{eqnarray}
which corresponds to the transformation invariance of the chiral
superfields:
\begin{eqnarray}
W_\alpha \rightarrow W_\alpha, ~~Q\rightarrow e^{i\beta}Q, ~~
\widetilde{Q}\rightarrow e^{i\beta}\widetilde{Q}.
\label{eq715a}
\end{eqnarray}
The anomaly of this current comes only from the one-loop quantum correction,
and can be read from the Konishi anomaly relation \cite{ref1p20},
\begin{eqnarray}
\overline{D}^2 \widetilde{K}= \overline{D}^2 Z\sum_{i=1}^{N_f}
\left(Q_i^{\dagger}e^VQ_i+\widetilde{Q}_ie^{-V}\widetilde{Q}^{\dagger}_i
\right)=\frac{N_f}{2\pi^2}\mbox{Tr}W^2,
\label{eq716}
\end{eqnarray}  
here and in what follows, $W^2=W^\alpha W_\alpha$. The superfield
\begin{eqnarray}
K=\sum_{i=1}^{N_f}\left(Q_i^{\dagger}e^VQ_i+\widetilde{Q}_ie^{-V}
\widetilde{Q}_i^{\dagger}\right)
\end{eqnarray}
is usually called Konishi supercurrent due to the Konishi
anomaly relation \cite{ref1p20}. 

The higher order effects of the $R_0$ anomaly can be easily inferred from
the operator anomaly equation of the supercurrent $J_{0\alpha\dot{\alpha}}$
found a long time ago \cite{ksv},
\begin{eqnarray}
\overline{D}^{\dot{\alpha}}J_{0\alpha\dot{\alpha}}=-\frac{1}{8}D_{\alpha}
\left[\frac{3N_c-N_f}{2\pi^2}\mbox{Tr}W^2+\gamma\overline{D}^2Z\sum_{i=1}^{N_f}
\left(Q_i^{\dagger}e^VQ_i+\widetilde{Q}_ie^{-V}\widetilde{Q}^{\dagger}_i
\right)\right],
\label{eq715}
\end{eqnarray}
with $\gamma$ being the anomalous dimension of the matter fields,
$\gamma =-\mu{\partial\ln Z}/{\partial \mu}=-{d\ln Z}/{d\ln\mu}$.
 The many loop effects are reflected in the term proportional 
to $\gamma$ in Eq.(\ref{eq715}). The combination of (\ref{eq716}) 
and (\ref{eq715}) gives the multi-loop anomaly equation for $R_0$,
\begin{eqnarray}
{\partial}^{\mu}R_{0\mu}=\frac{1}{16\pi^2}\left[3N_c-N_f(1-\gamma)\right]
F^{\mu\nu a}\widetilde{F}_{\mu\nu}^a.
\label{eq718}
\end{eqnarray}
The anomaly coefficient is proportional to the NSVZ beta function of 
the gauge coupling, given by Eq.(\ref{eq6.224x}). The $\theta^2$ 
component of the Konishi anomaly relation (\ref{eq716}) gives the 
axial anomaly of the Konishi current,
\begin{eqnarray}
\partial^{\mu}K_{\mu}=\frac{1}{16\pi^2}N_f
F^{\mu\nu a}\widetilde{F}_{\mu\nu}^a.
\label{eq719}
\end{eqnarray}
Eqs.\,(\ref{eq718}) and (\ref{eq719}) suggest that the anomaly-free 
$R$-current with the inclusion of the higher order effects should be 
the following combination
\begin{eqnarray}
R_{\mu}=R_{\mu}^0+\left(1-\frac{3N_c}{N_f}-\gamma\right)K_{\mu}.
\label{eq7110}
\end{eqnarray}

In the dual magnetic theory, complications will arise in 
constructing the anomaly-free $R$-current due to the cubic
superpotential (\ref{eq390}) involving the magnetic quarks 
$q$, $\widetilde{q}$ and the colour singlet $M$.
The magnetic Konishi current consists of a magnetic quark part
and a singlet field part,
\begin{eqnarray}
K_{\alpha\dot{\alpha}}&{\equiv}& K^q_{\alpha\dot{\alpha}}
+K^{\cal M}_{\alpha\dot{\alpha}},
\nonumber\\
K^{q}_{\alpha\dot{\alpha}}&=& \sum_i^{N_f}
\sum_{\widetilde{r}=1}^{N_f-N_c}\left( 
\psi_{q\dot{\alpha}}^{i\widetilde{r}\dagger}\psi^i_{q\widetilde{r}\alpha}+
\widetilde{\psi}_{q\dot{\alpha}}^{i\widetilde{r}\dagger}
\widetilde{\psi}_{q\widetilde{r}\alpha}^i\right),\nonumber\\
K^{\cal M}_{\alpha\dot{\alpha}}&=&
\sum_{k=1}^{N_f(N_f+1)/2}
\psi_{\cal M}^{k\dagger}\psi_{\cal M}^k.
\label{eq7112}
\end{eqnarray}
The $R_0$-current in the magnetic theory 
has the same form as in the electric theory, 
\begin{eqnarray}
\widetilde{R}_{0\alpha\dot{\alpha}}&=&\frac{2}{\widetilde{g}^2}
\mbox{Tr}(\widetilde{\lambda}_{\dot{\alpha}}\widetilde{\lambda}_{\alpha})
-K^q_{\alpha\dot{\alpha}}.
\label{eq7111}
\end{eqnarray}
The superpotential provides an additional classical source for current
non-conservation. The anomaly equations of the supercurrent superfield 
$\widetilde{J}$ and the Konishi relation are thus modifications of 
Eq.\,(\ref{eq715}) and (\ref{eq716}), respectively \cite{ksv},
\begin{eqnarray}
\overline{D}^{\dot{\alpha}}\widetilde{J}_{0\alpha\dot{\alpha}}
=D_{\alpha}\left\{\left(3{\cal W}-\sum_{i=1}^{N_f}\Phi^i
\frac{\partial{\cal W}}
{\partial\Phi^i}\right)-\left[\frac{\widetilde{\beta}_0}
{16\pi^2}\mbox{Tr}{\widetilde{W}}^2
+\frac{1}{8}\sum_{i=1}^{N_f}\gamma_iZ_i\overline{D}^2
\left(\Phi^{i\dagger}e^{\widetilde{V}}\Phi^i\right)
\right]\right\},
\label{eq7113}
\end{eqnarray}
\begin{eqnarray}
\frac{1}{8}\sum_{i=1}^{N_f}\gamma_iZ_i \overline{D}^2
\left(\Phi^{i\dagger}e^{\widetilde{V}}\Phi^i\right)
=\sum_{i=1}^{N_f}\left(\frac{1}{2}\Phi^i\frac{\partial {\cal W}}
{\partial \Phi^i}+\frac{C_i}{16\pi^2}\mbox{Tr}{\widetilde{W}}^2\right),
\label{eq7114}
\end{eqnarray}
where $\Phi^i$ is a certain general chiral superfield contained in the
superpotential ${\cal W}$, $\gamma_i$ and $Z_i$ are the anomalous 
dimension and the wave function renormalization for $\Phi_i$, 
$\widetilde{\beta}_0=3N_c-\sum_{i=1}^{N_f}C_i$ 
is the coefficient of the first order $\beta$-function, $C^i$ is defined 
by the normalization 
$\mbox{Tr}(\widetilde{T}^{\widetilde{a}}\widetilde{T}^{\widetilde{b}})
=C_i\delta^{\widetilde{a}\widetilde{b}}$ with $T^a$ being 
the matrix representation
to which the $\Phi^i$s belong, and ${\cal W}$ is a general 
classical superpotential.

 According to (\ref{eq7113}) and (\ref{eq7114}), 
the conserved $R$-current of the magnetic theory 
with the inclusion of the higher order quantum 
corrections should be the following combination \cite{ksv}:
\begin{eqnarray}
\widetilde{R}_\mu
&=&\widetilde{R}_{0\mu}+\left[1-\frac{3(N_f-N_c)}{N_f}-\gamma_q\right]
\left(K_\mu^q-2K_\mu^M\right)
-\left(2\gamma_q+\gamma_M\right)K_\mu^M\nonumber\\
&{\equiv}&\widetilde{R}^0_\mu+c_qK^q_\mu+c_MK^M_\mu,\nonumber\\
c_q&=&\frac{3N_c-2N_f}{N_f}-\gamma_q{\equiv}c_{q}^0-\gamma_q; 
~~~c_M=-2\frac{3N_c-2N_f}{N_f}-\gamma_M
=c_M^0-\gamma_M,
\label{eq7115}
\end{eqnarray} 
where $\gamma_q$ and $\gamma_M$ are the anomalous dimensions 
of the fields $q$, $\widetilde{q}$ and $M$, respectively. 
The meaning of the terms and their coefficients in (\ref{eq7115})
is as follows: the special combination $K_\mu^q-2K_\mu^M$ 
is classically conserved and its coefficient is the numerator of 
the NSVZ beta function of the magnetic gauge
theory; the coefficient of $K_\mu^M$ is proportional to 
the beta function $\beta_f=f(2\gamma_q+\gamma_M)$ of the Yukawa coupling
$f$ of the cubic superpotential
${\cal W}_f=fq_{\widetilde{r}}^{~i}M_{ij}\widetilde{q}^{j\widetilde{r}}$.
\footnote{The form of 
this beta function can be easily understood: due to the non-renormalization 
theorem of ${\cal W}_f$, the Yukawa vertex 
$f({\psi}_q\phi_M\psi_{\widetilde{q}}+\psi_q{\psi}_M\phi_{\widetilde{q}}+
\psi_M{\phi}_M\psi_{\widetilde{q}})$ is not renormalized, hence
the renormalization of the coupling constant $f$ is  determined by
the wave function renormalization constants, $f_R=Z_{q}^{-1}Z_M^{-1/2}f$.
Thus the beta function is $\beta_f=2\gamma_q+\gamma_M$.}

Although the higher order quantum corrections
in constructing the anomaly-free
R-current are considered,  
they actually do not affect 't Hooft's anomaly matching. 
 Since the 't Hooft anomaly is an external
gauge anomaly, the theory should be put in a background of some 
external gauge fields. It was explicitly demonstrated 
that all the higher order contributions to the external
anomalies of the $R$-current cancel exactly \cite{ksv}.
We first consider the $U_R(1)U_B(1)^2$ triangle diagram. 
Introducing the external $U_B(1)$ gauge 
field $G_{\mu}$ to couple to the baryon number current, one 
can easily calculate 
the anomaly of the $R_0$-current of the electric theory \cite{ksv},
\begin{eqnarray}
\partial_{\mu}R_0^\mu=-\frac{1}{48\pi^2}N_fN_c(1-\gamma)
G^{\mu\nu}\widetilde{G}_{\mu\nu},
\label{eq7117}
\end{eqnarray}
where $G_{\mu\nu}=\partial_{\mu}G_{\nu}-\partial_{\nu}G_{\mu}$ is
the external $U_B(1)$ gauge field strength. The anomaly for the
Konishi current $K_{\mu}$ in the external field background, like the 
internal anomaly, comes only from one-loop quantum corrections,
\begin{eqnarray}
\partial_{\mu}K^{\mu}=\frac{1}{48\pi^2}N_fN_c
G^{\mu\nu}\widetilde{G}_{\mu\nu}.
\label{eq7118}
\end{eqnarray}
(\ref{eq7110}), (\ref{eq7117}) and (\ref{eq7118}) lead to
\begin{eqnarray}
\partial_{\mu}R^{\mu}=-\frac{1}{16\pi^2}(-2N_c^2)
G^{\mu\nu}\widetilde{G}_{\mu\nu}.
\label{eq7119}
\end{eqnarray}
Thus the external anomaly of the $R$-current in the triangle diagram
$U_R(1)U_B(1)^2$ receives no higher order contributions.
This cancellation is actually attributed to the definition 
(\ref{eq7110}) of $R_{\mu}$, which ensures that the higher order 
contribution  of the triangle diagram containing $R_0$ is dismissed. 
Note that the anomalies we used to derive Eq.\,(\ref{eq7110}) 
are the internal ones. On the magnetic theory side, it can also 
easily be found from
(\ref{eq715}) that 
$\partial_{\mu}R^{\mu}=N_c^2/(8\pi^2)
G^{\mu\nu}\widetilde{G}_{\mu\nu}$ \cite{ksv}.
Thus the 't Hooft anomalies for the triangle diagram 
$U_R(1)U_B(1)^2$ match exactly as in the one-loop case. Similarly, 
the other two triangle diagrams containing the
$R$-current, the axial gravitational anomaly $U_R(1)$ and the triangle
diagram $U_R(1)SU(N_f)^2$, also match exactly as in the one-loop case.

There remains the $U_R(1)^3$ triangle diagram. 
This diagram only concerns the $U_R(1)$ current. It was argued from
the holomorphic dependence of the quantum effective action on the external
field that the external anomaly for this triangle diagram gets 
only contributions from one-loop quantum corrections \cite{ksv}. It is 
well known that the Wilson effective action of 
a pure supersymmetric Yang-Mills theory depend holomorphically 
on the gauge coupling and field strength \cite{ref6,ref7}.
In the presence of matter fields such as quarks and mesons, 
the Wilson effective action of a
supersymmetric gauge theory  with the inclusion of 
an external gauge field takes in Pauli-Villars regularization 
the following general form \cite{ref7,ksv},
\begin{eqnarray}
S_W(\mu)&=&\frac{1}{16\pi^2}\left[\frac{8\pi^2}{g_0^2}
-\left(3C_G\ln\frac{M_0}{\mu}-\sum_i
\ln\frac{M_i}{\mu}\right)\right]
\int d^4x d^2\theta\mbox{Tr}W^2\nonumber\\
&&+\frac{1}{16\pi^2}\left[\sum_i 
C_i^{\rm ext}\ln\frac{M_i}{\mu}\right]
\int d^4x d^2\theta\mbox{Tr}W^2_{\rm ext}
\nonumber\\
&&+\sum_i\frac{Z_i}{4}\int d^4xd^2\theta d^2\overline{\theta}
\Phi^{i\dagger}e^V\Phi^i
+\left[\frac{1}{2}\int d^4xd^2\theta {\cal W}(\Phi)+\mbox{h.c.}\right],
\label{eq7121}
\end{eqnarray}
where $g_0$ is the bare gauge coupling, $\mu$ is 
the renormalization scale, it also plays the role of infrared cut-off
in the Wilson effective action \cite{ref7}.
$M_0$ and $M_i$ are the Pauli-Villars regulator masses for the ghost
superfields and matter fields, respectively. The ghost fields arise
due to the (super-)gauge-fixing. $C_G$ is the $C_i$ in the adjoint
representation of gauge group; $W_{\rm ext}$ is the superfield strength
corresponding to the external gauge field; $C_i^{\rm ext}$ are
defined similar to $C_i$ for the generators when they appear in 
the interaction term with the external gauge field. 
${\cal W}(\Phi^i)$ is the superpotential,
and its presence or absence depends on the concrete model.

The holomorphy of the vector part of the Wilson effective action 
determines that the coefficients in front of $W^2$ and $W^2_{\rm ext}$ 
are saturated by only the one-loop quantum correction. 
Higher order quantum corrections only enter the wave 
function renormalization constant $Z$, and are absent 
for the coupling constant. Note that the Wilson effective 
action is actually an operator action. 
Its expectation value yields the usual quantum effective action, 
i.e. the generating functional of the 1PI Green functions,  
\begin{eqnarray}
\langle e^{iS_W(\mu)}\rangle=e^{i\Gamma (\mu)}.
\label{eq7122}
\end{eqnarray}
The one-loop contribution to the $R_0$ anomaly (including 
both the internal and external anomalies) 
can be obtained through the action of the operator
$M_0{\partial}/{\partial M_0}+\sum_iM_i{\partial}/{\partial M_i}$
on the vector field part of the one-loop quantum effective action,
\begin{eqnarray}
\Gamma^{\rm one-loop}&=&\frac{1}{16\pi^2}\left[\frac{8\pi^2}{g_0^2}
-\left(3C_G\ln\frac{M_0}{\mu}
-\sum_i\ln\frac{M_i}{\mu}\right)\right]
\int d^4x d^2\theta\mbox{Tr}W^2\nonumber\\
&&+\frac{1}{16\pi^2}\left[\sum_iC_i^{{\rm ext}}
\ln\frac{M_i}{\mu}\right]
\int d^4x d^2\theta\mbox{Tr}\left(W^2_{{\rm ext}}\right).
\label{eq7123}
\end{eqnarray}
The anomaly for the Konishi current is also given by the differentiation
$M_i\frac{\partial}{\partial M_i}\Gamma^{\rm one-loop}$.
This way of calculating the anomaly confirms that the $R$-currents 
constructed in Eqs.\,(\ref{eq7110}) and (\ref{eq7115}) 
indeed have no internal anomaly, since the $W^\alpha W_\alpha$ 
term of $\Gamma^{\rm one-loop}$ 
is invariant under the action of the operator
\begin{eqnarray}
M_0\frac{\partial}{\partial M_0}+\sum_iM_i\frac{\partial}{\partial M_i}
\left(1+c_i^{(0)}\right), ~~~~c_i^{(0)}=c_q^{(0)},  c_{\cal M}^{(0)}.
\label{eq7125}
\end{eqnarray} 
The non-invariance of the 
$W^2_{{\rm ext}}$ part
in (\ref{eq7123}) under the action
of (\ref{eq7125}) gives the 
external anomaly of the $R$-current. 
The higher order quantum effect is reflected  in 
the presence of the wave function renormalization 
factors $Z_i$, which can be included by the replacement \cite{ref6} 
\begin{eqnarray}
M_i {\longrightarrow} {\cal M}_i= \frac{M_i}{Z_i}, 
~~~~M_0{\longrightarrow} {\cal M}_0= \frac{M_0}{(g_0/g)^{2/3}}.
\label{eq7126}
\end{eqnarray}
Eq.\,(\ref{eq7126}) also implies that the role of $Z_i$ for 
the ghost regulator 
mass $M_0$ is played by the factor $(g_0/g)^{2/3}$. Consequently,
the multi-loop quantum effective action is
\begin{eqnarray}
\Gamma^{\rm multi-loop}&=& \frac{1}{16\pi^2}\left[\frac{8\pi^2}{g_0^2}
-\left(3C_G\ln\frac{M_0}{(g_0/g)^{2/3}\mu}
-\sum_i\ln\frac{M_i}{Z_i\mu}\right)\right]
\int d^4x d^2\theta\mbox{Tr}W^2\nonumber\\
&&+\frac{1}{16\pi^2}\left[\sum_iC_i^{{\rm ext}}
\ln\left(\frac{M_i}{Z_i\mu}\right)\right]
\int d^4x d^2\theta\mbox{Tr}W^2_{{\rm ext}}.
\label{eq7127}
\end{eqnarray}
The $W^2$ part of this multi-loop quantum effective action 
is invariant under the action of the modified operator
\begin{eqnarray}
{\cal M}_0\frac{\partial}{\partial {\cal M}_0}
+\sum_i{\cal M}_i\frac{\partial}{\partial {\cal M}_i}
\left(1+c_i^{(0)}\right).
\label{eq7128}
\end{eqnarray}
This means that the $R$-current remains internal anomaly-free 
in the presence of higher order quantum correction, as it should be, 
while the action of (\ref{eq7128}) on the 
$W^2_{\rm ext}$ part of $\Gamma^{\rm multi-loop}$ 
yields the external anomaly of the $R$-current.
With the relation (\ref{eq7126}) it can be easily found that 
\begin{eqnarray}
&&\left[{\cal M}_0\frac{\partial}{\partial {\cal M}_0}+\sum_i{\cal M}_i
\frac{\partial}{\partial {\cal M}_i}\left(1+c_i^{(0)}\right)\right]
\Gamma^{\rm multi-loop}\nonumber\\
&=&\left[M_0\frac{\partial}{\partial M_0}
+\sum_iM_i\frac{\partial}{\partial M_i}\left(1+c_i^{(0)}\right)\right]
\Gamma^{\rm one-loop}.
\label{eq7129}
\end{eqnarray}
This implies that the external anomaly of the $R$-current is exhausted
by the one-loop quantum correction, and thus the anomalies of the $U_R(1)^3$ 
triangle diagram match like in the one-loop case. 

The above discussion is a detailed analysis of the infrared behaviour 
of the $R$-current with the inclusion of the higher order quantum 
correction carried out in the operator form of the anomaly
equation. One can also start from the matrix element form of the anomaly 
equation \cite{ref10}. The matrix element
form of the anomaly equation for the Konishi current
has indeed revealed some more interesting 
features than its operator form. The operator anomaly 
equations for the Konishi current, Eqs.\,(\ref{eq719})
and (\ref{eq7118}), imply that its internal and external anomalies 
have only one-loop character, but from the matrix element form 
one sees that the external anomaly 
of the Konishi current is proportional to the $\beta$-function
and hence presents multi-loop character. Furthermore,
the matrix element form 
shows that the Konishi current must be renormalized and thus it receives
an anomalous dimension. This implies that the Konishi current remains 
anomalous at the critical point, despite the fact that the matrix element
of its operator anomaly equation is proportional to
the $\beta$-function, which vanishes
at the critical point. In the following we shall illustrate this
by working out two examples, supersymmetric QED and QCD.

For supersymmetric QED,  
its generating functional in the presence 
of an external vector superfield $V_{\rm ext}$ is  
\begin{eqnarray}
Z=\int {\cal D}V{\cal D}\Phi{\cal D}\widetilde{\Phi}
e^{iS(V,V_{\rm ext},\Phi,\widetilde{\Phi})}
=e^{\Gamma[V_{\rm ext}]},
\label{eq7131}
\end{eqnarray}
where $S[V,V_{\rm ext},\Phi,\widetilde{\Phi}]$ is 
the classical action of supersymmetric QED,
\begin{eqnarray}
S&=&\int d^4x\left\{\frac{1}{4g_0^2}
\left[\int d^2\theta W^2+\mbox{h.c.}\right]
+\int d^2\theta d^2\overline{\theta}
\left[\Phi^\dagger e^{-(V+V_{\rm ext})}\Phi
+\widetilde{\Phi}e^{-(V+V_{\rm ext})}\widetilde{\Phi}^{\dagger}
\right]\right\}\nonumber\\
&&+\mbox{ gauge fixing part},
\label{eq7132}
\end{eqnarray}
and $\Gamma$ is the quantum effective action. We consider the case 
that the external momentum $k {\gg}|W^2_{\rm ext}(k)|$,
then only the terms quadratic in $W_{\rm ext}$ survive in an expansion
in powers of $|W^2_{\rm ext}(k)|/|k|^3$. Thus 
the quantum effective action in this energy region is \cite{ans}
\begin{eqnarray}
\Gamma\left[V_{\rm ext}\right]
=\int d^4x d^2\theta \frac{1}{4g^2_{\rm eff}(k)}W^2_{\rm ext}+
\mbox{h.c.} \, .
\label{eq7133}
\end{eqnarray}
Eqs.\,(\ref{eq716}) and (\ref{eq715}) show that 
the operator form of the Konishi anomaly $\overline{D}^2K$ and 
the supercurrent anomaly $\overline{D}^{\dot{\alpha}}J_{\alpha\dot{\alpha}}$ 
are proportional to the vector superfield strength operators 
$W^2$ and $D_{\alpha}W^2$, respectively. Thus to determine 
the matrix element of the operator anomaly, one should first compute the 
expectation value $\langle W^2\rangle$. From Eqs.(\ref{eq7131})
and (\ref{eq7132}),   
this expectation value can be expressed in terms of the external field 
$V_{\rm ext}$, integrated over $d^4xd^2\theta$, and it is obtained by
taking the logarithmic derivative of $Z$ with respect to $1/g_0^2$, 
\begin{eqnarray} 
\int d^4x d^2\theta \langle W^2\rangle + \mbox{h.c.}=
-4i\frac{\partial}{\partial (1/g_0^2)}\ln Z. 
\label{eq7134}
\end{eqnarray}    
Omitting the integration over $x$ and $\theta$, (\ref{eq7133}) 
and (\ref{eq7134}) imply that the matrix element of the chiral 
superfield operator $W^2$ is directly connected to the external 
superfield strengths $W^2_{\rm ext}$ \cite{ref8,ref10},
\begin{eqnarray}
\langle W^2\rangle &=&\left[ \frac{\partial}{\partial (1/g_0^2)}
\frac{1}{g^2_{\rm eff}(k)}\right]W^2_{\rm ext}
=\frac{\partial g_0^2}{\partial (1/g_0^2)}
\frac{\partial g^2_{\rm eff}}{\partial g_0^2}
\frac{\partial}{\partial g_{\rm eff}^2}\left(\frac{1}{g^2_{\rm eff}}
\right)W^2_{\rm ext}\nonumber\\
&=&\frac{g_0^4}{g^4_{\rm eff}(k)}\frac{\beta[g^2_{\rm eff}(k)]}{\beta(g_0^2)}
W^2_{\rm ext}= \frac{\beta (g^2_{\rm eff})}{\beta_0(g^2_{\rm eff})}
\frac{\beta_0(g^2_0)}{\beta(g^2_0)}W^2_{\rm ext}. 
\label{eq7135}
\end{eqnarray}
Here $\beta_0=g^4/(2 \pi^2)$ is the one-loop $\beta$-function of 
supersymmetric QED \cite{ref8,ref10}. 
The factor $\beta_0 (g_0^2)/\beta (g_0^2)$ 
depends on the UV cut-off $\Lambda$, and hence it can be identified as
a renormalization factor of the operator $W^2$, i.e.  
\begin{eqnarray}
W^2=\frac{\beta_0(g_0^2)}{\beta (g_0^2)}W^2_{\rm ren}.
\label{eq7136}
\end{eqnarray}
Consequently, Eq.\,(\ref{eq7135}) becomes
\begin{eqnarray}
\langle W^2_{\rm ren}\rangle =
\frac{\beta (g^2_{\rm eff})}{\beta_0(g^2_{\rm eff})}W^2_{\rm ext}.
\label{eq7137}
\end{eqnarray}
(\ref{eq7137}) and the Abelian analogue of the 
operator anomaly equations, (\ref{eq716}) and (\ref{eq715}) \cite{ref7},
\begin{eqnarray}
\overline{D}^{\dot{\alpha}}J_{\alpha\dot{\alpha}}
=-\frac{\beta (g_0^2)}{6g_0^4}D_{\alpha}W^2, ~~~
D_{\alpha}\overline{D}^2K
=\frac{\beta_0 (g_0^2)}{g_0^4}D_{\alpha}W^2,
\label{eq7138}
\end{eqnarray}
lead to the renormalized matrix elements of the operator anomaly equation
 in the external background field $V_{\rm ext}$,
\begin{eqnarray}
\langle \overline{D}^{\dot{\alpha}}J_{\alpha\dot{\alpha}} \rangle
&=&-\frac{\beta (g_0^2)}{6g_0^4} \langle D_{\alpha} W^2 \rangle
=-\frac{1}{6(2\pi)^2}\frac{\beta (g_0^2)}{\beta (g_0^2)}
\langle D_{\alpha} W^2 \rangle
=-\frac{1}{6 (2\pi)^2} \langle D_\alpha W^2_{\rm ren} \rangle 
\nonumber\\
&=&-\frac{1}{6 (2\pi)^2}
\frac{\beta (g^2_{\rm eff})}{\beta_0 (g^2_{\rm eff})}
\langle D_\alpha W^2_{\rm ext}\rangle
=-\frac{\beta (g^2_{\rm eff})}{6g^4_{\rm eff}} D_{\alpha}W^2_{\rm ext}, 
\nonumber\\
\langle \overline{D}^2 K\rangle_{\rm ren}
&=&\frac{\beta (g^2_{\rm eff})}{g^4_{\rm eff}}W^2_{\rm ext}.
\label{eq7139}
\end{eqnarray}
(\ref{eq7135}) and (\ref{eq7139}) imply that the supersymmetry supercurrent
is not renormalized and it is conserved at the fixed point, whereas
the Konishi current must be renormalized with the renormalization factor 
$Z_K(g_0^2)={\beta (g_0^2)}/{\beta_0(g_0^2)}$,
and the  renormalized Konishi current is
\begin{eqnarray}
K_{\rm ren}=Z_K K.
\label{eq7141}
\end{eqnarray}
Therefore, the Konishi current will stay anomalous at the IR fixed point
despite the fact that the matrix element of its anomaly equation is 
proportional to the beta function.

For supersymmetric QCD, the analysis is not so simple as in the 
Abelian case because the external background
gauge field is charged with respect to the gauge group, 
otherwise it cannot participate in the interaction with the 
matter fields. Furthermore, the matrix element of a current
operator depends on gauge fixing. However, in the kinematic 
region where the currents carry zero momentum, 
their matrix elements can be reduced to a form
similar to (\ref{eq7139}) \cite{ans}. With the one-loop form of the 
Wilson effective action (\ref{eq7121})
\begin{eqnarray}
S_W[\mu]&=&\int d^4x d^2\theta \left[\frac{1}{2g_0^2}
+\beta_0\ln\left(\frac{\Lambda}{\mu}\right)\right]
\mbox{Tr}\left(W^2+W^2_{\rm ext}\right)\nonumber\\
&+&\int d^4x d^2\theta d^2\overline{\theta}
\sum_{i=1}^{N_f}\frac{Z_i}{4}
\left({Q}^{i\dagger}e^VQ^i+\widetilde{Q}^i
e^{-V}\widetilde{Q}^{i\dagger}\right)\nonumber\\
&=&\frac{1}{2g^2_{\rm one-loop}} \int d^4x d^2\theta
\mbox{Tr}\left(W^2+W^2_{\rm ext}\right)\nonumber\\
&+&\int d^4x d^2\theta d^2\overline{\theta}
\sum_{i=1}^{N_f}\frac{Z_i}{4}
\left({Q}^{i\dagger}e^VQ^i+\widetilde{Q}^i
e^{-V}\widetilde{Q}^{i\dagger}\right), 
\label{eq7142}
\end{eqnarray}
and the relation between $S_W[\mu]$ and $\Gamma [\mu]$, 
(\ref{eq7122}), the differentiation
of the 1PI effective action with respect to the one-loop coupling  
$1/g^2_{\rm one-loop}$ gives
\begin{eqnarray}
\langle \mbox{Tr}W^2\rangle =\frac{\beta (\alpha)}{\beta_0 (\alpha)}
\frac{\beta_0 (\alpha_0)}{\beta (\alpha_0)} \frac{1}{1-\alpha_0 N_c/(2\pi)}
\mbox{Tr}W^2_{\rm ext},
\label{eq7143}
\end{eqnarray}
where $\beta (\alpha)$ is the NSVZ $\beta$ function, and 
$\beta_0$ is the first order one.
 The matrix elements form  of the anomaly equations 
(\ref{eq716}) and (\ref{eq715}) 
at the renormalization scale $\mu$,
can be derived in the same way as in the QED case, 
\begin{eqnarray}
\langle \overline{D}^{\dot{\alpha}}J_{\alpha\dot{\alpha}} \rangle
=\frac{\beta (\alpha)}{24\pi\alpha^2}
\mbox{Tr}\left(D_{\alpha}W^2_{\rm ext}\right),~~~~
\langle \overline{D}^2K_{\rm ren}\rangle =\frac{N_f}{2\pi^2}
\frac{\beta (\alpha)}{\beta_1(\alpha)}\mbox{Tr}W^2_{\rm ext},
\label{eq7144}
\end{eqnarray}
where the renormalized Konishi current is defined as \cite{ans}
\begin{eqnarray}
K_{\rm ren}=\left(1-\frac{\alpha_0N_c}{2\pi}\right)Z_K(\alpha_0)K
\label{eq7145}
\end{eqnarray}
and the renormalization factor $Z_K(\alpha_0)
=\beta (\alpha_0)/\beta_0(\alpha_0)$.

Furthermore, the renormalization group invariance of
the two-point correlator of the Konishi current shows that its 
anomalous dimension is related to the slope of 
the beta function at the critical point. 
The scale dimension for a general
local operator $O(x)$ at the fixed point is the sum of its canonical 
dimension $d_0$ and anomalous dimension $\gamma (\alpha_\star)$ 
at the critical value of the coupling constant, which can be determined 
from the scaling behaviour of its two-point correlator 
$\langle O(x)O(y)\rangle$ at large distance 
($|x-y|{\rightarrow}\infty$). The Callan-Symanzik equation 
(in coordinate space)
\begin{eqnarray}
\left(|x-y|\frac{\partial}{\partial |x-y|}+2 d_0+2\gamma (\alpha) 
+\beta (\alpha)\frac{\partial}{\partial \alpha}\right)
\langle O(x)O(y)\rangle=0
\label{eq7146}
\end{eqnarray}
determines that the correlator should be of the form
\begin{eqnarray}
\langle O(x)O(y)\rangle 
=\left[Z_O(x-y)\right]^2\phi\left[\alpha (x-y)\right],   
\label{eq7147}
\end{eqnarray}
where $Z_O$ is the renormalization factor for the operator $O$, $\alpha (x-y)$
is the running coupling constant defined at the scale $\mu\sim 1/|x-y|$, 
$\phi [\alpha (x-y)]$ is an unknown function up to some space-time
or internal indices. At the critical point, $\beta (\alpha_\star)=0$,
$\gamma (\alpha)=\gamma_{\star}$, 
the correlator (\ref{eq7147}) behaves as 
\begin{eqnarray}
\langle O(x) O(y) \rangle {\sim}\frac{1}{|x-y|^{2d_0+2\gamma_\star}}.
\label{eq7148}
\end{eqnarray}

For the Konishi current, as discussed above, its 
renormalization factor depends on the beta function. 
It is enough to look at the two-point
correlator of its axial vector component,
\begin{eqnarray}
a_{\mu}{\sim}\left[\overline{D}_{\dot{\alpha}},D_\alpha\right]J|_{\theta =0}.
\label{eq7149}
\end{eqnarray}
(\ref{eq7141}), (\ref{eq7145}), the Lorentz covariance and 
the renormalization group equation for 
$\langle a_{\mu} (x) a_{\nu} (y)\rangle $ lead to \cite{ref10}
\begin{eqnarray}
\langle a_{\mu} (x) a_{\nu} (y)\rangle =
\left[\frac{\beta (\alpha)}{\beta_0(\alpha)}\right]^2 
\left[\frac{\phi_1(\alpha)g_{\mu\nu}}{|x-y|^6}
+\frac{(x-y)_{\mu} (x-y)_{\nu}\phi_2(\alpha)}{|x-y|^8}\right],
\label{eq7150}
\end{eqnarray}
where the form factors $\phi_{1,2}(\alpha)$ cannot
 be explicitly determined from
the Callan-Symanzik equation. At large distance, the running
gauge coupling $\alpha (|x-y|)$ flows to $\alpha_*$, i.e. the value at the 
IR fixed point. Without losing generality, we assume that the 
$\beta$-function has only a simple zero. Hence near the critical point,
\begin{eqnarray}
\beta (\alpha)=\beta^{\prime}(\alpha_*) \left(\alpha-\alpha_*\right).
\label{eq7151}
\end{eqnarray}
The definition of the $\beta$ function at the scale $\mu{\sim}{1}/{|x-y|}$,
 $\beta\left[\alpha (\mu)\right]
={d\alpha (\mu)}/{d\ln\mu}$,
gives
\begin{eqnarray}
\alpha_*-\alpha =|x-y|^{-\beta^{\prime} (\alpha^*)}, ~~~~\mbox{near}~
 |x-y|{\longrightarrow}\infty.
\label{eq7153}
\end{eqnarray}
In the case that $\phi_{1,2}$ have no pole at the critical point, 
the substitution (\ref{eq7153}) into (\ref{eq7150}) yields \cite{ref10},
\begin{eqnarray}
\langle a_{\mu}(x) a_{\nu}(y)\rangle {\sim} 
\frac{\phi_1(\alpha)g_{\mu\nu}}{|x-y|^{6+2\beta^{\prime}(\alpha_*)}}
+\frac{(x-y)_{\mu} (x-y)_{\nu}\phi_2(\alpha)}
{|x-y|^{8+2\beta^{\prime}(\alpha_*)}},
\label{eq7154}
\end{eqnarray}
Thus the anomalous dimension of the Konishi current is related to the 
slope of the beta function.

In the dual magnetic theory, similarly to the operator form, 
the matrix element form of the Konishi current anomaly equation
will become quite complicated due to the flavour interaction superpotential
among the magnetic quarks and the singlet fields. The conservation
of the Konishi current is spoiled by both the superpotential and 
the gauge anomaly.  To extract the matrix element form  of the Konishi
current anomaly, a technique was invented \cite{ref10} which
consists of first constructing an anomaly-free Konishi current 
for the Kutasov-Schwimmer model introduced in Sect.\,\ref{subsect43} 
\cite{kss,ref431,ref432}, then making the adjoint 
matter field decouple by introducing a large mass term for it. 
In this way, one can obtain the Konishi current in the usual 
supersymmetric QCD, which is thus called minimal supersymmetric QCD 
as in Refs.\,\cite{kss,ref431,ref432}. Recall that the 
electric theory side of the Kutasov-Schwimmer model allows the 
superpotential (\ref{eq4p328}), $W_{\rm el}=g_k\mbox{Tr}X^{k+1}$,
$X$ being the matter field in the adjoint representation
of the $SU(N_c)$ gauge group. The magnetic theory 
has a  superpotential (\ref{eq4p343})  
$W_{\rm mag}=\widetilde{g}_k\mbox{Tr}Y^{k+1}
+\sum_{p=1}^kt_p\widetilde{M}_p\widetilde{q}Y^{k-h}q$,
$g_k$, $\widetilde{g}_k$ and $t_p$ are the corresponding coupling 
constants, which are explicitly written out here. In the 
critical point of the model, the singlet field $\widetilde{M}_p$ 
is identical to the meson field $M_p$ in the electric theory 
defined by (\ref{eq4p329}) due to the duality.

When the Kutasov-Schwimmer model is at the critical point,
the various couplings including both the electric and  magnetic
gauge coupling and those appearing in the superpotentials $W_{\rm el}$ 
and $W_{\rm mag}$ take their critical values, $\alpha=\alpha_{\sharp}$,
$\widetilde{\alpha}=\widetilde{\alpha}_{\sharp}$, $s=s_{\sharp}$ etc.
The anomaly-free Konishi currents in both the electric and magnetic
theories are given by the Noether currents corresponding to 
the non-anomalous $U_A(1)$ transformations on the matter fields. 
According to Eqs.\,(\ref{eq4p313}) and (\ref{eq4p317}), 
the absence of the $U(1)$ anomalies in both the electric 
and magnetic theories requires that
\begin{eqnarray} 
&& N_fq_Q+N_c q_X=0, \nonumber\\
&& N_fq_q-(N_c-N_f)q_Y=0,
\label{eq7157}
\end{eqnarray}
where $q_Q(=q_{\widetilde{Q}})$, $q_X$, $q_q(=q_{\widetilde{q}})$, $q_Y$ 
are the corresponding $U(1)$ charges of $Q$, $\widetilde{Q}$, 
$X$, $q_q$, $q_{\widetilde{q}}$, 
respectively. At the same time,
the $U(1)$ invariance of the superpotentials $W_{\rm el}$  
and $W_{\rm mag}$ in both the electric and magnetic 
theories assign the coupling
$s$, $\widetilde{s}$, $t_p$ with the charges
\begin{eqnarray} 
&& q_s=-(k+1)q_X, ~~~q_{\widetilde{s}}=-(k+1)q_Y, \nonumber\\
&& q_{t_p}=\left(\frac{2N_c}{N_f}+1-p\right)q_X
+\left(p-k+2\frac{N_f-N_c}{N_f}\right)q_Y,~~p=1,2,\cdots,k,
\label{eq7158}
 \end{eqnarray}
where use was made of the first relation of (\ref{eq7157}) and the 
identification of the singlet fields $\widetilde{M}_p$ in 
the magnetic theory with the meson fields $M_p$ of the electric 
theory at the critical point. Furthermore, matching the 
coupling constant $t_p$ at the scale of 
decoupling one heavy flavour requires 
that the charges $q_{t_p}$ remain
identical under a deformation $N_f{\longrightarrow} N_f-1$. This
leads to \cite{ref10}
\begin{eqnarray}
q_X=q_Y, 
\label{eq7159}
\end{eqnarray}
and hence
\begin{eqnarray} 
q_s=q_{\widetilde{s}}, ~~~q_{t_p}=-q_s\frac{3-k}{k+1}.
\label{eq7160}
\end{eqnarray}
The relations (\ref{eq7158}), (\ref{eq7159}) and  (\ref{eq7160})   
determine that the non-anomalous Konishi supercurrent in the critical electric 
and magnetic Kutasov-Schwimmer model is a linear combination of the 
Konishi current in the minimal supersymmetric QCD and a current 
constructed from the adjoint matter fields \cite{ref10},
\begin{eqnarray}
K_{\rm el}&=&\sum_{i=1}^{N_f}\left(Q^{i\dagger}e^VQ^i
+\widetilde{Q}^ie^{-V}\widetilde{Q}^{i\dagger}\right)
-\frac{N_f}{N_c}\mbox{Tr}\left(X^{\dagger}e^VXe^{-V}\right), \nonumber\\
K_{\rm mag}&=&\sum_{i=1}^{N_f}\left(q^{i\dagger}e^Vq^i
+\widetilde{q}^ie^{-V}\widetilde{q}^{i\dagger}\right)
-\frac{N_f}{N_c}\mbox{Tr}\left(Y^{\dagger}e^V Ye^{-V}\right)\nonumber\\
&& +\left[2+\left(1-k\right)\frac{N_f}{N_c}\right]\mbox{Tr}
\left(M^{\dagger}M\right),
\label{eq7161}
\end{eqnarray}
where the $U(1)$ charges of the electric quark supermultiplet are chosen
as $1$ as shown in (\ref{eq715a}).

We decouple the adjoint field $X$  in the electric theory by 
introducing its mass term $m\mbox{Tr}X^2$. After $X$ is integrated out,
the theory returns to the minimal supersymmetric QCD with only quark
superfields $Q$, $\widetilde{Q}$. Furthermore, due to the duality map
of the chiral operators at the critical point of the Kutasov-Schwimmer
model \cite{kss,ref431,ref432}, $Y$ also 
acquires a mass term $m\mbox{Tr}Y^2$, thus 
in the magnetic theory the minimal supersymmetric QCD will be reproduced
in the low-energy limit with the matter fields $q$, $\widetilde{q}$ and $M$.
There are two delicate points to make clear in
realizing this: first, to precisely produce the minimal dual magnetic theory,
one must choose a phase to ensure that the magnetic 
gauge group $SU(kN_f-N_c-k)$
breaks down to $SU(N_f-N_c)$. This in principle should be under control. 
Second, despite starting from a critical 
Kutasov-Schwimmer model and decoupling the heavy field $X$ 
(in the electric theory )
or $Y$ (in the magnetic theory ), 
usually one does not get the minimal supersymmetric
QCD still at the critical point. This can be understood  
from another viewpoint: the heavy fields $X$ and $Y$ can be thought of as 
the regulator fields of the minimal supersymmetric QCD away from its 
critical point and their masses $m$ as the UV cut-off. Consequently, 
the resulting Konishi operators in the minimal theory should 
be thought of as bare operators defined at the scale given 
by $m$. Equivalently speaking, the Konishi
operators in the Kutasov-Schwimmer model may be thought of as 
operators of the minimal supersymmetric QCD regularized in a particular way. 
Therefore, the Konishi operators in the minimal (electric or magnetic) 
theories can be defined by directly dropping the fields $X$ and $Y$.
This immediately gives 
\begin{eqnarray}
K_{\rm el} &=& \sum_{i=1}^{N_f}\left(Q^{i\dagger}e^VQ^i
+\widetilde{Q}^ie^{-V}\widetilde{Q}^{i\dagger}\right); \nonumber\\
K_{\rm mag} &=& 2\mbox{Tr}\left(M^{\dagger}M\right)
+\sum_{i=1}^{N_f}\left(q^{i\dagger}e^Vq^i
+\widetilde{q}^ie^{-V}\widetilde{q}^{i\dagger}\right)
{\equiv}2J_M+\frac{N_f-N_c}{N_f}J_q.
\label{eq7163}
\end{eqnarray}
Eq.(\ref{eq7163}) shows that the electric Konishi current obtained 
in this way has its usual form, and hence is expected to flow in 
the infrared region
to the corresponding current in the critical minimal theory.

The magnetic Konishi current needs a delicate analysis. 
Since the $K_{\rm mag}$ given in (\ref{eq7163}) 
is obtained by the reduction of the 
magnetic Konishi current in the critical Kutasov-Schwimmer model, 
it is defined at the scale $m$. 
At this scale the various couplings of the magnetic minimal supersymmetric
QCD are equal to their critical values with the heavy
field $Y$ present, i.e. the critical values of the coupling constants in
Kutasov-Schwimmer model, $\widetilde{\alpha}=\widetilde{\alpha}_{\sharp}$, 
$\lambda=\lambda_{\sharp}$. Below the scale $m$, as stated above,
the low energy theory is the usual non-critical minimal supersymmetric QCD.
Thus the Konishi operator in (\ref{eq7163}) is actually
 defined on a particular 
renormalization group trajectory, which is a path
in the space parametrized by the couplings 
$(\widetilde{\alpha},\lambda)$ of the non-critical minimal 
supersymmetric QCD. After this point is clear, one can easily see  
how the Konishi current flows to the IR fixed point of 
the magnetic theory of the minimal supersymmetric QCD.
Note that the Konishi current operator is a composite operator, 
thus in general its renormalization 
will lead to mixing with operators of lower dimensions. 
However, before the decoupling of the heavy fields $X$ and $Y$, the 
electric and magnetic Konishi current operators are
identical in the critical Kutasov-Schwimmer model. Thus their
dynamical behaviour along the renormalization group trajectory, i.e.
at any distance (or energy scale), should be identical. 
This means that the magnetic Konishi current operator is renormalized 
multiplicatively, and hence it cannot mix with 
operators of lower dimension as soon as this is true for 
the electric Konishi current operator in the off-critical electric theory. Therefore,
the magnetic Konishi current operator can be rewritten as a linear combination
of renormalization group invariant operators in 
an off-critical minimal magnetic theory,
\begin{eqnarray}
K_{\rm mag}=AK_1+BK_2,
\label{eq7164}
\end{eqnarray}
where $K_1$ and $K_2$ are two fundamental
renormalization group invariant operators constituting a basis for 
any other renormalization group invariant operators. These 
two operators diagonalize the  anomalous dimensions matrix,  
i.e. that they do not mix with each other 
along the trajectory of renormalization group flow. $A$ and $B$ are two 
constants determined  by the original critical Kutasov-Schwimmer
model, thus they depend only on the critical values of the couplings, 
$\widetilde{\alpha}=\widetilde{\alpha}_\sharp$, $\lambda=\lambda_\sharp$. 

The decomposition (\ref{eq7164}) is explicitly suitable for analyzing the IR 
behaviour of the magnetic Konishi current near the fixed point. 
$K_1$ and $K_2$ can be constructed by inspecting the various $U(1)$
currents and their anomalies as well as their relations with the 
Konishi current operator. In the off-critical minimal supersymmetric 
QCD, one well known renormalization group invariant operator is the 
non-anomalous $R$-current. However, it does not mix with the Konishi current
since they belong to current supermultiplet with different 
superspins. Thus this $R$-current should be excluded from the
candidates for $K_1$ and $K_2$. Two other possible candidates
are combinations of $K^M$ and $K^q$ defined in (\ref{eq7115}).
One is $K_W{\equiv}K^q-2K^M$, whose 
divergence is given by the gauge anomaly, and the other current is
$K_{\rm sp}{\equiv}K^M$, whose divergence is 
only proportional to the superpotential.
The magnetic Konishi current (\ref{eq7163}) can be 
rewritten in the following form in terms of these two new currents,
\begin{eqnarray}
K_{\rm mag}=\frac{N_f-N_c}{N_c}K_W+2\frac{N_f}{N_c}K_{\rm sp}.
\label{eq7165}
\end{eqnarray}
Unfortunately, $K_W$ and $K_{\rm sp}$ mix under the renormalization
group flow, so they cannot play the roles of $K_1$ and $K_2$. 
It is found that $K_1$ is actually the following combination \cite{ref10}, 
\begin{eqnarray}
K_1=\frac{2N_f-3N_c+N_f\gamma_q}{48 \pi^2}K_W+\widetilde{\beta}_\lambda K_{\rm sp}
 =\widetilde{\beta}_{\widetilde{\alpha}}K_W
+\widetilde{\beta}_{\lambda} K_{\rm sp},
\label{eq7166}
\end{eqnarray}
where $\widetilde{\beta}_{\lambda}=\lambda (\gamma_M+2\gamma_q)/2$ 
is the beta function of $\lambda$, and $\gamma_q$, $\gamma_M$ are 
the anomalous dimensions of 
the fields $q$, $\widetilde{q}$ and $M$, respectively.
This form of $K_1$ explicitly remains invariant along the trajectory
of the renormalization group flow since 
its $D_\alpha\overline{D}^2$ divergence is proportional
to the anomaly of the supercurrent $J_{\alpha\dot{\alpha}}$,  
\begin{eqnarray}
D_\alpha\overline{D}^2 K_1=
\overline{D}^{\dot{\alpha}}J_{\alpha\dot{\alpha}}=
-\frac{2N_f-3N_c+N_f\gamma_q}{48 \pi^2} 
D_\alpha\overline{D}^2\mbox{Tr}W^2_{\rm mag} 
+\widetilde{\beta}_{\lambda} D_\alpha\overline{D}^2{\cal W},
\label{eq7167}
\end{eqnarray}
where $W_{\rm mag}$ is the gauge superfield strength of the magnetic gauge
theory and ${\cal W}$ is the interaction superpotential
among the magnetic quarks and the gauge singlet, 
${\cal W}=\lambda q_iM^i_{~j}\widetilde{q}^j$.
To find the renormalization group invariant combination $K_2$ 
requires determining the matrix $\widetilde{\Gamma}$ 
of anomalous dimensions. A similar procedure to the one used 
to derive (\ref{eq7141}) and (\ref{eq7145}) shows
that the entries of $\widetilde{\Gamma}$ 
are proportional to linear combinations of the beta functions
in the IR limit. Thus 
the magnetic Konishi current (\ref{eq7165}) 
in the minimal supersymmetric QCD can formally be written as
\begin{eqnarray}
K_{\rm mag}=A(\widetilde{\alpha}_{\sharp},\lambda_{\sharp}, m)K_1
+B (\widetilde{\alpha}_{\sharp},\lambda_{\sharp}, m)K_2. 
\label{eq7168}
\end{eqnarray}
In particular, $A(\widetilde{\alpha}_{\sharp},\lambda_{\sharp}, m)$ 
and $B(\widetilde{\alpha}_{\sharp},\lambda_{\sharp},m)$ 
have well defined non-vanishing
limits at $m{\longrightarrow}\infty$.
The form of Eq.\,(\ref{eq7168}) makes it possible to compute the 
anomalous dimension of the magnetic Konishi current. 
Like what was done in the electric theory, we consider the
two-point correlator 
$\langle\widetilde{a}_{\mu}(x)\widetilde{a}_{\nu}(y)\rangle$ 
of its axial vector component $\widetilde{a}_\mu$. The large 
distance behaviour of this correlator is
determined by the matrix of anomalous dimensions \cite{ref10}
and hence by the renormalization factors 
$\widetilde{Z}$. These renormalization factors take into account 
the mixing of the operators $\mbox{Tr}W^2_{\rm mag}$ and ${\cal W}$.
Since, as shown above, the entries of the $\widetilde{Z}$ matrix  
are given by a combination of the beta functions, 
the correlator $\langle\widetilde{a}_{\mu}(x)\widetilde{a}_{\nu}(y)\rangle$ 
at large distance must depend on 
the beta functions of the coupling constants taken at the scale
$\sim 1/|x-y|$. Assuming that the beta functions only have simple zeros at
the critical point, one has the following asymptotic expansion 
in the neighbourhood of the critical point, 
$\widetilde{\alpha}=\widetilde{\alpha}_\star$,
 $\widetilde{\lambda}=\widetilde{\lambda}_\star$, 
\begin{eqnarray}
\widetilde{\beta}_\alpha &=&
\widetilde{\beta}^{\prime}_{\alpha\alpha}
(\widetilde{\alpha}_\star,\lambda_\star)
(\widetilde{\alpha}-\widetilde{\alpha}_\star)+
\widetilde{\beta}^{\prime}_{\alpha\lambda}(\widetilde{\alpha}_\star,
\lambda_\star) (\lambda-\lambda_\star),\nonumber\\
\widetilde{\beta}_\lambda &=&
\widetilde{\beta}^{\prime}_{\lambda\alpha}
(\widetilde{\alpha}_\star,\lambda_\star)
(\widetilde{\alpha}-\widetilde{\alpha}_\star)+
\widetilde{\beta}^{\prime}_{\lambda\lambda}
(\widetilde{\alpha}_\star,\lambda_\star) (\lambda-\lambda_\star),
\label{eq7169}
\end{eqnarray}  
where the constants $\widetilde{\beta}^{\prime}_{\alpha\alpha}$, 
$\widetilde{\beta}^{\prime}_{\alpha\lambda}$,
 $\widetilde{\beta}^{\prime}_{\lambda\alpha}$
and $\widetilde{\beta}^{\prime}_{\lambda\lambda}$ are the 
elements of the matrix
$\widetilde{\beta}^{\prime}_{ij}
(\widetilde{\alpha}_\star,{\lambda}_\star)
=\partial \widetilde{\beta} (i)/\partial j$
$i,j=\widetilde{\alpha},\lambda$. A similar 
calculation as in the electric theory gives
\begin{eqnarray}
\langle \widetilde{a}_{\mu}(x)\widetilde{a}_{\nu}(y)\rangle
{\sim}\frac{1}{|x-y|^{2\widetilde{\beta}^{\prime}_{\rm min}+6}},
\label{eq7170}
\end{eqnarray} 
where $\widetilde{\beta}^{\prime}_{\rm min}$ is the minimal eigenvalue
of the matrix $(\widetilde{\beta}^{\prime}_{ij})$. 
The identification of the correlators
$\langle a_{\mu}(x) a_{\nu}(y)\rangle {\sim}\langle 
\widetilde{a}_{\mu}(x)\widetilde{a}_{\nu}(y)\rangle$ at 
the critical points yields
\begin{eqnarray}
\beta^{\prime}(\alpha_\star)_{\rm el}
=\widetilde{\beta}^{\prime}(\widetilde{\alpha}_\star,
 {\lambda}_\star)_{\rm mag},
\label{eq7171}
\end{eqnarray}
i.e. the anomalous dimensions of the electric and magnetic Konishi currents
are identical at the critical points and are given by the slope of the
beta functions evaluated at the critical points.  
This is one of the most important conclusions revealed by the matrix
element form of the anomaly equation of the Konishi current.

\subsection{Universality of operator product expansion}
\label{subsect72}
\renewcommand{\theequation}{6.2.\arabic{equation}}
\setcounter{equation}{0}
 
The operator product expansion (OPE) is an axiomatic-algebraic way to study 
conformal invariant quantum field theory. It had achieved great success 
in two dimensions \cite{ref12}, where the quantum conformal algebras,
 the Virasoro and Kac-Mody algebras, are derived with the use of 
the OPE, and all of the two-dimensional conformal field theories are 
classified according to the unitary representations of the conformal algebra.
Thus a natural idea is to look at the OPE in the four-dimensional case. 
Some new features of the four-dimensional
superconformal field theories have indeed been revealed by the OPE:  

   First, in contrast to the two-dimensional case, the OPE of products of  
energy-momentum tensors $T_{\mu\nu}$
does not form a closed algebra. 
A new operator $\Sigma$ with lower dimension arises.
An explicit calculation in a free superconformal field theory 
shows that $\Sigma$ is the Konishi current operator. 
In particular, $\Sigma$ develops an anomalous dimension. 
Thus, the discussions in Sect.\,\ref{subsect71} suggests
that even in an interacting four-dimensional superconformal field theory
$\Sigma$ may also be identified with the Konishi current operator \cite{ans}.
 
 Secondly, the OPE of $T_{\mu\nu}$'s and the OPE of
$\Sigma$'s  show that there are two central charges, 
$c$ and $c^{\prime}$, where $c$ is related to the gravitational trace 
anomaly of the theory. In addition, the OPE of $T_{\mu\nu}$ and 
$\Sigma$ implies that $\Sigma$ has a non-vanishing conformal dimension $h$.
These three numbers, $c$, $c^{\prime}$ and $h$ characterize
a four dimensional superconformal field theory. In the context of
a free four-dimensional field theory, an explicit 
calculation gives \cite{ref13}
 \begin{eqnarray}
c=\frac{1}{120 (4\pi)^2}\left(12 N_1+6N_{1/2}+N_0\right),
\label{eq721}
\end{eqnarray}
where $N_1$, $N_{1/2}$ and $N_0$ are the numbers of real vector
fields, Majorana spinor fields and real scalar fields. For a 
supersymmetric gauge field theory with gauge group $G$, there are
$N_v{\equiv}\mbox{dim}G$ components of the vector superfield
and $N_{\chi}{\equiv}\mbox{dim}T$ components of the chiral
superfield in the representation $T$. The central charge 
(\ref{eq721}) hence becomes  
\begin{eqnarray}
c=\frac{1}{24}(3\mbox{dim}G+\mbox{dim}T)=\frac{1}{24}(3N_v+N_{\chi}).
\label{eq722}
\end{eqnarray}
It was found that in a free supersymmetric field theory \cite{ref10},
$\Sigma =\Phi^{\dagger}\Phi$, $h=0$,  and 
\begin{eqnarray}
c^{\prime}=N_{\chi}=\mbox{dim}T.
\label{eq723}
\end{eqnarray}  
(\ref{eq721}), (\ref{eq722}) and (\ref{eq723}) mean that $c$ depends on 
the total number of degrees of freedom of the theory, while $c^{\prime}$ 
counts the degree of freedom of the chiral matter superfields. 

Thirdly, higher quantum corrections show that $c$ and $c^{\prime}$ are 
 universal, i.e. they are constants independent of the 
couplings, while $h$ is not. 

In the following we shall illustrate these features by working out several
examples of four dimensional conformal field theories such as free massless
scalar and spinor field theories and $N=1$ supersymmetric gauge theory
at the IR fixed point.

The simplest example of a four dimensional conformal field theory
is a massless free scalar field theory with the following Lagrangian 
and propagator in Euclidean space \cite{ref10},
\begin{eqnarray}
{\cal L}=\frac{1}{2}\partial^{\mu}\phi \partial_{\mu}\phi, 
~~~~\langle \phi (x) \phi (y) \rangle =\frac{1}{(x-y)^2},
\label{eq724}
\end{eqnarray}
and the energy-momentum tensor
\begin{eqnarray}
 T_{\mu\nu}(x)=\frac{2}{3}\partial_{\mu}\phi \partial_{\nu}\phi-
\frac{1}{6}\delta_{\mu\nu}(\partial_{\rho}\phi)^2-\frac{1}{3}
\phi \partial_{\mu} \partial_{\nu}\phi.
\label{eq725}
\end{eqnarray}
The correlator $\langle \phi (x) \phi (y) \rangle$
immediately leads to the OPE of $T_{\mu\nu}$'s,
\begin{eqnarray}
 T_{\mu\nu}(x)T_{\rho\sigma}(y)&=&-c\frac{1}{360}X_{\mu\nu\rho\sigma} 
\frac{1}{(x-y)^4}-\frac{1}{36}\Sigma (y)X_{\mu\nu\rho\sigma} \frac{1}{(x-y)^{4-h}}
+{\cdots}
\label{eq726}
\end{eqnarray}
with $c=1$ and $h=2$, where the omitted terms are less singular
terms, and the tensor operator $X$ is 
\begin{eqnarray}
X_{\mu\nu\rho\sigma}(x)&=&2\delta_{\mu\nu}\delta_{\rho\sigma}\Box^2
-3\left(\delta_{\mu\rho}\delta_{\nu\sigma}
+\delta_{\nu\rho}\delta_{\mu\sigma}\right)\Box^2
-2\left(\delta_{\mu\nu}\partial_{\rho}\partial_{\sigma}
+\delta_{\rho\sigma}\partial_{\mu}\partial_{\nu}\right)\Box \nonumber\\
&&-2\left(\delta_{\mu\rho}\partial_{\nu}\partial_{\sigma}+
\delta_{\mu\sigma}\partial_{\nu}\partial_{\rho}
+\delta_{\nu\sigma}\partial_{\mu}\partial_{\rho}
+\delta_{\nu\rho}\partial_{\mu}\partial_{\sigma}\right)\Box
-4\partial_{\mu}\partial_{\nu}\partial_{\rho}\partial_{\sigma}.
\label{eq727}
\end{eqnarray}
The OPE (\ref{eq726}) is fixed by the symmetry 
of $T_{\mu\nu}$, the conservation 
$\partial^{\mu}T_{\mu\nu}=\partial^{\nu}T_{\mu\nu}=0$ 
and the tracelessness $T^{\mu}_{~\mu}=0$. In particular, it 
 shows that a new operator $\Sigma=\phi^2$ arises. The OPE algebra is 
closed by the expansions,
\begin{eqnarray}
\Sigma (x) \Sigma (y)&=&\frac{2c^{\prime}}{(x-y)^{2h}}
+\frac{2}{(x-y)^h}\Sigma (y)+{\cdots}\, ,
\nonumber\\
T_{\mu\nu}(x)\Sigma (y)&=&-\frac{h}{3}\Sigma (y)
\partial_{\mu}\partial_{\nu}\frac{1}{(x-y)^2}
+\cdots
\label{eq728}
\end{eqnarray}
with $c^{\prime}=1$. Note that in this special case $c=c^{\prime}=1$, but
they are in general not equal.
The generalization to the 
case of $n$ free massless scalar fields $\phi^i$, 
$i=1,2,{\cdots},n$ is straightforward, where now 
$c=c^{\prime}=n$, $h=2$ and $\Sigma=\sum_i\phi^i\phi^i$.

Another familiar example of a four-dimensional conformal field theory 
is the free massless fermionic field (in Euclidean space),
\begin{eqnarray}
&&{\cal L}=\frac{1}{2}\left(\overline{\psi}/\hspace{-2mm}\partial\psi
-\partial_{\mu}\overline{\psi}\gamma_{\mu}\psi\right), 
~~~~\langle \psi(x) \overline{\psi}(y)\rangle 
=\frac{/\hspace{-2mm}x-/\hspace{-2mm}y}{(x-y)^4}, \nonumber\\
&&T_{\mu\nu}=\frac{1}{2}\left(\overline{\psi}\gamma_{\mu}\partial_{\nu}\psi+
\overline{\psi}\gamma_{\nu}\partial_{\mu}\psi-\partial_{\mu}\overline{\psi}
\gamma_{\nu}\psi-\partial_{\nu}\overline{\psi}\gamma_{\mu}\psi\right)
-g_{\mu\nu}{\cal L}.
\label{eq729}
\end{eqnarray}
The correlator, the symmetry and conservation of $T_{\mu\nu}$ 
and the parity determine that the form of the OPE of $T_{\mu\nu}$'s should be
of the form \cite{ref10},
\begin{eqnarray}
T_{\mu\nu}(x)T_{\rho\sigma}(y)&=&-c\frac{1}{360}X_{\mu\nu\rho\sigma}
\frac{1}{(x-y)^4}\nonumber\\
&+&\frac{1}{4}J^5_{\beta}(y)\left\{\left[\epsilon_{\mu\rho\alpha\beta}
\partial_{\nu}\partial_{\sigma}\partial_{\alpha}
+\left(\mu\leftrightarrow\nu\right)
\right] +\left[\rho\leftrightarrow\sigma\right]\right\}
\frac{1}{(x-y)^2}+{\cdots}
\label{eq7210}
\end{eqnarray}
with $c=6$. The $\Sigma$-term involves the axial vector current operator
$J_{\mu}^5=\overline{\psi}\gamma_5\gamma_\mu\psi$. This
is required by the conservation of parity since the parity odd tensor
$\epsilon_{\mu\nu\lambda\rho}$ appears in the OPE.
These two simple examples show that the non-closure of the OPE 
of $T_{\mu\nu}$ is a general feature of four-dimensional
conformal field theory and that the OPE should be the form
of (\ref{eq726}) and (\ref{eq728}) with $c$, $c^{\prime}$ 
and $h$ generic. 

Now we turn to the supersymmetric case. There 
is a long known four-dimensional
superconformal field theory, $N=4$ supersymmetric 
Yang-Mills theory. There also
exist some $N=2$ superconformal gauge theories 
with certain matter field contents. 
They all have identically vanishing beta functions, and their 
first central charges do not receive any higher order quantum corrections. 
Thus to reveal the special features of superconformal field theory we need to 
consider an $N=1$ supersymmetric gauge theory in the IR fixed point. 
Here we choose a general classical (four-component form) Lagrangian
including both supersymmetric 
QCD (\ref{eq163}) and a cubic superpotential 
$W=Y_{rst}Q^rQ^sQ^t/6$ \cite{ans1},  
\begin{eqnarray}
{\cal L}&=&\frac{1}{4}(F_{\mu\nu})^2+\frac{1}{2}
\overline{\lambda}/\hspace{-2.5mm}D\lambda+(D^{\mu}\Phi)^{\dagger}D_{\mu}\Phi+
\frac{1}{2}\overline{\Psi}/\hspace{-2.5mm}D\Psi\nonumber\\
&&+i\sqrt{2}g\left[\overline{\lambda}^a\Phi^{\star r}
\left(T^a\right)_r^{~s}(1-\gamma_5)\Psi_s-
\overline{\Psi}^r(1+\gamma_5)
\left(T^a\right)_r^{~s}\Phi_s\lambda^a\right]\nonumber\\
&&-\frac{1}{2}\left[\overline{\Psi}^r(1-\gamma_5)Y_{rst}\Phi^t\Psi^s+
\overline{\Psi}^r(1+\gamma_5)Y^{\star}_{rst}
\Phi^{\star t}\Psi^s\right]\nonumber\\
&&+\frac{1}{2}g^2\left[\Phi^{*r}\left(T^a\right)_r^{~s}\Phi_s\right]^2
+\frac{1}{4}Y_{rst}Y^{*rmn}\Phi^s\Phi^t\Phi^*_m\Phi_n^*,
\label{eq7211}
\end{eqnarray}
where $\Phi$ and $\Psi$ are the four-component form
of the quark superfield $Q^i_r$ and the flavour indices are
suppressed.
Since the energy-momentum tensor $T_{\mu\nu}$, the $R$-current
$R_{0\mu}$ and the supersymmetry current lie in the 
same supermultiplet,  it is enough to work out the OPE 
of the lowest component of 
this supermultiplet, i.e. the $R_0$-current,
$R_{0\mu}(x)=\overline{\lambda}^a\gamma_\mu\gamma_5\lambda^a/2
-\overline{\psi}\gamma_\mu\gamma_5\psi/2$. 
Then the OPE of the whole supermultiplet can be obtained through 
supersymmetry transformations.  
Note that at the IR fixed point the $R_0$-current is identical to the
anomaly-free $R$-current, $R_{\mu}(x)$.
The dimension and conservation of the $R$-current require that the 
OPE of $R_{\mu}(x)$ should be  \cite{ans1}
\begin{eqnarray}
R_\mu (x) R_\nu (y)|_{x\rightarrow y}&=&\frac{1}{3\pi^4}\left(\partial_{\mu}
\partial_{\nu}-\Box\delta_{\mu\nu}\right)\frac{c}{(x-y)^4}\nonumber\\
&+&\frac{2}{9\pi^2}\Sigma (y) \left(\partial_{\mu}
\partial_{\nu}-\Box\delta_{\mu\nu}\right)\frac{c}{(x-y)^{2-h}}
+{\cdots},
\label{eq7213}
\end{eqnarray}
where the new operator $\Sigma (x)$ is the lowest component of the 
real superfield $\Sigma (z)$ ($z=(x,\theta,\overline{\theta})$), and  is 
related to the renormalized Konishi operator by 
\begin{eqnarray}
\Sigma (z)=\rho (g,Y) K_{\rm ren}(z).
\label{eq7214}
\end{eqnarray}
$\rho (g,Y)$  is a function of the couplings that can be determined from
an explicit perturbative calculation \cite{ans1,jack}. It is
dimensionless since $K_{\rm rem}$ is 
a renormalized operator and carries the power
$\mu^h$ of the renormalization scale $\mu$. 
Similarly, the OPE of $\Sigma (x)$'s 
should be of the following form \cite{ans,jack},
\begin{eqnarray} 
\Sigma (x) \Sigma (y)|_{x\rightarrow y}=\frac{1}{16\pi^4}
\frac{c^{\prime}}{(x-y)^{2-h}}+{\cdots}\,,
\label{eq7215}
\end{eqnarray}
where the omitted parts in (\ref{eq7213}) and (\ref{eq7215}) 
denote less singular terms.

The second central charge $c^{\prime}$ and anomalous dimension $h$ can be
obtained by an explicit perturbative calculation. There are two
independent methods\cite{ans1}.
The first one is calculating
the connected four-point correlation function 
$\langle R_{\mu}(x)R_{\nu}(y)R_{\rho}(z)R_{\sigma}(w)\rangle$ in the
asymptotic region where $|x-y|, |z-w|{\ll} |x-z|, |y-w|$. However, this 
method cannot give the proportionality function $\rho (g,Y)$. 
 The second method is to consider the two- and three-point 
correlators $\langle K_{\rm ren}(x)K_{\rm ren}(y)\rangle$ and 
$\langle R_\mu(x) R_\nu (y) K_{\rm ren}(z)\rangle$, $K_{\rm ren}(x)$
being the lowest component of $K_{\rm ren}(z)$. 
Since for a conformal invariant field theory, the two- and three-point 
functions can be fixed up to a constant, 
the general forms of $\langle K_{\rm ren}(x)K_{\rm ren}(y)\rangle$ and 
$\langle R_\mu(x) R_\nu(y) K_{\rm ren}(z)\rangle$ can be easily written out. 
Concretely, scale and 
translation invariance lead to
\begin{eqnarray}
\langle K_{\rm ren}(x) K_{\rm ren}(y)\rangle=\frac{A}{16\pi^4(x-y)^4},
\label{eq7216}
\end{eqnarray}  
where $A=c^{\prime}/\rho^2$ due to (\ref{eq7214}). The tensor form
of $\langle R_\mu(x) R_\nu(y) K_{\rm ren}(z)\rangle$ 
is fixed by the conservation of the $R$-current and the correct 
transformation properties under inversion, 
$x^{\mu}\longrightarrow x^{\prime\mu}=x^{\mu}/x^2$, since any conformal
transformation can be generated by combining inversions with rotations 
and translations \cite{ref22}. Thus 
\begin{eqnarray}
\langle R_\mu(x) R_\nu(y) K_{\rm ren}(z)\rangle &=&\frac{B}{36\pi^6}
\frac{1}{(x-y)^{4-h}(x-z)^{2+h}(y-z)^{2+h}}
\left[\left(1-\frac{h}{4}\right)I_{\mu\nu}(x-y)\right. \nonumber\\
&&\left.-\frac{1}{2}\left(1+\frac{h}{2}\right)I_{\mu\rho}(x-z)I_{\rho\nu}(z-y)
\right],
\label{eq7217}
\end{eqnarray}
where $I_{\mu\nu}(x)=\partial x_{\mu}^{\prime}/\partial x^\nu
=\delta_{\mu\nu}-2x_{\mu}x_{\nu}/x^2$.  
 In the limit $x{\sim}y$, the most singular term is
\begin{eqnarray}
\langle R_\mu(x) R_\nu(y) K_{\rm ren}(z)\rangle{\sim}\frac{B}{72\pi^6 (h-2)}
\frac{1}{(y-z)^{4+2h}}\left(\partial_\mu\partial_\nu-\delta_{\mu\nu}\Box
\right)\frac{1}{(x-y)^{2-h}}.
\label{eq7218}
\end{eqnarray}
The comparison of (\ref{eq7216}) and (\ref{eq7218}) with the OPEs 
(\ref{eq7213}) and (\ref{eq7215}) leads to the relations
\begin{eqnarray}
c^{\prime}(g,Y)=\frac{B^2}{(h-2)^2A}, ~~~~~\rho=\frac{B}{(h-2)A}.
\label{eq7219}
\end{eqnarray}
$A$ and $B$ can be obtained through an explicit perturbative
calculation. We thus finally get \cite{ans1}:
\begin{eqnarray}
c^{\prime}=N_{\chi}+2\gamma^r_{~r}, 
~~~~h=\frac{3}{\pi^2N_{\chi}}Y_{rst}Y^{*rst}.
\label{eq7220}
\end{eqnarray}
The quantum corrections to the first central charge 
was calculated from a general renormalizable theory
containing vector, spinor and scalar fields in curved space-time \cite{jack}. 
Specialized to the $N=1$ supersymmetric gauge theory (\ref{eq7211}),
the result is \cite{ans1}
\begin{eqnarray}
c=\frac{1}{24}\left(3N_v+N_{\chi}+N_v\frac{\beta (g)}{g}-\gamma^r_{~r}\right).
\label{eq7221}
\end{eqnarray}
In (\ref{eq7220}) and (\ref{eq7221}), $N_v=\mbox{dim}G$ 
and $N_{\chi}=\mbox{dim}T$ as shown in (\ref{eq722}) and (\ref{eq723}). 
The gauge beta function $\beta (g)$ and the anomalous dimensions 
$\gamma^r_{~s}$ of the chiral matter superfields at 
the one-loop level of the model (\ref{eq7211}) are the following:
\begin{eqnarray}
16\pi^2\beta (g)&=& g^3\left[-3C(g)
+\frac{\mbox{Tr} C(T)}{\mbox{dim}G}\right],
\nonumber\\
\beta_{rst}&=&\frac{1}{3!}\left(Y_{mrs}\gamma^m_{~t}+Y_{mtr}\gamma^m_{~s}+
Y_{mst}\gamma^m_{~r}\right)\nonumber\\
16\pi^2\gamma^r_{~s}&=&\frac{1}{2}Y^{rmn}Y^*_{smn}-2g^2\left[C(T)\right]^r_{~s}, \nonumber\\
C(G)\delta^{ab}&=& f^{acd}f^{bcd}, ~~~\left[C(T)\right]^r_{~s}=(T^aT^a)^r_{~s}.
\label{eq7222}
\end{eqnarray}
The comparison of (\ref{eq7220}) and (\ref{eq7221}) with the free 
field case, (\ref{eq722}) and (\ref{eq723}), shows that the quantum 
corrections to $c$ and $c^{\prime}$ are proportional to a combination of 
anomalous dimension $\gamma^r_{~r}$ and the beta function 
$\beta (g)$. As we know, if the conformal symmetry is preserved at 
quantum level, the beta function and anomalous dimensions 
must vanish. Hence from (\ref{eq7220}) and (\ref{eq7221}), 
the first and second central charges $c$ and $c^{\prime}$ 
are identical to their classical values. Conversely, 
if both $c$ and $c^{\prime}$ receive no quantum correction, then 
the theory remains conformally invariant,
whereas the second relation of (\ref{eq7220}) 
shows that under these conditions the anomalous 
dimension $h$ remains non-vanishing if we consider the 
cubic superpotential composed of the chiral superfield. 
This means that $c$ and $c^{\prime}$ are independent of 
the couplings and hence are universal, while the anomalous 
$h$ is not. The above conclusion is the one-loop result.
From the NSVZ beta function (\ref{eq1p1}) 
and the relation between $\beta_{rst}$ and the anomalous dimension
$\gamma^r_{~s}$ given in (\ref{eq7222}), 
we conclude that to all orders in the couplings $g$ and
 $Y$ there exists a fixed surface of the renormalization group flow 
provided that the matter representation is chosen so that 
$3\mbox{dim}G C(G)-\mbox{Tr}C(T)=0$ and $g$, $Y$ are such that
$\gamma^r_{~s}=0$. Therefore, an important feature of four 
dimensional superconformal field theory has been revealed:
there exists a space of continuously 
connected conformal field theories and their central charges
are universal, i.e. $c$ and $c^{\prime}$ are constants, independent
of the couplings in this space. Such quantities are called
invariants of a four-dimensional superconformal field theory.  
The third quantity contained in the OPE, the anomalous dimension $h$ is,
as stated above, not an invariant of a four-dimensional superconformal 
field theory, it actually manifests the inequivalence of 
the continuously connected four-dimensional superconformal 
field theories in this space. These may be the essential
features of a superconformal field theory in four dimensions \cite{ans1}.

\subsection{Possible existence of a four-dimensional $c$-theorem}
\label{subsect73}
\renewcommand{\theequation}{6.3.\arabic{equation}}
\setcounter{equation}{0}

The $c$-theorem was proposed by Zamolodchikov in the context of
two-dimensional conformal quantum field theory \cite{zamo}. 
Its main idea is to define an appropriate function $c(g)$ on 
the space of the couplings, which monotonically decreases along the
trajectory of the renormalization group flow from the UV region 
to the IR one. At the fixed point $g=g_f$
of the beta function, $\beta (g_f)=0$, the $c$ function $c(g_f)$ should be 
equal to the central charge of the Virasoro algebra of the resulting
two-dimensional conformal field theory. 
Since this theorem can show how the dynamical degrees
of freedom are lost along the trajectory of 
the renormalization group flow from the UV region
to the IR region, a theorem, if it exists in four dimensions, 
 will be quite helpful to understand the dynamical mechanisms
of some non-perturbative phenomena 
such as confinement and chiral symmetry breaking. 
The search for a four-dimensional $c$-theorem
began immediately after the invention of 
the two-dimensional $c$-theorem. It was first suggested by Cardy  
that a $c$-function in the four-dimensional case should depend on the 
expectation value of the trace of the energy-momentum tensor of a quantum field
in a curved space-time, integrated  over a 4-dimensional 
sphere $S^4$ with constant radius \cite{car}
\begin{eqnarray}
C={\cal N} \int_{S^4}\langle T^{\mu}_{~\mu} \rangle \sqrt{g}d^4x.
\label{eq731}
\end{eqnarray}
Here ${\cal N}$ is a positive numerical factor and its value depends on
the normalization of the $c$-function. 
In a curved background space-time,
the trace $\langle T^{\mu}_{~\mu} \rangle$
takes the following general form,
\begin{eqnarray}
\langle T^{\mu}_{~\mu} \rangle&=&-aG+ bF+ d{\Box}R +e B,
\label{eq732}
\end{eqnarray}  
where $G=1/4({\epsilon}^{\mu\nu\sigma\rho}
{\epsilon}_{\alpha\beta\gamma\delta}R^{\alpha\beta}_{~~\mu\nu})^2
=(\widetilde{R}_{\mu\nu\lambda\rho})^2$ 
 is the Euler topological number density,
$R$ is the Riemann scalar curvature, 
$F=C^{\mu\nu\sigma\rho}C_{\mu\nu\sigma\rho}$ with 
$C_{\mu\nu\sigma\rho}$  the Weyl tensor, which in a general 
$n$-dimensional space-time can be expressed in terms of the Riemann curvature,
\begin{eqnarray}
C_{\mu\nu\sigma\rho}&=& R_{\mu\nu\sigma\rho}-\frac{2}{n-2}
\left(g_{\mu\sigma}R_{\nu\rho}-g_{\mu\rho}R_{\nu\sigma}-
g_{\nu\sigma}R_{\rho\mu}+g_{\nu\rho}R_{\sigma\mu}\right)\nonumber\\
&&+\frac{2}{(n-1)(n-2)}\left(g_{\mu\sigma}g_{\nu\rho}-
g_{\mu\rho}g_{\nu\sigma}\right)R.
\label{eq733}
\end{eqnarray} 
The last term in (\ref{eq732}) represents the contribution from
other external background fields. For example, in a background
gauge field $A_\mu^a$, $B=F_{\mu\nu}^aF^{a\mu\nu}$ \cite{duff}.
The coefficients $a$ and $b$ in (\ref{eq732}) 
have an universal meaning and they have been calculated \cite{osb1}, 
\begin{eqnarray}
a&=&\frac{1}{360 (4\pi)^2}\left(N_0+11N_{1/2}+62N_1\right),\nonumber\\
b&=&\frac{1}{120 (4\pi)^2}\left(N_0+6N_{1/2}+12N_1\right).
\label{eq734}
\end{eqnarray}  
One can see that the coefficient $b$ is just
the central charge $c$ given in (\ref{eq721}).
The coefficient $d$ has no universal meaning and its value
can be defined at will by introducing a local counterterm
proportional to the integral $R^2$, so it is renormalization 
scheme dependent coefficient. 

A four-dimensional $c$-theorem was first attempted  
in the $SU(N_c)$ gauge theory coupled to $N_f$ massless
Dirac fermions in the fundamental representation by choosing
the coefficient $a$ of the Euler number density to be 
the $c$-function \cite{car}. 
If the flavour number $N_f$
is sufficiently small, the theory is asymptotically free and
$g=0$ is the UV fixed point. 
At $g{\rightarrow}0$,
the particle spectrum contains $N_c^2-1$ gauge bosons and 
$N_cN_f$ fermions. In the low energy case, 
the theory will be strongly coupled,
$g{\rightarrow}\infty$ is the IR fixed point. 
The chiral symmetry $SU_L(N_f){\times}SU_R(N_f)$ breaks to
$SU_V(N_f)$ due to quark condensation, 
yielding $N_f^2-1$ Goldstone bosons. Therefore,  
at the UV fixed point
\begin{eqnarray}
c_{UV}=\lim_{g{\rightarrow}0}c(g)=
\frac{1}{360 (4\pi)^2}\left[62(N_c^2-1)+11N_cN_f\right],
\label{eq735}
\end{eqnarray}   
and at the IR fixed point
\begin{eqnarray}
c_{IR}=\lim_{g{\rightarrow}\infty}c(g)
=\frac{1}{360 (4\pi)^2}\left(N_f^2-1\right).
\label{eq736}
\end{eqnarray}   
 (\ref{eq735}) and (\ref{eq736}) 
show that the requirement of a $c$-theorem, $c_{UV}-c_{IR}>0$,
is not satisfied identically.

Later it was proposed by Osborn et al. \cite{osb1} that one can
directly choose a coefficient $a$ in the trace anomaly as 
a $c$-function even away from the fixed point 
and modify it to satisfy the renormalization
group flow equation. This scheme is an 
exact analogue of Zamolodchikov's procedure in two dimensions. 
This $c$-function can reduce to the coefficient $a$
of the trace anomaly when approaching the critical point \cite{duff}.
However, the monotonically decrease of the $c$-function  
along the trajectory of renormalization flow cannot be explicitly
shown. 

Another attempt was to define a $c$-function by using the spectral 
representation of the two-point correlator of the energy-momentum 
tensor and constructing a reduced spectral density for 
the spin-0 intermediate state \cite{cfl}. The concrete steps
are as follows. First work out the spectral representation 
of two-point correlator of energy-momentum tensors in $n$-dimensional
(Euclidean) space-time, 
\begin{eqnarray}
\hspace{-8mm} &&\langle T_{\mu\nu}(x)T_{\sigma\rho}(0) \rangle
={\langle} T_{\mu\nu}(x)T_{\sigma\rho}(0) {\rangle}_{s=0}
+{\langle} T_{\mu\nu}(x)T_{\sigma\rho}(0) {\rangle}_{s=2}
\nonumber\\
\hspace{-8mm}&&=
A_n\left[\int_0^{\infty}d\mu c^{(0)}(\mu)
\Pi^{(0)}_{\mu\nu,\sigma\rho}(\partial)G(x,\mu)
+\int_0^{\infty}d\mu c^{(2)}(\mu)
\Pi^{(2)}_{\mu\nu,\sigma\rho}(\partial)G(x,\mu)\right],
\end{eqnarray}
where 
\begin{eqnarray}
A_n=\frac{2\pi^{n/2}}{2^{n-1}\Gamma (n/2)(n+1)(n-1)^2},
\label{eq737}
\end{eqnarray}
and $s=0,2$ denotes spin.
\begin{eqnarray} 
G(x,\mu)=\int \frac{d^nx}{(2\pi)^n}
\frac{e^{ip{\cdot}x}}{p^2+\mu^2}
=\frac{1}{2\pi}\left(\frac{\mu}{2\pi|x|}\right)^{(n-2)/2}
K_{(n-2)/2}(\mu |x|)
\end{eqnarray} 
is the two-point correlator of the fields involved. 
Lorentz covariance and the conservation
of the energy-momentum tensor determine that there are only
two possible Lorentz structures for the intermediate states: $s=0$ and
$s=2$. The tensors $\Pi^{(s)}_{\mu\nu,\sigma\rho}$ are, respectively,  
\begin{eqnarray}
\Pi^{(2)}_{\mu\nu,\sigma\rho}(\partial)
&=&\frac{1}{\Gamma (n-1)}\left[\frac{n-1}{2}\left(\pi_{\mu\sigma}\pi_{\nu\rho}+
\pi_{\mu\rho}\pi_{\sigma\nu}+\pi_{\mu\nu}\pi_{\rho\sigma}\right)
-\pi_{\mu\nu}\pi_{\rho\sigma}\right],\nonumber\\
\Pi^{(0)}_{\mu\nu,\sigma\rho}(\partial)&=&\frac{1}{\Gamma (n)}\Pi_{\mu\nu}
\pi_{\sigma\rho}, ~~
\pi_{\mu\nu}=\partial_{\mu}\partial_{\nu}
-\Box{\delta}_{\mu\nu}.
\label{eq738}
\end{eqnarray}
Away from criticality, owing to the UV divergence, the spectral
density must be scale dependent, 
$c^{(s)}(\mu)=c^{(\mu)}(\mu,\Lambda)$.
 The analysis shows 
that the candidate for $c$-function should be the
``reduced'' spin-$0$ spectral density \cite{cfl},
$c(\mu,\Lambda) ={c^{(0)}(\mu,\Lambda)}/{\mu^{n-2}}$,
and this indeed satisfies 
\begin{eqnarray}
\lim_{\Lambda{\rightarrow}0}c(\mu,\Lambda)
=(c_{UV}-c_{IR})\delta (\mu){\equiv}
\Delta c^{(0)}\delta (\mu).
\label{eq7313}
\end{eqnarray}
This $c$-function  candidate indeed monotonically decreases 
along the trajectory of the renormalization group flow 
from short distance to short distance and becomes 
stationary at the fixed point,
but its physical meaning  at the fixed points is not clear. It has only
been checked that in free scalar and spinor field theories, 
$c(\mu,\Lambda)$ coincides with
the coefficients $a$ of the trace anomaly (\ref{eq732}). 

The above attempts at establishing a four dimensional $c$-theorem
imply that if the $c$-function exists it will reduce to the coefficients
$a$ or $b$ of the trace anomaly at the fixed points.

The new non-perturbative results make it possible to test a
$c$-theorem in four-dimensional supersymmetric gauge theory \cite{bast}.
The introduction in Sects.\,\ref{subsub6.3.4}, \ref{subsect7.1}
and \ref{subsect7.2}  shows that the particle spectrum
of $N=1$ supersymmetric QCD in the IR region can be deduced, 
hence a theorem can be explicitly checked for most of the 
ranges of $N_f$ and $N_c$. 
The analysis supports the existence of a 
four-dimensional $c$-theorem \cite{bast}.  
At the UV fixed point the theory is a free theory 
with a particle spectrum 
composed of $N_c^2-1$ vector
supermultiplets and $2N_fN_c$ scalar supermultiplets. Since for a free theory
the $c$-function reduces to the sum of the central charges carried by the 
free fields, one has in the UV fixed point
\begin{eqnarray} 
c_{UV}=(N_c^2-1)c_V+2N_fN_cc_S,
\end{eqnarray}
where $c_S$ and $c_V$ are central charges corresponding to 
the scalar and vector supermultiplets. 
At the IR fixed point, the low energy dynamics of the theory depends heavily
on the relative number of $N_f$ and $N_c$ and one must
perform the analysis according to different values of $N_f$ and $N_c$.
First, considering the $N_f=0$ case, 
the low-energy pure supersymmetric Yang-Mills theory is
strongly coupled, 
the gluons and gluoninos condensate into massive 
colour singlets and hence they develop a mass gap \cite{ref1p26}, 
so the IR theory contains only the vacuum state. Consequently, 
\begin{eqnarray}
c_{IR}=0.
\end{eqnarray}
For $0<N_f<N_c$, 
the dynamically generated  superpotential erases  all the vacua 
and the particle spectrum is not clear. For $N_f=N_c$, 
there exists a smooth moduli space described by the expectation values
of $N_f^2$ meson supermultiplet operators $M_{~j}^i$, 
the baryon $B$ and antibaryon superfield operators $\widetilde{B}$ with
the constraint (\ref{eq318}), $\det M-B\widetilde{B}=\Lambda^{2N_c}$, 
so that at the IR fixed point, there are $N_f^2+2-1$ massless scalar 
superfields, since the quantum fluctuations must satisfy the constraint. 
Thus we get
\begin{eqnarray}
c_{IR}=(N_f^2+1)c_S.
\end{eqnarray}
For $N_f=N_c+1$, the quantum moduli space has singularities associated 
with additional massless particles. 
The maximum number of massless particles is at the origin of the moduli space, 
there are $N_f^2$ free massless mesons and $2N_f$ free massless baryons. Their
contribution to the central charge is
\begin{eqnarray}
c_{IR}=(N_f^2+2N_f)c_S.
\end{eqnarray}
In the range $N_c+2{\leq}N_f{\leq}3/2N_c$,  
the low energy theory is effectively 
described by a magnetic $SU(N_f-N_c)$ gauge theory with
$N_f$ magnetic quarks, $N_f$ magnetic antiquarks and $N_f^2$ mesons.
All these particles are free at the IR fixed point and hence the 
central charge is
\begin{eqnarray}
c_{IR}=[(N_f-N_c)^2-1]c_V+[2N_f(N_f-N_c)+N_f^2]c_S.
\end{eqnarray}
In the range $3/2N_c{\leq}N_f{\leq}3N_c$, the IR region is described by a 
superconformal field theory, and can be parametrized equivalently
by the magnetic or electric variables. However, both versions of 
the theory are interacting theories, the one-loop results (\ref{eq732})
and (\ref{eq734}) for the
trace anomaly are not enough to extract the central charge $c_{IR}$. 
A technique of computing exactly the trace anomaly and hence
$c_{IR}$ for these interacting superconformal field theories will
be introduced in next section. In the range $N_f{\geq}3N_c$, the 
theory ceases to be asymptotically free,
the particle spectrum at the UV fixed point is not clear
and hence the central charge at the UV fixed
point cannot be calculated. The above known central charges at 
the IR and UV fixed points all give
\begin{eqnarray}
c_{UV}-c_{IR}>0.
\end{eqnarray}
This hints at the possible existence of a $c$-theorem for 
a supersymmetric gauge theory.  

More explicit and concrete evidence for the existence 
of a four dimensional $c$-theorem was provided by calculating exactly 
the renormalization group flows of the
central charges of $N=1$ supersymmetric QCD in the conformal window
$3/2N_c{\leq}N_f{\leq}3N_c$ \cite{afgj}. The explicit formula for the flows
of the trace anomaly coefficients were given. It was shown that 
the coefficient $a$ of the Euler number
density in the trace anomaly (\ref{eq732}) 
is always monotonically decreasing $a_{UV}-a_{IR}>0$.
 This is a strong support for the existence of a four-dimensional 
$c$-theorem in $N=1$ supersymmetric gauge theory. The following
section will give a detailed introduction. 

\subsection{Non-perturbative central functions and their renormalization
group flows}
\label{subsect74}
\renewcommand{\theequation}{6.4.\arabic{equation}}
\setcounter{equation}{0}

A quantitative investigation of the existence of a four dimensional
$c$-theorem was made by working out the explicit non-perturbative
formula for the renormalization group flow of the central functions.
The central functions are the central charges away from the conformal 
criticality \cite{jhep}. If an asymptotic gauge theory  has
a non-trivial infrared fixed point, at which the theory becomes
an interacting conformal field theory, then the quantum field
theory away from criticality can be regarded as a radiative interpolation 
between a pair of four dimensional conformal field theories. To extract
certain non-perturbative information about this quantum field theory,
one should first identify relevant physical quantities and observe
their renormalization group flow from the UV fixed points to 
the IR ones. 
As stated in previous sections, 
at the fixed points, the $c$-functions coincide with the central
charges. These central charges are the coefficients of the leading terms in
the operator product expansion of the various conserved quantities 
in the fixed points. In the conformal window
$3N_c/2<N_f<3N_c$ of $N=1$ supersymmetric QCD, 
the NSVZ $\beta$-function shows
that the theory has a non-trivial IR fixed point, 
where the theory admits an equivalent
dual magnetic description but with strong/weak coupling exchanged. 
In the following,
we shall review the derivation of the renormalization
group flow of the non-perturbative central functions associated with
the various conserved quantities developed in Ref.\,\cite{afgj}. The
explicit non-perturbative formula for the central functions
shows that the appropriate candidate for the
$c$-function should depend on the coefficient of the
Euler term in the trace anomaly.

In a supersymmetric gauge theory,
several facts help to determine non-perturbatively 
the renormalization group flow of the central functions.  First, 
the energy-momentum tensor
$T_{\mu\nu} (x)$ and the $R_0$-current $R_{0\mu}(x)$
lie in the same supermultiplet, so do the trace anomaly
$T^{\mu}_{~\mu}$ and the anomalous divergence
of the $R_0$-current, $\partial_{\mu}R_0^{\mu}$. This fact makes 
the coefficients of the trace anomaly and $\partial_{\mu}R_0^{\mu}$ 
relevant. Secondly, if one introduces external source 
fields for the flavour current $j_\mu (x)$ and the 
energy-momentum tensor $T_{\mu\nu}(x)$, the
trace anomaly has a close relation with the two-point 
correlators $\langle j_\mu (x) j_\nu (y)\rangle$ and 
$\langle T_{\mu\nu} (x) T_{\lambda\rho} (y)\rangle$
and their central charges.
 Finally, the anomalous divergence of the $R_0$-current can be 
exactly calculated
at an infrared fixed point through 't Hooft anomaly 
matching. As shown in Eq.\,(\ref{eq7110}),  
the $R_0$-current can be combined 
with the Konishi current to an (internal) 
anomaly-free $R$-current, and the combination coefficient of
the Konishi current is just the numerator of
the NSVZ $\beta$-function. Thus at the IR fixed point the
$R_0$-current will coincide with the (internal) anomaly-free
$R$-current, whereas the 't Hooft anomalies for this internal
anomaly-free $R$-currents can be calculated in the whole energy 
range from only the one-loop triangle diagram. Therefore, 
the 't Hooft anomalies of the $R_0$-current can be exactly 
determined at the IR fixed point.
In the UV region, $g=0$ must be the UV fixed point due to 
asymptotic freedom, so all the anomalies can be
computed in the context of a free field theory. We are going to show 
the non-perturbative derivation of the renormalization group flow of 
the central functions.

The $N=1$ supersymmetric QCD has 
the anomaly-free global symmetry $SU_L(N_f){\times}SU_R(N_f)$ 
${\times}U_B(1){\times}U_R(1)$. 
The fermionic parts of flavour currents corresponding to
 $SU_L(N_f){\times}SU_R(N_f)$ ${\times}U_B(1)$ given in 
Sect.\,\ref{sect6} are rewritten in four-component 
form \cite{afgj}, 
\begin{eqnarray}
j_{\mu}^A &=& \frac{1}{2}\overline{\psi}\gamma_\mu (1-\gamma_5)t^A\psi, ~~
\widetilde{j}_{\mu}^A = \frac{1}{2}\overline{\widetilde{\psi}}
\gamma_\mu (1-\gamma_5)\overline{t}^A\widetilde{\psi}, \nonumber\\
\nonumber\\
j_{\mu}^5 &=&\frac{1}{2N_c}\left(\overline{\psi}
\gamma_\mu \gamma_5\psi-\overline{\widetilde{\psi}}
\gamma_\mu \gamma_5\widetilde{\psi}\right).
\label{eqs1}
\end{eqnarray} 
If written in two-component form they are the $\theta\overline{\theta}$ 
component of the current superfields $Q^{\dagger}t^AQ$, 
$\widetilde{Q}\overline{t}^A\widetilde{Q}^{\dagger}$ and 
$(Q^{\dagger}Q+\widetilde{Q}\widetilde{Q}^{\dagger})$, respectively;  $t^A$ and
$\overline{t}^A$ being the generators of $SU(N_f)$ in the fundamental
and conjugate fundamental representation. The anomaly-free $R$-symmetry 
current is the combination (\ref{eq7110}) of the anomalous Konishi current
and the $R_0$-current, and the fermionic parts of their four-component 
forms are
\begin{eqnarray}
K_\mu &=& \frac{1}{2}\overline{\psi}
\gamma_\mu \gamma_5\psi+\frac{1}{2}\overline{\widetilde{\psi}}
\gamma_\mu \gamma_5\widetilde{\psi},
~~R_{0\mu} = \frac{1}{2}\overline{\lambda}^{a}\gamma_\mu\gamma_5\lambda^a
-\frac{1}{2}\left(\overline{\psi}
\gamma_\mu \gamma_5\psi+\overline{\widetilde{\psi}}
\gamma_\mu \gamma_5\widetilde{\psi}\right).
\label{eqs2}
\end{eqnarray}
Note that in (\ref{eqs1}) and (\ref{eqs2}) the flavour and colour
indices carried by the quarks are 
suppressed.

Let $j_{\mu}$ denote one of the flavour currents listed in 
(\ref{eqs1}). The conservation of the current, the dimensional
analysis and the renormalization effects
restrain the two-point correlator of 
$j_\mu$ to be of the form
\begin{eqnarray}
\langle j _\mu (x) j_\nu (y)\rangle =\frac{1}{(2\pi)^4}
\left(\partial_\mu\partial_\mu-\partial^2\delta_{\mu\nu}\right)
\frac{b[g(1/x)]}{x^4}.
\label{eqs3}
\end{eqnarray}
Since $j_\mu$ is conserved at quantum level, it has no
anomalous dimension. As a consequence, the Callan-Symanzik equation
\begin{eqnarray}
\left[\mu\frac{\partial}{\partial\mu}+\beta (g)\frac{\partial}{\partial g}
\right]\langle j_\mu (x) j_\nu (y)\rangle =0
\label{eqs4}
\end{eqnarray}
and the $\beta$-function
$\beta[g(\mu)]=\mu {dg(\mu )}/{d\mu}$
determine that the function $b$ depends only on 
the running coupling $g(1/x)$. At the UV and IR fixed points of the 
renormalization group flow, 
$g_{UV}$ and $g_{IR}$,  the following limits exist:
\begin{eqnarray}
b_{UV}&{\equiv}&\lim_{x{\rightarrow}0}b[g(1/x)]=b[g_{UV}],\nonumber\\
b_{IR}&{\equiv}&\lim_{x{\rightarrow}\infty}b[g(1/x)]=b[g_{IR}].
\label{eqs6}
\end{eqnarray}

To determine the renormalization flow $b_{UV}-b_{IR}$,
we first consider the correlation function (\ref{eqs3}) to any order in perturbation
theory, and then make reasonable arguments to extract the all orders result.
In perturbation theory, one unavoidable result is the
emergence of UV divergences. To regulate all the sub-divergences 
contained in $b[g(1/x)]$, we first expand the function
$b[g(1/x)]$ to any finite order of $g$,
\begin{eqnarray}
b[g(1/x)]=\sum_{n{\geq}0}b_n[g(\mu)]t^n, ~~~~~~t{\equiv}\ln (x\mu),
\label{eqs7}
\end{eqnarray}
where $b_n[g(\mu)]$ is a polynomial in $g(\mu)$. Evaluating Eq.(\ref{eqs7}) at
$x=1/\mu$ gives 
\begin{eqnarray}
b[g(\mu)]=b_0[g(\mu)].
\label{eqs8}
\end{eqnarray}
(\ref{eqs3}) and the Callan-Symanzik equation (\ref{eqs4}) yield 
\begin{eqnarray}
&&\mu\frac{\partial b[g(1/x)]}{\partial\mu}=0,\nonumber\\
&&\beta (g)\frac{db_n(g)}{dg}+(n+1)b_{n+1}(g)=0,\nonumber\\
&&b_{n+1}(g)=-\frac{\beta (g)}{n+1}\frac{db_n(g)}{dg}.
\label{eqs9}
\end{eqnarray}
Eq.(\ref{eqs9}) shows that all $b_n(g)$ with $n{\geq}1$ can be expressed 
in terms of $\beta (g)$, $b_0(g)$ and its derivatives. 
However, when inserting (\ref{eqs7}) into (\ref{eqs3}), one can easily see
that at short distance $t^n/x^4$ is too singular to have 
a Fourier transform in momentum space. This is 
actually the reflection of the UV divergence
in coordinate space. With a newly developed regularization method
in coordinate space called differential regularization \cite{jfl}, 
the overall divergence near $x=0$ can be regulated as follows \cite{afgj,jhep}:
\begin{eqnarray}
\frac{[\ln (x\mu)]^n}{x^4}=-\frac{n!}{2^{n+1}}\partial^2 \sum_{k=0}^n
\frac{2^kt^{k+1}}{(k+1)!x^2}-a_n\delta^{(4)}(x).
\label{eqs10}
\end{eqnarray}
The fully regulated form factor of the correlator (\ref{eqs3})
thus becomes
\begin{eqnarray}
\frac{b[g(1/x)]}{x^4}=-\sum_n b_n[g(\mu)]
\left[\frac{n!}{2^{n+1}}\partial^2 \sum_{k=0}^n
\frac{2^kt^{k+1}}{(k+1)!x^2}+a_n\delta^{(4)}(x)\right].
\label{eqs11}
\end{eqnarray}
Consequently, the scale derivative of $b[g(1/x)]/x^4$ can be written as the
sum of the local contribution of the $k=0$ term plus a non-local term 
proportional to $\beta (g)$ due to Eq.\,(\ref{eqs9}),
\begin{eqnarray}
\mu\frac{\partial}{\partial\mu}\frac{b[g(1/x)]}{x^4}
=2\pi^2\widetilde{b}[g(\mu)]\delta^{(4)}(x)+
\beta[g(\mu)]\partial^2\left[\frac{F(x)}{x^2}\right],
\label{eqs12}
\end{eqnarray}
where 
\begin{eqnarray}
\widetilde{b}[g(\mu)]=\sum_n b_n[g(\mu)]\frac{n!}{2^n},
\label{eqs13}
\end{eqnarray}
and $F(x)$ is the sum of all the terms with $k{\geq}1$ in the scale
derivative. In particular, with Eq.\,(\ref{eqs9}), 
the scale derivative of $\widetilde{b}[g(\mu)]$ yields a differential equation
for $\widetilde{b}[g(\mu)]$ \cite{afgj},
\begin{eqnarray}
\beta (g)\frac{d\widetilde{b}(g)}{dg}+2 \widetilde{b}(g)=2 b(g).
\label{eqs14}
\end{eqnarray}
Eq.\,(\ref{eqs14}) means that the functions $b[g(\mu)]$ and 
$\widetilde{b}[g(\mu)]$ coincide at the fixed points of the renormalization
group flow. This result is an important step towards the non-perturbative
determination of the renormalization flow of the $b$-function. It should be
emphasized that Eqs.\,(\ref{eqs9})---(\ref{eqs14}) were derived 
within perturbation theory, but they can be regarded as 
non-perturbative results \cite{afgj}.  

The function $\widetilde{b}[g(\mu)]$ turns out to be 
the coefficient $e$ of the external field part 
of the trace anomaly (\ref{eq732}). This can be easily shown by writing down
the generating functional for the current correlation function (\ref{eqs3}),
\begin{eqnarray}
e^{-\Gamma [B_\mu]}=\int [d\varphi]
e^{-S[\varphi]+i\int d^4x j^\mu (x)B_\mu(x)},
\label{eqs15}
\end{eqnarray}
where $\varphi$ denotes all the fields appearing in the 
functional integral
including the ghost fields associated with gauge fixing. 
$B_\mu$ is the external field coupled to the flavour 
current $j_\mu$ listed in (\ref{eqs1}).
Since the action of the scale derivative is equal to the insertion
of the trace of the energy-momentum tensor with 
zero momentum, i.e. the insertion 
of $\int d^4 x T^{\mu}_{~\mu}$ \cite{ref532},  
\begin{eqnarray}
\partial_\alpha d^{\alpha}=T^{\alpha}_{~\alpha}
=\beta (g)\frac{\partial\Gamma}{\partial g}
=\mu \frac{\partial\Gamma}{\partial\mu},
\label{eqs16}
\end{eqnarray}
we have
\begin{eqnarray}
\mu \frac{\partial}{\partial\mu}e^{-\Gamma}
=\int [d\varphi]e^{-S[\varphi]+i\int d^4x j^\mu (x)B_\mu(x)}\int d^4z\left[
-\frac{3N_c-N_f (1-\gamma)}{32\pi^2} (F^a_{\mu\nu})^2
+\frac{1}{4}q (B_{\mu\nu})^2\right],\nonumber\\
\label{eqs17}
\end{eqnarray}
where the general from of the trace anomaly of $N=1$ 
supersymmetric gauge theory plus an external anomaly was used:
\begin{eqnarray}
T^{\mu}_{~\mu}=
-\frac{3N_c-N_f (1-\gamma)}{32\pi^2} (F^a_{\mu\nu})^2
+\frac{1}{4}q (B_{\mu\nu})^2,
\label{eqs18}
\end{eqnarray}
$q$ being a coefficient needing to be determined.
On the other hand, the scale derivative of the flavour current 
correlator (\ref{eqs3}) gives
\begin{eqnarray}
&&\mu \frac{\partial}{\partial\mu}\langle j_\mu (x)j_\nu (y)\rangle
=\langle j_\mu (x)j_\nu(y)\int d^4z\theta^{\mu}_{~\mu}(z) \rangle
=\mu \frac{\partial}{\partial\mu}\left[\frac{\delta}{i\delta B_\mu (x)}
\frac{\delta}{i\delta B_\nu (y)}e^{-\Gamma}\right]\nonumber\\
&=&\frac{\delta^2}{\delta B_\mu (x)\delta B_\nu (y)}
\int [d\varphi]e^{-S[\varphi]+i\int d^4x j^\mu (x)B_\mu(x)}\int d^4z\left[
-\frac{3N_c-N_f (1-\gamma)}{32\pi^2} (F^a_{\mu\nu})^2
+\frac{1}{4}q (B_{\mu\nu})^2\right]\nonumber\\
&=&q(\partial_\mu \partial_\nu-\delta_{\mu\nu}\Box)\delta^{(4)} (x)
-\frac{3N_c-N_f(1-\gamma)}{32\pi^2}
\langle J_\mu (x)J_\nu (y)\int d^4z(F^a_{\mu\nu})^2 \rangle.
\label{eqs19}
\end{eqnarray}
Comparing the above equation with (\ref{eqs12}), one can immediately
identify the coefficients of their local terms up to contributions
${\cal O}(\beta [g(\mu)])$ which will vanish at the fixed points,
\begin{eqnarray}
q=\frac{1}{8\pi^2}\widetilde{b}[g(\mu)]+{\cal O}(\beta [g(\mu)]).
\label{eqs20}
\end{eqnarray}

The renormalization group flow of $\widetilde{b}[g(\mu)]$,
can be computed from the anomalous divergence
$\partial_\mu R_0^\mu$, which 
lies in the same supermultiplet with the trace
anomaly and hence their coefficients should be identical. 
In particular, as shown in Eq.\,(\ref{eq7110}), 
the $R_0^\mu$-current can be combined 
with the Konishi current to form an (internal) anomaly-free $R$-current. 
The (external) anomalous divergence $\langle \partial_\mu R^\mu\rangle$ 
can be calculated exactly from the one-loop triangle diagram
$\langle R_\mu (x)j_\nu (y) j_\rho (z)\rangle$.  
Note that 't Hooft anomaly matching holds only for the correlator
$\langle R_\mu (x)j_\nu (y) j_\rho (z)\rangle$, not for 
$\langle R_{0\mu} (x)j_\nu (y) j_\rho (z)\rangle$ or 
$\langle K_\mu (x)j_\nu (y) j_\rho (z)\rangle$. This is because
$R_{0\mu}(x)$ and $K_\mu(x)$ have internal
anomalies, so in the higher order triangle diagrams, 
there will arise so-called
``rescattering graphs'' \cite{ref8} containing an internal triangle
diagram in which the axial vector current will 
communicate with a pair of gluons
due to the internal anomaly coming from the sub-fermionic triangle
diagram. This will produce higher
order non-local contributions to the chiral anomaly.  In view of this,
the expectation values of the anomalous divergence of $R_\mu$, $R_{0\mu}$
and $K_\mu$ in the presence of the external field $B_\mu$ coupled to the
flavour current $j_\mu$ are as the following:
\begin{eqnarray}
&&\langle \partial^\mu R_\mu \rangle{\equiv}
\frac{1}{48\pi^2}sB^{\mu\nu}\widetilde{B}_{\mu\nu},\nonumber\\
&& \langle \partial^\mu R_{0\mu} \rangle =
-\frac{1}{48\pi^2}\widetilde{b}[g(\mu)]B^{\mu\nu}\widetilde{B}_{\mu\nu}
+{\cdots},
\nonumber\\
&& \langle \partial^\mu K_{\mu} \rangle =
-\frac{1}{48\pi^2}\widetilde{k}[g(\mu)]B^{\mu\nu}\widetilde{B}_{\mu\nu}
+{\cdots}.
\label{eqs21}
\end{eqnarray}
The omitted terms in $\langle \partial^\mu R_{0\mu}\rangle$
denote the non-local contributions proportional 
to the internal anomaly $\beta [g(\mu)]F_{\mu\nu}\widetilde{F}^{\mu\nu}$.
There is a similar non-local contribution 
to $\langle \partial^\mu K_{\mu} \rangle$,
which will cancel the corresponding terms of 
$\langle \partial^\mu R_{0\mu}\rangle$ in the linear combination 
(\ref{eq7110}) and leads to $\langle \partial^\mu R_\mu \rangle$.  
 In fact, these non-local contributions are irrelevant for 
the following analysis since their local terms
are of the order ${\cal O}\left(\beta [g(\mu)]\right)$; The quantity
$s$ is a constant independent 
of the renormalization scale.

The combination of (\ref{eq7110}) and (\ref{eqs21}) implies that 
the external anomaly coefficients satisfy the relation
\begin{eqnarray}
\widetilde{b}[g(\mu)]+\left(1-\frac{3N_c}{N_f}-\gamma [g(\mu)]\right)
\widetilde{k}[g(\mu)]=-s.
\label{eqs22}
\end{eqnarray}
This relation holds along the whole trajectory of the renormalization group
flow since  $s$ is independent of the renormalization scale. For an 
asymptotically free supersymmetric gauge theory, $g=0$ is the UV fixed point
and the anomalous dimension $\gamma (g_{UV})$  
vanish at the UV fixed point.
Consequently, $\widetilde{b}_{UV}$ and $\widetilde{k}_{UV}$ can be
calculated  from the one-loop triangle diagrams $\langle R_0JJ\rangle$
and $\langle KJJ\rangle$ in the context of a free field theory. 
The coincidence of Eq.\,(\ref{eqs22}) at scale $\mu$ and in the UV limit gives
\begin{eqnarray}
\widetilde{b}[g(\mu)]=b_{UV}+\gamma[g(\mu)]\widetilde{k}_{UV}-
\left(1-\frac{3N_c}{N_f}-\gamma [g(\mu)]\right)\left(\widetilde{b}[g(\mu)]
-\widetilde{k}_{UV}\right).
\label{eqs23}
\end{eqnarray}
As for the IR aspect, 
in the conformal window
$3N_c/2 <N_f<3N_c$,  the theory
flows to a non-trivial fixed point $g_\star$. The vanishing of the
NSVZ $\beta$-function gives the exact infrared limit value of the
anomalous dimension,
\begin{eqnarray}
\gamma_{IR}=1-\frac{3N_c}{N_f}.
\label{eqs24}
\end{eqnarray}
Then, at the IR fixed point, Eq.\,(\ref{eqs23}) becomes:
\begin{eqnarray}
b_{IR}-b_{UV}=\gamma_{IR}\widetilde{k}_{UV}.
\label{eqs25}
\end{eqnarray}
Furthermore, based on the observation that 
the gaugino does not contribute to the
flavour current correlator, and that the quark and antiquark
contribution to the combination $R_{0\mu}+K_\mu$ cancels, 
we obtain in the UV limit,
\begin{eqnarray}
\langle \partial^\mu R_{0\mu}\rangle 
+\langle \partial^\mu K_\mu\rangle =0.
\label{eqs26}
\end{eqnarray}  
This equation together with (\ref{eqs21}) yields
\begin{eqnarray}
b_{UV}=-\widetilde{k}_{UV}.
\label{eqs27}
\end{eqnarray} 
The calculation of the one-loop $\langle \partial^\mu R_\mu(x)
j^5_\nu(y)j^5_\rho (z)\rangle$ gives $b_{UV}=2N_f/N_c$. (\ref{eqs24}),
(\ref{eqs25}) and (\ref{eqs27}) lead to $b_{IR}=6$. 
The total renormalization group flow of the central function $b$ 
from the UV limit to the IR fixed point is thus obtained,
\begin{eqnarray}
b_{UV}-b_{IR}=6\left(\frac{N_f}{3N_c}-1\right).
\label{eqs28}
\end{eqnarray}
One can easily see that this flow is always negative in the whole 
conformal window, $3N_c/2<N_f<3N_c$, This contradicts the
$c$-theorem and hence the $b$-function is excluded from being a candidate
for a $c$-function.

In the magnetic theory, the four-component form of
 the flavour currents corresponding to the global symmetry
$SU(N_f)_q{\times}SU(N_f)_{\widetilde{q}}{\times}U_B(1){\times}U_R(1)$
is
\begin{eqnarray}
\widetilde{j}_{q\mu}^A 
&=& \frac{1}{2}\overline{\psi}_q\gamma_\mu (1-\gamma_5)\overline{t}^A\psi_q,
~~\widetilde{j}_{\widetilde{q}\mu}^A = \frac{1}{2}
\overline{\widetilde{\psi}}_{\widetilde{q}}
\gamma_\mu (1-\gamma_5)t^A\widetilde{\psi}_{\widetilde{q}}
\nonumber\\
\widetilde{j}_{\mu}^5 &=&\frac{1}{N_f-N_c}\left(\frac{1}{2}
\overline{\psi}_{\widetilde{q}}
\gamma_\mu \gamma_5\psi_{\widetilde{q}}
-\frac{1}{2}\overline{\widetilde{\psi}}_{\widetilde{q}}
\gamma_\mu \gamma_5\widetilde{\psi}_{\widetilde{q}}\right).
\label{eqs29}
\end{eqnarray}
The anomaly-free magnetic $R$-current is given by the combination 
(\ref{eq7115}), where the four-component form of the Konishi current 
and the original $R$-current are
\begin{eqnarray}
K_\mu &=&K_\mu^{q} +K_\mu^{M};
~~K_\mu^{q}=\frac{1}{2}\overline{\psi}_q
\gamma_\mu \gamma_5\psi_q+\frac{1}{2}
\overline{\widetilde{\psi}}_{\widetilde{q}}
\gamma_\mu \gamma_5\widetilde{\psi}_{\widetilde{q}},
\nonumber\\
K_\mu^{M}&=&\frac{1}{2}\overline{\psi}_M
\gamma_\mu \gamma_5\psi_M,
~~\widetilde{R}_{0\mu}
=\frac{1}{2}\overline{\lambda}^a\gamma_\mu\gamma_5\lambda^a-K_\mu.
\label{eqs30}
\end{eqnarray}
The flavour interaction superpotential ${\cal W}_f
=fq^{ri}M_{ij}\widetilde{q}^j_{~r}$ 
makes the analysis of the identification
of the external trace anomaly coefficient $e$ with the 
central function $b$ quite complicated, since now $b$ is a function
of both $g(1/x)$ and $f(1/x)$. However, the same analysis as in the 
electric theory shows that they indeed coincide at 
the fixed points \cite{afgj}. 

The derivation of the renormalization group flow of  
the central function  $\widetilde{b}$ in the magnetic flavour 
current correlator is 
similar to the electric theory, and the difference is only in the Konishi
current part. The combination (\ref{eq7115}) and the matrix elements 
of the external anomaly equations
\begin{eqnarray}
\langle\partial^\mu \widetilde{R}_\mu\rangle &=&\frac{1}{48\pi^2}
\widetilde{s}B^{\mu\nu}\widetilde{B}_{\mu\nu},\nonumber\\
\langle\partial^\mu \widetilde{R}_{0\mu}\rangle &=&-\frac{1}{48\pi^2}
\widetilde{b}[g(\mu),f(\mu)]B^{\mu\nu}\widetilde{B}_{\mu\nu}
+\cdots,
\nonumber\\
\langle\partial^\mu K_\mu^{q}\rangle&=&-\frac{1}{48\pi^2}
\widetilde{k}^{(q)}[g(\mu),f(\mu)]B^{\mu\nu}\widetilde{B}_{\mu\nu}
+\cdots,
\nonumber\\
\langle\partial^\mu K_\mu^{M}\rangle &=&-\frac{1}{48\pi^2}
\widetilde{k}^{M}[g(\mu),f(\mu)]B^{\mu\nu}\widetilde{B}_{\mu\nu}
+\cdots
\label{eqs33}
\end{eqnarray}
as well as the scale independence of $\widetilde{s}$, 
lead to the relation
\begin{eqnarray}
\widetilde{b}&=&\widetilde{b}_{UV}+\gamma_q\widetilde{k}^{(q)}_{UV}
+\gamma_M\widetilde{k}^{(M)}_{UV}
+\left(2\gamma_q+\gamma_M\right)\left(\widetilde{k}^{(M)}-
\widetilde{k}^{(M)}_{UV}\right)\nonumber\\
&&-\left[1-\frac{3(N_f-N_c)}{N_f}\right]\left(\widetilde{k}^{(q)}
-2\widetilde{k}^{(M)}-\widetilde{k}^{(q)}_{UV}+2\widetilde{k}^{(M)}_{UV}
\right),
\label{eqs34}
\end{eqnarray}
where, as in the case of the electric theory, a quantity 
with subscript $UV$ means that it is evaluated from the 
lowest order triangle diagrams at high energy, 
while the other quantities are defined at an arbitrary 
renormalization scale $\mu$.

At the IR fixed point, the vanishing of 
the $\beta$-functions for both the gauge
magnetic coupling and the Yukawa coupling gives
\begin{eqnarray}
\gamma^{IR}_q=-\frac{1}{2}\gamma^{IR}_M=1-\frac{3(N_f-N_c)}{N_f}.
\label{eqs35}
\end{eqnarray}
The above anomalous dimensions and the 
relation (\ref{eqs34}) lead to the renormalization group flow
\begin{eqnarray}
\widetilde{b}_{IR}-\widetilde{b}_{UV}=\gamma_q^{IR}\widetilde{k}_{UV}^{(q)}
+\gamma_M^{IR}\widetilde{k}_{UV}^{(M)}.
\label{eqs36}
\end{eqnarray} 
Eq.\,(\ref{eqs36}) is the non-perturbative formula for the flow of the
central function of the flavour currents in the magnetic theory. For
the baryon number current listed in (\ref{eqs29}), the 
one-loop triangle diagram $\langle \widetilde{j}_\mu^5 j_\nu j_\rho\rangle$ and
$\langle\partial^\mu \widetilde{R}_{0\mu}\rangle +\langle\partial^\mu K_\mu\rangle=0$
in the UV limit yield
\begin{eqnarray}
\widetilde{b}_{UV}=-\widetilde{k}_{UV}^{(q)}=\frac{2N_c}{N_f-N_c},~~~
\widetilde{k}_{UV}^{(M)}=0.
\label{eqs37}
\end{eqnarray}
(\ref{eqs36}) and (\ref{eqs37}) determine that $b_{IR}=6$. Therefore,
the flow of the central function of the baryon number current is
\begin{eqnarray}
\widetilde{b}_{IR}-\widetilde{b}_{UV}=6\left[1-\frac{N_f}{3(N_f-N_c)}\right]=
2\frac{2N_f-3N_c}{N_f-N_c}.
\label{eqs38}
\end{eqnarray}
This flow is again positive throughout the whole conformal window
of the magnetic theory, $3(N_f-N_c)/2<N_f<3(N_f-N_c)$ (i.e.
$3N_c/2<N_f<3N_c$). Moreover, the central functions in 
the electric and magnetic theories
have the same IR values, $b_{IR}=\widetilde{b}_{IR}=6$. 
The above results are actually 
an inevitable outcome of the electric-magnetic  
duality in the conformal window.  
They can be regarded as support for Seiberg's electric-magnetic duality
conjecture.

Having verified explicitly that the central functions
of the flavour currents 
are not suitable candidates for the $c$-function, 
we turn to the gravitational central functions \cite{jhep}. 
These central functions are contained in the operator product expansion 
of the energy-momentum tensor $T_{\mu\nu}$. The conservation
of $T_{\mu\nu}$ implies that the two-point correlator of the
energy-momentum tensor should be of the form \cite{jhep}:
\begin{eqnarray}
\langle T_{\mu\nu}(x)T_{\rho\sigma}(0)\rangle
=-\frac{1}{48\pi^4}\Pi_{\mu\nu\rho\sigma}\frac{c[g(1/x)]}{x^4}
+\pi_{\mu\nu}\pi_{\rho\sigma}\frac{f[\ln (x\mu), g(1/x)]}{x^4},
\label{eqs39}
\end{eqnarray}
where $\pi_{\mu\nu}=\partial_\mu\partial_\nu-\delta_{\mu\nu}\Box$
and $\Pi_{\mu\nu\rho\sigma}$ is the transverse traceless spin 2
projective tensor, the $n=4$ case given in Eq.\,(\ref{eq738})
\begin{eqnarray}
\Pi_{\mu\nu\rho\sigma}&=&2\pi_{\mu\nu}\pi_{\rho\sigma}
-3\left(\pi_{\mu\rho}\pi_{\nu\sigma}+\pi_{\mu\sigma}\pi_{\nu\rho}\right),
\nonumber\\
\Pi^{\mu}_{~\mu\rho\sigma}&=&\Pi_{\mu\nu\rho}^{~~~~\rho}=0, ~~~
\partial^\mu \Pi_{\mu\nu\rho\sigma}=\partial^\nu \Pi_{\mu\nu\rho\sigma}=
\partial^\rho \Pi_{\mu\nu\rho\sigma}=\partial^\sigma \Pi_{\mu\nu\rho\sigma}=0. 
\label{eqs40}
\end{eqnarray}
With a non-local redefinition \cite{jhep}
\begin{eqnarray}
{\cal T}_{\mu\nu}= T_{\mu\nu}(x)+\frac{1}{3}\frac{1}{\Box}
\pi_{\mu\nu}T^{\rho}_{~\rho},
\label{eqs41}
\end{eqnarray}
one can get the two-point correlator of the energy-momentum tensor
with an improved trace property and a simple correlator: 
\begin{eqnarray}
\langle {\cal T}_{\mu\nu}(x){\cal T}_{\rho\sigma}(0)\rangle 
= -\frac{1}{48\pi^4}\Pi_{\mu\nu\rho\sigma}\frac{c[g(1/x)]}{x^4}.
\label{eqs42}
\end{eqnarray}

As for the flavour current, the central function
$c[g(1/x)]$ is connected to the coefficient of the Weyl tensor part
of the gravitational anomaly. 
For $N=1$ supersymmetric QCD, 
we introduce the background metric $g_{\mu\nu}(x)$
for the energy-momentum tensor $T_{\mu\nu}(x)$ and the
external $U(1)$ gauge field $V_\mu$ for the $R_{0\mu}$-current.
According to the general form (\ref{eq732}) of the trace anomaly
in the  presence of external fields,
the trace anomaly at the fixed point is
\begin{eqnarray}
\langle T^{\mu}_{~\mu}\rangle 
=\frac{\widetilde{c}}{16\pi^2} C_{\mu\nu\rho\sigma}C^{\mu\nu\rho\sigma}
-\frac{a}{16\pi^2} \widetilde{R}_{\mu\nu\rho\sigma}
 \widetilde{R}^{\mu\nu\rho\sigma} 
+\frac{\widetilde{c}}{16\pi^2} V_{\mu\nu}V^{\mu\nu},
\label{eqs44}
\end{eqnarray}
where $V_{\mu\nu}=\partial_\mu V_\nu-\partial_\nu V_\mu$ is the field
strength of $V_\mu$. 
The coefficients of 
$(C_{\mu\nu\rho\sigma})^2$ and $(V_{\mu\nu})^2$ 
are identical since  $\partial^{\mu}R_{0\mu}$ and the trace
anomaly are in the same supermultiplet.
The coefficient of the Euler number density is an independent
constant \cite{ duff}. 

The renormalization group flow of the $c$-function and the $a$-function
can be determined using a similar technique as in 
the flavour current case. 
First, the $c$- and $a$-functions in the UV limit can be calculated 
in a free field theory due to the asymptotic freedom. 
In a free supersymmetric gauge theory
with $N_v$ vector and $N_{\chi}$ chiral supermultiplets, 
the central function $c$ at the UV fixed point is given
by (\ref{eq722}) and the $a$-function is \cite{duff,ref13}, 
\begin{eqnarray}
a_{UV}=\frac{1}{48}\left(9N_v+N_{\chi}\right).
\label{eqs45}
\end{eqnarray}
For $N=1$ supersymmetric QCD, $N_v=N_c^2-1$ and $N_{\chi}=2N_cN_f$.
Off criticality there will arise internal trace anomaly terms
in $\langle T^{\mu}_{~\mu}\rangle$, but they are proportional to
the $\beta$ function for gauge coupling, $\beta [g(\mu)]$, and hence give no
contribution to the flow. The central charges thus depend on the
running gauge coupling, $c=c[g(\mu)]$, $a=a[g(\mu)]$.  The next step is
still to resort to the $R_{0\mu}$ anomaly  
since the coefficient of the $\partial^\mu R_{0\mu}$ 
is connected to the trace anomaly coefficient. In particular, 
$\partial^\mu R_{0\mu}$ can be exactly evaluated in the IR region. 
In the presence 
of the background gravitational field and the external gauge field
coupled to the $R_0$-current, the external field part of the
$\partial^\mu R_{0\mu}$ anomaly is of the following general form: 
\begin{eqnarray}
\partial^\mu R_{0\mu}=\left(uc+va\right)R^{\mu\nu\rho\sigma}
\widetilde{R}_{\mu\nu\rho\sigma}+\left(rc+sa\right)V^{\mu\nu}
\widetilde{V}_{\mu\nu}, 
\label{eqs46}
\end{eqnarray} 
where $u$, $v$, $r$ and $t$ are universal (or model independent) coefficients,
and they can be calculated in the UV limit from free field theory
due to asymptotic freedom. 
With the observation that gauginos and quarks have opposite $R$-charges, 
$+1$ and $-1$, respectively
$uc_{UV}+va_{UV}$ and $rc_{UV}+sa_{UV}$ can be evaluated from
the triangle diagrams $\langle R_0TT \rangle$
and $\langle R_0R_0R_0\rangle$, respectively:
\begin{eqnarray}
uc_{UV}+va_{UV}=\frac{3N_v-N_{\chi}}{24\pi^2},\nonumber\\
rc_{UV}+sa_{UV}=\frac{27N_v-N_{\chi}}{24\pi^2}.
\label{eqs47}
\end{eqnarray}
With the values of $c_{UV}$ and $a_{UV}$ given in (\ref{eqs45}),
it can be easily found that
\begin{eqnarray}
 \langle \partial^\mu\left(\sqrt{g}R_{0\mu}\right)\rangle_{UV}=
\frac{c_{UV}-a_{UV}}{24\pi^2}R^{\mu\nu\rho\sigma}
\widetilde{R}_{\mu\nu\rho\sigma}
+\frac{5a_{UV}-3c_{UV}}{9\pi^2}V^{\mu\nu}\widetilde{V}_{\mu\nu} .
\label{eqs48}
\end{eqnarray}

The same procedure as for deriving Eq.\,(\ref{eqs20}) shows
that off criticality the coefficient $\widetilde{c}[g(\mu)]$ 
of $(W_{\mu\nu\rho\sigma})^2$ and $(V_{\mu\nu})^2$ in
the trace anomaly (\ref{eqs44}) is related to 
the central function $c[g(\mu)]$ as 
\begin{eqnarray}
\widetilde{c}[g(\mu)]=c[g(\mu)]+{\cal O}[\beta(g)].  
\label{eqs49}  
\end{eqnarray}
So they coincide at the fixed points of the renormalization group flow. 
The anomaly coefficients of the terms in 
$\partial^\mu R_{0\mu}$ appear as the combination $c-a$ and $5a-3c$. 
The renormalization group flow of $\widetilde{c}[g(\mu)]-a[g(\mu)]$
and $5a[g(\mu)]-3\widetilde{c}[g(\mu)]$ can be determined in the same way
as deriving Eq.\,(\ref{eqs28}). First, the triangle 
 axial gravitational anomalies $\langle RTT\rangle$, $\langle R_0TT\rangle$
and $\langle KTT\rangle$ give
\begin{eqnarray}
\langle \partial_\mu (\sqrt{g}R^\mu)\rangle &=&
\frac{1}{12\pi^2}s_1\epsilon^{\mu\nu\lambda\rho}R_{\mu\nu\sigma\delta}
R^{\sigma\delta}_{~~\lambda\rho},\nonumber\\
\langle \partial_\mu (\sqrt{g}R_0^\mu)\rangle &=&
\frac{1}{12\pi^2}\left(\widetilde{c}[g(\mu)]-a[g(\mu)]\right)
\epsilon^{\mu\nu\lambda\rho}R_{\mu\nu\sigma\delta}
R^{\sigma\delta}_{~~\lambda\rho}+{\cdots},\nonumber\\
\langle \partial_\mu (\sqrt{g}K^\mu)\rangle &=&
\frac{1}{12\pi^2}k[g(\mu)]\epsilon^{\mu\nu\lambda\rho}R_{\mu\nu\sigma\delta}
R^{\sigma\delta}_{~~\lambda\rho}+{\cdots},
\label{eqs50}  
\end{eqnarray} 
where the omitted are the non-local ${\cal O}[\beta(g(\mu))]$ terms
coming from the internal anomaly.
In the UV limit, 
the quantities $k[g(\mu)]$ and $\widetilde{c}[g(\mu)]-a[g(\mu)$ can be 
calculated exactly from the one-loop triangle diagrams, 
$\langle R_0TT\rangle_{UV}$ and
$\langle KTT\rangle_{UV}$ \cite{afgj},
\begin{eqnarray}
k_{UV}&=&-\frac{1}{16}N_{\chi}=-\frac{1}{8}N_fN_c, \nonumber\\
c_{UV}-a_{UV}&=&-\frac{1}{16}\left(N_v-\frac{1}{3}N_{\chi}\right)=
-\frac{1}{16}\left(N_c^2-1-\frac{2}{3}N_fN_c\right).
\label{eqs51}  
\end{eqnarray} 
The scale independence of $s_1$ and the combination Eq.\,(\ref{eq7110}) 
lead to
\begin{eqnarray}
\widetilde{c}[g(\mu)]-a[g(\mu)]&=&c_{UV}-a_{UV}+
\frac{1}{3}\gamma[g(\mu)]k_{UV}\nonumber\\
&&-\frac{1}{3}\left(1-\frac{3N_c}{N_f}
-\gamma[g(\mu)]\right)\left(k[g(\mu)]-k_{UV}\right).
\label{eqs52}  
\end{eqnarray}
The non-perturbative formula for $5a-3c$ can be obtained  
from the triangle diagram $\langle R_0R_0R_0\rangle$. Eq.\,(\ref{eqs2}) shows 
that the fermionic part
of the $R_0$-current is contributed by the
gaugino, the left- and right-handed quarks, all of which are Majorana spinors,
and these three contributions have the same form  
$\overline{\Psi}\gamma_\mu\gamma_5\Psi/2$.
The amplitude of the triangle diagram formed by three
identical axial currents composed of Majorana spinors has a Bose symmetry. 
A calculation in the UV limit yields \cite{ef}
\begin{eqnarray}
\frac{\partial}{\partial z_\rho}\langle J_\mu (x)J_\nu (y)J_\rho (z)
\rangle_{UV}&=&-\frac{1}{12\pi^2}\epsilon_{\mu\nu\lambda\rho}
\frac{\partial}{\partial x_\lambda}\frac{\partial}{\partial y_\rho}
\delta^{(4)}(x-z)\delta^{(4)}(y-z)\nonumber\\
&{\equiv}&\frac{16}{9}{\cal C}_{\mu\nu}(x,y,z),
\nonumber\\
{\cal C}_{\mu\nu}(x,y,z)&=&{\cal C}_{\nu\mu}(y,x,z).
\label{eqs53}  
\end{eqnarray}
With Eq.(\ref{eqs53}), rewriting the combination (\ref{eq7110}) as 
$R_{0\mu}=R_\mu+(\gamma-\gamma_{IR})K_{\mu}/3$
and considering the various Bose symmetric contributions of 
gaugino and quarks to the
anomalous divergence of the triangle $\langle R_0R_0R_0\rangle$, 
one can write down the following anomaly equations \cite{afgj}:
\begin{eqnarray}
&&\frac{\partial}{\partial z_\rho}\langle R_{0\mu}(x)R_{0\nu}(y)
R_{0\rho}(z)\rangle =\left(5a\left[g(\mu)\right]
-3c\left[g(\mu)\right]\right){\cal C}_{\mu\nu}(x,y,z),
\nonumber\\
&&\frac{\partial}{\partial z_\rho}\langle R_{\mu}(x)R_{\nu}(y)
R_{\rho}(z)\rangle = s_2{\cal C}_{\mu\nu}(x,y,z),
\nonumber\\
&&\frac{\partial}{\partial z_\rho}\left[\langle R_{\mu}(x)R_{\nu}(y)
K_{\rho}(z)\rangle + \langle K_{\mu}(x)R_{\nu}(y)
R_{\rho}(z)\rangle \right. \nonumber\\
&& \left. + \langle R_{\mu}(x)K_{\nu}(y)
R_{\rho}(z)\rangle\right]
=3k_1\left[g(\mu)\right]{\cal C}_{\mu\nu}(x,y,z),
\nonumber\\
&&\frac{\partial}{\partial z_\rho}\left[\langle R_{\mu}(x)K_{\nu}(y)
K_{\rho}(z)\rangle + \langle K_{\mu}(x)K_{\nu}(y)
S_{\rho}(z)\rangle 
\right.\nonumber\\
&& \left.+ \langle K_{\mu}(x)R_{\nu}(y)
K_{\rho}(z)\rangle \right]
=3k_2\left[g(\mu)\right]{\cal C}_{\mu\nu}(x,y,z),
\nonumber\\
&&\frac{\partial}{\partial z_\rho}\langle K_{\mu}(x)K_{\nu}(y)
K_{\rho}(z)\rangle = k_3\left[g(\mu)\right]{\cal C}_{\mu\nu}(x,y,z),
\label{eqs54}  
\end{eqnarray}
where the coefficient $s_2$ stays constant along the trajectory of the
renormalization group flow since the $R_\mu$-current is internal anomaly-free.
The above Bose symmetric anomaly equation and the combination (\ref{eq7110}) 
yield
\begin{eqnarray}
5a-3c=s_2+k_1 \left(\gamma-\gamma_{IR}\right)
+\frac{1}{3}\left(\gamma-\gamma_{IR}\right)^2k_2
+\frac{1}{27}\left(\gamma-\gamma_{IR}\right)^3k_3.
\label{eqs55}  
\end{eqnarray}
In the UV limit, the coefficients of the triangle anomaly 
involved in (\ref{eqs54}) can be exactly calculated in a free theory,
\begin{eqnarray}
\gamma_{UV}&=&0, ~k_{1UV}=\frac{9}{16}N_{\chi}\left(\frac{N_c}{N_f}\right)^2,
~k_{2UV}=-\frac{9}{16}N_{\chi}\frac{N_c}{N_f},\nonumber\\
 k_{3UV}&=&\frac{9}{16}N_{\chi}, ~N_{\chi}=2N_fN_c.
\label{eqs56}  
\end{eqnarray}
Thus Eq.\,(\ref{eqs56}) becomes
\begin{eqnarray}
5a_{UV}-3c_{UV}=s_2-k_{1UV}\gamma_{IR}+\frac{1}{3}\gamma_{IR}^2k_{2UV}
-\frac{1}{27}\gamma_{IR}^3k_{3UV}.
\label{eqs57}  
\end{eqnarray}
Eqs.\,(\ref{eqs55}) and (\ref{eqs57}) give 
\begin{eqnarray}
&&5a[g(\mu)]-3c[g(\mu)]=5a_{UV}-3c_{UV}+h, \nonumber\\
&&h{\equiv}\gamma_{IR}k_{1UV}
-\frac{1}{3}k_{2UV}\gamma_{IR}^2+\frac{1}{27}k_{3UV}\gamma_{IR}^3
+k_1\left[g(\mu)\right]\left(\gamma-\gamma_{IR}\right)\nonumber\\
&&+\frac{1}{3}k_2\left[g(\mu)\right] \left(\gamma-\gamma_{IR}\right)^2
+\frac{1}{27}k_3\left[g(\mu)\right] \left(\gamma-\gamma_{IR}\right)^3.
\label{eqs58}  
\end{eqnarray}
With Eqs.\,(\ref{eqs52}) and (\ref{eqs58}), 
the non-perturbative formula for the central functions turns out to be
\begin{eqnarray}
c&=&c_{UV}+\frac{5}{6}\left(\gamma-\gamma_{IR}\right)k\left[g(\mu)\right]
+\frac{5}{6}\gamma_{IR}k_{UV}+\frac{1}{2}h,\nonumber\\
a&=& a_{UV}+\frac{1}{2}\left(\gamma-\gamma_{IR}\right)k\left[g(\mu)\right]+
\frac{1}{2}\gamma_{IR}k_{UV}+\frac{1}{2}h.
\label{eqs60}  
\end{eqnarray}
Evaluating $c[g(\mu)]$ and $a[g(\mu)]$ 
at the IR fixed point, 
we get the renormalization group flow of the central charges,
\begin{eqnarray}
c_{IR}-c_{UV}&=&\frac{5}{6}\gamma_{IR}k_{UV}+
\frac{1}{2}\gamma_{IR}k_{1UV}-\frac{1}{6}\gamma_{IR}^2k_{2UV}+
\frac{1}{54}\gamma_{IR}^3k_{3UV} \nonumber\\
&=&\frac{N_cN_f}{48}\gamma_{IR}\left[9\left(\frac{N_c}{N_f}\right)^2
+3\frac{N_c}{N_f}-4\right], \nonumber\\
a_{IR}-a_{UV}&=&\frac{1}{2}\gamma_{IR}k_{UV}+
\frac{1}{2}\gamma_{IR}k_{1UV}-\frac{1}{6}\gamma_{IR}^2k_{2UV}+
\frac{1}{54}\gamma_{IR}^3k_{3UV} \nonumber\\
&=&-\frac{N_cN_f}{48}\gamma_{IR}^2\left(2+3\frac{N_c}{N_f}\right).
\label{eqs61}  
\end{eqnarray}
In deriving (\ref{eqs61}) we have used the values
listed in  (\ref{eqs51}) and (\ref{eqs56}). 

In fact, the values of the central functions
at both the UV and IR fixed points can be calculated directly.
Since the NSVZ $\beta$-function vanishes
at the IR fixed point,
the combination (7.1.10) shows that $R_{0\mu}$ and 
$R_{\mu}$ coincide at the IR fixed point. Because the
anomaly-free $R_{\mu}$ has one-loop exact external anomalies and the 
coefficients are scale independent, the IR values of the
external $R_\mu$ anomaly coefficients must be equal to the UV ones. This 
fact means
\begin{eqnarray}
\frac{\partial}{\partial z_\rho}\langle R_{0\mu}(x)R_{0\nu}(y)
R_{0\rho}(z) \rangle_{IR}
&=&\frac{\partial}{\partial z_\rho}\langle R_{\mu}(x)R_{\nu}(y)
R_{\rho}(z) \rangle_{IR}\nonumber\\
&=&\frac{\partial}{\partial z_\rho}\langle R_{\mu}(x)R_{\nu}(y)
R_{\rho}(z) \rangle_{UV}.
\label{eqs62}  
\end{eqnarray}
Using  $R^{UV}_{\mu}=R_{0\mu}+(1-3N_c/N_f)K_{\mu}$ with
$R_{0\mu}$ and $K_{\mu}$ given in (\ref{eqs2}) and considering
the relative contributions from the gaugino and quarks, one can find
\cite{afgj}
\begin{eqnarray}
5 a_{IR}-3c_{IR}=\frac{9}{16}\left[N_v-N_{\chi}\left(\frac{N_c}{N_f}\right)^3\right]
=\frac{9}{16}\left[N_c^2-1
-2N_cN_f\left(\frac{N_c}{N_f}\right)^3\right].
\label{eqs63}  
\end{eqnarray}  
Calculating $\partial/\partial z_{\rho}\langle R_\mu (x)
R_{\nu} (y) R_{\rho} (z)\rangle$ in the UV limit as in 
 free field theory, we obtain: 
\begin{eqnarray}
5 a_{UV}-3c_{UV}=\frac{9}{16}\left(N_c^2-1-\frac{2}{27}N_cN_f\right).
\label{eqs64}  
\end{eqnarray}  
Applying a similar procedure to the axial gravitational triangle
diagram $\langle TTR\rangle $, we find:
\begin{eqnarray}
c_{IR}-a_{IR}&=&\frac{1}{16}\left(N_c^2-1+2\right)= \frac{1}{16}\left(N_c^2+1\right);\nonumber\\
c_{UV}-a_{UV}&=&-\frac{1}{16}\left(N_c^2-1-\frac{2}{3}N_fN_c\right).
\label{eqs65}  
\end{eqnarray} 
The values of the central functions at the fixed points are thus 
fixed:
\begin{eqnarray}
c_{IR}&=&\frac{1}{16}\left[7 \left(N_c^2-1\right)-9N_cN_f
\left(\frac{N_c}{N_f}\right)^3+5 \right]
=\frac{1}{16}\left(7N_c^2-2-9 \frac{N_c^4}{N_f^2}\right), \nonumber\\
a_{IR}&=&\frac{3}{16}\left[2\left(N_c^2-1\right)-3N_cN_f
\left(\frac{N_c}{N_f}\right)^3+1 \right]
=\frac{3}{16}\left(2N_c^2-1-3\frac{N_c^4}{N_f^2}\right);\nonumber\\
c_{UV}&=&\frac{1}{24}\left[3(N_c^2-1)+2N_fN_c\right], ~~~
a_{UV}=\frac{1}{48}\left[9(N_c^2-1)+2N_fN_c\right].
\label{eqs66}  
\end{eqnarray} 
It can be easily checked that the flow equation (\ref{eqs61}) is satisfied
with the above explicit values for the central charges.
In a similar way, the flow of central charges in the dual magnetic theory
can be worked out:
\begin{eqnarray}
\widetilde{c}_{IR}-\widetilde{c}_{UV}
&=&\frac{1}{24}\left(1-\frac{3}{2}\frac{N_c}{N_f}\right)\left(
9\frac{N_c^3}{N_f}-6N_c^2+6N_f^2+N_cN_f\right),\nonumber\\
\widetilde{a}_{IR}-\widetilde{a}_{UV}
&=&-\frac{1}{12}\left(1-\frac{3}{2}\frac{N_c}{N_f}\right)^2
\left(3N_c^2+4N_cN_f+3N_f^3\right).
\label{eqs67}  
\end{eqnarray} 

 The result given in Eqs.\,(\ref{eqs61}) and (\ref{eqs67}) 
shows that the flow of the $c$-function can be both positive and negative. 
For example, in the electric theory, $c_{IR}-c_{UV}$ is
negative near the lower edge of the conformal window,
$N_f{\sim}3N_c/2$, but positive near the upper edge, $N_f{\sim}3N_c$,
while in the magnetic theory $c_{IR}-c_{UV}$ is positive in the entire
conformal window. Thus there is no $c$-theorem for this
$c$-function in supersymmetric gauge theory.
However, the flow of the central charge function $a$ always satisfies
$a_{IR}-a_{UV}<0$ for both the electric and magnetic theories in the
whole range of the conformal window. This fact seems to suggest the 
existence of an ``$a$-theorem", i.e. a suitable $c$-function would be  
the coefficient of the Euler term of the trace anomaly. 
In Sect.\,\ref{subsect73}  we have already introduced the
partial quantitative result obtained by Bastianelli that both 
$c_{IR}-c_{UV}$ and $a_{IR}-a_{UV}$ are negative \cite{bast}. Now the explicit
non-perturbative formula shows that the $c$-theorem is only applicable
to the $a$-function. This coincides with the initial choice 
made by Cardy \cite{car}.

\section{Concluding remarks}
\label{sect8n}
\renewcommand{\theequation}{7.\arabic{equation}}
\setcounter{equation}{0}

\subsection{A brief history of electric-magnetic duality}

The status of QCD as a true theory describing strong interaction has been
generally accepted for more than 20 years. Due to the property of asymptotic 
freedom, its predictions at small distances such  as deep inelastic scattering
processes can be calculated and coincide with the results of experimental tests. However, 
its description for the low energy dynamics of quarks is not clear yet. 
It is well known that the remarkable strong interaction phenomena at 
low energy are colour confinement and chiral symmetry breaking \cite{ref81}, 
but the mechanisms for both of them are still not fully understood. 
Although some phenomenological models have been proposed,  
the final truth should be determined by an exact solution of QCD. 
However, with the present methods, it is impossible to find 
an exact quantum action of QCD since it is a highly non-linear theory. 
Perturbative calculations become invalid at low-energy due to the strong coupling. 
Therefore, understanding the non-perturbative dynamics of strong interaction theory 
is a big problem awaiting to be solved.

Duality has provided a possibility to tackle this problem since it relates the
strong coupling in one theory to the weak coupling in another theory. 
The other feature of duality is that it interchanges fundamental quanta 
in the theory with the solitons of its dual theory. It has long been known
that such a duality relation really occurs in two-dimensional relativistic
quantum field theories. Based on Skyrme's conjecture, Coleman and Mandelstam 
found that the bosonic 
sine-Gordon model is completely equivalent to the 
massive Thirring model \cite{ref82,ref83}.
The strong and weak coupling in these two theories are exchanged and
the solitons of the sine-Gordon theory corresponds to the fundamental
fermions of the massive Thirring model.

The search for duality in four-dimensional gauge theory has a long history.
It was noticed early one, that classical electrodynamics, formulated
in terms of field strengths, possesses an electric-magnetic duality
symmetry, if magnetic charges and currents are introduced as sources. 
The quantum theory of magnetic charges was found by Dirac \cite{ref84}.
He found the famous quantization condition ensuring the consistency 
of the quantum mechanics of a magnetically charged particle moving
in an electromagnetic field. This quantization condition implies 
that the electric-magnetic duality, if it exists, exchanges strong 
and weak couplings.

However a consistent quantum field theory with magnetic 
monopoles was not found 
until after more than 40 years.
In 1974, 't Hooft \cite{monopole} and Polyakov \cite{monopolea} 
independently found finite energy classical solutions in the Georgi-Glashow
model and that they can be interpreted as monopoles. At large 
distance, this classical solution behaves as a Dirac monopole. 
Furthermore, 
another kind of soliton solutions carrying both electric charge 
and magnetic charge (dyon) was found by Julia and Zee \cite{dyons}. 
In the Prasad-Sommerfield limit, the explicit
analytic solutions for the magnetic monopole and the dyon were worked out. 
In this limit the classical masses of the magnetic monopole and dyon
saturate the Bogomol'nyi bound and hence such monopole and dyon solutions 
are called BPS solutions \cite{go}. 
A comparison of the classical masses and charges of the BPS 
monopoles with those of the fundamental quantum particles such as the massive
gauge bosons and Higgs particles produced by spontaneously symmetry breaking
in the Georgi-Glashow model \cite{go} shows that the whole particle spectrum
including the BPS magnetic monopoles is invariant under electric-magnetic  
duality provided that the BPS monopoles are interchanged with the massive
gauge vector bosons \cite{ref1p6}. 
This observation motivated the Montonen-Olive duality
conjecture that there should exist a dual description of the Georgi-Glashow 
model where the elementary gauge particles should be BPS magnetic 
monopoles and they should form a gauge triplet together 
with the photon, while the massive magnetic bosons 
should behave as ``electric monopoles". This conjecture 
was further reinforced by the fact that two very different 
calculations for the long-range force 
between the massive gauge bosons (done by computing the lowest order Feynman
diagrams contributed by photon and Higgs particle exchange \cite{ref1p6}),
and that between the BPS magnetic monopoles ( a calculation due to
 Manton \cite{ref810}) yielded identical results. Later Witten considered 
the strong CP violation effects (i.e. including a $\theta$-term) in 
the Georgi-Glashow model and found that 
the $\theta$-term shifts the allowed values of 
the electric charge in the monopole sector \cite{ref811}.
As a consequence, Montonen-Olive duality was extended from $Z_2$ to
$SL(2,Z)$ duality since the $\theta$-parameter and the coupling constant of the
theory can be combined into one complex  parameter.

However, the Montonen-Olive conjecture suffers from several serious drawbacks.
First, owing to the Coleman-Weinberg mechanism \cite{ref812}, 
a non-zero scalar potential will
be generated by quantum corrections even if the classical potential vanishes,
consequently, the classical mass formula will be modified. Thus there is no 
reason to believe that the electric-magnetic duality of the whole particle
spectrum is not broken by radiative corrections through a renormalization 
 of the Bogomol'nyi bound. Secondly, it is well known that the massive gauge
bosons have spin one, while the magnetic monopoles are spherical symmetric
solutions and should have spin zero. 
Thus although the mass spectrum is invariant 
under duality, the quantum states and quantum numbers in two dual theories
do not match. In addition, there is the difficulty to test this conjecture 
since the duality relates weak to strong coupling and  
we know very little about solutions of four dimensional 
strongly coupled theory. Fortunately 
supersymmetry provides a way to circumvent these problems.
   
The main physical motivation to introduce supersymmetry is to solve
the gauge hierarchy problem \cite{ref814}, To some extent, it prevents 
the generation of quantum corrections to the mass terms and  provides 
a naturalness to mass relations in quantum field
theory. In (extended) supersymmetric gauge theories with central charges, the 
supersymmetry has another effect: the Bogomol'nyi  bound  is a property of
some representations of the supersymmetry 
algebra. Due to the work of Witten and Olive \cite{ref815}, 
the physical meaning of the central charges was made clear: they
are precisely the electric and magnetic charges. 
Thus supersymmetry protects
the Bogomol'nyi bound against quantum corrections, guaranteeing that
the bound is true both at the classical and 
the quantum levels. The main reason for 
this is that the BPS mass formula is a necessary condition for the existence of
the short representations of the supersymmetry algebra. A 
typical example with a non-vanishing central charge
is $N=2$ supersymmetric Yang-Mills theory, which is obtained by the
dimensional reduction of $N=1$ supersymmetric Yang-Mills theory in
six dimensions and whose bosonic part is 
just the Georgi-Glashow model \cite{ref816}.
So the supersymmetric generalization of the BPS magnetic monopole and dyon 
solutions can be naturally obtained. Observing the massive supermultiplets
arising out of higgsing this model, one can see that they are only
four-fold degenerate and not sixteen-fold degenerate as usually
expected from the representation of supersymmetry algebra realized
on massive particle states. This fact indicates that the central charge
of $N=2$ supersymmetry saturates the bound. 
Nevertheless, this also shows that the states in 
the short massive supermultiplet have spin $1/2$ and
$0$, but not spin 1, or $1/2$ and $1$, but not $0$. 
Thus $N=2$ supersymmetric 
Yang-Mills theory is not an appropriate theory admitting 
Montonen-Olive duality. This obstacle is surmounted in $N=4$ supersymmetric
Yang-Mills theory, which can be obtained through a dimensional reduction
of ten-dimensional $N=1$ supersymmetric Yang-Mills theory \cite{ref1p7}. 
This model also admits the Higgs mechanism and has 
BPS monopole solutions. The massive vector multiplets
naturally saturate the Bogomol'nyi bound. However, in this model,
thanks to the larger supersymmetry, the short supermultiplet
containing the BPS states and the one containing the massive vector
bosons are isomorphic, both of them have spin $s=0$, $1/2$ and $1$ states.  
Further, the $N=4$ theory possesses another remarkable feature: the 
$\beta$-function vanishes identically \cite{ref819}. 
This means that the gauge coupling is a genuine constant
and does not run.
This property not only makes the relation between the couplings of the
electric and magnetic theories generally valid,
but also present a first non-trivial  
 superconformal field theory in four dimensions, since the trace
of the energy momentum tensor is measure of conformal symmetry
and it is proportional to the $\beta$ function \cite{ref1p15}.

This is far from the end of the duality story. After more
than a decade, Seiberg and Witten, in two papers \cite{ref1p1,ref1p1a}
 which already have
become classical, based on the general form of the low-energy Wilson
effective action of $N=2$ supersymmetric Yang-Mills theory \cite{ref1p2}, 
made the conjecture that there exists an effective electric-magnetic duality
in $N=2$ supersymmetric Yang-Mills theory. Further using this
conjecture they  
determined the explicit instanton coefficients appearing in the low-energy
effective action and the whole structure of quantum vacua. 
The results of Seiberg and Witten were 
supported by explicit instanton calculations
and hence this duality conjecture is convincing. As a consequence, 
Seiberg-Witten's work set off a new upsurge in the search for dualities 
in quantum field theory \cite{ref824}. 

In addition, in an exciting development Seiberg found that
if the numbers of colours and flavours satisfy $3N_c/2<N_f<3N_c$,
the $N=1$ supersymmetric $SU(N_c)$ QCD has a definite infrared fixed
point, where the theory becomes an interacting superconformal field
theory. A duality also then arises in the IR fixed point. Moreover,
 a series of non-perturbative results including those in $SO(N_c)$ and
$Sp(2n_c)$ gauge theories have been 
obtained \cite{se2,ref1p10,ref51,ref1p11}. 
$N=1$ supersymmetric theories have the possibility for
practical physical applications after breaking the supersymmetry. 
Therefore, a confirmation of $N=1$
duality would be remarkably significant. 

\subsection{Possible application of non-perturbative results
and duality of $N=1$ supersymmetric QCD to non-supersymmetric case}

To conclude this report,  we mention some possible
physical applications of Seiberg's $N=1$ duality.
A natural expectation is that we can get some 
understanding on the non-perturbative
aspects of ordinary QCD from the exact results of 
supersymmetric QCD deformed by breaking the 
supersymmetry. It was first investigated
in Ref.\,\cite{ahso}  how to extend Seiberg's exact results on
supersymmetric QCD to nonsupersymmetric 
models by considering the effects of soft supersymmetry
breaking terms. In the electric theory a soft breaking term
is composed of the mass terms of squark fields and the gaugino:
\begin{eqnarray}
{\cal L}_{\rm SB}=-m_Q^2({\phi}^{*r}_{Qi}{\phi}_{Qri}
+\phi^{*r}_{\widetilde{Q}i}\phi_{\widetilde{Q}ri})
+m_g\lambda^a
\lambda^a+m_g^{\star}\overline{\lambda}^a\overline{\lambda}^a.
\label{eq8p1}
\end{eqnarray}
This term can be rewritten in a superfield form:
\begin{eqnarray}
{\cal L}_{\rm SB}&=&
\int d^4{\theta}M_Q(Q^{\dagger}e^{gV}Q+\widetilde{Q}e^{-gV}
\widetilde{Q}^{\dagger})-\int d^2{\theta}M_g\mbox{Tr}(W^{\alpha}W_{\alpha})
\nonumber\\
&&-\int d^2\overline{\theta}M_g^*\mbox{Tr}(\overline{W}_{\dot{\alpha}}
\overline{W}^{\dot{\alpha}}),
\end{eqnarray}
where $M_Q$ is a vector superfield whose $D$-component equals 
$-m_Q^2$ and $M_g$ ($M_g^*$) is a chiral (anti-chiral) superfield
whose $F$-component equals $m_g$ ($m_g^*$). At the low energy,
there are many possible gauge invariant soft supersymmetry breaking 
terms that can be built from
the meson $M$ and baryons, $B$ and $\tilde{B}$. 
For the case $N_f{\leq}N_c+1$, 
the soft term (precisely speaking, 
the first order expansion of soft supersymmetry breaking terms
near the origin of moduli space) was proposed to be \cite{ahso},
\begin{eqnarray}
{\cal L}_{\rm SB}&=&\int d^4{\theta}\left[C_MM_Q\mbox{Tr}(M^{\dagger}M)+
C_BM_Q(B^{\dagger}B+\widetilde{B}^{\dagger}\widetilde{B})+C_{\cal M}{\cal M}(
M,B,\widetilde{B})+\mbox{h.c.}\right]\nonumber\\
&-&\left[\int d^2{\theta}M_g\langle \mbox{Tr}\left(W^{\alpha}W_{\alpha}\right) 
\rangle +\mbox{h.c.}\right],
\label{eq9.35}
\end{eqnarray}
where $C_M$, $C_B$ and $C_{\cal M}$ are normalization constant 
coefficients, ${\cal M}(M,B,\tilde{B})$
is a function of the composite chiral superfield and is invariant under
the global symmetry $SU_L(N_f)$ ${\times}SU_R(N_f)$ ${\times}U_B(1)$
${\times}U_R(1)$. The expectation value 
$\langle \mbox{Tr}\left(W^{\alpha}W_{\alpha}\right)
\rangle$ in (\ref{eq9.35}) should be understood as 
a combination of the various chiral superfields appearing
in the low energy effective Lagrangian which has the same
quantum numbers as $\mbox{Tr}\left(W^{\alpha}W_{\alpha}\right)$. 
The rationale in introducing this soft breaking term lies in that
${\langle}S{\rangle}{\equiv}-{\langle}\mbox{Tr}
\left(W^{\alpha}W_{\alpha}\right) \rangle$
exists in the low-energy effective Lagrangian of supersymmetric 
gauge theory \cite{veya}. As pointed out in Ref.\,\cite{ahso}, 
the soft breaking terms in (\ref{eq9.35}) are not the most
general terms that can be written down. 
There are 
term of higher order in the fields, suppressed by powers of 
$\Lambda$, of the form $\mbox{Tr}(M^{\dagger}M)^{2(n+1)}/\Lambda^{2n}$,
$\mbox{Tr}(B^{\dagger}B)^{2(n+1)}/\Lambda^{2n}$ and 
 $\mbox{Tr}(\widetilde{B}^{\dagger}\widetilde{B})^{2(n+1)}/\Lambda^{2n}$.
However, (\ref{eq9.35}) may be true near the 
origin of the moduli space
$\langle M\rangle=\langle B\rangle=\langle\widetilde{B}\rangle=0$, 
thus the vacuum under consideration
with the above soft breaking terms is the one 
near the origin of the moduli space.
For the range $N_f{\geq}N_c+2$, as seen in Sect.\,\ref{subsect7.1},  
the theory has a dual description and the composite superfield
is effectively replaced by the dual magnetic quarks. Correspondingly,
near the origin of the moduli space, 
$\langle M \rangle=\langle q \rangle=\langle \tilde{q} \rangle=0$, 
the soft breaking term is
\begin{eqnarray}
\widetilde{\cal L}_{\rm SB}
=\widetilde{C}_M m_M^2 \mbox{Tr}({\phi}_M^{\dagger}{\phi}_M)
+C_q m_q^2(\phi_q^{\dagger}\phi_q+
\phi_{\widetilde{q}}^{\dagger}\phi_{\widetilde{q}}),
\label{eq8p4}
\end{eqnarray}
where $\widetilde{C}_M$ and $C_q$ are normalization coefficients.
With these soft breaking terms at the fundamental and composite field levels, 
it is found that in the case of $N_f<N_c$, the standard 
vacuum of ordinary $QCD$ with both confinement and 
chiral symmetry breaking can be obtained. 
However, when $N_f{\geq}N_c$, some strange
phenomena arise: there appear new vacua with spontaneously 
broken baryon number for $N_f=N_c$ and a vacuum state with unbroken 
chiral symmetry for $N_f>N_c$, These exotic vacua  contain massless 
composite fermions and especially, in some cases dynamically generated 
gauge bosons. This indicates that it is not straightforward to obtain 
a complete understanding of non-perturbative QCD starting
from supersymmetric QCD. However one encouraging
result is that Seiberg's electric-magnetic duality seems 
to persist in the presence of small soft supersymmetry breaking.

In fact, due to the existence of the superpotential 
${\cal W}=q_iM^{i}_{~j}\widetilde{q}^j$ in the dual magnetic theory, 
the Lagrangian (\ref{eq8p4}) does not exhaust all 
the possible soft breaking terms.  Another trilinear 
interaction term (called $A$-term) 
is allowed by soft supersymmetry
breaking \cite{ref832} 
\begin{eqnarray}
\widetilde{L}^{\prime}_{\rm SB}
= h\phi_{qi}\phi^i_{Mj}{\phi}_{\widetilde{q}}^j,
\label{eq8p5}
\end{eqnarray}
where the trilinear coupling $h$ is introduced as a free parameter. 
Note that this $A$-term, like the gaugino mass term, breaks 
the $R$-symmetry. The effects of (\ref{eq8p5}) on 
the phases of the range $N_f{\geq}N_c+1$, the vacuum
structure and the fate of duality in the 
soft breaking $N=1$ supersymmetric QCD 
was investigated in Refs.\,\cite{ref833,ref834}. With the inclusion
of (\ref{eq8p5}), the whole scalar potential containing the soft 
breaking term is: 
\begin{eqnarray}
V(\phi_q, \phi_{\widetilde{q}},\phi_M)&=&
\frac{1}{k_T}\mbox{Tr}\left(\phi_q\phi^{\dagger}_q
\phi^{\dagger}_{\widetilde{q}}\phi_{\widetilde{q}}\right)
+\frac{1}{k_T}\mbox{Tr}\left(\phi_q\phi_M\phi_M^{\dagger}\phi_q^{\dagger}
+{\phi}^{\dagger}_{\widetilde{q}}\phi_M^{\dagger}\phi_M
{\phi}_{\widetilde{q}}\right)
\nonumber\\
&+&\frac{\widetilde{g}^2}{2}\left(\mbox{Tr}\phi_q^{\dagger}\widetilde{T}^a
\phi_q-\mbox{Tr}\phi_{\widetilde{q}}\widetilde{T}^a
\phi_{\widetilde{q}}^{\dagger}\right)^2
+m_q^2\mbox{Tr}(\phi_q^{\dagger}\phi_q)
+m_{\widetilde{q}}^2
\mbox{Tr}(\phi_{\widetilde{q}}^{\dagger}\phi_{\widetilde{q}})\nonumber\\
&+&m_M^2\mbox{Tr}(\phi_M^{\dagger}\phi_M)-
\left(h\mbox{Tr}\phi_{qi}\phi^i_{Mj}\phi_{\widetilde{q}}^j+\mbox{h.c.}\right),
\end{eqnarray}
where $\widetilde{T}^a$ 
is the generator of the magnetic gauge group $SU(N_f-N_c)$
and $\widetilde{g}$ is the gauge coupling constant.
The third term is the $D$-term of the magnetic gauge theory.
 $k_q$, $k_M$ are the normalization parameters for $q$,
$\widetilde{q}$ and $M$  so that their kinetic terms 
(K\"{a}hler potentials) take the canonical form \cite{ahso,ref832} 
\begin{eqnarray}
K(q,\widetilde{q},M)=k_q\mbox{Tr}\left(q^{\dagger}
e^{\widetilde{g}\widetilde{V}}q
+\widetilde{q}e^{-\widetilde{g}\widetilde{V}}
\widetilde{q}^{\dagger}\right)+k_M\left(M^{\dagger}M\right).
\end{eqnarray}
The phase structure can be revealed by analyzing the minimum
of the above scalar potential. With the assumption that $h$ is real,
the minimum of the potential can be obtained along the diagonal
direction
\begin{eqnarray}
&&\phi_{qi}^r=\left\{\begin{array}{ll}\phi_{q(i)}\delta^{r}_{~i} & 
                                      i,r=1,{\cdots}, N_f-N_c \\
                             0 & i=N_f-N_c+1, \cdots, N_f\end{array}\right., 
~\phi_{\widetilde{q}i}^r=\left\{\begin{array}{ll}
\phi_{\widetilde{q}(i)}\delta^{r}_{~i} &
i,r=1, {\cdots}, N_f-N_c\\
 0 & i=N_f-N_c+1, \cdots, N_f \end{array}\right. ,\nonumber\\
&&\phi_{Mj}^i=\left\{\begin{array}{ll}\phi_{M(i)}\delta^{i}_{~j}, 
 & i,j=1,{\cdots}, N_f-N_c\\
0 & i,j=N_f-N_c+1, \cdots, N_f\end{array}\right. .
\end{eqnarray} 
It is found that the  trilinear term  plays an important role in the
realization of the broken phase and leads to a rich  vacuum 
structure \cite{ref833,ref834}.
\begin{itemize}
\item In the direction $\phi_{q(i)}=q$ and $\phi_{\widetilde{q}(i)}=0$ (or 
$\phi_{q(i)}=0$ and $\phi_{\widetilde{q}(i)}=q$), the vacuum expectation value
$\langle \phi_{M(i)}\rangle =0$.  If $m_q^2 <0$ (or  $m_{\widetilde{q}}^2 <0$),
the scalar potential will be unbounded from below \cite{ref833} and
the theory becomes unphysical. 
If $m_q^2 >0$ (or  $m_{\widetilde{q}}^2 >0$), the scalar potential
has the minimum $V=0$ at $q=0$ and thus the theory is in the 
gauge and chiral symmetric phase.
\item In the flat direction of the $D$-term, 
$\phi_{q(i)}=\phi_{\widetilde{q}(i)}
=X_i$. A broken phase arises when the solution of the 
expectation value equation 
\begin{eqnarray}
G\left[\langle\phi_{M(i)}\rangle\right]
=\left(\frac{2}{k_q}\langle\phi_{M(i)}\rangle-h\right)
f(\langle\phi_{M(i)}\rangle)
-\frac{2}{k_M}m_M^2\langle \phi_{M(i)}\rangle=0
\label{eq8p9}
\end{eqnarray} 
satisfies the following inequalities 
\begin{eqnarray}
f(\langle\phi_{M(i)}\rangle){\equiv}\frac{2}{k_q}{\langle}
\phi_{M(i)}{\rangle}^2-2h{\langle}\phi_{M(i)}{\rangle}
+m_q^2+m^2_{\widetilde{q}}{\leq}0.
\label{eq8p10}
\end{eqnarray}
Otherwise, the theory will be in a (gauge and chiral) 
symmetric phase. (\ref{eq8p9}) 
requires that the soft breaking parameters should satisfy 
\begin{eqnarray}
h^2{\geq}\frac{2}{k_q}\left(m_q^2+m_{\widetilde{q}}^2\right),
\end{eqnarray}
having then a non-trivial solution:
\begin{eqnarray}
\frac{h-\sqrt{h^2-2(m_q^2+m_{\widetilde{q}}^2)/k_q}}{2/k_q}{\leq}\langle
\phi_{M(i)}\rangle {\leq}
\frac{h+\sqrt{h^2-2(m_q^2+m_{\widetilde{q}}^2)/k_q}}{2/k_q}.
\label{eq8p12}
\end{eqnarray}
The sufficient condition for the existence of a broken phase
is that the local maximum point of 
$G\left[\langle\phi_{M(i)}\rangle\right]$ should be within the
region (\ref{eq8p12}) and further at that point the value of
of $G\left[\langle\phi_{M(i)}\rangle\right]$ should not be negative. 
This requires 
\begin{eqnarray}
\left(\frac{1}{3}h^2+\frac{2m_M^2}{3k_M}-\frac{2}{3}
\frac{m_q^2+m_{\widetilde{q}}^2}{k_q}\right)^3
-\left(\frac{m_M^2}{k_M}h\right)^2{\geq}0.
\label{eq8p13}
\end{eqnarray}
Depending on the ratio $\rho{\equiv}
2m_M^2/(m_q^2+m_{\widetilde{q}}^2) k_q/k_M$, the phase structure in these
$D$-flat directions can present various patterns \cite{ref833}. 
For example, in the phase
diagram labelled by $(h^2, (m_q^2+m_{\widetilde{q}}^2)/2)$, when $\rho=1$,
the theory has only a chiral symmetry breaking phase 
and an unbroken phase if all $m_q^2$, $m_{\widetilde{q}}^2$ 
and $m_M^2$ are positive, whereas when $\rho=20$, e.g.,
the theory presents two unbroken phases and two kinds of 
broken chiral symmetry 
phases. In addition, in this direction, if all $m_q^2$, 
$m_{\widetilde{q}}^2$ and $m_M^2$ are negative, the scalar 
potential is still unbounded from below and hence the theory
becomes unphysical. 
\end{itemize}

 We still use the 't Hooft anomaly matching to check the survival
 of duality in the presence of the trilinear term. 
Since the trilinear term violates the $R$-symmetry 
explicitly, it can be checked that in the unbroken phase 
the 't Hooft anomalies $SU_{L(R)}(N_f)^3$ 
and $SU_{L(R)}(N_f)^2U_B(1)$ still match as in 
the supersymmetric limit. This seems to imply  
the existence of Seiberg's duality in this phase 
after soft supersymmetry breaking with the inclusion 
of the trilinear term. In the broken phase, the situation 
will become complicated since a large symmetry
breaking takes place. However, there is a simple case where conclusions
can be drawn. It should be first emphasized that
when adding the soft supersymmetry breaking terms we break
 the global flavour
symmetry $SU_L(N_f-N_c)\times SU_R(N_f-N_c)$ into
$SU_L(N_f-N_c-1)\times U_L(1)\times SU_R(N_f-N_c)\times U_R(1)$
breaking terms \cite{ref832}. Assume that
in the dual magnetic theory the first flavour 
has soft scalar masses
$m_{q1}$ and $m_{\widetilde{q}1}$, different 
from the other $N_f-1$ flavours,
$m_q$ and $m_{\widetilde{q}}$ and that only 
$m_{q1}$ and $m_{\widetilde{q}1}$
satisfy the conditions (\ref{eq8p12}) and 
(\ref{eq8p13}) for the broken phase. In this case
only the vacuum expectation values 
$\langle \phi_{q(1)}\rangle=\langle \phi_{\widetilde{q}(1)}\rangle= X_1$ and 
$\langle \phi_{M(1)}\rangle$ do not vanish. Consequently, the following
gauge symmetry breaking occurs,
\begin{eqnarray}
SU(N_f-N_c){\longrightarrow}SU(N_f-N_c-1).
\end{eqnarray}
Furthermore, the other $N_f-1$ flavours of the dual magnetic quarks and 
$(N_f-1)^2$ singlet fermion components $\psi_M$ remain 
massless. Thus the resulting theory has the global symmetry
$SU_L(N_f-1){\times}SU_R(N_f-1){\times}U_{B'}(1)$, under which massless
dual quarks, $\psi_q$, $\psi_{\widetilde{q}}$ and the singlet
fermion $\psi_{M}$ transform as $(\overline{N}_f,0, N_c/(N_f-N_c-1))$,
 $(0,N_f,-N_c/(N_f-N_c-1))$ and $(N_f,\overline{N}_f,0)$, respectively.

Now let us consider the corresponding electric theory. The soft breaking 
Lagrangian consisting only of the mass terms of squarks and gaugino 
is (\ref{eq8p1}). If at the supersymmetry breaking scale, 
the flavour symmetry is broken in the same 
way as in the magnetic theory, i.e. the remaining  global flavour symmetry is
$SU_L(N_f-1){\times}SU_R(N_f-1)$, then nothing can prevent the appearance 
of the superpotential
$W_1=M_1Q^1{\cdot}\widetilde{Q}_1$. 
Consequently, a soft breaking term corresponding to this superpotential
can be added to (\ref{eq8p1}). Especially, 
a $B$-term, $-M_B^2Q^1{\cdot}\widetilde{Q}_1$ 
can also arise as a soft breaking term. Thus the (mass$)^2$
 matrix of the first flavour of squarks, $\phi_{Q^1}$ 
and $\phi_{\widetilde{Q}_1}$ can be written 
as follows,
\begin{eqnarray}
M_{11}^2=\left(\begin{array}{cc}m_{Q^1}^2+M_1^2 & -M_B^2 \\
                                  -M_B^2 & m^2_{\widetilde{Q}_1}+M_1^2
\end{array}\right).
\end{eqnarray}
If $\det (M_{11}^2)>0$, the potential minimum corresponds to
$\langle \phi_{Q^1}\rangle= \langle\phi_{\widetilde{Q}_1}\rangle=0$ 
and the gauge symmetry $SU(N_c)$ remains unbroken. 
In this case it can be checked that the 't Hooft anomalies $SU_{L(R)}(N_f-1)^3$ 
and $SU_{L(R)}(N_f-1)^2U_B(1)$ match with those 
in the magnetic theory, $SU_{L(R)}(N_f-1)^3$ and
$SU_{L(R)}(N_f-1)^2U_{B'}(1)$. This seems to suggest that the
$N=1$ duality remains after supersymmetry breaking 
with a trilinear term even in the broken phase.

In the case that $\det (M_{11}^2)<0$ and 
$m^2_{Q^1}+m^2_{\widetilde{Q}_1}+2M_1^2>2|M_B^2|$, the squarks $\phi_{Q^1}$ and
$\phi_{\widetilde{Q}_1}$ will acquire the vacuum expectation values and the 
gauge symmetry is broken to $SU(N_c-1)$. This case seems to correspond 
to the dual magnetic theory with the scalar potential unbounded from below
and hence the duality disappears \cite{ref833}.  

It was further shown that the trilinear soft breaking term plays an 
important role in determining the vacuum structure in the 
cases $N_f{\leq}N_c+1$ \cite{ref834}. In particular, for the
range $N_f=N_c+1$, the trilinear term is just the flavour
interaction among the scalar components of the baryon and
mesons,
\begin{eqnarray}
\widetilde{\cal L}_{\rm SB}=h^{\prime}\phi_{Bi}\phi_{Mj}^i\phi_B^j.
\end{eqnarray}
The chiral symmetry can be broken if $|h^{\prime}|$ is sufficiently large.
This is completely different from the case that the soft 
breaking Lagrangian is only composed of mass terms of 
superpartners\cite{ahso,ref832}, where the chiral 
symmetry is always preserved for $N_f=N_c+1$. 
 
Overall, the duality should have physical applications 
in exploring non-perturbative dynamics,  but ahead of us 
there is still a long way to go. 

\vspace{3ex}

\centerline{\large \bf Acknowledgments:} 

\vspace{5mm}

The financial support of the Academy of Finland under the Project No. 37599 
and 163394 is greatly acknowledged. W.F.C is partially supported by
the Natural Sciences and Engineering  Council of Canada.

\end{document}